\documentclass[12pt,paper=a4,twoside=semi]{scrartcl}
\usepackage{mathtools}
\usepackage{multirow}
\usepackage[hidelinks]{hyperref} 
\usepackage[T1]{fontenc}        
\usepackage[utf8]{inputenc}	
\usepackage{ae,aecompl} 
\usepackage{amssymb}
\usepackage{mathrsfs}
\usepackage{amsmath}
\usepackage{amsthm}
\usepackage[english]{babel}   	
\usepackage{graphicx}	
\usepackage{color}	
\usepackage{epstopdf}   
\usepackage{url}	
\usepackage{listings}	
\usepackage[font=small,labelfont=bf]{caption}                            
\usepackage{booktabs}	
\usepackage{inputenc}
\usepackage{color,colortbl,xcolor}
\usepackage{physics}
\usepackage{dsfont}
\numberwithin{equation}{section}
\usepackage{bibgerm}	
\usepackage{subcaption}
\usepackage{enumitem}
\usepackage{diagbox} 
\usepackage{rotating}
\counterwithin{figure}{section}
\counterwithin{table}{section}
\newtheorem{theorem}{Theorem}[section]
\newtheorem{Remark}[theorem]{Remark} 
\newtheorem{lemma}[theorem]{Lemma}
\newtheorem{prop}[theorem]{Proposition}

\theoremstyle{definition}

\newtheorem{definition}[theorem]{Definition}

\newcommand{\re}{\operatorname{Re}}
\newcommand{\im}{\operatorname{Im}}
\newcommand{\Var}{\operatorname{Var}}
\newcommand{\Cov}{\operatorname{Cov}}
\newcommand{\Hp}{\mathcal{H}}
\newcommand{\Lw}{\mathscr{L}}

\newcommand{\rv}{\mathbf{r}}
\newcommand{\lv}{\mathbf{l}}
\newcommand{\Rv}{\mathbf{R}}

\newcommand{\GL}{\mathrm{GL}}
\DeclareMathOperator{\diag}{diag}
\DeclareMathOperator{\Pf}{Pf} 
\DeclareMathOperator{\erfc}{erfc} 
\DeclareMathOperator{\erfi}{erfi} 
\newenvironment{sproof}{\proof}{\endproof}

\def\boxit#1{%
	\smash{\color{gray}\fboxrule=1pt\relax\fboxsep=2pt\relax%
		\llap{\rlap{\fbox{\vphantom{0}\makebox[#1]{}}}~}}\ignorespaces
}
\begin{document}
\thispagestyle{empty}
\begin{center}

\centering

\vspace{5cm}

{\LARGE \textbf{Non-Hermitian Random Matrices:}
    
  }

\quad 
  
{\LARGE \textbf{Study of Beta-Ensembles and Symmetry Classes}
 
} 

\vspace{3cm}

\textbf{Doctoral Thesis}\\
submitted by\\
\textbf{Patricia Päßler}\\

\vspace{4cm}

Faculty of Physics\\

Bielefeld University\\

November 19, 2026

 \vspace{4cm}

Supervisor and 1st Examiner: Prof. Dr. Gernot Akemann

\quad 

2nd Examiner: Jun.-Prof. Dr. Jonas Jalowy

\vspace{1cm}

\end{center}

\cleardoublepage
\pagenumbering{roman}
\setcounter{page}{1} 
\section*{Abstract}
In this work, we study non-Hermitian random matrices with complex eigenvalues in a plenty of situations. We start with a review on Hermitian matrices, which have real eigenvalues, and discuss relations to the Dyson gas with inverse temperature $\beta>0$ as well as techniques used in the literature. We use this insight to study a tridiagonal matrix-valued process with stochastic resetting for arbitrary Dyson index $\beta$. Afterwards, we turn to non-Hermitian matrices, with complex eigenvalues, and search for possible matrix ensembles for a two-dimensional Coulomb gas ensemble at positive inverse temperature $\beta$. We discuss the difficulties of the construction of such a model in two dimensions compared to the situation for Hermitian matrices. Our first step is to revisit a simple ensemble of $2\times 2$ non-Hermitian matrices, which describes the local statistics of large matrices for small $\beta\in\left[0,0.5\right]$ well. We strengthen the importance of the random matrix model in demand by presenting a novel application onto the study of the distribution of nests of birds of prey in the Teutoburger Wald. It follows the introduction of the key random matrix ensembles in this work: We define and analyse three non-Hermitian tridiagonal matrices such that we find a Coulomb-type interaction of the eigenvalues, plus additional terms. One key term in the study of random matrices is the so-called universality of certain (statistical) properties. Among all 38 non-Hermitian symmetry classes it is conjectured that only three kinds of universal bulk statistics exist, which can be simplest represented by the classes A (complex Ginibre), AI$^\dagger$ (complex symmetric) and AII$^\dagger$ (complex self-dual). We find numerically an effective description of classes AI$^\dagger$ as a two-dimensional Coulomb gas at $\beta= 1.4$ and AII$^\dagger$ at $\beta=2.6$ due to comparison of (next-to-)nearest-neighbour spacing, while the correspondence of class A to a Coulomb gas at $\beta=2$ is well-established. As an alternative (numerical) tool for the study of the universal statistics, we discuss the complex spacing ratio and investigate the three classes in the bulk and at the edge of the spectrum. Additionally, we study the characteristic polynomials of those classes analytically and find distinct behaviour. We present the model of a quantum spin chain with dissipative disorder as one application of the non-Hermitian random matrix statistics.\noindent
\subsection*{Statement of Authorship}
I am familiar with the doctoral regulations of the Department of Physics at Bielefeld University from 10 January 2012 and 14 April 2022. I certify that this doctoral thesis has been composed by myself and describes my own work, unless otherwise acknowledged in the text. All references and verbatim extracts have been quoted and all sources of information have been specially acknowledged. It has not been accepted in any previous application for a degree. No AI has been utilized in the creation of this thesis.
\vspace{10mm}

\noindent
Bielefeld, November 19, 2026 \\
\hspace*{\fill}
\begin{tabular}{@{}l@{}}\hline
	\makebox[4cm]{Patricia Päßler}
\end{tabular}\clearpage
\section*{Publications}
The following publications should be considered as contribution of my dissertation.\vspace{0.5cm}\\
\begin{minipage}{0.2\textwidth}\vspace{-2.0cm}
	$\left[\text{AMP22}\right]$
\end{minipage}\begin{minipage}{0.8\textwidth}
G.~Akemann, A.~Mielke, and P.~Päßler, \textit{Spacing distribution in the 2D Coulomb gas: Surmise and symmetry classes of non-Hermitian random matrices at non-integer $\beta$}, Phys. Rev. E \textbf{106}: 014146:1-11 (2022), \href{https://arxiv.org/abs/2112.12624}{[arXiv:2112.12624]}.\vspace{0.5cm}
\end{minipage}\\
\begin{minipage}{0.2\textwidth}\vspace{-2.0cm}
	$\left[\text{ACKMOP}\right]$
\end{minipage}\begin{minipage}{0.8\textwidth}
	G.~Akemann, N.~Chakarov, O.~Krüger, A.~Mielke, M.~Ottensmann, and P.~Pässler, \textit{Interactions between different birds of prey as a random point process}, J. Stat. Mech.: Theor. Exp. \textbf{2024}: 053501:1-35 (2024), \href{https://arxiv.org/abs/2310.20670}{[arXiv:2310.20670]}.\vspace{0.5cm}
\end{minipage}\\
\begin{minipage}{0.2\textwidth}\vspace{-2cm}
	$\left[\text{ABCPRS}\right]$
\end{minipage}\begin{minipage}{0.8\textwidth}
 G.~Akemann, F.~Balducci, A.~Chenu, P.~Päßler, F.~Roccati, and R.~Shir, \textit{Two transitions in complex eigenvalue statistics: Hermiticity and integrability breaking}, Phys. Rev. Research \textbf{7}: 013098:1-14 (2025), \href{https://arxiv.org/abs/2409.10625}{[arXiv:2409.10625]}.\vspace{0.5cm}
\end{minipage}\\
\begin{minipage}{0.2\textwidth}\vspace{-2cm}
	$\left[\text{AAKP}\right]$
\end{minipage}\begin{minipage}{0.8\textwidth}
	G.~Akemann, N.~Aygün, M.~Kieburg, and P.~Päßler, \textit{Complex symmetric, self-dual, and Ginibre random matrices: Analytical results for three classes of bulk and edge statistics}, J. Phys. A: Math. Theor. \textbf{58}: 125204:1-50 (2025), \href{https://arxiv.org/abs/2410.21032}{[arXiv:2410.21032]}.\vspace{0.5cm}
\end{minipage}\\
\begin{minipage}{0.2\textwidth}\vspace{-1.0cm}
	$\left[\text{AMPT}\right]$
\end{minipage}\begin{minipage}{0.8\textwidth}
	G.~Akemann, F.~Mezzadri, P.~Päßler, and H.~Taylor, \textit{Logarithmic Spectral Distribution of a non-Hermitian  $\beta$-Ensemble}, SIGMA \textbf{22} (2026), 058, 50 pages, \href{https://arxiv.org/abs/2504.12120}{[arXiv:2504.12120]}.\vspace{0.5cm}
\end{minipage}\\
\begin{minipage}{0.2\textwidth}\vspace{-2.0cm}
	$\left[\text{AAAMPRW}\right]$
\end{minipage}\begin{minipage}{0.8\textwidth}
	G.~Akemann, G.~Angermann, N.~Aygün, A.~Mielke, P.~Päßler, C.~Raitzig, and T.~Winkler, \textit{Three non-Hermitian random matrix universality classes of complex edge statistics: Spacing ratios and distributions}, J. Phys. A: Math. Theor. \textbf{59}: 255201:1-32 (2026), \href{https://arxiv.org/abs/2603.28457}{[arXiv:2603.28457]}.\vspace{0.5cm}
\end{minipage}\\
\begin{minipage}{0.2\textwidth}\vspace{-2.0cm}
$\left[\text{AMP26}\right]$
\end{minipage}\begin{minipage}{0.8\textwidth}
G.~Akemann, S.~N.~Majumdar, and P.~Päßler, \textit{A tridiagonal matrix-valued process with stochastic resetting for arbitrary Dyson index $\beta>0$}, Submitted to J. Phys. A (2026), \href{https://arxiv.org/abs/2605.26945}{[arXiv:2605.26945]}.\vspace{0.5cm}
\end{minipage}\clearpage

\section*{Acknowledgements}
First, I would like to thank my supervisor Gernot Akemann for his constant support throughout my PhD. He introduced me to the subject of random matrices and pushed my work forward by many important remarks and questions. Moreover, I thank my co-authors for joined work and letting me learn from their expertise. Especially, I thank Adam Mielke, Francesco Mezzadri, Henry Taylor, Ruth Shir and Satya N. Majumdar for many insightful discussions.\\
\\
Furthermore, I would like to express my thanks to the organisers of the Bielefeld-Melbourne-Seoul random matrix theory seminar: Gernot Akemann, Sung-Soo Buyn, Peter J. Forrester, Lucas Hackl and Mario Kieburg. This seminar series broadened my perspective on the studies of random matrices and was a constant source of new input for me.\\
\\
The work in this thesis was supported by the 
Deutsche Forschungsgemeinschaft (DFG) grant SFB 1283/2 2021 – 317210226 "Taming uncertainty and profiting from randomness and low regularity in analysis, stochastics and their applications".\\
\\
Most important, I thank my family for their constant support, motivation and standing always behind me. Ein besonderer Dank geht an meine Eltern.
\cleardoublepage
\pagenumbering{arabic}
\setcounter{page}{1} 
\tableofcontents 
\clearpage 
\pagestyle{plain}
\section{Introduction}\label{sec-intro}
Let us start with a simple definition: A random matrix contains elements which follow a certain probability distribution. So far so good, but this concept opens an extremely wide range for studies both from a mathematical point of view as well as for applications in physics, biology, economy, etc.\\
In the study of the statistical distribution of level spacings of nuclei Wigner \cite{Wigner51} was the first to point out the relation to the spacing distribution of the real eigenvalues of Hermitian\footnote{A matrix $A$ is Hermitian, if $A^\dagger=A$, where $A^\dagger=\bar{A}^t$ denotes the complex conjugate transpose.} random matrices in 1957 \cite{Wigner57}. In the 1950s many data were available on the statistical properties of heavy nuclei and therefore average quantities as level-densities and spacing distributions came into the centre of interest \cite{Mehta60}, and thus attracted the description in a probabilistic sense. The work by Wigner \cite{Wigner57} was the spark to bring the study on random matrices and its applications to shine. It yields a wealth of further research areas, physical models and not least many researchers which contributed to the development of the subject which is now known as Random Matrix Theory (RMT). For a more detailed (historical) overview on the results and applications we refer the reader to the following books \cite{oxford handbook,mehta,haake}.\\
Now, we turn to the study of non-Hermitian matrices $A$, i.e. $A^\dagger\neq A$. While we could point on the physical motivation for the study of Hermitian random matrices above, the history is different here. In a highly influential paper in 1965 Ginibre \cite{ginibre} introduced and studied three ensembles of non-Hermitian matrices: complex, real and self-dual matrices. The simple the last sentence is, so big is the impact to the research area of random matrices. We would like to cite a comment made by Ginibre in his paper \cite{ginibre}: \textit{"Apart from the intrinsic interest of the problem, one may hope that the methods and results will provide further insight in the cases of physical interest or suggest as yet lacking applications."} This illustrates the ingenious mind of Ginibre starting to reach into a new (physically at that time not yet motivated) direction suspecting a bunch of new methods to arise. The next years will show that he was right.\\
Before we take a closer look into the studies by Ginibre, we would like to state a few applications to motivate the reader that it is a good idea to dive deeper in the non-Hermitian RMT. While a physicist might be sceptic about non-Hermitian Hamiltonians at first glance, it turns out, that there are a bunch of situations, where those studies are well-motived. The flux lines of superconductors can be described via the subject of localization of a random non-Hermitian Hamiltonian \cite{Hatano Nelson,abrikosov,serfaty supercond}. Dissipative quantum systems, i.e. quantum systems with damping, became of great interest and follow non-Hermitian statistics \cite{GHS}. Further applications are the theory of random neural networks, i.e. elements with random asymmetric coupling in a network \cite{SCS 88}, quantum field theory with non-zero quark chemical potential (such that the Dirac operator becomes non-Hermitian) \cite{QCD application} and the fractional quantum Hall effect in the study of electron gases \cite{FQHE}.\\
More applications can be found in biology, such as the spatial distribution of plants (investigating the "randomness") \cite{spatial dist plants} or the territorial behaviour of birds of prey, where several environmental influences contribute \cite{buzzards 1,ACKMOP}. The latter, will be discussed in greater detail in Section \ref{ssec-birds of prey} and was investigated by the author in \cite{ACKMOP}.\\
Furthermore, we present the study of a spin chain with imaginary disorder in Section \ref{ssec-application spin chain}, finding transitions in Hermiticity and integrability breaking. The latter is contained in the publication \cite{ABCPRS} of the author.\\
We turn back to the study of Ginibre. The following two ingredients are needed:
\begin{enumerate}
	\item A space (algebraic set) of matrices with complex, quaternion or real entries.
	\item A suitable probability measure on that space.
\end{enumerate}
\begin{Remark}\label{remark-Ginibre}
	Here is a good point to pause for a few remarks. First, Ginibre \cite{ginibre} points out that the choice of complex, quaternion or real matrix entries agree with the so called $Z$'s in Dyson's threefold way described in \cite{dyson62}.\footnote{We will give some more information on this in Section \ref{sec-universality symmetry classes}, when dealing with the non-Hermitian symmetry classification.} This touches an important issue, namely the study of symmetry classes in the subject of Hermitian and non-Hermitian random matrices. The latter will be discussed in detail in Section \ref{sec-universality symmetry classes} but it is important to note that this topic is the backbone of the Ginibre ensembles from the very beginning. Moreover, it is crucial for the understanding of what we call a \textit{suitable} probability measure. As discussed by Dyson \cite{dyson62} one starts from a set of matrices (possibly with symmetry properties) and defines a probability measure on them to find what we call a \textit{random matrix ensemble}. The starting point for the probability measure is the invariance under a group action.\footnote{One should be aware that the term invariance can be defined in different manners here. Especially, for non-Hermitian ensembles it is important to distinguish between "formal" and "physical" invariance, where the first is phrased in terms of the matrix itself and the second deals with scalar products \cite{dyson62}.} An excellent reference on which also Dyson based his work on is the book by Weyl \cite{book weyl}. 
\end{Remark}
\subsection{Introduction of the Ensembles}\label{ssec-intro models}
We start with a complex $N\times N$ matrix $J\neq J^\dagger$ and define a probability measure on its complex matrix elements $J_{i,j}=(J)_{i,j}$ as
\begin{align}
	\mu_L(J)=\prod_{i,j=1}^N dJ_{i,j}dJ^*_{i,j},
\end{align}
where $\mu_L(J)$ is the linear measure defined by the volume of the matrix elements. This probability measure turns out to be not suitable for a probabilistic interpretation as the measure of the whole space diverges. Recalling from Remark \ref{remark-Ginibre} that we are interested into invariance under a group action, we define
\begin{align}\label{eq-prob measure}
	d\mu(J)=\exp\left[-\Tr(JJ^\dagger)\right]d\mu_L(J).
\end{align}
This measure is invariant under unitary transformations $J\rightarrow UJU^\dagger$, where $U$ is unitary, i.e. $UU^\dagger=U^\dagger U=\mathds{1}_N$.\footnote{This is the action of the adjoint representation of $Z_U$ \cite{ginibre}.} Additionally, we observe that the distributions of the matrix elements of $J$ are independent of each other. Porter and Rosenzweig showed in \cite{Porter Rosenzweig} that the requirements of rotational invariance and independence imply that the matrix elements are normally distributed (in the case of Hermitian matrices but the arguments will carry through). We denote the normal distribution of mean $\mu\in \mathbb{R}$ and variance $\sigma^2\in \mathbb{R}_+$ by $\mathcal{N}(\mu,\sigma^2)$. The probability density of a variable $x\sim \mathcal{N}(\mu,\sigma^2)$ reads as \cite[Section 1.3.6.6.1.]{NIST statistics}
\begin{align}\label{eq-def normal dist}
	f(x)=\frac{1}{\sqrt{2\pi \sigma^2}}e^{-\frac{(x-\mu)^2}{2\sigma^2}}, \text{ for } x\in \mathbb{R}.
\end{align}
Complex matrices $J$, distributed with respect to \eqref{eq-prob measure} form the complex Ginibre ensemble, which is also denoted by GinUE due to the described unitary invariance. The next step is to determine the eigenvalue probability distribution. We consider the Schur decomposition of $J$,
\begin{align} \label{eq-schur decomp}
	J=U(Z+ T)U^{-1},
\end{align}
with $U$ unitary, $Z=\text{diag}(z_1,...,z_N)$ consisting of the $N$ distinct\footnote{The set of matrices with degenerate eigenvalues is of measure zero or can be mapped to a lower dimensional case \cite[Chapter 3]{mehta}.} complex eigenvalues and the complex strictly upper triangular matrix $T$. We can fix additional $N$ real degrees of freedom as this similarity transformation is invariant under $U\rightarrow U\text{diag}(e^{i\phi_1},...,e^{i\phi_N})
$ for $\phi_j\in[0,2\pi)$ $\forall j=1,\ldots, N$. Hence, we choose $U\in U(N)/U(1)^N$. Note that we will use this invariance in Section \ref{sssec-def tridiag jpdf} again to fix parameters in the eigenvectors. In order to find the joint probability density function (JPDF) in terms of the eigenvalues, the next steps are to calculate the Jacobian for this change of variables (from matrix entries of $J$ to the spectral coordinates encoded in $U,T,Z$) and integrate over $U$ and $T$. We refer the reader for the explicit calculation to \cite{ginibre} or \cite[Chapter 15]{mehta}, \cite[Chapter 15]{book forrester}. Nevertheless, we would like to draw the attention to some "\textit{miracles}" which occur here.\footnote{As a physicist or mathematician one should be careful to name something a miracle but we choose this phrase here to draw the reader's attention to some very important issues of the complex Ginibre ensemble.} First, it turns out that there is no explicit $U$-dependence in the Jacobian and the task reduces to the integration of the right coset space $U(N)/U(1)^N$. Second, using a clever reordering of the parameters, the Jacobian simplifies significantly and the integral over the $T$-matrix decouples into $N(N+1)/2$ Gauss-integrals. We refer the reader to \cite[A.33]{mehta} for an elegant proof.\\
The symmetrized JPDF $P(Z)$ of the eigenvalues of the complex matrix $J$ results to \cite[Eq. (1.35)]{ginibre}
\begin{align} \label{eq-jpdf complex ginibre}
	P(z_1,\ldots,z_N)=\frac{1}{\pi^N\prod_{j=0}^Nj!}e^{-\sum_{j=1}^N|z_j|^2}\underbrace{\prod_{1\leq i < j\leq N}|z_i-z_j|^2}_{=|\Delta_N(z_1,\ldots,z_N)|^2}.
\end{align}
Here, $\Delta_N(z_1,\ldots,z_N)$ denotes the Vandermonde determinant. We observe that $P(Z)$ consists of two contributions: a Gaussian confining potential preventing the eigenvalues (or particles) to escape to infinity and a pairwise repulsion between the eigenvalues. We can view this probability density in another way as:
\begin{align}\label{eq-jpdf complex ginibre CG}
	P(z_1,\ldots,z_N)=\frac{1}{\pi^N\prod_{j=0}^Nj!}\exp\left[-\sum_{j=1}^N|z_j|^2+2\sum_{1\leq i<j\leq N}\log|z_i-z_j|\right].
\end{align} 
This expression should now ring a bell in the head of a physicist: We recover the Hamiltonian of a Dyson gas with pairwise logarithmic interaction at inverse temperature $\beta=2$ for particles in two dimensions (2d). Some nice introductory material into this subject can be found for instance in \cite[Chapter 1]{book forrester}, \cite{serfaty}. The study of the two-dimensional Coulomb gas (2dCG) attracted a lot of researchers in the last couple of years. We are not in the position to state a complete list of contributions here, but we would like to highlight one conjectured feature, which makes contact to one application previously mentioned. Numerical studies show a transition of the density from constant to a triangular Abrikosov lattice in the large $N$-limit, around $\beta\approx 140$ (i.e. low temperatures) \cite{AB81,CC83}, see \cite{Abanov,Abanov 2} for recent works. This relates to the flux lines of superconductors, mentioned above.\\
Let us take a look (historically) back to the Hermitian random matrices with $J=J^\dagger$. The three classical ensembles are those of real, complex or quaternion matrices (all of them Hermitian) with probability densities invariant under orthogonal, unitary or symplectic conjugation respectively. This can be realised via the same probability measure as in Eq. \eqref{eq-prob measure} (and appropriate rescaling). The JPDF in terms of the $N$ distinct real eigenvalues $\lambda_1,\ldots,\lambda_N$ reads \cite{book forrester}
\begin{align}\label{eq-jpdf 1d Dyson}
	\begin{split}
	P(\lambda_1,\ldots,\lambda_N)&=\frac{1}{C_N(\beta)}e^{-\frac{1}{2}\sum_{j=1}^N\lambda_j^2}\prod_{1\leq i < j\leq N}|\lambda_i-\lambda_j|^\beta, \text{ with } \begin{cases}
	\beta=1 & \text{for GOE}\\
	\beta=2 & \text{for GUE}\\
	\beta=4 & \text{for GSE},
	\end{cases}\\
C_N(\beta)&=(2\pi)^{N/2}\frac{\prod_{j=1}^N\Gamma(1+\beta j/2)}{\left(\Gamma(1+N\beta/2)\right)^N}.
\end{split}
\end{align}
The value of $\beta$ counts the independent real variables per matrix entry, see also \cite[Chapter 3]{mehta}.
The Gaussian orthogonal/ unitary / symplectic ensemble (G O/U/S E) shares the probability density of a one-dimensional (1d) Dyson gas with Gaussian confining potential at the inverse temperature $\beta=1,2,4$, respectively. Having this correspondence at hand, it seems natural to ask for a random matrix ensemble for general real positive $\beta>0$. We come back to this question in Section \ref{sec-Hermitian beta ensembles}.
\begin{Remark}
	We would like to point out that this correspondence of GO/U/SE to 1d Dyson gases at $\beta=1,2,4$ respectively, does not carry through to the non-Hermitian ensembles of real, complex and symplectic Ginibre matrices. For the symplectic Ginibre ensemble (GinSE), with quaternion matrix elements, one can define a $2N\times 2N$ complex representation, as
	\begin{align}
		J=\begin{pmatrix}
			J_1 & J_2\\
			-\bar{J_2} & \bar{J_1}
		\end{pmatrix},
	\end{align}
where $J_1,J_2\in$GinUE($N$). The derivation of the JPDF works now similar to the one of the GinUE, with complex conjugate pairs of eigenvalues and is given in Ginibre's initial work \cite{ginibre}. With the replacement $N\rightarrow2N$ and $z_{j+N}\rightarrow\bar{z_{j}}$ for $j=1,\ldots,N$, the JPDF of the GinSE agrees with the one of the GinUE in Eq. \eqref{eq-jpdf complex ginibre} up to missing terms involving only differences in the $\left\{\bar{z_{j}}\right\}$.\footnote{It is interesting to note that the GinSE looks in that sense very similar to the GinUE, i.e. a 2dCG at $\beta=2$. We should keep this in mind and come back to this in Section \ref{sec-universality symmetry classes}, when discussing universality.} This is due to the counting of independent differentials in the quaternion elements, cf. \cite{book byun forrester,book forrester}.\\
The derivation of the JPDF of the orthogonal Ginibre ensemble (GinOE), consisting of independent real entries, took some more 25 years. Ginibre manifested in this work \cite{ginibre}, that he could not calculate the full JPDF with the same methods as for the GinUE and GinSE. A real Ginibre matrix has real eigenvalues as well as complex conjugate pairs of eigenvalues. Conditioning on the number of real eigenvalues, Lehmann and Sommers derived the JPDF in 1991 \cite{Lehmann Sommers 91}. An independent second proof was presented by Edelman in 1997 \cite{Edelman 97}. The density invokes again an amplitude of the differences between the eigenvalues but also a coupling of the complex eigenvalues to some charge on the real axis, cf. \cite[Chapter 15]{book forrester}.\\
The key message is here: While we suppress the explicit form of the JPDF's of the GinOE and GinSE, we make the strong point that correspondence of the value of $\beta=1,2,4$ to the number of independent matrix elements, does not result in a Coulomb repulsion at these inverse temperatures, as in the Hermitian ensembles. Moreover, we would like to stress that beyond the integrable point $\beta=2$, only few rigorous mathematical results exist. We comment on some of the difficulties along the lines.
\end{Remark}
\subsection{Structure of the Thesis}\label{ssec-intro overview}
In Section \ref{sec-Hermitian beta ensembles} we give an introduction into the subject of Hermitian $\beta$-ensembles. In a first Section \ref{ssec-Householder}, we discuss Householder reflections, which gave rise to the "first" tridiagonal $\beta$-ensemble in 1d, introduced by Dumitriu and Edelman in 2002 \cite{Dumitriu Edelman}. In the second Section \ref{ssec-Anderson Dixon}, we revisit an alternative proof strategy via the Anderson-Dixon integral invented by Forrester and Rains in 2005 \cite{Forrester Rains 05}. In Section \ref{ssec-tridiag with resetting} we use the insights from the previous sections to study a tridiagonal matrix-valued process with resetting. This section is based on the work \cite{AMP26} by the author. We give an overview over various existing Hermitian $\beta$-ensembles in Appendix \ref{app-overview Hermitian beta ensembles} for completeness.\\
We turn to non-Hermitian $\beta$-ensembles in Section \ref{sec-non Herm beta}, starting with the review of the Poisson point process in $D$ dimensions and important results on the complex Ginibre ensemble in Section \ref{ssec-point processes}. We introduce techniques for the analysis of numerics and comparison with analytics in Section \ref{ssec-numerics methods}. Next, we recall in Section \ref{ssec-surmise} a normal matrix ensemble of $2\times 2$ matrices with pairwise logarithmic interaction at the inverse temperature $\beta>0$ introduced by the author in \cite{masterthesis, AMP}. The ensemble yields a surmise for the local nearest-neighbour (NN) statistics of a 2dCG for small non-integer $\beta >0$ after appropriate adjustment. We present the study of the distribution of nest of birds of prey in the Teutoburger Wald as an application for the 2dCG and especially the surmise in the subsequent Section \ref{ssec-birds of prey}. The analysis of the data in \cite{ACKMOP} was contributed by the author. Next, we introduce three non-Hermitian tridiagonal matrix ensembles, which include a Coulomb gas type interaction at inverse temperature $\beta \in \mathbb{R}_+$ for general matrix dimension $N$ in Section \ref{ssec-tridiag beta}. We derive the JPDF and the global density via the study of the characteristic polynomial and discuss their local statistics numerically. The extensive calculations both analytical and numerical serve as a main contribution to this thesis and are published in \cite{AMPT}. \\
Section \ref{sec-universality symmetry classes} deals with non-Hermitian symmetry classes. We revisit the symmetry classification derived in the literature \cite{Bernard LeClair,non-h classes} and discuss the suggested universality in level statistics in Section \ref{ssec-intro symmetry classes}. The classes AI$^\dagger$ of complex symmetric, A of complex Ginibre, and AII$^\dagger$ of complex self-dual matrices serve as the three simplest representatives amongst the 38 non-Hermitian symmetry classes \cite{hamazaki2019non}. For two (classes AI$^\dagger$ and AII$^\dagger$) of the three classes, we lack an explicit form of the JPDF until today.\footnote{Recently, progress on the spectral density was made for complex symmetric matrices in \cite{complex sym Gernot}.} Therefore, we turn to a numerical study of the statistics for these two classes. We find an effective description as 2dCG with non-integer $\beta$ for the NN and next-to-nearest (NNN) distributions numerically. While in a first work \cite{AMP} the NN was studied, the effective description was extended to the NNN statistics in \cite{ABCPRS} by the author. We investigate the complex spacing ratio as another measure of the statistics both in the bulk and at the edge of the spectrum in Section \ref{sssec-complex spacing ratio} and find distinct behaviour between the three studied symmetry classes AI$^\dagger$, A, and AII$^\dagger$ following \cite{AAAMPRW}, which was contributed by the author. In Section \ref{ssec-expectation of characteristic polynomials}, we investigate the distinction of the three representatives in the study of expectation values of characteristic polynomials further. We present results on the product of one pair of characteristic polynomials in detail, which were contributed by the author in \cite{AAKP}. After collecting these results on the numerical and analytical distinction of the three conjectured universality classes of non-Hermitian matrices, we turn again to an application. We present the study of the non-Hermitian statistics of a spin chain with imaginary disorder, investigating comparison with various (non-)Hermitian random matrix statistics, such as the surmise introduced in Section \ref{ssec-surmise}, the symmetry class AI$^\dagger$ of the spin chain Hamiltonian, and the $D$-dimensional Poisson process, based on \cite{ABCPRS}. The analysis was mainly contributed by the author. \\
In the last Section \ref{sec-summary}, we summarize the main results of this thesis and discuss possible topics for further research. \clearpage
\section{Hermitian $\boldsymbol{\beta}$-Ensembles}\label{sec-Hermitian beta ensembles}
One might wonder why, we start with a section on Hermitian $\beta$-ensembles, while the title of this thesis deals with non-Hermitian matrices. There are several reasons: First, it is natural to follow the historical process on the study of matrix ensembles, which describe a certain (physical) model, just as we did in the introduction Section \ref{sec-intro}. Moreover, we wish to point out that we can learn a lot for the study of non-Hermitian matrices by revisiting the Hermitian results. We will present the ideas of the proofs for a Hermitian $\beta$-ensemble, consisting of tridiagonal matrices which share its eigenvalue distribution with a Dyson gas in 1d. We follow in Section \ref{ssec-Householder} the strategy of Dumitriu and Edelman in \cite{Dumitriu Edelman}, using Householder reflections to reduce a general Hermitian matrix to a tridiagonal one. We observe that the Householder reflections can be defined very generally which makes them applicable also to non-Hermitian matrices. We will briefly revisit this in the next section on non-Hermitian $\beta$-ensembles.\\
We present a second view on the derivation of Hermitian $\beta$-ensembles in Section \ref{ssec-Anderson Dixon}. Using rank-one perturbations and making contact to the Anderson-Dixon integral Forrester and Rains \cite{Forrester Rains 05} gave an alternative way of understanding the rise of the tridiagonal matrix ensembles.\\
After explaining these two proof techniques on the matrix ensembles of a 1d Dyson gas at inverse temperature $\beta$, we present how the understanding of the proof techniques for the Hermitian $\beta$-ensembles enabled us to study the following Hermitian matrix ensemble: We define and analyse a time-dependent version of the Dumitriu-Edelman ensemble under stochastic resetting in Section \ref{ssec-tridiag with resetting}. This work is based on \cite{AMP26}, where the author mainly contributed the introduction of the matrix ensemble and the numerical analysis.\\
For completeness, we present an overview over all Hermitian $\beta$-ensembles, which were invented step by step after the opening of this field by Dumitriu and Edelman in Appendix \ref{app-overview Hermitian beta ensembles}.
\subsection{The Role of Householder Reflections}\label{ssec-Householder}
We recall from the introduction Section \ref{sec-intro} that the theory of Hermitian random matrices came into play in the 1950s probing statistics on level spacings of nuclei. At the same time the possibility to study possibly large matrices on computers came within reach with the invention of the programming language Fortran \cite[Chapter 2]{parlett}. Therefore, the research on computationally efficient processes was of main interest, cf. \cite{parlett,horn,Golub van Loan}. In 1958, Householder introduced the following unitary matrix operation, which can be defined as \cite{Householder}:
\begin{definition}\label{def-housholder}
	Let $\mathbf{w}=(w_1,\ldots,w_N)\in \mathbb{C}^N$ be a complex vector, with non-zero norm $\|\mathbf{w}\|_2:=\sqrt{\sum_{i=1}^{N}|w_i|^2}\neq 0$. The \textit{Householder reflector} with respect to the \textit{Householder vector} $\mathbf{w}$ is defined as
	\begin{align}\label{eq-def householder}
		U_{\mathbf{w}}:=\mathds{1}_N-2\frac{\mathbf{w}\mathbf{w}^\dagger}{\|\mathbf{w}\|_2^2}\, .
	\end{align} 
\end{definition}
\begin{lemma}\label{lem-Householder}
	The Householder reflector $U_{\mathbf{w}}$ in Eq. \eqref{eq-def householder}, with respect to $\mathbf{w}\in \mathbb{C}^N$, is \begin{itemize}
		\item Hermitian: $U_{\mathbf{w}}=U_{\mathbf{w}}^\dagger$ and
		\item unitary: $U_{\mathbf{w}}=U_{\mathbf{w}}^{-1}$.
	\end{itemize}
	Hence, $U_{\mathbf{w}}^2=\mathds{1}_N$ and $U_{\mathbf{w}}$ is a projector.
\end{lemma}
\begin{proof}
	The matrix $U_{\mathbf{w}}$ is Hermitian by definition, cf. Eq. \eqref{eq-def householder}. Also, the unitarity follows straight forward from its definition in \eqref{eq-def householder}.
\end{proof}
The introduction of this matrix operation was of great impact, and we would like to point out, why. Therefore, let us study the action of $U_{\mathbf{w}}$ on some complex vector $\mathbf{x}=(x_1,\ldots,x_N)\in \mathbb{C}^N$. Assume that we want to project this vector $\mathbf{x}$ onto the axis in $\mathbf{e}_1$-direction.\footnote{The set $\{\mathbf{e}_j\}_{j=1}^N$ denotes the canonic basis vectors with $(\mathbf{e}_j)_i=\delta_{i,j}$. Here, $\delta_{i,j}$ is the Kronecker delta, i.e. $\delta_{i,i}=1$ for all $i=1,\ldots,N$ and $\delta_{i,j}=0$ for $i\neq j$.} This can be achieved which the choice $\mathbf{w}:=\mathbf{x}+e^{i\Theta}\|\mathbf{x}\|_2\mathbf{e}_1$, with $x_1=e^{i\Theta}|x_1|$, cf. \cite{francesco}:
\begin{align}
	\begin{split}
	U_{\mathbf{w}}\mathbf{x}
	&=\mathbf{x}-2\frac{(\mathbf{x}+e^{i\Theta}\|\mathbf{x}\|_2\mathbf{e}_1)(\|\mathbf{x}\|_2^2+|x_1|\|\mathbf{x}\|_2)}{2(\|\mathbf{x}\|_2^2+|x_1|\|\mathbf{x}\|_2)}\\
	&=\|\mathbf{x}\|_2e^{i\Theta}\mathbf{e}_1
	\end{split}
\end{align}
Hence, we can project an arbitrary complex vector $\mathbf{x}\in\mathbb{C}^N$ onto the first unit vector $\mathbf{e}_1$ and the transformed vector is characterized by $\|\mathbf{x}\|_2$ and the angular part of the $x_1$-component. In the study of real vectors, the Hessenberg reflectors turn out to be symmetric ($U_{\mathbf{w}}=U_{\mathbf{w}}^t$) and orthogonal $U_{\mathbf{w}}^t=U_{\mathbf{w}}^{-1}$. The Householder transformations became of great impact as they present a computationally efficient way to reach a QR-decomposition.\footnote{The QR-decomposition of a matrix $A$ into the product of an orthogonal matrix $Q$ and a triangular matrix $R$, i.e. $(R)_{i,j}\neq$ for $i\leq j$, is the starting point of the QR-algorithm for computation of eigenvalues of the matrix $A$, cf. \cite{Golub van Loan, horn}.} The latter allows to calculate the eigenvalues of a possibly large matrix efficiently \cite{Golub van Loan}. The strategy is to reflect the first column of a given $N\times N$ matrix onto $\mathbf{e}_1\in \mathbb{C}^N$ in a first iteration step. Then, the lower $(N-1)$-dimensional part of the second column is mapped onto $\mathbf{e}_1\in \mathbb{C}^{N-1}$ and so on.\\
Besides this very important impact to numerical implementation of eigenvalue calculus, we want to examine the subject of Householder reflectors from the random matrix perspective. Shortly after the introduction in 1958, Householder and Bauer studied an application to real symmetric matrices in \cite{Householder Bauer,Bauer}. See also \cite[Chapter 1.9.1]{book forrester}.
\begin{prop}\label{prop-tridiagonalization Householder}
	Let $A=(A_{i,j})_{i,j=1,...,N}$ be a real symmetric matrix. Then there exist symmetric real orthogonal matrices $U^{(j)}$ such that
	\begin{align}\label{eq-def Householder tridiag}
	B^{(N-2)}:=	U^{(N-2)}U^{(N-3)}\ldots U^{(1)}AU^{(1)}U^{(2)}\ldots U^{(N-2)}
	\end{align}
	is a symmetric tridiagonal\footnote{A tridiagonal matrix has only non-vanishing entries on the diagonal and the sub- and super-diagonals, i.e. $(B^{(N-2)})_{i,j}=0$ for $j\neq i-1,i,i+1$.} matrix with $U^{(j)}$ defined as in Eq. \eqref{eq-def householder} with respect to the real vector $\mathbf{u}^{(j)}\in \mathbb{R}^N$. The matrix $B^{(N-2)}$ has the same eigenvalues as $A$.
\end{prop}
\begin{sproof}
	The idea is, to stepwise tridiagonalize the matrix $A$. We choose $U^{(1)}$ such that it reflects the lower $(N-1)$-dimensional part of the first column of $A$ onto the $\mathbf{e}_1\in \mathbb{R}^{N-1}$ direction. Using the symmetry of $A$, we note that the operation $AU^{(1)}$ maps also the $(N-1)$-dimensional part of the first row of $A$ onto the direction of the first unit vector. We proceed with the lower $(N-1)\times (N-1)$ sub-matrix of $A$ in the very same way and iterate the procedure.\\
	The properties of the matrices $U^{(j)}$ follow from the definition in Eq. \eqref{eq-def householder}, with $(\mathbf{u}^{(j)})^\dagger$ replaced by $(\mathbf{u}^{(j)})^t$ in the real setting.\\
	The product $U^{(N-2)}U^{(N-3)}\ldots U^{(1)}$ of symmetric real orthogonal matrices $U^{(j)}$ is again real orthogonal and we conclude that the definition of $B^{(N-2)}$ in Eq. \eqref{eq-def Householder tridiag} is a similarity transformation of $A$, i.e. $B^{(N-2)}=UAU^{-1}$, with $U:=U^{(N-2)}U^{(N-3)}\ldots U^{(1)}$ and $U^{-1}=U^t$. Hence, $B^{(N-2)}$ and $A$ share the same eigenvalues, cf. \cite{parlett,horn}.
\end{sproof}
Talking about random matrices means to consider statistics of the matrix elements. As we pointed out, the amount of a vector mapped to the direction of the first unit vector, is explicitly given and fully determined (in the real setting) by the 2-norm of the initial vector. In 1984, Trotter \cite{Trotter 84} studied the consequence of a Householder reduction applied to a GOE matrix, i.e. a real symmetric matrix with Gaussian distributed entries, all independent up to the inner symmetry of the matrix. Using Proposition \ref{prop-tridiagonalization Householder}, he concluded that an $N\times N$ matrix $A$ from the GOE shares the eigenvalue distribution, with the tridiagonal matrix $B^{(N-2)}$, with
\begin{align}\label{eq-Housholder tridiag GOE}
	B^{(N-2)}\sim \begin{pmatrix}
		\mathcal{N}(0,1)& \frac{1}{\sqrt{2}}\chi^{}_{(N-1)}& & &\\
		\frac{1}{\sqrt{2}} \chi^{}_{(N-1)}&\mathcal{N}(0,1)& \frac{1}{\sqrt{2}}\chi^{}_{(N-2)}& &\\
		&\ddots&\ddots&\ddots&\\
		 & &\frac{1}{\sqrt{2}}\chi^{}_{2}&  \mathcal{N}(0,1) & \frac{1}{\sqrt{2}}\chi^{}_1\\
		 & & &\frac{1}{\sqrt{2}}\chi^{}_1&\mathcal{N}(0,1)
	\end{pmatrix},
\end{align}
where $\mathcal{N}(0,1)$ is the normal distribution with zero mean and unit variance, defined in Eq. \eqref{eq-def normal dist}. $\chi_k$ is the Chi-distribution with parameter $k$. The probability density of a variable $y\sim \chi_k$ defined as \cite[Section 1.3.6.6.6.]{NIST statistics}
\begin{align}\label{eq-def chi dist}
		f_k(y)=\begin{cases}
			\frac{2^{1-k/2}}{\Gamma \left(\frac{k}{2}\right)}e^{-y^2/2}y^{k-1}\quad & \text{ for } x\geq 0,\  k>0,\\
			0\quad &\text{ otherwise.}
		\end{cases}
\end{align}
Here, $\Gamma(m)$ is the Gamma function \cite[Section 8.31]{gradshteyn}. The squared Chi-distribution $\chi_k^2$ is the probability distribution of the sum of the squares of $k$ independent $\mathcal{N}(0,1)$ random variables. It is a special case of the Gamma-distribution, which we will encounter below in Section \ref{ssec-tridiag with resetting}. The Chi-distribution follows from the reflection of an $(N-k)$-dimensional vector of independent normal elements onto its 2-norm for $k=1,\ldots,N-1$. The factor $1/\sqrt{2}$ in the off-diagonals comes from the fact that the variance of the off-diagonal elements of $A$ is $2$ due to the inner symmetry. One important point is that the distribution of the $(N-k)\times(N-k)$-dimensional sub-matrix of $A$ in the $k$th iteration step is invariant under the Householder transformation \eqref{eq-def Householder tridiag}. Recall, that the density of $A$ in GOE is proportional to $\exp(-\Tr(AA^\dagger))$, which is invariant under $A\rightarrow UAU^{-1}$ with $U$ unitary. The impact of the Householder reduction to other random matrix ensembles was studied for example by Silverstein in 1985 \cite{Silverstein 85}. See also \cite{edelman raj rao} for a review.\\
What do we have so far? Using the Householder reflections, introduced in Eq. \eqref{eq-def householder}, we found that Hermitian (respectively symmetric) matrices can be reduced to a tridiagonal form (Proposition \ref{prop-tridiagonalization Householder}) and the distributions of the resulting matrix elements are explicitly known. Especially, we find non-negative off-diagonal elements. As the Householder transformation in Eq. \eqref{eq-def Householder tridiag} is a similarity transformation, we can study the eigenvalue distribution of the tridiagonal matrix instead of the full matrix. Recall, that the GO/U/SE can be viewed as a 1d Dyson gas at the inverse temperature $\beta=1,2,4$ respectively. For the tridiagonalization of a symmetric matrix, i.e. an element of GOE, we found Eq. \eqref{eq-Housholder tridiag GOE}. What happens for an element of GUE? The matrix is now Hermitian, i.e. the off-diagonal elements are complex with real- and imaginary-part independently normal distributed. We can view this as a vector with $2$ components and the number of real parameters in a certain sub-column of the initial matrix is doubled. This yields to a replacement $\chi_k\rightarrow \chi_{2k}$ compared to the GOE result. Similarly, we argue that for the four real normal distributed variables in each off-diagonal entry of a matrix from the GSE. Hence, the tridiagonal reduction can be found from the GOE result with $\chi_k\rightarrow \chi_{4k}$. Note that the factor in the shift of the Chi-distribution is exactly the corresponding $\beta=1,2,4$. Using this insight, Dumitriu and Edelman \cite{Dumitriu Edelman} introduced in 2002 a tridiagonal matrix model, with the elements distributed according to 
\begin{align}\label{eq-DE model}
	H_\beta \sim \frac{1}{\sqrt{2}}\begin{pmatrix}
		\mathcal{N}(0,2)& \chi^{}_{(N-1)\beta}& & &\\
		\chi^{}_{(N-1)\beta}&\mathcal{N}(0,2)& \chi^{}_{(N-2)\beta}& &\\
		&\ddots&\ddots&\ddots&\\
		& &\chi^{}_{2\beta}&  \mathcal{N}(0,2) & \chi^{}_\beta\\
		& & &\chi^{}_\beta&\mathcal{N}(0,2)
	\end{pmatrix}, \text{ for } \beta\in \mathbb{R}_+.
\end{align}
They proved that the eigenvalue distribution of this matrix $H_\beta$ gives the one of a 1d Dyson gas at inverse temperature $\beta>0$, cf. Eq. \eqref{eq-jpdf 1d Dyson}.\\
The strategy of the proof in \cite{Dumitriu Edelman} is as follows: We start from the tridiagonal matrix $H_\beta$, defined in Eq. \eqref{eq-DE model}, for which the correspondence to the GO/U/SE at $\beta=1,2,4$ respectively, is known. Next, the eigenvalue decomposition of $H_\beta$ as $H_\beta=Q\Lambda Q^{-1}$, with $QQ^t=\mathds{1}_N$ orthogonal and $\Lambda=\diag(\lambda_1,\ldots,\lambda_N)$ is studied. It turns out that the first row $\mathbf{q}$ of $Q$ is independent of $\Lambda$. The main task of the proof is now to calculate the Jacobian for the change of variables from the matrix variables of $H_\beta$ to the spectral parameters $\Lambda,\mathbf{q}$. For more details, we refer the reader to \cite{Dumitriu Edelman,book forrester}. We will perform a similar proof in Section \ref{sssec-def tridiag jpdf} for the three tridiagonal non-Hermitian random matrix ensembles introduced there.

\subsection{Anderson-Dixon Integral} \label{ssec-Anderson Dixon}
We present an alternative approach to the tridiagonal representation \eqref{eq-DE model}. Similar as in the previous Section \ref{ssec-Householder}, we start with the study of the GOE. The strategy is to study minor processes, i.e. recursively build matrices of larger dimension $N$ from smaller ones. The idea is that we know the eigenvalues of the smaller, i.e. $(N-1)\times(N-1)$-dimensional matrix and calculate the eigenvalues of the $N\times N$ matrix, conditioned on these $(N-1)$ real eigenvalues. The probability distribution of this conditional process turns out to be the Anderson-Dixon density \cite{Forrester Rains 05}. After exploring this for GOE (i.e. $\beta=1$), we can reach a 1d Dyson gas description for general $\beta>0$.\\
Let $M\in$ GOE($N-1$) be a real symmetric $(N-1)\times (N-1)$ matrix. We build an $N\times N$ matrix $\widetilde{M}$ from $M$ as
\begin{align}
	\widetilde{M}=\begin{pmatrix}
		M & \mathbf{b}\\
		\mathbf{b}^t & a
	\end{pmatrix}.
\end{align}
With the choice $a\sim \mathcal{N}(0,1)$ and $b_i\sim \chi_{1/2}$ for $i=1,\ldots,N-1$ it follows, that $\widetilde{M}\in$ GOE($N$). The characteristic polynomial of $\widetilde{M}$ is given as
\begin{align}
	\Phi_{N}(\lambda):&=\det(\lambda\mathds{1}_{N}-\widetilde{M})\nonumber \\
	&=\det(\lambda\mathds{1}_{N-1}-M)\det((\lambda-a)\mathds{1}_{N-1}-\mathbf{b}^t(\lambda-M)^{-1}\mathbf{b}). \label{eq-anderson dixon p n}
\end{align}
It follows that the distribution of the eigenvalues of $\widetilde{M}$ is invariant under the replacement $M\rightarrow \diag(a_1,\ldots,a_{N-1})$, where $\{a_i\}_{i=1}^{N-1}$ are the real eigenvalues of $M$ \cite{Forrester Rains 05, oxford handbook}. The eigenvalues of $\widetilde{M}$ are determined by the zeros of the rational function in the second factor in Eq. \eqref{eq-anderson dixon p n}. The idea is now, to benefit from the ordering of the real eigenvalues of $\widetilde{M}$ and its minor process $M$ with respect to each other. Without loss of generality, we impose $a_1>a_2>\ldots >a_{N-1}$ and find the interlacing property, i.e.
\begin{align}\label{eq-interlacing}
	\infty>\lambda_1>a_1>\lambda_2>\ldots>a_{N-1}>\lambda_{N}>-\infty,
\end{align}
where $\{\lambda_i\}_{i=1}^{N}$ are the eigenvalues of $\widetilde{M}$.\footnote{We assume non-degeneracy of the eigenvalues, which holds with probability one.} The result by Forrester and Rains \cite[Corollary 1]{Forrester Rains 05} reads as follows, cf. \cite[Proposition 4.3.3]{book forrester}.
\begin{prop}\label{prop-Forrester Rains}
	Let $M_N$ be a sequence of random matrices
	\begin{align}\label{eq-recur M n Forrester Rains}
		M_{N}=\begin{pmatrix}
			\diag M_{N-1} & \mathbf{b}\\
			\mathbf{b}^t & a\\
		\end{pmatrix},
	\end{align}
with $a\sim \mathcal{N}(0,1)$ and $b_j\sim \chi_{\beta /2}$ for $j=1,\ldots, N-1$. The JPDF of $M_N$ agrees with the JPDF of the Gaussian-$\beta$-ensemble \eqref{eq-jpdf 1d Dyson}.
\end{prop}
Let us briefly review the strategy of the proof here, as we did in Section \ref{ssec-Householder} for the work by Dumitriu and Edelman. The key ingredient is the computation of the conditional probability density $G_N(\{\lambda_i\}_{i=1}^N,\{a_j\}_{j=1}^{N-1})$ of $M_N$ given the eigenvalues of the minor process $M_{N-1}$ (cf. \cite[Eq. 20.3.6]{oxford handbook}).
\begin{align}
	\begin{split}\label{eq-Anderson Dixon special case}
	G_N(\{\lambda_i\}_{i=1}^N,\{a_j\}_{j=1}^{N-1})\!=\!\frac{1}{\sqrt{2\pi}}&\frac{1}{(\Gamma(\beta/2))^{N-1}}\frac{\prod_{1\leq j<k \leq N}(\lambda_j-\lambda_k)}{\prod_{1\leq j < k\leq N-1}(a_j-a_k)^{\beta-1}}\prod_{\ell=1}^N\prod_{m=1}^{N-1}\left|\lambda_\ell-a_m\right|^{\frac{\beta}{2}-1}\\
	&\times \exp\left(-\frac{1}{2}\left(\sum_{j=1}^N\lambda_j^2-\sum_{k=1}^{N-1}a_k^2\right)\right)
	\end{split}
\end{align}
The support of $G_N(\{\lambda_i\}_{i=1}^N,\{a_j\}_{j=1}^{N-1})$ is given in Eq. \eqref{eq-interlacing}. Integration over $\{a_j\}_{j=1}^{N-1}$ yields Proposition \ref{prop-Forrester Rains}. Eq. \eqref{eq-Anderson Dixon special case} is a special case of the so-called Anderson-Dixon integral. The initial appearance of these conditional probabilities was in the study of the Selberg-integral. The latter is defined as \cite{Selberg 44, importance of selberg}
\begin{align}
	S_N(\alpha,\beta,\gamma):&=\int_0^1\! dt_1\cdots \int_0^1\!dt_N\ \prod_{i=1}^Nt_i^{\alpha-1}(1-t_i)^{\beta-1}\prod_{1\leq j<k\leq N}\left|t_j-t_k\right|^{2\gamma}\label{eq-selberg integral}\\
	&=\prod_{j=0}^{N-1}\frac{\Gamma(\alpha+j \gamma)\Gamma(\beta+j\gamma)\Gamma(1+(j+1)\gamma)}{\Gamma(\alpha+\beta+(N+j-1)\gamma)\Gamma(1+\gamma)},\label{eq-result selberg}
\end{align}
for complex parameters $\alpha,\ \beta,\ \gamma$, with
\begin{align}
	\Re(\alpha)>0,\ \Re(\beta)>0,\ \Re(\gamma)>-\min\left(\frac{1}{N},\frac{\Re(\alpha)}{N-1} , \frac{\Re(\beta)}{N-1}\right).
\end{align}
We notice the pairwise interaction term in Eq. \eqref{eq-selberg integral}, which we know from the Dyson gas repulsion. It turns out, that the Selberg-integral is related to the normalization constant $C_N(\beta)$ from Eq. \eqref{eq-jpdf 1d Dyson} as \cite{importance of selberg}
\begin{align}
	\lim_{L\rightarrow\infty}2^{L^2}(2L)^{N+N(N-1)\gamma}S_N(L^2/2,L^2/2,\gamma)=\frac{C_N(\gamma)}{(2\pi)^{N/2}}.
\end{align}
The studies on the Selberg-integral are diverse, see \cite{importance of selberg}, for a review. There exist several methods, to prove Eq. \eqref{eq-result selberg}, see \cite[Chapter 4]{book forrester} for the discussion of four of them. One of those techniques uses the conditional probability \eqref{eq-Anderson Dixon special case}. Interestingly, this density function was studied already 39 years before Selberg published his paper \cite{Selberg 44} by Dixon in the theory of Legendre's formula \cite{Dixon}. In 1991, Anderson \cite{Anderson} adapted it directly to the Selberg-integral and thus we refer to Eq. \eqref{eq-Anderson Dixon special case} as (a special case of) the Anderson-Dixon integral today.\footnote{While this seems to be a short note on the Selberg-integral, we would like to emphasize the importance of the result. We will repeatedly be in touch with it throughout this thesis.}\\
Before, we introduce and study a tridiagonal matrix-valued process with resetting in the next Section \ref{ssec-tridiag with resetting}, we would like to have another look onto the recursive construction of $M_N$ in Eq. \eqref{eq-recur M n Forrester Rains}. The rational function, which determines the conditional eigenvalues of $\widetilde{M}$ in Eq. \eqref{eq-anderson dixon p n}, occurs also in the following rank one projection, cf. \cite{Forrester Rains 05,book forrester}. Let $\widetilde{M}\in$ GOE($N$) with eigenvalues $\lambda_1>\lambda_2>\ldots>\lambda_N$, and $\mathbf{x}\in\mathbb{R}^N$ a normalized vector of normal distributed entries. We define
\begin{align}
	M':=\Pi \widetilde{M} \Pi, \text{  where } \Pi=\mathds{1}_N-\mathbf{x}\mathbf{x}^\dagger. 
\end{align}
The projection $\Pi$ onto the subspace orthogonal to the vector $\mathbf{x}$ is of codimension 1. It follows, that the eigenvalues of $M'$, which are non-zero and different from $\{\lambda_i\}_{i=1}^N$, are determined by a rational function, which is of the same form as in Eq. \eqref{eq-anderson dixon p n}. Hence, we can view the described procedure of this section as a (recursive) projection onto a sub-matrix. The eigenvalue density is determined via a conditional probability.\footnote{This method enables the study of minor processes, cf. \cite{oxford handbook}.}
\begin{Remark} \label{rem-perturbations}
	The described procedure can also be viewed in terms of rank-one perturbations of the smaller matrix $M_{N-1}$, cf. \cite{forrester rank one} and references therein. It is well known that non-Hermitian matrices are much more unstable under perturbations than Hermitian matrices and there exists various works probing additive or multiplicative non-Hermitian rank-one perturbations on classical Hermitian-$\beta$-ensembles, showing a significant change in the eigenvalue JPDF, cf. \cite{kozhan additive,kozhan multiplicative}.   
\end{Remark}
\subsection{Application: A Tridiagonal Matrix-Valued Process with Stochastic Resetting for Arbitrary $\boldsymbol{\beta}>0$}\label{ssec-tridiag with resetting}
We can benefit from the insight into the Hermitian $\beta$-ensembles. Namely, we modify the Dumitriu-Edelman ensemble \eqref{eq-DE model} such that we allow for a time-dependent variance $\sigma \rightarrow \sigma(t)$, finding the Dyson gas at inverse temperature $\beta$ as stationary solution for $t\rightarrow \infty$. We study this time-dependent matrix-valued process with stochastic resetting, where the matrix elements are reset to the origin independently and compare to the case of simultaneous resetting which was studied earlier in \cite{BMS2025} without the need of an explicit matrix model. \\
We follow \cite{AMP26} closely, where the author mainly contributed to the introduction of the matrix ensemble and the numerical analysis. We study the following time-dependent process: Let $N$ point particles on a line, with coordinates 
$\mathbf{x}(t)\equiv \{x_1(t),x_2(t),\ldots, x_N(t)\}$, evolve stochastically
with time $t$ via the overdamped Langevin equation
\begin{align}
	\frac{dx_i}{dt}= - \mu\,x_i + D\, \sum_{j(\ne i)} \frac{1}{x_i-x_j} + \sqrt{\frac{2D}{\beta}}\, \eta_i(t),
	\label{eq-dbm0.def}
\end{align}
where the $\eta_i(t)$ are independent zero-mean Gaussian white noises with correlator
\begin{align}
	\langle \eta_i(t)\eta_j(t')\rangle = \delta_{i,j}\delta(t-t')\, .
	\label{eq-noise.0}
\end{align}
Here, $\delta(t-t')$ denotes the Dirac delta function, i.e. the continuous version of $\delta_{i,j}$.
In \eqref{eq-dbm0.def}, the quantities $(\mu, D, \beta)$ are just three positive parameters. The first term on the right-hand-side (RHS) represents a force due to a confining harmonic potential $V(x_i)= \mu x_i^2/2$ acting on each particle that tries to drive the particle towards the origin. The second term on the RHS corresponds to a pairwise repulsion term between the particles that tries to push them apart. We saw these two terms earlier in the introduction of the 1d Dyson gas in Section \ref{ssec-intro models}. The third term on the RHS corresponds to thermal fluctuations induced by a heat bath with inverse temperature $\beta$ to which the particles are connected. The stochastic process in Eq. \eqref{eq-dbm0.def}, introduced by Dyson in 1962 \cite{Dyson1962a,Dyson1962b}, is referred to as the $\beta$-Dyson Brownian motion ($\beta$-DBM).
For simplicity, we assume that the Langevin equation \eqref{eq-dbm0.def} starts from the initial condition where all the particles are localised close to the origin, e.g., over an interval $(-\epsilon,\epsilon]$ with
\begin{align}
	x_i(0)= \epsilon\, \left(-1 +\frac{2i}{N}\right)\, , \quad i=1,2,\ldots N\, ,
	\label{eq-init.0}
\end{align}
and eventually the limit $\epsilon\to 0$ is taken. However, irrespective of the initial condition, it is easy to see that at long times $t\to \infty$, the JPDF $P_0(\mathbf{x},t)$ of the positions of the particles approaches
a stationary (st) form which is simply given by the equilibrium Gibbs-Boltzmann distribution
\begin{align}
	P_0^\textrm{st}(\mathbf{x})= P_0\left(\mathbf{x}, t\to \infty\right)\propto e^{-\beta\, E(\mathbf{x})}\, ,
	\label{eq-gibbs.0}
\end{align}
where the energy $E(\mathbf{x})$ of the gas of particles is given by
\begin{align}
	E(\mathbf{x})= \frac{\mu}{2D}\, \sum_{i=1}^N x_i^2 -\frac{1}{2}\sum_{i\ne j} \ln |x_i-x_j|\, .
	\label{eq-energy.0}
\end{align}
We encounter exactly the Dyson gas discussed Section \ref{ssec-intro models}. It follows that the stationary distribution \eqref{eq-gibbs.0} agrees with the JPDF \eqref{eq-jpdf 1d Dyson} in terms of the eigenvalues of the GO/U/SE for $\beta=1,2,4$, respectively the Hermitian $\beta$-ensemble \eqref{eq-DE model}.

At any finite time $t$, for the three special values $\beta=1,\,2$ and $4$, Dyson showed \cite{Dyson1962a,Dyson1962b} that the positions $\{x_i(t)\}$ in \eqref{eq-dbm0.def} can be interpreted as $N$ real eigenvalues of a time-dependent $N\times N$ Gaussian matrix $H(t)$, which is real symmetric ($\beta=1$), complex Hermitian $(\beta=2)$ and quaternionic self-dual ($\beta=4$). The entries of this matrix are evolving stochastically in a `fictitious' time $t$ by performing independent Ornstein-Uhlenbeck (OU) processes. Such a time-dependent stochastic matrix $H(t)$ will be referred to as a \textit{matrix-valued process}. Dyson demonstrated, considering the evolution of $H(t)$ from $t$ to $t+\Delta t$ and using a second order perturbation theory in $\Delta t$, that for $\beta=1,2$, and $4$, if one diagonalizes this matrix $H(t)$ at any instant $t$, then its $N$ real instantaneous eigenvalues $\mathbf{x}(t)$ evolve via Eq. \eqref{eq-dbm0.def}. However, the stochastic process \eqref{eq-dbm0.def} is well-defined
for any $\beta>0$. This $\beta$-DBM process is interesting in its own right (irrespective of its connection to random matrix theory) and has been studied extensively in many different applications \cite{book forrester,KT04,Bor09,ABG12,ABMV13,BBMP2014,AD2015,LY2017,GMS21,MM21,TLS23,BMS2025,FLO2025}. 
A natural question then arises whether, for arbitrary $\beta>0$, one can find an appropriate underlying matrix-valued process $H(t)$ with stochastically evolving entries such that its eigenvalues at any time $t$ evolve by the $\beta$-DBM process \eqref{eq-dbm0.def}. This subject was studied earlier in various works without resetting \cite{Li beta process, Holcomb Paquette,Yabuoko,Edelman Brownian}. While our finding of the matrix-valued process agrees with parts of results by Li \cite{Li beta process}, we derive the ensemble independent and especially via a different proof technique. We comment on the parallels and difference on the lines. Moreover, we are the first to revisit this process in the case of resetting. 

The introduction of a time-dependent process $H(t)$ becomes particularly relevant in the context of stochastic resetting in many-body classical systems. Stochastic resetting simply means interrupting the natural dynamics (stochastic or deterministic) of a system at random times and restarting the dynamics from the same initial condition \cite{EM2011,EM2011_b}. We provide a brief introduction to stochastic resetting later in Section \ref{sssec-intro resetting}. One of the principal effects of stochastic resetting, amongst others, is to drive the system at long times to a nonequilibrium stationary state (NESS), since the stochastic interruption typically violates detailed balance \cite{EM2011}. Recently, the $\beta$-DBM process in Eq. \eqref{eq-dbm0.def} in the presence of a simultaneous stochastic resetting
with rate $r$ was studied in reference \cite{BMS2025}. More precisely, one starts the $\beta$-DBM process from the initial condition \eqref{eq-init.0} and evolves it up to a random time $\tau_1$ drawn from an exponential distribution
$p(\tau)= r\, e^{-r \tau}$. At this time $\tau_1$, all the particles are simultaneously and instantaneously reset to their initial positions in \eqref{eq-init.0} and they again evolve up to a random time $\tau_2$ drawn
again from $p(\tau)$. Then they are reset and the process continues. 
This process was referred to as $\beta$-RDBM (resetting Dyson Brownian motion).
It was proved in reference \cite{BMS2025} that
the JPDF of the positions of particles $P_r(\mathbf{x}, t)$, at long times $t\to \infty$, approaches a NESS 
$P_r^\textrm{st}(\mathbf{x})$ given explicitly, for arbitrary $\beta>0$, by
\begin{align}
	P_r^\textrm{st}(\mathbf{x})= \frac{r}{C_N(\beta)}\,\int_0^{\infty}\! \frac{d\tau\, e^{-r\, \tau}}{
		\left[\sigma(\tau)\right]^{N+ \beta N (N-1)/2}}\,
	e^{-\frac{1}{2\, \sigma^2(\tau) }\sum_i x_i^2} \, \prod_{i> j}^N
	|x_i-x_j|^{\beta}\,  ,
	\label{eq-RDBM_st.0}
\end{align}
where 
\begin{align}
	\sigma^2(\tau)= \frac{D}{\beta\, \mu}\, \left(1- e^{-2\, \mu\, \tau}\right)\, 
	\label{eq-sigma2_def.0}
\end{align}
and $C_N(\beta)$ is given in Eq. \eqref{eq-jpdf 1d Dyson}.
In the limit $r\to 0$, the JPDF $P_r^\textrm{st}(\mathbf{x})$ in Eq. \eqref{eq-RDBM_st.0} reduces again to the JPDF \eqref{eq-jpdf 1d Dyson} of the 1d Dyson gas with $\mathbf{x}\equiv \boldsymbol{\lambda}$. This can be easily seen by implementing the change of variables $r\tau \to y$ in Eq. \eqref{eq-RDBM_st.0}.

For $\beta=1, 2$ and $4$, it follows that the $\beta$-RDBM process describes the evolution of the eigenvalues of a resetting matrix-valued process whose
OU entries undergo simultaneous resetting to the origin with rate $r$ \cite{BMS2025}. Consequently, for these three special values of $\beta=1, 2$ and $4$, this resetting matrix-valued process also reaches a NESS as $t\to \infty$, i.e., a nonequilibrium ensemble (no longer Gaussian, but still rotationally invariant) whose eigenvalues are jointly distributed via the JPDF $P_r^\textrm{st}(\mathbf{x})$ in Eq. \eqref{eq-RDBM_st.0}. It is then natural to ask if, for arbitrary $\beta>0$ (not necessarily $\beta=1,2$ and $4$), one can construct a resetting matrix-valued process $H(t)$ such that the matrix ensemble, driven by simultaneous resetting of its entries, approaches a NESS and its eigenvalues are jointly distributed via Eq. \eqref{eq-RDBM_st.0} with general $\beta>0$. We construct precisely such a general-$\beta$ resetting matrix-valued process $H(t)$ using the insight of the proof techniques for the stationary matrix ensemble and thus generalise the results of reference \cite{BMS2025}.

The rest of this section is organized as follows. In Section \ref{sssec-intro resetting}, we provide a brief description of stochastic resetting to be self-containing. We then recall in Section \ref{sssec-RDBM} some properties of the $\beta$-DBM process in Eq. \eqref{eq-dbm0.def} and of the $\beta$-RDBM process once the simultaneous resetting with rate $r$ is switched on.
In Section \ref{sssec-CIR}, we introduce the CIR process of a single particle and state its exact time-dependent position distribution at all times $t$. This result is crucially needed in the next Section \ref{sssec-TMP}, where we introduce the matrix-valued $\beta$-TMP process for arbitrary $\beta>0$ re-deriving parts of an earlier result by Li \cite{Li beta process} using different techniques.
In Section \ref{sssec-RTMP}, we investigate the resetting matrix-valued processes
$\beta$-SRTMP (simultaneous resetting of entries) and $\beta$-IRTMP (independent resetting of entries) which was not considered in the literature before. We prove that the JPDF of the eigenvalues of the $\beta$-SRTMP ensemble at long times is given precisely by
Eq. \eqref{eq-RDBM_st.0} for arbitrary $\beta>0$. We also present 
numerical results for the average density of eigenvalues in the $\beta$-IRTMP stationary ensemble and compare it to the analytical results for the corresponding $\beta$-SRTMP stationary ensemble, demonstrating that they are quite different from each other.

\subsubsection{Stochastic Resetting}
\label{sssec-intro resetting}

We briefly outline here the main ideas behind stochastic resetting to strengthen the wide applicability and the importance of it. It is remarkable that this topic, which was introduced by Evans and Majumdar in \cite{EM2011} in 2011, has seen an explosion of activities both theoretically and experimentally, spanning across disciplines such as physics, chemistry, biology, mathematics and computer science (for reviews see \cite{EMS2020,GJ22,PKR22}). The main idea 
behind stochastic resetting is very simple. Imagine a process evolving under its own natural dynamics (deterministic or stochastic) starting
from a fixed initial condition. Stochastic resetting simply means interrupting the natural dynamics at random times and re-starting the process from the same initial condition. The original motivation for studying this process came from random search problems \cite{EM2011,EM2011_b}, where it proved to be very useful in designing random search algorithms with a stochastic resetting component leading to several important practical applications \cite{EMS2020,PKR22}.

There is also a second aspect of stochastic resetting on which we will mainly focus here. 
Stochastic resetting typically drives a system
to a nonequilibrium stationary state (NESS) at long times by eliminating long wandering trajectories which can lead the system far away from a stationary state. Nevertheless, this steady state is `nonequilibrium' as the resetting move violates detailed balance: there is always a probability current from any position to the initial position via instantaneous resetting move, but there is no reverse move. Hence, there is always a non-zero probability current in configuration space even in the steady state \cite{EM2011,EMS2020}. 

The physical properties in the resetting driven NESS have since been studied in both single particle and multiple particles (with or without interactions) systems, with a host of interesting phenomena. We refer the reader to \cite{AMP26}, where we give a detailed list of references.

For simultaneous resetting in a generic many-body system, there exists a very simple general relation, based on a renewal argument, that connects the probability distribution in the configuration space in the presence and in the absence of resetting. This argument is very general and holds for any process. 
Consider a system with any number of degrees of freedom and let us denote a configuration of the system by $\mathcal{S}$. We will view $\mathcal{S}$ below either as the configuration of positions ($\equiv$ eigenvalues) or as a given (random) matrix. The system undergoes its natural dynamics in continuous time starting from a fixed configuration $\mathcal{S}_0$ and let $r$ denote the resetting rate at which the system returns instantaneously from a current configuration to its initial configuration $\mathcal{S}_0$. The resetting events thus form a Poisson process with rate $r$. Let $P_r(\mathcal{S},t)$ denote the probability that the system is in configuration $\mathcal{S}$ at time $t$, starting at $\mathcal{S}_0$ at time $t=0$. The subscript $r$ denotes the presence of a non-zero resetting rate $r$. 
This is actually the propagator $P_r(\mathcal{S},t)= G_r(\mathcal{S},t|S_0,0)$ from $\mathcal{S}_0$ at $t=0$ to $\mathcal{S}$ at time $t$, but we omit the explicit dependence on $\mathcal{S}_0$ for brevity. Similarly, $P_0(\mathcal{S},t)$ denotes the probability that the reset-free system ($r=0$) is in configuration $\mathcal{S}$ at time $t$, starting from $\mathcal{S}_0$
at time $t=0$. These two probabilities are simply related to each other by the following renewal relation \cite{GMS2014,EMS2020}
\begin{align}
	P_r(\mathcal{S},t)= \underbrace{e^{-r t}\, P_0(\mathcal{S},t)}_{\text{no resetting in } [0,t]} + \underbrace{r\, \int_0^t \! d\tau\, e^{-r\, \tau}\, P_0(\mathcal{S},\tau)}_{\text{resetting}}\, .
	\label{eq-renewal_gen.1}
\end{align}
The interpretation of the two terms on the RHS of Eq. \eqref{eq-renewal_gen.1} is very simple. Consider all possible trajectories of the system going from $\mathcal{S}_0$ at time $t=0$ to the configuration $\mathcal{S}$ at time $t$. 
There are two possibilities: (i) no resetting events in $[0,t]$ and (ii) one or more resetting events occur in $[0,t]$. The probability for event (i) is simply $e^{-r t}$ since the resetting events are Poissonian with rate $r$.
In this case, the probability to reach $\mathcal{S}$ at time $t$ (starting at $\mathcal{S}_0$) is simply the bare (reset-free) probability $P_0(\mathcal{S}, t)$ (also starting at $\mathcal{S}_0$).
Hence, multiplying $P_0(\mathcal{S},t)$ by $e^{-r t}$ gives the joint probability
of reaching $\mathcal{S}$ at time $t$ and that there is no resetting event in $[0,t]$. For the second term in Eq. \eqref{eq-renewal_gen.1}, where we consider having one or more resettings in $[0,t]$, it is useful to first consider the epoch $t_l$ at which the last resetting before $t$ occurred in the trajectory. Let $\tau=t-t_l$. This is useful because conditioned on the last resetting at $t_l$ before $t$, the process evolves freely (without resetting) during the interval $\tau$ from $t_l$ to $t$. The probability for the process to reach $\mathcal{S}$ at time $t$ (starting from $\mathcal{S}_0$), conditioned on $t_l$, is then $P_0(\mathcal{S}, \tau=t-t_l)$ where we assumed that the propagator of the reset-free process depends only on the time difference $\tau=t-t_l$. Now, the probability of having the last resetting event at $t_l$ is simply the probability of a reset at $t_l$ (which occurs with density $r$) followed by no resetting during
$\tau=t-t_l$ that occurs with probability $e^{-r \tau}$. Hence, multiplying all factors and integrating $\tau$ over $[0,t]$ gives the joint probability of reaching $\mathcal{S}$ at time $t$ and having one or more resettings in $[0,t]$. The probability $P_r(\mathcal{S}, t)$ is normalized to unity when summed over all configurations $\mathcal{S}$. Thus, the renewal equation \eqref{eq-renewal_gen.1} 
allows one to compute the probability $P_r(\mathcal{S}, t)$ of the process exactly at any time $t$, provided one knows the probability $P_0(\mathcal{S}, t)$ of the underlying
reset-free process \textit{at all times}. 

Now, we consider the renewal equation \eqref{eq-renewal_gen.1} in the limit $t\to \infty$. The first term on the RHS drops out and we get the stationary state
\begin{align}
	P_r^\textrm{st}(\mathcal{S})= P_r(\mathcal{S}, t\to \infty)= r \int_0^{\infty}\! d\tau\, e^{-r \tau}\, P_0(\mathcal{S}, \tau)\, ,
	\label{eq-stat_gen.1}
\end{align}
provided, of course, the integral in Eq. \eqref{eq-stat_gen.1} exists. Thus the NESS is fully determined if one knows the probability $P_0(\mathcal{S},t)$ of the underlying reset-free process \textit{at all times}, and not just at late times. In many situations, it is easy to determine the late time behaviour of the reset-free propagator $P_0(\mathcal{S},t)$, but is much harder to compute $P_0(\mathcal{S},t)$ \textit{at all times} $t$ which is needed to determine the stationary
probability $P_r^\textrm{st}(\mathcal{S})$ of the process with reset.

\subsubsection{Dyson Brownian Motion and Underlying Matrix-Valued Process}
\label{sssec-RDBM}

In this section, we briefly recall the $\beta$-DBM process \eqref{eq-dbm0.def} and its underlying matrix-valued process, first without resetting and then in the presence of a simultaneous resetting with rate $r>0$. We consider first the standard $\beta$-Dyson Brownian motion ($\beta$-DBM) defined by the Langevin equation \eqref{eq-dbm0.def}. As discussed above, for three special values of the parameter $\beta=1,2$ and $4$,  
Dyson showed \cite{Dyson1962a,Dyson1962b} that the set of positions $\{x_i(t)\}$ can be interpreted as the $N$ real eigenvalues of a time-dependent $ N\times N $ Gaussian matrix $H(t)$ (real symmetric, complex Hermitian and quaternion self-dual for $\beta=1,2,4$ respectively), whose elements are evolving via the Ornstein-Uhlenbeck (OU) process. 
For instance, for $\beta=1$, the real symmetric matrix entry $H_{j,k}(t)$ evolves via
\begin{align}
	\frac{d H_{j,k}(t)}{dt}= - \mu\, H_{j, k}(t) + \sqrt{\left(1+ \delta_{j,k}\right)\, D}\, \xi_{j, k}(t)\, ,
	\label{eq-OU.1}
\end{align}
where the $\xi_{j,k}(t)$ are again independent (for $j\geq k$ subject to symmetry) zero-mean Gaussian white noises with correlator
$\langle \xi_{j,k}(t)\xi_{j', k'}(t')\rangle= \delta_{j, j'}\, \delta_{k,k'}\, \delta(t-t')$. A similar construction can be made for $\beta=2$ and $\beta=4$ with complex and quaternionic matrix entries, subject to the Hermiticity and self-duality condition, respectively. The JPDF of the matrix entries, starting from $H_{j,k}(0)=0$ (for simplicity), then evolve as a Gaussian process (up to an overall normalization constant)
\begin{align}
	P_0(H,t)= \textrm{Prob.}\left[H(t)=H\right]\propto \frac{1}{\left[\sigma(t)\right]^{N+\beta\, N(N-1)/2}}\, 
	\exp\left[- \frac{1}{2\, \sigma^2(t)}\, \textrm{Tr}(H^2) \right]\ , 
	\label{eq-matrix_OU.1}
\end{align}
with $\sigma(t)$ given in \eqref{eq-sigma2_def.0}.
Notice that when writing out the invariant probability distribution of matrix elements \eqref{eq-matrix_OU.1}, it factorises into the product of 
$N+\beta\, N(N-1)/2$ independent real Gaussian factors, with different variances for the diagonal and off-diagonal matrix elements, due to symmetry. For example, for $\beta=1$, we have for the 
$N(N+1)/2$ independent real matrix elements $H_{l\geq k}$
\begin{align}
	\textrm{Prob.}\left[H(t)=H\right]_{\beta=1}\propto \frac{1}{\left[\sigma(t)\right]^{N(N+1)/2}}\, 
	\exp\left[- \frac{1}{2\, \sigma^2(t)}\, \sum_{j=1}^N\, H_{j,j}^2 - \frac{1}{\sigma^2(t)}\, \sum_{j>k}^N
	H_{j,k}^2 \right]\ .
	\label{eq-matrix_OU.1b1}
\end{align}

At late times, the JPDF \eqref{eq-matrix_OU.1} of the matrix entries then converges to the standard, stationary Gaussian ensembles for $\beta=1,\, 2,\, 4$ given by
\begin{align}
	P_0^\textrm{st}[H]= \lim_{t\to \infty}\textrm{Prob.}\left[H(t)=H\right]\propto \exp\left[- 
	\frac{\beta\, \mu}{2D}\, \textrm{Tr}(H^2)\right]\, .
	\label{eq-GE_st}
\end{align}
It is natural to ask how the JPDF of the eigenvalues $ P_0(\mathbf{x},t)$ 
at time $t$ looks for $\beta=1,2,4$? This can be answered using a simple rescaling approach, without doing any explicit computation, as shown below.

Rescaling the matrix $H(t)$ by $\sigma(t)$, i.e. defining a matrix $\tilde{H}(t)= H(t)/\sigma(t)$ with $\sigma(t)$ defined in Eq. \eqref{eq-sigma2_def.0}, one sees from
Eq. \eqref{eq-matrix_OU.1} that at any instant $t$, the JPDF of the matrix $\tilde{H}(t)$ has a time-independent form
\begin{align}
	\textrm{Prob.}\left[\tilde{H}(t)=\tilde{H}\right]\propto 
	\exp\left[- \frac{1}{2}\, \textrm{Tr}({\tilde{H}}^2)\right]\, .
	\label{eq-matrix_OU.2}
\end{align}
Consequently, it follows that the eigenvalues $\tilde{\mathbf{x}}$ of $\tilde{H}$ are also time-independent with a JPDF given by
\begin{equation}
	P_0\left(\tilde{\mathbf{x}}\right)=
	\frac{1}{C_N(\beta)}\, e^{-\frac{1}{2}\sum_{i=1}^N \tilde{x}_i^2} \, \prod_{ i>j}^N
	|\tilde{x}_i-\tilde{x}_j|^{\beta} \, ,
	\label{st_jpdf.2}
\end{equation}
Reverting back to the unscaled eigenvalues $x_i(t)= \sigma(t)\, \tilde{x}_i$, one sees immediately that the PDF of the original unscaled eigenvalues $\mathbf{x}(t)$ at any time $t$ is given by \cite{book forrester}
\begin{equation}
	P_0\left(\mathbf{x}, t\right)=
	\frac{1}{C_N(\beta)}\, \frac{1}{[\sigma(t)]^{N+ \beta N(N-1)/2}}\, 
	e^{-\frac{1}{2 \sigma^2(t) }\sum_{i=1}^N x_i^2} \, \prod_{i> j}^N
	|x_i-x_j|^{\beta} \, , 
	\label{eq-dbm_jpdf.2}
\end{equation}
where we recall that $\sigma(t)$ is given in Eq. \eqref{eq-sigma2_def.0}. The overall factor $[\sigma(t)]^{-N-\beta N(N-1)/2}$ comes from the change of variables in the Vandermonde determinant to the power $\beta$, but crucially $C_N(\beta)$ does not depend on $\sigma(t)$. In deriving the result in Eq. \eqref{eq-dbm_jpdf.2}, we assumed that $\beta=1,\, 2$ or $4$ so that we could exploit the underlying matrix-valued process and its rotational symmetries for $\beta=1,\,2,\, 4$. We will see below that the result in \eqref{eq-dbm_jpdf.2} holds for arbitrary $\beta>0$ \cite{BMS2025}, and not necessarily for $\beta=1,\,2,\, 4$.
\\
By solving explicitly the Fokker-Planck equation associated to the Langevin equation \eqref{eq-dbm0.def}, it was shown recently \cite{BMS2025} that actually, the time-dependent JPDF in Eq. \eqref{eq-dbm_jpdf.2} is exact for any $t$ and any $\beta>0$, not necessarily restricted to $\beta=1,2,4$. In this derivation, it was assumed that the Langevin process \eqref{eq-dbm0.def} start from the initial condition \eqref{eq-init.0} with $\epsilon\to 0$. Hence the JPDF \eqref{eq-dbm_jpdf.2}, for arbitrary $\beta>0$, can also be interpreted as the
propagator of the $\beta$-DBM process from $\mathbf{x} (0)\equiv \{x_1(0)\to0, x_2(0)\to 0,\ldots, x_N(0)\to 0\}$ to $\mathbf{x} \equiv \{x_1,x_2,\ldots, x_N\}$ in time $t$.

We recall from Section \ref{sssec-intro resetting}, that the knowledge at all times is indeed needed to determine the exact NESS of this process when it is subjected to simultaneous resetting of all the particles to the initial condition with rate $r$. Substituting the explicit form \eqref{eq-dbm_jpdf.2} of $P_0(\mathbf{x}, \tau)$ into the renewal equation \eqref{eq-stat_gen.1} for $\mathcal{S}=\mathbf{x}$, one then gets in the limit $t\rightarrow \infty$ the stationary JPDF of $\beta$-RDBM process in Eq. \eqref{eq-RDBM_st.0} in the reset induced NESS for arbitrary $\beta>0$ \cite{BMS2025}. Similarly, we can express the probability distribution $P_0(H,t)$ in Eq. \eqref{eq-matrix_OU.1} of the matrix $H(t)$ in the presence of resetting with rate $r$ using $\mathcal{S}=H$ in the renewal equation \eqref{eq-renewal_gen.1}. In particular at late times, the JPDF of the resetting matrix-valued process approaches a stationary distribution
\begin{align}
	P_r^\textrm{st}(H) \propto r\, \int_0^{\infty}\! d\tau\, e^{-r\, \tau}\, \frac{1}{\left[\sigma(\tau)\right]^{N+
			\beta\, N(N-1)/2}}\,
	\exp\left[- \frac{1}{2\, \sigma^2(\tau)}\, \textrm{Tr}(H^2) \right]\,,  
	\label{eq-matrix_reset_st.1}
\end{align}
with $\sigma(t)$ from Eq. \eqref{eq-sigma2_def.0}. Here, the proportionality constant just normalizes the matrix PDF. We conclude that if a matrix $H$ is distributed via Eq. \eqref{eq-matrix_reset_st.1} with $\beta=1,2,4$, then its eigenvalues $\{x_i\}$'s are distributed via Eq. \eqref{eq-RDBM_st.0}. In other words, the $\beta$-RDBM process, for $\beta=1,2,4$, has a stationary distribution \eqref{eq-RDBM_st.0} which can be interpreted as the JPDF of the eigenvalues drawn from \eqref{eq-matrix_reset_st.1}. Such a deformed Gaussian matrix ensemble is 
called super-statistical Gaussian ensemble and has been studied in the literature in other contexts \cite{AM2005,BCP2008,AMAV2009}.

Again, we ask: What happens when $\beta\neq 1,\, 2,\, 4$? In Section \ref{sssec-TMP}, we will see that indeed, for arbitrary $\beta>0$, one can construct a matrix-valued resetting process $H(t)$ that reaches a stationary ensemble at late times and the eigenvalues of that stationary ensemble are indeed distributed via Eq. \eqref{eq-RDBM_st.0}. 
However, to construct explicitly such a matrix-valued process, we need to first remind the reader about another stochastic process known as the Cox-Ingersoll-Ross (CIR) process, which we do in the next section.

\subsubsection{Cox-Ingersoll-Ross Process}
\label{sssec-CIR}

We consider a stochastic process $x(t)$ defined on the semi-infinite line $x\ge 0$ that evolves via the Langevin equation
\begin{align}
	\frac{dx}{dt}= \mu_0\, (\theta-x(t))+ \sigma_0\, \sqrt{x(t)}\, \eta(t)\, ,
	\label{eq-cir.1}
\end{align}
starting from $x_0\ge 0$. Here $\eta(t)$ is a zero-mean Gaussian white noise with correlator $\langle \eta(t)\eta(t')\rangle= \delta(t-t')$. The multiplicative term in Eq. \eqref{eq-cir.1} is interpreted in the It$\hat{\textrm{o}}$ sense.\footnote{Taking only information up to time $t$ into account.} The process is characterized by three parameters
$\mu_0>0$, $\theta> 0$ and $\sigma_0>0$. It is a particular case of the celebrated Feller process \cite{Feller51} that describes an overdamped particle in the presence of a linear drift and a space dependent diffusion constant. The process \eqref{eq-cir.1} has been studied extensively and has found numerous applications across disciplines: from neurobiology \cite{CR73,CR74},
finance \cite{CIR85,Hull,Heston93,DY2002} and branching process with death \cite{MR2025}.
In finance, the process in Eq. \eqref{eq-cir.1} goes by the name of CIR (Cox-Ingersoll-Ross) process \cite{CIR85}. The first-passage properties and the extreme value statistics of the CIR process have been investigated more
recently \cite{MP2012,Somrita2022,MR2025}.

Let $P_0(x,t|x_0)$ denote the probability density for the process to be at $x$ at time $t$, starting from $x_0>0$ at $t=0$. 
The exact result for $P_0(x,t|x_0)$ is known \cite{CR73,CIR85} and reads
\begin{align}
	P_0(x,t|x_0,0) \equiv P_0(x,t)= C(t)\, e^{-u-v} \left(\frac{v}{u}\right)^{\gamma/2}\, I_\gamma\left(2\, \sqrt{u\, v}\right)\, ,
	\label{eq-full_solx.1}
\end{align}
where
\begin{align}
	C(t)= \frac{2\mu_0}{\sigma_0^2 \left(1- e^{-\mu_0\, t}\right)}\, , \quad\quad  u= C(t)\, x_0\, e^{-\mu_0\, t}\, ,
	\quad\quad v= C(t)\, x\, , \quad \textrm{and}\quad  \gamma= \frac{2\mu_0\theta}{\sigma_0^2}-1\, .
	\label{eq-def_cons}
\end{align}
For the interested reader, we present its derivation in \cite{AMP26}.\\
In the limit $t\to \infty$, the probability density $P_0(x,t)$ approaches
a stationary solution, $P_0(x,t\to \infty) \to P^\textrm{st}_0(x)$. To derive the stationary solution from Eq. \eqref{eq-full_solx.1},
we note from Eq. \eqref{eq-def_cons} that, to leading order as $t\to \infty$,
\begin{align}
	C(t)\to \frac{2\mu_0}{\sigma_0^2}\, , \quad\quad u\to \frac{2\mu_0 x_0}{\sigma_0^2}\, e^{-\mu_0\, t}\, ,
	\quad\quad v\to \frac{2\mu_0\, x}{\sigma_0^2}\, .
	\label{eq-cons_st.1}
\end{align}
Since $u\to 0$ in Eq. \eqref{eq-full_solx.1} when $t\to \infty$, we can use the small argument behaviour of $I_\gamma(z)$ \cite[Eq. 8.440]{gradshteyn}
\begin{align}
	I_\gamma(z) \approx \frac{2^{-\gamma}}{\Gamma(\gamma+1)}\, z^{\gamma}\, \quad \textrm{as} \quad z\to 0\,.
	\label{eq-BesselI_smallz}
\end{align}
Substituting this behaviour along with Eq. \eqref{eq-cons_st.1}, one gets from Eq. \eqref{eq-full_solx.1} the limiting behaviour as $t\to \infty$, namely
\begin{align}
	P^\textrm{st}_0(x)= P_0(x, t\to \infty)= \left(\frac{2\mu_0}{\sigma_0^2}\right)^{\gamma+1}\, \frac{x^{\gamma}}{\Gamma(\gamma+1)}\, 
	e^{-2\mu_0\, x/\sigma_0^2}\, , 
	\label{eq-stat_sol.1}
\end{align}
with $\gamma=\frac{2\mu_0\theta}{\sigma_0^2}-1$. It is easy to check that this stationary solution is normalized to unity, i.e., $\int_0^{\infty}\! dx \, P^\textrm{st}_0(x)=1$, and of course, it is independent of the initial position $x_0$. The stationary solution $P^\textrm{st}_0(x)$ 
in Eq. \eqref{eq-stat_sol.1} can be expressed as PDF of a Gamma distributed random variable. A random variable $\Gamma(q,l)\ge 0$ is called Gamma distributed if its PDF can be written as
\begin{align}
	f(x, q, l)= \textrm{Prob.}\left[\Gamma(q,l)=x\right] = \frac{1}{\Gamma(q)\, l^q}\, x^{q-1}\, e^{-x/l}\, , 
	\quad \textrm{for}\quad x\ge 0\,  ,
	\label{eq-Gamma_pdf.1}
\end{align}
where the shape factor $q\ge 0$ and the scale factor $l$ parametrize this PDF. Comparing Eq. \eqref{eq-Gamma_pdf.1} to the exact stationary solution $P_0^\textrm{st}(x)$ in Eq. \eqref{eq-stat_sol.1}, we see immediately that 
\begin{align}
	P^\textrm{st}_0(x)= f\left(x,\, q=\frac{2\mu_0\theta}{\sigma_0^2},\, l= \frac{\sigma_0^2}{2\mu_0}\right)\, , 
	\label{eq-stat_sol.2}
\end{align}
where $f(x,q,l)$ is given in Eq. \eqref{eq-Gamma_pdf.1}. Thus 
$P^\textrm{st}_0(x)=\textrm{Prob.}[y_\textrm{st}=x]$ is the PDF of a Gamma distributed random variable $y_\textrm{st}$
\begin{align}
	y_\textrm{st}=\Gamma\left(q=\frac{2\mu_0\theta}{\sigma_0^2},\, l= \frac{\sigma_0^2}{2\mu_0}\right)\, .
	\label{eq-Gamma_random.1}
\end{align}

For a generic initial condition $P_0(x, t=0)= \delta(x-x_0)$ with $x_0\ge 0$, the full time-dependent solution \eqref{eq-full_solx.1} is somewhat complicated. However, for the special case when $x_0=0$, i.e., the process
starts at the origin, the full time-dependent solution takes a very simple self-similar form which will be useful later. To derive this form, we note from Eq. \eqref{eq-def_cons} that as $x_0\to 0$, we have $u\to 0$. Consequently, we can again use the small argument asymptotics of $I_\gamma(z)$ given in
Eq. \eqref{eq-BesselI_smallz} to take the limit $x_0\to 0$ in Eq. \eqref{eq-full_solx.1}. This gives, for any $t\ge 0$,
\begin{align}
	\begin{split}
	P_0(x,t|x_0=0) &= C(t)\, \frac{[C(t)\, x]^{\gamma}}{\Gamma(\gamma+1)}\, e^{-C(t)\, x}\, ,\\ \textrm{where}\quad
	C(t)&=\frac{2\mu_0}{\sigma_0^2 \left(1- e^{-\mu_0\, t}\right)}\, , \quad \textrm{and}\quad
	\gamma=\frac{2\mu_0\theta}{\sigma_0^2}-1\, .
	\end{split}
	\label{eq-origin_init.1}
\end{align}
This solution, for any $t\ge 0$ and $x\ge 0$, can be expressed in terms of the PDF $f(x,q,l)$ of a Gamma distributed random variable, i.e.,
\begin{align}
	P_0(x,t|x_0=0)= f\left(x, \,  q=\frac{2\mu_0\, \theta}{\sigma_0^2},\,  l=\frac{1}{C(t)}=
	\frac{\sigma_0^2 \left(1- e^{-\mu_0\, t}\right)}{2\mu_0}\right)\, ,
	\label{eq-sol_final.1}
\end{align}
where $f(x,q,l)$ is defined in Eq. \eqref{eq-Gamma_pdf.1}. Thus, $p_0(x,t|x_0=0)=\textrm{Prob.}[y(t)=x]$ represents the PDF of a Gamma distributed random variable $y(t)$
\begin{align}
	y(t)= \Gamma\left(q=\frac{2\mu_0\, \theta}{\sigma_0^2},\,  l=
	\frac{\sigma_0^2 \left(1- e^{-\mu_0\, t}\right)}{2\mu_0}\right)\, .
	\label{eq-yt_def.1}
\end{align}
Of course, as $t\to \infty$, one recovers the stationary solution, i.e., $y(t\to \infty)\to y_\textrm{st}$
with the latter given in Eq. \eqref{eq-Gamma_random.1}.
In summary, for the special initial condition $x_0=0$, the PDF $P_0(x,t|x_0=0)$ of the CIR process is given exactly, \textit{at all times} $t\ge 0$, by the PDF of a Gamma distributed random variable defined in Eq. \eqref{eq-yt_def.1}.
\begin{Remark}
	We note that the tridiagonal $\beta$-ensemble \eqref{eq-DE model} by Dumitriu and Edelman \cite{Dumitriu Edelman} incorporates also Gamma distributions in the off-diagonal matrix entries, as we recover the Chi-distribution for $l=1/2$ in \eqref{eq-Gamma_pdf.1}.
\end{Remark}
\begin{Remark}
	We commented already on earlier results on a matrix-valued process for the $\beta$-DBM. In the tridiagonal matrix process, introduced by Li in \cite{Li beta process} the off-diagonal elements evolve via a generalized Bessel process rather than a CIR process. The two processes show main similarities, yielding to the finding of the same eigenvalue JPDF we derive below in Eq. \eqref{eq-tmp_jpdf.1}. For a comparison of the processes, we refer to the studies in \cite{CIR Bessel 1,CIR Bessel 2,CIR Bessel 3}. While squared Bessel processes display stochastic instability, the CIR process is ergodic and stable.
\end{Remark}
\subsubsection{Definition of a Tridiagonal Matrix-Valued Process}
\label{sssec-TMP}

Having gathered the necessary background on the OU and the CIR process, we now proceed to construct a matrix-valued process \`a la Dyson. Again, we note that this $\beta$-DBM for the Gaussian-$\beta$-ensemble was studied in earlier works \cite{Li beta process,Holcomb Paquette,Yabuoko,Edelman Brownian}. However, the proof techniques show main differences. Li in \cite{Li beta process} derives a tridiagonal matrix-valued process via the study of the transition probability density for a class of stationary Markov processes which are the steady stationary state of the tridiagonal Dumitriu Edelman ensemble \eqref{eq-DE model}. Here, comes the generalized Bessel process as Markov process corresponding to the Chi-distribution into play. We comment on the used proof technique below. The references \cite{Holcomb Paquette,Yabuoko} deal instead with the processes of the matrix entries itself, study conditions for the existence of the corresponding flows of matrix entries as well as the differential equations for the eigenvalues. A time-dependent matrix representation incorporating $\sigma(t)$ as we give below in \eqref{eq-H-matrix}, is not given in \cite{Holcomb Paquette,Yabuoko}.

We consider a real, symmetric $N\times N$ tridiagonal matrix $H_{k,k'}(t)$ that evolves stochastically in a fictitious time $t$. Since the matrix is tridiagonal, we have $H_{k,k'}(t)=0$ for all $t$, for $k' \ne k-1, k, k+1$. In addition, since the matrix is symmetric, we have $H_{k,k+1}(t)=H_{k+1,k}(t)$ for
$k=1,2,\ldots, N-1$. All non-zero matrix entries (not related by symmetry) are assumed to be statistically independent of each other at all times $t$ and evolve by the following dynamics:

\begin{itemize}
	
	\item{\textbf{Diagonal elements $\mathbf{H_{k,k}(t)}$.}} The $k$th diagonal entry is assumed to evolve as an OU process, i.e., 
	\begin{align}
		\frac{d H_{k,k}(t)}{dt}= -\mu\, H_{k,k}(t) + \sqrt{\frac{2D}{\beta}}\, \eta_k(t)\, \text{, for }k=1,\ldots,N,
		\label{eq-TMP_diag.1}
	\end{align}
	where $\eta_k(t)$'s are zero-mean Gaussian white noises with the correlator $\langle \eta_k(t)\eta_{k'}(t')\rangle= \delta_{k,k'}\, 
	\delta(t-t')$. This process has three parameters $(\mu, \, D,\, \beta)$ which are arbitrary for the moment. We assume
	that the process, for each $k=1,2\ldots, N$, starts at the origin, i.e., $H_{k,k}(t=0)=0$. The probability density of the process
	$H_{k,k}(t)$ at time $t$ is given by the Gaussian distribution
	\begin{align}
		\textrm{Prob.}[H_{k,k}(t)=x]= \frac{1}{\sqrt{2\,\pi\,\sigma^2(t)}}\, e^{-\frac{x^2}{2\sigma^2(t)} }\, ,    
		\label{eq-OU_pdf.1}
	\end{align}
	where
	\begin{align} \label{eq-sigma_TMP}
		\sigma^2(t)=\frac{D}{\beta\, \mu}\, \left(1- e^{-2\,\mu\, t}\right)\, .
	\end{align}
	Thus, $H_{k,k}(t)= {\cal N}\left(0, \sigma^2(t)\right)$ is just a zero-mean Gaussian normal variable with variance $\sigma^2(t)$.
	For later convenience, let us express this after rescaling by a factor $1/\sqrt{2}$ as
	\begin{align}
		H_{k,k}(t)=\frac{1}{\sqrt{2}}\, {\cal N}\left(0, 2\sigma^2(t)\right)\, ,
		\label{eq-Hdiag.1}
	\end{align}
	with $\sigma(t)$ given in Eq. \eqref{eq-sigma_TMP}.
	
	\item{\textbf{Off-diagonal elements $\mathbf{H_{k,k+1}(t)}$.}} The element $H_{k,k+1}(t)$ with $k=1,2,\ldots, N-1$ equals $H_{k,k+1}(t)= \sqrt{Y_k(t)/2}$, where $Y_k(t)$ represents a CIR process in Eq. \eqref{eq-cir.1} starting at the origin $Y_k(t=0)=0$
	with parameters $(\mu_0,\, \theta_k,\, \sigma_0)$. From Eq. \eqref{eq-yt_def.1}, $Y_k(t)$ is a Gamma distributed random variable
	and hence we have\footnote{We choose a different labelling of the matrix elements $H_{j,k}$ here, compared to \cite{Dumitriu Edelman}.}
	\begin{align}
		H_{k,k+1}(t)=\sqrt{\frac{\Gamma\left(q=\frac{2\mu_0\, \theta_k}{\sigma_0^2},\,  l=
				\frac{\sigma_0^2 \left(1- e^{-\mu_0\, t}\right)}{2\mu_0}\right)}{2}}\, ,
		\label{eq-Hup.1}
	\end{align}
	where we are at liberty to choose the parameters $(\mu_0, \theta_k, \sigma_0)$. We will make the following choice for the parameters
	\begin{align}
		\mu_0=2\, \mu\, , \quad\quad \theta_k= \frac{D}{\mu}(N-k)\, , \quad\quad \textrm{and}\quad \sigma_0^2= \frac{8D}{\beta}\, ,
		\label{eq-para_choice}
	\end{align}
	where $(\mu,\, D, \, \beta)$ are the parameters of the OU process characterizing the diagonal entries. With this choice of parameters, we then have from Eq. \eqref{eq-Hup.1}
	\begin{align}
		H_{k,k+1}(t)= \sqrt{ \frac{\Gamma\left(\frac{\beta(N-k)}{2}\, ,\,  2\sigma^2(t)\right)}{2}}\, , 
		\label{eq-Hup.2}
	\end{align}
	with $\sigma(t)$ given in Eq. \eqref{eq-sigma_TMP}. The probability density of the off-diagonal process $\sqrt{2}H_{k,k+1}(t)$ at time $t$ is given by the (generalized) Gamma distribution
	\begin{align}
		\textrm{Prob.}\left[\sqrt{2}H_{k,k+1}(t)=x\right]= \frac{2^{1-\frac{\beta(N-k)}{2}}}{\Gamma\left(\frac{\beta(N-k)}{2}\right)\sigma^{\beta(N-k)}(t)}x^{\beta(N-k)-1}e^{-\frac{x^2}{2\sigma^2(t)}} , \text{ for } x>0.
		\label{eq-CIR_pdf.1}	
	\end{align}
\end{itemize}
This symmetric tridiagonal matrix $\frac{1}{\sqrt{2}}H(t)$ (with parameters $(\mu,\, D\, ,\beta)$) reads
{\small
\begin{align}
	\begin{pmatrix}
		{\cal N}(0,2\sigma^2(t)) & \sqrt{\Gamma\left(\frac{\beta(N-1)}{2},2\sigma^2(t)\right)}& & &\\
		\sqrt{\Gamma\left(\frac{\beta(N-1)}{2},2\sigma^2(t)\right)} & {\cal N}(0,2\sigma^2(t))& \sqrt{\Gamma\left(\frac{\beta(N-2)}{2},2\sigma^2(t)\right)}& & \\
		\ddots & \ddots &\ddots&\\
		&  \sqrt{\Gamma\left(\frac{2\beta}{2},2\sigma^2(t)\right)} &
		{\cal N}(0,2\sigma^2(t))&\sqrt{\Gamma\left(\frac{\beta}{2},2\sigma^2(t)\right)}\\	
		& & \sqrt{\Gamma\left(\frac{\beta}{2},2\sigma^2(t)\right)}& {\cal N}(0,2\sigma^2(t))
	\end{pmatrix}.\!
	\label{eq-H-matrix}
\end{align}}
The JPDF of the matrix elements of $H(t)$ is determined via Eqs. \eqref{eq-OU_pdf.1}, \eqref{eq-CIR_pdf.1}
\begin{align}
	P_0(H,t)=\frac{1}{c_N(\beta)}\frac{2^{-\beta N(N-1)/4}}{[
		\sigma(t)]^{N+ \beta N(N-1)/2}} \prod_{j=1}^N\,
	e^{-\frac{H_{j,j}^2}{2\sigma^2(t)}}\,
	\prod_{k=1}^{N-1}\, H_{k,k+1}^{\beta(N-k)-1}\, e^{-\frac{H_{k,k+1}^2}{\sigma^2(t)}} \, , 
	\label{eq-JPDF_H}	
\end{align}
where
\begin{align}
	c_N(\beta)=
	\frac{(2\pi)^{N/2}\prod_{k=1}^{N-1}\Gamma\left(\frac{\beta k}{2}\right)}{2^{N-1}}\, ,
	\label{eq-cnbeta_def}
\end{align}
and $\sigma(t)$ is given in Eq. \eqref{eq-sigma_TMP}. The symmetric tridiagonal matrix $H(t)$ is a generalization of the tridiagonal matrix model introduced by Dumitriu and Edelman \cite{Dumitriu Edelman}. The Chi-distribution of the off-diagonal model in the Dumitriu-Edelman matrix ensemble is a special case of the Gamma-distribution when the variance $\sigma^2(t)$ is time-independent and has unit value. In the present case the variance $\sigma^2(t)$ in Eq. \eqref{eq-JPDF_H}, however, depends explicitly on time $t$. The JPDF in terms of the real eigenvalues of $H(t)$ follows very similarly as in the time-independent model in \cite{Dumitriu Edelman}. 
The exponential parts in Eq. \eqref{eq-JPDF_H} can be rewritten as
\begin{align}
	\exp\left(-\sum_{k=1}^N\frac{H_{k,k}^2}{2\sigma^2(t)}-\sum_{k=1}^{N-1}\frac{H_{k,k+1}^2}{\sigma^2(t)}\right)=\exp \left(-\frac{1}{2\sigma^2(t)}\textrm{Tr}(H^2)\right).
\end{align}
It follows, that the JPDF of $H(t)$ is of the same functional form as in the time-independent model $H(0)$ (cf. Proof of Theorem 2.12 in \cite{Dumitriu Edelman}) up to the factor $1/\sigma^2(t)$ in the exponential 
and an overall normalization term incorporating $\sigma(t)$. To see this, 
we briefly review all ingredients for the proof. We impose a spectral decomposition of $H(t)$ as
\begin{align}
	H(t)=QX(t)Q^t,
\end{align}
where $X(t)=\textrm{diag}(x_1(t),x_2(t),\ldots,x_N(t))$ contains the real eigenvalues of $H(t)$ and $Q$ is a real orthogonal. We apply \cite[Lemma 
2.5]{Dumitriu Edelman} and find that the first row $\mathbf{q}$ of $Q$ and the eigenvalues uniquely determine $Q$ and $H(t)$. Next, we use \cite[Lemma 2.7]{Dumitriu Edelman} to find a rewriting of the Vandermonde determinant in terms of the off-diagonal elements $H_{k,k+1}(t)$ ($H_{k,k+1}(t)$ 
corresponds to $b_{N-k}$ in \cite{Dumitriu Edelman}) and the spectral parameters $\mathbf{q}$. The computation of the Jacobian in \cite[Lemma 2.9]{Dumitriu Edelman} for the change of variables from the matrix entries to the spectral parameters $(X(t),\mathbf{q})$ is unchanged, as the proof uses 
only the tridiagonal symmetric form of $H(t)$.

Hence, the JPDF of the eigenvalues $\mathbf{x}(t)$ at any instant $t$, and for any $\beta>0$, is given exactly by
\begin{align}
	P_0\left(\mathbf{x}, t\right)=
	\frac{1}{C_N(\beta)}\, 
	\frac{1}{\left[\sigma(t)\right]^{N+ \beta\, N(N-1)/2}}\, 
	e^{-\frac{1}{2\,\sigma^2(t)}\, \sum_{i=1}^N x_i^2} \, \prod_{i> j}^N
	|x_i-x_j|^{\beta} \, ,
	\label{eq-tmp_jpdf.1}
\end{align}
with $\sigma(t)$ from Eq. \eqref{eq-sigma_TMP}.
The normalizing constant $C_N(\beta)$ is given in Eq. \eqref{eq-jpdf 1d Dyson} and most crucially, it does not depend on $t$.

This result \eqref{eq-tmp_jpdf.1} coincides precisely with the JPDF of the $\beta$-DBM process in Eq. \eqref{eq-dbm_jpdf.2}. Hence, this construction shows that for arbitrary $\beta>0$, the $\beta$-DBM process corresponds to the eigenvalues of an underlying tridiagonal symmetric time-evolving matrix $H(t)$
whose diagonal entries perform independent OU processes starting at the origin and the off-diagonal entries perform independent CIR processes (with a rank dependent shape parameter $q$ in Eq. \eqref{eq-Hup.2}) starting at the origin. We will call this tridiagonal matrix-valued process as $\beta$-TMP.

We note that our derived tridiagonal matrix-valued process \eqref{eq-H-matrix} coincides with the process proposed earlier in \cite[Eqs. (3.1), (3.11)]{Li beta process}. However, we would like to point out two main differences: First, we incorporate the CIR process in the off-diagonal elements rather than the stochastic unstable generalized Bessel process. Second, our proof strategy shows main differences. We follow the idea by Dumitriu and Edelman in \cite{Dumitriu Edelman} and ask, what happens to a full Hermitian matrix with matrix elements having a time-dependent standard deviation $\sigma(t)\neq 1$. Following the arguments of the Householder reduction, we find that the off-diagonal elements in the tridiagonalised matrix are distributed as the square root of Gamma variables \eqref{eq-yt_def.1}. Next, we consider $\sigma(t)$ \eqref{eq-sigma_TMP} following a DBM, yielding to the matrix processes in the diagonal, respectively off-diagonal elements. The studies of the $\beta$-DBM for the Gaussian-$\beta$-ensemble in \cite{Li beta process,Holcomb Paquette,Yabuoko} view the problem from a different perspective. The shared goal of finding a time-dependent solution, with stationary solution represented by the DBM associated to the Dumitriu-Edelman ensemble, is reached via the study of the Markovian transition processes or the time-dependent flows there. We add the subject of resetting to the analysis of this tridiagonal matrix-valued process below.
\subsubsection{Tridiagonal Matrix-Valued Resetting Processes}
\label{sssec-RTMP}

\begin{figure}[t]
	\hspace{50pt}{$\beta=1$} \hspace{120pt}{$\beta=2.6$}\hspace{120pt}{$\beta=3.1$} \hfill\\
	\begin{turn}{90}\hspace{30pt}{$\mu=1$}\end{turn}
	\includegraphics[width=0.3\linewidth]{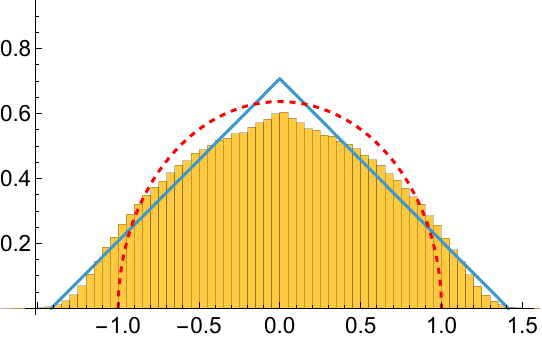} 
	\includegraphics[width=0.3\linewidth]{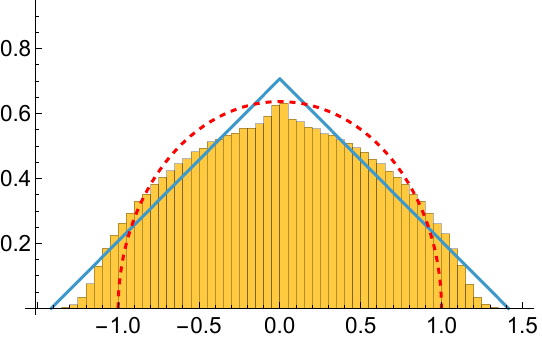}
	\includegraphics[width=0.3\linewidth]{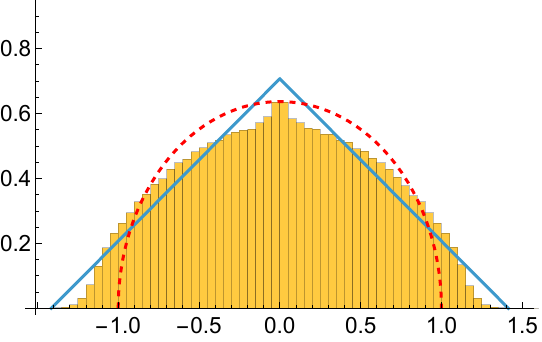}  \\
	\begin{turn}{90}\hspace{20pt}{$\mu=1/2$}\end{turn}
	\includegraphics[width=0.3\linewidth]{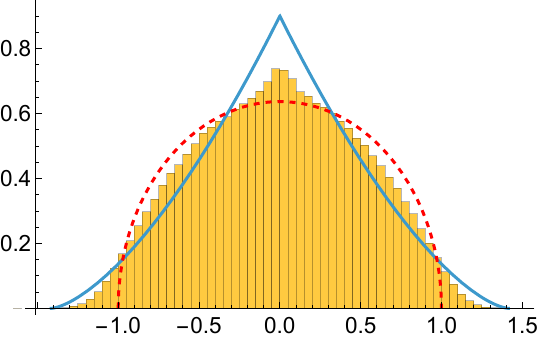} 
	\includegraphics[width=0.3\linewidth]{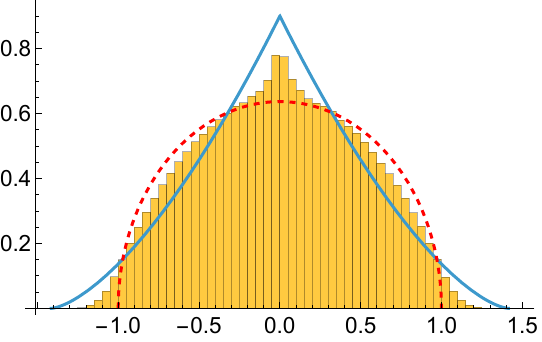}
	\includegraphics[width=0.3\linewidth]{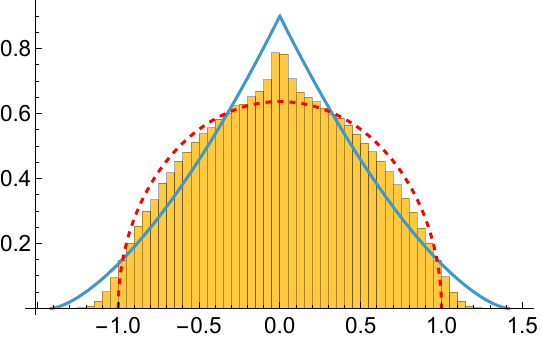}\\  
	\caption[Figure 1 from \cite{AMP26}, \textit{Study of the eigenvalue density for the stationary ensemble of a tridiagonal matrix valued process with resetting and parameter-dependence on $\beta$}]{We generated numerically the stationary ensemble of the $\beta$-IRTMP process in Eq. \eqref{eq-prob.Htridiag} by drawing independently the diagonal entries from Eq. \eqref{eq-diag_pdf.1} and off-diagonal entries from Eq. \eqref{eq-off_diag_pdf.1} using the inverse CDF method described in the text. We then diagonalized this $N\times N$ random matrix to obtain the histogram of its $N$ eigenvalues. We plot the rescaled average eigenvalue density (rescaled by a factor $\sqrt{N}$, respectively $\sqrt{2N}$) for $N=20$ (the histogram averaged over $50000$ samples). We chose parameters values $\beta=1,\, 2.6,\, 3.1$ (from left to right) and $\mu=1,\, 1/2$ (top, bottom row) respectively. We have set $r=D=1$ everywhere. We compare this eigenvalue density of this $\beta$-IRTMP
	process with the known analytical density for the $\beta$-SRTMP process
	(shown by blue curves, cf. Eqs. \eqref{eq-main-density}, \eqref{eq-main-density-special}), where we rescaled the density by $1/\sqrt{N}$ in the top row and by $1/\sqrt{2N}$ in the bottom row. For comparison we give also the semi-circle \eqref{eq-semicircle} (red dashed curve), which is the limiting density without resetting. \cite[Figure 1]{AMP26}}
	\label{fig-n20 nmtr50000}
\end{figure}

Now, we subject the $\beta$-TMP process to Poissonian resetting with rate $r$. While the process was engaged in the literature before, it was until today not studied under resetting. The resetting move can be introduced in two alternative ways: (i) $\beta$-SRTMP process where all the matrix entries \textit{simultaneously} reset to the origin with rate $r$ and (ii) $\beta$-IRTMP process where the matrix entries reset \textit{independently} with rate $r$ to the origin. Thus, in $\beta$-IRTMP process, the matrix entries (not connected by symmetry) remain independent at all times, while in $\beta$-SRTMP process the matrix entries get 
correlated with time due to the shared history of simultaneous resetting. We consider the two processes separately below.
\begin{enumerate}
	\item \textbf{Tridiagonal matrix with simultaneous resetting of entries ($\beta$-SRTMP).}
	Our starting point is the exact time-dependent JPDF of the matrix entries $P_0(H(t))$ in Eq. \eqref{eq-JPDF_H}. If now the matrix entries are simultaneously reset to the origin with rate $r$, the JPDF of the resetting matrix ensemble at time $t$ is given by the general renewal equation \eqref{eq-renewal_gen.1} with $\mathcal{S}=H$. In particular, as $t\to \infty$, the $\beta$-SRTMP process converges to a stationary JPDF \eqref{eq-matrix_reset_st.1}, which reads explicitly
	\begin{equation}
		P_r^\textrm{st}(H)= \frac{r\, 2^{-\beta N(N-1)/4}}{c_N(\beta)}\, \int_0^{\infty}\! \frac{d\tau\, 
			e^{-r\tau}}{\left[\sigma(\tau)\right]^{N+\beta N(N-1)/2}}\,\prod_{j=1}^N e^{-\frac{H_{j,j}^2}{2\sigma^2(\tau)}} \prod_{k=1}^{N-1} H_{k,k+1}^{\beta(N-k)-1}e^{-\frac{H_{k,k+1}^2}{\sigma^2(\tau)}}\, ,
		\label{eq-SRTMP_stat.2}
	\end{equation}
	where $c_N(\beta)$ and $\sigma^2(t)$ are defined in \eqref{eq-cnbeta_def}, \eqref{eq-sigma_TMP} respectively. Given this distribution of the matrix ensemble, one can then ask how are its eigenvalues distributed? When the matrix entries are simultaneously reset with rate $r$, it follows that its eigenvalues also undergo simultaneous resetting with rate $r$. Now, at any given instant $t$, the JPDF of the eigenvalues of the $\beta$-TMP process is given by Eq. \eqref{eq-tmp_jpdf.1}. We again apply the general renewal property \eqref{eq-stat_gen.1} in the limit $t\to \infty$, but now to the JPDF of the eigenvalues rather than the matrix itself, i.e. for $\mathcal{S}=\mathbf{x}$. This gives the JPDF of the eigenvalues $\mathbf{x}\equiv 
	\{x_1,x_2,\ldots, x_N\}$ of the $\beta$-SRTMP stationary ensemble in Eq. \eqref{eq-SRTMP_stat.2} as
	\begin{align}
		P_r^\textrm{st}(\mathbf{x})= \frac{r}{C_N(\beta)}\,\int_0^{\infty}\! \frac{d\tau\, e^{-r\, \tau}}{
			\left[\sigma(\tau)\right]^{N+ \beta N (N-1)/2}}\,
		e^{-\frac{1}{2\, \sigma^2(\tau) }\sum_i x_i^2} \, \prod_{i> j}^N
		|x_i-x_j|^{\beta}\,
		\label{eq-SRTMP_eigen_st.0}
	\end{align}
	where $C_N(\beta)$ is given in Eq. \eqref{eq-jpdf 1d Dyson} and $\sigma(t)$ in \eqref{eq-sigma_TMP}. Note that Eq. \eqref{eq-SRTMP_eigen_st.0} coincides with the JPDF of the $\beta$-RDBM process in Eq. \eqref{eq-RDBM_st.0}. This then answers one of our motivations raised above that there is indeed an underlying matrix-valued process, here the $\beta$-SRTMP process, such that its eigenvalues are distributed according to \eqref{eq-SRTMP_eigen_st.0} 
	for arbitrary $\beta>0$.
	
	The JPDF of the eigenvalues in Eq. \eqref{eq-SRTMP_eigen_st.0} for arbitrary $\beta>0$ was studied in detail in reference \cite{BMS2025} and several interesting exact results in the limit of large $N$ were derived. For instance, the average density of eigenvalues in the limit of large $N$ reads \cite{BMS2025}
	\begin{align}
		\label{eq-main-density}
		\rho_N(x | r) \simeq \sqrt{\frac{\mu}{N D}} \;
		g\left( x \sqrt{\frac{\mu}{N D}} , \, \frac{\mu}{r} \right) \;,
	\end{align}
	where the scaling function $g(z, \nu= \mu/r)$ is symmetric around $z=0$ and
	is independent of $\beta$. It has a finite support over $z \in [-\sqrt{2}, \sqrt{2}]$ and is normalized to unity, i.e. $\int_{-\sqrt{2}}^{\sqrt{2}}\! dz \, g(z, \nu) = 1$. Its explicit expression reads \cite{BMS2025}
	\begin{align} \label{eq-main-f}
		g(z, \nu) = \frac{\Gamma\left( 1 + \frac{1}{2 \nu} \right)}{\sqrt{2 \pi}
			\Gamma\left( \frac{3}{2}+\frac{1}{2 \nu}\right)}
		\left( 1 - \frac{z^2}{2} \right)^{\frac{1 + \nu}{2 \nu}}\, \,
		_2\,{F}_1 \left[1, \frac{1}{2 \nu}, \frac{3}{2} +
		\frac{1}{2\nu}, 1 -\frac{z^2}{2} \right] \;,
	\end{align}
	where $\,_2\,{F}_1[a, b, c, z]$ is the hypergeometric function.
	For some special values of $\nu=\mu/r$ the hypergeometric function simplifies, for example for $\nu = 1$ and $\nu = 1/2$, where one obtains \cite{BMS2025}
	\begin{align}
		\label{eq-main-density-special}
		g(z, \nu= 1) = \frac{1}{\sqrt{2}} - \frac{|z|}{2} \mbox{~~and~~}
		g(z, \nu = 1/2) = \frac{2}{\pi} \left( \sqrt{2 - z^2} - |z| \,
		\textrm{ArcCos} \left[ \frac{|z|}{\sqrt{2}} \right] \right) \; ,
	\end{align}
	see blue full lines in Figure \ref{fig-n20 nmtr50000} top respectively bottom row. In the $\beta$-SRTMP process, the matrix elements are highly correlated as seen in Eq. \eqref{eq-SRTMP_stat.2}, as the $\tau$-dependent variance $\sigma^2(\tau)$, which is common to all independent matrix elements, is integrated over, creating correlations. Hence it is not easy to generate this matrix numerically. However, the JPDF of the eigenvalues of this ensemble can be explicitly expressed as in Eq. \eqref{eq-SRTMP_eigen_st.0} which allows
	for analytical computations.  
	
	\item \textbf{Tridiagonal matrix with independent resetting of entries ($\beta$-IRTMP).} We now consider the $\beta$-IRTMP matrix process where each entry of the tridiagonal $\beta$-TMP process resets independently to $0$ with rate $r$. Thus, in this case, the matrix elements remain independent at all times, including in the stationary state. We now apply the general renewal equation \eqref{eq-renewal_gen.1} in the limit $t\to \infty$ for the evolution of each matrix entry separately, i.e. $\mathcal{S}=H_{i,j}$ for $j=i-1,i,i+1$ and $i=2,\ldots, N-1$. 
	It follows that in the $t\to \infty$ limit, the elements of the tridiagonal matrix $H(t)$ approach a stationary state, where the diagonal and the off-diagonal entries are distributed as follows.
	
	\begin{itemize}
		\item{\textbf{Diagonal elements.}} From Eqs. \eqref{eq-OU_pdf.1} and \eqref{eq-stat_gen.1}, it follows that the diagonal element $H_{k,k}(t)$, in the limit $t\to \infty$ and in the presence
		of $r>0$, is distributed via the stationary PDF
		\begin{align}
			\lim_{t\to \infty} \textrm{Prob.}[H_{k,k}(t)=x]= r\, \int_0^{\infty}\! d\tau\, e^{-r\, \tau}\, \frac{1}{\sqrt{2\,\pi\,\sigma^2(\tau)}}\, 
			e^{-\frac{x^2}{2\sigma^2(\tau)} }\, ,
			\label{eq-diag_pdf.1}
		\end{align}
		with $\sigma(t)$ from Eq. \eqref{eq-sigma_TMP}. The entries for different $k$ are independent of each other. 
		
		\item{\textbf{Off-diagonal elements.}} The matrix remains symmetric and tridiagonal. From Eqs. \eqref{eq-Hup.1} and \eqref{eq-stat_gen.1} it follows that the element $H_{k,k+1}(t)$, in the limit $t\to \infty$ and $r>0$, can be expressed as
		\begin{align}
			H_{k,k+1}(t\to \infty)= \sqrt{\frac{Y_k(r)}{2}}\, ,
			\label{eq-off_diag_element.1}
		\end{align}
		where the random variable $Y_k(r)$ is Gamma distributed via the stationary PDF
		\begin{align}
			\textrm{Prob.}\left[Y_k(r)=y\right] = r\, \int_0^{\infty}\! d\tau\, e^{-r\, \tau}\, 
			\frac{y^{q-1}}{\Gamma(q)\, \left[2\, \sigma^2(\tau)\right]^q}\, e^{-y/2\sigma^2(\tau)} ,
			\label{eq-off_diag_pdf.1}
		\end{align}
		with $q= \frac{\beta}{2}\,(N-k)$ and $\sigma(t)$ given in Eq. \eqref{eq-sigma_TMP}.
	\end{itemize}
	
	The JPDF of matrix entries of this $\beta$-IRTMP in the stationary state is then given by the product of all 
	independent entries of matrix $H(t)$ in the limit $t\to \infty$, 
	\begin{align}
		\textrm{Prob.}\left[ H, t\to \infty\right] = \lim_{t\to \infty}\, \left[ \prod_{j=1}^N \textrm{Prob.}\left[ 
		H_{j,j}(t)\right]\, 
		\prod_{k=1}^{N-1} \textrm{Prob.}\left[ H_{k,k+1}(t)\right] \right].
		\label{eq-prob.Htridiag}
	\end{align}
	This can be compared for example with the distribution \eqref{eq-matrix_OU.1b1} of independent elements for $\beta=1$ in the $t\to \infty$ limit. Notice, however, that here the JPDF in \eqref{eq-prob.Htridiag} \textit{cannot} be written in an invariant way, in terms of the trace of $H(t\to \infty)$ squared. Compared to the stationary JPDF with simultaneous resetting in \eqref{eq-SRTMP_eigen_st.0}, where we integrate over the variable $\tau$ of the product of distributions of independent matrix elements with $\tau$-dependent variance $\sigma^2(\tau)$, we have in \eqref{eq-prob.Htridiag} the product of integrals in the case of the independent resetting. In this sense, this $\beta$-IRTMP stationary ensemble is like the Wigner matrices, rather than rotationally invariant matrices in random matrix theory \cite{book forrester}. In fact, given the stationary JPDF of the matrix entries of $\beta$-IRTMP ensemble in Eq. \eqref{eq-prob.Htridiag}, it is not easy to compute the JPDF of its eigenvalues, unlike in the $\beta$-SRTMP process in Eq. \eqref{eq-SRTMP_eigen_st.0}. In contrast, it is much easier to generate this matrix numerically since its entries are independent.
\end{enumerate}

\noindent
We compute numerically the average density of eigenvalues in the
$\beta$-IRTMP process and compare it with the exact analytical result in Eqs. \eqref{eq-main-density} and \eqref{eq-main-f} for the $\beta$-SRTMP process. We find that they are quite different as expected.\\
We study the stationary ensemble, defined via Eqs. \eqref{eq-diag_pdf.1} and \eqref{eq-off_diag_pdf.1} numerically. For the off-diagonal elements $H_{k,k+1}(t)$ the probability distributions follow from Eqs. 
\eqref{eq-off_diag_element.1}, \eqref{eq-off_diag_pdf.1} with $\textrm{Prob.}\left[H_{k,k+1}=y\right] 
=\textrm{Prob.}\left[Y_k(r)=2y^2\right] 4|y|$. We use the so called \textit{inverse CDF method} to generate the random matrix $H$, cf. \cite{Simulation_book}. Let $\mathcal{X}$ be a random variable distributed with respect to some density $f(z)$. We calculate its cumulative distribution function (CDF) $F(t)=\int_{-\infty}^t \! dz\  f(z)$ for several values $t$ and list this as pairs $\{F(t),t\}$, which gives data points of the inverse CDF of $\mathcal{X}$. Next, we interpolate this function linearly and evaluate it at a uniform distributed random variable in $\left[0,1\right]$. Here it is important to note that the inverse CDF can diverge around 0 or 1. Thus, a random variable in $\left[0.0001,1-0.0001\right]$ is chosen, where we checked that this cut-off is a good choice. 

In the computation of the CDF two integrals arise in each matrix variable, i.e. the $\tau$-integration for $\tau\in[0,\infty)$ and the $x$-integration with $x\in (-\infty,t)$ with $t\in\mathbb{R}$ for the diagonal elements, respectively the $y$-integration with $y\in[0,t)$ with $t\in \mathbb{R}_+$ for the off-diagonal elements, cf. Eqs. \eqref{eq-diag_pdf.1} and \eqref{eq-off_diag_pdf.1}. In order to evaluate these integrals, we have to consider cut-offs for the integration domains, where the result is not significantly changed by the respective choice. We consider 
$\tau\in[0,10]$ and $t\in [-2.5,2.5]$ for the diagonal elements, respectively $t\in[0,6]$ for the off-diagonal elements. We observe that especially the integrands for the off-diagonal elements become highly oscillatory for 
large $N$, as $q=\frac{\beta}{2}(N-k)$ will be large. We set the parameters $D=r=1$ and vary $\beta=1,\, 2.6,\, 3.1$ as well as $\mu=1,\, 1/2$.

It is important to note that in this described procedure for the simulation of the tridiagonal matrix $H(t\rightarrow \infty)=H$, the resetting is at independent times for each matrix elements rather than simultaneous. Hence this average density in the $\beta$-IRTMP ensemble is expected to be different from the exact result in Eq. \eqref{eq-main-density} for the $\beta$-SRTMP process. For comparison, we provide the analytical results on the limiting density of eigenvalues in the $\beta$-SRTMP ensemble in Figure \ref{fig-n20 nmtr50000} as the blue curves. We plot here the rescaled density in the $\beta$-SRTMP ensemble in Eq. \eqref{eq-main-density-special} for $\mu=1$ and $\mu=1/2$ (with $r=D=1$). Additionally, we plot the semi-circle as red dashed curve in Figure \ref{fig-n20 nmtr50000}, which is the limiting density without resetting, i.e.
\begin{align}
	\rho^{\text{sc}}(x)=\frac{2}{\pi}\sqrt{1-x^2}, \text{ with } x\in[-1,1].\label{eq-semicircle}
\end{align}
We observe in Figure \ref{fig-n20 nmtr50000} that the stationary states of the models of independent (histograms), respectively simultaneous resetting (blue curves) are very different. While the agreement of the histogram data for independent resetting with the blue analytic curve for simultaneous resetting is relatively good in the tail for small $\beta$, it becomes worse for increasing $\beta$ (from left to right). At the origin, we observe the opposite behaviour, i.e. the average density becomes more peaked for large $\beta$. Computing analytically the average density in this $\beta$-IRTMP ensemble remains an interesting open problem.

We conclude that using the insight into the Dumitriu-Edelman matrix ensemble, we present an independent introduction of a symmetric tridiagonal matrix-valued process ($\beta$-TMP) $H(t)$ whose diagonal entries $H_{k,k}(t)$ evolve independently via an Ornstein-Uhlenbeck process starting at the origin and the off-diagonal entries $H_{k,k+1}(t)$ evolve independently via the Cox-Ingersoll-Ross (CIR) process, starting at the origin and with parameters that depend on the row index $k$. We find agreement with earlier results without resetting, cf. \cite{Li beta process,Holcomb Paquette,Yabuoko,Edelman Brownian}. The joint distribution of the entries of the matrix can be computed exactly at all times and moreover, the joint distribution of its $N$ real eigenvalues can 
also be computed exactly at all times and is given by the $\beta$-ensemble of Dumitriu and Edelman up to a simple time-dependent rescaling factor with arbitrary Dyson index $\beta>0$. We then add the following new study. We subject this time-evolving matrix valued process to stochastic resetting with rate $r$ in two different settings: (i) when the matrix entries are simultaneously reset to the origin with rate $r$ ($\beta$-SRTMP process) and (ii) when the matrix entries are independently reset to the origin with rate r ($\beta$-IRTMP process). These studies, using the $\beta$-TMP were not present in the literature. In the $\beta$-SRTMP process, the matrix elements get strongly correlated in the stationary state, while in the $\beta$-IRTMP process the matrix elements remain independent at all times, including in the stationary state. While such an underlying resetting matrix-valued process was known previously \cite{BMS2025} only for three special values of the Dyson index $\beta=1,2$ and $4$, our construction of the $\beta$-SRTMP process here generalises the resetting matrix-valued process to 
arbitrary $\beta>0$.
\cleardoublepage
\section{Non-Hermitian $\boldsymbol{\beta}$-Ensembles}\label{sec-non Herm beta}
Having the impressive results from the previous Section \ref{sec-Hermitian beta ensembles} in mind, we start here with the statement of our goal:\\
We are looking for a non-Hermitian matrix ensemble, whose complex eigenvalues $\{z_i\}_{i=1}^N$ show pairwise-logarithmic interaction at inverse temperature $\beta>0$, i.e. have a JPDF
\begin{align}\label{eq-jpdf 2dCG}
	P(z_1,\ldots,z_N)\propto \prod_{1\leq i < j\leq N}|z_i-z_j|^\beta \prod_{k=1}^Ne^{-V(z_k)},
\end{align}
with the potential $V(z)$. We will focus on the rotationally symmetric potential $V(z)=|z|^2$ in the following.\\
We recall from the introduction Section \ref{sec-intro}, that we saw already one non-Hermitian matrix ensemble which can be viewed as a Dyson gas in 2d: The complex Ginibre ensemble. We repeat its definition for convenience: Let $J\neq J^\dagger$ be a complex $N\times N$ matrix with elements distributed according to $\propto \exp\left[-\Tr(JJ^\dagger)\right]$, i.e. $(J)_{ij}\sim \mathcal{N}_\mathbb{C}(0,1)$ $\forall i,j=1,\ldots, N$ identically and independently sampled from the complex normal distribution. The latter is defined via the following probability density, with mean $\mu\in \mathbb{C}$ and variance $\sigma^2\in \mathbb{R}_+$, cf. \cite{tao vu}
\begin{align}
	f(z)=\frac{1}{\pi \sigma^2}e^{-\frac{|z-\mu|^2}{\sigma^2}},\text{ for } z\in \mathbb{C}.
\end{align}
Comparing with the definition of the real normal distribution in Eq. \eqref{eq-def normal dist}, we find that a complex variable $Z\sim\mathcal{N}_\mathbb{C}(0,1)$ can be sampled from two independent normal variables as $Z\sim\mathcal{N}(0,1/2)+i\mathcal{N}(0,1/2)$. One natural question is, what $J$ looks like, if we apply a Householder-transformation as in Section \ref{ssec-Householder}. Therefore, let us briefly recall the main points on this transformation: One constructs a unitary and Hermitian (cf. Definition \ref{def-housholder}, Lemma \ref{lem-Householder}) projector such that a column vector is mapped to the first unit vector times the norm of the initial vector. The definition, we state in Section \ref{ssec-Householder} is very general and holds also for complex vectors/ matrices. However, we saw that if the matrix, we want to transform has inner symmetry, e.g. is Hermitian, we find a similarity transformation of unitary matrices onto a tridiagonal matrix. Removing the symmetry of the initial matrix, we are no longer able to reduce transpose column and row vectors with the same transformation. Hence, we can reduce only "one side" of the matrix, without affecting the eigenvalues (cf. Remark \ref{remark-Housholder nonH} below). It is well known that the reduction of a complex Ginibre matrix yields the following form (cf. \cite[Proposition 51]{tao vu}, \cite[Section 2]{krishnapur virag})
\begin{align}\label{eq-Hessenberg GinUE}
	J_\beta \sim \begin{pmatrix}
		\mathcal{N}_\mathbb{C}[0,1]& ...&...&\mathcal{N}_\mathbb{C}[0,1]\\
		\frac{1}{\sqrt{\beta}}\chi_{\beta(N-1)}&\mathcal{N}_\mathbb{C}[0,1]&...&\mathcal{N}_\mathbb{C}[0,1]\\
		...&...&...&...\\
		0&...&\frac{1}{\sqrt{\beta}}\chi_{\beta}& \mathcal{N}_\mathbb{C}[0,1]\\
	\end{pmatrix}
\end{align}
with $\beta=2$ and $\chi_k$ denoting the Chi-distribution defined in Eq. \eqref{eq-def chi dist}. A matrix $J$ with elements non-zero only above the sub-diagonal, i.e. $(J)_{i,j}\neq 0$ for $i>j+1$ is named a Hessenberg matrix. Again, it is important to note that only for $\beta=2$, we find a correspondence to a 2d Dyson gas. Moreover, it is (until today) not possible to perform a direct computation of the Jacobian from the matrix $J_\beta$ in Eq. \eqref{eq-Hessenberg GinUE} to its eigenvalues. One attempt, using the recursive structure of the determinant of a Hessenberg matrix is studied in \cite{krishnapur virag} via asymptotic analysis of Pólya type recursion. We will comment on the difficulties along the line in the proof of the Jacobian for the tridiagonal matrix model defined in the later Section \ref{ssec-tridiag beta}.
\begin{Remark}\label{remark-Housholder nonH}
	It is important to note that one can reduce a generally non-Hermitian matrix $J$ to a tridiagonal form using Householder transformations. However, this requires different unitary transformations from the left and the right, i.e. $J\rightarrow U_1JU_2$ is tridiagonal with $U_1,U_2$ unitary matrices. Let $J=RZR^{-1}$ be the spectral decomposition of $J$, with $Z=\diag(z_1,\ldots,z_N)$ containing the complex eigenvalues. We observe that the transformation $J\rightarrow U_1JU_2$ will affect the eigenvalues, whereas the eigenvalues of $JJ^\dagger$ (also known as the squared singular values of $J$) are unchanged. One main advantage of the reduction to tridiagonal form is the three-term recurrence relation of the resulting characteristic polynomial, which enabled Dumitriu and Edelman to perform a direct proof of the Jacobian to the eigenvalue-decomposition in \cite{Dumitriu Edelman}. In the Hermitian setting, the same eigenvalue distribution is captured. There exists a $\beta$-ensemble, where the singular values, i.e. the eigenvalues of $LL^t$ (in the real setting) are captured: the Wishart-$\beta$-ensemble, see the overview in Appendix \ref{app-overview Hermitian beta ensembles}.
	\end{Remark}
Shortly, after the introduction of the $\beta$-ensembles by Dumitriu and Edelman in 2002, Killip and Nenciu studied the Circular-$\beta$-ensemble in \cite{Killip Nenciu}. Let us give an overview on the main features below.
\begin{itemize}
	\item Also known as the Jacobi-$\beta$-ensemble, see also Appendix \ref{app-overview Hermitian beta ensembles}
	\item Consists of unitary (non-Hermitian) $N\times N$ matrices. The eigenvalues lie on the unit circle in the complex plane and can be parametrized with eigenphases as $\{e^{i\vartheta_k}\}_{k=1}^N$. The JPDF of the eigenvalues reads for $\beta>0$
	\begin{align*}
		P(e^{i\vartheta_1},\ldots,e^{i\vartheta_N})&=\frac{1}{Z_{\beta}^{\textrm{circ}}}\prod_{1\leq k < j\leq N}|e^{i\vartheta_k}-e^{i\vartheta_j}|^\beta\\
		Z_{\beta}^{\textrm{circ}}&=(2\pi)^{N} \frac{\Gamma\left(\frac{\beta}{2}N+1\right)}{\Gamma\left(\frac{\beta}{2}+1\right)^N}.
	\end{align*}
	\item Parametrization via a unitary Hessenberg matrix, i.e. $H_{i,j}=0$ for $i>j+1$ due to \cite{Killip Nenciu}\footnote{We present the parametrization of the Hessenberg matrix element-wise (cf. \cite[Chapter 2.8]{book forrester}), which results from the block-wise parametrization introduced in \cite{Killip Nenciu}.}
	\begin{align*}
		H_{i,j}=\begin{cases}
			-\alpha_{i-1}\bar{\alpha}_i & \text{ for } i=j\\
			\rho_{i-1}& \text{ for } j=i-1\\
			-\alpha_{i-1}\bar{\alpha}_j\prod_{\ell=i}^{j-1}\rho_\ell & \text{ for } i<j\\
			0 & \text{ otherwise}
		\end{cases},
	\end{align*}
	where $\{\alpha_{j-1}\}_{j=0}^{N-1}$ are complex numbers with $|\alpha_j|<1$, $\alpha_{-1}:=-1$ and $\rho_j=\sqrt{1-|\alpha_j|^2}$. They are distributed as
	\begin{align*}
		\alpha_k\sim \Theta_{\beta(N-k-1)+1}.
	\end{align*}
	The probability distribution $\Theta_\nu$ of a complex random variable has the density
	\begin{align*}
		f_\nu(z)=\frac{\nu-1}{2\pi}(1-|z|^2)^{(\nu-3)/2}, \text{ for } z\in \mathbb{C},\ |z|=1.
	\end{align*}
	\item Proof via direct computation of Jacobian in \cite{Killip Nenciu,Forrester Rains unitary}
\end{itemize}
It is remarkable that this model consisting of Hessenberg matrices can be studied via direct computation of the Jacobian. One major ingredient to the proof is that although the eigenvalues are no longer real, they can be fully determined via real phases in $[0,2\pi)$, which have again interlacing properties. Hence, the problem can be mapped to a one-parameter space and similar techniques as in the Hermitian-$\beta$-models apply.\\
Before, we start to study different non-Hermitian $\beta$-ensembles in the subsequent subsections, we discuss how to identify/ characterize a certain quantum system via its local statistics. Let us recall from the introduction Section \ref{sec-intro}, where the origin of the first application of RMT lies, i.e. in the analysis of the distribution of level spacing of nuclei \cite{Wigner57}. It was a well-known fact in the 1950's, that nuclei as well as eigenvalues of Hermitian matrices show repulsion. This was for instance already 1929 investigated by von Neumann and Wigner \cite{von Neumann Wigner}, see also \cite{Wigner collection} for a comprehensive collection of the works by Wigner and collaborators. Hence, it is a natural question how to quantify this repulsion. In \cite{Wigner57} Wigner studied a real symmetric $2\times 2$ matrix, computed schematically its real eigenvalues $\lambda_1,\lambda_2$ and derived the distribution of the spacing $s=|\lambda_1-\lambda_2|$ between the eigenvalues. He surmised that this distribution captures the spacing distribution between nearest eigenvalues also for large systems $N\gg 2$.\footnote{We do not deprive the reader a nice anecdote on the evolution of Wigner's surmise: As mentioned by Wigner \cite{Wigner57}, he came along with it at a conference on Neutron Physics by Time-of-Flight, held at the Oak Ridge National Laboratory in 1956, answering a question about the possible
shape of the distribution of the spacings of energy levels in a heavy nucleus at the blackboard, cf. \cite{introduction rmt}.} While this might seem like an optimistic wish at a first glance, the experiments gave it a truth: The $N=2$ spacing between real eigenvalues describes the distribution of the NN eigenvalues in a large $N\gg 2$ ensemble accuratly but not exactly, cf. \cite{haake}. While this result is quite impressive and extensively used in the literature, it is not the only reason why to investigate the local statistics of RMT. As discussed in the introduction Section \ref{sec-intro}, the three Hermitian random matrix ensembles GO/U/SE show invariance with respect to orthogonal/ unitary/ symplectic transformations. This can also be viewed as the invariance under symmetry operations. In that sense, the random matrix models serve as representatives of Hermitian symmetry classes. The classification was extended to 10 classes by Altland and Zirnbauer in 1997 \cite{Altland Zirnbauer}. Starting in the late 1970's, the level spacings of quantum systems were conjectured to follow those of random matrix ensembles: Berry and Tabor \cite{Berry Tabor} conjectured this for closed integrable quantum systems, which correspond to the statistics of independent identical particles described by the Poisson process in 1d. The Bohigas-Giannoni-Schmit-conjecture (BGS) \cite{BGS,BGS 1}, cf. \cite{CGV}, extends this to closed fully chaotic quantum systems, where the NN spacing distribution turns out to be a good measure to distinguish symmetry classes of the Hamiltonians corresponding with different RMT ensembles. We will not detail the analysis of symmetry classes here and come back to this topic in Section \ref{sec-universality symmetry classes}. However, it turns out that among the 10 Hermitian symmetry classes only three different NN distributions appear in the bulk of the spectrum: the simplest representatives of these universality classes of local statistics are exactly the GO/U/SE. Let us record here, that BT and BGS conjectures turn out to hold in many settings, see \cite{review BGS BT} for a review on the applications. Turning our view back to non-Hermitian RMT, we find applications to open quantum systems, which are not perfectly isolated, i.e. they interact with its environment, under the assumption that the time scale of the latter this much shorter than the systems one and the Born approximation (the loss of the energy is much smaller than the total energy of the system). Some examples \cite{Hatano Nelson, GHS} were mentioned in the introduction Section \ref{sec-intro}, more recent applications can be found for instance in \cite{MMCN 19,NAJM 19,RPCB 22}. We will discuss the model of a spin chain with dissipative disorder in the later Section \ref{ssec-application spin chain}. Grobe, Haake and Sommers (GHS) \cite{GHS} conjectured that also in this non-Hermitian setting of open quantum systems, the spectral properties such as the NN spacing, can be used to characterize chaotic or integrable structures. Numerical evidence has been found in many applications, e.g. \cite{non-h classes,AKMP,complex sym pandey,Rubio22} and references therein. There exist other measures for integrability, which became of increasing interest over the past years. One is the complex spacing ratio, which we will review in Section \ref{sssec-complex spacing ratio} below. While its use is well established for Hermitian matrices \cite{OH,Atas_et_al}, it has been recently extended to the non-Hermitian setting \cite{SaRibeiroProsen}. Another measure is the so called Krylov complexity, where we refer the reader to the review \cite{krylov complexity} and references therein.

A first natural guess could be that also here the three classical Ginibre ensembles of real, complex, respectively quaternion matrix elements, serve as possible representatives of local universality classes. However, this is not true, and we will again come back to this subject in Section \ref{sec-universality symmetry classes} in detail. Another first attempt would be to ask whether there is a simple formula (having the $2\times 2$-Wigner surmise in mind) describing the universality classes in 2d?\\
Before we turn our attention to this question, we have to lay the foundations for the analysis. Therefore, we state in Section \ref{ssec-point processes} results on some point processes, which will serve as main references in the following studies: the Poisson point process in $D$ dimensions and the complex Ginibre ensemble. Here the NN and NNN distributions are known analytically.\\
Thereafter, we describe the subject of unfolding a spectrum and comparing numerics possibly with analytics or within each other in Section \ref{ssec-numerics methods}, following the discussion in \cite{ABCPRS} which was contributed by the author. We would like to point the reader's attention to this section, as these settings allow us to draw conclusions in the subsequent applications.\\
On the way to find a non-Hermitian $\beta$-ensemble of random matrices, we start in the next Section \ref{ssec-surmise} with the introduction of an ensemble of normal $2\times 2$ matrices, which are exactly tractable. The local statistics are compared with numerical generated data for 2d Coulomb gases (2dCG) at inverse temperature $\beta$ for large $N$. Fitting to these yields an effective $\beta$, which provides a good effective description of the local statistics for small $\beta>0$ at large $N$. We follow the results published in \cite{AMP} closely.\\
We present an application of this surmise in Section \ref{ssec-birds of prey}. Here, the 2dCG at inverse temperature $\beta$ serves as a simple one-parameter model for the study of repulsive interaction of birds of prey in the Teutoburger Wald. Data on occupied nests from over 20 years of observation, are analysed in \cite{ACKMOP}. The parameter $\beta$ can be interpreted as a measure of the strength of interaction within and between species under consideration.\\
We introduce three tridiagonal $\beta$-ensembles of non-Hermitian random matrices in Section \ref{ssec-tridiag beta}. Extending a previously studied ensemble by Mezzadri and Taylor \cite{mezzadri taylor}, we generalise their method and provide the explicit calculation of the Jacobian. The JPDF of the complex eigenvalues contains a pairwise logarithmic interaction at $\beta$ as well as an additional coupling to the eigenvectors for each ensemble. While the latter makes a study of the spectral density in the limit $N\rightarrow \infty$ in general hard, we introduce a mechanism to perform the calculation in the low temperature limit $\beta\gg 1$. Therefore, we first study the characteristic polynomial in this limit, finding explicit expressions on the coefficients. Secondly, we apply a result from free probability on the distribution of zeros of polynomials with random coefficients to the characteristic polynomial of our ensembles. The results are in good agreement with extensive numerical data. Numerically, the local NN statistics show accordance with the behaviour of independent particles in the complex plane, i.e. the 2d Poisson process, independent of $\beta$. The results presented in this section serve as a main contribution to this thesis and are published in \cite{AMPT}.
\subsection{Point Processes in 2 Dimensions and Beyond}\label{ssec-point processes}
We discussed above, how local statistics of non-Hermitian point processes can be used to distinguish integrable and chaotic open quantum systems. Therefore, we collect several results on the $D$-dimensional Poisson point process, of independent particles as well as on the complex Ginibre ensemble, which describes a fully chaotic system in 2d. In Figure \ref{fig-scatterplots Poi GinUE} below, we give scatter plots of the eigenvalues from a 2d Poisson point process of $N$ independent normally distributed random variables (left) and for the complex Ginibre ensemble (right), both for $N=100$. Due to normalization, both processes are supported within the unit disc (in the limit $N\rightarrow \infty$). Having a closer look, we observe that eigenvalues from the 2d Poisson process can lie arbitrary close, while we have a nearly uniform distribution on the RHS. This illustrates very well the absence/ existence of repulsion between the eigenvalues. We follow the discussion in \cite{ABCPRS}, contributed by the author.\\
\begin{figure}[t]
	\centering
	\hspace{0.0\linewidth}Poi \hspace{0.3\linewidth} GinUE \hfill\\
	\includegraphics[width=0.38\linewidth,angle=0]{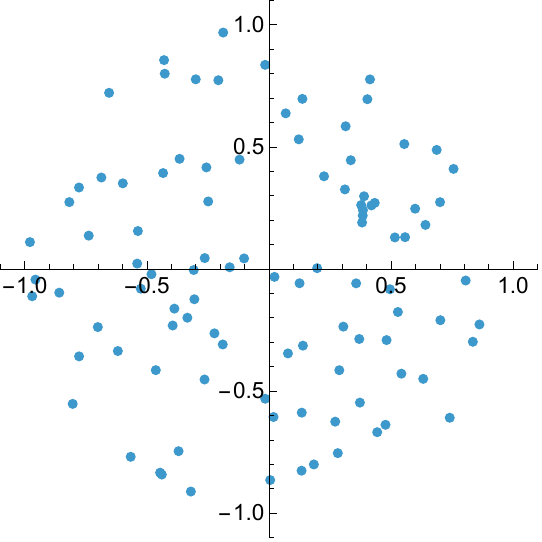}
	\includegraphics[width=0.38\linewidth,angle=0]{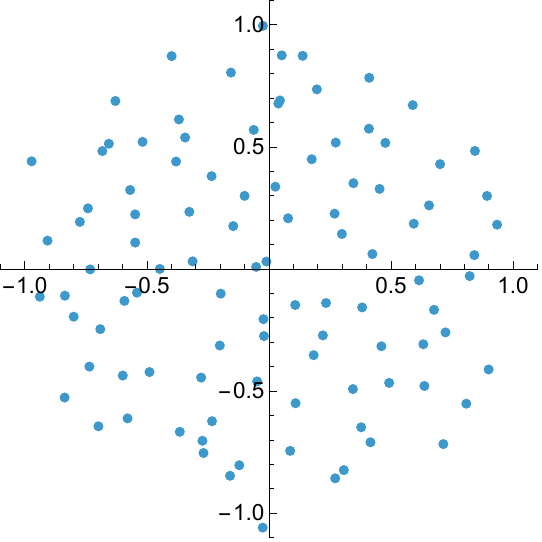}
	\caption[Mathematica plots, \textit{Scatter plots for a Poisson point process in the unit circle and eigenvalues of the complex Ginibre ensemble, both for $N=100$}]{Comparison of distributions of points in the case of independent particles in the complex unit circle (left, Poi) with the complex eigenvalues of the complex Ginibre ensemble (right, GinUE), both for $N=100$. We normalized the eigenvalues of the latter to be supported in the unit circle in the limit $N\rightarrow \infty$. On the left, we find that points can cluster together while we observe an even spread on the right.
	}
	\label{fig-scatterplots Poi GinUE}
\end{figure}\noindent
Let $z_1,\ldots,z_N$ be the complex points (eigenvalues) of a process under consideration. We define the radial\footnote{While in 1d, i.e. point processes with real particles, a relative ordering between the points can be introduced, we choose this radial distance in 2d.} \textit{nearest-neighbour} (NN) spacing of the eigenvalue $z_i$ as 
\begin{align}\label{eq-def nn}
	s_i^\textrm{NN}=\min_{j\neq i} |z_i-z_j|.
\end{align}
Let the minimum in Eq. \eqref{eq-def nn} being realised for some $j=i_0$. One can also study the \textit{next-to-nearest-neighbour} (NNN) spacing of the eigenvalue $z_i$:
\begin{align}\label{eq-def nnn}
	s_i^\textrm{NNN}=\min_{j\neq i,i_0} |z_i-z_j|.
\end{align}
The $D$-dimensional \textit{Poisson point process} consists of $N$ identically and independently distributed random variables in a $D$-dimensional ball of radius $R$, keeping the density of points fixed by setting $N/R^D=1$. Numerically, one generates a diagonal matrix with its entries distributed as describes, i.e. the JPDF of the $N$ particles factorizes completely.\\
The NN and NNN distributions can be derived analytically in the large $N$-limit \cite{SaRibeiroProsen}:
\begin{align}
	\label{eq-Poid-NN}
	p^{\textrm{(NN)}}_{\textrm{Pois},D}(s) &= D\,\Gamma\left(1+\frac{1}{D}\right)^D s^{D-1} \, e^{-\Gamma\left(1+\frac{1}{D}\right)^D s^D},\\
	\label{eq-Poid-NNN}
	p^{\textrm{(NNN)}}_{\textrm{Pois},D}(s) &= D\,\Gamma\left(1+\frac{1}{D}\right)^{2D} s^{2D-1} \, e^{-\Gamma\left(1+\frac{1}{D}\right)^D s^D},
\end{align}
where each distribution is normalized. For convenience, also the first moment of $p^{\textrm{(NN)}}_{\textrm{Pois},D}(s)$, i.e. $\int_0^\infty\! ds \ sp^{\textrm{(NN)}}_{\textrm{Pois},D}(s)$ is rescaled to unity and $p^{\textrm{(NNN)}}_{\textrm{Pois},D}(s)$ becomes rescaled with respect to the very same factor. For $D=2$, these are given for example in \cite{haake,SakhrNieminen}, with
\begin{align}
	\label{eq-Poi2-NN}
	p^{\textrm{(NN)}}_{\textrm{Pois},2}(s) &= \frac{\pi}{2} s \, e^{-\frac{\pi}{4} s^2}\sim s\\
	\label{eq-Poi2-NNN}
	p^{\textrm{(NNN)}}_{\textrm{Pois},2}(s) &= \frac{\pi^2}{8} s^{3} \, e^{-\frac{\pi}{4} s^2} \sim s^3,
\end{align}
where we give the leading coefficient for small $s\rightarrow 0$ too.\\
A second point process, which frequently occurred in this thesis already, is the 2-dimensional \textit{Coulomb gas} (2dCG). Let us repeat, that it consists of particles, that show pairwise logarithmic interaction at inverse temperature $\beta$ in a confining potential $V(z)$, cf. Eq. \eqref{eq-2dCoulomb} below. While the parameter $\beta$ is a measure of the strength of the repulsion within the particles (eigenvalues), we find that in the absence of the interaction, i.e. $\beta=0$, we recover the Poisson point process. On the other hand, for $\beta=2$, we saw already in the introduction Section \ref{ssec-intro models}, that this CG interaction describes the complex Ginibre ensemble (cf. Eq. \eqref{eq-jpdf complex ginibre CG}). We repeat that a 2dCG at real $\beta>0$ is of great interest but so far a random matrix realisation is not known. We comment on the difficulties throughout the following lines. 
\begin{Remark}
	Before we move on and state some explicit results for the spacing statistics of the complex Ginibre ensemble, we would like to motivate (again), why the Coulomb gas is of such a great interest. The key ingredient is here the \textit{universality}. While other interactions than the pairwise logarithmic originating from the Coulomb force exist (cf. \cite{Lewin} and references therein), it turns out that the CG is highly applicable to plenty of models.
\end{Remark}
The NN and NNN spacing distributions of an $N\times N$ complex Ginibre matrix are given in \cite{GHS} for NN and \cite{APS} for NNN:
\begin{align}\label{eq-GinibreSpacingNN}
	p^{\textrm{(NN)}}_{\textrm{Gin}}(s) &=
	E_{\textrm{Gin}}^\textrm{(NN)}(s)
	\sum_{j=1}^{N-1} \frac{2s^{2j+1}e^{-s^2}}{\Gamma(1 + j, s^2)}
	\sim s^3, 
	\\
	p^{\textrm{(NNN)}}_{\textrm{Gin}}(s) &=
	E_{\textrm{Gin}}^\textrm{(NN)}(s)
	\sum_{j,k=1; k \neq j}^{N-1}
	\frac{\gamma(1+j, s^2)}{\Gamma(1 + j, s^2)}
	\frac{2s^{2k+1}e^{-s^2}}{\Gamma(1 + k, s^2)}	\sim s^5, \text{ with }\label{eq-GinibreSpacingNNN}\\
	E_{\textrm{Gin}}^\textrm{(NN)}(s)&= \prod_{j=1}^{N-1}\frac{\Gamma(1 + j, s^2)}{j!}\ .
\end{align}
Here $\gamma(1 + k, s^2)=\int^{s^2}_0\! dt\ t^k e^{-t}$ is the lower incomplete
Gamma function, and we give again the behaviour for small $s\rightarrow 0$. These results extend to the limit $N\rightarrow \infty$. The first moment of $p^{\textrm{(NN)}}_{\textrm{Gin}}(s)$ still needs to be normalised to unity, by $s\rightarrow 1.1429 \cdot s$, which can be calculated numerically \cite{haake}. The spacing distributions \eqref{eq-GinibreSpacingNN} and \eqref{eq-GinibreSpacingNNN} hold throughout the bulk\footnote{We comment on the terms bulk and edge of a spectrum in the next Section \ref{sssec-unfolding}.} of the spectrum. Moreover, they apply to a much larger class of confining potentials $V(|z|)$ than Gaussian \cite{Hedenmalm}, which proves again universality.  
\\
We present the spacing distributions (NN and NNN) of the Poisson process, with dimension $1\leq D\leq 2$ varying in steps $0.1$ (red--grey) in Figure \ref{fig-2d-statistics} below left. Notice that in both cases the maximum moves from right to left when decreasing dimension $D$. On the right, we present the numerically generated NN and NNN distributions of the 2dCG for $\beta\in [0,2]$, increasing in steps $0.2$ (grey--blue). The lowest curves correspond to 2d Poisson with $\beta=0$, i.e. Eqs. \eqref{eq-Poid-NN} and \eqref{eq-Poid-NNN} for $D=2$ respectively, while the blue ones with the highest maxima are given by Eqs. \eqref{eq-GinibreSpacingNN} and \eqref{eq-GinibreSpacingNNN}, i.e. $\beta=2$.
\begin{figure}[t]
	\includegraphics[width=\linewidth]{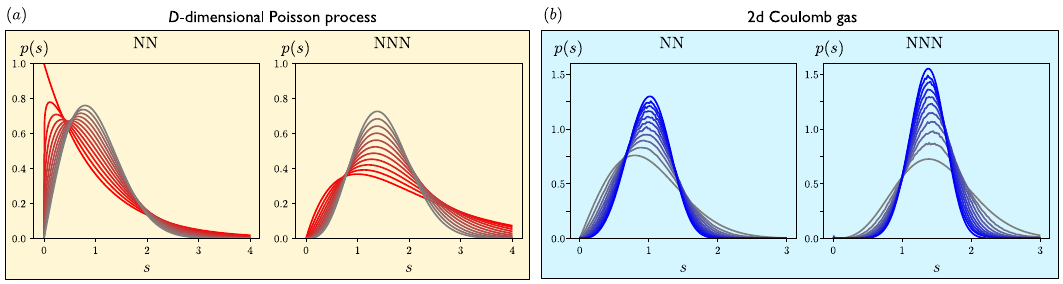}
	\caption[Figure 2 from \cite{ABCPRS}, \textit{NN and NNN spacing distributions of the $D$-dimensional Poisson process and the 2dCG at $\beta\in[0,2]$}]{\textbf{(a) Left:} Spacing distributions of the Poisson process, with dimension $1\leq D\leq 2$ varying in steps $0.1$ (red to grey). Notice that in both cases the maximum moves from right to left when decreasing dimension $D$. \textbf{(b) Right:} The numerically generated NN and NNN spacing distributions of the 2dCG for $\beta\in [0,2]$, increasing in steps $0.2$ (grey to blue). This also includes analytic formulas for 2d Poisson with $\beta=0$, \eqref{eq-Poid-NN} and \eqref{eq-Poid-NNN} with the lowest maxima, and the complex Ginibre ensemble with $\beta=2$, \eqref{eq-GinibreSpacingNN} and \eqref{eq-GinibreSpacingNNN}, with highest maxima. Note that the grey curves for $D=2$ in (a) are the same as the grey curves for $\beta=0$ in (b), while they are plotted on different scales. \cite[Figure 2]{ABCPRS}  
		\label{fig-2d-statistics}}
\end{figure}
\subsection{Analysing Numerics: Methodologies}\label{ssec-numerics methods}
The comparison with numerical data from either random matrix ensembles or data sets from applications plays an important role in the research and therefore, we introduce two concepts which will be used throughout this thesis. Again, we follow the discussion in \cite{ABCPRS} by the author closely.
\subsubsection{Unfolding Spectra in 2d}\label{sssec-unfolding}
We pointed out that the study of the local statistics of a set of particles (eigenvalues) in the complex plane, is a main part in the analysis of a certain system. In order to treat only the local statistics, one has to extract them from a global, possibly changing, background density. Let $\rho(z)^\textrm{ave}$ be the mean background density, which in general varies on a global scale, i.e. on the support of the point process. The rescaling of it to unity is called \textit{unfolding}. In 1d this procedure is well understood, and possibly different methods yield same results in the most cases under consideration \cite[Chapter 4.8]{haake}, \cite{GMW}. However, in 2d, without a total ordering of the particles, the unfolding is more complicated and several different methods have been developed, cf. \cite{QCD application,AKMP,Hamzaki et al,unfolding Li et al,unfolding Suthar et al,unfolding Garcia et al}. In the applications, presented in this thesis in the subsequent Sections \ref{ssec-birds of prey}, \ref{ssec-application spin chain}, we will apply the method proposed in \cite{AKMP}. We repeat the latter briefly. The density $\rho(z)$ is a superposition of the mean spectral density $\rho(z)^\textrm{ave}$ and a fluctuating part $\rho(z)^\textrm{fluc}$
\begin{align}
	\rho(z)=\rho(z)^\textrm{ave}+\rho(z)^\textrm{fluc},
\end{align}
where $\rho(z)^\textrm{ave}\sim \mathcal{O}(1)$ for an ensemble of $N$ particles and $\rho(z)^\textrm{fluc}\sim 1/\sqrt{N}$ (typically) in 2d for geometrical reasons. The exact density of states of $N$ eigenvalues $z_1,\ldots,z_N$ in the complex plane is determined by the sum of delta functions at these positions, i.e.
\begin{align}
	\rho(z)=\frac{1}{N}\sum_{j=1}^N \delta^{(2)}(z-z_j).
\end{align}
If the explicit form of $\rho(z)$ is not known analytically, the method in \cite{AKMP} is to approximate the later by a superposition of smeared out Gaussian distributions with variance $\Sigma$,
\begin{align}
	\rho(z)^\textrm{ave}=\frac{1}{2\pi\Sigma^2 N}\sum_{j=1}^N e^{-\frac{|z-z_j|^2}{2\Sigma^2}}.
\end{align}
Here, the variance is $\Sigma=\alpha \bar{s}$, where $\bar{s}$ is the mean NN spacing and $\alpha$ is a constant of $\mathcal{O}(1)$, which is chosen properly to rescale $\rho(z)^\textrm{ave}$ to unit on its support. In \cite{AKMP}, it was shown that $\alpha=4.5$ is a good choice and we will use this value in the applications described below, too.
\begin{Remark}
	In the analysis of eigenvalue spectra, it is important to distinguish between regions which are called bulk, respectively edge. The eigenvalues are supported with respect to a certain domain which can be chosen to be compact via rescaling. Hence, possible repulsion of the eigenvalues interacts with the effect of the confining potential near the edge of the support. The spectral density in the bulk shows often a significantly different behaviour than at the edge, often also the scaling dimensions are distinct. Therefore, it is essential to distinguish between eigenvalues in the bulk of the spectrum, where boundary effects can be neglected and the edge, where this is no longer the case. We can see from above that also the unfolding procedure must be adjusted, if one moves to the edge regime as the mean NN spacing and therefore $\alpha$ will change. In the most parts of this thesis, we will consider the bulk of the spectrum. We come back to the discussion of the choice of bulk and edge regions in Section \ref{sec-universality symmetry classes}. 
\end{Remark}
\subsubsection{Comparison of Numerical Data}\label{sssec-comparison numerical data}
After mentioning comparison of random matrix statistics and applications already several times, the natural question arises how to quantify differences or agreement between either analytics and a data set or two data sets. We briefly recall two measures, which will be used in the applications \cite{ACKMOP,ABCPRS}, presented in the subsequent Sections \ref{ssec-birds of prey} and \ref{ssec-application spin chain}, following the discussion (contributed by the author) in \cite{ABCPRS} closely. Let $p_1(s)$, $p_2(s)$ be two distributions, where at least one is based only on numerical data and the second one might be given by an analytical expression or by numerical data as well. The \textit{standard deviation} $\sigma$, is defined as
\begin{align}\label{eq-standard deviation}
	\sigma=\left(\frac{1}{n}\sum_{j=1}^n (p_1(s_j)-p_2(s_j))^2\right)^{\frac{1}{2}},
\end{align}
where $n$ is the number of bins in the data and $p_1(s_j)$, respectively $p_2(s_j)$ are the number of counts in the $j$th bin at its midpoint $s_j$. Hence, one first determines a fine binning of the data into a histogram and uses these histogram points $(s_j)_{j=1}^n$ for the comparison. If the second distribution $p_2(s_j)$ is analytically known, one chooses a binning on the numerical data of $p_1(s_j)$ and evaluates $p_2(s)$ at $s=s_j$. For data with an exponentially suppressed tail at large $s$, a cut off at $s = 3$, is convenient, when the first moment is normalized to unity as described above. See also Figure \ref{fig-comparison n=2 GUE GinUE} for illustration.\\
The second measure is the \textit{Kolmogorov-Smirnov distance}, which is defined as
\begin{align}\label{eq-KS distance}
	d=\max_{x\geq 0}|F_1(x)-F_2(x)|\in [0,1],
\end{align}
between the cumulative distributions $F_1$ and $F_2$ of distributions $p_1$ and $p_2$, i.e. $F_i(x)=\int_0^x \! ds\  p_i(s)$ for $i=1,2$. It is independent of the binning into histograms. The functions $F_k(x)$ are either determined numerically or analytically. However, the calculations take much longer than the ones for the standard deviation.\\
Having these two measures on hand, we need a reference point, when to speak about a good agreement between two ensembles under consideration. Therefore, we compare the numerical generated complex Ginibre ensemble, with the 2dCG data at $\beta=2$. The latter was generated via the Metropolis-Hastings algorithm \cite{MetropolisHastings,Metropolis}, using the 2dCG potential, cf. \cite{AKMP,buzzards 1} for details. We obtain for the matrix size $N=5000$,
\begin{align}
	\begin{split}\label{eq-reference sigma d}
	\sigma_\textrm{ref}&= 0.44 \cdot 10^{-2}\\
	d_\textrm{ref}&=0.20 \cdot 10^{-2}.
	\end{split}
\end{align}
\subsection{First Attempt: A Surmise}\label{ssec-surmise}
We study a surmise, which is inspired from the Wigner surmise discussed above, here for normal non-Hermitian $2\times 2$ matrices. Before the reader gets too enthusiastic on this presumably simple idea, we have to insist: It is well-known that the NN statistics of the complex Ginibre ensemble at $N=2$ do not describe the large $N$ distribution well. For illustration, we reprint Figure 1 from \cite{Wigner surmise pic} in Figure \ref{fig-comparison n=2 GUE GinUE} below. On the left, the Wigner surmise \eqref{eq-pWigner} (red dashed) is shown compared with the exact large $N$ result (blue full curve), cf. \cite{Wigner surmise}. We observe excellent agreement. On the right, the spacing distribution \eqref{eq-GinibreSpacingNN} of the complex Ginibre ensemble for $N=2,3,4,20$ (from red left bottom to blue right top) is given. We find significant differences between $N=2$ and $N=20$.
\begin{figure}[t]
	\centering
	\includegraphics[width=\linewidth,angle=0]{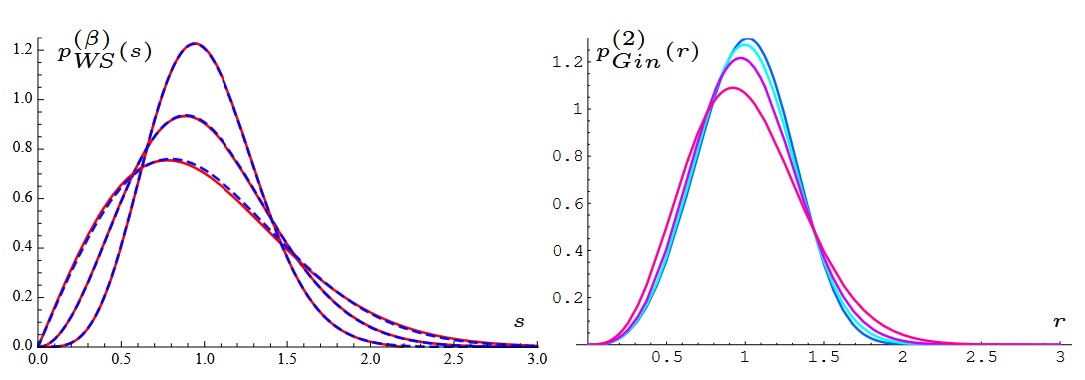}
\caption[Figure 1 from \cite{Wigner surmise pic}, \textit{Comparison NN statistics for large $N$ with the $N=2$ analytics for the Wigner surmise and the GinUE}]{\textbf{Left: }Comparison of the NN statistics $N=2$ Wigner surmise \eqref{eq-pWigner} (red dashed) and the exact large $N$ result (blue full curve), cf. \cite{Wigner surmise}. We observe excellent agreement. \textbf{Right:} Spacing distribution \eqref{eq-GinibreSpacingNN} of the complex Ginibre ensemble for $N=2,3,4,20$ (from red left bottom to blue right top). We find significant differences between $N=2$ and $N=20$. \cite[Figure 1]{Wigner surmise pic}
	}
	\label{fig-comparison n=2 GUE GinUE}
\end{figure}\noindent
\begin{Remark}\label{rem-stability N=2}
	We pause here to make an important observation: It is well-known that small $N$ results represent often good approximations for large $N$-limits in the case of Hermitian matrices. Similarly, it is established that this "surmise"-behaviour breaks down for non-Hermitian matrices. One can view this from several perspective. First, observables like Green functions (Feynman propagators), which are fundamental objects in quantum field theory (see \cite{PS} for an introduction), are integrated with respect to the measure $d^Dk/k^2$ in $D$ dimensions in the phase space. For $D=1$, this yields $\mathcal{O}(1/N)$ as leading correction, while it results in significant $\mathcal{O}(\log(N))$-corrections for $D=2$, with $d^2k/k^2=dk/k$ exploiting polar coordinates. Therefore, one can argue that it is due to geometrical reasons that in 1d small $N$ results are in better agreement with the large $N$ results than in 2d \cite{satya private}. Secondly, one can view this in terms of the different scales of the condition number \cite{Edelman}, i.e. stability under perturbation. We come back to this subject in Section \ref{sssec-numerics}.\\
	We refer the reader to a very readable discourse by Wigner which is called "The unreasonable effectiveness of mathematics in the natural sciences" \cite{wigner effect} and deals with similar questions.
\end{Remark}
We follow \cite{AMP}, which was mainly contributed by the author and partly covered in \cite{masterthesis}, closely and start by revisiting the in the meanwhile often referred to Wigner's surmise.
\begin{align}
	\label{eq-pWigner}
	\begin{split}
	&p_\textrm{Wigner}^\textrm{1d}(s)=a_\beta s^\beta \exp[-b_\beta s^2], \\
	&b_\beta=\frac{\Gamma[(\beta+2)/2]^2}{\Gamma[(\beta+1)/2]^2},\ \ a_\beta=2b_\beta^{(\beta+1)/2}. 
	\end{split}
\end{align}
The $N=2$ NN distribution $p_\textrm{Wigner}^\textrm{1d}(s)$ works very well also for large $N\gg 2$, cf. \cite{Wigner surmise}. While it becomes a more accurate approximation at increasing values of $\beta=1,2,4$, it fails closer to Poisson in 1d,  
\begin{align}
	\label{eq-Poisson1d}
	p_\textrm{Poisson}^\textrm{1d}(s)=\exp[-s],
\end{align}
corresponding to $\beta=0$. Several phenomenological interpolations between $\beta=0$ and 1 exist, cf. \cite{GMW}. Investigating a result for continuous $\beta\in[0,2]$, more recently, a $2\times2$ invariant matrix representation of the 1d Dyson gas has been used to derive an integral representation \cite{PierSatya}. We will follow this approach and extend it to a 2dCG at $\beta\geq0$, staying well below the crystallisation transition at $\beta_c\approx 140$ \cite{Abanov}, where the Coulomb gas is conjectured to attain a triangular crystal structure.\\
In 2d closed form expressions for the local statistics exist only for the spacing for Poisson ($\beta=0$), see Eqs. \eqref{eq-Poi2-NN}, \eqref{eq-Poi2-NNN} above, and the complex Ginibre ensemble (GinUE) at $\beta=2$, see Eqs. \eqref{eq-GinibreSpacingNN}, \eqref{eq-GinibreSpacingNNN}. Consequently, a detailed comparison to the spacing in the intermediate regime previously relied on numerical simulations of the 2dCG, cf. \cite{AKMP}. 
We start by briefly recalling the construction of the matrix model used in \cite{AMP}. The goal is to find a model, whose complex eigenvalues are distributed with respect to the following JPDF (cf. Eq. \eqref{eq-jpdf complex ginibre CG}).
\begin{align}
	\label{eq-2dCoulomb}
	P_{\beta,N}(z_1,\ldots,z_N) \propto
	e^{\beta\sum_{i> j}^N\log|z_i-z_j|
		-{\sum}_{l=1}^N V(z_l,z_l^*)
	}.
\end{align}
In order to be able to take the limit $\beta\to0$ of uncorrelated points, cf. \cite{ABy}, we have rescaled the potential to remove its dependence on $\beta$. Let us construct an ensemble of $N\times N$ complex normal non-Hermitian random matrices $J\neq J^\dag$, with $[J,J^\dag]=0$, that has the same JPDF of complex eigenvalues as the 2dCG \eqref{eq-2dCoulomb}, for arbitrary $\beta>0$. 
Following \cite[Chapter 15]{mehta} the Vandermonde determinant 
\begin{align}
	\label{eq-Vandermonde}
	\Delta_N(Z)= \prod_{N\geq i> j\geq 1}(z_i-z_j) 
\end{align}
of complex eigenvalues of $J$, that provides the logarithmic Coulomb interaction in \eqref{eq-2dCoulomb}, can be written as follows, after taking the modulus squared,
\begin{align}
	\label{eq-Delta-rewrite}
	|\Delta_N(Z)|^2
	= \det_{1\leq i,j\leq N}\left[\Tr(J^{i-1}J^{\dag\,j-1})\right].
\end{align}
Here, we used that the normal matrix $J$ can be diagonalised via a unitary transformation, i.e. $J=UZU^\dag$ with $U\in U(N)$ and $Z=$diag$(z_1,\ldots,z_N)$, which yields 
\begin{align}
	\Tr (J^kJ^{\dag\,l})=\sum_{i=1}^Nz_i^k{z}^{*l}_i, \quad k,l=0,1,...,N-1.
\end{align}
We obtained the modulus square of the Vandermonde determinant expressed in rotational invariant terms. Note that this invariance is also one of the main ingredients discussed in Section \ref{ssec-intro models} and it follows that we can insert this term, raised to an arbitrary power, in the JPDF.
Because the Jacobian for the above stated diagonalisation of complex normal matrices $J$ is known \cite{Oleg} to be proportional to 
$|\Delta_N(Z)|^2$, we obtain the following distribution
\begin{align}
	P_{\beta,N}(J)\propto&\frac{\exp[-\Tr V(J,J^\dag)]}{\det_{1\leq i,j\leq N}\left[\Tr(J^{i-1}J^{\dag\,j-1})\right]^\xi}
	\nonumber\\
	P_{\beta,N}(Z)\propto& |\Delta_N(Z)|^{2-2\xi}
	e^{-{\sum}_{l=1}^N V(z_l,z_l^*)},
	\label{eq-betaRM}
\end{align}
for a non-Hermitian $\beta$-ensemble. 
It agrees with \eqref{eq-2dCoulomb} when identifying $\beta=2-2\xi$. 
The integrals converge for $\beta\geq0$ or $\xi\leq1$ (however, cf. \cite{ABy} when $\beta=2c/N$ for $c> -2$).
This allows us to interpolate between the Poisson point process at $\beta=0$ and the GinUE at $\beta=2$, or larger values of $\beta$, when considering $\xi<0$. Although this $\beta$-ensemble \eqref{eq-betaRM} is invariant and well defined for any $N$, we have so far only been able to obtain analytical results for $N=2$ with Gaussian potential $V(J,J^\dag)=JJ^\dag$, following \cite{PierSatya} closely.\\
\begin{Remark}
	Let us pause a moment to summarize the main advantage of this model: Using the fact, that we consider \textit{normal} random matrices, we are able to impose a diagonalisation via unitary matrices. Comparing with the Schur-decomposition of a generally non-Hermitian matrix, the upper triangular matrix $T$ in Eq. \eqref{eq-schur decomp} vanishes. As mentioned in the previous Section \ref{ssec-intro models}, the integration of this triangular part as well as the possible coupling to the other spectral parameters is in general highly non-trivial. Here, its absence enables us to impose a rotationally invariant weight, which depends on $\Tr (J^kJ^{\dag\,l})$ raised to some non-integer power $\xi$. While this might convince the reader at a first glance, that the normality condition is a "good" choice, we have to deal with its restrictions. While normal random matrices are studied for many years, cf. \cite{Oleg} and references therein, the lack of the Selberg-integral, which determines the normalization of the JPDF in terms of the eigenvalues, is one main difficulty. While we outlined in Section \ref{ssec-Anderson Dixon}, that several methods exist to compute the Selberg-integral \eqref{eq-selberg integral} over real variables, the result is until today not known for general real $\beta>0$ over the complex plane \cite{importance of selberg}. In order to solve at least for the normalization of this model for general $N>2$, we have to deal with the business of the complex Selberg-integral which is still an open problem. 
\end{Remark}
\begin{prop}[\cite{AMP,masterthesis}]
	Let $J$ be a $2\times 2$ normal matrix from the ensemble defined in Eq. \eqref{eq-betaRM}, with $V(z,z^*)=|z|^2$. The NN spacing distribution $p_{\beta,2}(s)$, defined by 
	\begin{align}
		p_{\beta,2}(s)=\int_\mathbb{C}\! d^2z_1 \int_\mathbb{C}\! d^2z_2 ß P_{\beta,2}(z_1,z_2)\delta(s-|z_2-z_1|)
	\end{align}
	reads 
	\begin{align}
		\label{eq-peta}
		{p}_{\beta,2}(s)=\frac{2\alpha^{1+\beta/2}}{\Gamma[1+\beta/2]}\,s^{1+\beta} e^{-\alpha s^2},\  \alpha=\frac{\Gamma[(3+\beta)/2]^2}{\Gamma[1+\beta/2]^2}.
	\end{align}
	Here, $p_{\beta,2}(s)$ and its first moment $\int_0^s \! ds \  sp_{\beta,2}(s)$ are normalized to unity both.
\end{prop}
\begin{figure}[t!]
		\centering
		\includegraphics[width=0.49\linewidth,angle=0]{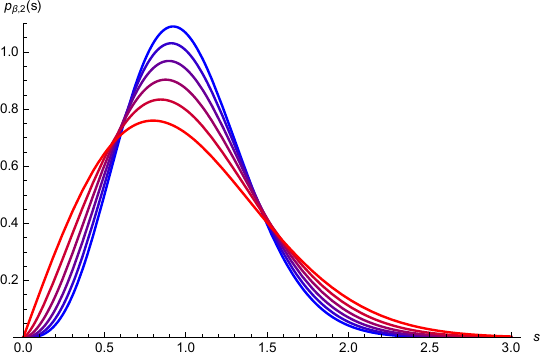}
		\includegraphics[width=0.49\linewidth,angle=0]{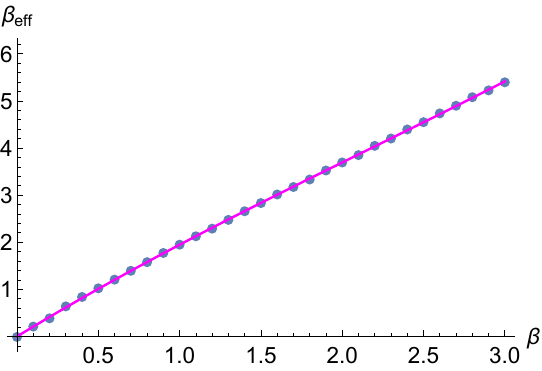}
		\caption[Part of Figure 1 and Figure 2 from \cite{AMP}, \textit{Spacing distribution of the surmise \eqref{eq-pbetasurmise} and fit of the effective $\beta$ \eqref{eq-cubic}}]{\textbf{Left:} Spacing distribution $p_{\beta,2}(s)$ \eqref{eq-peta} at $N=2$ for several values of $\beta=0,0.4,0.8,1.2,1.6,2$, where the blue top curve corresponds to $\beta=2$ (GinUE) and the red bottom curve to $\beta=0$ (2d Poisson). \textbf{Right:} Fit of $\beta_\textrm{eff}(\beta)$ with a polynomial of degree three from \eqref{eq-cubic} (full line), compared to the best fit $\beta_\textrm{eff}$ as a function of $\beta$, in steps of 0.1 (points). \cite[Cf. Figures 1, 2]{AMP}}
		\label{fig-spacing, cubic beta n=2}
	\end{figure}
We refer to \cite{AMP,masterthesis} for the proof. We plot $p_{\beta,2}(s)$ in Figure \ref{fig-spacing, cubic beta n=2} left for several values of $\beta$. It correctly reproduces the 2d Poisson distribution \eqref{eq-Poi2-NN} (red bottom curve) when setting $\beta=0$, which is valid for an infinite number of independent particles, cf. \cite{haake}. Second, by construction it agrees with the spacing distribution of the GinUE for $N=2$ at $\beta=2$ (blue top curve) \cite[Eq. (8.8.28)]{haake}.
Moreover, we observe, that the maxima of the curves become sharper and move to the right within increasing $\beta$, i.e. we find qualitatively a similar behaviour as discussed for the large $N$ statistics in Section \ref{ssec-point processes}, Figure \ref{fig-2d-statistics} right. We recall from Figure \ref{fig-comparison n=2 GUE GinUE} that this $N=2$ result is \textit{not} a good approximation of the large $N$ behaviour in the GinUE.\\
The spacing \eqref{eq-peta} can be modified by fitting it to the numerically generated spacings of the 2dCG at a given $\beta$ in \cite{AKMP}, where $\beta$ in $p_{\beta,2}(s)$ serves as a fit parameter. The resulting effective inverse temperatures $\beta_\textrm{eff}$ are shown in Figure \ref{fig-spacing, cubic beta n=2} right as blue dots. Although the curve is reasonably well approximated by a straight line, see discussion in \cite{AMP}, we use a polynomial of degree 3, 
\begin{align}
	\beta_{\text{eff}}(\beta)=2.108\beta-0.190\beta^2+0.030\beta^3 ,
	\label{eq-cubic}
\end{align}
shown in Figure \ref{fig-spacing, cubic beta n=2} above (red full line).
We conclude that\footnote{Correcting a typo in Eqs. (23), (27) in \cite{AMP}.}
\begin{align}
	\label{eq-pbetasurmise}
	{p}^\textrm{(NN)}_{\textrm{sur},\beta_\textrm{eff}}(s)&=
	\frac{2\alpha_\textrm{eff}^{1+\beta_\textrm{eff}}/2}{\Gamma[1+\beta_\textrm{eff}/2]}
	\,s^{1+\beta_\textrm{eff}} \exp[-\alpha_\textrm{eff} s^2]\\
	\alpha_\textrm{eff}&=\frac{\Gamma[(3+\beta_\textrm{eff})/2]^2}{\Gamma[1+\beta_\textrm{eff}/2]^2}\nonumber
\end{align}
is a \textit{surmise} for the spacing distribution in the 2dCG or non-Hermitian $\beta$-ensemble. Although it does not capture the expected repulsion $\sim s^{1+\beta}$ at very small values of $s\ll1$, it describes the overall spacing distribution very well, including its maximum and tail.\\
Using the Kolmogorov-Smirnov-distance \eqref{eq-KS distance} in comparison to data, we state the CDF of \eqref{eq-pbetasurmise}, cf. \cite{ACKMOP},
\begin{align}
	\label{eq-CDFsurmise}
	{E}^\textrm{(NN)}_{\textrm{sur},\beta_\textrm{eff}}(s) &= 1 - \frac{\Gamma\left(1 + \frac{\beta_\textrm{eff}}{2}, \alpha_\textrm{eff} s^2\right)}{\Gamma\left(1 + \frac{\beta_\textrm{eff}}{2}\right)},
\end{align}
where $\Gamma(n+1, x)=\int_x^\infty \! dt\  t^n e^{-t}$ is the upper incomplete Gamma function.\\ 
We refer to \cite{AMP,masterthesis} for details on the analysis. We conclude that the $N=2$ result of the spacing distribution of a 2dCG at inverse temperature $\beta$, can be adjusted to a good large $N$ description for small $\beta\in\left[0,0.5\right]$ via fitting to CG data. The main advantage is here, that Eq. \eqref{eq-pbetasurmise} is a rather simple formula, which describes the large $N$ CG spacing well enough to get a first insight of the interaction strength $\beta$. In contrast to the Wigner surmise in 1d \eqref{eq-pWigner}, which becomes a better approximation for increasing values of $\beta=1$ to $\beta=4$ but fails when getting closer to 1d Poisson \eqref{eq-Poisson1d} at lower $\beta\to0$, we observe the opposite behaviour here. The closer we get to 2d Poisson \eqref{eq-Poi2-NN}, the better is the approximation, cf. \cite{AMP,masterthesis}. Thus, when quantifying the transition between integrable (Poissonian) behaviour, fully chaotic random matrix statistics in the respective symmetry class, and potential intermediate transitions, this surmise serves as a good parametrisation. We present an application to birds of prey in the preceding section and later in Section \ref{ssec-application spin chain} to a quantum spin chain with dissipative disorder.

\subsection{Application: Birds of Prey}\label{ssec-birds of prey}
We discuss an application of the surmise from the previous section, which was studied by the author and collaborators in \cite{ACKMOP}. We recall from the beginning of Section \ref{sec-non Herm beta} that the local NN statistics are a good measure to distinguish integrable and chaotic systems using the GHS conjecture \cite{GHS}. We follow \cite{ACKMOP} closely. First, we start in Section \ref{sssec-motivation and biology} with a motivation, why a fitting with 2dCG data is a good approach to quantify the repulsion between highly territorial birds of prey. While we made the point, that one really should care about the 2dCG, already several times in this thesis, we emphasize its role in applications for systems, where an underlying theory is possibly not known (yet). We discuss the biological setting and the data collection to set the stage.\\
In Section \ref{sssec-interact}, we present the main results, i.e. the study of interaction within a species and between two different species, quantified by fits of the NN and NNN spacing distributions to a 2dCG. The outcomes agree with the known degree of aggressive behaviour amongst the different species of birds of prey.\\
Although we highlighted the universality of the CG interaction, we probe in Section \ref{sssec-variations rpp} the variation of the underlying random point process. First, we compare with the $D$-dimensional Poisson point process, where $D<2$ is fitted. The motivation here is that the birds cannot nest everywhere in a 2d plane, but only on (fractal) regions which show the right biological conditions for nesting. We discuss the subject of "independence" of the occupied nests in consecutive years, by quantifying the reusage of nests within one or all species. The results in Section \ref{sssec-interact} yield different (decreasing) $\beta$'s for the NN and NNN spacing distribution for the same setting, pointing to a shorter correlation length than Coulomb. This raises the question to study point processes with varying correlation length. This was investigated in \cite{ACKMOP} via a two-parameter fit of the Yukawa interaction in 2d. As this model did not resolve the discrepancy of the fit parameters of the NN and NNN statistics, we restrict ourselves to the 2dCG here and refer to \cite[Section 4.2]{ACKMOP} for details on the Yukawa interaction. \\ 
We present supplementary material in Appendix \ref{app-birds of prey}. In Appendix \ref{apps-random displacement}, we study the effect of random displacement to probe whether the observed repulsion can be captured by this only, which turns out to be not the case. We give the exact numbers of nests for the Common Buzzards, Goshawks, Eagles Owls and the split of the Eagle Owls per year each in Table \ref{tab-numbersbirds} in Appendix \ref{apps-number of nests}.
\subsubsection{The Setup}\label{sssec-motivation and biology}
\begin{figure}[t]
	\centering
		\includegraphics[width=0.32\linewidth,angle=0]{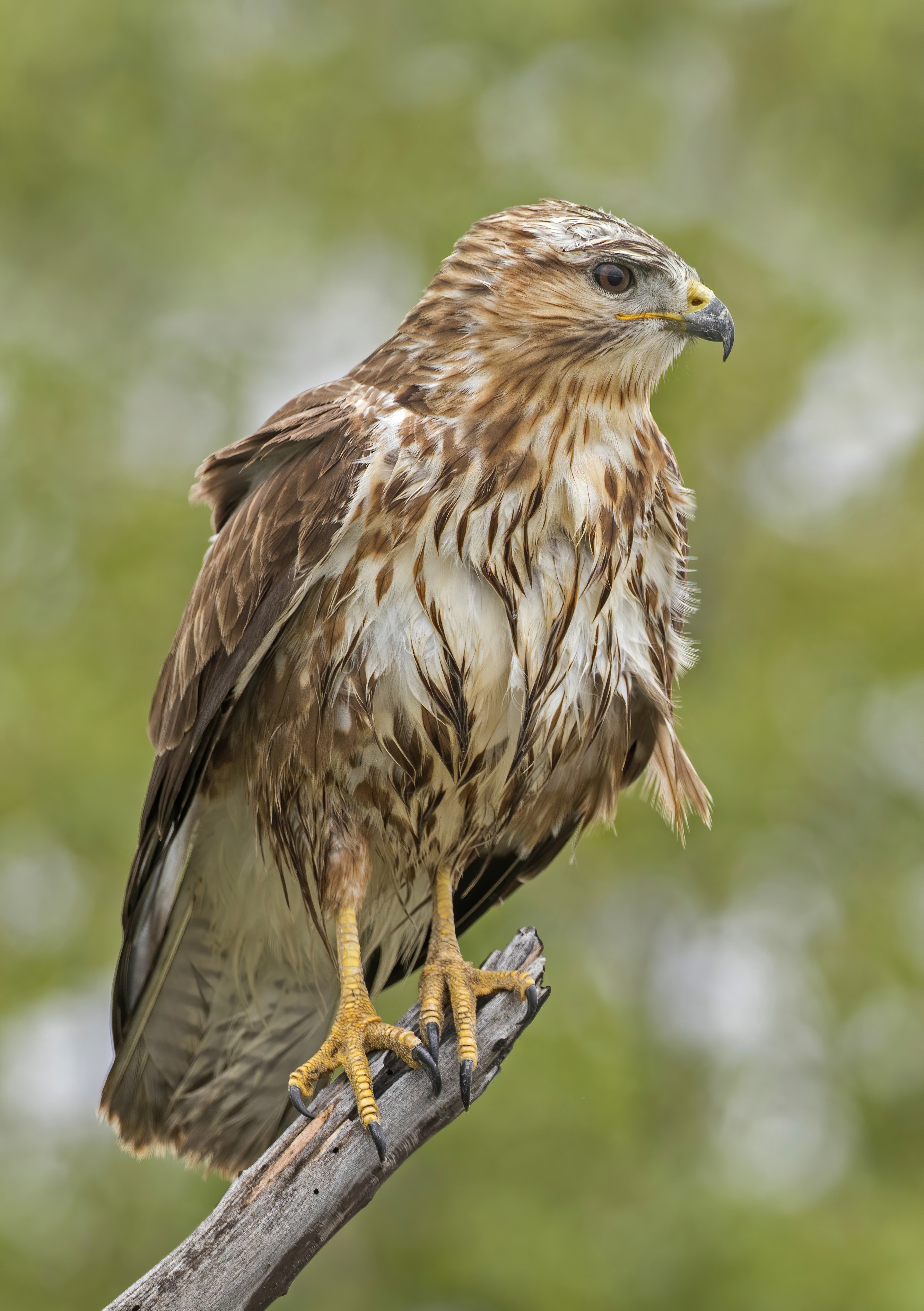}	
		\includegraphics[width=0.32\linewidth,angle=0]{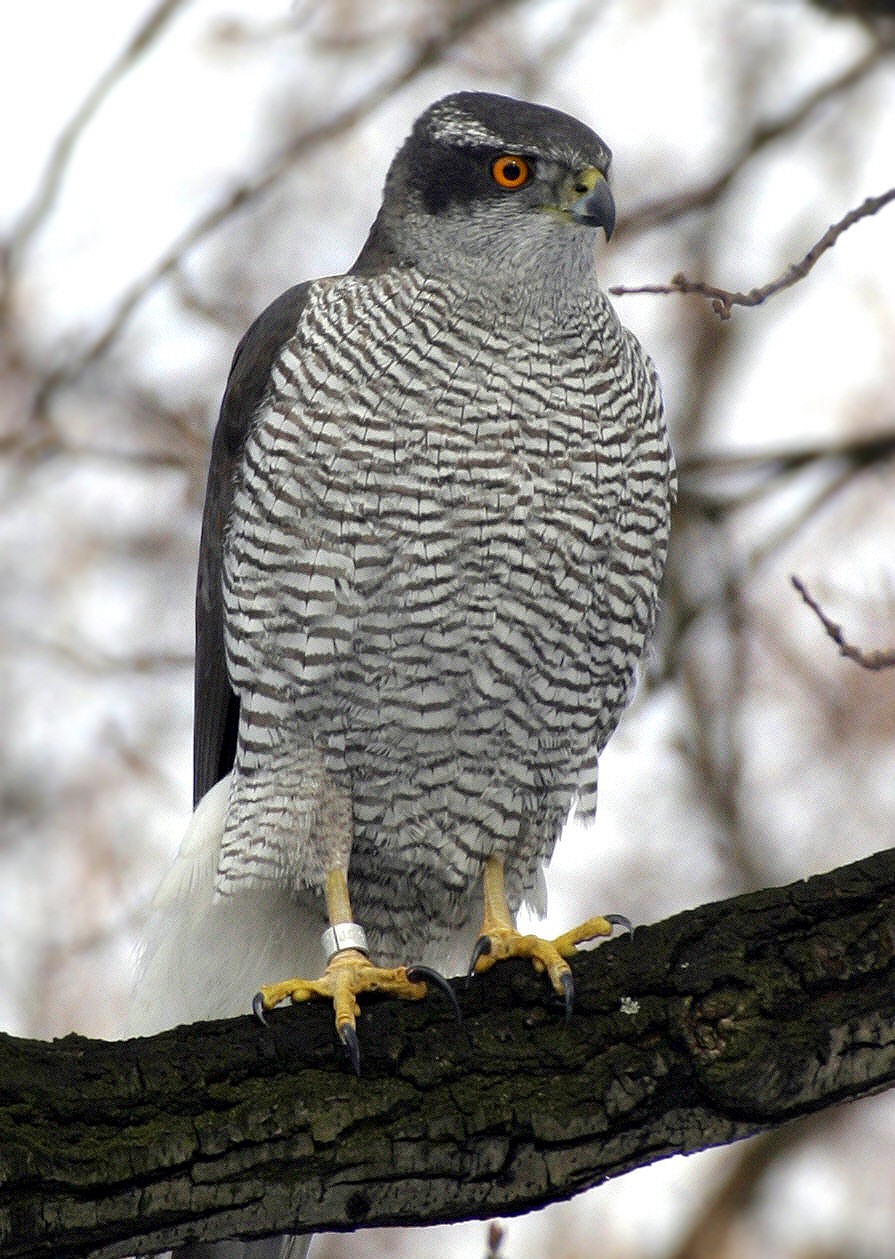}
		\includegraphics[width=0.32\linewidth,angle=0]{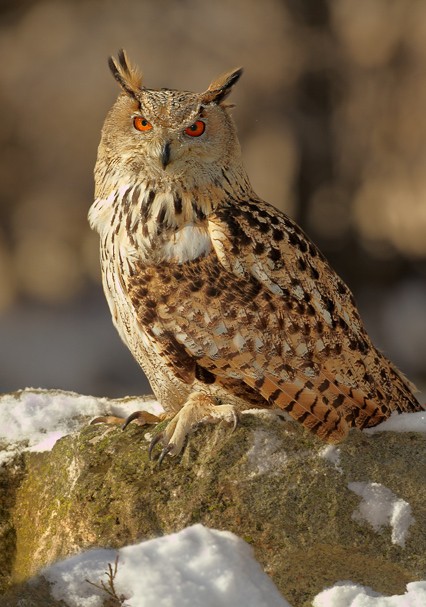}
	\caption[Pictures from \cite{picBuzzard,picGoshawk,picEagleOwl}, \textit{Pictures of the birds of prey: Common Buzzard (Buteo buteo L.), Goshawk (Accipiter gentilis) and Eagle Owl (Bubo bubo)}]{
		We give pictures of the three birds of prey for which we investigate the distribution of nests. We have from left to right: Common Buzzard (Buteo buteo L.), Goshawk (Accipiter gentilis) and Eagle Owl (Bubo bubo). While the Common Buzzard (males 525–1180 g, females 625–1364 g) and the Goshawk (males 517–1170 g, females 820–1590 g) are of comparable medium size, the Eagle Owl (males 1500–2800 g; females 1750–4200 g) is significant larger \cite{MCHK}. \cite{picBuzzard,picGoshawk,picEagleOwl}		
	}
	\label{fig-picBirds}
\end{figure}\noindent
Many point processes found in nature show correlations, and in particular repulsion between points in a characteristic, \textit{universal} way. This universal behaviour should now rings a bell to the reader: We pointed out that universality of local statistics is a feature of models from statistical mechanics, which is frequently studied. Hence, it seems reasonable to describe those point processes in nature via statistical models and then, we are only one step away from random matrices. What kind of understanding could be expected from such an approach to biological systems, which typically have a high degree of complexity?

A central question in biology is to understand the distribution of animals or plants in space and time. Furthermore, one would like to disentangle the effect of direct interaction within one species, between different species, and the effect of the environment. Such an understanding is of great value for conservation etc. 
Competition is one of the most ubiquitous features of ecology \cite{Begon}. Animals compete for limited resources with individuals of their own (intraspecific competition) or another species (interspecific competition), see \cite{Krueger2002a}. One of the easiest and hence most often used estimates of competition is the NN distance \cite{Clark}. It has been shown to be a useful measure, with the underlying assumption that a too near neighbour of the same or another species ultimately leads to decreased reproduction or survival \cite{CK10,MCHK}. 

Two examples where 2d point processes have been applied in biology are the spatial distribution of trees, see \cite{spatial dist plants} for a review, and the distribution of nests of birds of prey in space and time in a previous work \cite{buzzards 1}. Using data from \cite{KCNLGS-JM,CBK,MCHK} on the yearly locations of occupied nests of the Common Buzzard in an area of 300 km$^2$ in the Teutoburger Wald, Germany, the authors of \cite{buzzards 1} draw the following conclusions. The distribution of their nests differs significantly from 2d Poisson statistics of independent particles. The repulsion between the locations of nests of these highly aggressive birds of prey is quantified by fitting the spacing distribution between NN and NNN in radial distance to those of a static 2dCG of point charges in a confining (Gaussian) potential. It turns out that this single parameter fit yields a plausible description of the distributions of nests of Common Buzzards, where the inverse temperature $\beta$ serves as a measure for the coupling strength. Moreover, this simple parametrisation enables the authors to identify a change in time of the interaction strength measured by $\beta$, over the observed period of 20 years, as a function of the increasing population density. 
It is important to note that the fitted value of $\beta$ does not have a direct biological meaning but merely serves as a measure to quantify the repulsion, particularly the deviation from 2d Poisson statistics at $\beta=0$.
\begin{Remark}
	Let us comment, why these findings are highly attractive, while we do not have an underlying microscopic model that would lead to a NN spacing distribution between the nests of birds of prey of (approximately) 2dCG type. We turn our view back to a 1d point process, where a statistical mechanics description worked out very well. In \cite{Seba} a model for the spacing between parked cars and birds on a power line has been developed, leading to a Gamma-distribution. 
	Based on independently distributed symmetric spacings, it provides a link to the perception of space by humans and birds. Other applications are discussed in the review \cite{GMW}, such as the study of 740 spacings in the Sinai billiard, and 1726 spacings from nuclear scattering data. We state here the explicit amount of available data, to strengthen our analysis of the in a statistical sense rather small, but in the effort of collection huge, data set of the nests of birds of prey in the Teutoburger Wald. Over the course of the study, 3377 breeding attempts of Common Buzzard have been documented, over 423 for Goshawks and over 174 for Eagle Owls. While in the mentioned 1d examples, the development of a complete analytical picture based on a semi-classical expansion has taken over 25 years, the universality arguments set the course, cf. \cite{haake} and references therein.\\
	Because the absolute ordering into consecutive spacings between cars (birds) in 1d is absent in 2d, a simple generalisation of such an approach is not possible. We recall from the introduction Section \ref{sec-intro}, that the spacing distribution of the 2dCG has been observed to apply in several very different examples in physics due to universality.
	This is of course a strong assumption for the biological system, but it has worked surprisingly well for the Common Buzzard \cite{buzzards 1}.
	A plethora of models have been formulated to explain intra- and interspecific competition and density dependence in ecological time series, and they commonly make several ecological assumptions. Here, we test how powerful a simple approach from statistical mechanics might be, that does not carry any ecological assumptions with it.
\end{Remark}
The successful study of the Common Buzzard via this 2dCG approach in \cite{buzzards 1} inspired us to extend this analysis for the three species of competing birds of prey living in the observed area, the Common Buzzard, Goshawk and Eagle Owl, although less statistics are available for the latter two species, cf. Table \ref{tab-numbersbirds}. Annual data for the locations of their nests over the same area and period of time have been collected \cite{CBK,MCHK,CPK,Mueller}. We present pictures of the birds under consideration in Figure \ref{fig-picBirds} to allow the reader to get a feel about the subject we study. While the typical size of the Common Buzzard (males 525–1180 g, females 625–1364 g) and the Goshawk (males 517–1170 g, females 820–1590 g) is comparable, the Eagle Owl (males 1500–2800 g; females 1750–4200 g) is significant larger \cite{MCHK}. This extended study allows us to try to quantify the interspecies interaction between each two species, and to compare it to the intraspecies interaction found within each species individually. \begin{figure}[t]
	\centering
	\begin{minipage}{0.47\linewidth}
		\includegraphics[width=\linewidth,angle=0]{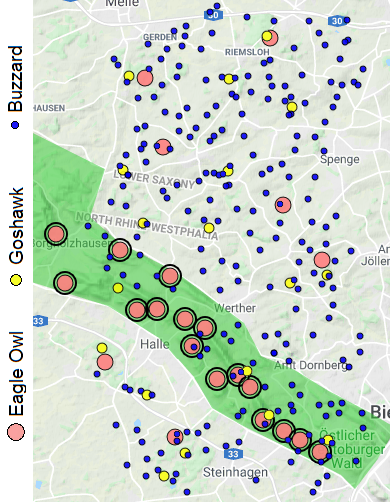}	
	\end{minipage}
	\hspace{10pt}
	\begin{minipage}{0.44\linewidth}
		\includegraphics[width=\linewidth,angle=0]{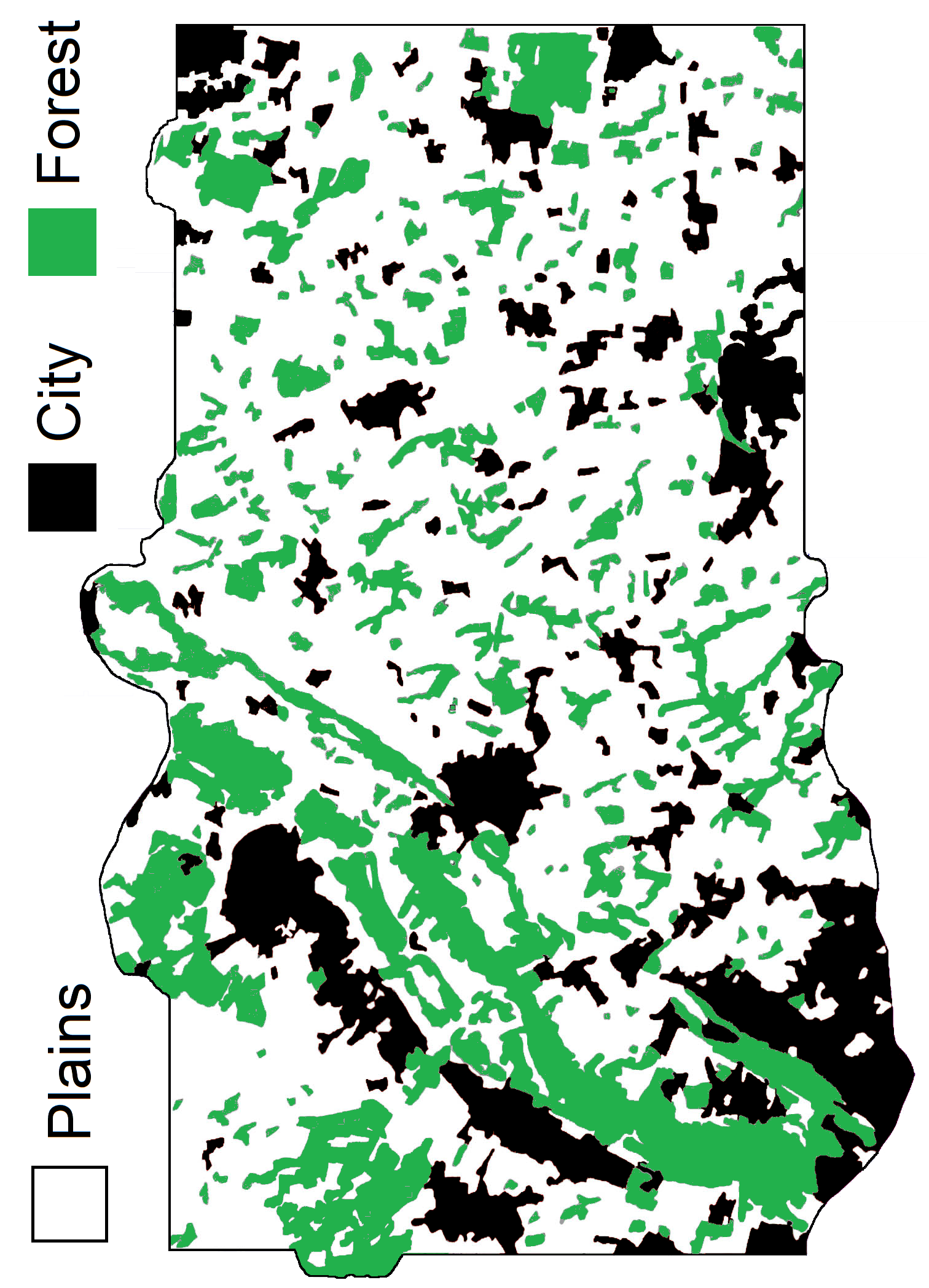}
	\end{minipage}
	\hspace{10pt}
	\caption[Figure 1 from \cite{ACKMOP}, \textit{Distribution of occupied bird nests in 2020 and classification of the observed area}]{
		\textbf{Left:} Distribution of all observed occupied bird nests in the year 2020: Common Buzzards (dark blue points), Goshawks (yellow points) and Eagle Owls (large pink points). The green shaded area marks the approximate extend of the Teutoburger Wald based on the roads on either side, cf. right plot. Eagle Owl nests within this green area are distinguished by a black ring around the large pink points. The extent of the monitored area is roughly $12\times25$ km. \textbf{Right:} A classification of the observed area in three types of terrain: City (black), forest (green) and cultivated land in the plains (white). \cite[Figure 1]{ACKMOP}		
	}
	\label{fig-PlotBirds}
\end{figure}\noindent
\\
\\
\textbf{Data Collection}\\
\\
\noindent
Let us know turn to the explicit description of the collection of data we use and the geographical setup. The locations of occupied nests of three kinds of birds were collected in late winter and early spring during the years 2000-2020 in an area in and around the Teutoburger Wald close to Bielefeld. The three species are the Common buzzard (Buteo buteo L.), Goshawk (Accipiter gentilis) and Eagle Owl (Bubo bubo) (shown in Figure \ref{fig-picBirds} from left to right). The area of 300 km$^2$ (8$^\circ$25'E and 52$^\circ$6'N) is located in Eastern Westphalia. It consists of two 125 km$^2$ grid squares and 50 km$^2$ edge areas, see Figure \ref{fig-PlotBirds} above for illustration. 
The main habitat of these birds of prey is the Teutoburger Wald and a cultivated landscape to the north and south of it. This is a low mountain region of height up to 315 m above sea level, and we will treat the area as approximately flat, that is 2d. 
All three species nest in forest patches. Their size varies from rows of trees to large patches, of more than 10 km$^2$. In total approximately 17\% of the study site is forested. The area has been intensively monitored for birds of prey, and the resulting spatial data have been published and used before extensively \cite{KCNLGS-JM,CBK,MCHK}.

The forest patches were visited in March and April each year to look for incubating birds, and if a nest was occupied in both months, the pair was classified as breeding. The locations of these nests were marked in large scale maps or using GPS-devices to monitor the spatial distribution of occupied nests. In Figure \ref{fig-PlotBirds} left, a snapshot for the year 2020 is shown, with occupied nests of the three species marked with coloured points, see caption. The classification of the monitored area into forest (green), city (black) or plains (white) is shown in Figure \ref{fig-PlotBirds} right.

The example year 2020 chosen has amongst the highest population for all three kinds of birds. We collect a few features that emerge already on eyesight. 
\begin{itemize}
	\item Common Buzzards (dark blue points) are much more abundant than Goshawks (yellow points) and Eagle Owls (large pink points). Whereas Common Buzzards and Goshawks spread relatively evenly over the monitored area, the density of Eagle Owls inside the Teutoburger Wald is much higher than outside.
	Because of this difference in distribution of Eagle Owl nests, we decided to split the group of Eagle Owls in two: Those nesting in the forest area (F) in green in the left plot (nests marked by a black rings around the pink dot), and those in the plains (P) (nests marked by a pink dot only). Notice that the plains were populated by Eagle Owls only after 2010, see Figure \ref{fig-EagleOwlGroup} bottom right below and Appendix \ref{apps-number of nests} Table \ref{tab-numbersbirds} for their annual populations. The marking of the forest area in green in Figure \ref{fig-PlotBirds} left is of course approximate. In particular, one might argue whether the two outliers of Eagle Owls south of the Teutoburger Wald should be counted as (F) or (P), in view of the two large patches of forest in Figure \ref{fig-PlotBirds} right. Although living in large forest patches, they are separated from the main forest by cities and are thus counted as (P) here.
	Such a choice does not change the quantitative picture, with the low statistics for Eagle Owls being difficult anyhow, cf. Table \ref{tab-numbersbirds}.
	\item There are holes in the population densities around cities (marked in black in Figure \ref{fig-PlotBirds} right). 
	Although effects of the edge of a density are known in point processes, we will disregard them here and treat all points as bulk points, see Remark \ref{rem-edge birds} below. 
	\item The mean spacing between nests within one species is very different for Common Buzzards, Goshawks, Eagle Owls (F) and Eagle Owls (P). It is approximately given by the inverse density (for a constant density). 
	The mean spacing in kilometres, e.g. $m_B=0.63$ km for the Common Buzzard and $m_G=2.63$ km for the Goshawk in 2020, is very different and encodes important biological information about the range of interaction within one (or different) species. Goshawks are among the most territorial of all European birds of prey and have large territory sizes of commonly between 5 and 50 km$^2$ and regularly make foraging trips of 5-10 km distance \cite{Kenward}. 
	Interactions between neighbouring territories are rather aggressive and intruders into the territory are regularly killed \cite{Kenward}. As the mean NN distance in our study site 
	$m_G=2.63$ km is very low compared to other study sites due to our high population density \cite{KCNLGS-JM}, regular antagonistic interaction between the territorial pairs is exceedingly likely.\\
	However, we are interested in comparing with (universal) features from spatial statistics in simple 2d point processes. Therefore, we have to apply unfolding explained in Section \ref{sssec-unfolding} above.
\end{itemize}
\begin{Remark}\label{rem-edge birds}
	We come back to the question how to handle the (inner) edges of the supporting area in Figure \ref{fig-PlotBirds}. In the Ginibre ensemble at $\beta=2$, the complex eigenvalue correlations at the edge of the droplet have also been investigated. They are universal and agree with the correlations at an inner edge, that is a droplet with a hole, that can be studied in the induced Ginibre ensemble \cite{Fischmann}. In our data there is no outer edge in Figure \ref{fig-PlotBirds}, it is simply given by the area of observation, and there are birds nesting outside the area containing dots. However, we do observe inner edges, as the cities of Bielefeld, Halle or Werther show up as holes in Figure \ref{fig-PlotBirds}. We have not been able to address such edge correlations, mainly due to lack of statistics, but also because it is not so clear where to draw the boundaries of the inner edges. In our analysis we have thus treated all points (nests) as bulk points.
\end{Remark}

\subsubsection{Interaction of Species: Study via 2dCG} \label{sssec-interact}
In this section, we study the NN and NNN spacing distribution of the birds of prey for each species and within each other. In order to quantify the repulsion, we fit to the approximate NN spacing ${p}^\textrm{(NN)}_{\textrm{sur},\beta_\textrm{eff}}(s)$, Eq. \eqref{eq-pbetasurmise}, from the surmise \cite{AMP}. As the surmise was derived from a $2\times 2$ matrix ensemble, cf. Section \ref{ssec-surmise}, we cannot treat the NNN distribution, which requires at least three eigenvalues. Secondly, we compare with the numerically determined NN and NNN spacings from \cite{AKMP} as functions of $\beta$, in steps of $0.1$. We refer to \cite{AKMP} for details about the numerics. Around 200000 spacings have been used there to generate these numerical distributions.\\
In a first step we repeat the analysis from \cite{buzzards 1} for Goshawks and Eagle Owls individually. This includes the dependence on the respective change in population over 20 years, ranging between 12-29 and 1-23 pairs, respectively. 
Because the spacing distribution for Eagle Owls differs considerably, whether they nest in the Teutoburger Wald (F) or in the adjacent plains (P), we have further split them into two groups. Due to a comparably low statistics, we have been unable to address the time dependence for the interspecific interaction between different species so far.\clearpage
\begin{figure}[t]
	\centering
	\begin{minipage}{0.02\linewidth}
		\begin{turn}{90}\hspace{-20pt} Eagle Owls (F) \hspace{15pt} Goshawks \hspace{40pt} Buzzards \hspace{30pt}\end{turn}
	\end{minipage}
	\begin{minipage}{0.92\linewidth}
		\centering
		\hspace{0.05\linewidth}2020 \hspace{0.2\linewidth} 2011-2020 \hspace{0.15\linewidth} 2000-2020\\
		\includegraphics[width=0.3\linewidth,angle=0]{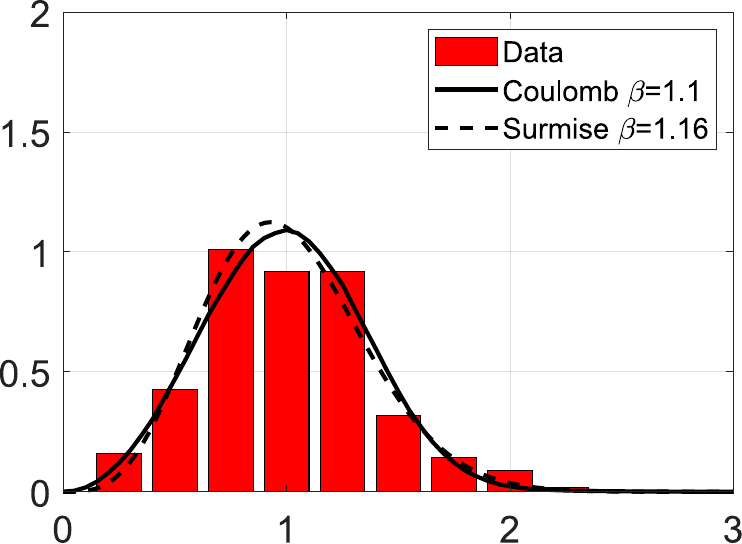}
		\includegraphics[width=0.3\linewidth,angle=0]{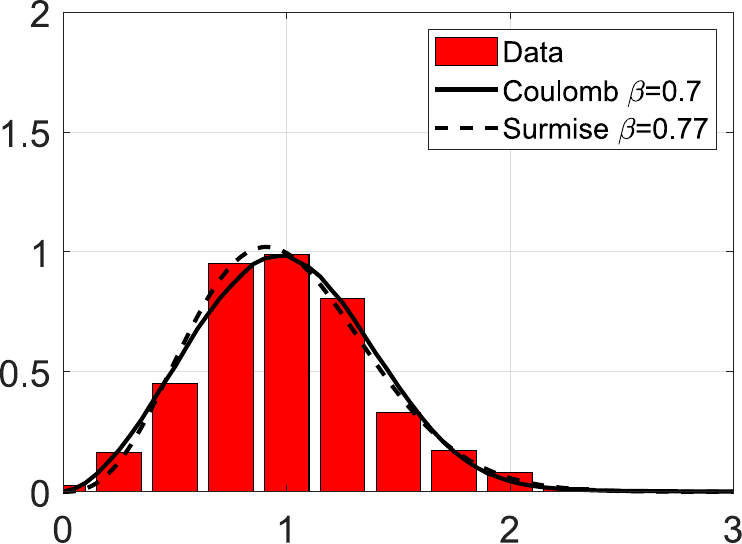}
		\includegraphics[width=0.3\linewidth,angle=0]{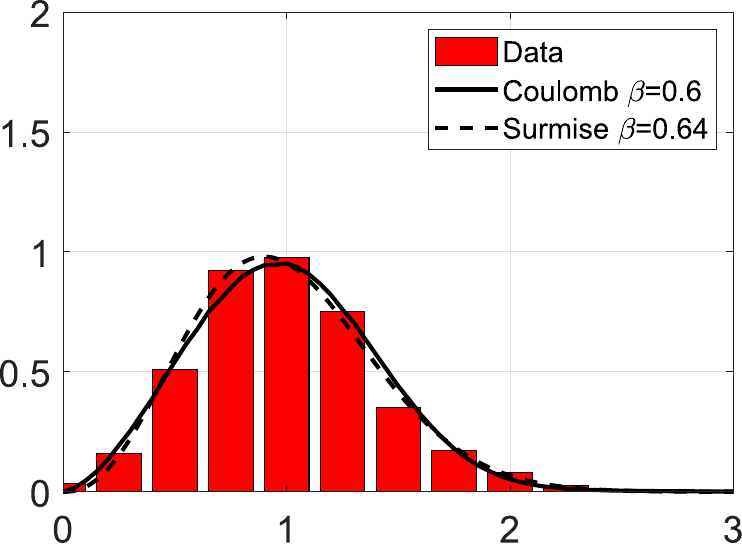}
		\\
		\includegraphics[width=0.3\linewidth,angle=0]{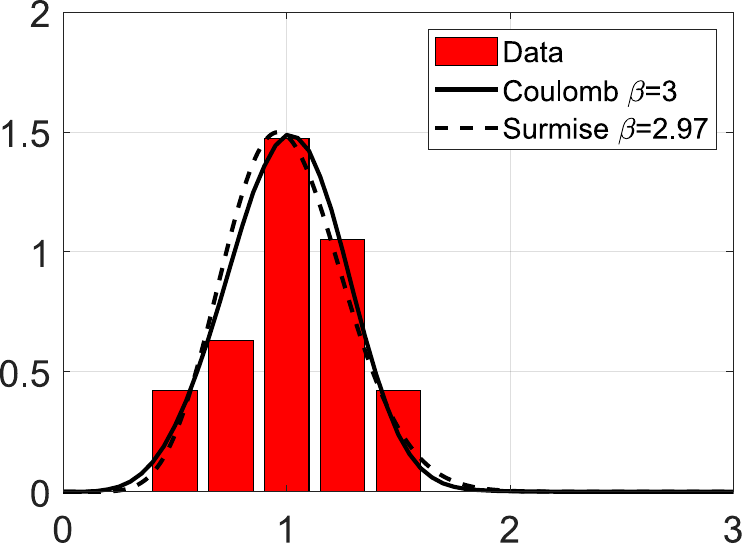}
		\includegraphics[width=0.3\linewidth,angle=0]{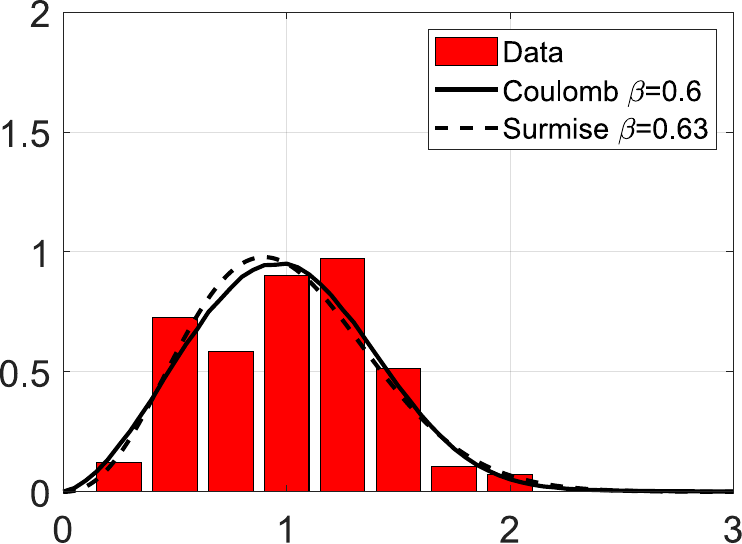}
		\includegraphics[width=0.3\linewidth,angle=0]{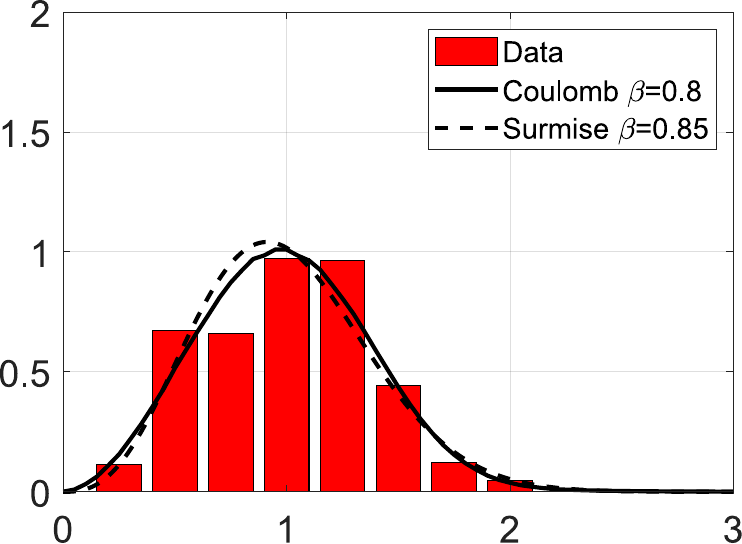}
		\\			
		\includegraphics[width=0.3\linewidth,angle=0]{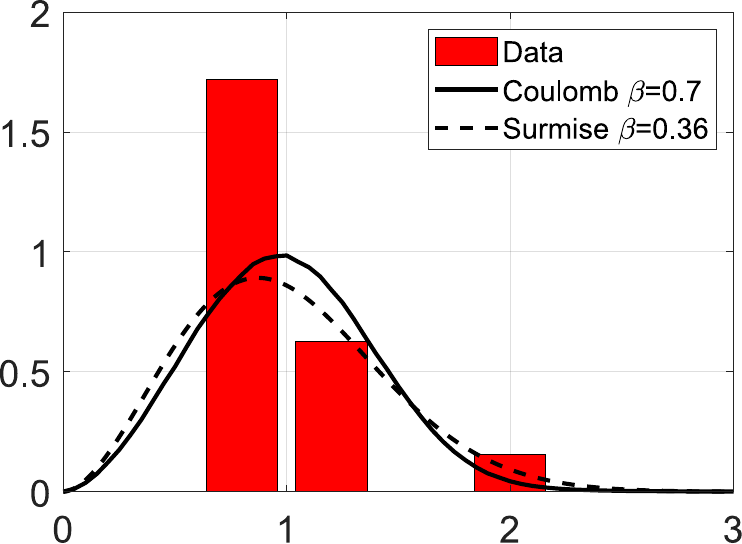}
		\includegraphics[width=0.3\linewidth,angle=0]{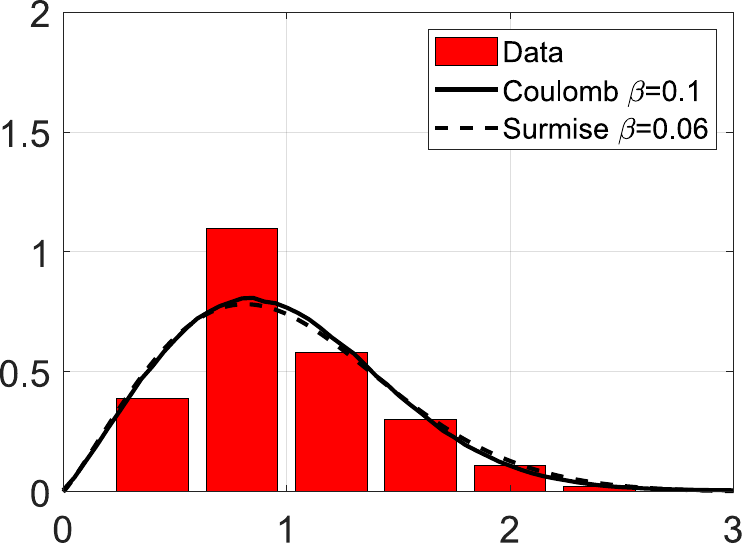}
		\includegraphics[width=0.3\linewidth,angle=0]{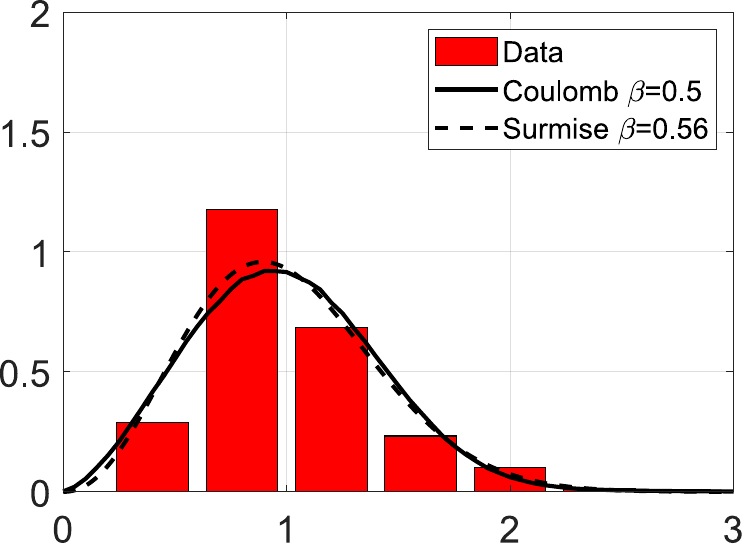}\\
	\end{minipage}
	\caption[Figure 3 from \cite{ACKMOP}, \textit{Comparison of the NN spacing distributions of birds of prey with 2dCG and the surmise \eqref{eq-pbetasurmise}}]{Comparison of the fitted $\beta$-values from the NN spacing distribution from 2dCG in steps of $0.1$ (full line), respectively from the surmise \eqref{eq-pbetasurmise} with continuous $\beta$ (dashed line). We present examples for fits for a single year 2020 (left column), an average over 10 consecutive years from 2011-2020 (middle column), and over all years (right column). Clearly when moving from individual years to time moving averages, the number of spacings for Goshawks and Eagle Owls (F) leads to an improved statistics and quality of the fits. \cite[Figure 3]{ACKMOP}}
	\label{fig-GroupingYears}
\end{figure}
\noindent
\textbf{Interaction within One Species: Common Buzzard, Goshawk and Eagle Owl}\\ \\ \noindent
Let us describe the procedure initiated in \cite{buzzards 1} for Common Buzzards, to quantify the supposedly repulsive interaction between these birds of prey, by fitting the NN and NNN spacing distributions between occupied nests. Each year is observed for each species and treated as an ensemble. In principle, for each year a value of $\beta$ can be determined for each species, as it is done in Figure \ref{fig-GroupingYears} left column for the NN spacing distribution in the year 2020 as an example. As it can be seen, the quality of the fit decreases rapidly with the amount of data available, in particular for the Eagle Owls (bottom left plot in Figure \ref{fig-GroupingYears}). 
In all figures we show both the fit using the CG (full line), with the spacings determined numerically in \cite{AKMP} in discrete steps of size $0.1$ for $\beta$, and the explicit formula \eqref{eq-pbetasurmise} for the surmise (dashed line), with $\beta$ varying continuously. We remark that the surmise \eqref{eq-pbetasurmise} was not available in the first study of the Common Buzzards \cite{buzzards 1}.\\  
The fits are done using the CDF, i.e. Eq. \eqref{eq-CDFsurmise} for the surmise, to avoid any dependence on the choice of binning (cf. Kolmogorov-Smirnov-distance \eqref{eq-KS distance}). To guide the eye, we nevertheless show the spacing distribution compared to histograms of the data.\\
The population of pairs of Common Buzzards and Goshawks per year is shown as red crosses in Figure \ref{fig-AllBirds} bottom left, respectively right, and for Eagle Owls in Figure \ref{fig-EagleOwlGroup} bottom in the forest (F, left) and plains (P, right), cf. Table \ref{tab-numbersbirds} in Appendix \ref{apps-number of nests}. Between Common Buzzards and Goshawks there is a factor of 10 in abundance, and typically another factor of 2 between Goshawks and Eagle Owls, with the latter being completely absent in the plains up to the year 2010. To remedy the resulting low statistics, we have introduced a time moving average, averaging over 10 consecutive ensembles (years) for all three species. The correspondingly averaged population is shown as a black line in the population plots in Figures \ref{fig-AllBirds} and \ref{fig-EagleOwlGroup}, respectively. For comparison, because of the better statistics for the Common Buzzards, in \cite{buzzards 1} a time moving average of 5 years was chosen, that gave a better resolution of the time dependence of the fitted $\beta$ as a measure of repulsion. 
In \cite{buzzards 1}, the fits for all individual years (compare Figure 4 left in \cite{buzzards 1}) are presented for the Common Buzzard. This result is very noisy and makes it difficult to see an overall trend in time, which is why we do not present such plots here.
We collect findings in Figure \ref{fig-GroupingYears}.
\begin{itemize}
	\item The fitted $\beta$-values vary over a wide range. For example, in the year 2020 in Figure \ref{fig-GroupingYears} \textbf{left column}, we find the strongest repulsion among Goshawks ($\beta=3$), followed by the Common Buzzard ($\beta=1.1$) down to Eagle Owls (F) with $\beta=0.7$. Apart from the latter plot with very low statistics, the surmise and CG fit agree well. We will introduce the time moving average for this purpose below. Notice that the continuous respectively discrete fit in $\beta$ adds to the discrepancy between the two, cf. \cite[Appendix B]{ACKMOP} for more details.
	\item All fitted values are far from 2d Poisson statistics at $\beta=0$. Let us emphasise again, that the fitted parameter $\beta$ does not have a biological meaning, but that it is the relative strength that allows us to compare between different species of birds, or their interaction in the next part of the study.
	\item The fitted $\beta$-value for the time averaged (10 consecutive years 2011-2020, including the single year 2020 from the left column) for each species is shown in Figure \ref{fig-GroupingYears} \textbf{middle column}. It is striking to see that in all three cases the ensemble average leads to a much lower $\beta$-value. This is especially true for the Goshawks ($\beta=0.6$), which now show a repulsion comparable to the Common Buzzards ($\beta=0.7$). 
	There is a clear trend that the fitted $\beta$ value goes down for the Goshawks over the years, see Figure \ref{fig-AllBirds} below. Notice that their population is peaked in 2012 (cf. Table \ref{tab-numbersbirds} in Appendix \ref{apps-number of nests}). 
	For the Eagle Owls (F) we obtain a fitted value $\beta=0.1$, being very close to 2d Poisson in this window of time average. Also, here the averaged fitted $\beta$ goes down over time, see Figure \ref{fig-EagleOwlGroup} top left.
	\item Finally, if we abandon any time resolution and make an ensemble average over all available years, we obtain closer values for $\beta$ for all three species, as shown in Figure \ref{fig-GroupingYears} \textbf{right column}. The strongest repulsion is again amongst Goshawks ($\beta=0.8$), followed by Common Buzzards ($\beta=0.6$), and almost equally among Eagle Owls (F) with $\beta=0.5$. 
	The particularly high $\beta$ for intraspecific repulsion in Goshawk makes a lot of ecological sense, as the Goshawk is one of the most territorial species in Europe \cite{KL01}. The Goshawk preys on larger prey and needs exclusive access to hunting areas \cite{KS96}. Therefore, intruding individuals are always met with high aggression and are regularly killed by the territory owners \cite{CS80,KL01,K07}, 
	cf. Figure \ref{fig-PlotBirds} for their relatively large spacings.
	All values are significantly above 2d Poisson statistics ($\beta=0$).
	Because of the strong growth in population for Common Buzzards, and for Eagle Owls (F), see Figures \ref{fig-AllBirds} and \ref{fig-EagleOwlGroup}, respectively, and Table \ref{tab-numbersbirds} in Appendix \ref{apps-number of nests}, the later years are weighted much higher here in the average over all years, because they contribute to more spacings per year.
\end{itemize}
	\begin{figure}[t]
		\centering
		\hspace{0.05\linewidth} $\beta$-fit Buzzards \hspace{0.2\linewidth}
		$\beta$-fit Goshawks\\
		\includegraphics[width=0.4\linewidth,angle=0]{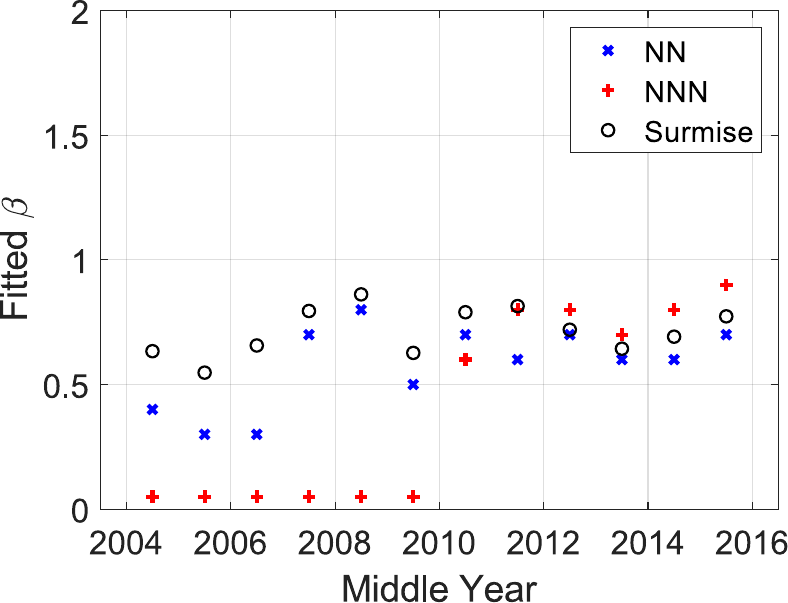}
		\hspace{10pt}
		\includegraphics[width=0.4\linewidth,angle=0]{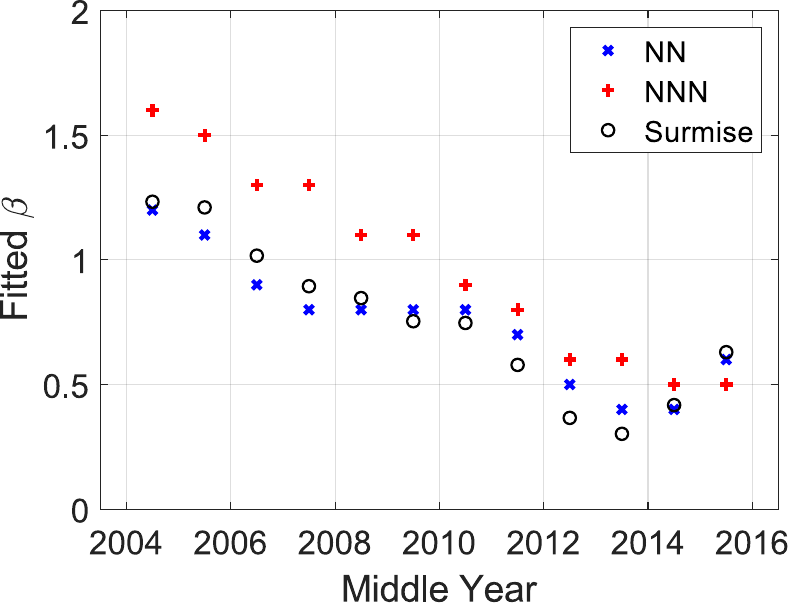}
		\\
		\hspace{0.05\linewidth} Common Buzzard population \hspace{0.14\linewidth}
		Goshawk population\\
		\includegraphics[width=0.4\linewidth,angle=0]{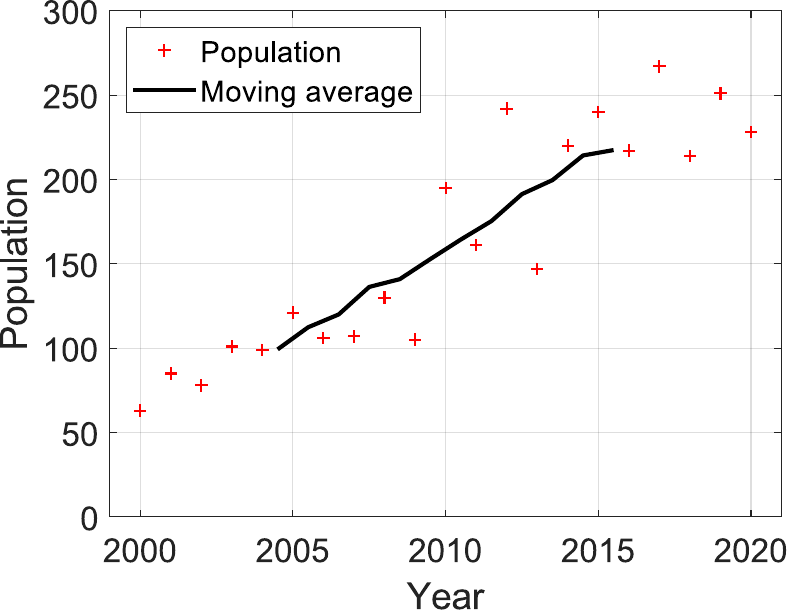}\hspace{10pt}
		\includegraphics[width=0.4\linewidth,angle=0]{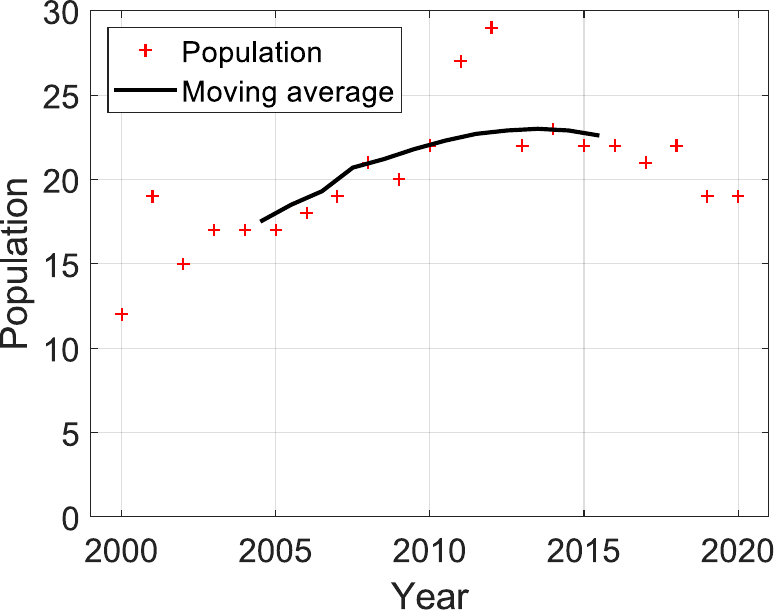}
		\caption[Figure 4 from \cite{ACKMOP}, \textit{Results on fitted $\beta$ for the NN and NNN distribution from the 2dCG and the surmise for Common Buzzards and Goshawks and population per year with time moving average}]{\textbf{Top row:} Fit of $\beta$ for the NN spacing for Common Buzzards (left) and Goshawks (right) from the 2dCG (blue crosses) and surmise (black circles), as well as for the NNN spacings distribution (red crosses). We use a time moving average of 10 years, where the middle year is indicated on the x-axis. \textbf{Bottom row:} Population of birds per year (red crosses) for Common Buzzards (left) and Goshawks (right), over the entire period of time 2000-2020. The resulting average populations using a time moving average of 10 years is shown in comparison (black line), where again the middle year is chosen as a label, as in the top plots. For a better comparison these points are connected by the line of the moving average. \cite[Figure 4]{ACKMOP}}
		\label{fig-AllBirds}
	\end{figure}
\noindent
Two general observations can be made from the results in Figures \ref{fig-GroupingYears}, \ref{fig-AllBirds}, and \ref{fig-EagleOwlGroup} for the $\beta$-values fitted to the time moving average. First, the values obtained using the surmise agree quite well with the NN spacings form the 2dCG, apart from the years with low statistics for the Eagle Owls (F). 
In particular, they follow the same trend as seen from the NN CG fits and thus can be used as a very simple approximate, but analytical tool. Second, the $\beta$-values obtained from NN and NNN fits do not agree in general, sometimes lying systematically below (Common Buzzards, Eagle Owls (F)) or above (Goshawks) the NN value. This discrepancy may be used to compare the observed interaction range to that of the Coulomb interaction. 
The same observation was already made in \cite{buzzards 1} and motivated us to introduce and study an interaction with shorter range, the 2d Yukawa interaction, see Section \ref{sssec-variations rpp} below. It turned out, however, that also with such a two-parameter family (with $\beta$ and $\gamma$) the phenomena of obtaining two different values for $\beta$ from NN and NNN remains, see the discussion there. Therefore, we kept the simpler 2d Coulomb interaction as a measure for repulsion.
\begin{figure}[t]
	\hspace{0.05\linewidth} $\beta$-fit Eagle Owls (F) \hspace{0.17\linewidth}
	$\beta$-fit Eagle Owls (P)\\
	\includegraphics[width=0.4\linewidth,angle=0]
	{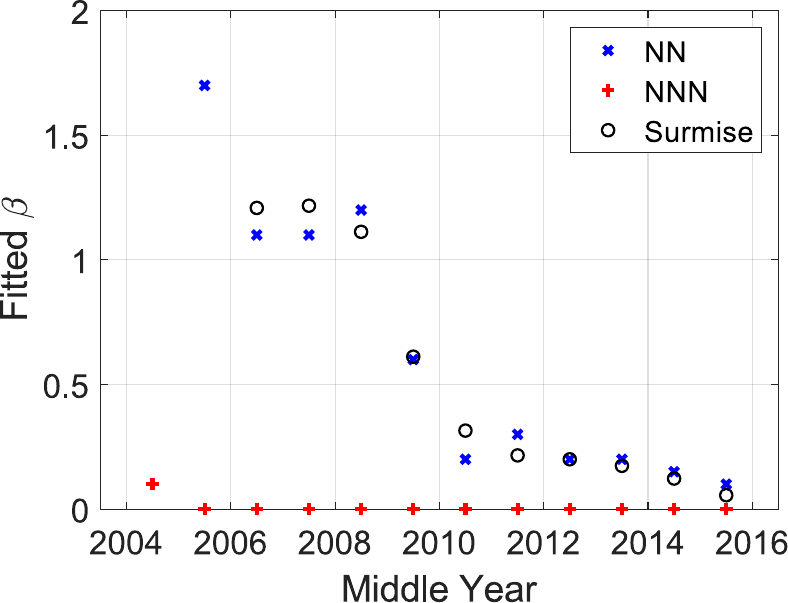}
	\raisebox{7pt}{
		\includegraphics[width=0.4\linewidth,angle=0]{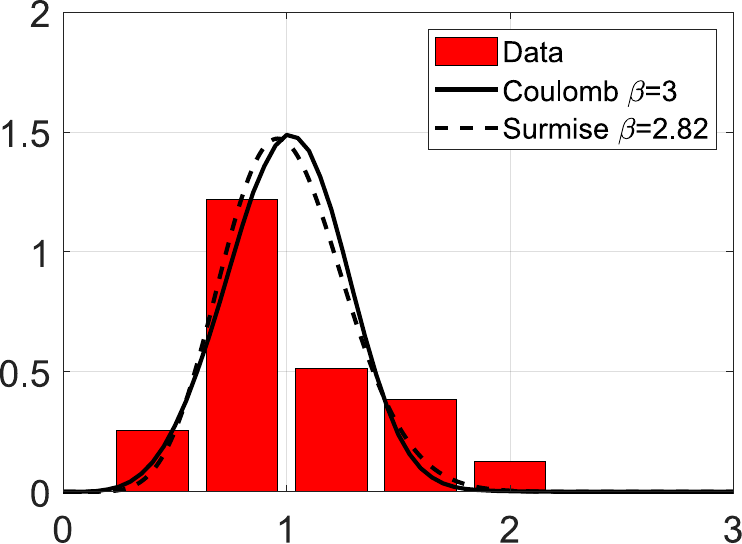}}
	\\
	\hspace*{0.05\linewidth}Eagle Owl (F) population \hspace{0.10\linewidth}
	Eagle Owl (P) population\\
	\includegraphics[width=0.4\linewidth,angle=0]{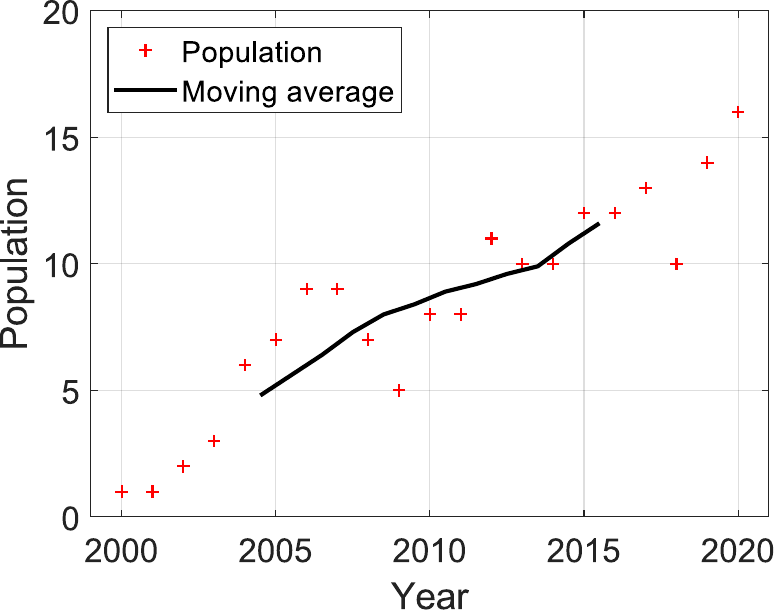}
	\includegraphics[width=0.4\linewidth,angle=0]{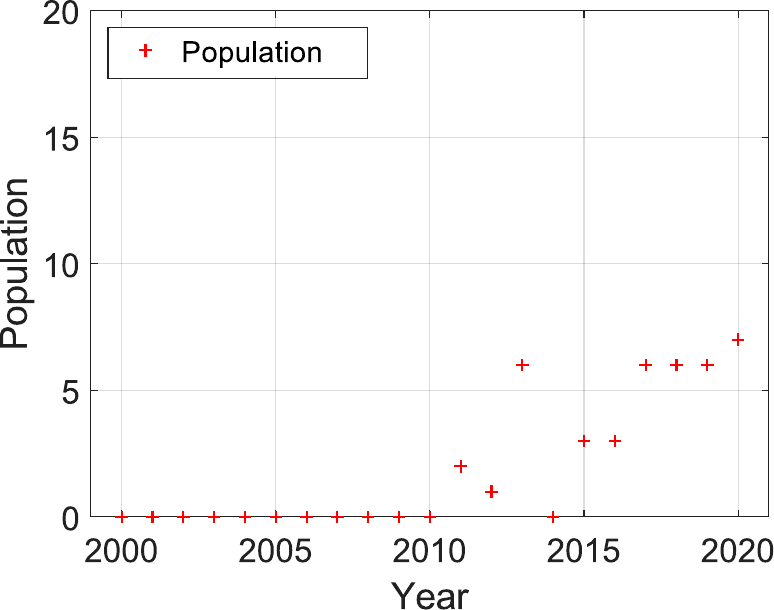}
	\caption[Figure 5 from \cite{ACKMOP}, \textit{Results on fitted $\beta$ for the NN and NNN distribution from the 2dCG and the surmise for Eagle Owls (F) and for Eagle Owls (P) averaged over all years and population per year (with time moving average)}]{The same plots as in Figure \ref{fig-AllBirds} for the Eagle Owls (F) are presented in the left column, with the corresponding $\beta$-fits for time moving averages (top left), and corresponding population with time moving averages in black (bottom left). Values above $\beta=2$ are not shown. Due to the lack of data for the Eagle Owls (P), see the corresponding population on the bottom right, we only present a fit over all years containing 40 spacings (top right), cf. Table \ref{tab-numbersbirds}. Notice the lack of nests in the plains (right) until 2010. \cite[Figure 5]{ACKMOP}
	}
	\label{fig-EagleOwlGroup}
\end{figure}\noindent
In Figures \ref{fig-AllBirds} and \ref{fig-EagleOwlGroup} we do not give error bars for the fitted values of $\beta$. In the previous studies in \cite{buzzards 1} an attempt to estimate the error was made by fitting $\beta$ to approximately the same number of points from a 2dCG simulation, that is $N=200$ for the Common Buzzards. This estimate (which is not related to the data of occupied nests) seemed to overestimate the fluctuations, e.g. when comparing to a linear fit for the trend in $\beta$. Because in the ensembles of Goshawks and Eagle Owls (F) the number of nests per year is smaller by (at least) a factor of 10, we refrain from giving such rough estimates for the errors.  

Let us summarize the findings from Figures \ref{fig-AllBirds}, \ref{fig-EagleOwlGroup} on the (evolution of the) NN and NNN statistics for the three birds of prey under consideration.
\begin{itemize}
	\item The results for the \textbf{Common Buzzards} from Figure \ref{fig-AllBirds} \textbf{left} agree with the previous studies in \cite{buzzards 1}. The NN values vary in range between $\beta=0.3$ and 0.8 over the observed period. Although they still fluctuate considerably for this time moving average, an increasing trend can be identified. In contrast, the NNN values remain close to $\beta=0$ for half of the observed period and then jump approximately to the NN values from the mid-year 2010 onwards. In comparison, the population growth is approximately linear from around 100 pairs on average to above 200 (10-year average given by the black curve in Figure \ref{fig-AllBirds} bottom left). Apparently below a certain population threshold the interaction range is rather limited to NN and then increases in range to be more 2d Coulomb like up to NNN. 
	Notice that because of unfolding the data, the mean spacing is the same in all years. Thus, the trivial effect of having a smaller spacing for a larger density is removed here.
	\item For the \textbf{Goshawks} we observe in Figure \ref{fig-AllBirds} \textbf{right} that there is a clear trend for the fitted $\beta$-value to go down from approximately 1.2 to 0.4 for NN and from 1.6 to 0.5 for NNN. In contrast to the Common Buzzards, the NNN values are systematically above the NN values, apart from the last middle year. This means that the interaction range is longer than 2d Coulomb, which goes in line with the high aggressiveness of this species discussed earlier. The repulsion decreases over time, and the population development is also different from the Common Buzzards when comparing the bottom rows in Figure \ref{fig-AllBirds}. Notice the difference in abundance of a factor of 5-10 between the two species. Although the Goshawk population fluctuates between 12 and 29 pairs, the average only slightly goes up from 17.5 to 22.5, stabilises and then slightly goes down again. Thus, a different factor than the Goshawk population seems to be at work here. If it lies in the interaction between the two other species, it will be answered (negatively) in the next paragraph below.
	\item We recall that the \textbf{Eagle Owls} have the lowest statistics, see Figure \ref{fig-EagleOwlGroup} and Table \ref{tab-numbersbirds}, about a factor of 2-10 in abundance below the population of Goshawks. This makes the interpretation of the findings very difficult, and the results should be treated with caution, especially for the much smaller group nesting in the plains (P). 
	For the nesting in the forest (F), the $\beta$ value from NN clearly decreases from large values above 1.5 down to 0.2-0.1, which is almost comparable with 2d Poisson at $\beta=0$. 
	In contrast, the NNN values remain consistent with $\beta=0$ over the entire period. A comparison with the population development from 5 pairs on average to 12 shows an approximate linear increase on average, see Figure \ref{fig-EagleOwlGroup} bottom left. The effect is thus opposite to the Common Buzzards, the Eagle Owls (F) show a decreasing repulsion, with short range interaction, despite an increase or stabilisation in population, respectively. It should be noted that for relatively few spacings of the order 20, the distribution is strongly peaked, often leading to larger values for the fitted $\beta$. For a better resolution the first NN value at $\beta=3$ is suppressed in Figure \ref{fig-EagleOwlGroup} top left.
\end{itemize}
\noindent
The time dimension of the strength of the repulsion also reflects the species' biology as well as population trends in the study area. For the Common Buzzard, we see an increase of $\beta$ over time, most likely due to the increase in the population density until carrying capacity might have been approached in the last five years. The strength of intraspecific competition is expected to increase with increasing population density. With regard to the Goshawk, a decreasing $\beta$ over time is mirrored by a decrease in population density over the last five years, most likely due to displacement by Eagle Owls \cite{MCHK}. Why $\beta$ shows a decrease for Eagle Owls, the population of which has increased very rapidly over the last 20 years, is difficult to explain. It could be that the population is still not approaching carrying capacity and hence progressively smaller repulsions measured through NN spacings are observed.

The reader might wonder whether the finding $\beta \approx 0$ indicates a complete absence of repulsion and this is consistent with the described aggressive behaviour of the birds. We investigate this point in the subsequent Section \ref{sssec-variations rpp}, taking the fact into account that the birds nest only in a (fractal) area in the 2d plane shown in Figure \ref{fig-PlotBirds}.
\\
\\
\textbf{Interaction Among Two Species}\\
\\
We turn our view now to the repulsion between all combinations of two different species of birds, while keeping the two groups of Eagle Owls (F) and (P) separate, see Figure \ref{fig-PlotBirds} left. Following the 2dCG analogy, it would be perhaps natural to associate to each species a different charge, according to their observed (average) repulsion strength. However, such a multi-component 2dCG would depend on the ratio of the different charges, which may change over the years. 
If one charge is very abundant, we may consider the others to be screened, but this is also not always the case. The multi-parameter fit is therefore not easy to do, and we stick to the simple one-parameter fits for each pair of species instead, see Figure \ref{fig-Interspecies} below. We thus fit one $\beta$-value to the spacings between species $A$ and $B$. 

	\begin{figure}[h]
		\begin{minipage}{0.02\linewidth}
			\begin{turn}{90}\hspace{5pt} Eagle Owls (P) \hspace{10pt} Eagle Owls (F)\end{turn}
		\end{minipage}
		\begin{minipage}{0.52\linewidth}
			\hspace{0.0\linewidth} vs. Common Buzzards
			 \hspace{0.1\linewidth} 
				vs. Goshawks \hspace{0.0\linewidth} \hspace{0.1\linewidth}\\
			\includegraphics[width=0.48\linewidth,angle=0]{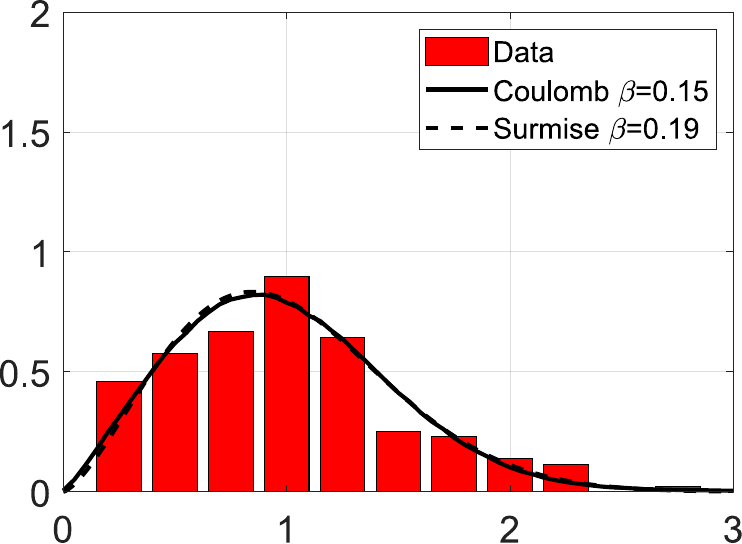}
			\includegraphics[width=0.48\linewidth,angle=0]{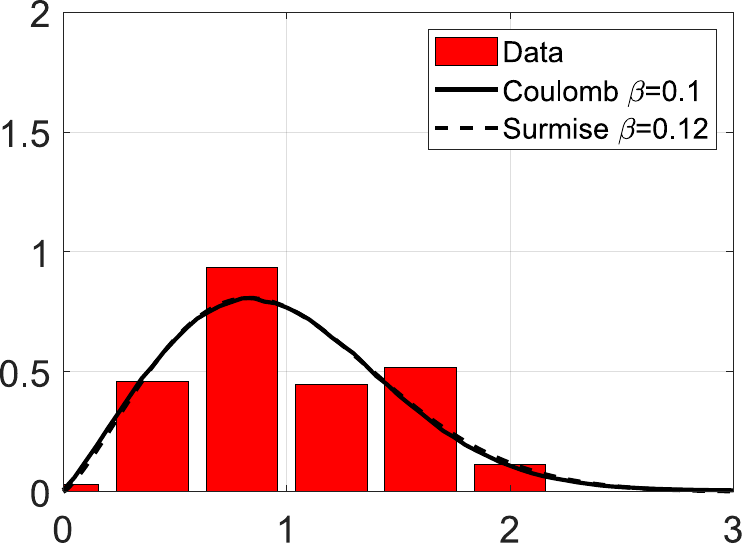}\\
			
			\includegraphics[width=0.48\linewidth,angle=0]{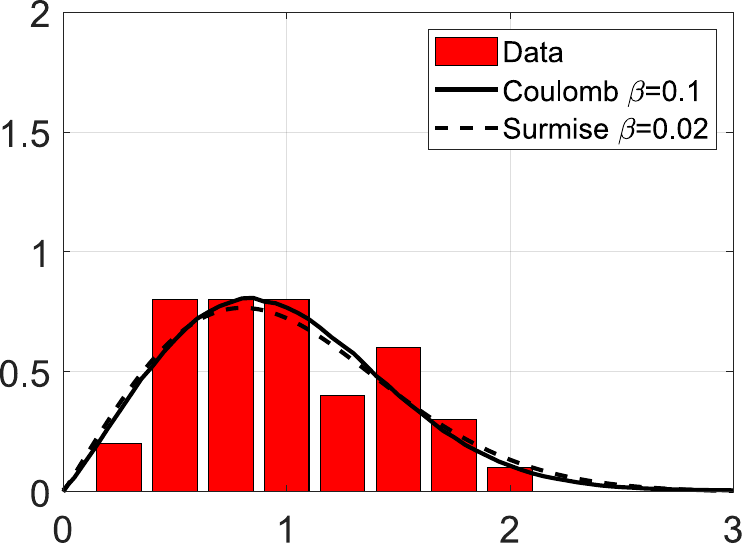}
			\includegraphics[width=0.48\linewidth,angle=0]{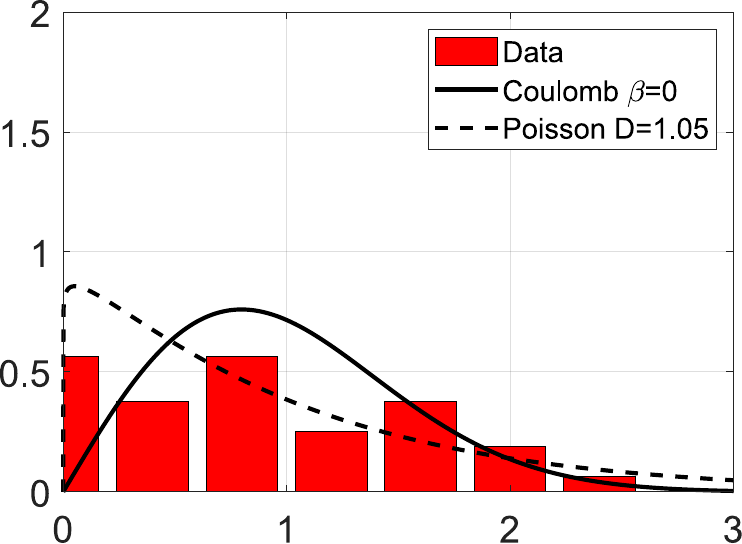}\\
			\centering
		\end{minipage}
		\begin{minipage}{0.02\linewidth}
			\begin{turn}{90}\hspace{5pt} Goshawks 
		\end{turn}	\end{minipage}
		\begin{minipage}{0.31\linewidth}
			vs. Common Buzzards \vspace{2pt}\\
			\includegraphics[width=0.96\linewidth,angle=0]{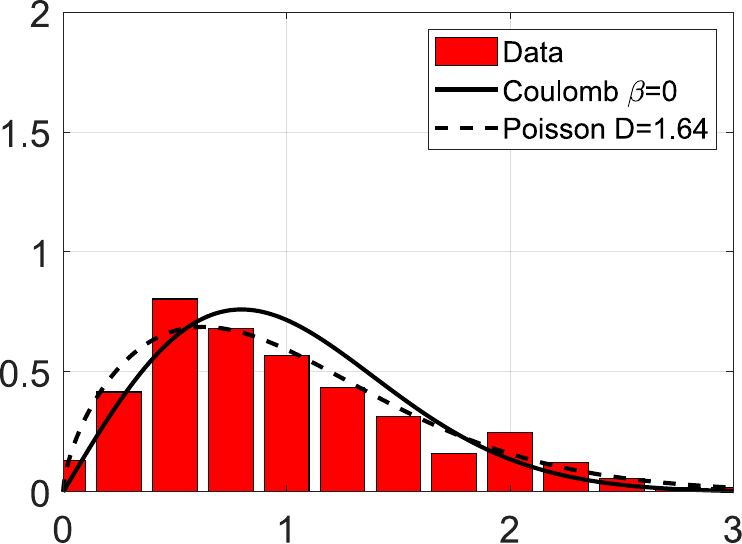}\\
			\centering
		\end{minipage}
		\caption[Figure 6 from \cite{ACKMOP}, \textit{2dCG and surmise fits of the NN distributions between all pairs of bird of prey}]{2dCG and surmise fits for the $\beta$-values of the NN spacing distributions (full respectively dashed line) between all pairs of species for all years, cf. Table \ref{tab-numbersbirds}, where for Eagle Owls we distinguish (F) and (P) in the top row and bottom row, respectively. When the best fit to the 2dCG gives $\beta=0$ (bottom middle and right), we fit instead to the Poisson distribution with $D<2$ from Eq. \eqref{eq-Poid-NN} (dashed line), with the resulting value for $d$ given in the inset. \cite[Figure 6]{ACKMOP}}
		\label{fig-Interspecies}
	\end{figure}

In order to avoid an over counting of spacings, we always go through the number of points (nest) of the less abundant species between the two and find its NN to the more abundant species in a given year, defining an ensemble. 
Because in all years the number of pairs of Common Buzzards is larger than the pairs of Goshawks, which in turn is larger than the number of pairs of Eagle Owls ((F) or (P)), the ordering is clear. For example, for the Common Buzzard - Goshawk interaction we go through all Goshawk nests in a given year and find its NN Common Buzzard nest. 
Our statistic is thus limited by the population of Goshawks or Eagle Owls, respectively. 
For that reason, we choose to make an average over all ensembles of 21 years in this analysis. For the comparison with Eagle Owls (F) we have 174 spacings (top left and middle plots in Figure \ref{fig-Interspecies}), and for Eagle Owls (P) we have 40 spacings (bottom left and middle plots in Figure \ref{fig-Interspecies}). In the comparison Goshawk to Buzzard (right column Figure \ref{fig-Interspecies}) we have 423 spacings available, cf. Table \ref{tab-numbersbirds}. Furthermore, we restrict ourselves to NN spacings only. The unfolding of the data has been made using all occupied nests from all three species per year, to get an approximate mean global density. This certainly represents a simplification but does not introduce any extra bias.

The following picture emerges from Figure \ref{fig-Interspecies}. All fitted values are closer to $\beta=0$, with the largest repulsion observed between Common Buzzards and Eagle Owl (F) at $\beta=0.15$. Thus, the 2dCG and surmise values are very close. The NN spacing between Goshawks and Eagle Owls (P) is not well fitted by Poisson at $D\leq 2$, nor by a 2dCG at $\beta>0$.

Because we average over all years, we have to compare with the values in Figure \ref{fig-GroupingYears} right column, for the repulsion within each species for all years: $\beta=0.6$ for Common Buzzards, $\beta=0.8$ for Goshawks, and $\beta=0.5$ for Eagle Owls (F). 
Clearly, the repulsion measured in $\beta$ is much weaker between different species than within one species. 
The strongest interspecies repulsion measured in this way is between Common Buzzards and Eagle Owls (F) with $\beta=0.15$, followed by that between Goshawks and Eagle Owls (F), and Common Buzzards and Eagle Owls (P) at $\beta=0.1$ each. Do these low values of $\beta$ close to zero indicate, that there is almost no repulsion between different species, as in a random Poisson point process in 2d? As previously mentioned, we investigate in Section \ref{sssec-variations rpp} if the random point process on the forest patches reduces the dimension, finding $D=1.66$ as effective dimension. 
This seems to indicate that even low $\beta$-values close to zero represent a certain repulsion. For the interaction between Common Buzzards and Goshawks we (perhaps coincidentally) find that the best fit is obtained by the Poisson distribution \eqref{eq-Poid-NN} close to this effective dimension at $D=1.64$, and in that sense there is no repulsion also according to this measure between the two. 
The plot with the lowest statistics between Goshawks and Eagle Owls (P) could be interpreted in the same way, beside the low quality of the fit. 

The finding of a small repulsion between the three species is in line with empirical findings that intraspecific competition is commonly stronger than interspecific competition in avian predators \cite{Krueger2002a,Krueger2002b}. The effect of Eagle Owls is rather different and has been shown to be clearly negative for both Common Buzzard and Goshawk \cite{CBK,MCHK}. These interactions are, however, very dynamic at a very small scale \cite{CBK,MCHK}, and hence the 
simplistic 2dCG picture (using the time averaging for better statistics) employed here is perhaps at its limits in detecting these interactions.

\subsubsection{The Poisson Point Process Revisited} \label{sssec-variations rpp}
In the discussion of the results in Section \ref{sssec-interact}, we raised the following two points: First, we noted that the repulsion strength $\beta$ does not agree between the NN and NNN fit, cf. Figures \ref{fig-GroupingYears} (top) and \ref{fig-EagleOwlGroup} (top left). This can be interpreted in the sense that the correlation length of the actual system is shorter ($\beta_\textrm{NNN}$<$\beta_\textrm{NN}$) / longer ($\beta_\textrm{NNN}$>$\beta_\textrm{NN}$) than in the 2dCG. Therefore, we study in \cite{ACKMOP} the 2d Yukawa interaction, which allows for a varying correlation length. However, we find that also in this two-parameter model, the NN and NNN distributions do not agree in the parameters and conclude that we can stick to the "simpler" 2dCG fit with only one parameter $\beta$. We refer the reader to \cite[Section 4.2]{ACKMOP} for details.\\ A second question, that was raised in Section \ref{sssec-interact} is, whether resulting values of $\beta\approx 0$ show the absence of interaction. While we studied the CG in 2d in the previous section (including the 2d Poisson point process at $\beta=0$), we now make use of the general $D$-dimensional Poisson process (cf. Section \ref{ssec-point processes}) and allow for a $D<2$. Finding an effective dimension of $D=1.66$ for the forest patches in Figure \ref{fig-PlotBirds}, we conclude that results for $\beta\approx 0$ in Section \ref{sssec-interact} represent a repulsion through. After this extensive discussion of the Poisson point process, which consists per definition of independent particles, we have to ask whether this "independence" is given here? From the data collection, it is known that "old" nests are reused in consecutive years. We investigate this reusage below.\\
\\
\textbf{Application of the Poisson Point Process in Varying Dimension $\boldsymbol{D}$}\\
\\
We investigate if the 2d Poisson point process used so far indeed describes the situation of nests placed as independent random variables in the plane. 
It has been observed that all three species of birds of prey invariably breed in forest patches. 
A look on Figure \ref{fig-PlotBirds} right thus poses the question, if the forest represents a lower dimensional, fractal domain in the full 2d area. Notice that we completely ignore the elevation of the terrain, by treating it as 2d. It varies from about 70 m to 300 m in height above sea level, compared to the dimension of roughly $12\times25$ km extent of the monitored area.

We will try to answer this question by generating a Poisson point process solely on the forested area, determine its NN spacing distribution numerically, and compare it to the analytic result for the NN spacing distribution of a Poisson point process for general dimension $D$, by fitting $D$ as a free parameter. We use the well-known distribution \eqref{eq-Poid-NN}.

\begin{figure}[t]
	\centering
	\includegraphics[width=0.4\linewidth,angle=0]{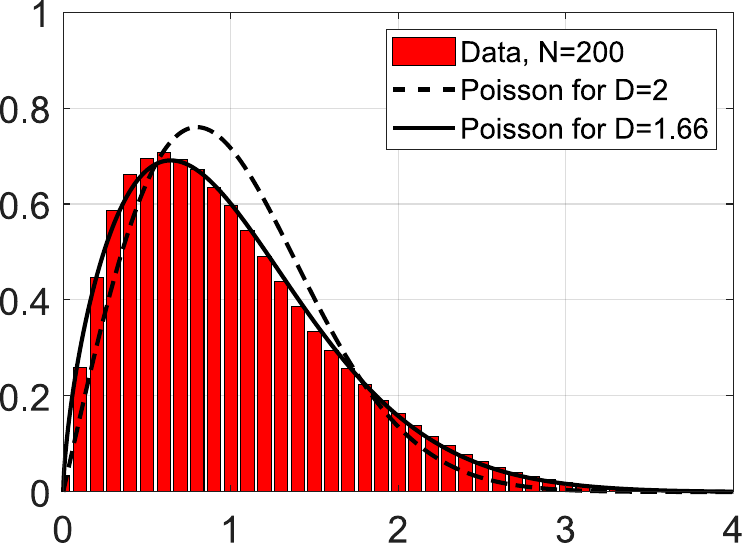}
	\caption[Part of Figure 7 from \cite{ACKMOP}, \textit{Fit of the dimension $D$ of a Poisson point process to the forest area}]{Fit of the dimension $D$ in \eqref{eq-Poid-NN} (solid curve) to the random point process generated on the forested area as described in the main text (histograms). From the fit we obtain an effective dimension of $D=1.66$. For comparison, the NN spacing distribution in $D=2$ is also shown (dashed curve). \cite[Cf. Figure 7]{ACKMOP}}
	\label{fig-PoissonForest}
\end{figure}

The Poisson point process on the forest is generated as follows. A certain number $N$ of points is distributed independently in the monitored area. If they fall onto a green forest patch in Figure \ref{fig-PlotBirds} right, they are accepted, else they are rejected. Note that we accept all areas with trees, i.e., not just the main forest depicted as a green band in Figure \ref{fig-PlotBirds}. 
In the limit $N\gg1$ with a fixed area, we would recover a collection of two-dimensional patches, given the number of points per patch is large enough. In order to capture roughly the same length scale as the distances between occupied nests, we stop after having $N=200$ accepted points, which is about the number of Common Buzzard pairs in the entire area. For the fit we use a Kolmogorov-Smirnov-distance \eqref{eq-KS distance}, see Figure \ref{fig-PoissonForest} above for the result.

\noindent

We find that the Poisson point process generated as described indeed leads to slight reduction of dimension from $D=2$ to an effective dimension of $D=1.66$. Consequently, we may conclude that a fit of data to a CG with $\beta=0$, or equivalently the Poisson point process with $D=2$, still reflects a small repulsion, compared to the process on the forested area.
Consequently, this makes the repulsion found for small $\beta$ more pronounced.

In Section \ref{sssec-interact}, where the repulsion between different species is quantified by fitting to the NN spacing of the 2dCG with $\beta\geq0$, we encounter the situation that the fit to $\beta=0$ is apparently still not satisfactory, because the maximum of the NN data is further to the left. In that cases we rather fit the dimension $D$ of the Poisson NN distribution \eqref{eq-Poid-NN}, that has a maximum further to the left, see Figure \ref{fig-Interspecies} and the discussion there. 
\\
\\
\textbf{Independence of Points: Reuse of Nests}\\
\\
We revisited the Poisson point process in $D$-dimensions in the paragraph above. The latter consists by definition of independent points. For the data, where we find agreement with the Poisson statistics in Section \ref{sssec-interact}, we investigate, whether the positions of the nests can be treated as "independent" in consecutive years. The reason for this question is that some of the nests are reused, not only within one species but also between different species, and we will present data about this fact below. Such an analysis is possible because all old and new nests were marked precisely with GPS coordinates, up to an error of about 10 m. In this paragraph we do not split the group of Eagle Owls into two.

\begin{figure}[t]
	\centering
	\hspace{0.02\linewidth} Buzzards \hspace{0.18\linewidth}
	Goshawks \hspace{0.17\linewidth}
	Eagle Owls\\
	\begin{turn}{90}\hspace{30pt}\tiny Probability of Reuse\end{turn}
	\includegraphics[width=0.3\linewidth,angle=0]{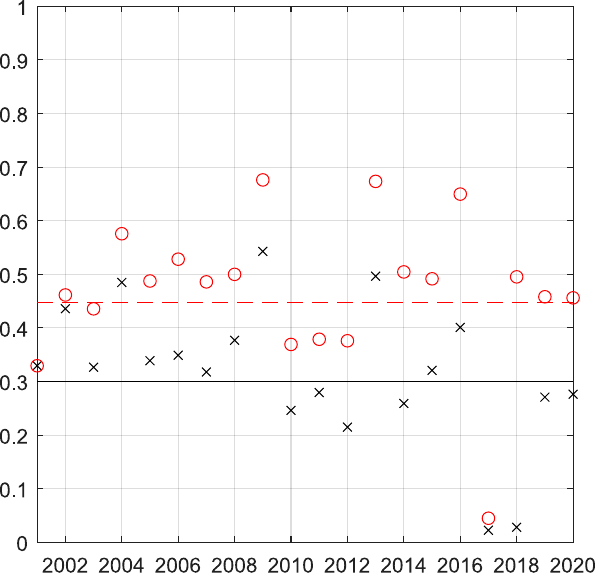}
	\includegraphics[width=0.3\linewidth,angle=0]{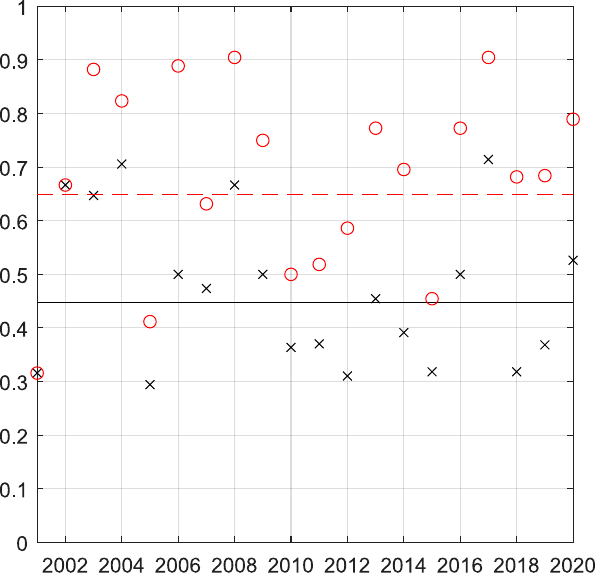}
	\includegraphics[width=0.3\linewidth,angle=0]{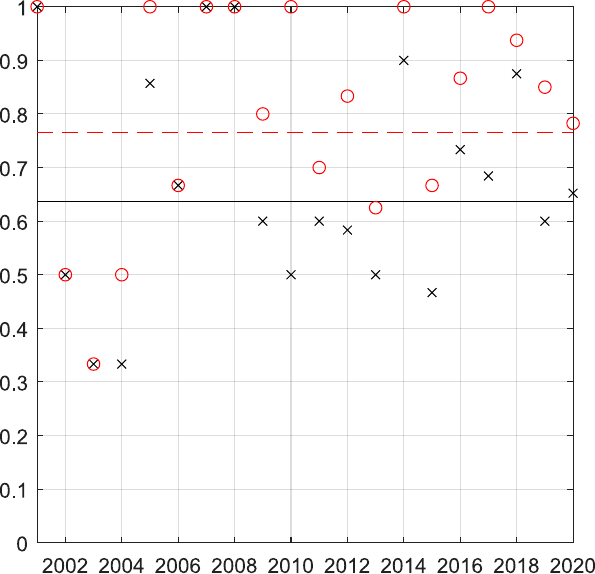}
	\caption[Figure 9 from \cite{ACKMOP}, \textit{Study of the reuse of nests per year by each species of birds of prey}]{Percentage of nests that have been reused per year by each species. The black crosses indicate the fraction of nests that were reused from the previous year by the same species, i.e. Common Buzzards reusing a Common Buzzard nest in the left plot. The black full line indicates the mean over the entire period of observation. The red circles indicate the fraction of nests that one species reused from all nests occupied by any of the three species in any previous year. The red dashed line again indicates the mean value. Notice that both percentages may sometimes coincide, as observed for the Eagle Owls. \cite[Figure 9]{ACKMOP}
	}
	\label{fig-Reuse}
\end{figure}

The amount of reusage of occupied nests from one year to the next year varies between 30\% (Common Buzzards), 45\% (Goshawks) and 65\% (Eagle Owls (F) + (P)) within one species, on average. The reuse of any nest from any previous year is about 10-20\% higher. 
The frequent reuse of nest sites over years and even decades has been documented previously for the studied area and is typical of avian predators \cite{Krueger2002b}. Habitat heterogeneity in many dimensions is commonly observed in these species \cite{KL01}, and individual preferences for certain habitat features are also common \cite{Krueger2002b}, leading to frequent reuse of nests. In other species, the same nest has been in use for centuries \cite{Ellis09} and even millennia \cite{BBN09}.

What is the consequence for our data analysis? In Section \ref{sssec-interact}, we treated every set of occupied nests per year as an independent ensemble, in order to compare it to a statistical mechanics picture, that represents an equilibrium configuration.  
The above data show that this is a simplification, that is only partly justified. The fact that a substantial fraction of nests is reused, both for the repulsion within one species or among two different species, means that not all NN and NNN positions are new. 

In order to improve our statistics, we have then made an ensemble average over 10 consecutive years (in \cite{buzzards 1} over 5 years). We have not analysed if after two or more years the positions become more independent or randomised, but it seems reasonable to assume that. Our situation could be compared to ensemble generation in statistical mechanics. If a new configuration is generated from an old one, e.g. with the Metropolis-Hastings algorithm used to generate the 2d Coulomb ensembles in application, one must monitor the autocorrelation time before a new configuration can be accepted as independent. In the generation of points, we could simply let enough iterations go by, but for the observed data we cannot wait one or more years, until the next set of occupied nests is more independent, without losing much information (and statistics). This is because the repulsion within one or more species can change over time, as we have observed.

There are ecological reasons why certain locations of nests are more popular than others, which is why they get reused. Plus, it takes time and effort to build a new nest. Certain nest sites are associated with higher reproductive success, independent of the individuals using these nests \cite{Krueger2002b}. The proximity of high-quality hunting habitat, less exposure to predators and parasites, and less disturbance from humans are just some of the key factors \cite{Krueger2002b,CBK}.

The fact that there is a large number of used nests in the observed area seems to guarantee at least to some extent, that a kind of "random sampling" can take place between consecutive years.
There is an estimate on the total number of 500-600 reusable nests in the area for birds of prey.

As a final remark, there seems to be no correlation between the population growth and the percentage of reusage of nests. It is approximately constant in Figure \ref{fig-Reuse} and largely fluctuates around the mean. As we have seen in Figure \ref{fig-AllBirds} (and Table \ref{tab-numbersbirds}) for the Common Buzzard, their population has grown from 63 to 267 pairs, and from a single pair to 23 for all Eagle Owls in total, see Figure \ref{fig-EagleOwlGroup} and Table \ref{tab-numbersbirds}.
The Goshawk population has also grown but more slowly and then declined again. 
Together with the high number of empty nests available each year this could be argued to be in favour of a certain independence of the locations of nests each year, despite the high fraction of reusage.

This study has shown that a simple, one-parameter dependent approach from statistical mechanics, without any biological assumptions, has the potential to explain important features in this three-species guild of avian predators.
We have shown that the intraspecific repulsion measured through $\beta$ clearly deviates from Poisson statistics of independent particles. The relative ranking in repulsion strength amongst the three species, averaged over all years, has been found compatible with experience from ecology. 

The correlation between $\beta$ and the time dependence of the population have been found plausible for at least two of the three species analysed. For the Common Buzzard the steep rise in population from 63 to 267 pairs has been paralleled by a clear rise in $\beta$, in particular in the NNN spacing, showing also an increased interaction range. For the Goshawk the population has been more stable around 20 pairs, with a decline after a peak around 10 years ago. Both NN and NNN show a clear decrease in the fitted $\beta$. In contrast for the Eagle Owls, with the population increasing from a single pair to over 23, we have seen a steep decrease in $\beta$ for NN and no repulsion between NNN. This relation is not clear to us and may be due to the limitation of our approach, or an artefact due to very low statistics. The change in $\beta$ over time might hence be a fruitful further topic for investigation.

At the same time, the results also clearly indicate some limitations when it comes to interspecific interactions. 
We have found the repulsion between Eagle Owls in the forest (F) on the one hand, and the Common Buzzard respectively Goshawk on the other hand to be significantly smaller than the respective intraspecific repulsion, which is plausible. Both interspecies repulsions have values of $\beta=0.15-0.1$ above Poisson at $\beta=0$. Taken together with the effect of the terrain, effectively lowering the dimension of the Poisson distribution, this makes the repulsion found more pronounced and significant. 
For the Eagle Owls in the plains (P) the same order of repulsion was only observed with the Common Buzzard, and no repulsion was seen for the Goshawk, neither among the latter two. 
In view of the continuing growth of population in Common Buzzards and Eagle Owls, notably populating the plains for the latter, and the population of Goshawks being under pressure, this seems to be a limitation of our approach (and perhaps present statistics).

We have shown that a simple formula based on a surmise from random matrix theory allows to quantify and discuss the intra- and interspecies NN repulsion, instead of using the numerically generated spacings distributions from a full 2dCG, that also include NNN. We quantified the effect of the terrain via a fit to a Poisson point process with $D=1.66$. The extensive reusage of nests, especially among Eagle Owls, has put the assumption of independence of ensembles as realised by consecutive years under scrutiny. However, at least from a random matrix perspective we know that there, spectral statistics show a certain robustness or universality under such perturbations.
In Appendix \ref{apps-random displacement}, we present as a further cross check a numerical experiment for two examples of year with medium to low statistics. Introducing a random displacement of the original locations of the nests by an amount comparable to the mean spacings between the nests, the displacement destroys the repulsion we have seen in fitting $\beta$ to the NN spacing and rapidly leads to a Poisson distribution. 
It would be very valuable to find an underlying microscopic model that leads at least approximately to the kind of spacing distribution of a 2dCG, as the one we have applied in our comparison with the data.
We hope that this example of transfer between disciplines has been an insightful exercise for the reader and probes the wide applicability of random matrix theory.
It remains to be seen whether interspecific interactions might be better captured with another decade of data with increased statistics, 
and with all three species at their ecological carrying capacity.
\subsection{Tridiagonal $\boldsymbol{\beta}$-Ensembles}\label{ssec-tridiag beta}
Let us summarize, where we are at the moment: We started this Section \ref{sec-non Herm beta} with the formulation of our goal, i.e. finding a matrix ensemble, which has a pairwise interaction among its complex eigenvalues proportional to $|\Delta_N(z)|^\beta$. We motivated this via the 2dCG with many possible applications of the latter. Let us now have a look into the literature and summarize, what is known/ conjectured so far. 
\begin{itemize}
	\item For $\beta=0$, we find the 2d Poisson process, discussed in Section \ref{ssec-point processes}, which is fully chaotic and well-studied, cf. \cite{haake}.
	\item In the vicinity of $\beta\approx0$ of distance $1/N$, the local eigenvalue correlations become Poisson, studied via probabilistic methods in \cite{Lambert}.
	\item For small $\beta=0,\ldots,0.5$ the surmise \eqref{eq-pbetasurmise}, introduced in \cite{AMP,masterthesis} describes the NN distribution of a 2dCG well.
	\item Perturbative results on the density exist for $\beta$ around 2, cf. \cite{Can et al}. 
	\item At $\beta=2$ we have the determinantal point process given by the GinUE, cf. \cite{book byun forrester} for a survey on the numerous results.
	\item Around $\beta=2$ strong oscillations of the density close to the edge of the spectrum are observed for $\beta>2$ \cite{Can et al}. These results are based on numerics and non-rigorous, perturbative calculations, cf. \cite{Abanov, Abanov 2} for the most recent works.
	\item At $\beta\approx 140$ a crystallisation towards the triangular Abrikosov lattice is conjectured \cite{AB81,CC83}, see again \cite{Abanov} for recent work.
\end{itemize}
We emphasize that in 1d, much more results exist and also for 2d further results on mesoscopic scales, e.g. \cite{leble}, are available. We are not in the position to give a complete list here but rather present the most common techniques used so far in the macroscopic study of the 2dCG:
Why is the development of a $\beta$-ensemble in 2d so challenging? We discussed in Section \ref{sec-Hermitian beta ensembles} in detail how the Hermitian $\beta$-ensembles arise. Let us recall that one way to reach the tridiagonal matrix representation is due to Householder reflections (cf. Section \ref{ssec-Householder}). However, we discussed in the beginning of this section that it is a well-known fact that a general complex Ginibre matrix can only be reduced to a Hessenberg matrix $J_\beta$ cf. Eq. \eqref{eq-Hessenberg GinUE} with $\beta=2$, i.e. $(J_\beta)_{i,j}=0$ for $i>j+1$, without affecting the eigenvalues. While the computation of the Jacobian for a diagonalization of such a Hessenberg matrix worked out for the case of unitary matrices in \cite{Killip Nenciu}, the $N$-term recurrence of the corresponding characteristic polynomial makes it in general, i.e. with eigenvalues in (a part of) the complex plane rather than the unit circle, difficult to perform similar proof techniques. We saw a second proof strategy in the Hermitian setting in Section \ref{ssec-Anderson Dixon}, via minor processes which appear also in the study of the Selberg integral. Nevertheless, as pointed out already several times, the Selberg integral is not known for the complex setting and is an open problem for decades now. One can ask for different similarity transformations which might reduce a complex Ginibre matrix to tridiagonal form. One possible candidate could be Givens rotations, which transform selective elements to zero, e.g. one component of a row/ column in a given matrix is mapped to zero via an orthogonal or unitary transformation, cf. \cite[Chapter 5.1]{Golub van Loan}. It exists an explicit algorithm for the computation of the parameters in the applied Givens rotation, which can yield quite complicated expressions. Consequently, this mechanism serves in general not as a good alternative to Householder reflections in the non-Hermitian case, we would like to emphasize the importance of Givens rotations in sparse-matrix models, see for instance \cite{Givens rot,iteration methods}.

A different ansatz is the following: Let us start from a non-Hermitian tridiagonal random matrix, whose distributions of matrix entries contain a positive real parameter $\beta$. Can we calculate the Jacobian for the eigenvalue decomposition of this matrix? If so, how should we choose the distributions of the matrix elements to find a Coulomb-type repulsion of the eigenvalues? The idea is to turn our view to the spectral decomposition of a tridiagonal matrix, where we can use similar techniques as in the Hermitian tridiagonal ensembles, rather than to the reduction of a full non-Hermitian matrix to this form. The reader might insist, that this ansatz is an over-simplification of the problem. We would like to recall that, while the possible reduction of the GO/U/SE to tridiagonal form was known already in 1984 by Trotter \cite{Trotter 84}, it took nearly 20 years until the proof of the Jacobian and therefore the possible generalisation for $\beta$-ensembles was developed by Dumitriu and Edelman in 2002 \cite{Dumitriu Edelman}. Hence, we would like to emphasize that the generalisation of the proof techniques, used frequently in Hermitian $\beta$-ensembles (cf. \cite{Dumitriu Edelman,Killip Nenciu,Forrester Rains unitary,Dumitriu Forrester}), is an important step towards non-Hermitian $\beta$-ensembles, respectively a matrix parametrisation of a 2dCG at general inverse temperature $\beta>0$.

Recently, Mezzadri and Taylor introduced
an ensemble of particular complex non-symmetric tridiagonal matrices, where the inverse temperature $\beta$ is a continuous parameter in \cite{mezzadri taylor}. 
They constructed the joint density of complex eigenvalues, by establishing a one-to-one correspondence between the random variables on the diagonal and lower diagonal of a particular tridiagonal matrix on the one hand, and the complex eigenvalues and the first row of the diagonalising similarity transformation on the other hand. The latter set of variables completely determines the diagonalisation.  
The construction allowed them to calculate the Jacobian and joint eigenvalue distribution explicitly. The proof strategy has main parallels to the one performed by Dumitriu and Edelman \cite{Dumitriu Edelman}. The key point is that the distributions of the matrix elements are chosen in such a way that the joint eigenvalue distribution contains a logarithmic Colomb repulsion at inverse temperature $\beta>0$. While this property is a strong argument for the importance of this matrix ensembles, as well as the explicit known normalisation constant (corresponding to a Selberg type integral in the complex plane), the increased complexity of this non-Hermitian matrix ensemble yields an additional function in the joint eigenvalue distribution. This term contains a non-trivial integral representing a coupling between the eigenvalues and eigenvector components.\footnote{The subject of evolved (compared to the Hermitian case) eigenvector statistics for non-Hermitian matrices is well-known and is studied for instance in the eigenvector overlaps of left- and right-eigenvectors, to be defined below, initiated by \cite{Chalker Mehlig}.} We would like to emphasize that this ensemble of \cite{mezzadri taylor} does not yield the complex Ginibre ensemble for $\beta=2$. The described additional factor in the joint eigenvalue distribution is the reason not only for this but also for making the computation of eigenvalue correlation functions challenging.

In the joined work \cite{AMPT}, which is the basis of this section, we introduce two new ensembles of general, respectively symmetric complex tridiagonal matrices. We will show that these share many of the attractive features and several open questions of the first ensemble \cite{mezzadri taylor} that we just described. Moreover, we develop a novel\footnote{in the study of non-Hermitian matrices} method for the computation of the limiting spectral density of eigenvalues for general centred Jacobi matrices, i.e. tridiagonal matrices with vanishing diagonal. In order to reach this matrix form, we perform a low temperature limit in $\beta\gg 1$, where we can again follow proofs from the Hermitian $\beta$-ensembles \cite{DE02B}.

In Section \ref{sssec-def tridiag jpdf}, we define the three (the non-symmetric one by \cite{mezzadri taylor} and two novel) tridiagonal matrix ensembles and derive the JPDF in terms of the complex eigenvalues for each of the three ensembles. They show a logarithmic Coulomb repulsion at arbitrary positive inverse temperature $\beta>0$ in the eigenvalues. Additionally, we find a function which depends on the integral of eigenvector components and eigenvalues contributing to this JPDF. We investigate the role of this function in a later Section \ref{sssec-numerics} studying the NN statistics of the ensembles.\\
In the next Section \ref{sssec-low temperature}, we study the three ensembles in the low temperature limit $\beta\gg1$ analytically. We numerically do not find a crystallization of the eigenvalues, which is conjectured for the 2dCG above $\beta>140$ \cite{AB81,CC83} for tested $\beta$ up to $5000$. Nevertheless, the matrices reduce to centred Jacobi-form in this limit, which makes the calculation of the limiting spectral density for $N\rightarrow \infty$ valuable. We comment on the role of this low temperature limit in the Hermitian models on the lines.\\
The reached centred Jacobi-form enables us to apply a novel technique in the study of the eigenvalue distribution in the limit $N\rightarrow \infty$ using the characteristic polynomial in Section \ref{sssec-characteristic polynomial}. While similar results are known for real orthogonal polynomials, we establish a proof for general complex polynomials here.\\
In Section \ref{sssec-equilibrium density}, we derive the limiting spectral density for each of the three tridiagonal matrix ensembles. Therefore, we slightly generalise results from free probability on the limiting density of zeros of a random polynomial and apply these to the characteristic polynomials of our matrix ensembles.
It turns out that the spectral density of the general ensemble
is much more stable with respect to an expansion in large matrix size $N$ \textit{and} large $\beta$. In that limit the global spectral density of the general and symmetric ensemble agrees. The previously introduced ensemble \cite{mezzadri taylor} is less stable in this limit, and we will show that to leading order the spectral density becomes a Dirac delta at the origin. In the more general ensemble, the spectral density is absolutely continuous and very simple, depending on the logarithm of the modulus of the eigenvalue plus a constant. This limiting law seems to be a novel feature for non-Hermitian random matrix ensembles.
In Section \ref{sssec-numerics}, we present extensive numerical studies on the eigenvalue density, confirming the results from the previous Section \ref{sssec-equilibrium density} for the general and the symmetric ensembles. Moreover, we investigate the local NN spacing distribution numerically and observe that the general and symmetric tridiagonal ensembles share the statistics with a 2d Poisson point process of independent particles. It turns out that the local statistics are independent of $\beta$. For the complex non-symmetric tridiagonal ensemble, we find a $\beta$-dependence of the NN distribution which deviates from 2d Poissonian statistics. We currently do not have an understanding of this difference.\\
We present results on the case $N=2$ in Appendix \ref{app- tridiag N=2}.
\subsubsection{Definition of the Tridiagonal Matrix Ensembles and JPDF of the Complex Eigenvalues}
\label{sssec-def tridiag jpdf}
We introduce a complex tridiagonal non-Hermitian matrix $J=\widetilde{T},S,T$ of dimension $N$ as 
\begin{align}
	\label{eq-tridiag}
	J= \begin{pmatrix}
		a_N & c_{N-1} &  &  &  \\
		{b}_{N-1} & a_{N-1} & c_{N-2}   & & \\
		& \ddots & \ddots &   \ddots&\\
		&  &  {b}_2 & a_2& c_1\\ 
		&  & & {b}_1 & a_1 
	\end{pmatrix},
\end{align}
with $a_j,b_j,c_j\in\mathbb{C}$, $b_j,c_j \neq 0$, and all other matrix elements vanishing. For the three tridiagonal complex ensembles of non-symmetric, symmetric and general matrices, we use the labels $\widetilde{T}$, $S$ and $T$ respectively. The matrix elements are assumed to be all independent and follow the distributions given in Table \ref{tab-dists tridiag} above. Here, "$\sim $" is to be understood as "distributed as". We recall that $\mathcal{N}(0,1)$ denotes the normal distribution (cf. Eq. \eqref{eq-def normal dist}), $\mathcal{U}(0,2\pi)$ the uniform distribution on $[0,2\pi)$, and $\chi_\ell$ is the Chi-distribution (cf. Eq. \eqref{eq-def chi dist}).\\ 
\begin{table}[t]
	{ 
		\begin{tabular}{|c|c|c|c|c|c|c|}
			\hline
			\rule{0pt}{2ex}  $J$ & $\re(a_j)$ & $\im(a_j)$ & $|b_k|$ & $\arg(b_k)$ & $|c_k|$ & $\arg(c_k)$\\
			\hline
			\rule{0pt}{3ex}  $\widetilde{T}$  &\multirow{3}{*}{$\sim\mathcal{N}(0,1)$}& \multirow{3}{*}{$\sim\mathcal{N}(0,1)$} & $\sim\chi_{\beta k/2}$ & \multirow{3}{*}{$\sim\mathcal{U}(0,2\pi)$}&1&0\\ 
			$S$& & &$\sim\frac{1}{\sqrt{2}}\chi_{\beta k}$& & $=|b_k|$& $=\arg(b_k)$\\
			$T$ & & & $\sim \chi_{\beta k/2}$& & $\sim \chi_{\beta k/2}$&  $\sim\mathcal{U}(0,2\pi)$\\
			\hline			
			
		\end{tabular}
		\caption[\textit{Distributions of the matrix elements for the three tridiagonal non-Hermitian matrix ensembles $\widetilde{T},S,T$}]{We give the distributions of the matrix elements $a_j$ for $j=1,\ldots,N$ and $b_k,c_k$ for $k=1,\ldots,N-1$ of the tridiagonal matrix $J=\widetilde{T},S,T$ \eqref{eq-tridiag} for $\beta>0$. While the distributions of the diagonal elements and the angle of the sub-diagonal $\arg(b_k)$ agree in all three ensembles, we have scaling differences in $|b_k|$ and distinct settings for the super-diagonal elements $c_k$.}
		\label{tab-dists tridiag}}
\end{table}
We notice, that we have $2N+2(N-1)=4N-2$ real degrees of freedom for the non-symmetric matrix $\widetilde{T}$, introduced in \cite{mezzadri taylor}, as well as $4N-2$ real degrees of freedom for the symmetric matrix $S$ and $2N+4(N-1)=6N-4$ real parameters in the general case of matrix $T$. The diagonal elements $a_j$ and the angular part of the sub-diagonal elements $\arg(b_k)$ are drawn from the same distribution for all three ensembles. The main differences are in the scalings of the Chi-distributions in the radial part of $b_k$ and the distribution of the super-diagonal elements $c_k$. In the non-symmetric model, they are deterministic $c_k=1$, while they are equal to $b_k=c_k$ in the symmetric model. In the general model they are independently chosen from the same distribution, i.e. $b_k\sim c_k$. In the following, we will only consider matrices which are diagonalisable and have pairwise distinct non-vanishing eigenvalues.\\
Let us compare these tridiagonal non-Hermitian random matrices, with the ensemble introduced by Dumitriu and Edelman for the Gaussian-$\beta$-matrices in Eq. \eqref{eq-DE model}. We observe that the distributions in Table \ref{tab-dists tridiag} are "natural" complexifications of this Hermitian model, i.e. the normal distributions on the diagonal are replaced by complex normals with independent real- and imaginary-part, while the uniform distribution of the angles of the off-diagonal elements is used to extend into the complex plane. In terms of the scaling in the Chi-distribution, the symmetric non-Hermitian ensemble (second row in Table \ref{tab-dists tridiag}) agrees with the Gaussian-$\beta$-ensemble due to its symmetry. The exponential part of the distributions in Table \ref{tab-dists tridiag} can be written as $\propto\exp\left[-\Tr(JJ^\dagger)\right]$ as for the Hermitian matrix $H_\beta$ \eqref{eq-DE model}.

The question could be raised, why we study these three different kinds of tridiagonal matrix ensembles, defined via \eqref{eq-tridiag} and Table \ref{tab-dists tridiag}. Moreover, why are they distinct? It is a well-known fact, that the eigenvalues of a tridiagonal matrix $J$ \eqref{eq-tridiag}, with $b_kc_k\neq 0$ for all $k=1,\ldots,N-1$, depend only on the diagonal elements $a_j$ for $j=1,\ldots,N$ and the products of the off-diagonal elements $b_kc_k$ \cite{cullum willoghby}. There exists a similarity transformation between a general tridiagonal matrix $T$ and a symmetric one $S$, via the diagonal matrix 
\begin{align}\label{eq-Cdef}
	C=\begin{pmatrix}
		1 & &  & \\
		& d_2 & & \\
		& & \ddots &\\
		&  & & d_{N-1}
	\end{pmatrix},
\end{align}
with $d_i=d_{i-1}\sqrt{\frac{c_i}{b_i}}$ for $i>1$ and $d_1=1$. Then, $S=CTC^{-1}$ \cite{cullum willoghby vol 1}. In \cite{mezzadri taylor} it was observed that the similarity transformation $\widetilde{T} = \widetilde{C} T\widetilde{C}^{-1}$ brings $T$ to the form
\begin{align}
	\label{eq-T_tilde_def}
	\widetilde{T} = \begin{pmatrix}
		a_N & 1 &  &  &  \\
		b_{N-1}c_{N-1} & a_{N-1} & 1 &  &  \\
		& \ddots & \ddots & \ddots &  \\
		& & b_2c_2 & a_2 & 1 \\
		&  &  & b_1c_1 & a_1 
	\end{pmatrix}, 
\end{align}
with the diagonal matrix
\begin{align}
	\label{eq-Ctildedef}
	\widetilde{C} = \begin{pmatrix} 1 & &  & \\
		& c_{N-1} & & \\
		& & \ddots &\\
		&  & & c_{N-1}\dotsm c_1
	\end{pmatrix}.
\end{align}
Hence, there exist similarity transformations between all three ensembles from \eqref{eq-tridiag} and Table \ref{tab-dists tridiag}. However, it is important to note that these ensembles are defined exactly via both, its tridiagonal form in \eqref{eq-tridiag} and an explicit choice of distribution of matrix elements in Table \ref{tab-dists tridiag}. The similarity transformations defined via $C$ \eqref{eq-Cdef} respectively $\widetilde{C}$ \eqref{eq-Ctildedef} are not necessarily unitary and will therefore not leave the Gaussian part $\propto\exp\left[-\Tr(JJ^\dagger)\right]$ of the distributions invariant. Thus, we consider truly distinct random matrix ensembles. One can view this as follows: On the level of matrices, there exist similarity transformations between non-symmetric, symmetric and general tridiagonal matrices. However, these affect the distributions of the matrix elements which will no longer correspond to the choice we made in Table \ref{tab-dists tridiag}. Nevertheless, from an algebraical view, these transformations are important and we will discuss their role in later proofs repeatedly.
\begin{Remark}
In order to illustrate this important point, let us consider the similarity transformation between $\widetilde{T}$ and $T$ via $\widetilde{C}$. The independent distribution $\chi_{j\beta/2}$ for the moduli $|b_k|$ and $|c_k|$ in $T$ results in the distribution $g_j(y)$ for the sub-diagonal matrix elements of $\widetilde{T}$, given by the product of matrix elements of matrix $T$, with
	\begin{align}
		g_j(y)=\frac{1}{2^{\frac{j\beta}{4}-2}\Gamma\left(\frac{j\beta}{4}\right)}\ y^{j\beta-2}\exp[-y^4], \ y>0,
	\end{align}
This is \textit{not} the Chi-distribution, which is imposed in the first row in Table \ref{tab-dists tridiag}. 
\end{Remark}

We are interested in the distribution of complex eigenvalues $z_j\in\mathbb{C}$, $j=1,\dotsc,N$ of matrix $J=\widetilde{T},S,T$, and thus in the diagonalisation 
\begin{align}
	\label{eq-diagonal}
	J=RZ R^{-1}
\end{align}
with $Z= \diag(z_1,\dotsc,z_N)$ and $R \in \mbox{GL}(N,\mathbb{C}) $.\footnote{Here, $\mbox{GL}(N,\mathbb{C})=\{M\in \mathbb{C}^{N\times N}|\det M\neq 0\}$ is the general linear group.} The columns of $R$ contain the right-eigenvectors, and the rows of $R^{-1}$ contain the left-eigenvectors. These are in general distinct.\footnote{Notice however, that for symmetric matrices any right-eigenvector is also a left-eigenvector \cite{cullum willoghby}.} This is in stark contrast to Hermitian matrices, where left- and right-eigenvectors agree. The diagonalising matrix $R$ is unique up to permutations of the eigenvalues and (right) multiplication by an invertible complex diagonal matrix. We will express main parts of the proofs in terms of this tridiagonal matrix $J=\widetilde{T},S,T$ \eqref{eq-tridiag}, i.e. for all three ensembles in parallel. This follows the proofs given in \cite{mezzadri taylor} for the non-symmetric ensemble $\widetilde{T}$ and the results in \cite{AMPT} on the general tridiagonal matrix $T$, in a unified way. However, it is important to notice that the spectral parameters, i.e. the matrices $R$, $Z$ and $R^{-1}$ are different within the distinct ensembles.

Let $\Lw(N)$ denote the set of matrices $J$, of the form given in Eq. \eqref{eq-tridiag}, with the eased constraints $a_j,b_j,c_j\in\mathbb{C}$ for all $j$, such that $\det(J)\neq0$ and the spectrum of $J$ is non-degenerate. Note that the probability distributions in Table \ref{tab-dists tridiag}, define an absolutely continuous measure on the set of tridiagonal matrices $J=\widetilde{T},S,T$ each. Consequently, the exceptional set of matrices with degenerate spectra forms a set of measure zero and is therefore omitted.
We define:
\begin{align}
	\label{eq-def D(n)}D(N)  & = \left\{\diag(z_1,\dotsc,z_N)\colon z_j \in \mathbb{C}, \quad  z_j \neq 0, \quad
	j=1,\dotsc,N \right\}, \\
	\label{eq-def Z(n)}Z(N) & =  \left\{ \diag(z_1,\dotsc,z_N)\colon z_j,z_k \in \mathbb{C}, 
	\quad z_j \neq z_k \neq 0, \quad 1 \le j < k \le N \right\},\\
	R^J(N)  & = \left\{R \in \GL(N,\mathbb{C}) \colon J = 
	RZ R^{-1},\quad J \in \Lw(N),\quad  Z \in Z(N)\right\}.\label{eq-R(n)def}
\end{align}
When $J \in \Lw(N)$, the spectral decomposition of $J$ is given by
\begin{align}\label{eq-eigdecompgeneral}
	J=RZ R^{-1},
\end{align}
where $Z\in Z(N)$ and $R\in R^{J}(N)$. This decomposition is unique up to a permutation of the eigenvalues and the right multiplication $R\mapsto RD$, where $D\in D(N)$. We observed a similar invariance in Section \ref{ssec-intro models} in the Schur decomposition.

Consequently, the spectral decomposition in Eq. \eqref{eq-eigdecompgeneral} defines a bijection
\begin{align}\label{eq-sing_valued_map_general}
	\mathcal{F}^{J}: \Lw(N)\rightarrow\mathcal{L}(N)\times\mathcal{R}^{J}(N),
\end{align}
where
\begin{align}\label{eq-def curlyL(n)}
	\mathcal{R}^J(N)= R^J(N)/D(N),  \quad  \mathcal{L}(N) = Z(N)/\mathfrak{S}_N,
\end{align}
and $\mathfrak{S}_N$ is the symmetric group of order $N$. We restrict the domain of $\mathcal{F}^{J}$ to $\Lw'(N)=\Lw(N)\setminus \widetilde{B}^J$, where
\begin{align}
	\widetilde{B}^{\widetilde{T},S}&=\{\{\widetilde{T},S\}\in \Lw(N):b_j=0\hspace{2mm}\text{for some}\hspace{2mm} j\in\{1,\dots,N\}\},\\
	\widetilde{B}^T&=\{T\in \Lw(N):b_j=0\hspace{2mm}\text{or}\hspace{2mm} c_j=0\hspace{2mm}\text{for some}\hspace{2mm} j\in\{1,\dots,N\}\}.
\end{align}
Since the probability measures in Table \ref{tab-dists tridiag}, are absolutely continuous, we note that the set of matrices not in $\Lw'(N)$ has measure zero and is therefore omitted from our analysis.
\begin{Remark}\label{rem-parameters 0}
For the general case $J=T$, we can have the case that both $c_{N-j}=0$ and $b_{N-j}=0$ for any $j$. In order to consider this, we perform a direct sum decomposition. Specifically, we have
	\begin{align}\label{eq-decomp2}
		T=T_j\oplus U_j,
	\end{align}
	where $T_j$ is the principal sub-matrix obtained by keeping the first $j$ rows and columns of $T$. In contrast, $U_j$ denotes the principal sub-matrix obtained by keeping the last $j$ rows and columns of $T$. Thus, we can carry out our analysis on these two smaller dimensional matrices, namely $T_j$ and $U_j$, which are still tridiagonal.
\end{Remark}
Let
\begin{align*}
	\widetilde{R}^{J}(N)  = \left\{R \in \GL(N,\mathbb{C}) \colon\quad J = 
	RZ R^{-1},\quad J \in \Lw'(N),\quad  Z \in Z(N)\right\},  
\end{align*}
and
\begin{align}
	\widetilde{\mathcal{R}}^{J}(N)= \widetilde{R}^{J}(N)/D(N).
\end{align}
It was pointed out above that the matrix ensembles consisting of $J=\widetilde{T}$, $S$, $T$ with definitions in Table \ref{tab-dists tridiag}, are not equivalent in a probabilistic sense. Nevertheless, there exists the similarity transformations $C$ \eqref{eq-Cdef}, respectively $\widetilde{C}$ \eqref{eq-Ctildedef}. We use this one-to-one mapping between the matrices (not in distributions) in this section. Therefore, we can inherit the proofs from \cite{mezzadri taylor} with the replacement $b_j\rightarrow b_jc_j$. Another way to see, why this is the right mapping, is to look into the three-term recurrence relation of the characteristic polynomials. Again, one finds the same recurrence, which is used in the proofs \cite{mezzadri taylor}, for the general tridiagonal matrix after the suggested replacement. For completeness, we will repeat the proofs from \cite{mezzadri taylor} here in the general setting.
\begin{lemma}\label{lem-r_diff_from_zero}
	Let $\mathbf{r}^t=(r_1,r_2,\dots,r_N)$ and $\mathbf{v}^t=(v_1,\dots,v_N)$ be the first and last rows of matrix $R\in \widetilde{R}^{J}(N)$, respectively. We have
	\begin{align}
		r_j\neq0,\quad v_j\neq0\quad\text{for all}\quad j\in\{1,\dots,N\}.
	\end{align}
	Furthermore, the subset of $\widetilde{R}^{J}(N)$ such that 
	\begin{align}\label{eq-rgeneral}
		r_1+r_2+\dots+r_N=1,
	\end{align}
	spans a set of representatives in $\widetilde{\mathcal{R}}^{J}(N)$ for each $J=\widetilde{T},S,T$.
\end{lemma}
\begin{proof}
We follow the proof of Lemma $2.1$ in \cite{mezzadri taylor} closely and present its application to a general tridiagonal matrix $J$ \eqref{eq-tridiag}.\footnote{We use a slightly different notation as in \cite{mezzadri taylor}.} Let $J\in \Lw'(N)$ and let 
\begin{align}
	\label{eq-tridiag k}
	J_k= \begin{pmatrix}
		a_N & c_{N-1} &  &  &  \\
		{b}_{N-1} & a_{N-1} & c_{N-2}   & & \\
		& \ddots & \ddots &   \ddots&\\
		&  &  {b}_{N-k+2} & a_{N-k+2}& c_{N-k+1}\\ 
		&  & & {b}_{N-k+1} & a_{N-k+1} 
	\end{pmatrix},
\end{align}
be the principal sub-matrix of $J$, after removing the last $(N-k)$ rows and columns. We have $J_N=J$. Let $z_j^{(k)}$, $j=1,\ldots,k$, denote the eigenvalues of $J_k$. Let $\Phi_k(z)$ denote the characteristic polynomial of $J_k$, i.e.
\begin{align}\label{eq-det charac poly}
	\Phi_k(z)=\det(z\mathds{1}_k-J_k)=\prod_{j=1}^k(z-z_j^{(k)}).
\end{align}
The tridiagonal structure of the matrix $J_k$ implies a three-term recurrence relation of $\Phi_k$
\begin{align}\label{eq-3 term rec}
	\Phi_k(z)=(z-a_{N-k+1})\Phi_{k-1}(z)-b_{N-k+1}c_{N-k+1}\Phi_{k-2}(z),\ k=1,\ldots,N,
\end{align}
with the initial conditions $\Phi_0(z)=1$ and $\Phi_{-1}(x)$=0 and $b_N,c_N$ undetermined. It follows from Eq. \eqref{eq-det charac poly} for $k=N$ that $z_j=z_j^{(N)}$ is an eigenvalue of $J$ and the corresponding right-eigenvector is
\begin{align}\label{eq-r eigenvec}
	\begin{pmatrix}
		\Phi_0(z_j)\\
		\vdots\\
		\Phi_{N-1}(z_j)
	\end{pmatrix}.
\end{align}
Hence, the matrix $R$, which contains the right-eigenvectors as $JR=RZ$, can be written as
\begin{align}
	R=\begin{pmatrix}
		\Phi_0(z_1) & \ldots & \Phi_0(z_N)\\
		\Phi_1(z_1) & \ldots & \Phi_1(z_N)\\
		\ldots & \ldots & \ldots\\
		\Phi_{N-1}(z_1) & \ldots & \Phi_{N-1}(z_N)
	\end{pmatrix},
\end{align}
which belongs to $\widetilde{R}^{J}(N)$. As we assume that the spectrum is non-degenerate, the eigenvectors are unique up to multiplication by a constant. With the initial condition $\Phi_0(z_j)=r_j=1$, it follows that $r_j\neq0$ for $j=1,\ldots,N$. This proves the first part of Lemma \ref{lem-r_diff_from_zero}. We also need to show that the $v_j\neq 0$. Here, we cannot use the initial condition and perform a proof via contradiction instead. Suppose that $\Phi_{N-1}(z_j)=0$ for some $j\in\{1,\ldots,N\}$. As $\Phi_N(z_j)=v_j=0$ and $b_jc_j\neq 0$ for all $j=1,\ldots, N-1$, this yields by induction that all entries of the right-eigenvector in Eq. \eqref{eq-r eigenvec} vanish. Hence, $\det R=0$, which is a contradiction to the assumption that the spectrum is simple. \\
Next, we need to show that the set $\left\{R\in \widetilde{\mathcal{R}}^{J}(N): r_1+\cdots +r_N=1\right\}$ is a set of representatives of $\widetilde{\mathcal{R}}^{J}(N)$. Following the steps above using the principal sub-matrix of $J$ with removed first $(N-k)$ rows and columns, one can prove that the elements of the first column of $R^{-1}$ are all non-zero. We use now the freedom of the property $RR^{-1}=\mathds{1}_N$ under right multiplication $R\rightarrow RD$, where $D\in D(N)$ to rescale all elements in the first column in $R^{-1}$ to 1. Then, $RR^{-1}=\mathds{1}_N$ yields Eq. \eqref{eq-rgeneral}.\\
For the matrix ensembles of non-symmetric $\widetilde{T}$, respectively symmetric $S$ matrices, it is sufficient to prove the condition on the $r_j$'s, as we have only $4N-2$ real degrees of freedom from the matrix entries. However, for the general case $T$, where we have $6N-4$ real degrees of freedom, we will need more spectral parameters for a one-to-one mapping and therefore need also non-vanishing $v_j$'s.
 \end{proof}
Next, we aim to show that the bijection in Eq. \eqref{eq-sing_valued_map_general} induces the injective map
\begin{align}\label{eq-induced_map_abc}
	\mathcal{G}^{S,\widetilde{T}}&: \Lw'(N)\rightarrow\mathcal{L}(N)\times\mathcal{H}_N,\\
	\mathcal{G}^{T}&: \Lw'(N)\rightarrow\mathcal{L}(N)\times\mathcal{H}_N\times \mathcal{V}_N,
\end{align}
where 
\begin{align}
	\label{eq-hyperplane}
	\Hp_N = \left\{ \rv \in \mathbb{C}^N \colon r_1 + \dotsb + r_N =1, \quad r_j \neq 0, 
	\quad\text{for}\quad j=1,\dotsc,N \right\},
\end{align}
and
\begin{align}
\mathcal{V}_N=\{\mathbf{v}^t=(v_1,\dotsc,v_N)^t\in\mathbb{C}^N:v_j\neq0\quad\text{for}\quad j=1,\dotsc,N\}.
\end{align}

From Lemma \ref{lem-r_diff_from_zero} and Eq. \eqref{eq-sing_valued_map_general}, it follows that $\mathcal{G}^{J}$ is single-valued. To prove the injectivity of $\mathcal{G}^J$, we let 
\begin{align}\label{eq-def Rgenmodel}
	R=\begin{pmatrix}
		\rv_1^t\\\vdots\\\rv^t_N
	\end{pmatrix}=(\Rv_1,\dotsc,\Rv_N)\quad\text{and}\quad R^{-1}=\begin{pmatrix}
		\mathbf{L}_1^t\\\vdots\\\mathbf{L}_N^t
	\end{pmatrix}=(\lv_1,\dotsc,\lv_N),
\end{align}
where $\mathbf{R}_j$ denotes the $j$th column of $R$ and $\mathbf{r}^t_j$ denotes its $j$th row, with $\mathbf{r}_1^t=\mathbf{r}^t$ and $\mathbf{r}_N^t=\mathbf{v}^t$ from Lemma \ref{lem-r_diff_from_zero}. Similarly, let $\mathbf{L}_j^t$ denote the $j$th row of $R^{-1}$ and $\mathbf{l}_j$ denote the $j$th column.\footnote{One should note that $\mathbf{R}_j$ and $\mathbf{L}_j^t$ denote the right- and left-eigenvectors in $R$ and $R^{-1}$ respectively.} 

Furthermore, as in the previous proofs, we will write the biorthogonality as
\begin{align}\label{eq-biorthogonality}
	\rv^t_j \lv_{k} = r_{j1}l_{1k}+ \dotsb + r_{jN}l_{Nk} = \delta_{jk}, \quad\text{for}\hspace{2mm} 
	j,k=1,\dotsc,N,
\end{align}
where the notation on the left-hand-side (LHS) represents matrix multiplication between the row vector $\rv_j^t$ and the column vector $\lv_k$, rather than the usual scalar product in $\mathbb{C}^N$. Finally, let $(\mathbf{x})_1$ denote the first entry of the vector $\mathbf{x}$.
\begin{theorem}\label{thm-roughbijection}
	Let $Z\in Z(N)$, $\mathbf{r}^t=(r_1,r_2,\dots,r_N)\in\mathcal{H}_N$ and $\mathbf{R}=(R_1,R_2,\dotsc,R_N)^t\in\mathcal{V}_N$. Then, there exists a unique $R\in \widetilde{R}^{J}(N)$ whose 
	\begin{itemize}
		\item first row is $\mathbf{r}^t$ for $\mathbf{J=\widetilde{T},S,T}$
		\item and first column is $\mathbf{R}$ for $\mathbf{J=T}$,
	\end{itemize}
	and a unique $J\in\Lw'(N)$ such that $J=RZ R^{-1}$ for each ensemble $J=\widetilde{T},S,T$, except for $\rv^t$ belonging to an exceptional set $M\subset\mathcal{H}_N$ of Lebesgue measure zero.
\end{theorem}
\begin{proof}
	We start the proof by taking $R$ from the set of representatives of $\widetilde{\mathcal{R}}^{J}(N)$ defined in Lemma \ref{lem-r_diff_from_zero}.
	Let us write the matrix equations
	\begin{align*}
		JR=RZ\quad\text{and}\quad
		R^{-1}J=Z R^{-1},
	\end{align*}
	as a set of vector equations:
	\begin{align}\label{eq-genmodelrecrelation}
		&\mathbf{r}_j^tZ=b_{N-j+1}\mathbf{r}^t_{j-1}+a_{N-j+1}\mathbf{r}_j^t+c_{N-j}\mathbf{r}^t_{j+1},\\
		&Z\mathbf{l}_j=c_{N-j+1}\mathbf{l}_{j-1}+a_{N-j+1}\mathbf{l}_j+b_{N-j}\mathbf{l}_{j+1}\label{eq-genmodelrecrelation1},
	\end{align}
	for $j=1,2,\dots,N$, where we have adopted the notation from Eq. \eqref{eq-def Rgenmodel}. Let\footnote{Lemma \ref{lem-r_diff_from_zero} implies that the entries of $\mathbf{R}_1$ are all different from zero. Thus, $\mathbf{R}_1\in\mathcal{V}_N$.} $\mathbf{R}=\mathbf{R}_1$ denote the first column of $R$ and $\mathbf{r}_1^t=\mathbf{r}^t$. We want to show that given $Z$, $\mathbf{r}_1^t$ and $\mathbf{R}_1$, with $\mathbf{r}_1^t\mathbf{l}_1=1$, where
	\begin{align*}
	\mathbf{l}_1=\begin{pmatrix}
		1\\\vdots\\1
	\end{pmatrix},
	\end{align*}
	we can reconstruct $J$, $R$ and $R^{-1}$ uniquely from the recurrence relations in Eqs. \eqref{eq-genmodelrecrelation} and \eqref{eq-genmodelrecrelation1}, with boundary conditions $\mathbf{l}_0=\mathbf{r}_{0}^t=\mathbf{0}$ and $b_0=c_0=0$. This construction depends on the matrix ensemble $J=\widetilde{T},S,T$ under consideration. We need the condition $c_0=0$ only for the general model $T$, as $c_j=1$ for all $j=1,\ldots,N-1$ in $\widetilde{T}$ and $b_j=c_j$ in the symmetric ensemble $S$. 
	For $j=1$, in Eqs. \eqref{eq-genmodelrecrelation} and \eqref{eq-genmodelrecrelation1}, we thus have
	\begin{align}
		\label{eq-cnot1recrelation}
		&\mathbf{r}^t_1Z=a_N\mathbf{r}^t_1+c_{N-1}\mathbf{r}^t_2,\\
		&Z\mathbf{l}_1=a_N\mathbf{l}_1+b_{N-1}\mathbf{l}_2\label{eq-cnot1recrelation1}.
	\end{align}
	We construct the tridiagonal matrix $J$ iteratively. Here, we give constructions specialised to each ensemble and discuss the differences.
	\begin{enumerate}
		\item[I.] The first step reads for $\mathbf{J=\widetilde{T}}$ \footnote{Note again a slightly different notation compared to \cite{mezzadri taylor}.}
	\begin{subequations}\label{eq-first recurrence step T tilde}
	\begin{align}
		a_N&=\mathbf{r}_1^tZ\mathbf{l}_1,\label{eq-a_1 T tilde}\\
		\label{eq-r_2 T tilde}\mathbf{r}_2^t&=\mathbf{r}_1^tZ-a_N\mathbf{r}_1^t,\\
		b_{N-1}&=\mathbf{r}_2^tZ\mathbf{l}_1,\\
		\mathbf{l}_2&=\frac{1}{b_{N-1}}(Z\mathbf{l}_1-a_N\mathbf{l_1})\label{eq-l_2 T tilde}.
	\end{align}
	\end{subequations}
\item[II.] For $\mathbf{J=S}$, we find
\begin{subequations}\label{eq-first recurrence step S}
	\begin{align}
		a_N&=\mathbf{r}_1^t Z \mathbf{l}_1, \label{eq-recurrence a_n for S}\\
		\mathbf{r}_2^t&=\frac{1}{b_{N-1}}(\mathbf{r}_1^t Z-a_N\mathbf{r}_1^t) \label{eq-recurrence r_2 S}\\
		b_{N-1}&=\mathbf{r}_2^t Z \mathbf{l}_1 \label{eq-recurrence b_n-1 by vector eqs S} \\
		\mathbf{l}_2&=\frac{1}{b_{N-1}}(Z \mathbf{l}_1-a_N\mathbf{l}_1) \label{eq-recurrence l_2 S}\\
		\Rightarrow b_{N-1}&= \sqrt{(\mathbf{r}_1^t Z-a_N\mathbf{r}_1^t)Z \mathbf{l}_1}. \label{eq-recurrence b_n-1 S}
	\end{align}
\end{subequations}
Here, it is important to note that we find a recursive iteration due to the symmetry of $S$, i.e. the parameter $b_{N-1}$ appears both in Eq. \eqref{eq-recurrence r_2 S} and Eq. \eqref{eq-recurrence b_n-1 by vector eqs S}, which allows us to substitute \eqref{eq-recurrence b_n-1 by vector eqs S} in \eqref{eq-recurrence r_2 S} and find \eqref{eq-recurrence b_n-1 S}. Hence, the matrix parameters are recursively defined by the spectral parameters $\mathbf{r}_1^t,\ \mathbf{l}_1,\ Z$ via Eqs. \eqref{eq-recurrence a_n for S} and \eqref{eq-recurrence b_n-1 S}. The spectral parameters $\mathbf{r}_2^t$ and $\mathbf{l}_2$ are defined by Eqs. \eqref{eq-recurrence r_2 S} and \eqref{eq-recurrence l_2 S}. We note that the ordering of the recursion is here slightly different to the non-symmetric model, where the parameters are determined as $a_N$, $\mathbf{r}_2^t$, $b_{N-1}$ and $\mathbf{l}_2$ in Eqs. \eqref{eq-first recurrence step T tilde}.
\item[III.] For the general tridiagonal matrix $\mathbf{J=T}$, we find 
\begin{subequations}\label{eq-first recurrence step T}
	\begin{align}
		a_N&=\mathbf{r}_1^tZ\mathbf{l}_1,\label{eq-a_1 T}\\
		c_{N-1}&=\frac{(\mathbf{r}_1^tZ)_1-a_N(\mathbf{r}^t_1)_1}{(\mathbf{r}_2^t)_1},\label{eq-c_n-1 T}\\
		\label{eq-r_2 T}\mathbf{r}_2^t&=\frac{1}{c_{N-1}}(\mathbf{r}_1^tZ-a_N\mathbf{r}_1^t),\\
		b_{N-1}&=\mathbf{r}_2^tZ\mathbf{l}_1,\\
		\mathbf{l}_2&=\frac{1}{b_{N-1}}(Z\mathbf{l}_1-a_N\mathbf{l_1})\label{eq-l_2 T},
	\end{align}
\end{subequations}
where in the above equations we have $(\rv_j^t)_1=R_j$. Comparison with the non-symmetric model, where $c_{N-1}=1$ yields that there, Eq. \eqref{eq-c_n-1 T} is not needed and hence the spectral parameter $(\mathbf{r}_2^t)_1=R_1$. The recursive definitions agree otherwise. Especially, the ordering of the iterative mapping agrees within the non-symmetric and the general ensemble, while being different to the structure for the symmetric ensemble, discussed above.\\
In the case where $c_{N-1}=b_{N-1}=0$, the constructions in Eqs. \eqref{eq-r_2 T} and \eqref{eq-l_2 T} fail. Therefore, we refer to Remark \ref{rem-parameters 0} and decompose $T$ into two smaller tridiagonal matrices. Specifically, we express $T$ as the direct sum 
\[
T=T_1\oplus U_{N-1},
\]
where $T_1$ and $U_{N-1}$ are defined as in Remark \ref{rem-parameters 0}. This decomposition reduces the problem to two lower-dimensional matrices, which can be handled in two separate proofs. In general, if we encounter a $j$ for which $c_{N-j}=b_{N-j}=0$, we use the decomposition in Eq. \eqref{eq-decomp2} and carry out separate proofs for $T_j$ and $U_{N-j}$. The proofs will follow a similar structure to the one presented here, as $T_j$ and $U_{N-j}$ are tridiagonal matrices, albeit with lower degrees of freedom. Essentially, we restate the theorem for these lower-dimensional matrices and then perform the proofs for $T_j$ and $U_{N-j}$ independently. Therefore, for the remainder of the proof, we assume that $c_{N-j}$ and $b_{N-j}$ are never simultaneously zero for any $j$, thereby ensuring that our constructions avoid any singularities.
Additionally, since the probability distributions of the matrix elements, cf. Table \ref{tab-dists tridiag}, are absolutely continuous, for any fixed $Z$ the set of exceptional points $\rv_1^t$ such that $b_{N-1}=0$ or $c_{N-1}=0$ has measure zero in $\Hp_N$ and can therefore be neglected.
\end{enumerate}
For all three matrices $J=\widetilde{T}, S, T$ we successively express the LHS (cf. Eqs. \eqref{eq-first recurrence step T tilde}, \eqref{eq-first recurrence step S}, \eqref{eq-first recurrence step T} respectively) in terms of known quantities from the previous equations. Each of the Eqs. \eqref{eq-first recurrence step T tilde}, \eqref{eq-first recurrence step S}, \eqref{eq-first recurrence step T} respectively, solve the recurrence relations in Eqs. \eqref{eq-cnot1recrelation} and \eqref{eq-cnot1recrelation1} for the matching choice of $c_{N-1}=1$, $b_{N-1}=c_{N-1}$ and $b_{N-1}\neq c_{N-1}$ respectively. Now, we check that the solutions are consistent with the equation $RR^{-1}=\mathds{1}_N$. From the equations for $a_N$ and $\mathbf{r}_2^t$ one can deduce that $\mathbf{r}_2^t\mathbf{l}_1=0$. Then, from the equation for $a_N$ and $\mathbf{l}_2$ one can show that $\mathbf{r}_1^t\mathbf{l}_2=0$. Furthermore,
\begin{align*}
	\mathbf{r}_2^t\mathbf{l}_2&=\begin{cases}
		\frac{1}{b_{N-1}}\mathbf{r}_2^t(Z\mathbf{l}_1-a_N\mathbf{l_1}) & \text{, for } J=\widetilde{T}, T,\\
		\frac{1}{b_{N-1}^2}(\mathbf{r}_1^t Z-a_N\mathbf{r}_1^t)(Z\mathbf{l}_1-a_N\mathbf{l}_1)& \text{, for } J=S
	\end{cases}\\
&=1.
\end{align*}
Note that combining the orthogonality relations $\mathbf{r}_1^t\mathbf{l}_2=0$ and $\mathbf{r}_2^t\mathbf{l}_2=1$ with \eqref{eq-r_2 T tilde}, \eqref{eq-recurrence r_2 S}, \eqref{eq-r_2 T} respectively gives
\begin{align*}
	\mathbf{r}_1^tZ\mathbf{l}_2=\begin{cases}
		1& \text{, for } J=\widetilde{T},\\
		b_{N-1}& \text{, for } J=S,\\
		c_{N-1} & \text{, for } J=T,
	\end{cases}
\end{align*}
as expected. Furthermore, by the orthogonality relations $\mathbf{r}_2^t\mathbf{l}_2=1$ and $\mathbf{r}_2^t\mathbf{l_1}=0$ with \eqref{eq-l_2 T tilde}, \eqref{eq-recurrence l_2 S}, \eqref{eq-l_2 T} (which agree for all three matrix ensembles) respectively, we have
\begin{align*}
	\mathbf{r}_2^tZ\mathbf{l}_1=b_{N-1}
\end{align*}
for each $J=\widetilde{T}, S, T$ as expected. When $1<j<N$ take 
\begin{align*}
	\mathbf{r}_{j-1}^t,\, \mathbf{r}_j^t,\, \mathbf{l}_{j-1},\, \mathbf{l}_j,
\end{align*}
subject to the conditions
	\begin{subequations}\label{eqs-Gen Cases}
		\begin{align}
			&\mathbf{r}_j^t\mathbf{l}_j=\mathbf{r}^t_{j-1}\mathbf{l}_{j-1}=1,\\
			&\mathbf{r}_j^t\mathbf{l}_{j-1}=\mathbf{r}^t_{j-1}\mathbf{l}_j=0,
		\end{align}
	\end{subequations} for all three ensembles $J=\widetilde{T}, T, S$. For the off-diagonal elements, we have to distinguish between them, i.e. for $J=S$ we need further
\begin{align}\label{eq-Gen Cases b S}
	b_{N-j+1}=\sqrt{(\mathbf{r}_{j-1}^tZ-a_{N-j+2}\mathbf{r}_{j-1}^t-b_{N-j+2}\mathbf{r}_{j-2}^t)Z\mathbf{l}_{j-1}}\neq 0.
\end{align}
For $J=\widetilde{T}, T$, the condition on the sub-diagonal elements agree with 
\begin{align}\label{eq-Gen Cases b T T tilde}
	b_{N-j+1}=\mathbf{r}_j^tZ\mathbf{l}_{j-1}\neq0
\end{align}
and we have one further restriction for the construction of the general matrix $T$, with
\begin{align}\label{eq-Gen Cases c T}
	c_{N-j+1}=\frac{(\mathbf{r}_{j-1}^tZ)_1-b_{N-j+2}(\mathbf{r}_{j-2}^t)_1-a_{N-j+2}(\mathbf{r}_{j-1}^t)_1}{(\mathbf{r}_{j}^t)_1}\neq0.
\end{align}
Now, we define the $j$th recursion step for each ensemble.
\begin{enumerate}
	\item[I.] We start again with $\mathbf{J=\widetilde{T}}$
\begin{subequations}\label{eqs-Gen cases T tilde}
		\begin{align}
			a_{N-j+1}&=\mathbf{r}_j^tZ\mathbf{l}_j,\\
			\mathbf{r}_{j+1}^t&=\mathbf{r}_j^tZ-a_{N-j+1}\mathbf{r}_j^t-b_{N-j+1}\mathbf{r}^t_{j-1},\\
			b_{N-j}&=\mathbf{r}_{j+1}^tZ\mathbf{l}_j,\\
			\mathbf{l}_{j+1}&=\frac{1}{b_{N-j}}(Z\mathbf{l}_j-a_{N-j+1}\mathbf{l}_j-\mathbf{l}_{j-1}).
		\end{align}
	\end{subequations}
\item[II.] For the symmetric matrix $\mathbf{J=S}$, we give the recursive definition in the accurate order. Note that we illustrated in the first recursion step in Eqs. \eqref{eq-first recurrence step S} how to find a definition for $b_{N-1}$ and give the definitions there in a slightly different order. We set
\begin{subequations}\label{eqs-Gen cases S}
	\begin{align}
		a_{N-j+1}&=\mathbf{r}_j^tZ\mathbf{l}_j,\\
		b_{N-j}&=\sqrt{(\mathbf{r}_j^t Z-a_{N-j+1}\mathbf{r}_j^t-b_{N-j+1}\mathbf{r}_{j-1}^t)Z\mathbf{l}_j},\\
		\mathbf{r}_{j+1}^t&=\frac{1}{b_{N-j}}(\mathbf{r}_j^tZ-a_{N-j+1}\mathbf{r}_j^t-b_{N-j+1}\mathbf{r}^t_{j-1}),\\
		\mathbf{l}_{j+1}&=\frac{1}{b_{N-j}}(Z\mathbf{l}_j-a_{N-j+1}\mathbf{l}_j-b_{N-j+1}\mathbf{l}_{j-1}).
	\end{align}
\end{subequations}
\item[III.] For the general matrix $\mathbf{J=T}$, we define
\begin{subequations}\label{eqs-Gen cases T}
	\begin{align}
		&a_{N-j+1}=\mathbf{r}_j^tZ\mathbf{l}_j,\\
		&c_{N-j}=\frac{(\mathbf{r}_j^tZ)_1-b_{N-j+1}(\mathbf{r}_{j-1}^t)_1-a_{N-j+1}(\mathbf{r}_j^t)_1}{(\mathbf{r}_{j+1}^t)_1},\\
		&\mathbf{r}_{j+1}^t=\frac{1}{c_{N-j}}(\mathbf{r}_j^tZ-b_{N-j+1}\mathbf{r}^t_{j-1}-a_{N-j+1}\mathbf{r}_j^t),\\
		&b_{N-j}=\mathbf{r}_{j+1}^tZ\mathbf{l}_j,\\
		&\mathbf{l}_{j+1}=\frac{1}{b_{N-j}}(Z\mathbf{l}_j-c_{N-j+1}\mathbf{l}_{j-1}-a_{N-j+1}\mathbf{l}_j).
	\end{align}
\end{subequations}
\end{enumerate}
We note again that the definitions for $\widetilde{T}$ in \eqref{eqs-Gen cases T tilde} and $T$ in \eqref{eqs-Gen cases T} agree for $c_{N-j}=1$. In contrast to this, we have a significant different definition of the off-diagonal elements $b_{N-j}$ in the symmetric matrix $S$ and also the recursive structure needs a different ordering here. Combining Eqs. \eqref{eqs-Gen Cases}, \eqref{eq-Gen Cases b S}, \eqref{eq-Gen Cases b T T tilde}, \eqref{eq-Gen Cases c T} with corresponding Eqs. \eqref{eqs-Gen cases T tilde}, \eqref{eqs-Gen cases S}, \eqref{eqs-Gen cases T} respectively and proceeding in a similar way too, when $j=1$, we prove that
	\begin{align*}
		\mathbf{r}_k^t\mathbf{l}_j=0 \text{ , for }k=j-2,j-1,j+1,j+2,
	\end{align*}
	alongside
	\begin{align*}
		\mathbf{r}_{j+1}^t\mathbf{l}_{j+1}&=1,\\
		\mathbf{r}_j^tZ\mathbf{l}_{j+1}&=\begin{cases}
			1 & \text{, for } J=\widetilde{T},\\
			b_{N-j} & \text{, for } J=S,\\
			c_{N-j} & \text{, for } J=T,\\
		\end{cases}\\
		\mathbf{r}_{j+1}^tZ\mathbf{l}_j&=b_{N-j}.
	\end{align*}
	The orthogonality relations
\begin{align*}
	\mathbf{r}_{j+1}^t\mathbf{l}_k=0,\quad\mathbf{r}_k^t\mathbf{l}_{j+1}=0,
\end{align*}
	for $k=1,2,\dots,j-2$, follow by induction and from the fact that by construction $\rv_j^tZ\lv_k=0$ if $|j-k|>1$. The boundary conditions $b_0(=c_0)=0$ terminate the recurrence relations.
\end{proof}

\begin{Remark}
	It is a well-known fact \cite{Chalker Mehlig} that in non-Hermitian random matrix theory there exists a residual symmetry in rescaling the respective right- and left-eigenvectors simultaneously as follows: 
	\begin{align}
		\mathbf{r}_j^t \to d_j \mathbf{r}_j^t,\quad \mathbf{l}_j\to d_j^{-1} \mathbf{l}_j, \ 0\neq d_j\in\mathbb{C}, \quad j=1,\ldots,N\ .
		\label{eq-invariance}
	\end{align}
	This transformation does not change the biorthogonality of left- and right-eigenvectors in \eqref{eq-biorthogonality}, nor does it change the eigenvalues. 
	It is this rescaling that fixes the $N-1$ additional parameters $c_1,\ldots,c_{N-1}$ in ensemble $T$ given in Table \ref{tab-dists tridiag} last row, compared to ensemble $\tilde{T}$ given in Table \ref{tab-dists tridiag} top. The matrix elements of the diagonal matrix $\widetilde{C}$ in \eqref{eq-Ctildedef} relating the two ensembles provide the rescaling in \eqref{eq-invariance}, with $d_1=1,d_2=c_{N-1}, d_3=c_{N-1}c_{N-2}, \ldots, d_N= c_{N-1}\cdots c_1$. It is not difficult to see that this rescaling together with $b_j\to b_jc_j$ maps the recurrence \eqref{eqs-Gen cases T} above to the recurrence \eqref{eqs-Gen cases T tilde}.\\
	For complex symmetric matrices it is well known that the left- and right-eigenvectors are transpose to each other. Note that we construct the right- and left-eigenvectors component-wise in Eqs. \eqref{eq-recurrence r_2 S}, \eqref{eq-recurrence l_2 S} and \eqref{eqs-Gen cases S}.
\end{Remark}

The generalisation of the volume element calculation to the spectral decomposition of $T\in\Lw(n)$ from the non-symmetric matrix calculations, as seen in \cite{mezzadri taylor}, is more computationally involved. We need to introduce another $(N-1)$ complex parameters, requiring a more careful adaption of the previous theorems. For the symmetric ensemble $S$, where we have $4N-2$ real parameters as in the non-symmetric matrix $\widetilde{T}$, we do not require these additional spectral parameters. We define the matrix parameters
\begin{align}\label{eq-abc-var}
	\mathbf{a}:=(a_N,\dotsc,a_1),\quad\mathbf{b}:=(b_{N-1},\dotsc,b_1),\quad\mathbf{c}:=(c_{N-1},\dotsc,c_1),
\end{align}
where $\mathbf{c}$ can be suppressed for $J=\widetilde{T}, S$. We proved an injective correspondence with the $4N-2$, respectively $6N-4$ real spectral parameters
\begin{align}\label{eq-lrR-var}
	\mathbf{z}:=(z_1,\dotsc,z_N),\quad\rv:=(r_1,\dotsc,r_N)\quad\text{and}\quad\mathbf{R}_1:=(R_1,\dotsc,R_N),
\end{align}
when $r_1+\dotsc+r_N=1$ and again $\mathbf{R}$ can be suppressed for $J=\widetilde{T}, S$. We would like to repeat here that while we perform the proofs using a unified notation for all three tridiagonal matrix ensembles under consideration, the actual parameters agree only in their names.

\begin{lemma}\label{lem-jac1}
	Let $J=RZ R^{-1}$ be the spectral decomposition of $J\in \Lw(N)$, where $R$ is defined in \eqref{eq-def Rgenmodel} and
	$Z = \diag(z_1, \dotsc,z_N)$. For each $J=\widetilde{T},S, T$ we have
	\begin{align}\label{eq-Vandermonde determinant reduced}
		\Delta_N(\boldsymbol{z})^2=\frac{\prod_{j=1}^{N-1}(b_jc_j)^j}{\prod_{j=1}^{N}r_{j}},
	\end{align}
	where $\rv = (r_1,\dotsc,r_N)$ is the first row of $R$ with $r_1 + \dotsb + r_N =1$ and $c_j=1$ for $J=\widetilde{T}$, respectively $b_j=c_j$ for $J=S$.
\end{lemma}
We repeat the proof of \cite[Lemma $3.3$]{mezzadri taylor} when the substitution $b_j\to b_jc_j$ is made for completeness below. First, we state some results on the characteristic polynomial given and proved for the general matrix $J$ in \cite[Proposition $3.2$]{mezzadri taylor}. We recall that $\Phi_k(z)$ in Eq. \eqref{eq-det charac poly} is the characteristic polynomial of the sub-matrix $J_k$ \eqref{eq-tridiag k}, obtained after removing the last $(N-k)$ rows and columns of $J$. Let $\mathcal{P}_j[\mathbb{C}]$ be the set of monic polynomials of degree $j$ over $\mathbb{C}$ and $\mathbf{e}_j$ be a canonical basis vector in $\mathbb{C}^N$, i.e $\mathbf{e}_j^t
=\overset{j}{(0,\dotsc,1,\dotsc,0)}$.
\begin{prop}[\text{\cite[Proposition $3.2$]{mezzadri taylor}}]
	\label{prop-char poly}
	Let $j=1,\dotsc,N-1.$ The following statements hold:
	\begin{enumerate}
		\item $\bigl|b_{N-1}b_{N-2}\dotsm b_{N-j}\bigr|=
		\bigl \|\Phi_j(J)\mathbf{e}_1\bigr \|=\underset{\psi \in \mathcal{P}_j[\mathbb{C}]}
		{\min}\bigl \|\psi(J)\mathbf{e}_1\bigr \|$;
		\item $\Phi_j(J)\mathbf{e}_1=\mathbf{e}_{j+1}b_{N-1}b_{N-2}\dotsm b_{N-j}$;
		\item $|c_{N-1}c_{N-2}\dotsm c_{N -j}|=
		\bigl \|\mathbf{e}_1^t\Phi_j(J)\bigr \|=\underset{\psi \in \mathcal{P}_j[\mathbb{C}]}
		{\min}\bigl \|\mathbf{e}_1^t\psi(J)\bigr\|$;
		\item $\mathbf{e}_1^t\Phi_j(J)= \mathbf{e}_{j+1}^tc_{N-1}c_{N-2}\dotsm c_{N-j}$.
	\end{enumerate}
\end{prop}
\begin{proof} We present a proof of Lemma \ref{lem-jac1} for general $J$, which follows the one of \cite[Lemma $3.3$]{mezzadri taylor} closely. Proposition \ref{prop-char poly} yields
	\begin{subequations}
		\begin{align}
			\Phi_j(J)\mathbf{e}_1 & =\mathbf{e}_{j+1}b_{N-1}b_{N-2}\dotsm b_{N-j},\\
			\mathbf{e}_1^t\Phi_j(J) &= \mathbf{e}_{j+1}^tc_{N-1}c_{N-2}\dotsm c_{N-j},
		\end{align}
	\end{subequations}
	for $j=1,2,\dotsc,N-1$. Together with the spectral decomposition $J = RZ R^{-1}$ and the definition of $R$ given in \eqref{eq-def Rgenmodel}, we find
	\begin{align*}
		\Phi_j(Z)R^{-1}\mathbf{e}_1 & = \Phi_j(Z) \mathbf{l}_1 = R^{-1}\mathbf{e}_{j+1}b_{N-1}b_{N-2}\dotsm b_{N-j} \nonumber \\
		& = \mathbf{l}_{j+1}b_{N-1}b_{N-2} \dotsm b_{N-j},\\
		\mathbf{e}_1^tR\Phi_j(Z) &= \mathbf{r}_1^t\Phi_j(Z) = \mathbf{e}_{j+1}^t R  c_{N-1}c_{N-2}\dotsm c_{N-j}\\
		& = \mathbf{r}_{j+1}^t c_{N-1}c_{N-2}\dotsm c_{N-j}.
	\end{align*}
	Multiplying these equations side by side gives 
	\begin{align*}
	\mathbf{r}_1^t\Phi_j(Z) \Phi_k(Z)\mathbf{l}_1 = 
	\mathbf{r}_{j+1}^t  \mathbf{l}_{k+1}\prod_{m=1}^k b_{N-m}c_{N-m},
	\end{align*}
	which can be rewritten as 
	\begin{align}\label{eq-lem jac tridiag}
		\sum_{l=1}^N r_l\Phi_j(z_l)\Phi_k(z_l)=
		\begin{cases} 1 & \text{if $j=k=0$,} \\
			\delta_{jk}\prod_{m=1}^k b_{N-m}c_{N-m}, & \text{if $j \neq 0$ or $k \neq 0$,}
		\end{cases}
	\end{align}
	for $j,k=0,\dotsc, N-1$, where we have used $\mathbf{l}_1^t = (1,\dotsc,1)$ and the fact that by definition $\Phi_0(z)=1$.
	
	We define
	\begin{equation*}
		W = \begin{pmatrix} 1 & \Phi_1(z_1) & \dots &\Phi_{N-1} (z_{1}) \\
			1 & \Phi_1(z_2) & \dots &\Phi_{N-1} (z_{2}) \\
			\hdotsfor[2]{4}\\
			1 & \Phi_1(z_N) & \dots &\Phi_{N-1} (z_{N}) \\
		\end{pmatrix}                        
	\end{equation*}
	and
	\begin{equation*}
		Q = \begin{pmatrix} r_1 & & & \\
			& r_2 & &  \\
			& & \ddots& \\
			& &  & r_N 
		\end{pmatrix},  \quad 
		K = \begin{pmatrix} 1 & & & \\
			& b_{N-1} c_{N-1}& &  \\
			& & \ddots& \\
			& &  & \prod_{m=1}^{N-1} b_{N-m} c_{N-m}
		\end{pmatrix}.
	\end{equation*}
	Using these, Eq. \eqref{eq-lem jac tridiag} can be rewritten as the matrix product
	\begin{equation*}
		W^t Q W = K.
	\end{equation*}
	 Taking the determinant of both sides of this equation and using $\det(W) = \Delta_N(\mathbf{z})$ completes the proof.
\end{proof}
This expression of the Vandermonde determinant in terms of the off-diagonal elements and eigenvector-components is the key which enables us to find a Coulomb-type repulsion in the JPDF of the complex eigenvalues. A very similar relation was found in the Hermitian $\beta$-ensembles, cf. \cite{Dumitriu Edelman,book forrester}. Now, we have to perform a change of parameters from the matrix elements to the spectral parameters, and we will find that the appearing factor can be partly rewritten in terms of the Vandermonde determinant. The following theorem gives the relation between the differentials of the matrix elements and the spectral parameters (cf. \cite[Theorem $3.1$]{mezzadri taylor} for $J=\widetilde{T}$):
\begin{theorem}\label{thm-generalised Jacobian}
	We have for\begin{itemize}
		\item $\mathbf{J=\widetilde{T}}$ \begin{align}
			\prod_{j=1}^N d^2a_j \prod_{k=1}^{N-1}d^2b_k
		 = \left|\frac{\prod_{j=1}^{N-1}b_j}{\prod_{k=1}^N r_k}\right|^2
			\prod_{k=1}^{N-1}d^2r_k\prod_{j=1}^N d^2z_j,
		\end{align}
		\item $\mathbf{J=S}$ \begin{align}
			\prod_{j=1}^N d^2a_j \prod_{k=1}^{N-1}d^2b_k = \left|\frac{\prod_{j=1}^{N-1}b_j }{\prod_{k=1}^N r_k}\right|^2
			\prod_{k=1}^{N-1}d^2r_k\prod_{j=1}^N d^2z_j,
		\end{align}
		\item and $\mathbf{J=T}$ \begin{align}
			\prod_{j=1}^N d^2a_j \prod_{k=1}^{N-1}d^2b_k
			d^2c_k = \left|\frac{\prod_{j=1}^{N-1}b_jc_j }{\prod_{k=1}^N r_k\prod_{w=2}^NR_w}\right|^2
			\prod_{k=1}^{N-1}d^2r_k\prod_{w=2}^Nd^2R_w\prod_{j=1}^N d^2z_j.
		\end{align}
	\end{itemize}
	
The parameter domains are $\{a_j\}_{j=1}^N\in \mathbb{C}^N$, $\{b_j\}_{j=1}^{N-1}\in \mathbb{C}^{N-1}$, $\{c_j\}_{j=1}^{N-1}\in \mathbb{C}^{N-1}$ for the matrix parameters subject to the distributions given in Table \ref{tab-dists tridiag} for each $J=\widetilde{T},S,T$ and $\{z_j\}_{j=1}^N\in \mathbb{C}^N$, $\{r_j\}_{j=1}^{N}\in \mathbb{C}^{N-1}$ with $\sum_{j=1}^N r_j=1$ and $\{R_j\}_{j=2}^N\in \mathbb{C}^{N-1}$ as spectral parameters. Again, we suppress the parameters $\{c_j\}_{j=1}^{N-1}$ and $\{R_j\}_{j=2}^N$ for the non-symmetric and symmetric matrices $\widetilde{T},S$.
\end{theorem}
\begin{Remark}
	Before we proceed with the proof of this theorem, we would like to draw the reader's attention to the comparison of the results for the three matrix ensembles. In the previous discussion, we treated them at the same time by the replacement $c_j=1$ for $\widetilde{T}$ and $b_j=c_j$ for $S$. However, for the results in this theorem, even in the product over the off-diagonal elements, this does no longer holds. The reason is that we are now considering differentials and hence the appearing $b_j^2$ in the three-term recurrence of the characteristic polynomial of the symmetric matrix $S$ compared to $b_j$ in $\widetilde{T}$ and $b_jc_j$ in $T$ cause additional factors from an inner derivative.
\end{Remark}
\begin{proof}
We present the proof again for the general matrix $J$ \eqref{eq-tridiag} and explain the specialisations to the various matrix ensembles throughout the lines. Let $J=RZ R^{-1}$ denote the spectral decomposition of $J$. Denote by $\mathbf{R}_1$ the first column of $R$ and let its $j$th component be denoted by $R_j$. We note that the first right-eigenvector of $J$ satisfies the equation
\begin{align*}
	J\mathbf{R}_1=z_1\mathbf{R}_1.
\end{align*}
Hence, if we equate row $j$ of this equation then
	\begin{align}\label{eq-firstrighteigenvector}
		b_{N-j+1}R_{j-1}+a_{N-j+1}R_j+c_{N-j}R_{j+1}=z_1R_j,
	\end{align}
	for $1\leq j\leq N$ with boundary conditions $R_0=c_0=0$. The differential of this equation is
	\begin{align}\label{eq-diff}
		R_{j-1}&db_{N-j+1}+R_jda_{N-j+1}+R_{j+1}dc_{N-j}\nonumber\\&=R_jdz_1+dR_j[z_1-a_{N-j+1}]-b_{N-j+1}dR_{j-1}-c_{N-j}dR_{j+1}.
	\end{align}
We will need this equation only for the general tridiagonal matrix $T$ to determine the $2N-2$ additional parameters compared to $\widetilde{T}$ and $S$. Nevertheless, these equations hold true also for the latter two ensembles with $c_j=1$, respectively $b_j=c_j$, while $R_j$ do not serve as independent parameters here. The remaining $4N-2$ conditions are obtained in a very similar way as in the proof performed by Dumitriu and Edelman \cite{Dumitriu Edelman} or Forrester \cite{book forrester} on the Gaussian $\beta$-ensemble. We consider the first element of the von Neumann series
	\begin{align}\label{eq-resolvent}
		((\mathds{1}_N-zJ)^{-1})_{11}=\sum_{j=1}^N\frac{r_j}{1-zz_j},
	\end{align}
	where RHS follows from the diagonalisation $J=RZ R^{-1}$. We would like to pause here to explain the power of this equation \eqref{eq-resolvent}: On the RHS, we find only spectral parameters which we identified as elementary in the one-to-one-mapping in Theorem \ref{thm-roughbijection}, i.e. $\{r_j\}_{j=1}^{N-1}$ and $\{z_j\}_{j=1}^N$ in each of the ensembles $J=\widetilde{T}, S, T$. On the other hand, we have on the LHS only matrix parameters. The crucial point is now, that the Taylor expansion in powers of $z$ yields a recursive appearance of these in the expansion due to the tridiagonal structure of the initial matrix $J$. This is the main point, why the proof by Dumitriu and Edelman in 2002 and also here, will work out. One can also see from here, that for a Hessenberg matrix (to which we can reduce a complex Ginibre matrix, cf. Eq. \eqref{eq-Hessenberg GinUE}) which has in general an $N$-term recurrence of the characteristic polynomial will not share this attractive feature.\\
	Taking the Taylor expansion of both sides of the equation and equating the coefficients of the powers of $z$ up to $z^{2N-1}$, we find the needed $4N-2$ real conditions mentioned above. For the general matrix $J$ an interlacing with the rows of the eigenvector equation \eqref{eq-firstrighteigenvector} is needed to complete the equation set. We achieve:\clearpage
	\begin{subequations}
		\label{eqs-Taylor_exps_1}
		\begin{align}
			\label{eq-ord_0_1}
			O(z^0):&\hspace{3mm}1=\sum_{j=1}^Nr_j,\\
			\label{eq-ex1_1}
			O(z^1):&\hspace{3mm}a_N=\sum_{j=1}^Nr_jz_j,\\
			&\nonumber\\j=1:&\hspace{3mm}a_NR_1+c_{N-1}R_2=z_1R_1,\\
			&\nonumber\\O(z^2):&\hspace{3mm}
			a_N^2+c_{N-1}b_{N-1}=*\;+c_{N-1}b_{N-1}=\sum_{j=1}^Nr_jz_j^2
			\label{eq-231d},\\
			O(z^3):&\hspace{3mm}*\;+a_{N-1}b_{N-1}c_{N-1}=\sum_{j=1}^Nr_jz_j^3,\\
			&\nonumber\\j=2:&\hspace{3mm}
			*\:+c_{N-2}R_3=z_1R_2,\\
			&\nonumber\\O(z^4):&\hspace{3mm}*\; +b_{N-1}c_{N-1}b_{N-2}c_{N-2}=\sum_{j=1}^Nr_jz_j^4,\\
			O(z^5):&\hspace{3mm}*\;+a_{N-2}b_{N-2}c_{N-2}b_{N-1}c_{N-1}=\sum_{j=1}^Nr_jz_j^5,\\
			\nonumber \vdots \hspace{.6cm} & \hspace{1.8cm}\vdots \hspace{2.5cm}\vdots\\
			&\nonumber\\
			j=N-1:&\hspace{3mm}
			*\:+c_{1}R_N=z_1R_{N-1},\\
			&\nonumber\\O(z^{2N-2}):&\hspace{3mm}*\;+b_1c_1\dots b_{N-1}c_{N-1}=\sum_{j=1}^Nr_jz_j^{2N-2},\\
			\label{eq-ex2n-1_1}
			O(z^{2N-1}):&\hspace{3mm}*\; +a_1b_{1}c_{1}\dots b_{N-1}c_{N-1}=\sum_{j=1}^Nr_jz_j^{2N-1}.
		\end{align}
	\end{subequations}
	Here, $*$ denotes terms that have appeared in the previous equations, as exemplified in \eqref{eq-231d}.\footnote{Note, for every even integer power $2j$ of $z$ we introduce $b_{N-j}$ and $c_{N-j+1}$. To counteract adding two new variables each even integer line we interlace in the rows of the eigenvector equation. This provides us with a formula for the $b_j$.}
	
	We choose the ordering $a_N, c_{N-1}, b_{N-1}, a_{N-1}, c_{N-2}, b_{N-2},\dots$ for $J=T$, respectively $a_N, b_{N-1}, a_{N-1}, b_{N-2},\dots$ for $J=\widetilde{T},S$, such that the matrix of coefficients for the differentials of the LHS is triangular. Hence, taking the wedge product line by line gives
	\begin{enumerate}
		\item[I.] for $\mathbf{J=\widetilde{T}}$ with $c_j=1$
	\begin{align}\label{eq-leftderivative T Tilde}
		(-1)^{(N-1)(N-2)/2}\prod_{w=1}^{N-1}(b_{w})^{2w-1}\hspace{1mm}d\mathbf{a}\wedge d\mathbf{b}.
	\end{align}
\item[II.] For $\mathbf{J=S}$ with $b_j=c_j$, we find
	\begin{align}\label{eq-leftderivative S}
		(-1)^{(N-1)(N-2)/2}\prod_{w=1}^{N-1}(b_{w})^{4w-1}\hspace{1mm}d\mathbf{a}\wedge d\mathbf{b}
	\end{align}
\item[III.] For $\mathbf{J=T}$ we have
	\begin{align}\label{eq-leftderivative T}
		(-1)^{(N-1)(N-2)/2}\prod_{j=1}^{N-1}R_{j+1}\prod_{w=1}^{N-1}(b_{w}c_{w})^{2w-1}\prod_{i=1}^{N-1}c_i\hspace{1mm}d\mathbf{a}\wedge d\mathbf{b}\wedge d\mathbf{c},
	\end{align}
	whilst noting that we have re-arranged Eq. \eqref{eq-diff} such that the LHS of the differentials of the eigenvector Eqs. \eqref{eqs-Taylor_exps_1} is
	\begin{align}\label{eq-lhsderivative}
		R_{j-1}db_{N-j+1}+R_jda_{N-j+1}+R_{j+1}dc_{N-j}.
	\end{align}
	\end{enumerate} 
Comparison within the three ensembles shows that, while the overall structure is very similar, the symmetric model differs from the general one with replacement $b_j=c_j$ due to a factor $\prod_{w=1}^{N-1}b_w$, which comes from the inner derivatives of the factors $b_w^2$ in Eqs. \eqref{eqs-Taylor_exps_1}. For $J=S$, the result \eqref{eq-leftderivative S} agrees with the calculations in the Hermitian set up (cf. \cite[Proof of Proposition $1.9.3$]{book forrester}) with the replacement of real variables by complex ones.\\ 
Differentiating the RHS of Eqs. \eqref{eqs-Taylor_exps_1}, without the introduction of the rows of the eigenvector equations, i.e. for $J=\widetilde{T},S$, leads to 
	\begin{align}\label{eq-newjacdiffs}
		d\Big(\sum_{k=1}^Nr_kz_k^j\Big)=\sum_{k=1}^{N-1}dr_k(z_k^j-z_N^j)+j\sum_{k=1}^Nr_kz_k^{j-1}dz_k,
	\end{align}
	where we have used the differential relation
	\begin{align*}
	dr_N=-\sum_{j=1}^{N-1}dr_j,
\end{align*}
	as a substitution for $dr_N$. For the matrix ensembles of non-symmetric $\widetilde{T}$, respectively symmetric $S$, matrices, no further differentiation is needed. For the general matrix $T$, we note that the RHS of the differential of Eq. \eqref{eq-firstrighteigenvector} is fixed by Eq. \eqref{eq-lhsderivative} to be
	\begin{align}\label{eq-firsteigendiff}
		R_jdz_1+dR_j[z_1-a_{N-j+1}]-b_{N-j+1}dR_{j-1}-c_{N-j}dR_{j+1}.
	\end{align}
Taking the wedge product, line by line, of the RHS differentials of Eq. \eqref{eqs-Taylor_exps_1} produces
\begin{enumerate}
	\item[I.] for $\mathbf{J=\widetilde{T},S}$ (where the RHS of Eq. \eqref{eqs-Taylor_exps_1} agree)
\begin{align}\label{eq-Jacrearrange1 T tilde S}
	\left(\prod_{k=1}^Nr_k \right)
	\det\bigg[\left(z_k^j-z_N^j\right)_{
		\begin{subarray}{l}
			j=1,\dotsc,2N-1 \\ k=1,\dotsc,N-1
	\end{subarray}}
	\left(jz_k^{j-1}\right)_{
		\begin{subarray}{l}
			j=1,\dotsc,2N-1 \\ k=1,\dotsc,N
	\end{subarray}}
	\bigg] 
	d\mathbf{r}\wedge d\mathbf{z}.
\end{align}
\item[II.] For $\mathbf{J=T}$, we find
	\begin{align}\label{eq-Jacrearrange1 T}
		\prod_{j=1}^{N-1}c_j\left(\prod_{k=1}^Nr_k \right)
		\det\bigg[\left(z_k^j-z_N^j\right)_{
			\begin{subarray}{l}
				j=1,\dotsc,2N-1 \\ k=1,\dotsc,N-1
		\end{subarray}}
		\left(jz_k^{j-1}\right)_{
			\begin{subarray}{l}
				j=1,\dotsc,2N-1 \\ k=1,\dotsc,N
		\end{subarray}}
		\bigg] 
		d\mathbf{r}\wedge d\mathbf{R}_1\wedge d\mathbf{z}.
	\end{align}
\end{enumerate}
	The determinant in Eqs. \eqref{eq-Jacrearrange1 T tilde S}, \eqref{eq-Jacrearrange1 T} is calculated explicitly in the book by Forrester \cite[Eq. (1.175)]{book forrester}.
	We have
	\begin{gather}
		(-1)^{(N-1)(N-2)/2}\det\bigg[\left(z_k^j-z_N^j\right)_{
			\begin{subarray}{l}
				j=1,\dotsc,2N-1 \\ k=1,\dotsc,N-1
		\end{subarray}}
		\left(jz_k^{j-1}\right)_{
			\begin{subarray}{l}
				j=1,\dotsc,2N-1 \\ k=1,\dotsc,N
		\end{subarray}}
		\bigg] \nonumber \\
		\label{eq-crazy_det}
		= \prod_{1 \le j < k \le N}\left(z_k - z_j\right)^4 = \Delta_N(\mathbf{z})^4.
	\end{gather}
	Substituting Eq. \eqref{eq-crazy_det} into Eq. \eqref{eq-Jacrearrange1 T tilde S}, \eqref{eq-Jacrearrange1 T} respectively, we produce 
	\begin{enumerate}
		\item[I.] for $\mathbf{J=\widetilde{T},S}$
	\begin{align}\label{eq-rhsderivative T tilde S}
		&(-1)^{(N-1)}(-1)^{(N-1)(N-2)/2}\left(\prod_{k=1}^Nr_k \right)\Delta_N(\mathbf{z})^4
		d\mathbf{r}\wedge d\mathbf{z},
	\end{align}
\item[II.] and for $\mathbf{J=T}$
	\begin{align}\label{eq-rhsderivative T}
		&(-1)^{(N-1)}(-1)^{(N-1)(N-2)/2}\prod_{j=1}^{N-1}c_j\left(\prod_{k=1}^Nr_k \right)\Delta_N(\mathbf{z})^4
		d\mathbf{r}\wedge d\mathbf{R}_1\wedge d\mathbf{z}.
	\end{align}
	\end{enumerate}
	Note, to achieve the above results one needs to utilise that the wedge product is anti-commutative.
	Combining the Eqs. \eqref{eq-leftderivative T Tilde} and \eqref{eq-rhsderivative T tilde S} we obtain 
	\begin{enumerate}
		\item[I.] for $\mathbf{J=\widetilde{T}}$
	\begin{align*}
	\prod_{i=1}^{N-1}(b_{i})^{2i-1}d\mathbf{a}\wedge d\mathbf{b}=(-1)^{(N-1)}\left(\prod_{k=1}^Nr_k \right)\Delta_N(\mathbf{z})^4
	d\mathbf{r}\wedge  d\mathbf{z},\\
	\Rightarrow d\mathbf{a}\wedge d\mathbf{b} =(-1)^{N-1} \frac{\left(\prod_{k=1}^N r_k\right) \Delta_N(\mathbf{z})^4}{\prod_{j=1}^{N-1}(b_{j})^{2j-1}} d\mathbf{r}\wedge  d\mathbf{z}.
	\end{align*}
\item[II.] Similar, we find with Eqs. \eqref{eq-leftderivative S} and \eqref{eq-rhsderivative T tilde S} for $\mathbf{J=S}$
\begin{align*}
	\prod_{i=1}^{N-1}(b_{i})^{4i-1}d\mathbf{a}\wedge d\mathbf{b}=(-1)^{(N-1)}\left(\prod_{k=1}^Nr_k \right)\Delta_N(\mathbf{z})^4
	d\mathbf{r}\wedge  d\mathbf{z},\\
	\Rightarrow d\mathbf{a}\wedge d\mathbf{b} =(-1)^{N-1} \frac{\left(\prod_{k=1}^N r_k\right) \Delta_N(\mathbf{z})^4}{\prod_{j=1}^{N-1}(b_{j})^{4j-1}} d\mathbf{r}\wedge  d\mathbf{z}.
\end{align*}
\item[III.] For $\mathbf{J=T}$, we conclude with Eqs. \eqref{eq-leftderivative T} and \eqref{eq-rhsderivative T}
\begin{align*}
	\prod_{j=1}^{N-1}R_{j+1}\prod_{i=1}^{N-1}(b_{i}c_{i})^{2i-1}d\mathbf{a}\wedge d\mathbf{b}\wedge d\mathbf{c}=(-1)^{(N-1)}\left(\prod_{k=1}^Nr_k \right)\Delta_N(\mathbf{z})^4
	d\mathbf{r}\wedge d\mathbf{R}_1\wedge  d\mathbf{z},\\
	\Rightarrow d\mathbf{a}\wedge d\mathbf{b}\wedge d\mathbf{c} =(-1)^{N-1} \frac{\left(\prod_{k=1}^N r_k\right) \Delta_N(\mathbf{z})^4}{\prod_{j=1}^{N-1}(c_{j}b_{j})^{2j-1}\prod_{w=2}^{N}R_{w}} d\mathbf{r}\wedge d\mathbf{R}_1\wedge  d\mathbf{z}.
\end{align*}
	\end{enumerate}
	Substitution of Lemma \ref{lem-jac1}, with the replacement $c_j=1$ for $\widetilde{T}$ and $b_j=c_j$ for $S$, into the above equations completes the proof.
\end{proof}
We are now in the position to state and prove the following first main statement in this section.
\begin{theorem}\label{thm-JPDF tridiag}
	 The JPDF of the eigenvalues of the random matrix ensemble $J=\widetilde{T},S,T$ is given by 
	\begin{align}
		\label{eq-jpdf eig tridiag}
		P_J(Z) = \frac{1}{Z_\beta^J} \prod_{j=1}^N
		\exp\left(-\frac{|z_j|^2}{2}\right)\prod_{1\le j < k \le N}|z_k - 
		z_j|^\beta\  f_J(Z),
	\end{align}
	where $\beta\in \mathbb{R}_+$, $Z=\diag(z_1,\ldots,z_N)$,  $Z_\beta^J$ normalization constant with 
	\begin{subequations}\label{eqs-partition functions tridiag}
	\begin{align}
		Z_\beta^{\widetilde{T}}&=\pi^{2N-1}2^{\frac{N((N-1)\beta+8)}{8}}\prod_{k=1}^{N-1}\Gamma\left(\frac{\beta k}{4}\right),\\
		Z_\beta^{S}&=\pi^{2N-1}2^{\frac{3N-1}{2}+\frac{N(N-1)\beta}{4}}\prod_{k=1}^{N-1}\Gamma\left(\frac{\beta k}{2}\right),\\
		Z_\beta^{T}&=\pi^{3N-2}2^{\frac{N((N-1)\beta+4)}{4}}\prod_{k=1}^{N-1}\Gamma\left(\frac{\beta k}{4}\right)^2.
	\end{align}
\end{subequations}
The function $f_J(Z)$ contains a coupling between eigenvalues and eigenvectors.
\begin{itemize}
	\item For $\mathbf{J=\widetilde{T},S}$: $f_J(Z)$ depends on the first element $\mathbf{r} =(r_1,\dotsc,r_{N}) \in \mathbb{C}^{N}$ of each right-eigenvector of $J$, with $1 = r_1 +\ldots+ r_{N}$, as 
	\begin{align}\label{eq-factor f S T tilde}
		f_J(Z) =
		\int_{\mathbb{C}^{N-1}}\!\prod_{k=1}^{N-1}d^2r_k\ e^{-g_J(Z,\mathbf{r})}\prod_{j=1}^{N}|r_j|^{\frac{\beta}{2}-2}.
	\end{align}
\item For $\mathbf{J=T}$: $f_T(Z)$ depends on the first element $\mathbf{r}_1 =(r_1,\dotsc,r_{N}) \in \mathbb{C}^{N}$ of each right-eigenvector of $T$, with $1 = r_1 +\ldots+ r_{N}$, 
and $\mathbf{R}_1=(R_1,\dotsc,R_N)^t$ denoting the first right-eigenvector of $T$ 
\begin{align}\label{eq-factor f T}
	f_T(Z) =\int_{\mathbb{C}^{(N-1)}}\int_{\mathbb{C}^{N-1}\backslash \{0\}}\prod_{i=2}^N|R_i|^{-2}d^2R_id^2r_{i-1}\ 
	e^{-g_{T}(Z,\mathbf{r},\mathbf{R}_1)}\prod_{k=1}^{N}|r_k|^{\frac{\beta}{2}-2}.
\end{align}
\end{itemize}
The exponent is positive with
\begin{align}\label{eq-def g J}
	g_J(Z,\mathbf{r},\mathbf{R}_1) =  \frac{1}{2}\Tr(JJ^\dagger)
	- \frac{1}{2}\sum_{j =1}^N |z_j|^2,
\end{align}
where the $\mathbf{R}_1$-dependence applies only for $J=T$. It explicitly depends on the respective random variables $a_j$, $j=1,\ldots,N$, $b_k$ and $c_k$, $k=\ldots,N-1$ due to the one-to-one-correspondence established in Theorem \ref{thm-roughbijection}. Again, the $c_k$-dependence holds only for $J=T$. Furthermore, it is rotationally invariant, 
$g_{J}(Z,\mathbf{r},\mathbf{R}_1)=g_{J}(Z e^{i\phi},\mathbf{r},\mathbf{R}_1)$ for all $\phi\in[0,2\pi)$ and has the asymptotics:
\begin{align} 
	\label{g-asymptotic}
	g_{J}(Z,\mathbf{r},\mathbf{R}_1)=\begin{cases}
		O\left(|z_j|^4\right) & \text{, for } J=\widetilde{T},\\
		O\left(|z_j|^2\right) & \text{, for } J=S,T
	\end{cases}
	\end{align}
as $z_j\rightarrow\infty$ for $j=1,\ldots,N$
\end{theorem}
\begin{Remark}
	We would like to pause and discuss these results another time before proving them below. What did we achieve with the definitions of this three tridiagonal non-Hermitian matrices $\widetilde{T},S, T$ in Eq. \eqref{eq-tridiag} and Table \ref{tab-dists tridiag}? Due to "clever" choice of the distributions of the matrix elements, we encounter a Coulomb-type interaction term in the JPDF \eqref{eq-jpdf eig tridiag}, i.e. a factor $|\Delta_N(Z)|^\beta$. Additionally, we find a non-trivial function $f_J(Z)$, which represents a coupling between eigenvalues and eigenvector-components. The latter depends on the Frobenius norm of the triangular part of the Schur decomposition. The dependence on the matrix variables follows from the one-to-one-mapping established in Theorem \ref{thm-roughbijection} and is therefore, while explicit, still challenging.\\
	Another question might be: Do we recover the complex Ginibre ensemble for $\beta=2$? The answer is: No, because of the additional factor $f_J(Z)$. We recall from the discussion on the Ginibre ensemble that there the triangular part from the Schur decomposition decouples and can be integrated out. This is not the case here. Moreover, we would like to stress that we could not suspect to find GinUE from the beginning. We introduced tridiagonal matrices with complex normal, Chi- or uniformly distributed elements. This choice was not motivated from a Householder reduction as in the Hermitian situation. We discussed that a GinUE reduces via Householder reflections only to a Hessenberg-form, where again we find complex normal and Chi-distributed elements and is it not clear how this should be reduced further to a tridiagonal structure and will (to the best of our knowledge) not result in Chi-distributed (plus uniform angles) on the off-diagonals.
\end{Remark}
\begin{proof}
	The JPDF of the matrix elements of matrices $J\in\Lw'(N)$, distributed according to Table \ref{tab-dists tridiag} reads
	\begin{enumerate}
		\item[I.] for $\mathbf{J=\widetilde{T}}$
	\begin{align}
		\label{eq-jpdf_mat_el_T tilde}
		p_{\beta}^{\widetilde{T}}(\mathbf{a},\mathbf{b}) = \frac{1}{Z_\beta^{\widetilde{T}}}\prod_{j=1}^{N-1}\abs{b_j}^{\frac{\beta j}{2} -2} 
		\exp\left[ -\frac12\left(\sum_{j=1}^N \abs{a_j}^2 + \sum_{j=1}^{N-1} 
		\abs{b_j}^2+(N-1)\right)\right],
	\end{align}
\item[II.] for $\mathbf{J=S}$ 
\begin{align}
	\label{eq-jpdf_mat_el_S}
	p_{\beta}^S(\mathbf{a},\mathbf{b}) = \frac{1}{Z_\beta^{S}}\prod_{j=1}^{N-1}\abs{b_j}^{\beta j -2} 
	\exp\left[ -\frac12\left(\sum_{j=1}^N \abs{a_j}^2 + 2\sum_{j=1}^{N-1} 
	\abs{b_j}^2\right)\right],
\end{align}
\item[III.] and for $\mathbf{J=T}$
	\begin{align}
		\label{eq-jpdf_mat_el_T}
		p_{\beta}^T(\mathbf{a},\mathbf{b},\mathbf{c}) = \frac{1}{Z_\beta^{T}}\prod_{j=1}^{N-1}\abs{b_jc_j}^{\frac{\beta j}{2} -2} 
		\exp\left[ -\frac12\left(\sum_{j=1}^N \abs{a_j}^2 + \sum_{j=1}^{N-1} 
		\abs{b_j}^2+\sum_{j=1}^{N-1}|c_j|^2\right)\right],
	\end{align}
	\end{enumerate}	
where the normalising partition functions can be read off from the normalizations of the matrix elements, resulting in \eqref{eqs-partition functions tridiag}. Notice one extra inverse power in the exponent $\frac{\beta j}{2}-2$, respectively $\beta j-2$, compared to the Chi-distribution in Table \ref{tab-dists tridiag}, recalling its definition in Eq. \eqref{eq-def chi dist}. This is because when going to polar coordinates, the differentials $d^2b_j(d^2c_j)$ will contribute an extra positive power in the radii.
	By Lemma \ref{lem-jac1} and Theorem \ref{thm-generalised Jacobian}, we obtain
	\begin{enumerate}
		\item[I.] for $\mathbf{J=\widetilde{T}}$
	\begin{align}\label{eq-mes T tilde}
		\prod_{j=1}^{N-1}\abs{b_j}^{\frac{\beta j}{2} -2}  \prod_{w=1}^Nd^2a_w\prod_{k=1}^{N-1}d^2b_k = \abs{\Delta_N(\mathbf{z})}^\beta \prod_{j=1}^{N}\abs{r_j}^{\frac{\beta }{2}-2}d^2r_{k-1}\prod_{w=1}^Nd^2z_w,
	\end{align}
\item[II.] for $\mathbf{J=S}$
\begin{align}\label{eq-mes S}
	\prod_{j=1}^{N-1}\abs{b_j}^{\beta j -2}  \prod_{w=1}^Nd^2a_w\prod_{k=1}^{N-1}d^2b_k = \abs{\Delta_N(\mathbf{z})}^\beta \prod_{j=1}^{N}\abs{r_j}^{\frac{\beta }{2}-2}d^2r_{k-1}\prod_{w=1}^Nd^2z_w,
\end{align}
\item[III.] and for $\mathbf{J=T}$
\begin{align}\label{eq-mes T}
	\begin{split}
	&\prod_{j=1}^{N-1}\abs{b_jc_j}^{\frac{\beta j}{2} -2}  \prod_{w=1}^Nd^2a_w\prod_{k=1}^{N-1}d^2b_kd^2c_k\\ =& \abs{\Delta_N(\mathbf{z})}^\beta \prod_{j=1}^{N}\abs{r_j}^{\frac{\beta }{2}-2}\prod_{k=2}^N|R_k|^{-2}d^2R_kd^2r_{k-1}\prod_{w=1}^Nd^2z_w.
	\end{split}
\end{align}
	\end{enumerate}
The exponents in Eqs. \eqref{eq-jpdf_mat_el_T tilde}, \eqref{eq-jpdf_mat_el_S}, \eqref{eq-jpdf_mat_el_T} respectively, can be expressed in terms of the squared Frobenius norm of $J$. Specifically,
	\begin{align}
		\label{eq-Frobenius_exp}
		\norm{J}_{\mathrm{F}}^2= \Tr(J^\dag J)=\sum_{j=1}^N \abs{a_j}^2 + \sum_{j=1}^{N-1}\abs{b_j}^2+\sum_{j=1}^{N-1}\abs{c_j}^2,
	\end{align}
with $c_j=1$ for $J=\widetilde{T}$ and $c_j=b_j$ for $J=S$.	Next, we use the Schur decomposition to simplify the exponent further. Let
	\begin{align}
		\label{eq-schur1}
		J = Q\left(Z + K\right)Q^\dagger,
	\end{align}
	where $Z \in Z(N)$, $Q$ is a unitary matrix and $K$ strictly upper triangular. Combining Eq. \eqref{eq-schur1} with Theorem \ref{thm-roughbijection}, we write
	\begin{align}
		\label{eq-exponent1}
		\norm{J}_{\mathrm{F}}^2 =
		\sum_{j=1}^N \abs{z_j}^2 + 2g_{J}(Z,\rv,\Rv_1),
	\end{align}
	where $g_{J}(Z,\rv,\Rv_1)$ is a positive function defined by Eq. \eqref{eq-def g J}. Combining Eqs. \eqref{eq-jpdf_mat_el_T tilde}, \eqref{eq-jpdf_mat_el_S}, respectively \eqref{eq-jpdf_mat_el_T}, with the corresponding equation out of Eqs. \eqref{eq-mes T tilde}, \eqref{eq-mes S}, \eqref{eq-mes T} together with Eq. \eqref{eq-exponent1} yields \eqref{eq-jpdf eig tridiag} for each $J=\widetilde{T},S,T$ after integration over the eigenvector components.
For the general matrix $T$, we note an apparent logarithmic singularity in the integration over the radii $|R_k|$ when going into polar coordinates, with the JPDF being proportional to $1/|R_k|$ for each $k=2,\ldots,N$. However, when expressing the variables in the trace \eqref{eq-Frobenius_exp} defining $g_{T}(Z,\rv,\Rv_1)$ in terms of variables $R_j$, $r_j$ and $z_j$, in particular the $|c_j|^2$, it becomes apparent from Eq. \eqref{eqs-Taylor_exps_1} that for all $j=1,\ldots,N-1$ we have $|c_{N-j}|^2\sim 1/|R_{j+1}|^2$. Therefore, we have an essential singularity from the exponent that makes the integral over the radii $|R_k|$ well-behaved at the origin.
The rotational symmetry
\begin{align*}
	g_{J}(Z e^{i\phi},\rv,\Rv_1)=g_{J}(Z,\rv,\Rv_1),\hspace{2mm}\text{for all}\hspace{2mm}\phi\in[0,2\pi),
\end{align*}
follows from the observation that 
\begin{align*}
	|a_{N-j+1}|=|\mathbf{r}_jZ\mathbf{l}_j|,\quad|c_{N-j}|=|\mathbf{r}_jZ\mathbf{l}_{j+1}|,\quad|b_{N-j+1}|=|\mathbf{r}_{j+1}Z \mathbf{l}_j|,
\end{align*}
for $j=1,\ldots,N$ from the proof of Theorem \ref{thm-roughbijection} with $c_j=1$ for $J=\widetilde{T}$ and $b_j=c_j$ for $J=S$. Furthermore, the mapping $Z\rightarrow Z e^{i\phi}$ leaves the JPDF of the eigenvalues invariant under rotation of the eigenvalues. To obtain the asymptotic behaviour of $g_J$, we study the dependence on $z_k$ in the recursion in Theorem \ref{thm-roughbijection} and find for $\mathbf{J=\widetilde{T}}$
	\begin{align}
		\begin{split}
			a_{N-j+1}=O\left(z_k^2\right),\hspace{2mm}   b_{N-j+1}=O\left(z_k^2\right),
		\end{split}
	\end{align}
and for $\mathbf{J=S,T}$
\begin{align}
	\begin{split}
		a_{N-j+1}=O\left(z_k\right),\hspace{2mm}  c_{N-j}=O\left(z_k\right),\hspace{2mm} b_{N-j+1}=O\left(z_k\right),
	\end{split}
\end{align}
	as $z_k\rightarrow\infty$ and $j=1,\ldots,N$. This completes the proof.
\end{proof}
Let us summarize what we achieved so far: We introduced three different non-Hermitian tridiagonal matrix ensembles, starting with the non-symmetric ensemble $\widetilde{T}$ studied by Mezzadri and Taylor in \cite{mezzadri taylor}, generalising to a symmetric version $S$ and a general tridiagonal matrix $T$. In all ensembles, the matrix elements are distributed according to complex normal, Chi- or uniform distributions with slightly different choices of the parameters, cf. Table \ref{tab-dists tridiag}. We construct a one-to-one-mapping to spectral parameters in Theorem \ref{thm-roughbijection}, calculate the Jacobian for the change of variables in Theorem \ref{thm-generalised Jacobian} and find the JPDF in terms of the complex eigenvalues, which contains a Coulomb-type interaction $|\Delta_N(\mathbf{z})|^\beta$ for each ensemble. While the matrix distribution was constructed in the way to find this repulsion term, we obtain also a coupling between eigenvalues and eigenvectors $f_J(Z)$ which prevent us to obtain further explicit results from this directly.

\subsubsection{Low Temperature Expansion $\boldsymbol{\beta\gg1}$}\label{sssec-low temperature}
It turns out to be rather challenging to calculate the limiting spectral density of these tridiagonal matrix ensembles via standard techniques such as energy minimization via the CG analogy (cf. \cite[Chapter $1.4$]{book forrester},\cite[Chapter $6$]{oxford handbook}) or the replica method for instance (cf. \cite{Kanzieper replica,Nishigaki Kamenev,feinberg zee}). However, for large $\beta\gg1$, the method of steepest descent, i.e. a saddle point approximation works out for $N=2$, see Appendix \ref{app- tridiag N=2} for the explicit calculations. This was a first reason to investigate the low temperature limit $\beta\gg 1$ for general $N$. Another motivation can be found again in the Hermitian ensembles, where Dumitriu and Edelman showed in \cite{DE02B} that as $\beta\rightarrow\infty$, the eigenvalues of the $\beta$-Hermite and $\beta$-Laguerre ensembles crystallise, fixing their positions as the roots of the Hermite and Laguerre polynomials, respectively. In this low temperature regime, the randomness of the eigenvalues is effectively removed.\\
However, in contrast to the 2dCG we have not observed numerically a condensation of complex eigenvalues at large values of $\beta$, in the three ensembles under consideration. 
Nevertheless, the question can be asked if a simplification also occurs for our non-Hermitian $\beta$-ensembles at large $\beta$.\\
Throughout this section if $B$ denotes an $N\times N$ complex matrix let its eigenvalues be given by $z_1(B),z_2(B),\ldots,z_N(B)$, labelled such that 
\begin{align}
	|z_N(B)|\leq|z_{N-1}(B)|\leq\dots\leq|z_1(B)|.
\end{align}
We study the asymptotics of the eigenvalues of the three non-Hermitian $\beta$-ensembles $\widetilde{T},S,T$, focusing on fixed matrix dimension $N$ and the limit $\beta\rightarrow\infty$, known as the low temperature limit.

To obtain asymptotic results for the eigenvalues of the non-Hermitian $\beta$-ensembles in this low temperature limit, we first introduce two lemmas. The first, a perturbation theory lemma \cite[Chapter 4.3]{Demmel}, is crucial for our analysis. A particular form of this lemma, applicable when $A$ and $B$ are symmetric, will also be used and is detailed in \cite[Lemma $2.1$]{DE02B}.
\begin{lemma}\label{lem-perturb}
	Let $A$ and $B$ be $N\times N$ matrices and let $\epsilon>0$. Suppose the eigenvalues of $A$ are distinct. Define $M=A+\epsilon B+o(\epsilon)$, where $o(\epsilon)$ defines a matrix in which every entry goes to zero faster than $\epsilon$ goes to zero. Let $z_i(A)$ denote the $i$th eigenvalue of $A$ for $1\leq i \leq N$ and let $\mathbf{x}_i$ and $\mathbf{y}_i^\dagger$ denote the right- and left-eigenvectors associated with $z_i(A)$, respectively. Then, we have
	\begin{align*}
	\lim_{\epsilon\rightarrow0}\frac{1}{\epsilon}(z_i(M)-z_i(A))=\mathbf{y}_i^\dagger B\mathbf{x}_i.
	\end{align*}
	Equivalently, for every $1\leq i\leq N$, the eigenvalue of $z_i(M)$ can be expanded as
	\begin{align*}
	z_i(M)=z_i(A)+\epsilon \mathbf{y}_i^\dagger B\mathbf{x}_i.
\end{align*}
\end{lemma}
Succeeding this we detail a second lemma \cite[Lemma $2.3$]{DE02B} in which one considers a Chi-distribution where the degree of freedom $r$ is growing.
\begin{lemma}\label{lem-chilimit}
	Let $r>0$ and $X$ be a random variable with distribution $\chi_r$. Then, as $r\rightarrow\infty$ the distribution of $X-\sqrt{r}$ converges to a normal distribution of mean $0$ and variance $1/2$.
\end{lemma}
	
Let $J_{\beta}$ be a random matrix from one of the three $N\times N$ matrix ensembles $J=\widetilde{T},S, T$ defined in \eqref{eq-tridiag} and Table \ref{tab-dists tridiag}, scaled by $\frac{1}{\sqrt{2N\beta}}$.
\begin{theorem}\label{thm-largebetagen}
	Let $z_i(J_{\beta})$ be the $i$th eigenvalue of $J_{\beta}$ for $1\leq i\leq N$. Then,
	\begin{align*}
	z_i(J_{\beta})\rightarrow\frac{1}{\sqrt{2N}}z_i(D_{N}^J)+o\left(\frac{1}{\sqrt{\beta}}\right),
	\end{align*}
	in distribution as $\beta\rightarrow\infty$. The matrix $D_{N}^J$ depends on the matrix ensemble under consideration, i.e. we find 
	\begin{itemize}
		\item for $\mathbf{J=\widetilde{T}}$
	\begin{align}\label{eq-largebetas T tilde}
		D_{N}^{\widetilde{T}}=\begin{pmatrix}
			0 & \frac{1}{\sqrt{\beta}} &  &  &  \\
			\sqrt{\frac{N-1}{2}}e^{i\theta_{N-1}} & 0 & \ddots &  &  \\
			& \ddots & \ddots & \ddots &  \\
			& & \sqrt{\frac{2}{2}}e^{i\theta_{2}} & 0 & \frac{1}{\sqrt{\beta}} \\
			&  &  & \sqrt{\frac{1}{2}}e^{i\theta_1} & 0 
		\end{pmatrix},
	\end{align}
	\item for $\mathbf{J=S}$
	\begin{align}\label{eq-largebetas S}
		D_{N}^S=\frac{1}{\sqrt{2}}\begin{pmatrix}
			0 & \sqrt{N-1}e^{i\theta_{N-1}} &  &  &  \\
			\sqrt{N-1}e^{i\theta_{N-1}} & 0 & \ddots &  &  \\
			& \ddots & \ddots & \ddots &  \\
			& & \sqrt{2}e^{i\theta_{2}} & 0 & \sqrt{1}e^{i\theta_1} \\
			&  &  & \sqrt{1}e^{i\theta_1} & 0 
		\end{pmatrix},
	\end{align}
	\item and for $\mathbf{J=T}$
	\begin{align}\label{eq-largebetas T}
		D_{N}^T=\frac{1}{\sqrt{2}}\begin{pmatrix}
			0 & \sqrt{N-1}e^{i\phi_{N-1}} &  &  &  \\
			\sqrt{N-1}e^{i\theta_{N-1}} & 0 & \ddots &  &  \\
			& \ddots & \ddots & \ddots &  \\
			& & \sqrt{2}e^{i\theta_{2}} & 0 & \sqrt{1}e^{i\phi_1} \\
			&  &  & \sqrt{1}e^{i\theta_1} & 0 
		\end{pmatrix}.
	\end{align}
	\end{itemize}
	Here $\theta_i$, $\phi_i$ are independent random variables distributed uniformly, $\theta_i,
	\phi_i\sim\mathcal{U}(0,2\pi)$ for $1\leq i \leq N-1$. All other elements vanish.\footnote{Similar as before, the angular variables share (partly) the same names within different ensembles but are other than this independent of each other.} Let
	\begin{align}
		\Phi_k^{D,J}(z):=\det(z\mathds{1}_k-D_{k}^J),
	\end{align}
	denote the characteristic polynomials of $D_{k}^J$ for $1\leq k\leq N$.
	Then, \begin{itemize}
		\item for $\mathbf{J=\widetilde{T}}$
	\begin{align*}
		\begin{split}
			&\sqrt{\beta}\left(z_i(\widetilde{T}_{\beta})-\frac{1}{\sqrt{2N}}z_i(D_{N}^{\widetilde{T}})\right)\\
			&=\frac{1}{\sqrt{2N}}\frac{\sum_{l=0}^{N-1}\Phi_{l}^{D,\widetilde{T}}(z_i)\Phi^{D,\widetilde{T}}_{l}(z_i^*)M_{l+1}+\sum_{l=1}^{N-1}\Phi^{D,\widetilde{T}}_{l-1}(z_i^*)\Phi_l^{D,\widetilde{T}}(z_i)U_{l}}{\sum_{l=0}^{N-1}\Phi^{D,\widetilde{T}}_l(z_i)\Phi^{D,\widetilde{T}}_l(z_i^*)}+o(1),
		\end{split}
	\end{align*}
	\item for $\mathbf{J=S}$
	\begin{align*}
		\begin{split}
			&\sqrt{\beta}\left(z_i(S_{\beta})-\frac{1}{\sqrt{2N}}z_i(D_{N}^S)\right)\\
			&=\frac{1}{\sqrt{2N}}\frac{\sum_{l=0}^{N-1}\Phi^{D,S}_{l}(z_i)^2M_{l+1}+2\sum_{l=1}^{N-1}\left[\Phi^{D,S}_{l-1}(z_i)\Phi^{D,S}_l(z_i)Y_{l}\right]}{\sum_{l=0}^{N-1}\Phi^{D,S}_l(z_i)^2}+o(1),
		\end{split}
	\end{align*}
	\item and for $\mathbf{J=T}$
	{\small
	\begin{align*}
		\begin{split} 
			&\sqrt{\beta}\left(z_i(T_{\beta})-\frac{1}{\sqrt{2N}}z_i(D_{N}^T)\right)\\
			=&\frac{1}{\sqrt{2N}}\frac{\sum_{l=0}^{N-1}\Phi^{D,T}_{l}(z_i)\Phi^{D,T}_{l}(z_i^*)M_{l+1}+\sum_{l=1}^{N-1}\left[\Phi^{D,T}_{l-1}(z_i^*)\Phi^{D,T}_l(z_i)U_{l}+\Phi^{D,T}_{l}(z_i^*)\Phi^{D,T}_{l-1}(z_i)V_{l}\right]}{\sum_{l=0}^{N-1}\Phi^{D,T}_l(z_i)\Phi^{D,T}_l(z_i^*)}\\
			&+o(1),
		\end{split}
	\end{align*}}
\end{itemize}
	in distribution as $\beta\rightarrow\infty$ 
	for the independent random variables $\re(M_j), \im(M_j)\sim\mathcal{N}(0,1)$, $|U_w|,|V_w|\sim\mathcal{N}(0,\frac{1}{2})$, $Y_w\sim\mathcal{N}(0,\frac{1}{4})$ and $\arg(U_w),\arg(V_w),\arg(Y_w)\sim\mathcal{U}(0,2\pi)$ for $1\leq w\leq N-1$ and $1\leq j\leq N$.
\end{theorem}

\begin{proof}
	Let $D_{N}^J$ 
	denote an $N\times N$ non-Hermitian tridiagonal matrix defined by Eqs. \eqref{eq-largebetas T tilde}, \eqref{eq-largebetas S}, \eqref{eq-largebetas T} for $J=\widetilde{T},S,T$ respectively. We restrict ourselves to the set of matrices $D_{N}^J$ with non-vanishing determinant and all eigenvalues distinct. As discussed in Section \ref{sssec-def tridiag jpdf}, the exceptional set of matrices with degenerate spectra forms a set of measure zero. Associated to the $i$th eigenvalue $z_i:=z_i(D_{N}^J)$, the matrix $D_{N}^J$ has right- and left-eigenvectors given by
	\begin{align}\label{eq-right and left eigenvectors D}
		\mathbf{u}_i=\begin{pmatrix}
			\Phi^{D,J}_{N-1}(z_i)\\
			\Phi^{D,J}_{N-2}(z_i)\\
			\vdots\\
			\Phi^{D,J}_0(z_i)
		\end{pmatrix}\quad\text{and}\quad
		\mathbf{v}_i^\dagger=\begin{pmatrix}
			\Phi^{D,J}_{N-1}(z_i^*),&\Phi^{D,J}_{N-2}(z_i^*),&\dotsm,&\Phi^{D,J}_{0}(z_i^*)
		\end{pmatrix}.
	\end{align}
For $J=S$, we have $\mathbf{u}_j^t=\mathbf{v_j}^\dagger$ due to the symmetry of $S$.
	Using Lemma \ref{lem-chilimit}, it follows that
	\begin{align}\label{eq-limbeta}
		\lim_{\beta\rightarrow\infty}\left[\sqrt{2N\beta}J_{\beta}-\sqrt{\beta}D_{N}^J\right]=Z^{D,J},
	\end{align}
	almost surely, where $Z^{D,J}$ is a random $N\times N$ complex tridiagonal matrix with independently distributed entries.
	\begin{enumerate}
		\item[I.] For $\mathbf{J=\widetilde{T}}$, we find
	\begin{align}\label{eq-Zabc T tilde}
		Z^{D,\widetilde{T}}=\begin{pmatrix}
			M_N & 1  &  &  \\
			U_{N-1} & \ddots & \ddots &  \\
			& \ddots & M_2 & 1 \\
			&  & U_1 & M_1 
		\end{pmatrix},
	\end{align}
\item[II.] for $\mathbf{J=S}$
\begin{align}\label{eq-Zabc S}
	Z^{D,J}=\begin{pmatrix}
		M_N & Y_{N-1}  &  &  \\
		Y_{N-1} & \ddots & \ddots &  \\
		& \ddots & M_2 & Y_1 \\
		&  & Y_1 & M_1 
	\end{pmatrix},
\end{align}
\item[III.] and for $\mathbf{J=T}$
	\begin{align}\label{eq-Zabc T}
		Z^{D,J}=\begin{pmatrix}
			M_N & V_{N-1}  &  &  \\
			U_{N-1} & \ddots & \ddots &  \\
			& \ddots & M_2 & V_1 \\
			&  & U_1 & M_1 
		\end{pmatrix}.
	\end{align}
	\end{enumerate}
	The matrix entries are distributed according to the following distributions: $\re(M_j)\sim\mathcal{N}(0,1)$ and $\im(M_j)\sim\mathcal{N}(0,1)$ for $1\leq j\leq N$, $|U_k|,|V_k|\sim\mathcal{N}(0,\frac{1}{2})$, $|Y_k|\sim\mathcal{N}(0,\frac{1}{4})$ and $\arg(U_k),\arg(V_k),\arg(Y_k)\sim\mathcal{U}(0,2\pi)$ for $1\leq k \leq N-1$.
	
	Moreover, we find
	\begin{align}\label{eq-TbetaPertExpanTabc}
		J_{\beta}=\frac{1}{\sqrt{2N}}D_{N}^J+\frac{1}{\sqrt{2N\beta}}Z^{D,J}+o\left(\frac{1}{\sqrt{\beta}}\right),
	\end{align}
	in distribution as $\beta\rightarrow\infty$.
	
	Using this, we express the eigenvalues of $J_{\beta}$ as
	\begin{align*}
	z_i(J_{\beta})=z_i\left(\frac{1}{\sqrt{2N}}D_{N}^J+\frac{1}{\sqrt{2N\beta}}Z^{D,J}+o\left(\frac{1}{\sqrt{\beta}}\right)\right),
	\end{align*}
	in distribution for any $1\leq i\leq N$. 
	
	Using Lemma \ref{lem-perturb}, we expand the $i$th eigenvalue of $J_{\beta}$ as
	\begin{align}\label{eq-eigpertgen}
		z_i(J_{\beta})=\frac{1}{\sqrt{2N}}\lambda_i(D_{N}^J)+\frac{1}{\sqrt{2N\beta}}\frac{\mathbf{v}_i^\dagger Z^{D,J}\mathbf{u}_i}{\mathbf{v}_i^\dagger\mathbf{u}_i}+o\left(\frac{1}{\sqrt{\beta}}\right),
	\end{align}
	where $1\leq i\leq N$, in distribution as $\beta\rightarrow\infty$. 
	
	Substituting the definition of $Z^{D,J}$ from Eqs. \eqref{eq-Zabc T tilde}, \eqref{eq-Zabc S}, \eqref{eq-Zabc T} respectively, into Eq. \eqref{eq-eigpertgen} concludes the proof. For the symmetric matrix $S$, we used the symmetry of the characteristic polynomial $\Phi_k^{D,S}(z)=\Phi_k^{D,S}(z^*)$ and the symmetry of the left- and right-eigenvectors.
\end{proof}

\begin{Remark}
We note that the difference in the parameters of the Chi-distribution within the three tridiagonal ensembles (cf. Table \ref{tab-dists tridiag}) results in a rescaling of the off-diagonal elements in $Z^{D,J}$ above.\\
In the case of the non-symmetric tridiagonal ensemble, we apply the low temperature limit to the continuous density function of the diagonal and sub-diagonal elements to illustrate a similar limiting behaviour as in the general and symmetric ensembles. However, we keep the order $\mathcal{O}(1/\sqrt{\beta})$ on the super-diagonal at the same time. The reason is the following: If we neglect this order at this stage, we find a matrix with all zero elements up to the sub-diagonal which yields directly only zero eigenvalues.
\end{Remark}

What we gained in this section is the following: For each rescaled ensemble $J_\beta=\widetilde{T}_\beta, S_\beta, T_\beta$ we obtain a matrix $D_{N}^J$ with the same eigenvalues in the low temperature limit $\beta\rightarrow \infty$. Up to leading order, we can now study the eigenvalue statistics of this centred Jacobi matrix, i.e. a tridiagonal matrix with vanishing diagonal entries. While it seems at a first glance that this does not reduce the complexity of the problem significantly, it turns out in the next section, that we are able to find an explicit form for the characteristic polynomial of such a matrix. 

\subsubsection{Characteristic Polynomials of Centred Jacobi Matrices}\label{sssec-characteristic polynomial}

In this section, we study the characteristic polynomials of the random matrices we derived in the low temperature limit $\beta\gg1$ in the previous section. As we find only non-zero entries on sub- and super-diagonal in each of the matrix ensembles under consideration, we keep the discussion general and treat the characteristic polynomial of the general centred matrix $J|_{a_j=0}$. Define the polynomial of degree $N$
\begin{align}\label{eq-Characteristic Polynomial}
	\Phi_N(z):=\det\left(z\mathds{1}_N-J|_{a_j=0}\right)=z^N+\sum_{\ell=1}^{\lfloor \frac{N}{2}\rfloor}\kappa_\ell^{(N)}z^{N-2\ell}.
\end{align}
We also define $\Phi_0(z)=1$. 
These polynomials have the following properties, where we define sums
$\sum_{l=j}^k$ to vanish for $k<j$.
\begin{prop}\label{prop-formula coefficients P}
	The polynomials defined in \eqref{eq-Characteristic Polynomial} are either even or odd, depending on the parity of their degree $N$. They obey the following three-term recurrence relation
	\begin{align}
		\label{eq-3 term recurrence}
		\Phi_{N+1}(z)= z \Phi_{N}(z)-\widetilde{b}_{N}\Phi_{N-1}(z) \quad \mbox{for}\ \ N=1,2,\ldots, 
	\end{align}
	with $\widetilde{b}_j = c_j b_j$. The coefficients $\kappa_\ell^{(N)}$ defined in \eqref{eq-Characteristic Polynomial} are given by
	\begin{equation}\label{eq-formula coefficients P}
		\kappa_\ell^{(N)}=(-1)^\ell \sum_{\gamma_1=2\ell-1}^{N-1}\widetilde{b}_{\gamma_1} \sum_{\gamma_2=2\ell-3}^{\gamma_1-2}\widetilde{b}_{\gamma_2} 
		\cdots \sum_{\gamma_\ell=1}^{\gamma_{\ell-1}-2}\widetilde{b}_{\gamma_\ell} ,
	\end{equation}
	for $\ell=1,2,\ldots, \lfloor \frac{N}{2}\rfloor$. Note that $\kappa_\ell^{(N)}$ consists of the product of $\ell$ sums.	All sums can be extended to begin with $\gamma_j=1$ for all $j=1,\ldots,\ell$ as the additional terms vanish.
\end{prop}
\begin{Remark}
	We observe that the coefficients of the characteristic polynomial $\Phi_N(z)$ depend only on the product of the sub- and super-diagonal entries. This matches the well-known fact that the eigenvalues of a tridiagonal matrix depend only on the diagonal elements and the product of the sub- and super-diagonal elements (cf. \cite{cullum willoghby}) and the study of Jacobi matrices \cite{GK05}.
\end{Remark}
The exact form of the coefficients is known \cite{PS06} when the polynomials \eqref{eq-Characteristic Polynomial} constitute orthogonal polynomials on $\mathbb{R}$. The same objects are also studied in \cite{LRR91} without proof. The latter authors found the following expression
\begin{align}\label{eq-coeff LRR91}
	\kappa_{\ell}^{(N)}=(-1)^\ell \sum _{\{i_N\}\in \{0,1,\ldots,N-1\}}^\prime \widetilde{b}_{i_1}\cdots \widetilde{b}_{i_\ell},
\end{align}
where the prime on the sum indicates that no two adjacent indices ever occur within ${i_N}$. Note that we adapted their result to our notation. The sum is defined in the following way: We start to choose $i_1$ in the full range $i_1=0,1,\ldots,N-1$ and choose the next summation index such that it runs only up to $i_1-2$ to have disjoint indices. Similar to the earlier remark, the first summand will vanish for $i_2>i_1$. From this, it is straightforward to see that Eq. \eqref{eq-formula coefficients P} and \eqref{eq-coeff LRR91} agree.
Let us emphasize, however, that in general the coefficients in \eqref{eq-3 term recurrence} are complex, $\widetilde{b}_j\in\mathbb{C}$. In contrast, for real orthogonal polynomials with respect to some density, the coefficients must be positive, as they represent the ratio of the squared norms of the monic orthogonal polynomials. This property is used in \cite{PS06}; thus, we present a detailed proof below without this assumption. A similar result on the moments of a specific tridiagonal matrix with vanishing diagonal and the off-diagonal elements being real and positive is given in \cite{AKvI00} Eq. (5) with $p=1$, which can be related to the coefficients of the characteristic polynomial. However, we would like to emphasize that these proofs in terms of orthogonal polynomials rely massively on the fact that, due to Favard's theorem (cf. \cite[Theorem 4.4]{Chihara}), the existence of a unique probability measure for polynomials satisfying a three-term recurrence is guaranteed if and only if the elements of the $\widetilde{b}_j$ are real and positive for all $j=1,\ldots ,N-1$. 
 
\begin{proof}
	The three-term recurrence \eqref{eq-3 term recurrence} follows from the form of $J$ with vanishing diagonal elements $a_j$ and expansion of the determinant twice, for instance first with respect to the first row and second with respect to the first column.\\
	Recall, that we set $\Phi_0(z):=1$. The first characteristic polynomials for $N=0,1,2,3$ can be calculated straightforwardly from determining the determinant and read
	\begin{align}\label{eq-characteristic polynomial small n examples}
		\begin{split}
			\Phi_0(z)&=1,\\
			\Phi_1(z)&=z,\\
			\Phi_2(z)&=z^2-\widetilde{b}_1, \\
			\Phi_3(z)&=z^3-(\widetilde{b}_2+\widetilde{b}_1)z.
		\end{split}
	\end{align}
	From these explicit forms of the characteristic polynomials for small $N$ and the three-term recurrence \eqref{eq-3 term recurrence} without diagonal term it follows directly that the polynomials $\Phi_N(z)$ are either even or odd, depending on the parity of their degree $N$. 
	
	Let us emphasize again that the polynomials $\Phi_N(z)$ are not orthogonal polynomials on $\mathbb{R}$ as the coefficients $\widetilde{b}_j$ are complex and do not correspond to some real positive norm of orthogonal polynomials.
	
	We start with the observation that the three-term recurrence \eqref{eq-3 term recurrence} yields a recurrence relation for the coefficients. Inserting the notation from Eq. \eqref{eq-Characteristic Polynomial} for $\Phi_{N+1}(z)$, $\Phi_{N}(z)$ and $\Phi_{N-1}(z)$ into \eqref{eq-3 term recurrence} gives
	{\small
	\begin{align}
	z^{N+1}+\!\sum_{\ell=1}^{\lfloor \frac{N+1}{2} \rfloor}\!\kappa_\ell^{(N+1)}z^{N+1-2\ell}=z^{N+1}+\!\sum_{\ell=1}^{\lfloor \frac{N}{2} \rfloor}\!\kappa_\ell^{(N)}z^{N+1-2\ell}-\widetilde{b}_{N}z^{N-1}-\widetilde{b}_{N}\!\sum_{\ell=1}^{\lfloor \frac{N-1}{2} \rfloor}\!\kappa_\ell^{(N-1)}z^{N-1-2\ell}. \label{eq-3 term recurrence with coeff} 
	\end{align}}
	Comparing the coefficients in orders $z^{N+1-2\ell}$ for $\ell=1,\ldots, \lfloor \frac{N+1}{2}\rfloor-1$ yields for the coefficients of $\Phi_{N+1}(z)$
	\begin{align}\label{eq-recursion kappas}
		\kappa_\ell^{(N+1)}=\begin{cases}
			\kappa_1^{(N)}-\widetilde{b}_{N} & \text{ for }\ell=1,\\
			\kappa_\ell^{(N)}-\widetilde{b}_{N}\kappa_{\ell-1}^{(N-1)} & \text{ for } \ell=2,\ldots, \lfloor \frac{N+1}{2}\rfloor-1,\\
			\kappa_m^{(N)}-\widetilde{b}_{N}\kappa_{m-1}^{(N-1)} & \text{ for }\ell= \lfloor \frac{N+1}{2}\rfloor=m,\ \ \ \ \ \ \text{ with}\ N+1=2m+1\ \text{ odd},\\
			\quad \quad -\widetilde{b}_{N}\kappa_{m}^{(N-1)} & \text{ for }\ell= \lfloor \frac{N+1}{2}\rfloor=m+1,\ \text{ with}\ N+1=2m+2\ \text{ even}.
		\end{cases}
	\end{align}
	We will treat the coefficient of the lowest order $z^0$ resulting from the last equation separately below. From \eqref{eq-characteristic polynomial small n examples} we have that Eq. \eqref{eq-formula coefficients P} is true for $N=0,1,2,3$. We prove via induction that this holds also for some $N+1$. Therefore, we insert Eq. \eqref{eq-formula coefficients P} in the RHS of Eq. \eqref{eq-recursion kappas} and find for $\ell=1$
	\begin{align*}
		\kappa_1^{(N+1)}&=-\sum_{\gamma=1}^{N-1}\widetilde{b}_\gamma-\widetilde{b}_{N} =-\sum_{\gamma=1}^{N}\widetilde{b}_\gamma 
	\end{align*}
	as claimed. We have for $\ell=2,\ldots, \lfloor \frac{N+1}{2}\rfloor-1$ for $N+1$ even, and for $\ell=2,\ldots, \lfloor \frac{N+1}{2}\rfloor$ for $N+1$ odd
	\begin{align*}
		\kappa_\ell^{(N+1)}&=
		(-1)^\ell \sum_{\gamma_1=2\ell-1}^{N-1}\widetilde{b}_{\gamma_1} \sum_{\gamma_2=2\ell-3}^{\gamma_1-2}\widetilde{b}_{\gamma_2} 
		\cdots \sum_{\gamma_\ell=1}^{\gamma_{\ell-1}-2}\widetilde{b}_{\gamma_\ell}\ -\ \widetilde{b}_{N}(-1)^{\ell-1} \sum_{\gamma_1=2\ell-3}^{N-2}\widetilde{b}_{\gamma_1} 
		\cdots \sum_{\gamma_{\ell-1}=1}^{\gamma_{\ell-2}-2}\widetilde{b}_{\gamma_{\ell-1}}  \\
		&=(-1)^\ell \left(\sum_{\gamma_1=2\ell-1}^{N-1}\widetilde{b}_{\gamma_1}+\widetilde{b}_n\right)\sum_{\gamma_2=2\ell-3}^{\gamma_1-2}\widetilde{b}_{\gamma_2} 
		\cdots \sum_{\gamma_\ell=1}^{\gamma_{\ell-1}-2}\widetilde{b}_{\gamma_\ell}\\
		&=(-1)^\ell \sum_{\gamma_1=2\ell-1}^{N}\widetilde{b}_{\gamma_1} \sum_{\gamma_2=2\ell-3}^{\gamma_1-2}\widetilde{b}_{\gamma_2} 
		\cdots \sum_{\gamma_\ell=1}^{\gamma_{\ell-1}-2}\widetilde{b}_{\gamma_\ell}, 
	\end{align*}
	which proves Eq. \eqref{eq-formula coefficients P} via induction for all 
	coefficients resulting from order in $z$ down to $z^2$ for $N+1=2m+2$ even, and down to $z^1$ for $N+1=2m+1$ odd.  
	For the coefficient of the lowest order $z^0$, we have to treat even $N+1=2m+2$ differently, as the recursion is different.
	Here, we obtain
	\begin{align*}
		\kappa_{m+1}^{(2m+2)}&=-\widetilde{b}_{2m+1} \kappa_{m}^{(2m)}.
	\end{align*} 
	Iterating this recurrence, using the initial condition $\kappa_1^{(2)}=-\widetilde{b}_1$, and induction assumption
	\begin{align}
		\kappa_{m}^{(2m)}=(-1)^m\prod_{\ell=1}^{m} \widetilde{b}_{2\ell-1},
	\end{align}
	we complete the proof of Eq. \eqref{eq-formula coefficients P}.
	We can further simplify the expression as follows: We relax each lower bound of the nested sums to $1$. One intermediate step reads as
	\begin{align*}
		\sum_{\gamma_p=2\ell-2p+1}^{\gamma_{p-1}-2}\!\widetilde{b}_{\gamma_p}\sum_{\gamma_{p+1}=2\ell-2p-1}^{\gamma_p-2}\!\widetilde{b}_{\gamma_{p+1}}=	\sum_{\gamma_p=1}^{\gamma_{p-1}-2}\widetilde{b}_{\gamma_p}\sum_{\gamma_{p+1}=2\ell-2p-1}^{\gamma_p-2}\!\widetilde{b}_{\gamma_{p+1}}-	\sum_{\gamma_p=1}^{2\ell-2p}\widetilde{b}_{\gamma_p}\sum_{\gamma_{p+1}=2\ell-2p-1}^{\gamma_p-2}\!\widetilde{b}_{\gamma_{p+1}}
	\end{align*}
	for $p=1,2,\ldots,\ell$ with $\gamma_{0}=N+1$. Following our convention for the sums stated before Proposition \ref{prop-formula coefficients P}, the second term vanishes as the upper bound of the inner sum is always smaller than the lower bound, i.e.
	\begin{align*}
		\gamma_p-2<2\ell -2p-1\  \text{ for }\  1\leq \gamma_p \leq 2\ell-2p.
	\end{align*}
\end{proof}
We emphasize that Proposition \ref{prop-formula coefficients P} is another main result which has its importance independent of the key it gives us to compute the limiting spectral density of the eigenvalues of the tridiagonal matrix ensembles under consideration. It provides an explicit expression for the characteristic polynomial of a (possibly) non-Hermitian centred Jacobi matrix. In the next section, we will use a result from free probability to calculate the limiting spectral density of these centred Jacobi matrices via the distribution of zeros of the characteristic polynomial. Therefore, we need to take a closer look into the coefficients of the latter.
\begin{lemma} \label{lem-mean and variance kappa's}
	Assume that the random variables $\{\widetilde{b}_i \}_{i=1}^{N-1}$ are independent of each other and have all zero mean. It follows that the coefficients $\kappa_\ell^{(N)}$ \eqref{eq-formula coefficients P} of the polynomial $\Phi_N(z)$ are pairwise-uncorrelated of each other, have zero mean and variance
	\begin{align}\label{eq-variance kappas}
		\Var(\kappa_\ell^{(N)})=\sum_{\gamma_1=1}^{N-1}\Var(\widetilde{b}_{\gamma_1})\sum_{\gamma_2=1}^{\gamma_1-2}\Var(\widetilde{b}_{\gamma_2})\cdots \sum_{\gamma_\ell=1}^{\gamma_{\ell-1}-2}\Var(\widetilde{b}_{\gamma_\ell}),
	\end{align}
	$\forall \ell=1,2,\ldots, \lfloor \frac{N}{2}\rfloor$.
\end{lemma}
\begin{proof}
	\begin{enumerate}
		\item The random variables $\{\widetilde{b}_i \}_{i=1}^{N-1}$ are independent by assumption. Let us recall the notions of independent and uncorrelated variables. For independent (complex) variables, the joint probability distribution factorizes, especially all moments (of possibly complex conjugate powers) factorize. On the other hand, for uncorrelated variables only the covariance vanishes. The latter is defined as
			\begin{align}
				\Cov(X,Z):=\mathbb{E}(X,Z^*),
			\end{align} 
			for complex random variables $X,Z$. We show that $\Cov(\kappa_{\ell}^{(N)},\kappa_{\ell'}^{(N)})=0$ with $\ell<\ell'$. From the derived explicit form of the $\kappa_{\ell}^{(N)}$'s in Eq. \eqref{eq-formula coefficients P}, we find that it consists of $\ell$ nested sums, with each of the intersection of index sets of the sums being empty. Hence, each coefficient contains sums of a product of $\ell$ different $\widetilde{b}_j$'s. As the latter are independent complex random variables with zero mean, the expectation of them can only yield a non-zero contribution if we find contributions $|\widetilde{b}_j|^2$. We observe that the expectation value $\mathbb{E}(\kappa_{\ell}^{(N)},\kappa_{\ell'}^{(N)})$ for $\ell<\ell'$ will vanish as we have always one factor of a $\widetilde{b}_j$ without a matching part $\widetilde{b}_j^*$.
		\item The random variables $\{\widetilde{b}_i \}_{i=1}^{N-1}$ have zero mean by assumption. 
		\item Next, we study the sum $Z=Z_1+Z_2$ of two independent complex random variables with vanishing mean. The mean of $Z$ is given by the sum of the means of $Z_1$ and $Z_2$ and, hence vanishes. For the variance we find 
		\begin{align*}
			\Var(Z)&=\Var(Z_1+Z_2)\\
			&=\mathbb{E}(|Z_1|^2)+\mathbb{E}(|Z_2|^2).
		\end{align*}
		\item Let $Z_1$ and $Z_2$ be two independent complex random variables with zero mean. The mean of the product $Z=Z_1 Z_2$ vanishes and the variance is given by
		\begin{align*}
			\Var(Z)=\Var(Z_1)\Var(Z_2).
		\end{align*}
	\end{enumerate}
	Hence, the mean of $\kappa_\ell^{(N)}$ vanishes for all $\ell$ and the variance is given by \eqref{eq-variance kappas}.
\end{proof}
We will see in the next section how we can make use of the explicit formula for the coefficients of the characteristic polynomials, using the results from Section \ref{sssec-low temperature} in the low temperature limit to derive the equilibrium density of our non-Hermitian tridiagonal matrix ensembles.

\subsubsection{Equilibrium Density via Free Probability for $\boldsymbol{\beta \gg 1}$ and $\mathbf{N\rightarrow \infty}$} \label{sssec-equilibrium density}
We determine the equilibrium density of all three tridiagonal matrix ensembles under consideration in the low temperature limit $\beta \gg 1$. We find that the limiting density of the eigenvalues agrees for the general and the symmetric tridiagonal matrix ensemble, leading to a logarithmic density, whereas the leading order contribution for the non-symmetric ensemble is given by a Dirac delta. For this, we study their characteristic polynomials and apply the following theorem from free probability, which determines the equilibrium density of the zeros of polynomials with random coefficients.
\begin{Remark}
	The application of free probability to the study of random polynomials is well-established, see \cite{book free prob} for an introduction to the subject and \cite{jalowy I, jalowy II, campbell} for related works on polynomials with real or non-positive zeros. For polynomials with both complex coefficients and complex zeros the results are much rarer. Free probability was used for non-Hermitian matrices for instance to investigate the support of the complex eigenvalues \cite{free prob non h support} or the spectral edge \cite{free prob non h edge}.
\end{Remark} We slightly rephrase a result from \cite[Theorem 1.2]{Barbarino Noferini} for the case of random variables instead of random matrices as coefficients, replacing the condition of independent random variables by pairwise-uncorrelated ones.
\begin{theorem}\label{thm-Barbarino Noferini}
	Suppose that for any $N$ we have a vector of pairwise-uncorrelated random variables $X_0^{(N)}$, $X_1^{(N)},\ldots,X_N^{(N)}$ with zero mean, unit variance and continuous distributions with densities uniformly bounded by a constant $M>0$ not depending on $N$. Let also $\alpha_0^{(N)},\alpha_1^{(N)},\ldots, \alpha_N^{(N)}$ be sequences of complex numbers.
	We consider the polynomial of degree $N$
	\begin{align}\label{eq-polynomial thm Barbarino Noferini}
		\Phi_N(z)=\sum_{j=0}^N \alpha_j^{(N)}X_j^{(N)}z^j.
	\end{align}
	Then, the empirical spectral distribution of zeros of the random polynomials $\Phi_N$ converges almost surely to a probability measure $\mu$ as $N\rightarrow \infty$, when the coefficients $\alpha_j^{(N)}$ satisfy the following conditions:
	\begin{itemize}
		\item $\alpha_N^{(N)}\neq 0$ for every $N$,
		\item for almost every $z$, let
		\begin{align}\label{eq-Qfunc}
			Q_N(z)=\sum_{i=0}^N\frac{|\alpha_i^{(N)}|^2}{|\alpha_N^{(N)}|^2}z^i,
		\end{align}
		such that the limit $U(z)=\lim_{N\rightarrow \infty} \frac{1}{2N} \ln \left(Q_N(|z|^2)\right)$ exists as a locally integrable function or a distribution,
		\item the probability measure $\mu$ satisfies
		\begin{align*}
			\mu=\frac{1}{2\pi}\Delta_z U(z) \mathbf{1}_\textrm{sup},
		\end{align*}
		in the sense of distributions, on the support (sup) of $\mu$. Here, $\Delta=\partial_z\partial_z^*$ denotes the two-dimensional Laplacian and $\mathbf{1}_\textrm{sup}$ is the indicator function, which gives 1 for $z\in $sup and vanishes elsewhere.
	\end{itemize}
\end{theorem}
\begin{proof}
		We relaxed the stronger condition of independence of the coefficients $X_k^{(N)}$ to pairwise-uncorrelatedness. In Theorem 1.2 in \cite{Barbarino Noferini}, the authors allow for random $k\times k$ matrices $C_j^{(N)}$ in the polynomials which contain independent copies of the random variables $X_j^{(N)}$. Hence, the condition of independence appears twice in their theorem, i.e. that the complex random variables $X_j^{(N)}$ are independent and the random matrices $C_j^{(N)}$ are build up by independent copies of the variables each. As we need only the case of $k=1$ (in the notation of \cite{Barbarino Noferini} $k$ and $N$ are interchanged) we briefly revisit the key points of their proof.\\
		We can assume without loss of generality that $|\alpha_N^{(N)}|=1$ for each $N$. If we fix $z$ and analyse $\Phi_N(z)$ from \eqref{eq-polynomial thm Barbarino Noferini}, we find that it can be treated as a random variable with mean 0 and variance $\sigma^2=\sum_{i=0}^N|\alpha_i^{(N)}|^2|z|^{2i}$. This follows from the fact that the $X_j^{(N)}$'s are pairwise-uncorrelated as
		\begin{align}
			\Var\left(\sum_{i=0}^N\alpha_i^{(N)}X_i^{(N)}z^i\right)&=\sum_{i=0}^N\Var\left(\alpha_i^{(N)}X_i^{(N)}z^i\right)+2\sum_{i<j}^N\Cov\left(\alpha_i^{(N)}X_i^{(N)}z^i,\alpha_j^{(N)}X_j^{(N)}z^j\right)\\
			&=\sum_{i=0}^N|\alpha_i^{(N)}|^2|z|^{2i},
		\end{align}
		where we used that the $X_j^{(N)}$'s have unit variance. Next, we need to prove that the random variable $\Phi_N(z)/\sigma$, which has zero mean and unit variance, has a bounded density function. We have that $\Phi_N(z)/\sigma$ is of the type
		\begin{align}
			Y_N=X_N^{(N)}\frac{\alpha_N^{(N)}z^N}{\sigma}+X_{N-1}^{(N)}\frac{\alpha_{N-1}^{(N)}z^{N-1}}{\sigma}+\ldots+X_1^{(N)}\frac{\alpha_1^{(N)}z}{\sigma}+X_0^{(N)}\frac{\alpha_0^{(N)}}{\sigma},
		\end{align}
		where all $X_i^{(N)}$ have a density function uniformly bounded by a constant $M>0$ (by assumption). Here comes now the crucial change in the proof for uncorrelated variables. In the work of Barbarino-Noferini in \cite{Barbarino Noferini}, where $Y_N$ is the sum of independent uniformly bounded random variables with zero mean and unit variance, they applied a result by Bobkov-Chistyakov \cite{Bobkov-Chistyakov} to prove that $Y_N$ is bounded by the very same constant $M$. Then, a matrix built up by independent copies of this $Y_N$ can be further studied. However, we can study this complex random variable directly and do not need this step here. In order to complete the proof, the following two bounds are needed
		\begin{align}
			\mathbb{P}\left(\sigma(\Phi_N(z)/\sigma)\leq u\right)&\leq c u^2,\\
			\mathbb{P}\left(\lVert \Phi_N(z)/\sigma\rVert\geq v\right)&\leq \frac{1}{v^2}\quad \forall u,v>0,
		\end{align}
		where $\sigma(A)$ denotes the largest singular value of the matrix $A$.
		Let us recall that we deal here with a much simpler case, i.e. $\Phi_N(z)/\sigma$ is a complex random variable for each fixed $z$ rather than a random matrix. The first inequality holds as we replace the squared singular value of $\Phi_N(z)/\sigma$ in this case of a $1\times 1$ matrix by the absolute square. The second relation follows from Chebyshev's inequality. The rest of the proof stays the same as in the case discussed in \cite{Barbarino Noferini}. 
\end{proof}
The remarkable point is that the coefficients of the polynomial $\Phi_N(z)$ in Theorem \ref{thm-Barbarino Noferini} are not necessarily identically distributed. The condition of unit variance can be full-filled by pulling the non-unit variance of the coefficients into the complex coefficients $\alpha_j^{(N)}$. We apply this theorem on random polynomials to the characteristic polynomials of the non-Hermitian tridiagonal matrix ensembles, to obtain the limiting distribution of the eigenvalues of the ensembles which equals the distribution of the zeros of their characteristic polynomials. In the low temperature limit $\beta \rightarrow \infty$ we determined an explicit expression of the characteristic polynomials in the previous Section \ref{sssec-characteristic polynomial} and derived in Lemma \ref{lem-mean and variance kappa's} the variance and mean of the coefficients. Using these results one can in principle calculate the equilibrium density of the zeros of the characteristic polynomial $\Phi_N(z)$, which correspond to the eigenvalues of the centred random Jacobi matrix $J|_{a_j=0}$ (full-filling the conditions of Lemma \ref{lem-mean and variance kappa's}) via Theorem \ref{thm-Barbarino Noferini} by choosing $|\alpha_j^{(N)}|^2=\Var(b_jc_j)$. However, as the above expression of the variances still contains the product of nested sums, it is difficult to proceed in this general setting. Therefore, we will now turn to the explicit low temperature limits we derived in Section \ref{sssec-low temperature}. The limiting centred Jacobi matrices are given in Theorem \ref{thm-largebetagen} Eqs. \eqref{eq-largebetas T tilde}, \eqref{eq-largebetas S}, \eqref{eq-largebetas T}, which we repeat below for convenience. We have for $\mathbf{J=\widetilde{T}}$
\begin{align}\label{eq-largebetas T tilde rescaled}
	D_{N}^{\widetilde{T}}=\begin{pmatrix}
		0 & \frac{1}{\sqrt{\beta}} &  &  &  \\
		\sqrt{N-1}e^{i\theta_{N-1}} & 0 & \ddots &  &  \\
		& \ddots & \ddots & \ddots &  \\
		& & \sqrt{2}e^{i\theta_{2}} & 0 & \frac{1}{\sqrt{\beta}} \\
		&  &  & e^{i\theta_1} & 0 
	\end{pmatrix},
\end{align}
for $\mathbf{J=S}$
\begin{align}\label{eq-largebetas S rescaled}
	D_{N}^S=\begin{pmatrix}
		0 & \sqrt{N-1}e^{i\theta_{N-1}} &  &  &  \\
		\sqrt{N-1}e^{i\theta_{N-1}} & 0 & \ddots &  &  \\
		& \ddots & \ddots & \ddots &  \\
		& & \sqrt{2}e^{i\theta_{2}} & 0 & \sqrt{1}e^{i\theta_1} \\
		&  &  & \sqrt{1}e^{i\theta_1} & 0 
	\end{pmatrix},
\end{align}
and for $\mathbf{J=T}$
\begin{align}\label{eq-largebetas T rescaled}
	D_{N}^T=\begin{pmatrix}
		0 & \sqrt{N-1}e^{i\phi_{N-1}} &  &  &  \\
		\sqrt{N-1}e^{i\theta_{N-1}} & 0 & \ddots &  &  \\
		& \ddots & \ddots & \ddots &  \\
		& & \sqrt{2}e^{i\theta_{2}} & 0 & \sqrt{1}e^{i\phi_1} \\
		&  &  & \sqrt{1}e^{i\theta_1} & 0 
	\end{pmatrix},
\end{align}
where $\theta_\ell,\phi_\ell\sim \mathcal{U}(0,2\pi)$ are independent and identically distributed random variables. To apply Theorem \ref{thm-Barbarino Noferini} we incorporated a factor of $\sqrt{2}$ in front of $D_{N}^S$, $D_{N}^T$ and a factor of $\sqrt{2}$ in the sub-diagonal elements of $D_{N}^{\widetilde{T}}$. In particular, we need a factor of $\sqrt{2}$ in front of each random variable of the form $e^{i\theta}$, with $\theta\sim \mathcal{U}(0,2\pi)$. This is because after applying the low temperature limit, where we effectively performed the one-dimensional limit of the Chi-distribution while keeping the angular component fixed, we are left with a one-dimensional random variable in each sub- and super-diagonal entry. However, due to their explicit form, these lead to complex-valued entries. Thus, we effectively have a one-dimensional random variable, in complex representation as a phase. According to the methodology of the authors of \cite{Barbarino Noferini}, this should be interpreted as a one-dimensional random variable multiplied by a complex coefficient. However, we cannot do this because the real- and imaginary-parts of the variable $e^{i\theta}$, with $\theta\sim \mathcal{U}(0,2\pi)$, are not independent. Therefore, we interpret the matrix entries as complex random variables, whose real- and imaginary-parts contribute equally, i.e. each contributes to half of the total variance. By multiplying with the factor $\sqrt{2}$, we take this into account.

Let $\Phi_N^{D,J}(z)$ denote the characteristic polynomial of $D_{N}^J$ for $J=\widetilde{T},S,T$.
\begin{lemma}\label{lem-var coeff P D}
	The coefficients $\kappa_\ell^{(N),D,J}$ of the polynomial $\Phi_N^{D,J}(z)$ are pairwise uncorrelated\footnote{We correct a typo in \cite{AMPT} by replacing the independence of the coefficients $\kappa_\ell^{(N),D,J}$ by pairwise uncorrelatedness.} and have zero mean.
	\begin{itemize}
		\item For $\mathbf{J=\widetilde{T}}$ the variance is given by
		\begin{align}\label{eq-var coeff P D T tilde}
			\Var(\kappa_\ell^{(N),D,\widetilde{T}})= \frac{1}{\beta^\ell}\sum_{\gamma_1=1}^{N-1}\gamma_1\sum_{\gamma_2=1}^{\gamma_1-2}\gamma_2\cdots \sum_{\gamma_\ell=1}^{\gamma_{\ell-1}-2}\gamma_\ell,
		\end{align}
		for all $\ell=1,2,\ldots,\lfloor \frac{N}{2}\rfloor$.
		\item For $\mathbf{J=S,T}$ the variance agrees and reads 
		\begin{align}\label{eq-var coeff P D S T}
			\Var(\kappa_\ell^{(N),D,J})= \sum_{\gamma_1=1}^{N-1}\gamma_1^2\sum_{\gamma_2=1}^{\gamma_1-2}\gamma_2^2\cdots \sum_{\gamma_\ell=1}^{\gamma_{\ell-1}-2}\gamma_\ell^2,
		\end{align}
		for all $\ell=1,2,\ldots,\lfloor \frac{N}{2}\rfloor$.
	\end{itemize} 
	
\end{lemma}
\begin{proof}
	First, we check that the coefficients $\kappa_\ell^{(N),D,J}$, cf. Eq. \eqref{eq-formula coefficients P}, of the polynomial $\Phi_N^{D,J}(z)$ have zero mean for all $J=\widetilde{T},S,T$. We have from Eqs. \eqref{eq-largebetas T tilde rescaled}, \eqref{eq-largebetas S rescaled}, \eqref{eq-largebetas T rescaled} and Eq. \eqref{eq-Characteristic Polynomial} $\widetilde{b}_j=\sqrt{\frac{j-1}{\beta}}e^{i\theta_{j-1}}$ for $J=\widetilde{T}$, $\widetilde{b}_j=(j-1)e^{2i\theta
	_{j-1}}$ for $J=S$ and $\widetilde{b}_j=(j-1)e^{i(\theta_{j-1}+\phi_{j-1})}$ for $J=T$, where $\theta_j,\phi_j\sim \mathcal{U}(0,2\pi)$ for all $j=1,\ldots,N-1$ independent. In all three ensembles, the random variable is represented by uniformly distributed angles. Therefore, we study a complex random variable of the form $Z:=ae^{im\theta}$, with $a\in\mathbb{R}$, $\theta\sim \mathcal{U}(0,2\pi)$ and $m\in\mathbb{N}$. We assume that $m=1$ without loss of generality as the exponential function is periodic. The mean of $Z$ vanishes since
	\begin{equation}
		\mathbb{E}[Z]=\frac{a}{2\pi}\int_{0}^{2\pi}\! d\theta \ e^{i\theta}=0.
	\end{equation}
	Next, note the variance of a complex random variable $W$ is defined by:
	\begin{align}
		\Var(W)=\mathbb{E}[|W|^2]-|\mathbb{E}[W]|^2.
	\end{align}
	Applying this definition to the random variable $Z$, we find:
	\begin{align}
		\Var(Z)&=\mathbb{E}[|ae^{i\theta}|^2]-|\mathbb{E}[ae^{i\theta}]|^2=a^2.
	\end{align}
	Hence, we find $\Var(\widetilde{b}_j)=\frac{j}{\beta}$ for $J=\widetilde{T}$ and $\Var(\widetilde{b}_j)=j^2$ for $J=S,T$ for all $j=1,\ldots,N-1$. Utilising Lemma \ref{lem-mean and variance kappa's} allows us to conclude the results stated in Eqs. \eqref{eq-var coeff P D T tilde}, \eqref{eq-var coeff P D S T} as desired. 
\end{proof}
Remarkably, we find that the variance of the coefficients of $\Phi_N^{D,J}(z)$ agrees within the symmetric and the general tridiagonal matrix ensembles. Using Theorem \ref{thm-Barbarino Noferini}, we conclude that the limiting spectral density of the complex eigenvalues also agrees. Therefore, we distinguish in the following between the non-symmetric and the symmetric/ general ensemble only.
As established in the previous sections, the random coefficients of the characteristic polynomial $\Phi_N^{D,J}(z)$ satisfy the necessary conditions for applying Theorem \ref{thm-Barbarino Noferini}. To simplify the polynomial $Q_N^{D,J}(x)$ \eqref{eq-Qfunc}, constructed from the coefficients $|\alpha_\ell^{(N)}|^2=\Var(\kappa_\ell^{(N),D,J})$, we analyse these polynomials using generating functions. Consequently, we state the following lemma:
\begin{lemma}\label{lem-Q D polynomials with var kappa}
	The polynomials
	\begin{align}
		Q_N^{D,J}(x)=\sum_{\ell=0}^{\lfloor \frac{N}{2}\rfloor}\Var(\kappa_\ell^{(N),D,J})x^{N-2\ell},
	\end{align}
	with $\Var(\kappa_\ell^{(N),D,J})$ given by Eqs. \eqref{eq-var coeff P D T tilde}, \eqref{eq-var coeff P D S T} respectively, for $\ell=1,\ldots,\left\lfloor \frac{N}{2}\right\rfloor$ and $\Var(\kappa_0^{(N),D,J})=1$, satisfy
	\begin{itemize}
		\item for $\mathbf{J=\widetilde{T}}$ the following three-term recurrence relation
		\begin{align}\label{eq-3 term recurrence Q D T tilde}
			Q_{N+1}^{D,\widetilde{T}}(x)=xQ_N^{D,\widetilde{T}
			}(x)+\frac{N}{\beta}Q_{N-1}^{D,\widetilde{T}}(x), \quad N\geq1.
		\end{align}
	\item For $\mathbf{J=S,T}$, we find 
	\begin{align}\label{eq-3 term recurrence Q D S T}
		Q_{N+1}^{D,J}(x)=xQ_N^{D,J
		}(x)+N^2Q_{N-1}^{D,J}(x), \quad N\geq1.
	\end{align}
	\end{itemize} 

\end{lemma}
\begin{proof}
	From the variance at $\ell=0$ we have $Q_0^{D,J}(x)=1$ and $Q_1^{D,J}(x)=x$. The rest 
	follows from Proposition \ref{prop-formula coefficients P} with $\Var(-b_\ell)=\ell/\beta$ for $J=\widetilde{T}$, respectively $\Var(-b_\ell c_\ell)=\ell^2$ for $J=S,T$, for $\ell=1,\ldots,N-1$, where we have $b_\ell=c_\ell$ for $J=S$.
\end{proof}
\begin{prop}\label{prop-explicit form Q H}
	The polynomials $Q_N^{D,J}(x)$ can
	\begin{itemize}
		\item for $\mathbf{J=\widetilde{T}}$ be rewritten as
		\begin{align}\label{eq-polynomial Q H T tilde}
			Q_N^{D,\widetilde{T}}(x)=\sum_{j=0}^{\lfloor\frac{N}{2}\rfloor}\frac{N!}{j!(N-2j)!(2\beta)^j}\ 
			x^{N-2j}.
		\end{align}
	\item For $\mathbf{J=S,T}$ we find
	\begin{align}\label{eq-polynomial Q H S T}
		Q_N^{D,J}(x)=\sum_{k=0}^N\frac{(-1)^{N-k}(N!)^22^k}{(k!)^2(N-k)!}\left(\frac{1}{2}+\frac{x}{2}\right)_k,
	\end{align}
	where $(a)_b:=\Gamma(a+b)/\Gamma(a)$ denotes the Pochhammer symbol.
	\end{itemize}
	
\end{prop}
\begin{proof}
\begin{enumerate}
	\item[I.] We start with the proof for the non-symmetric ensemble $\mathbf{J=\widetilde{T}}$ and begin by simplifying the expression for the nested summation of
	$\Var(\kappa_k^{(N),D,\widetilde{T}})$ defined in Eq. \eqref{eq-var coeff P D T tilde}.
	
	Let $B_{N}$ denote an $N\times N$ Jacobi matrix defined as follows:
	\begin{align*}
	B_N:=\frac{1}{\sqrt{2}}\begin{pmatrix}
		0 & \sqrt{N-1} &  &  \\
		\sqrt{N-1} & 0 & \ddots &  \\
		& \ddots & \ddots & \sqrt{1} \\
		&  & \sqrt{1} & 0 
	\end{pmatrix}.
	\end{align*}
	Utilising the notation of Eq. \eqref{eq-Characteristic Polynomial}, the characteristic polynomial of $B_N$ is given by: 
	\begin{align}\label{eq-CharacteristicB}
		\Phi_N^B(x)=\sum_{m=0}^{\lfloor\frac{N}{2}\rfloor}\kappa_m^{(N),B}x^{N-2m},
	\end{align}
	where
	\begin{align}\label{eq-closedformB}
		\kappa_m^{(N),B}=\frac{(-1)^m}{2^m}\sum_{\gamma_1=1}^{N-1}\gamma_1\dots\sum_{\gamma_m=1}^{\gamma_{m-1}-2}\gamma_m.
	\end{align}
	Next, we note that the probabilist's Hermite polynomials $\mbox{He}_N(x)$, defined in \cite[Eq. 18.7.11]{NIST}, satisfy the three-term recurrence relation:
	\begin{align}
		\mbox{He}_{N+1}(x)=x\mbox{He}_N(x)-N\mbox{He}_{N-1}(x).
	\end{align}
	Identifying the recurrence coefficients $\tilde{b}_N$
	in Eq. \eqref{eq-3 term recurrence} 
	for Hermite polynomials and using Proposition \ref{prop-formula coefficients P}, alongside the explicit form of the Hermite polynomials \cite[Eq. 18.5.13]{NIST}, we obtain:
	\begin{align}\label{eq-ChebyshevPoly}
		\mbox{He}_N(x)=\sum_{m=0}^{\lfloor\frac{N}{2}\rfloor}\kappa_{m}^{(N),\textrm{He}}x^{N-2m}=N!\sum_{m=0}^{\lfloor\frac{N}{2}\rfloor}\frac{(-1)^m}{m!(N-2m)!2^m}\ x^{N-2m},
	\end{align}
	for
	\begin{align}\label{eq-ChebyshevCoeff}
		\kappa_{l}^{(N),\textrm{He}}=(-1)^l\sum_{\gamma_1=1}^{N-1}\gamma_1\dotsc\sum_{\gamma_l=1}^{\gamma_{l-1}-2}\gamma_{l}.
	\end{align}
	Now, we can derive an expression for $\kappa_m^{(N),D,\widetilde{T}}$ by comparing the closed forms of the characteristic polynomial in \eqref{eq-CharacteristicB} and the Hermite polynomial. Specifically, by equating the Eqs. \eqref{eq-CharacteristicB} and \eqref{eq-closedformB} with \eqref{eq-ChebyshevPoly} and \eqref{eq-ChebyshevCoeff}, we find:
	\begin{align*}
	\frac{1}{2^{2m}}\frac{N!}{m!(N-2m)!}=\frac{1}{2^m}\sum_{\gamma_1=1}^{N-1}\gamma_1\sum_{\gamma_2=1}^{\gamma_1-2}\gamma_2\dots\sum_{\gamma_m=1}^{\gamma_{m-1}-2}\gamma_m.
	\end{align*}
	Thus, we can express $\Var(\kappa_l^{(N),D,\widetilde{T}})$ as 
	\begin{align}
		\Var(\kappa_l^{(N),D,\widetilde{T}})=\frac{1}{\beta^l}\sum_{\gamma_1=1}^{N-1}\gamma_1\sum_{\gamma_2=1}^{\gamma_1-2}\gamma_2\dots\sum_{\gamma_l=1}^{\gamma_{l-1}-2}\gamma_l=\frac{1}{2^{l}\beta^l}\frac{N!}{l!(N-2l)!},
	\end{align}
	as desired.
	\item[II.] For the symmetric and the general matrices $\mathbf{J=S,T}$, we use the method of generating functions. The latter is defined by
	\begin{align}\label{eq-generating function Q D}
		q(z,x)=\sum_{N=0}^\infty \frac{Q_N^{D,J}(x)}{(N!)^2}z^N.
	\end{align}
	We study a differential equation for $q(z,x)$, with differential operator $D_z:=\partial_z z \partial_z=z\partial_z^2+\partial_z$,
	\begin{align*}
		D_zq(z,x)&=x\sum_{N=0}^\infty \frac{Q_N^{D,J}(x)}{N!N!}z^N+\sum_{N=0}^\infty \frac{Q_{N-1}^{D,J}(x)}{(N-1)!(N-1)!}z^N\\
		&=xq(z,x)+z q(z,x)
	\end{align*}
	The definition
	$q(z,x)=e^{-z}r(z,x)$ leads to
	\begin{align}\label{eq-diff eq gen fct Q H}
		z\partial_z^2 r(z,x)+(1-2z)\partial_z r(z,x)-(x+1)r(z,x)=0,
	\end{align}
	which is Kummer's differential equation in variable $2z$, cf. \cite[Eq. 13.2.1]{NIST}. We can thus immediately write down the two independent solutions
	\begin{align}\label{eq-gen sol q(z,x)}
		q(z,x)=c_1e^{-z}U\left(\frac{1}{2}+\frac{x}{2},1;2z\right)+c_2e^{-z}
		M\left(\frac{1}{2}+\frac{x}{2},1;2z\right)	,
	\end{align}
	where $U(a,b;y)$ denotes the confluent or Kummer hypergeometric function and $M(a,b;y)$ Tricomi's hypergeometric function.
	The initial conditions $q(0,x)=1$ and $\partial_zq(z,x)|_{z=0}=x$ follow from the definition \eqref{eq-generating function Q D} and the two lowest order polynomials which determine the constants $c_1,c_2$ in the solution of the differential equation. 
	Because $U(a,1,z)$ has a logarithmic singularity for $z\rightarrow 0$ \cite[Eq. 13.2.19]{NIST}, we have to set $c_1=0$ and $c_2=1$, as $M(a,b;y)={}_1F_1(a;b;y)$ is unity at the origin. The second boundary condition can be easily checked from the derivative of 
	$M(a,b;y)$ at the origin using \cite{NIST}.
	
	Finally, the explicit form \eqref{prop-explicit form Q H} follows from writing out the hypergeometric function ${}_1F_1(a,b;z)$ as a series and using the Cauchy product
	\begin{align*}
		e^{-z}{}_1F_1\left(\frac{1}{2}+\frac{x}{2},1;2z\right)&=\sum_{k=0}^\infty \frac{(-1)^kz^k}{k!}\sum_{N=0}^\infty \frac{(a)_N2^N}{N!N!}z^N\\
		&=\sum_{N=0}^\infty \sum_{k=0}^N (-1)^{N-k}\frac{(a)_k2^{k}}{k!k!(N-k)!}z^N,
	\end{align*}
	with $a=\frac{1}{2}+\frac{x}{2}$. 
\end{enumerate}
\end{proof}
We observe that for $J=S,T$ the polynomials $Q_N^{D,J}(x)$ in Proposition \ref{prop-explicit form Q H} are the \textit{Meixner-Pollaczek polynomials}, denoted by $P_N^{(\lambda)}(y;\phi)$ in its hypergeometric representation \cite[Chapter 9.7]{Hypergeometric Orthogonal Polynomials}
\begin{align}
	P_N^{(\lambda)}(y;\phi)=\frac{(\lambda)_N}{N!}e^{-iN\phi}{}_2F_1\left( -N, \lambda+iy;2\lambda;1-e^{-2i\phi}\right).
\end{align}
We obtain the polynomials $Q_N^{D,J}(y)$ with the choice $\lambda=\frac{1}{2}$, $\phi=-\frac{\pi}{2}$ and $y=-\frac{i}{2}$. The Meixner-Pollaczek polynomials appeared already in another context of random matrices, namely the study of moments and hypergeometric orthogonal polynomials in \cite{Cunden et al}.\\
Combining all results, we are now in the position to compute the equilibrium eigenvalue density of eigenvalues of all three complex tridiagonal matrices $\widetilde{T}_\beta, S_\beta,T_\beta$ in the low temperature limit.

\begin{theorem}\label{thm-density beta}
	The spectral density $\rho^{J}(z)$ in the limit $N\rightarrow \infty$ and low temperature regime $\beta \gg 1$ is given 
	\begin{itemize}
		\item for the non-symmetric ensemble $\mathbf{J=\widetilde{T}}$ to leading order by 
		\begin{equation}\label{eq-limiting density T tilde}
			\rho^{\widetilde{T}}(z)=\delta^{(2)}(z),
		\end{equation}
		which only has a singular component at the origin. 
		\item For the symmetric and the general ensembles $\mathbf{J=T,S}$ we find
		\begin{equation}\label{eq-limiting density T,S}
			\rho^{J}(z)=\frac{e^1}{2\pi}(\ln(2)-1-\ln(|z|^2)),
		\end{equation} 
		for $|z|\in [0,r_0]$ and zero otherwise, with $r_0=\sqrt{2/e}$. 
	\end{itemize}
\end{theorem}
\begin{proof}
We start again, with some general results, which are valid for all three tridiagonal ensembles $J=\widetilde{T},S,T$ and distinguish the two cases $J=\widetilde{T}$, respectively $J=S,T$ afterwards.	
We derived Lemma \ref{lem-var coeff P D} for the variances of the coefficients of the characteristic polynomial of matrix $D_{N}^J$. To apply Theorem \ref{thm-Barbarino Noferini}, we observe that we found an expression for the characteristic polynomial of $D_{N}^J$ in each ensemble $J$, such that the coefficients of this polynomial in $z$ are pairwise uncorrelated,\footnote{, where we again correct a typo in \cite{AMPT}} have zero mean and the variance is given by \eqref{eq-var coeff P D T tilde}, respectively \eqref{eq-var coeff P D S T}. Moreover, we proved in Lemma \ref{lem-Q D polynomials with var kappa} that the expressions for the variances of the coefficients $\kappa_k^{(N),D,J}$ are the coefficients of the polynomial $Q_N^{D,J}(x)$. We define the parameters $\alpha_k^{(N),J}$ in Theorem \ref{thm-Barbarino Noferini} by the square root of the variances $\sqrt{\Var(\kappa_k^{(N),D,J})}$ and find that we have to study the polynomial, whose coefficients are the variances of the coefficients of the characteristic polynomial of $D_{N}^J$ as a polynomial in $|z|^2$, i.e. we study $Q_N^{D,J}(|z|^2)$. In order to find the corresponding limit for the matrix $J_\beta$, we have to take the scaling due to Theorem \ref{thm-largebetagen} into account. We conclude that the two characteristic polynomials $\Phi^J_N(z)$ of $J_\beta$ and $\Phi^{D,J}_N(z)$ of $D_{N}^J$ from Eqs. \eqref{eq-largebetas T tilde rescaled}, \eqref{eq-largebetas S rescaled}, \eqref{eq-largebetas T rescaled} respectively, are related due to Theorem \ref{thm-largebetagen} as
\begin{align}
	\Phi^J_N(z)&=\prod_{i=1}^N(z-z_i(J_\beta))\nonumber\\
	&=\prod_{i=1}^N\left(z-\frac{1}{\sqrt{2N}}z_i(D_{N}^J)+o\left(\frac{1}{\sqrt{\beta}}\right)\right)\label{eq-prop char poly D}\\
	&=\frac{1}{(2N)^{\frac{N}{2}}}\Phi^{D,J}_N(\sqrt{2N}z) +o\left(\frac{1}{\sqrt{\beta}}\right).
	\label{eq-charac poly with beta correction}
\end{align}
We will only take the leading order of the low temperature limit $\beta \gg 1$ into account and comment on possible corrections later. Recall that for calculating the limiting density we have to take the variance of the coefficients of the characteristic polynomials into account which yields the following result due to Proposition \ref{prop-explicit form Q H} and Theorem \ref{thm-Barbarino Noferini}:
\begin{align}\label{eq-U(z) in proof}
	U(z)&=\lim_{N\rightarrow \infty}\frac{1}{2N}\ln \left(\frac{1}{(2N)^{N}}Q_N^{D,J}\left((2N)|z|^2+o\left( \frac{1}{\beta}\right)\right)\right).
\end{align}
Now, we have to distinguish the two possible forms of $Q_N^{D,J}(z)$ from Proposition \ref{prop-explicit form Q H}.
\begin{enumerate}
\item[I.] We start with $\mathbf{J=\widetilde{T}}$ and study the argument of the logarithm using Eq. \eqref{eq-polynomial Q H T tilde}.
\begin{align}
	\frac{1}{(2N)^{N}}Q_N^{D,\widetilde{T}}\left(2N|z|^2\right)
	&=\sum_{j=0}^{\lfloor\frac{N}{2}\rfloor}\frac{N!}{j!(N-2j)!}\frac{|z|^{2(N-2j)}}{2^{3j}\beta^jN^{2j}}\nonumber\\
	&=
	\frac{1}{\left(2iN\sqrt{\beta}\right)^N}\mbox{He}_{N}\left(2iN\sqrt{\beta}|z|^2\right).
	\label{eq-RelationtoHermite}
\end{align}
However, the difficulty is that the scaling parameter $N$ appears in both the argument and the coefficient of the Hermite polynomial in \eqref{eq-RelationtoHermite}. Therefore, before taking the limit of the logarithm
we apply the multiplication theorem for Hermite polynomials \cite[Eq. 18.18.13]{NIST}, applied to \eqref{eq-RelationtoHermite} to produce:
\begin{align}
	\frac{1}{\left(2iN\sqrt{\beta}\right)^N}\mbox{He}_{N}\left(2iN\sqrt{\beta}|z|^2\right)&=\sum_{j=0}^{\lfloor\frac{N}{2}\rfloor}\left(1+\frac{1}{4\beta N^2}\right)^j\frac{N!}{j!(N-2j)!2^j}\mbox{He}_{N-2j}(|z|^2)\nonumber\\
	&\approx \sum_{j=0}^{\lfloor\frac{N}{2}\rfloor}\frac{N!}{j!(N-2j)!2^j}\mbox{He}_{N-2j}(|z|^2).
\end{align}
In the second step, we are neglecting higher order terms in $1/\beta$. For the latter sum, there is the following identity \cite[Eq. 18.18.20]{NIST}:
\begin{align}
	|z|^{2N}=\sum_{j=0}^{\lfloor\frac{N}{2}\rfloor}\frac{N!}{j!(N-2j)!2^j}\mbox{He}_{N-2j}(|z|^2),
\end{align}
and thus we arrive at the following approximation for the limit
\begin{equation}\label{eq-U(z)Hermiteform}
	U(z)\approx\lim_{N\rightarrow\infty}\frac{1}{2N}\log( |z|^{2N})=\log|z|.
\end{equation}
It is well known that the logarithm is the Green's function for the Laplacian in two dimensions, and thus we obtain from Theorem \ref{thm-Barbarino Noferini},
\begin{align}
	\widetilde{\rho}^{\widetilde{T}}(z)=\frac{1}{2\pi}\Delta_z U(z)=\frac 14\delta^{(2)}(z).
\end{align}
After appropriately normalising this is Eq. \eqref{eq-limiting density T tilde}, in coordinates $z=x+iy$. 
\item[II.]
Now, we turn to the case $\mathbf{J=S,T}$. Application of Eq. \eqref{eq-polynomial Q H S T} in \eqref{eq-U(z) in proof} yields
\begin{align}
	U(z)&=\lim_{N\rightarrow \infty}\frac{1}{2N}\ln \left(\frac{1}{(2N)^{N}}\sum_{k=0}^N\frac{(-1)^{N-k}(N!)^22^k}{(k!)^2(N-k)!}\left(\frac{1}{2}+N|z|^2\right)_k+o\left( \frac{1}{\beta}\right)\right)\nonumber\\
	&=\lim_{N\rightarrow \infty}\frac{1}{2N}\ln \left((-1)^N \frac{N!}{(2N)^N}\ {}_2F_1\left(-N,\frac{1}{2}+N|z|^2;1;2\right)+o\left( \frac{1}{\beta}\right)\right). \label{eq-expression via 2F1 symm model}
\end{align}
In the second step, we have replaced the sum by Gauss' hypergeometric function which truncates. Notice that the prefactor in \eqref{eq-charac poly with beta correction} gets squared as we now consider the polynomial $Q_N^{D,J}((2N)|z|^2)$ of the variances.
To compute the limit of the hypergeometric function, we assume that we can first expand in the first argument as it diverges faster than the second argument, because we have $|z|<1$ in the corresponding scaling limit. We apply the large $N$ asymptotics given in \cite[Eq. 15.7.3]{Abramowitz Stegun} with $\arg(-2N)=\pi$ for $N\in \mathbb{N}$:
\begin{align*}
	{}_2F_1\left(-N,\frac{1}{2}+N|{z}|^2;1;2\right)
	=&\frac{ie^{i\pi N|{z}|^2}}{\Gamma(\frac{1}{2}-N|{z}|^2)}(-2N)^{-\frac{1}{2}-N|{z}|^2}
	\left(1+\mathcal{O}\left(\frac{1}{2N}\right)\right)\\
	+&\frac{1}{\Gamma(\frac{1}{2}+N|{z}|^2)}e^{-2N}(-2N)^{-\frac{1}{2}+N|{z}|^2}\left(1+\mathcal{O}\left(\frac{1}{2N}\right)\right).
\end{align*}
Using the Stirling formula to approximate $\Gamma(x)$ for large $x$ as \cite[Eq. (8.327)]{gradshteyn}
\begin{align}
	\Gamma(x)\sim x^{x-1/2}e^{-x}\sqrt{2\pi}\left(1+ \mathcal{O}\left(\frac{1}{x}\right)\right),
\end{align}
we conclude
{\small
\begin{align}
	&{}_2F_1\left(-N,\frac{1}{2}+N|z|^2;1;2\right)\nonumber\\
	=& \frac{i}{\sqrt{2\pi}}e^{i\pi N|z|^2}(-N|z|^2)^{N|z|^2}e^{-N|z|^2} 
	(-2N)^{-\frac{1}{2}-N|z|^2}
	\left(1+\mathcal{O}\left(\frac{1}{2N}\right)\right)\nonumber\\
	&+\frac{1}{\sqrt{2\pi}}(N|z|^2)^{-N|z|^2}e^{N|z|^2}e^{-2N} (-2N)^{-\frac{1}{2}+N|z|^2}\left(1+\mathcal{O}\left(\frac{1}{2N}\right)\right)\nonumber\\
	 \approx&\frac{i}{\sqrt{2\pi}}e^{N|z|^2(i\pi-1)}(|z|^2/2)^{N|z|^2}(-2N)^{-\frac12}
	+\frac{1}{\sqrt{2\pi}}e^{N(-2+|z|^2)}(-2/{|z|^2})^{N|z|^2}(-2N)^{-\frac12}.\label{eq-hypergeomsimp}
\end{align}}
To calculate the limit $U(z)$, we have to find an expression for the logarithm of this approximation. We use the following rewriting of the logarithm of the sum of two complex variables $a,b$ with $\frac{|b|}{|a|}\gg 1$:
\begin{align}\label{eq-logsimp}
	\ln(a+b)&=\ln(|a+b|)+i\arg(a+b)
	\approx\ln |b|+i\arg(a+b),
\end{align}
for $\frac{|b|}{|a|}\gg 1$, where $\arg(a+b)\in [0,2\pi)$ is the argument of the complex variable $a+b$. It vanishes in the limit $N\rightarrow \infty$ because of the factor $1/(2N)$. We proceed by implementing the approximation of the hypergeometric function into the logarithm and we set
\begin{align}\label{def-aandb}
	a&=\frac{i}{\sqrt{2\pi}}e^{N|z|^2(i\pi-1)}2^{-N|z|^2}(|z|^2)^{N|z|^2}\frac{1}{\sqrt{-2N}}\\
	b&=\frac{1}{\sqrt{2\pi}}e^{N(-2+|z|^2)}(|z|^2)^{-N|z|^2}2^{N|z|^2}\frac{1}{\sqrt{-2N}},\nonumber
\end{align}
and observe that
\begin{align}\label{eq-ratio}
	\frac{|a|}{|b|}=e^{-2N|z|^2+2N}(|z|^2/2)^{2N|z|^2}\ll 1,\text{ for } N\rightarrow \infty.
\end{align}
Note that the value of $\left|\frac{a}{b}\right|$ is not always close to zero for all values of $z\in\mathbb{\mathbb{C}}$ as $N\to\infty$. This is because the exponent in Eq. \eqref{eq-ratio} can be positive or negative, depending on the value of $z$, for example, when $z$ is close to zero.

For values of $z$ where the exponent is positive, we instead consider $\left|\frac{b}{a}\right|$, which decays to zero as $N\to\infty$. In this case, we rewrite Eq. \eqref{eq-logsimp} as
\begin{align}\label{eq-ratio1}
	\ln(a+b)&\approx\ln(|a|)+i\arg(a+b),
\end{align}
and carry on with the proof using this approximation.

To proceed, we therefore need to consider two cases, depending on the specific value of $z$, since the sign of the exponent in Eq. \eqref{eq-ratio} depends on $z$. This will determine whether $\left|\frac{a}{b}\right|$ or $\left|\frac{b}{a}\right|$ decays to zero. These two cases are:
\begin{enumerate}
	\item For values of $z$ where the exponent in Eq. \eqref{eq-ratio} is negative, $\left|\frac{a}{b}\right|\ll1$ as $N\to\infty$. Therefore, we use Eq. \eqref{eq-ratio} to approximate the logarithm of Eq. \eqref{eq-hypergeomsimp}.
	\item For values of $z$ where the exponent in Eq. \eqref{eq-ratio} is positive, $\left|\frac{b}{a}\right|\ll1$ as $N\to\infty$. Therefore, we use Eq. \eqref{eq-ratio1} to approximate the logarithm of Eq. \eqref{eq-hypergeomsimp}.
\end{enumerate}
However, in both cases, Eq. \eqref{eq-U(Z)unsimp} below will lead to the same spectral density in Eq. \eqref{eq-final spec dens}, regardless of whether we take $a$ or $b$ in Eq. \eqref{def-aandb}. Thus, for the remainder of the proof, we focus on the case described in Eq. \eqref{eq-ratio}, substituting Eq. \eqref{eq-logsimp} into Eq. \eqref{eq-U(Z)unsimp}. The proof for the other case follows analogously and is therefore omitted. Nevertheless, more care must be taken because Eqs. \eqref{eq-M def} and \eqref{eq-eta def} are slightly different, as the signs of $m_1$ and $m_2$ will be flipped.

Next, we approximate the prefactor of the hypergeometric function using the Stirling formula again,
\begin{align*}
	\frac{(-1)^NN!}{(2N)^N}\approx 
	&\sqrt{2\pi N}\frac{(-1)^Ne^{-N}}{2^N}.
\end{align*}
Bringing all together, we find
\begin{align}\label{eq-U(Z)unsimp}
	U(z)&=\lim_{N\rightarrow \infty}\frac{1}{2N}\ln\left(\frac{(-1)^NN!}{(2N)^N}{}_2F_1(-N,\frac{1}{2}+N|z|^2;1;2)\right)\nonumber\\
	&\approx \lim_{N\rightarrow \infty}\frac{1}{2N}\ln\left((-1)^{N}e^{N(-3+|z|^2)}2^{N(|z|^2-1)}(|z|^2)^{-N|z|^2}\right)\nonumber\\
	&=m_0+m_1|z|^2+m_2|z|^2\ln(|z|^2),
\end{align}
where we obtain
\begin{align}\label{eq-M def}
		m_0=\frac{i\pi}{2}-\frac{3}{2}-\frac{1}{2}\ln(2),\quad
		m_1=\frac{1}{2}+\frac{\ln(2)}{2},\quad
		m_2=-\frac{1}{2}.
\end{align}
Calculating the Laplacian yields
\begin{align}
	\Delta_z U(z)\approx m_1+2m_2+m_2\ln(|z|^2).
\end{align}
We conclude that the equilibrium density of the general and the symmetric tridiagonal matrix ensemble is given by
\begin{align}\label{eq-functional form rho S T}
	\widetilde{\rho}^J(z)=\eta_0 +\eta_1\ln(|z|^2),
\end{align}
valid on its support to be determined, where the constants $\eta_0, \eta_1$ are defined by 
\begin{align}\label{eq-eta def}
	\eta_0= m_1+2m_2=-\frac{1}{2}+\frac{\ln 2}{2},\quad 
	\eta_1= m_2=-\frac{1}{2}.
\end{align}
The function $\widetilde{\rho}^J(z)$ is a density as long as $\widetilde{\rho}^J(z)\geq 0$. We define the radius of the complex variable $z$ by $r:=|z|$, and determine the support also called droplet via the following condition
\begin{align}
	\eta_0+\eta_1\log(r^2)\geq 0
	\Leftrightarrow	r&\leq \exp\left(-\frac{\eta_0}{2\eta_1}\right),
\end{align}
where we used that $\eta_1<0$. We denote
\begin{align}
	r_0:=\exp\left(-\frac{\eta_0}{2\eta_1}\right)=\sqrt{\frac{2}{e}}\approx 0.858.
\end{align}
The support $[0,r_0]$ is determined by the ratio of the two constants $\eta_0, \eta_1$ and therefore independent of the normalisation.
We still have to normalise the density properly. Therefore, we define
\begin{align*}
	A:=&\int_{\mathbb{C}}\! d^2z \ \widetilde{\rho}^J(z)=
	2\pi	\int_0^{r_0}\! dr \ r\widetilde{\rho}^J(z)
	=2\pi	\int_0^{r_0}\! dr\  r(\eta_0+\eta_1\ln(r^2))\\
	&= -\pi\eta_1e^{-\frac{\eta_0}{\eta_1}}=\frac{\pi}{e}.
\end{align*}
Our final result is the following normalised equilibrium density
\begin{align}\label{eq-final spec dens}
	\rho^J(z)&= \frac{e}{\pi}\widetilde{\rho}^J(z)=
	\frac{e^1}{2\pi}\left(\ln(2)-1-\ln(|z|^2)\right)\\
	&\approx -0.133-0.433\ln(|z|^2).\nonumber
\end{align}
\end{enumerate}
\end{proof}

\begin{Remark}
	As we calculated the leading order of the equilibrium density in the low temperature limit $\beta \gg 1$, we will comment on the order of the sub-leading terms. 
	We start from \eqref{eq-prop char poly D} and perform the product as
	\begin{align*}
		\Phi_N^J(z)
		=&\frac{1}{(2N)^{\frac{N}{2}}}\Phi_N^{D,J}(\sqrt{2N}z)+o\left(\frac{1}{\sqrt{\beta}}\right)\left(\sum_{j=1}^N\prod_{k\neq j}(z-\frac{1}{\sqrt{2N}}z_i(D_{N}^J))\right)\\
		&+o\left(\frac{1}{\beta}\right)\left(\sum_{i=1}^N\sum_{j=1}^N\prod_{k\neq j,i}(z-\frac{1}{\sqrt{2N}}z_i(D_{N}^J))\right)+...
	\end{align*}
	We observe that we can express the lower order terms in terms of derivatives of the characteristic polynomial $\Phi_N^{D,J}(\sqrt{2N}z)$ as
	\begin{align}\label{eq-expansion in order of beta}
		\Phi_N^J(z)=\frac{1}{(2N)^{\frac{N}{2}}}\Phi_N^{D,J}(\sqrt{2N}z)+\frac{1}{(2N)^{\frac{N}{2}}}\sum_{i=1}^N o\left(\frac{1}{\beta^{\frac{i}{2}}}\right)\partial_z^i \Phi_N^{D,J}(\sqrt{2n}z).
	\end{align}
	We normalised each matrix $J_\beta=\widetilde{T}_\beta,S_\beta,T_\beta$ by $1/\sqrt{2N\beta}$ such that the eigenvalues are supported in a compact disc. Therefore, we expect the characteristic polynomial $\Phi_N^J(z)$ to be bounded on the support of $J_\beta$. Hence, the quantity $\frac{1}{(2N)^{\frac{N}{2}}}\Phi_N^{D,J}(\sqrt{2N}z)$ is bounded, too. Furthermore, we use the fact that the derivative of the characteristic polynomial $\Phi_N^{D,J}(\sqrt{2N}z)$ is given by the sum of the characteristic polynomials of all $(N-1)\times (N-1)$ dimensional sub-matrices of $D^J$. The latter are normalised with respect to a factor $1/(2N-1)^{\frac{N-1}{2}}$ (by an induction hypothesis, this normalisation is of the same order for all $N$). Notice that we get another factor of $\sqrt{2N}$ for each derivative as we act on the variable $z'=\sqrt{2N}z$.\\
	It follows that the first correction reads 
	\begin{align}
		o\left(\frac{1}{\sqrt{\beta}}\right)\frac{\sqrt{2N}}{(2N)^{\frac{N}{2}}}(2N-1)^{\frac{N-1}{2}}\times \text{norm.}=o\left(\frac{1}{\sqrt{\beta}}\right)\left(1-\frac{1}{2N}\right)^{\frac{N-1}{2}}\times \text{norm.},
	\end{align}
	where "norm." denotes the fixed norm of the characteristic polynomial of $D_{N}^J$. We observe that the leading order is independent of $N$. For the $i$th order correction in $\beta$, we find
	\begin{align}
		o\left(\frac{1}{\beta^{\frac{i}{2}}}\right)\left(1-\frac{i}{2N}\right)^{\frac{N-i}{2}}\times \text{norm.},
	\end{align}
	for $i=1,\ldots,N$. For the calculation of the equilibrium density, we need the polynomial built by the variances of the coefficients of $\Phi_N^J(z)$. Therefore, we have to estimate the correction to each coefficient separately and calculate its variance. For $z^{N-1}$ we have only one contribution coming from the first correction term in \eqref{eq-expansion in order of beta}. Using the independence of the matrix elements of $D^J$, as we did previously, it is clear that the variance splits and we obtain a correction of order $o\left(\frac{1}{\beta}\right)$ for the second highest coefficient in $z$. Hence, we conclude that
	\begin{align}
		U(z)=\lim_{N\rightarrow \infty}\frac{1}{2N}\ln \left(\frac{1}{(2N)^N}Q_N^{D,J}(2N|z|^2)+\frac{1}{(2N)^N}\sum_{i=1}^N o\left(\frac{1}{\beta^i}\right)Q_{N-i}^{D_{N-i}^J,J}(2N|z|^2)\right),
	\end{align}
	where the $Q_{N-i}^{D_{N-i}^J,J}(2N|z|^2)$ is the polynomial of the variances of all $(N-i)\times (N-i)$ dimensional sub-matrices of $D_J$ for $i=1,\ldots,N$. We estimate the leading order in $N$ by
	\begin{align}
		U(z)&=\lim_{N\rightarrow \infty}\frac{1}{2N}\ln \left(\frac{1}{(2N)^N}Q_N^{D,J}(2N|z|^2)(1+\sum_{i=1}^N o\left( \frac{1}{\beta^i} \right) \right)\\
		&=\lim_{N\rightarrow \infty}\frac{1}{2N}\ln \left(\frac{1}{(2N)^N}Q_N^{D,J}(2N|z|^2)\right)+\lim_{N\rightarrow \infty}\frac{1}{2N}\ln \left(1+\sum_{i=1}^N o\left( \frac{1}{\beta^i} \right) \right),
	\end{align}
	where we find that the first correction term in orders of $\beta$ is massively suppressed in the large $N$-limit. This is the reason that the radial density of the general and the symmetric tridiagonal ensemble shows universality in $\beta$ in the large $N$-limit, cf. Figure \ref{fig-radial dist T and T tilde beta 1/2,2,1000 n=5000} right. This contrasts with the behaviour for $N=2$, which is discussed in Appendix \ref{app- tridiag N=2}.
	
	In the non-symmetric ensemble, we have that the coefficients of the polynomial $Q^{D,\widetilde{T}}_N(|z|^2)$ still contain an explicit $\beta$-dependence (cf. Lemma \ref{lem-var coeff P D}, Eq. \eqref{eq-var coeff P D T tilde}) which makes the study of the correction terms much more evolved.
\end{Remark}

We conclude by determining the moments of our radially symmetric logarithmic density. 
\begin{lemma}
	The radial moments of the rotational symmetric density $\rho^{J}(z)$ for the general and symmetric ensemble $\mathbf{J=T,S}$ are defined by
	\begin{align}\label{eq-def rad moments}
		m_k:=\int_0^{r_0}\! dr\  r^{2k+1}\rho^{J}(z),
	\end{align}
	for $k=1,2,\ldots$ with $|z|=r$.
	They are given as
	\begin{align}\label{eq-rad moments}
		m_k=
		\frac{e^1 }{4\pi }\frac{1}{(k+1)^2}r_0^{2k+2}= \frac{1}{2\pi(k+1)^2}\left(\frac{2}{e}\right)^k.
	\end{align}
\end{lemma}
\begin{proof}
	Due to the radial symmetry of density $\rho^{J}(z)$ we study only the non-vanishing radial moments defined in Eq. \eqref{eq-def rad moments}. We parametrize $|z|=r$ and insert Eq. \eqref{eq-limiting density T,S}, which leads to \eqref{eq-rad moments} using elementary integrals.
\end{proof}
The moments seemingly do not correspond to any interesting combinatorial objects, unlike the semi-circle, which relates to the Catalan numbers.

\subsubsection{Numerical Simulations, Pseudospectrum, and Local NN Spacing}\label{sssec-numerics}
\begin{figure}[ht]
	\begin{subfigure}{0.49\textwidth}
		\includegraphics[width=1\linewidth]{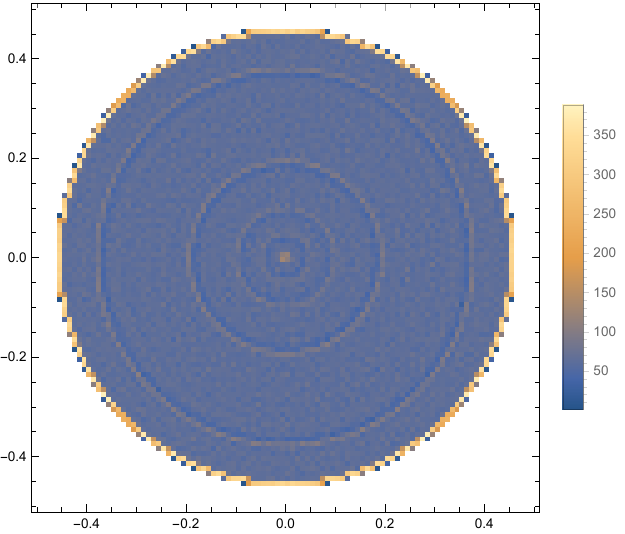}
		\caption{$\widetilde{T}_\beta$}\label{fig-scatterplot T tilde n5000}
	\end{subfigure}
\begin{subfigure}{0.49\textwidth}
	\includegraphics[width=1\linewidth]{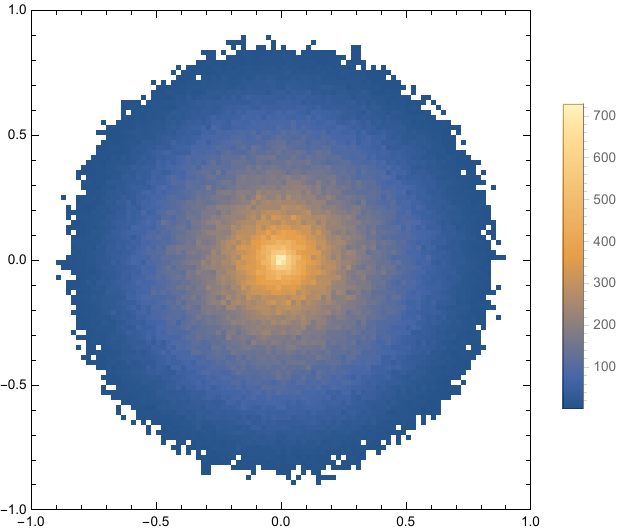} 
	\caption{$T_\beta$}
\end{subfigure}
	\caption[Cf. Figure 1 from \cite{AMPT}, \textit{Spectral density of the complex eigenvalues of the tridiagonal matrix ensembles $\widetilde{T}_\beta$ and $T_\beta$}]{Spectral density of the complex eigenvalues of $\widetilde{T}_\beta$ (left) and $T_\beta$ (right), for $100$ matrices of dimension $N=5000$, both with $\beta=6$. The spectral density is normalised by $\sqrt{2N\beta}$. The light yellow indicates here a high density, and dark blue regions are less dense. \cite[Cf. Figure 1]{AMPT}\label{fig-scatter plots tridiag models n5000}}
\end{figure}
In this section, we study the tridiagonal matrix ensembles numerically and compare the equilibrium densities for $\widetilde{T}_\beta$ \eqref{eq-limiting density T tilde} and $T_\beta, S_\beta$ \eqref{eq-limiting density T,S} derived in the previous section in the limit of low temperature $\beta \gg 1$ and large matrices $N\rightarrow \infty$ with the numerics. As the limiting spectral density of the symmetric and the general matrices agree, we focus on the numerics of matrix ensembles $T_\beta$.

We begin with two scatter plots of the complex eigenvalues of the tridiagonal random matrices $\widetilde{T}_\beta$ and $T_\beta$ in Figure \ref{fig-scatter plots tridiag models n5000} left respectively right, for 100 matrices of dimension $N=5000$ and $\beta=6$. In both cases the spectrum is rotationally invariant, as expected. For this reason, below we will only present plots for the radial density of the modulus $|z|$. 
However, already for a single value of $\beta$ the ensembles show very different features. 
While the spectrum of $T_\beta$ in Figure \ref{fig-scatter plots tridiag models n5000} right is smooth, with a strong peak at the origin, that of $\widetilde{T}_\beta$ looks almost flat, with pronounced concentric rings, including at the origin and at the edge.

\begin{figure}[h]
	\begin{subfigure}{0.49\textwidth}
		\includegraphics[width=\textwidth]{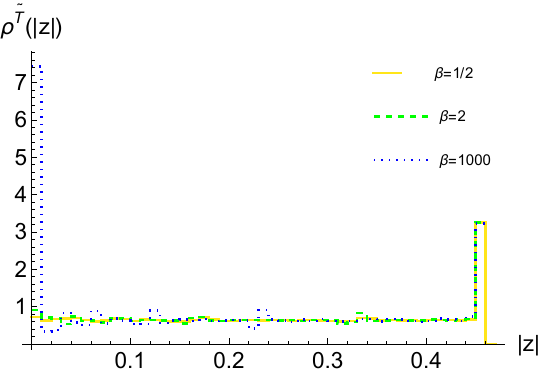}
		\subcaption{$\widetilde{T}_\beta$\label{fig-radial dist T tilde n=5000}}	
	\end{subfigure}
\begin{subfigure}{0.49\textwidth}
	\includegraphics[width=\textwidth]{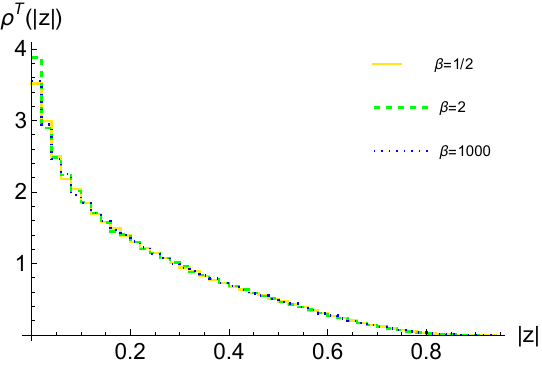}
	\subcaption{$T_\beta$ \label{fig-radial dist T n=5000}}	
\end{subfigure}
	\caption[Cf. Figure 2 from \cite{AMPT}, \textit{Radial density plots for the tridiagonal matrix ensembles $\widetilde{T}_\beta$ and $T_\beta$ for various $\beta$}]{Histograms of the radial density of the complex eigenvalues for $100$ matrices of size $N=5000$ of the ensembles $\widetilde{T}_\beta$ (left) and $T_\beta$ (right), at $\beta=1/2$ (yellow full curve), $\beta=2$ (green dashed curve) and $\beta=1000$ (blue dotted curve).\\ \cite[Cf. Figure 2]{AMPT} \label{fig-radial dist T and T tilde beta 1/2,2,1000 n=5000}}
\end{figure}

In Figure \ref{fig-radial dist T and T tilde beta 1/2,2,1000 n=5000} below, the radial density $\rho^J(|z|)$ of the complex eigenvalues $z$ is plotted for $J=\widetilde{T},T$ for different values of $\beta=1/2,2,1000$ in different colours (yellow full, green dashed, blue dotted respectively), for the same size and number of realisations as in Figure \ref{fig-scatter plots tridiag models n5000}. For the general tridiagonal model $T_\beta$ (right) we observe that the histograms of the radii $|z|$ look very similar for the plotted $\beta$-values over a range of three orders of magnitude. Only near the origin small differences can be seen. We conclude that the spectral density of the general tridiagonal ensemble $T_\beta$ seems to be universal for a large range of $\beta$ in the limit of large matrix size $N$. This shows that although we are only able to compute the limiting spectral density for matrices $T_\beta$ in the limit of low temperature $\beta \gg 1$, due to universality the result will be a good approximation also for small $\beta$ down to $\beta=1/2$, when $N$ is large, too.

For the radial density of the ensemble $\widetilde{T}_\beta$ the situation is different. 
We observe that all three curves for different values of $\beta=1/2,2,1000$ share a 
flat density and a peak at the edge of the spectrum, which also appears to be universal.
However, at large $\beta$ - which is the limit in which we derived the global density - a strong peak at the origin develops, in addition to small peaks within the spectrum.
We conclude for the non-symmetric tridiagonal model the equilibrium density $\rho^{\widetilde{T}}(z)=\delta(x)\delta(y)$ \eqref{eq-limiting density T tilde} 
completely dominates the spectrum in the limit of large $\beta$ and $N$, and that we cannot resolve the sub-leading, potentially universal features of the density in this limit. 
The tails of the peak at the edge may be different for small to intermediate $\beta$ too, but such a local analysis is currently not feasible. 
We will discuss below what could be the reason of this singular behaviour, introducing the notation of a pseudo-spectrum, which plays an important role here.

In Figure \ref{fig-comparison T,S beta} below we give a visual (left) and quantitative comparison (right) between the radial distribution of the numerically generated general tridiagonal ensemble $T_\beta$ and symmetric ensemble $S_\beta$ at $\beta=100$, keeping the same matrix and ensembles size. The left plot shows coloured histograms in blue and red, respectively, versus the analytical solution \eqref{eq-limiting density T,S}, which we derived in the low temperature limit for $N\rightarrow \infty$ (black curve). We observe an excellent agreement within both matrix ensembles. Moreover, the distributions of the two ensembles show only small departures from each other and from the limiting spacing density. To estimate the systematic error between the analytical result \eqref{eq-limiting density T,S} and the numerics of the two matrix ensembles $T_\beta$ and $S_\beta$, 
we calculate the Kolmogorov-Smirnov distance $d$ (cf. Eq. \eqref{eq-KS distance}) for four different values of $\beta$ in units of $10^{-2}$ in the table of Figure \ref{fig-comparison T,S beta} (right).\clearpage
\begin{figure}[t]
	\begin{minipage}{0.45\textwidth}
		\includegraphics[width=\linewidth,angle=0]{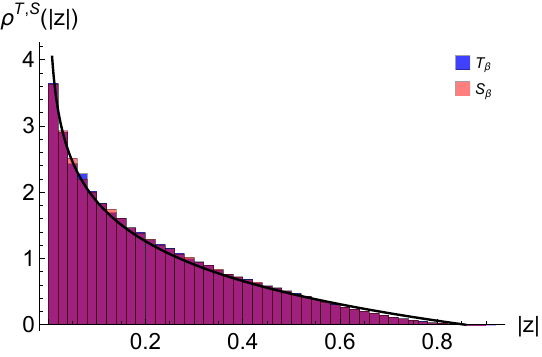} 
	\end{minipage}
	\begin{minipage}{0.45\textwidth}
		\begin{tabular}{|c|l|c|c|c|c|}
			\hline
			&\diagbox{$N(m)$}{$\beta$} & $\frac{1}{2}$ & 2 & 100 & 1000 \\
			\hline
			\multirow{3}{*}{$T_\beta$}&100 (5000) & 7.60 & 4.57&  10.70& 11.73\\
			\cline{2-6}
			&1000 (1000) & 3.25 & 1.35 & 3.60& 4.02\\
			\cline{2-6}
			&5000 (100) & 1.38 & 1.41&  1.39& 1.38\\ \hline \hline
			\multirow{3}{*}{$S_\beta$}&100 (5000) & 9.84 & 5.74&10.62 & 11.82\\
			\cline{2-6}
			&1000 (1000) & 4.10 & 2.24&3.59 & 4.15\\
			\cline{2-6}
			&5000 (100) & 2.22 & 1.38& 1.37& 1.34\\
			\hline
			
		\end{tabular}
		\label{tab-KS dist T,S beta}
	\end{minipage}
	\caption[Figure 3 from \cite{AMPT}, \textit{Comparison of the radial density of the tridiagonal matrix ensembles $S_\beta, T_\beta$ with the analytics: Plot and Table of Kolmogorov-Smirnov-distances}]{Comparison between analytics and numerics for $T_\beta$ and $S_\beta$. The histogram ({left}) shows the distribution of the radii for an ensemble of $100$ matrices of size $N=5000$ for the general ensemble $T_\beta$ (blue) and symmetric ensemble $S_\beta$ (light red) for $\beta=100$. The region where both distributions agree is purple. The black curve gives the analytical expression for the density \eqref{eq-limiting density T,S}. 
		We also give the Kolmogorov-Smirnov distances ({right}) in units $10^{-2}$ between the numerics and the analytical result \eqref{eq-limiting density T,S} for both ensembles for various matrix sizes $N$, ensemble sizes $m$ and $\beta$ values. \cite[Figure 3]{AMPT}}\label{fig-comparison T,S beta}
\end{figure}

\noindent
We recall from Section \ref{ssec-numerics methods}, that we give as a reference, Kolmogorov-Smirnov distance between the statistics of the complex Ginibre ensemble and a 2dCG at $\beta=2$ with the same ensemble size $N=5000$ and realisations $m=100$, which produces $d=0.17\cdot 10^{-2}$. Note that the range of error for the equilibrium density $\rho^{T,S}(z)$ of both ensembles is always comparable and shows the same systematic when increasing $\beta$, $N$ and $m$. We observe that the best agreement between the derived analytical result and numerics is given for large $N=5000$. However, the Kolmogorov-Smirnov distance is still by one order larger than for the same ensemble of Ginibre matrices.
For the largest $N$, where we see the best agreement in Kolmogorov-Smirnov distance, we can go to even larger $\beta= 5000, 7000, 50000$ and find that $d$ is not reduced further, i.e. we have $d=1.37, 1.39, 1.37$ in units $10^{-2}$, respectively for the symmetric ensemble $S_\beta$.

\begin{Remark}
	As a technical remark, we would like to point out that for the calculation of the Kolmogorov-Smirnov distances, the data is normalised with respect to a flat measure, which gives a much simpler numerical computation of the cumulative distribution. Therefore, we compare with the limiting density, which is also normalised with respect to the flat measure, i.e.
	\begin{align}
		1=\int_0^{r_0}\! dr\  \rho^{J,\text{flat}}(z).
	\end{align}
	Since this only affects the overall scaling of the function, both for analytics and numerics, the calculated Kolmogorov-Smirnov distances still provide a good estimate of the systematic error.
\end{Remark}

\clearpage
\noindent\textbf{Contours of $\boldsymbol{\epsilon}$-Pseudospectra for Non-Hermitian Tridiagonal Matrices}
\\
\\
It is well known that the numerically determined eigenvalues of a non-normal matrix do not always provide a complete picture of the matrix properties, particularly its resonances and stability. In many cases, the spectrum alone is insufficient to capture the sensitivity of the matrix to perturbations. This limitation motivates the study of the pseudospectrum, see \cite{Trefethen Embree} for an introduction. The pseudospectrum extends the analysis of the eigenvalue spectrum to incorporate the effects of perturbations and the behaviour of the resolvent. The pseudospectrum can be formally defined as follows:
\begin{definition}[\text{\cite[Eq. (2.1)]{Trefethen Embree}}]
	Let $A\in \mathbb{C}^{N\times N}$ and $\epsilon >0$. The \textit{$\epsilon$-pseudospectrum} $Z_\epsilon(A)$ is the set of $z\in \mathbb{C}$ such that
	\begin{align}
		||(z\mathds{1}_N-A)^{-1}||>\epsilon^{-1}.
	\end{align}
\end{definition}	
The matrix $(z\mathds{1}_N-A)^{-1}$ is known as the resolvent of $A$, and the spectrum $\sigma(A)$ consists of those $z\in\mathbb{{C}}$ where the resolvent does not exist (\emph{i.e.} the eigenvalues of $A$). Following the convention in \cite{Trefethen Embree}, for $z\in\sigma(A)$, we set
\begin{align}
	||(z\mathds{1}_N-A)^{-1}||=\infty.
\end{align}
An alternative definition of the $\epsilon$-pseudospectrum is given in terms of the $2$-norm and the smallest singular value of $(z\mathds{1}_N-A)$, denoted by $s_{\min}(z\mathds{1}_N-A)$. Specifically, when $\|.\|=\|.\|_2$, it holds that 
\[
s_{\min}(z\mathds{1}_N-A)=\frac{1}{\|(z\mathds{1}_N-A)^{-1}\|_2}.
\]
\begin{definition}\label{def-singpseudo}
	For $\|.\|=\|.\|_2$, the $\epsilon$-pseudospectrum $Z_\epsilon(A)$ is the set of all $z\in\mathbb{C}$ such that
	\begin{align}\label{eq-def min sv}
		s_{\textrm{min}}(z\mathds{1}_N-A)=\inf_{\|\mathbf{x}\|_2=1}\|(z\mathds{1}_N-A)\mathbf{x}\|_2< \epsilon.
	\end{align}
\end{definition}

For normal matrices (matrices which satisfy $AA^\dagger=A^\dagger A$), the behaviour of the resolvent is relatively straightforward: the $\epsilon$-pseudospectrum consists of the union of balls of radius $\epsilon$ centred at each eigenvalue in the complex plane. This reflects that the eigenvalues of normal matrices are stable under small perturbations (cf. Remark \ref{rem-perturbations} for behaviour of non-normal/ non-Hermitian matrices), and the matrix properties can be understood purely from the spectrum. We note that we touched this stability also in Remark \ref{rem-stability N=2} discussing the small $N$ behaviour of (non-)Hermitian matrices.

However, for non-normal matrices, the situation is more complicated. Even when $z$ is far from the spectrum, the norm of the resolvent $\|(z\mathds{1}_N-A)^{-1}\|$ can be large. This occurs because non-normal matrices can have eigenvalues that are extremely sensitive to small perturbations. As a result, the $\epsilon$-pseudospectrum of a non-normal matrix does not simply form $\epsilon$-balls around the eigenvalues. This behaviour is described in \cite[Theorem 2.2]{Trefethen Embree}.

A significant result that connects the spectrum and the pseudospectrum is the Bauer-Fike Theorem, which provides bounds on how eigenvalues behave under perturbations:
\begin{theorem}[Bauer-Fike Theorem \text{\cite[Theorem $2.3$]{Trefethen Embree}}]\label{thm-Bauer-Fike}
	Let $A\in\mathbb{C}^{N\times N}$ be diagonalisable, such that $A=RZ R^{-1}$. Then, for any $\epsilon>0$, with $\|.\|=\|.\|_2$,
	\begin{align}
		\sigma_A+\mathcal{B}_\epsilon\subseteq Z_\epsilon(A)\subseteq \sigma_A+\mathcal{B}_{\epsilon\kappa(R)},
	\end{align}
	where
	\[
	\mathcal{B}_\epsilon=\{z\in\mathbb{C}:|z|<\epsilon\}\hspace{2mm}\text{and}\hspace{2mm}\kappa(R)=\frac{\|R\|}{\|R^{-1}\|}\hspace{2mm}\text{is the condition number of}\ R.
	\]
\end{theorem}
This result shows that for diagonalisable matrices, the pseudospectrum $Z_\epsilon(A)$ is contained within a neighbourhood controlled by the condition number $\kappa(R)$ of the eigenvector matrix $R$. In the case of non-normal matrices, where $\kappa(R)$ can be large (see Table \ref{tab-condition number} below), the pseudospectrum provides an enlarged region around the spectrum.

\begin{figure}[t]
	\begin{subfigure}{0.49\textwidth}
		\includegraphics[width=\textwidth]{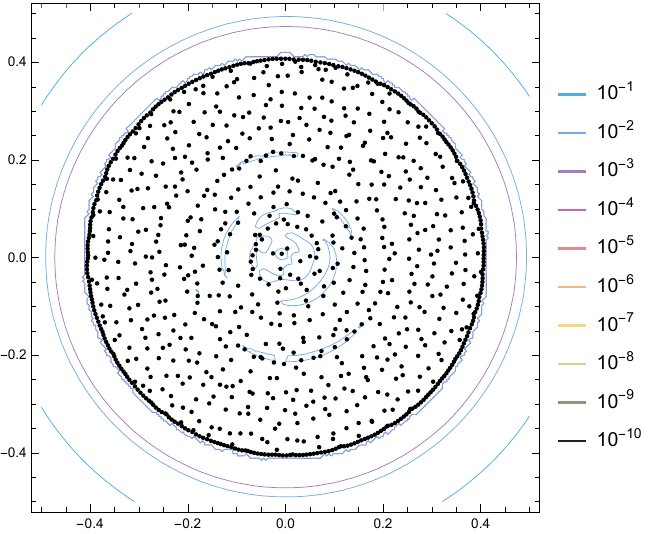}
		\subcaption{$\widetilde{T}_\beta$}
	\end{subfigure}
	\begin{subfigure}{0.49\textwidth}
	\includegraphics[width=\textwidth]{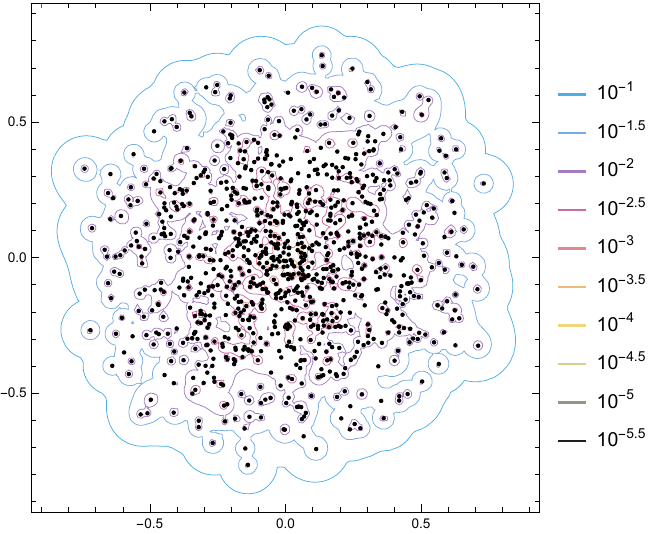}
	\subcaption{$T_\beta$ }	
	\end{subfigure}
	\caption[Cf. Figure 4 from \cite{AMPT}, \textit{$\epsilon$-pseudospectra of the tridiagonal matrices $\widetilde{T}_\beta$ and $T_\beta$}]{The plots show the $\epsilon$-pseudospectra of $\widetilde{T}_{\beta}$ (left) and $T_{\beta}$ (right), for $N=1000$, compared to its eigenvalues (dots) for $\beta=2$. The contour scale represents $10^{-\epsilon}$. \cite[Cf. Figure 4]{AMPT}}
	\label{fig-pseudospectrum}
\end{figure}
\begin{prop}
	For arbitrary $\epsilon>0$ the $\epsilon$-pseudospectrum is given as 
	\begin{align}\label{eq-pseudo}
		Z_\epsilon(J_\beta)=\left\{z\in\mathbb{C}:\left|z-\frac{a_1}{\sqrt{2N\beta}}\right|^2+\frac{|c_1|^2}{2N\beta}<\epsilon^2\right\},\text{ with} \begin{cases}
			c_1=1 &\text{, for }  J_\beta=\widetilde{T}_\beta\\
			|c_1|\sim \frac{1}{\sqrt{2}}\chi_{\beta}  &\text{, for }  J_\beta=S_\beta\\
			|c_1|\sim \chi_{\beta/2}  &\text{, for }  J_\beta=T_\beta
		\end{cases}
	\end{align} and $\re(a_1),\im(a_1)\sim\mathcal{N}(0,1)$.

\end{prop}
\begin{proof}
We present the proof for the $J_\beta=\widetilde{T}_\beta,S_\beta, T_\beta$ and specialise with respect to the probability distributions of the matrix elements later. Using the Definition \ref{def-singpseudo} of $\epsilon$-pseudospectrum, we start analysing the 2-norm of the vector $(z\mathds{1}_N-J_\beta)\mathbf{e}_N$:
	\begin{align}
		||(z\mathds{1}_N-J_\beta)\mathbf{e}_N||_2&=\left\lVert \begin{pmatrix}
			0,&
			\dotsc&
			0,&
			\frac{c_1}{\sqrt{2N\beta}},&
			z-\frac{a_1}{\sqrt{2N\beta}}
		\end{pmatrix}^t \right\rVert_2.
	\end{align}
	This yields
	\begin{equation}
		||(z\mathds{1}_N-J_\beta)\mathbf{e}_N||_2=\sqrt{\frac{|c_1|^2}{2N\beta}+|z|^2-\frac{2|za_1^*|}{\sqrt{2N\beta}}\cos(\arg(za_1^*))+\frac{|a_1|^2}{2N\beta}},
	\end{equation}
with $\re(a_1),\im(a_1)\sim\mathcal{N}(0,1)$ for all three matrices $J_\beta=\widetilde{T}_\beta, S_\beta, T_\beta$ and $c_1=1$ for $J_\beta=\widetilde{T}_\beta$, $|c_1|\sim \frac{1}{\sqrt{2}}\chi_\beta$ for $J_\beta=S_\beta$, respectively $|c_1|\sim \chi_{\beta/2}$ for $J_\beta=T_\beta$ from Table \ref{tab-dists tridiag}.\\
Next, we must establish that the vector $\mathbf{e}_N$ minimises this expression over all unit vectors $\mathbf{x}\in\mathbb{C}^N$. To do so we use the Cauchy-Schwartz inequality since $\Var(|c_j|)\geq\Var(|c_1|)$ for $j=2,\dots,N-1$ for $S_\beta, T_\beta$, as $|c_j|\sim\frac{1}{\sqrt{2}}\chi_{\beta j}$, respectively $|c_j|\sim\chi_{\beta j/2}$. Note that the entries on the sub- and super-diagonals are of the same order for the symmetric and general tridiagonal ensemble. For the non-symmetric matrix $\widetilde{T}_\beta$, with $c_j=1$ for all $j=1,\ldots,N-1$, we have that the super-diagonal elements are much smaller than the sub-diagonal elements (cf. low temperature limit of the rescaled matrix Eq. \eqref{eq-largebetas T tilde rescaled}). Assuming we have a vector $\mathbf{x}^t=(x_1,\ldots, x_N)$ with $\|\mathbf{x}\|_2=1$ one obtains:
	\begin{align}
		||(z\mathds{1}_N-J_\beta)\mathbf{x}||_2
		&\geq \sqrt{\left|(z-\frac{a_1}{\sqrt{2N\beta}})x_N\right|^2+2\Re \left(\frac{c_1}{\sqrt{2N\beta}}x_{N-1}(\frac{a_1^*}{\sqrt{2N\beta}}-z^*)x_N^*\right)}.
	\end{align}
	We can choose a representative, such that the variable $\Re (\frac{c_1}{\sqrt{2N\beta}}x_{N-1}(\frac{a_1^*}{\sqrt{2N\beta}}-z^*)x_N^*)$ is positive, without loss of generality. Therefore,
	\begin{align}
		||(z\mathds{1}_N-J_\beta)\mathbf{x}||_2&
		\geq \sqrt{\left|(z-\frac{a_1}{\sqrt{2N\beta}})x_N\right|^2},
	\end{align} 
	which implies that $\mathbf{e}_N$ is the infimum.
\end{proof}

As $N\rightarrow\infty$ or $\beta\rightarrow\infty$, the $\epsilon$-pseudospectrum converges to $\epsilon$-balls centred at the origin for all three matrix ensembles. However, for small $N$ and $\beta$ the randomness displayed in $|c_1|$ altering the radius of the contours results in a significantly different situation for the general and symmetric tridiagonal ensemble, cf. Eq. \eqref{eq-pseudo}. In the non-symmetric ensemble $\widetilde{T}_\beta$, we have a deterministic $c_1=1$. We present the $\epsilon$-pseudospectra for $N=100$ at $\beta=2$ of $\widetilde{T}_\beta$ and $T_\beta$ in Figure \ref{fig-pseudospectrum} above left, respectively right. We observe that the contours are much more sensitive to individual eigenvalues for the general ensemble $T_\beta$ than for the non-symmetric one $\widetilde{T}_\beta$.\clearpage


\noindent\textbf{Measures of Non-Normality}\\
\\
As discussed above, the $\epsilon$-pseudospectrum becomes more important as the degree of non-normality grows. However, the latter is too complex to be determined by one parameter completely. There are several possible measures of non-normality, which are discussed in \cite[Chapter 48]{Trefethen Embree}. To show that the degree of non-normality is much larger for the non-symmetric tridiagonal matrices than for general/ symmetric tridiagonal matrices, we consider the condition number of the diagonalizing matrix as a measure. This quantity can be easily derived numerically.\\
As previously, we assume that the matrix $J_\beta=\widetilde{T}_\beta, S_\beta, T_\beta$, can be diagonalized by matrix $R$ as
\begin{align*}
	J_\beta=RZ R^{-1},
\end{align*}
where $Z$ is a diagonal matrix containing the complex eigenvalues of $J_\beta$.\footnote{We suppress the explicit dependence of $Z,\ R$ on the ensemble $J_\beta$ as before.} The \textit{condition number} of the diagonalizing matrix $R$ (which contains the eigenvectors) is also defined as
\begin{align}
	\kappa(R)=\frac{s_{\textrm{max}}(R)}{s_{\textrm{min}}(R)},
\end{align}
where $s_{\textrm{max}}(R)$ and $s_{\textrm{min}}(R)$ are the maximal and minimal singular value of $R$. Recall, that the diagonalizing matrix $R$ is not determined uniquely and one has to fix a representative. The condition number is bounded by $1<\kappa(R)<\infty$, where the condition number $\kappa(R)=1$ is possible if and only if the matrix $J_\beta$ is normal. Therefore, we will give the condition number of the diagonalizing matrix as a measure of non-normality for matrix ensembles of $m$ matrices of size $N=100,500,1000$ and values $\beta=1/2,2,100,1000$ each in the Table \ref{tab-condition number}. We observe that the condition numbers are of order at least $10^{12}$ times larger for the non-symmetric ensemble compared with the general and the symmetric one. This confirms the observations we have made thus far, as the spectrum of the general tridiagonal matrix ensemble is quite stable for large $N$ and different $\beta$ and we calculated the equilibrium density which describes the numerics for a large range of ensemble sizes. For a better comparison, we also give the condition numbers for the same ensemble sizes of complex Ginibre matrices (normalised such that the droplet is the unit disk) in the last three rows of Table \ref{tab-condition number}.
\begin{table}[h]\centering
	\caption[Table 1 from \cite{AMPT}, \textit{Condition numbers of the matrix of eigenvectors for the tridiagonal matrix ensembles $\widetilde{T}_\beta, S_\beta,T_\beta$ and the complex Ginibre ensemble}]{We give the condition numbers of the matrices of eigenvectors for matrix ensembles of $m$ matrices of size $N=100,500,1000$ and $\beta=1/2,2,100,1000$ each for the non-symmetric matrix $\widetilde{T}_\beta$, the symmetric matrix $S_\beta$, the general tridiagonal matrix $T_\beta$ and the Ginibre ensemble corresponding to $\beta=2$. \cite[Table 1]{AMPT}
	}
	\begin{tabular}{|c|c|c|c|c|c|}
		\hline
		&\diagbox{$N(m)$}{$\beta$} & 1/2 & 2  & 100 & 1000\\
		\hline
		\hline 
		\multirow{3}{*}{$\widetilde{T}_\beta$}&100 (1000) & 2.57$\cdot10^{13}$ & 3.12$\cdot 10^{13}$ &  3.10$\cdot 10^{13}$ &  2.99$\cdot 10^{13}$\\
		\cline{2-6}
		&500 (500) & 3.90$\cdot 10^{13}$ & 3.87$\cdot 10^{13}$ &3.87$\cdot 10^{13}$ & 3.79$\cdot 10^{13}$\\
		\cline{2-6}
		&1000 (100) & 4.06$\cdot 10^{13}$ & 4.06$\cdot 10^{13}$  & 4.12$\cdot 10^{13}$ & 4.08$\cdot 10^{13}$\\
		\hline
		\hline
		\multirow{3}{*}{$S_\beta$}&
		100 (1000) & 26.47 & 36.33 &  42.40 & 41.70\\
		\cline{2-6}
		&500 (500) & 77.95 & 101.37 & 114.83 &111.19\\
		\cline{2-6}
		&1000 (100) & 118.67&154.42 & 188.21 &177.21\\
		\hline
		\hline    

		\multirow{3}{*}{$T_\beta$}&
		100 (1000) &38.83 & 42.09 &  41.20 & 41.67\\
		\cline{2-6}
		&500 (500) &105.81 & 116.81 & 124.03 &114.04\\
		\cline{2-6}
		&1000 (100) & 156.14& 177.47 &  178.90 &181.97\\
		\hline
		\hline
		
		\multirow{3}{*}{Ginibre}&
		100 (1000)& & 98.13 & & \\
		\cline{2-6}
		&500 (500)& & 405.45 & &\\
		\cline{2-6}
		&1000 (100)& &  704.47 & &\\
		
		\hline
	\end{tabular}
	\label{tab-condition number}
\end{table} 

It is known that Ginibre matrices are not normal, but their scatter plots can be described very well by the spectrum. We observe that the values of the condition numbers are of the same order as for the general and the symmetric tridiagonal matrix ensemble. Moreover, Chalker and Mehlig \cite{Chalker Mehlig} derived that the eigenvalue condition numbers scale as $\mathcal{O}(\sqrt{N})$ as $N\rightarrow \infty$, which indicates a mild degree of non-normality (cf. discussion in \cite[Chapter 35]{Trefethen Embree}). Edelman proved in \cite{Edelman} that the slope of the linearly growing condition number of a dense random matrix in the dimension is given by $C_E=2e^{\gamma /2}$, with Euler's constant $\gamma$. However, we want to emphasize that this condition number, studied by Edelman and others, refers directly to the random matrix $X$, rather than to the matrix of eigenvectors. We refer the reader to the literature \cite{Edelman} for a formal definition. We checked that the calculation of this measure of non-normality yields qualitatively the same picture as we have seen in Table \ref{tab-condition number}, i.e. the values of this quantity are of the same order for the general and the symmetric tridiagonal ensemble while the results for the non-symmetric model are of orders of magnitudes larger.\\
There exist also other measures of non-normality such as the departure of normality (introduced by Henrici \cite{Henrici}, see also \cite[Chapter 48]{Trefethen Embree}). We observed that the computation of this quantity is in the case of our matrix ensembles extremely sensitive to the machine precision and we therefore cannot observe large differences between the tridiagonal matrix ensembles. 
\\
\\
\noindent\textbf{Local NN Spacing}\\
\\
We discussed in the beginning of this Section \ref{sec-non Herm beta} that the NN spacing distribution is an important measure to study local statistics of matrix ensemble which might show universal features. Therefore, it is natural to study this quantity also for the three introduced tridiagonal matrix ensembles. 
\begin{figure}[h]
	\hspace{50pt}{$\beta=1/2$} \hspace{100pt}{$\beta=2$}\hspace{100pt}{$\beta=1000$} \hfill\\
	\begin{turn}{90}\hspace{40pt}{$\widetilde{T}_\beta$}\end{turn}    
	\includegraphics[width=0.3\textwidth,angle=0]{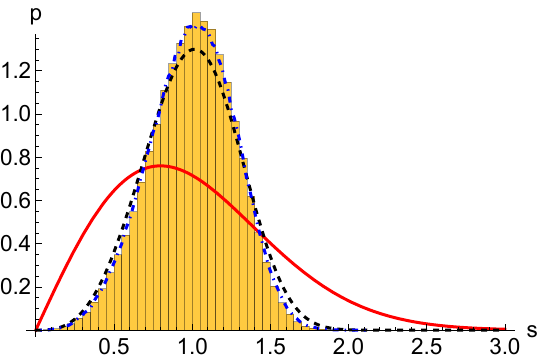}
	\includegraphics[width=0.3\textwidth,angle=0]{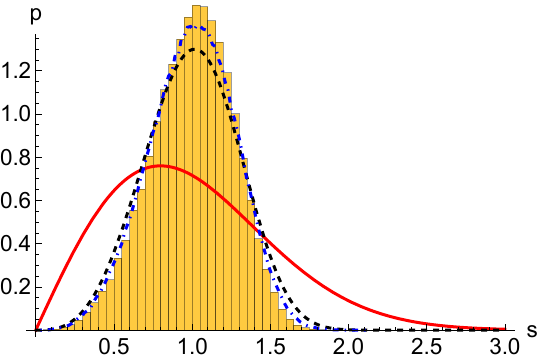}
	\includegraphics[width=0.3\textwidth,angle=0]{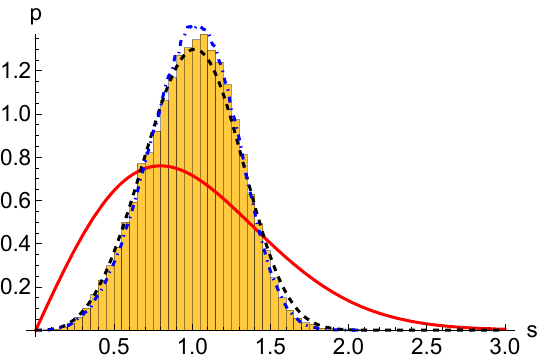}\\
	\begin{turn}{90}\hspace{40pt}{$S_\beta$}\end{turn}    
	\includegraphics[width=0.3\textwidth,angle=0]{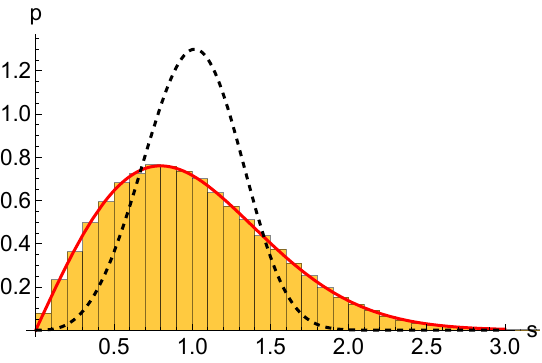}
	\includegraphics[width=0.3\textwidth,angle=0]{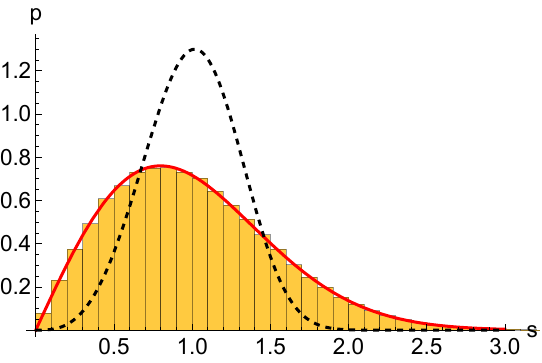}
	\includegraphics[width=0.3\textwidth,angle=0]{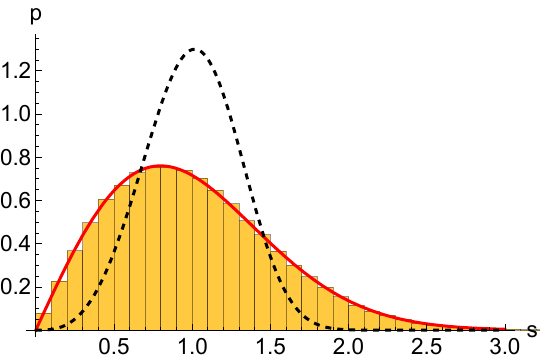}\\
	\begin{turn}{90}\hspace{40pt}{$T_\beta$}\end{turn}	
	\includegraphics[width=0.3\textwidth,angle=0]{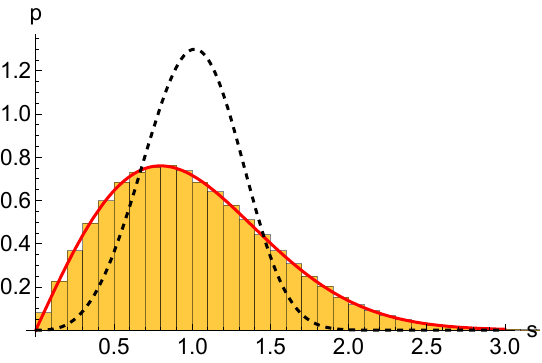}
	\includegraphics[width=0.3\textwidth,angle=0]{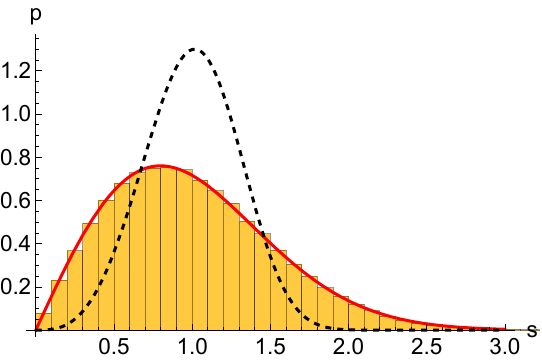}
	\includegraphics[width=0.3\textwidth,angle=0]{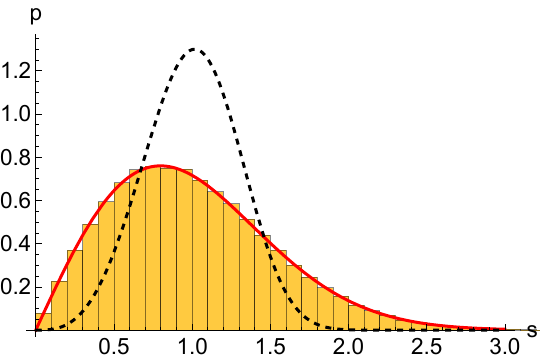}
	\caption[Cf. Figure 5 from \cite{AMPT}, \textit{Comparison of the NN statistics of the tridiagonal matrix ensembles $\widetilde{T}_\beta, S_\beta,T_\beta$ for various $\beta$ with the 2d Poisson and complex Ginibre analytics (and the statistics of class AII$^\dagger$)}]{We present histograms of the NN statistics for the ensembles $\widetilde{T}_\beta$, $S_\beta$, and $T_\beta$ from top to bottom row. For all ensembles we consider $\beta=1/2,2,1000$ in the plots from left to right. Additionally, we give the analytical known results for 2d Poisson ($\beta=0$) \eqref{eq-Poi2-NN} (red full) and for complex Ginibre ($\beta=2$) \eqref{eq-GinibreSpacingNN} (black dashed). We observe an agreement of the NN statistics of the ensembles $T_\beta$ and $S_\beta$ with 2d Poisson independently of $\beta$. The ensemble $\widetilde{T}_\beta$ shows a $\beta$-dependence, and it overshoots the maximum of the complex Ginibre ensemble for $\beta=1/2,2$, whereas it is in better agreement for $\beta=1000$. In the first row, we also plot the numerical distribution of the symmetry class AII$^\dag$ (dot-dashed blue) which agrees well with $\beta=1/2$ for $\widetilde{T}_\beta$. \cite[Cf. Figure 5]{AMPT}}
	\label{fig-nn}
\end{figure}
\noindent
We consider 500 matrices of size $N=5000$ for each of the three discussed tridiagonal matrix ensembles. For the general and symmetric matrix ensemble, we study the bulk of the spectrum, i.e. radii in the interval $ \left[0.1,0.6\right]\subset\left[0,\sqrt{2/e}\right]$ with $\sqrt{2/e}\approx0.858$ to stay away from the edge (cf. Figure \ref{fig-radial dist T n=5000}) and unfold the spectrum with the square root of the analytically known spectral density \eqref{eq-limiting density T,S} (cf. Section \ref{sssec-unfolding}). In both cases we find that the local NN spacing distribution (after rescaling the first moment to unity) looks for various $\beta=1/2,2,1000$ similar to the statistics of a 2d Poisson process of independent particles ($\beta=0$), see Eq. \eqref{eq-Poi2-NN} (red full curve in Figure \ref{fig-nn}). This is clearly not the NN spacing of a 2dCG at the inverse temperature $\beta$ (cf. Figure \ref{fig-2d-statistics} right) and we neither expected this nor understand this behaviour at the moment. For the non-symmetric model, we cut off the large peaks at the origin and the edge of the spectrum, i.e. consider radii in $\left[0.02,0.44\right]$ (cf. Figure \ref{fig-radial dist T tilde n=5000}) and find a different behaviour compared to the other ensembles. The maximum of the distribution is shifted to the right with respect to Poisson, and we observe $\beta$-dependence of the distributions. For comparison we plot the analytically known NN distribution \eqref{eq-GinibreSpacingNN} of the complex Ginibre ensemble ($\beta=2$) as a black dashed line. We observe a convergence towards the Ginibre statistics for increasing $\beta$ in the top row ($\widetilde{T}_\beta$), which we do not understand at the moment. Hence, the coupling of the eigenvectors plays an important role and needs further study in the future. In the first row, we also plot the numerical distribution of the symmetry class AII$^\dag$ (dot-dashed blue) which agrees well with $\beta=1/2$ for $\widetilde{T}_\beta$. We will come back to this comparison with the symmetry class of complex self-dual matrices in Section \ref{sec-universality symmetry classes}, where we discuss universality classes of non-Hermitian matrices.\\
Let us summarize the main results of this Section \ref{ssec-tridiag beta}. We study three tridiagonal non-Hermitian matrix ensembles, starting from the non-symmetric one $\widetilde{T}$ introduced by Mezzadri and Taylor in \cite{mezzadri taylor}, and generalise their proof techniques to symmetric $S$ and general tridiagonal $T$ matrix ensembles, see Theorems \ref{thm-roughbijection}, \ref{thm-JPDF tridiag}. The motivation behind the chosen matrix probability distributions is in each case to find a repulsion term of the form $|\Delta_N(\mathbf{z})|^\beta$ in terms of the complex eigenvalues and with some real positive parameter $\beta$. Due to the non-Hermiticity of the matrices, we have an additional contribution in the JPDF of eigenvalues, which represents the coupling to eigenvector components which cannot be integrated out so far. By application of the low temperature limit $\beta\gg 1$ and revisiting results from free probability, we calculate the limiting spectral density for each ensemble, see Theorem \ref{thm-density beta}. For the symmetric and the general tridiagonal matrices, we find a novel logarithmic density \eqref{eq-limiting density T,S}. For the less stable non-symmetric matrix $\widetilde{T}$, the leading contribution to the limiting spectral density is a Dirac-delta function at the origin \eqref{eq-limiting density T tilde}, which does not describe the full statistics, we observe from the numerics, cf. Figures \ref{fig-scatterplot T tilde n5000}, \ref{fig-radial dist T tilde n=5000}. We discuss these findings in view of the pseudospectrum in Section \ref{sssec-numerics}. Lastly, we numerical study the NN distribution for all tridiagonal matrix ensembles under consideration, finding a surprisingly a good agreement with the 2d Poisson statistics for the symmetric and the general ensemble. In the non-symmetric case, we find an overshooting of the maximum of the complex Ginibre statistics. We will revisit this behaviour after the discussion of universality classes of local statistics in non-Hermitian symmetry classes in the next Section \ref{sec-universality symmetry classes}.\cleardoublepage

\section{Three Universality Classes Amongst Non-Hermitian Symmetry Classes}\label{sec-universality symmetry classes}
After discussing the importance of (non-Hermitian) $\beta$-ensembles in the last sections, emphasizing the universality of random matrix statistics in the role of applications repeatedly, it seems natural to state the following questions: What are the non-Hermitian symmetry classes? Moreover, which of them share the same universal (local) statistics? \\
Hence, let us start this section again with the statement of a goal: Study the (simplest) representatives of non-Hermitian local statistics.\\
In Section \ref{ssec-intro symmetry classes} we start with an introduction into the study of (non-Hermitian) symmetry classes. We discuss the various symmetry operations and repeat the finding of 38 non-Hermitian classes by Kawabata et al \cite{non-h classes}. Next, we investigate the question of universality classes of local statistics. Following numerics and heuristic arguments Hamazaki et al \cite{Hamzaki et al} find three universal NN statistics in the bulk of the spectrum. We discuss the split of the 38 classes into these three conjectured universality classes.\\
In the next Section \ref{ssec-numerics symmetry classes} we study the three simplest representatives of the local statistics from the previous section in detail. They are class A of complex Ginibre, class AI$^\dagger$ of complex symmetric and class AII$^\dagger$ of complex self-dual matrices, respectively. We present known analytical results, which are richest for class A, recently expanded significantly for class AI$^\dagger$ and very rare for class AII$^\dagger$. Numerically, we study the NN and NNN statistics of all three classes in the bulk of the spectrum, confirming clearly distinct behaviours. While class A corresponds exactly to a 2dCG with $\beta=2$, as discussed previously, we establish a very good agreement of local statistics of class AI$^\dagger$ with the numerics of a 2dCG at $\beta=1.4$ and similarly for class AII$^\dagger$ with $\beta=2.6$. The introduction of this effective description was mainly contributed by the author providing the necessary comparison in \cite{AMP,ABCPRS}. It can be used to interpret the local repulsion of the three symmetry classes. We present also another numerical tool in the study of open quantum systems: The so-called complex spacing ratio and discuss its findings in the bulk and the edge of the spectrum. As an analytical tool to investigate the distinct behaviour of the three representatives, we study the expectation value of characteristic polynomials. Using supersymmetric techniques allows for explicit results while missing the JPDF for classes AI$^\dagger$ and AII$^\dagger$. Here, we will focus on the $k=1$ results, which were mainly contributed by the author.\\
We end with the discussion of applications for the universal statistics of the non-Hermitian symmetry classes: First, we present a spin chain model with dissipative disorder which shows both integrability and Hermiticity breaking at different scales. The comparison between the eigenvalue data of the model with the possible non-Hermitian random matrix ensembles such as the surmise from Section \ref{ssec-surmise}, numerical generated 2dCG data and the symmetry class AI$^\dagger$ of the Hamiltonian was mainly contributed by the author. Second, we revisit the NN statistics of the tridiagonal $\beta$-ensembles from the Section \ref{sssec-numerics} under the view of universal behaviour in 2d. We leave the interpretation of the findings for future research.
\subsection{Introduction to the Study of Non-Hermitian Symmetry Classes}\label{ssec-intro symmetry classes}
We introduce the classification of non-Hermitian matrices/ Hamiltonians raised by Kawabata et al \cite{non-h classes}. The study of symmetry classes has a long history both from a mathematical and physical point of view. The reason is very simple and is the backbone of each (mathematical) physicist: We use symmetries of a model/ space to simplify our description of it. It is this principle which is taught in a physics study first: Look for symmetries of your system and use them to make your calculations/ equations simpler. Another way to view this is to find (equivalence) classes with respect to some (symmetry) action, allowing for a simultaneous description of these larger sets. We recall that the invariance under a group action is one main ingredient for the construction of matrix ensembles, as discussed in the introduction Section \ref{sec-intro}. We will present here an overview of the non-Hermitian symmetry classes, revisiting the Hermitian ones. While the Cartan classification has its origin in works by Cartan in 1926 and 1927 \cite{Cartan 1926,Cartan 1927}, also the study of symmetry classes in quantum mechanics and random matrices has a rich history, cf. the books by Wigner \cite{book Wigner} and Weyl \cite{book weyl}. We recall from the introduction Section \ref{sec-intro} that random matrices became of increasing interest due to the application of the spacing distribution of the real eigenvalues of a Hermitian random matrix to level spacings of nuclei in 1957 by Wigner \cite{Wigner57}. Shortly thereafter, Dyson proposed in 1962 what is nowadays known as \textit{Dyson's threefold way} \cite{dyson62}. The key message is the following: Dyson identifies the time-reversal symmetry (TRS) (based on the Frobenius theorem over the real number field), which we will define below, as key to find distinct local statistics.
\begin{Remark}
	We distinguished so far between Hermitian matrices with $H=H^\dagger$ and non-Hermitian matrices with $J\neq J^\dagger$. From these definitions it follows that $H$ has a priori an inner symmetry of the matrix elements while in $J$ the elements are independent. However, this leaves still space for further symmetry relations which are compatible with the (non-)Hermiticity condition. We will see that the missing inner symmetry of non-Hermitian matrices enables us to find 38 symmetry classes, while we are restricted to 10 different classes for Hermitian matrices.
\end{Remark}
\begin{Remark}
	We remark that Kawabata et al \cite{non-h classes} were not the first investigate into the classification of non-Hermitian symmetries in 2019. Until that date the work by Bernard and Le Clair from 2002 \cite{Bernard LeClair} was accepted as the standard reference on this subject with a finding of 43 symmetry classes. Shortly after \cite{non-h classes} appeared, including a detailed list of the over- and undercounting of the classes in the initial work \cite{Bernard LeClair}, Bernard and LeClair published a revised version agreeing with the total number of 38 symmetry classes. However, the two works are (to the best of our knowledge) still not in agreement/ same language of symmetry classification: While Bernard and LeClair state in \cite{Bernard LeClair} that a misprint in the tables yield to the number of 43 classes, the group of Kawabata et al argue in \cite{non-h classes} that in the initial work of Bernard and LeClair 10 classes are overcounted while 5 classes are not considered. We will not judge these disagreeing arguments, but we would like to point the reader to the fact that there is still a discrepancy in the two classifications.
\end{Remark}\clearpage
\noindent
We start now with a list of the various possible symmetries for a (non-Hermitian) Hamiltonian, following \cite{non-h classes} closely. We restrict ourselves to the mathematical formulation of the symmetry relations rather than the physical interpretation. However, simply from the naming of the operations it is clear that strong relations to symmetries of physical systems exist. Let $J$ be a generally non-Hermitian Hamiltonian/ matrix of dimension $N\times N$.
\begin{itemize}
	\item \textbf{Particle Hole Symmetry (PHS)}: Let $\mathcal{C}_-, \mathcal{T}_-$ be unitary matrices, i.e. $\mathcal{C}_-\mathcal{C}_-^\dagger=\mathcal{C}_-^\dagger\mathcal{C}_-=\mathds{1}_N$ and $\mathcal{T}_-\mathcal{T}_-^\dagger=\mathcal{T}_-^\dagger\mathcal{T}_-=\mathds{1}_N$. 
	\begin{itemize}
		\item We define the \textbf{PHS} of $J$ as
		\begin{align} \label{eq-PHS}
			\mathcal{C}_-^{-1}J^t\mathcal{C}_-=-J,\ \ \mathcal{C}_-\mathcal{C}_-^*=\pm \mathds{1}_N.
		\end{align}
	\item We define the \textbf{PHS$^{\boldsymbol{\dagger}}$} of $J$ as\footnote{Note that this is the standard definition of the PHS symmetry for Hermitian Hamiltonians, cf. \cite{Altland Zirnbauer}. We follow the naming used in \cite{non-h classes}, which differs from the convention used in physics literature.}
	\begin{align}\label{eq-PHS dag}
		\mathcal{T}_-^{-1}J^*\mathcal{T}_-=-J,\ \ \mathcal{T}_-\mathcal{T}_-^*=\pm \mathds{1}_N.
	\end{align}
\end{itemize}
We have PHS=PHS$^\dagger$ for a Hermitian $J=J^\dagger$. However, the non-Hermiticity of $J\neq J^\dagger$ yields to an in-equivalence of the two relations \eqref{eq-PHS} and \eqref{eq-PHS dag}. We observe that for both, we have the three cases that the symmetry is absent (0) or is fulfilled with $\mathcal{C}_-\mathcal{C}_-^*$, respectively $\mathcal{T}_-\mathcal{T}_-^*$ equal (+1) or (-1) multiple of the unity matrix $\mathds{1}_N$, which we will suppress in this identification.
\item \textbf{Time Reversal Symmetry (TRS)}: Let $\mathcal{C}_+, \mathcal{T}_+$ be unitary matrices, i.e. $\mathcal{C}_+\mathcal{C}_+^\dagger=\mathcal{C}_+^\dagger\mathcal{C}_+=\mathds{1}_N$ and $\mathcal{T}_+\mathcal{T}_+^\dagger=\mathcal{T}_+^\dagger\mathcal{T}_+=\mathds{1}_N$. 
\begin{itemize}
	\item We define the \textbf{TRS} of $J$ as
	\begin{align} \label{eq-TRS}
		\mathcal{T}_+^{-1}J^*\mathcal{T}_+=J,\ \ \mathcal{T}_+\mathcal{T}_+^*=\pm \mathds{1}_N.
	\end{align}
	\item We define the \textbf{TRS$^{\boldsymbol{\dagger}}$} of $J$ as
	\begin{align}\label{eq-TRS dag}
		\mathcal{C}_+^{-1}J^t\mathcal{C}_+=J,\ \ \mathcal{C}_-\mathcal{C}_-^*=\pm \mathds{1}_N.
	\end{align}
	\end{itemize}
	Again, we indicate the symmetry of a Hamiltonian with respect to TRS or TRS$^\dagger$ as 0, +1, -1.

\item \textbf{Chiral Symmetry (CS)}: Let $\Gamma, \mathcal{S}$ be unitary matrices, i.e. $\Gamma \Gamma^\dagger=\Gamma^\dagger \Gamma=\mathds{1}_N$ and $\mathcal{S}\mathcal{S}^\dagger=\mathcal{S}^\dagger \mathcal{S}=\mathds{1}_N$. 
\begin{itemize}
	\item We define the \textbf{CS} of $J$ as
	\begin{align} \label{eq-CS}
		\Gamma^{-1}J^\dagger \Gamma=-J,\ \ \Gamma^2= \mathds{1}_N.
	\end{align}
	\item We define the \textbf{CS$^{\boldsymbol{\dagger}}$=SLS} of $J$ as
	\begin{align}\label{eq-CS dag}
		\mathcal{S}^{-1}J\mathcal{S}=-J,\ \ \mathcal{S}^2= \mathds{1}_N.
	\end{align}
\end{itemize}
For CS, we indicate the respective symmetry by 0,1. The chiral symmetry CS$^\dagger$ can also be viewed as a sublattice symmetry (SLS) yielding new symmetry classes respective to the (anti-)commutation relations between the unitary operations defining a symmetry relation.
\end{itemize}
We have that the non-Hermiticity of $J$ ramifies symmetry operations $S\neq S^\dagger$ for $S=\{\text{PHS, TRS, CS}\}$. However, it also happens that we find a unification of symmetry relations: Let $J$ respect the TRS \eqref{eq-TRS} with unitary matrix $\mathcal{T}_+$ and PHS$^\dagger$ \eqref{eq-PHS dag} with unitary matrix $\mathcal{T}_-$. The matrices $\mathcal{T}_+$ and $\mathcal{T}_-$ are a priori independent of each other. Although, if $J$ respects TRS, it follows directly that $iJ$ respects PHS$^\dagger$ with $\mathcal{T}_+=\mathcal{T}_-$ due to the complex conjugation in the symmetry relations \eqref{eq-TRS}, \eqref{eq-PHS dag}. We conclude that we have for non-Hermitian $J\neq J^\dagger$: \textbf{TRS=PHS$^{\boldsymbol{\dagger}}$}. This is not true for PHS and TRS$^\dagger$, where transposition is employed.\\ The 16 different symmetry classes, which are determined by the symmetries TRS, PHS, CS, TRS$^\dagger$ and PHS$^\dagger$ are listed in Table \ref{tab-AZ} below. The first 2+8=10 classes, which invoke only TRS, PHS and CS are equivalent to the 10 symmetry classes for Hermitian Hamiltonians due to the classification by Altland and Zirnbauer \cite{Altland Zirnbauer}. The classes AI and D$^\dagger$, respectively AII and C$^\dagger$ are equivalent as TRS=PHS$^\dagger$, indicated by the grey boxes around D$^\dagger$ and $C^\dagger$, which do not contribute to the counting of independent symmetry classes.
\begin{table}[t]
	\centering
	\caption[Table I from \cite{non-h classes}, \textit{AZ and AZ$^\dagger$ non-Hermitian symmetry classes incorporating TRS, PHS, TRS$^\dagger$, PHS$^\dagger$ and CS}]{We give a list of the AZ and AZ$^{\dag}$ non-Hermitian symmetry classes incorporating TRS, PHS, TRS$^\dagger$, PHS$^\dagger$ and CS. We count 10 classes for complex and real AZ and another 6 for real AZ$^\dagger$, where the classes D$^\dagger$ and C$^\dagger$ which are equivalent to AI, respectively AII are boxed as they do not count as independent classes. Moreover, we highlight classes with different values of the TRS$^\dagger$ as: bold for \textbf{0}, red for \textcolor{red}{+1} and blue for \textcolor{blue}{-1}, finding 10, 3, 3 classes respectively. \cite[Table I]{non-h classes}}
	\label{tab-AZ}
	\begin{tabular}{ccccccc} \hline 
		\multicolumn{2}{c}{~Symmetry class~} & TRS & PHS & TRS$^\dagger$ & PHS$^\dagger$ & CS \\ \hline
		\multirow{2}{*}{Complex AZ} 
		& A & 0 & 0 & \textbf{0} & 0 & 0 \\
		& AIII & 0 & 0 & \textbf{0} & 0 & 1 \\ \hline
		\multirow{9}{*}{Real AZ} 
		& AI & +1 & 0 & \textbf{0} & 0 & 0 \\
		& BDI & +1 & +1 & \textbf{0} & 0 & 1 \\
		& D & 0 & +1 & \textbf{0} & 0 & 0 \\
		& DIII & -1 & +1 & \textbf{0} & 0 & 1 \\
		& AII & -1 & 0 & \textbf{0} & 0 & 0 \\
		& CII & -1 & -1 & \textbf{0} & 0 & 1 \\
		& C & 0 & -1 & \textbf{0} & 0 & 0 \\
		& CI & +1 & -1 & \textbf{0} & 0 & 1 \\  \hline 
		\multirow{9}{*}{Real AZ$^{\dag}$}  
		&AI$^{\dag}$ & 0 & 0 & \textcolor{red}{+1} & 0 & 0 \\
		& BDI$^{\dag}$ & 0 & 0 &\textcolor{red}{+1} & +1 & 1 \\
		& \boxit{2.75in} D$^\dagger$ & 0 & 0  & 0  & +1  & 0  \\
		& DIII$^{\dag}$ & 0 & 0 & \textcolor{blue}{-1} & +1 & 1 \\
		& AII$^{\dag}$ & 0 & 0 & \textcolor{blue}{-1} & 0 & 0 \\
		& CII$^{\dag}$ & 0 & 0 &\textcolor{blue}{-1} & -1 & 1 \\
		& \boxit{2.75in} C$^{\dag}$ & 0  & 0  & 0  & -1  & 0 \\
		& CI$^{\dag}$ & 0 & 0 & \textcolor{red}{+1} & -1 & 1 \\ \hline 
	\end{tabular}
\end{table}
The remaining 22 classes arise by considering additionally the SLS from \eqref{eq-CS dag}. We remark that one can equivalently choose pseudo-Hermiticity as an additional internal symmetry, with condition
\begin{align}
	\eta J^\dagger \eta^{-1}=J,\ \ \eta^2=\mathds{1}_N,
\end{align}
with $\eta$ unitary and Hermitian, i.e. $\eta \eta^\dagger=\eta^\dagger \eta=\mathds{1}_N$ and $\eta=\eta^\dagger$. We refer to \cite[Table XIV]{non-h classes} for details on this and discuss SLS as additional symmetry here. The remaining symmetry classes are defined with respect to the (anti-)commutation relations between matrices generating the symmetries of the corresponding class. For the complex AZ from Table \ref{tab-AZ}, we consider by $\Gamma \mathcal{S}_\pm=\pm \mathcal{S}_\pm \Gamma$ the commutation (+) respectively anti-commutation (-) relation with the CS generator $\Gamma$. We find three new classes in Table \ref{tab-complex AZ + SLS}, where we find one new class for class A in the absence of CS symmetry and two new ones for class AIII. As these three classes will have also TRS$^\dagger$=0, we highlight them in bold for later purposes. Next, we consider the real AZ from Table \ref{tab-AZ} with additional SLS. The subscript of $\mathcal{S}_{\pm}$ specifies again the commutation (+), respectively anti-commutation (-) relation with the TRS or PHS respected by the corresponding symmetry class. If a class respects both TRS and PHS, the first subscript of $\mathcal{S}$ specifies the relation to the TRS generator and the second to the PHS one. Not all of these 24 classes are independent of each other. The classes AI, BDI and CII with SLS anti-commuting with TRS are equivalent to classes AII, DIII, and CI, respectively, with SLS respect to the same anti-commutation relations. Hence, we find 19 new classes in Table \ref{tab-real AZ + SLS}. Alternatively, one can also consider the real AZ$^\dagger$ symmetry classes from Table \ref{tab-AZ} and add SLS. The equivalence between the real AZ classes with SLS and the real AZ$^\dagger$ classes with SLS is displayed in Table \ref{tab-equiv AZ AZdag with SLS} following the discussion in \cite[Appendix A]{non-h classes}.\\
We conclude that the 38 non-Hermitian symmetry classes due to Kawabata et al \cite{non-h classes} are given by the 16 inequivalent classes in Table \ref{tab-AZ}, the three additional classes for complex AZ with SLS in Table \ref{tab-complex AZ + SLS} and the 19 additional classes for real AZ/ AZ$^\dagger$ with SLS given in Tables \ref{tab-real AZ + SLS}, \ref{tab-equiv AZ AZdag with SLS} respectively.

\begin{figure}[h]
		\centering
		\hspace{-20pt} NN \hspace{210pt} NNN \hfill\\
		\includegraphics[width=0.49\linewidth,angle=0]{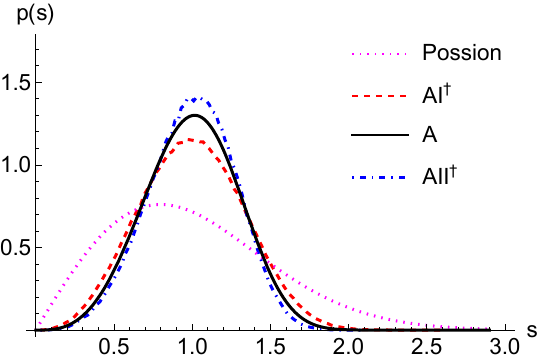}
		\includegraphics[width=0.49\linewidth,angle=0]{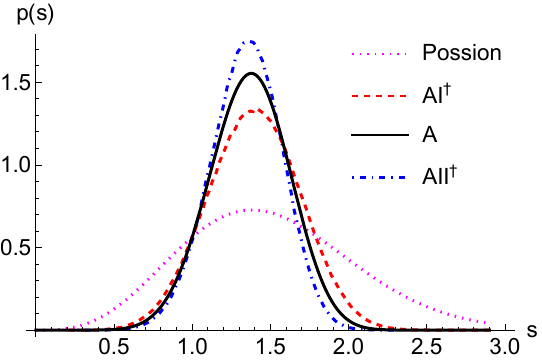}
		\caption[Figure 6 from \cite{AMPT} and Mathematica Plot, \textit{Comparison of the NN and NNN statistics of symmetry classes A, AI$^\dagger$, AII$^\dagger$ and a 2d Poisson process}]{We compare the NN (left) and NNN (right) distributions of the numerically generated statistics of 500 matrices of size $N=5000$, respectively $2N=5000$ of classes AI$^\dag$ (red dashed), respectively AII$^\dag$ (blue dot-dashed) with the analytically known results of class A (black full, Eqs. \eqref{eq-GinibreSpacingNN}, \eqref{eq-GinibreSpacingNNN}) and a 2d Poisson process (magenta dotted, Eqs. \eqref{eq-Poi2-NN}, \eqref{eq-Poi2-NNN}) in the bulk. We normalized the distribution with respect to the first moment of the NN statistics each. \cite[Cf. Figure 6]{AMPT}
		}
		\label{fig-NN NNN symmetry classes}
	\end{figure}

\begin{table}[t]
	\centering
	\caption[Table XI from \cite{non-h classes}, \textit{Complex AZ symmetry classes with additional SLS}]{We give the list of possible types [t=0,1 (mod 2)] of the complex AZ symmetry classes with additional SLS \eqref{eq-CS dag}, respect to the (anti-)commutation relation of the CS and SLS=CS$^\dagger$ generators. \cite[Table XI]{non-h classes}}
	\label{tab-complex AZ + SLS}
	\small
	\begin{tabular}{ccc} \hline 
	 AZ class & $t=0$ & $t=1$ \\ \hline
	 A & & $\boldsymbol{\mathcal{S}}$ \\
		 AIII & $\boldsymbol{\mathcal{S}_{+}}$ & $\boldsymbol{\mathcal{S}_{-}}$ \\ \hline
	\end{tabular}
\end{table}
\begin{table}[h]
	\centering
	\caption[Table XII from \cite{non-h classes}, \textit{Real AZ symmetry classes with additional SLS}]{We give the list of possible types [t=0,1,2, 3 (mod 4)] of the real AZ symmetry classes with additional SLS \eqref{eq-CS dag}, respect to the (anti-)commutation relation of the TRS or/ and PHS generators with SLS. Elements in grey boxes do not count as inequivalent classes. \cite[Table XII]{non-h classes}}
	\label{tab-real AZ + SLS}
	\small
	\begin{tabular}{ccccc} \hline 
		~AZ class~ & ~$t=0$~ & ~$t=1$~ & ~$t=2$~ & ~$t=3$~ \\ \hline
	 AI &  & $\mathcal{S}_{-}$ &  & $\mathcal{S}_{+}$ \\
		 BDI & $\mathcal{S}_{++}$ & $\mathcal{S}_{-+}$ & $\mathcal{S}_{--}$ & $\mathcal{S}_{+-}$ \\
		 D &  & $\mathcal{S}_{+}$ &  & $\mathcal{S}_{-}$ \\
		 DIII &\boxit{0.9in} $\mathcal{S}_{--}$ & $\mathcal{S}_{-+}$ & $\mathcal{S}_{++}$ & $\mathcal{S}_{+-}$ \\
		  AII &  & \boxit{0.2in}$\mathcal{S}_{-}$ &  & $\mathcal{S}_{+}$ \\
	 CII & $\mathcal{S}_{++}$ & $\mathcal{S}_{-+}$ & $\mathcal{S}_{--}$ & $\mathcal{S}_{+-}$ \\
		 C &  & $\mathcal{S}_{+}$ &  & $\mathcal{S}_{-}$ \\
		 CI & \boxit{0.9in}$\mathcal{S}_{--}$ & $\mathcal{S}_{-+}$ & $S_{++}$ & $\mathcal{S}_{+-}$ \\ \hline 
	\end{tabular}
\end{table}
\begin{table}[h]
	\centering
	\caption[Table XIII from \cite{non-h classes}, \textit{Equivalence between real AZ symmetry classes with SLS and real AZ$^\dagger$ classes with SLS}]{We present the equivalence between real AZ symmetry classes with SLS and real AZ$^\dagger$ classes with SLS, respect to the commutation (+) or anti-commutation (-) relation of SLS to TRS/TRS$^\dagger$ or/ and PHS/PHS$^\dagger$. In the case of two subscripts, the first specifies the relation to TRS/TRS$^\dagger$. Elements in grey boxes do not count as inequivalent classes. Classes which respect TRS$^\dagger$ with \textcolor{blue}{+1}/ \textcolor{red}{-1} are marked in blue/ red respectively. \cite[Table XIII]{non-h classes}}
	\label{tab-equiv AZ AZdag with SLS}
	\small
	\begin{tabular}{ccccccc} \hline 
		AZ$^\dagger$ class & $\mathcal{S}_+$ & $\mathcal{S}_-$ & $\mathcal{S}_{++}$ & $\mathcal{S}_{+-}$ & $\mathcal{S}_{-+}$& $\mathcal{S}_{--}$\\ \hline
		AI$^\dagger$ & \textcolor{blue}{D + $\mathcal{S}_+$} & \textcolor{red}{C + $\mathcal{S}_-$} &  & & & \\
		BDI$^\dagger$ &  &  & \textcolor{blue}{BDI + $\mathcal{S}_{++}$} & \boxit{0.77in}DIII + $\mathcal{S}_{-+}$& \textcolor{blue}{CI + $\mathcal{S}_{+-}$} &\textcolor{red}{CII + $\mathcal{S}_{--}$}\\
		D$^\dagger$ & \textcolor{blue}{AI + $\mathcal{S}_{+}$} & \textcolor{red}{AII + $\mathcal{S}_{-}$} &  & & &\\
		DIII$^\dagger$ & & & \textcolor{red}{CI + $\mathcal{S}_{++}$} & \textcolor{blue}{CII + $\mathcal{S}_{-+}$}   & \textcolor{red}{BDI + $\mathcal{S}_{+-}$} & \textcolor{blue}{DIII + $\mathcal{S}_{--}$} \\
		AII$^\dagger$ & \textcolor{red}{C + $\mathcal{S}_{+}$} & \textcolor{blue}{D + $\mathcal{S}_{-}$} &  & & & \\
		CII$^\dagger$ &  &  & \textcolor{red}{CII + $\mathcal{S}_{++}$} & \boxit{0.68in}CI + $\mathcal{S}_{-+}$ & \textcolor{red}{DIII + $\mathcal{S}_{+-}$} & \textcolor{blue}{BDI + $\mathcal{S}_{--}$}\\
		C$^\dagger$ & \boxit{0.66in}AII + $\mathcal{S}_{+}$ & \textcolor{red}{AI + $\mathcal{S}_{-}$} &  & & &\\
		CI$^\dagger$ &  &  & \boxit{0.77in}DIII + $\mathcal{S}_{++}$ &\textcolor{red}{BDI + $\mathcal{S}_{-+}$} & \textcolor{blue}{CII + $\mathcal{S}_{+-}$} & \boxit{0.68in}CI + $\mathcal{S}_{--}$ \\ \hline 
	\end{tabular}
\end{table}
Having the 38-fold classification of non-Hermitian Hamiltonian by hand, we can now raise the question of universality. To be more precise: We are looking for universal local statistics of the eigenvalues of these Hamiltonians/ matrices.\footnote{Recalling the first applications of random matrices to physics, where scattering data of nuclei were considered, it is rather natural to try to find universal behaviour on the local repulsion of eigenvalues of random matrices.} We discussed in the beginning of Section \ref{sec-non Herm beta} that the NN spacing distribution is a good measure to distinguish symmetry classes due to the BGS and GHS conjectures \cite{BGS,BGS 1, GHS}. Indeed, shortly after the classification by Kawabata et al appeared, Hamazaki et al \cite{Hamzaki et al} investigate the question of universal local statistics in the bulk of the spectrum for 9 of the 38 symmetry classes, finding three distinct behaviours. Remarkably, the split into these three universality classes can (so far) not be understood in a similar way to the splitting for Hermitian symmetry classes. Here the real, complex and quaternion Hermitian matrices are representatives of the three irreducible representations respect to unitary symmetries, a fact which is widely known in representation theory, cf. \cite{dyson62}, \cite[English version, Chapter 26]{book Wigner} and references therein. Without going into the details of the analysis for the Hermitian case, we would like to highlight, that the universality here is determined by the TRS which is in a one-to-one-correspondence to these irreducible representations. Dyson made the connection to the GO/U/SE (see Eq. \eqref{eq-jpdf 1d Dyson}) and their statistics. For the case of non-Hermitian matrices, the case is more evolved. First, we note what Dyson did already 1962, i.e. that the Frobenius theorem (which provides a splitting into real, complex and quaternion elements correspondingly) does not apply to the complex (non-Hermitian) setting \cite{dyson62}. Therefore, let us revisit the heuristic argument by Hamazaki et al \cite{Hamzaki et al} for a universality of local statistics in non-Hermitian symmetry classes. The authors state that the latter is determined by TRS$^\dagger$ rather than the TRS as in the Hermitian case for the following reason. In the Dyson's three-fold way the TRS yields to three universality classes. However, turning to the Ginibre ensembles they share the same statistics in the bulk as the TRS creates only non-local correlations between complex-conjugate pairs of eigenvalues due to the employed complex conjugation in Eq. \eqref{eq-TRS}, but does not change the repulsion of neighbouring eigenvalues away from the real axis. The authors of \cite{Hamzaki et al} argue that in non-Hermitian matrices the complex-conjugate symmetry TRS$^\dagger$ determines the repulsive interaction of neighbouring eigenvalues, i.e. the TRS$^\dagger$ imposes symmetry of left- and right-eigenvectors, whereas the TRS imposes symmetry of complex-conjugate pairs of eigenvectors. Using this interpretation, we can count the numbers of symmetry classes in the three proposed universality classes, determined via $TRS^\dagger=\textbf{0},\textcolor{red}{+1}, \textcolor{blue}{-1}$. We count the members of each class in Tables \ref{tab-AZ}, \ref{tab-complex AZ + SLS} and \ref{tab-equiv AZ AZdag with SLS}, where we need to consider the equivalence between the real AZ and real AZ$^\dagger$ symmetry classes with SLS, as only the latter incorporate TRS$^\dagger$. We conclude that we have \textbf{13} classes with TRS$^\dagger=\textbf{0}$, \textcolor{red}{13} with TRS$^\dagger=\textcolor{red}{+1}$ and \textcolor{blue}{12} with TRS$^\dagger=\textcolor{blue}{-1}$. The three Ginibre ensembles of real/ complex/ quaternion elements share the same TRS$^\dagger=\textbf{0}$, which agrees with the shared local NN statistics of them. Hamazaki et al revisit the 9 symmetry classes (A, AI, AIII, C, D, AI$^\dagger$, AII$^\dagger$ and AIII out of Tables \ref{tab-AZ}, \ref{tab-complex AZ + SLS}) which are symmetrized by at most one symmetry \cite{Hamzaki et al}, finding only three distinct NN statistics in the bulk of the spectrum, cf. Figure \ref{fig-NN NNN symmetry classes} left. The three simplest representatives are:
\begin{itemize}
	\item Class A of complex Ginibre matrices, with
	\begin{align}\label{eq-def class A}
		J\neq J^\dagger.
	\end{align} 
	\item Class AI$^\dagger$ of complex symmetric matrices, with 
	\begin{align}\label{eq-def class AI dag}
		J=J^t.
	\end{align}
	\item Class AII$^\dagger$ of complex self-dual matrices, with 
	\begin{align}\label{eq-def class AII dag}
		J=\Sigma J^t\Sigma \text{ , where }  \Sigma=\begin{pmatrix}
			0& -i\mathds{1}_{N}\\
			i\mathds{1}_{N} & 0
		\end{pmatrix},
	\end{align} for $J\in \mathbb{C}^{2N\times 2N}$.
\end{itemize}
We display the NN (left) and NNN (right) statistics in the bulk for the three classes A (black full, analytics), AI$^\dagger$ (red dashed) and AII$^\dagger$ (blue dot-dashed) as well as the 2d Poisson statistics (magenta dotted, analytics) in Figure \ref{fig-NN NNN symmetry classes}. They show a clear distinction both in the NN and NNN distributions. Moreover, we observe a relative ordering of the position and the height of the maxima, with the lowest and left-most maximum for Poisson ($\beta=0$), followed by class AI$^\dagger$, A and AII$^\dagger$.
\begin{Remark}
	The ensemble of complex self-dual matrices AII$^\dagger$ was first studied by Hastings in 1999 \cite{hastings}, where one motivation was the double degeneracy of the eigenvalues compared to ensembles of anti-Hermitian matrices, where the eigenvalues come in pairs of $\pm z$. This degeneracy of self-dual matrices is a well-known fact, which was used for instance also by Mehta \cite[Chapter 3.2]{mehta}. In view of the non-Hermitian symmetry classes, the degeneracy arise from respecting TRS$^\dagger$ \eqref{eq-TRS dag} with $\mathcal{C}_+\mathcal{C}_+^*=-\mathds{1}_N$ as follows (cf. \cite{non-h classes}): Let $J=\mathcal{C}_+J^t\mathcal{C}_+^{-1}$ with $\mathcal{C}_+$ unitary and $\mathcal{C}_+\mathcal{C}_+^*=-\mathds{1}_N$. We denote the complex eigenvalue $z$ of $J$ corresponding to the right- (left-)eigenvector $\mathbf{R}$ ($\mathbf{L}^t$), i.e. $J\mathbf{R}=z\mathbf{R}$ and $\mathbf{L}^tJ=\mathbf{L}^tz$. Using the symmetry of $J$, we find
	\begin{align*}
		J\mathcal{C}_+ \mathbf{L}=\mathcal{C}_+J^t\mathbf{L}=\mathcal{C}_+E\mathbf{L}=E\mathcal{C}_+ \mathbf{L},
	\end{align*}
i.e. $\mathcal{C}_+\mathbf{L}$ is also an eigenvector of $J$ to the same eigenvalue $z$. Using the property TRS$^\dagger$=-1, i.e. $\mathcal{C}_+^t=-\mathcal{C}_+$, we have
\begin{align*}
	\mathbf{L}^t\mathcal{C}_+\mathbf{L}=\mathbf{L}^t\mathcal{C}_+^t\mathbf{L}=-\mathbf{L}^t\mathcal{C}_+\mathbf{L},
\end{align*}
where we used the transposition symmetry of this scalar in the first step. It follows that $\mathbf{L}^t$ and $\mathcal{C}_+\mathbf{L}$ are bi-orthogonal to each other and linearly independent. Hence, the eigenvalues are all at least double degenerate.\footnote{Similarly, one finds for the symplectic Ginibre ensemble in class AII, that the eigenvalues come in complex conjugate pairs.}
\end{Remark}
\begin{figure}[h]
	\includegraphics[width=0.75\linewidth]{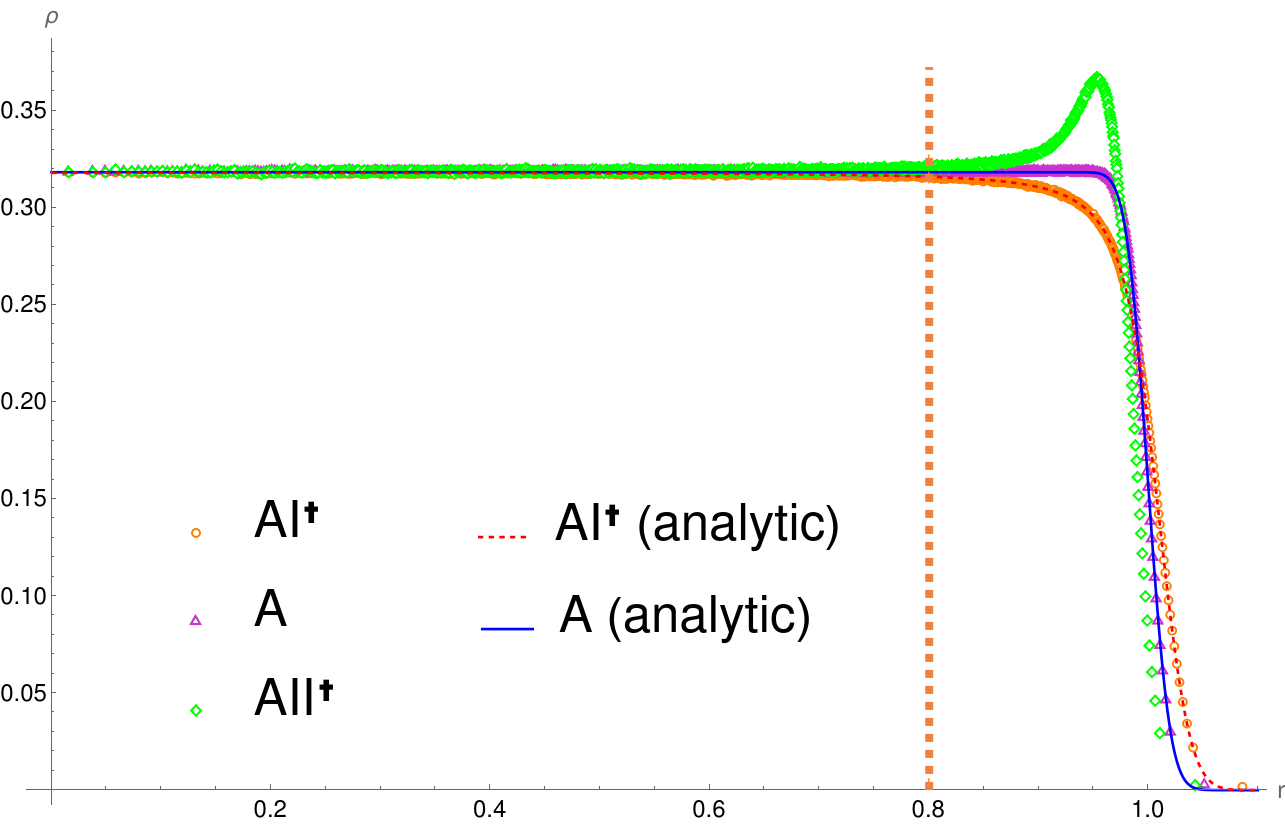}
	\caption[Part of Figure 4 from \cite{AAAMPRW}, \textit{Radial spectral density of the symmetry classes A, AI$^\dagger$ and AII$^\dagger$}]{The radial spectral density of class A \eqref{eq-R1A} (blue full curve: analytical; purple triangle: numerical), class AI$^\dagger$ \eqref{eq-R1AI+} (red dashed curve: analytical; orange circle: numerical), both at $N=1024$, and of class AII$^\dagger$ (green diamond: numerical only) with $2N=1024$, with $720\,000$ samples for all ensembles.
	For later purposes, the orange dotted vertical line marks the radius 
	 $r_b=0.8$. \cite[Cf. Figure 4]{AAAMPRW}
	}
	\label{fig-radial density plot symmetry classes}
\end{figure}\noindent
We can now benefit from the previous Section \ref{sec-non Herm beta}: First, we recall the idea of testing $N=2$ results to large $N$ data (cf. Section \ref{ssec-surmise}). Finding, that the exact $N=2$ results are (as for the complex Ginibre class A) are not good approximations (cf. \cite{AMP} for details), we turn our view back to the 2dCG description, which is exact for class A with $\beta=2$. Moreover, we recall from Section \ref{ssec-point processes}, Figure \ref{fig-2d-statistics} that NN distributions for the 2dCG model show a movement of the maximum of the distribution from left to right as well as an increase of its height in $\beta$. Comparing to Figure \ref{fig-NN NNN symmetry classes} it seems to be worth a try to compare AI$^\dagger$ with a 2dCG with $\beta<2$ and AII$^\dagger$ with a 2dCG for $\beta>2$. Before we investigate this in the next Section \ref{ssec-numerics symmetry classes}, we collect what is known on spectral density and the NN and NNN distributions of the three classes A, AI$^\dagger$ and AII$^\dagger$ analytically. We follow here the work \cite{AAAMPRW} closely, which was contributed by the author. We repeat from Section \ref{ssec-tridiag beta} that complex eigenvalues in class A represent a determinantal point process, which is well described by the universal exponential Ginibre kernel in the bulk, and the complementary error function kernel (also called Faddeeva plasma kernel) at the edge of the spectrum, both known at finite $N$ and in the limit of infinite matrix dimension $N\to\infty$. Closed form expressions exist for the gap probabilities in terms of the Fredholm determinant of these two kernels, which are most explicitly known in the bulk \cite{GHS}. This leads to explicit expressions for the NN and NNN spacing distribution \cite{GHS}, which we presented earlier in Eqs. \eqref{eq-GinibreSpacingNN} and \eqref{eq-GinibreSpacingNNN}. For convenience of the reader, we give here the spectral density \cite{ginibre}
\begin{align}
	R^{\textrm{A}}_{N}(|z|=r)=\frac{1}{\pi} e^{-Nr^2}\sum_{j=0}^{N-1}\frac{(Nr)^{2j}}{j!}=\frac{\Gamma(N,Nr^2)}{\pi \Gamma(N)},
	\label{eq-R1A}
\end{align} which is normalised to integrate to unity. For a recent review on the Ginibre and related ensembles we refer again to \cite{book byun forrester}.\\
In classes AI$^\dag$ and AII$^\dag$ no such determinantal expressions are known, or even expected, that would at least in principle allow to formulate NN spacing statistics in terms of Fredholm determinants or Pfaffians. In class AI$^\dag$ the spectral density was obtained most recently \cite{complex sym Gernot} and reads 
\begin{align}
	R^{\textrm{AI}^\dag}_{N}(|z|=r)&=
	\frac{\sqrt{2N}}{\pi r \Gamma(N+2)}\left[\sqrt{\frac{Nr^2}{2}}\left( (Nr^2)^Ne^{-Nr^2}+(N-\frac{Nr^2}{2})\Gamma[N,Nr^2]
	\right) \right.
	\nonumber\\
	&+\left.2^{\frac{N}{2}}\gamma\left(\frac{N+3}{2},\frac{Nr^2}{2}\right)\left(
	(Nr^2)^{\frac{N}{2}}e^{-Nr^2/2} +\frac{(N-1-Nr^2)\Gamma(N,Nr^2)}{2(Nr^2)^{\frac{N}{2}}e^{-Nr^2/2}}
	\right)
	\right].\quad  
	\label{eq-R1AI+}
\end{align}
Note that compared to \cite{complex sym Gernot} we have rescaled the edge to unity and given the density without radial factor $r$ from the radial measure $d r\,r$,
corresponding to the circular (and not triangular) law here. They are based on the Kac--Rice formalism \cite{SommersIida}, established rigorously in this context recently \cite{Yan25}. For class AII$^\dag$, based on universality arguments the limiting density at the edge is conjectured \cite{OK}. It is important to note that for classes AI$^\dagger$ and AII$^\dagger$ the JPDF in terms of the eigenvalues (for a Gaussian weight of the matrix elements) is not known until today. The densities \eqref{eq-R1A} and \eqref{eq-R1AI+} are shown in Figure \ref{fig-radial density plot symmetry classes} and compared to the numerics. Here we consider $720\,000$ samples for all ensembles, with class A \eqref{eq-R1A} (blue full curve: analytical; purple triangle: numerical), class AI$^\dagger$ \eqref{eq-R1AI+} (red dashed curve: analytical; orange circle: numerical), both at $N=1024$, and class AII$^\dagger$ (green diamond: numerical only) with $2N=1024$. All three ensembles share a constant density in the bulk and show differences at the edge. Moreover, we find an excellent agreement of the analytics and numerics for classes A and AI$^\dagger$.\\
In the next Section \ref{ssec-numerics symmetry classes}, we study the three conjectured universality classes A, AI$^\dagger$ and AII$^\dagger$ numerically using different techniques: Firstly, we compare the NN and NNN statistics in the bulk of the classes AI$^\dagger$ and AII$^\dagger$ with 2dCG data, finding a very good effective description of them corresponding to $\beta= 1.4$, respectively $\beta= 2.6$. We use here, that the spectra are flat in the bulk of the spectrum, removing the need of unfolding. As a second method, we discuss the complex spacing ratio and present results on the universality here in the bulk and at the edge. It is supposed that no unfolding is needed for the computation of this quantity in the bulk and we investigate whether this holds true at the edge of the spectrum.
\subsection{Numerical Study of the Classes A, AI$\boldsymbol{^\dagger}$, and AII$\boldsymbol{^\dagger}$}\label{ssec-numerics symmetry classes}
\subsubsection{NN and NNN Distributions}\label{sssec-NN NNN classes}
\begin{figure}[t]
	\centering
	\begin{turn}{90}\hspace{80pt} \large AI$^\dagger$ \end{turn}
	\includegraphics[width=0.9\linewidth,angle=0]{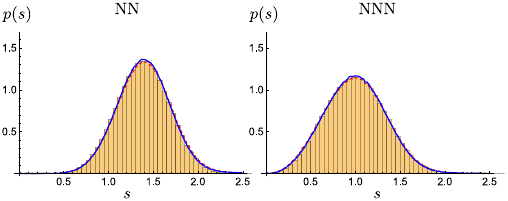}\\
	\begin{turn}{90}\hspace{75pt} \large AII$^\dagger$ \end{turn}
	\includegraphics[width=0.9\linewidth,angle=0]{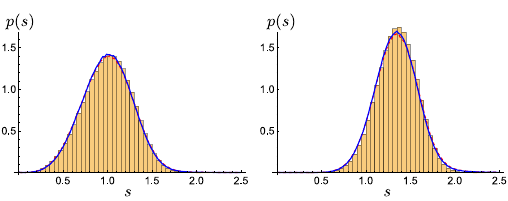}
	\caption[Figures 11 and A2 from \cite{ABCPRS}, \textit{Comparison of the NN and NNN statistics of symmetry classes AI$^\dagger$ and AII$^\dagger$ with 2dCG data}]{The NN (left) and NNN (right) AI$^\dag$/ AII$^\dagger$ (top/ bottom) spacing distribution generated numerically (histograms) for 500 matrices of size $N=5000$, respectively $2N=5000$. Also shown are comparisons to the NN and NNN spacing distribution of a 2dCG, with fitted values $\beta=1.3$/ $\beta=2.5$ (top/ bottom) (red dashed) and $\beta=1.4$/ $\beta=2.6$ (top/ bottom) (blue full curve). \cite[Figures 11, A2]{ABCPRS} 
	}
	\label{fig-AIdag AIIdag NN NNN}
\end{figure}
Let us draw the starting point for this section: After the revisited symmetry classification \cite{non-h classes} appeared, Hamazaki et al \cite{Hamzaki et al} conjectured based on local NN statistics that the three classes A, AI$^\dagger$ and AII$^\dagger$ represent the three different universality classes of local behaviour in the bulk. Here it is important that all three ensembles share a constant density in the bulk of the spectrum (see Figure \ref{fig-radial density plot symmetry classes}) which removes the need of unfolding. 
\begin{Remark}
	We would like to point the reader's attention again to the key word: Universality. As mentioned throughout this thesis, this subject is very rich and can be viewed from several perspectives. Therefore, it should be noted that Hamazaki et al \cite{Hamzaki et al} investigate the universality in the distribution of matrix elements in their work too. They show that random matrices with normal distributed elements share the same statistics as random matrices with properly distributed Bernoulli random variables. The universality with respect to the distribution is well-known \cite{tao vu}. It allows us to choose Gaussian (normal) distributions of the matrix elements for simplicity.
\end{Remark}
We come back to the intuition discussed above: We observe in Figure \ref{fig-NN NNN symmetry classes} that the maximum of the NN/ NNN distribution is lower/ higher compared to class A for class AI$^\dagger$/ AII$^\dagger$. Recalling the behaviour of the NN spacing distributions of 2dCG which we discussed in Section \ref{ssec-point processes} Figure \ref{fig-2d-statistics} right: The height of the maximum of the NN spacing distribution of a 2dCG increases with increasing $\beta>0$ and its position moves from left to right. Comparing the numerics of the classes AI$^\dagger$ and AII$^\dagger$ with the NN/ NNN statistics of numerically generated 2dCG (cf. Section \ref{ssec-surmise}), we find a good agreement of NN \textbf{and} NNN statistics of class AI$^\dagger$ with the 2dCG statistics at $\beta= 1.4$ and correspondingly at $\beta= 2.6$ for class AII$^\dagger$, see Figure \ref{fig-AIdag AIIdag NN NNN} above. While the comparison with the NN statistics in the bulk was established in earlier works by the author \cite{masterthesis,AMP}, this effective description of the two symmetry classes by non-integer 2dCG statistics was extended to the NNN statistics in \cite{ABCPRS} by the author. This shows that the effective description assigning a weaker/ stronger local repulsion of the eigenvalues compared to class A at $\beta=2$, extends from the local NN statistics to the NNN statistics at the very same values of inverse temperature $\beta$. We present the comparison of the NN (left) and NNN (right) distributions for the classes AI$^\dagger$ (top row) and AII$^\dagger$ (bottom row) in Figure \ref{fig-AIdag AIIdag NN NNN}, where the statistics for the symmetry classes are given by the histograms. Additionally, we give the best fit 2dCG statistics, i.e. $\beta= 1.4$/ $\beta= 2.6$ as blue full curves as well as the curve of slightly smaller $\beta=1.3$, respectively $\beta=2.5$ (red dashed). We observe that the curves for different $\beta$ show only few differences. Moreover, we find that the fits for class AII$^\dagger$ are less good than for class AI$^\dagger$ with a decrease of agreement for the NNN distributions. This is also displayed in the standard deviations and Kolmogorov-Smirnov distances (cf. Section \ref{ssec-numerics methods}, Eqs. \eqref{eq-standard deviation}, \eqref{eq-KS distance}) in Table \ref{tab-NN and NNN AI dag AII dag 2DC} below, where we mark the best fit in bold. We find a slight variation between the best fit via the standard deviation and the Kolmogorov-Smirnov distance in the NNN fits. From Figure \ref{fig-AIdag AIIdag NN NNN}, we have that the deviation between $\beta=1.3$ and $\beta=1.4$, respectively $\beta=2.5$ and $\beta=2.6$ is relatively small. Moreover, we stress that the finding serves as a tool for an effective description of the repulsion in the ensembles AI$^\dagger$ and AII$^\dagger$ and we do not expect the classes to share the JPDF of a 2dCG at the corresponding $\beta$. We observe in \cite{AAAMPRW} that the NN spacings in small arguments show cubic repulsion (as expected, cf. \cite{haake}) for all three classes AI$^\dagger$, A and AII$^\dagger$ and not a behaviour $\sim s^{1+\beta}$ for classes AI$^\dagger$ and AII$^\dagger$.
\begin{table}[h!]\centering
	\caption[Tables IV and VI from \cite{ABCPRS}, \textit{List of standard deviations and Kolmogorov-Smirnov distances for the comparison of the NN and NNN distributions of classes AI$^\dagger$, AII$^\dagger$ with 2dCG statistics}]{List of standard deviations $\sigma$ and Kolmogorov-Smirnov distances $d$ (both in units $ 10^{-2}$) between the NN and NNN distributions of AI$^\dagger$ (top), respectively AII$^\dagger$ (bottom), and 2dCG, with $\beta\in[1.1,1.6]$ (top), respectively $\beta\in[2.3,2.8]$ (bottom), in steps $0.1$. \cite[Tables IV, VI]{ABCPRS}}\vspace{3mm}
	\begin{tabular}{|c|c|c|c|c|c|c|c|c|}
		\hline
		&fitted &$\beta$  & 1.1 & 1.2 & 1.3 & 1.4 & 1.5& 1.6\\
		\hline
		\hline
		\multirow{4}{*}{AI$^\dagger$}&\multirow{2}{*}{NN}&	$\sigma$ & 3.3 & 2.2 & 1.1 & \textbf{0.9} & 1.7 & 2.7 \\
		\cline{3-9}
	&	&$d$ & 1.7 & 1.2 & 0.6 & \textbf{0.3} & 0.7 & 1.2 \\
		\cline{2-9}
		\cline{2-9}
	&	\multirow{2}{*}{NNN}&$\sigma $& 3.4 & 2.0 & \textbf{0.8} & 1.4 & 2.7 & 4.0 \\
		\cline{3-9}
	&	&$d$ & 3.0 & 1.8 & 0.8 & \textbf{0.6} & 1.7 & 2.7\\
		\hline
		\hline
	&	fitted &$\beta$  & 2.3 & 2.4 & 2.5 & 2.6 & 2.7& 2.8\\
		\hline
		\hline
	\multirow{4}{*}{AII$^\dagger$}	&\multirow{2}{*}{NN}&	$\sigma$ & 3.4 & 2.8 & 2.4 & \textbf{2.3} & 2.4 & 3.0 \\
		\cline{3-9}
	&	&$d$ & 1.4 & 1.1 & 1.0 & \textbf{0.9} & 1.0 & 1.2 \\
		\cline{2-9}
		\cline{2-9}
		&\multirow{2}{*}{NNN}&$\sigma $& 5.5 & 4.6 & 4.1 & \textbf{3.9} & 4.3 & 4.9 \\
		\cline{3-9}
	&	&$d$ & 2.8 & 2.1 & \textbf{1.6} & 1.9 & 2.4 & 2.8\\
		\hline
	\end{tabular}
	\label{tab-NN and NNN AI dag AII dag 2DC}
\end{table}
Are these effective values of $\beta$ for the local statistics consistent with the global shape of the spectral density in Figure \ref{fig-radial density plot symmetry classes}? Therefore, we revisit the study of 2dCG for $\beta \neq 2$ briefly, where we follow the discussion in \cite{AAKP}. For $\beta<2$ the authors of \cite{Abanov} found a damping of the spectral density at the edge compared to $\beta=2$, see Figure 2 therein. We observe a similar damping of the density for class AI$^\dagger$ (red dashed curve, orange circles) in Figure \ref{fig-radial density plot symmetry classes}. Perturbation theoretic methods \cite{Can et al} as well as recent numerical simulations \cite{Abanov, Abanov 2} show that for $\beta > 2$ the spectral density of the 2dCG overshoots the $\beta=2$ result in the neighbourhood of the edge and eventually starts to oscillate when increasing $\beta$ even further, cf. \cite[Figure 1]{Can et al} and \cite[Figure 4]{Abanov}. \\
\\
Chronologically, there started intensive research on the two rather "unexplored" classes AI$^\dagger$ and AII$^\dagger$. Recall that at this stage analytical results were available only for the $N=2$ NN spacing distributions, cf. \cite{Hamzaki et al,complex sym pandey}. Nevertheless, using the universality in the distribution of matrix elements in making the choice of a Gaussian distribution, computations of expectation values of products of characteristic polynomials came into flavour \cite{AAKP,Liu Zhang,Kulkarni Kawabata Ryu,Forrester24}. While the computed quantities are not in direct relation to the spectral density for the classes AI$^\dagger$ and AII$^\dagger$, a first analytical comparison for finite and large $N$ was reached. The results lead to extend the conjecture of different local statistics from the bulk to the edge. We will discuss some of these results, which were derived at earliest by the author and collaborators in the next Section \ref{ssec-expectation of characteristic polynomials}. Before, turning to these analytical studies, we introduce and discuss another quantity which became of great interest over the last years as a measure to distinguish different ensembles from each other: The complex spacing ratio.
\subsubsection{Complex Spacing Ratio}\label{sssec-complex spacing ratio}
We investigate the numerical distinguishing of local statistics for the three symmetry classes A, AI$^\dagger$ and AII$^\dagger$, via the complex spacing ratio. We follow \cite{AAAMPRW} closely, were the author mainly contributed the interpretation of the numerical findings and the supervision of the implementation of the numerics subject of \cite{BScCR}.\\
So far, we revisited the NN (and NNN) spacings as a tool to distinguish integrable and chaotic behaviour, cf. Section \ref{ssec-numerics methods}. A second popular tool besides NN spacings are complex spacing ratios and its moments. They have been extended to non-Hermitian RMT in \cite{SaRibeiroProsen} and have the advantage of not relying of unfolding the local spectral data, which is much more involved in two dimensions, cf. Section \ref{sssec-unfolding}. The authors of \cite{SaRibeiroProsen} find that the complex spacing ratio shows significant differences within the three representatives of non-Hermitian symmetry classes A, AI$^\dagger$ and AII$^\dagger$ in the bulk of the spectrum. An application of the study of universality in those classes to QCD-like theories, is discussed in \cite{KanazawaWettig}. As the complex spacing ratio is the main object of our current studies, let us give a clear definition below. Let again $z_1,\ldots,z_N$ denote the complex eigenvalues of an $N\times N$ matrix. The complex spacing ratio $\lambda_k$ of the $k$th eigenvalue $z_k$, where no ordering is implied, is
\begin{align}\label{eq-def spacing ratio}
	\lambda_k=\frac{z_k^{\textsc{NN}}-z_k}{z_k^{\textsc{NNN}}-z_k},
\end{align}
where $z_k^{\textsc{NN}}$ denote the NN of $z_k$ in radial distance, while $z_k^{\textsc{NNN}}$ is eigenvalue with second smallest radial distance to $z_k$. We observe that $\lambda_k$ is a complex variable. We can assign a radius $r_k$ and an angle $\phi_k$ to it. As the NNN is always equal or further away than the nearest one, we have $r_k\leq 1$. One attractive feature of the spacing ratio is that no unfolding is needed. As $z_k^{\textsc{NN}}$, and $z_k^{\textsc{NNN}}$ feel a similar averaging of the underlying spectral density, the unfolding will be (almost) the same in the numerator and denominator of Eq. \eqref{eq-def spacing ratio} and therefore, drops out. It is remarkable that even in class A only approximate expressions exist for the complex spacing ratio \cite{SaRibeiroProsen,DusaWettig}.
They include an $N$-dependent determinantal expression and a surmise at $N=3$, which is not very accurate. This is in stark contrast to the spacing ratio in Hermitian RMT introduced by \cite{OH,Atas_et_al} in the bulk of the real spectrum, which is very well approximated by $N=3$.\footnote{We note that again for Hermitian RMT the small $N$ results yield a good approximation for the large $N$ behaviour, similar to the Wigner surmise discussed in Section \ref{ssec-surmise}.} Exact analytic limiting results are scarce for the spacing ratio, see however \cite{Nishigaki,Buijsman}. Higher order spacing ratios have been considered \cite{Bhosale,Ruth surmise}, as well as transitions between different symmetry classes \cite{KKSG}, and more generally $\beta$-ensembles of a 1d Dyson gas \cite{Atas2}.\\
In \cite{AAAMPRW}, we derive analytical expressions for the complex spacing ratio at finite $N$ within class A, including the parameter-dependent elliptic or Ginibre-Girko ensemble (eGinUE) \cite{Girko}, which interpolates the Hermitian GUE and the non-Hermitian GinUE. While we will not present the results here, we want to highlight, one key ingredient in this study: Following the idea of Grobe, Haake and Sommers \cite{GHS}, we consider the point process with one eigenvalue conditioned at the origin for general $N$. Together with the known translational invariance of the GinUE in the large $N$ bulk limit, this explains why the approximation proposed in \cite{SaRibeiroProsen,DusaWettig} converges well. Moreover, we compare the NN and NNN spacing distributions to the numerics for all three classes in \cite{AAAMPRW}. Here, an unfolding procedure at the edge of the spectrum is established, where the mean density varies on the scale of the mean level spacing. We find distinct statistics for all three classes of the NN and NNN spacing at the edge and in the bulk. In contrast, for 2d Poisson bulk and edge statistics agree after unfolding. We refer to \cite{AAAMPRW} for more details. We will now start with study of the complex spacing ratio. \\
For simplicity, we will choose a Gaussian distribution of matrix elements for all three ensembles as follows. Let $J$ be an $N\times N$ (respectively $2N\times 2N$ in the case of AII$^\dagger$) random matrix, where we assign a Gaussian weight $\propto \exp\left[-N\Tr (JJ^\dagger) \right]$ with it. Here, the Gaussian weight is normalised such that the limiting eigenvalue density will be supported on the unit disc. The classes A, AI$^\dagger$ and AII$^\dagger$ are defined with respect to the symmetry relations \eqref{eq-def class A}, \eqref{eq-def class AI dag}, \eqref{eq-def class AII dag} respectively. It is convenient to parametrise the $2N\times 2N$ matrix $J$ of class AII$^\dag$ via three non-Hermitian $N\times N$ matrices $A,B,C$, where the matrices $B$ and $C$ are anti-symmetric, i.e. $B=-B^t$ and $C=-C^t$ as
\begin{align}\label{eq-para class AII dag}
	J=\begin{pmatrix}
		A & B\\
		C & A^t
	\end{pmatrix},
\end{align}
and $A$ has no symmetry. Here, it is important to scale the variance of the elements of the $A,B,C$ to unity (up to symmetries for $B,C$), as the variance of $J$ cannot be fixed by an overall factor as for the other matrix ensembles under consideration, cf. \cite[Eq. (2.9)]{AAKP}. The distribution for 2d Poisson can be formulated in the same manner in matrix form, with $J=$diag$(z_1,\ldots,z_N)$ being diagonal. \\
We provide our data for the complex eigenvalues of the random matrices from the Poisson statistics, respectively the symmetry classes AI$^\dagger$, A, and AII$^\dagger$ for
720~000 ensembles each and $N=1024$, respectively $2N=1024$ for AII$^\dagger$ in \cite{data repository}. We refer to \cite{AAAMPRW} for details on the implementation.
\begin{figure}[t]
	\includegraphics[width=0.49\linewidth]{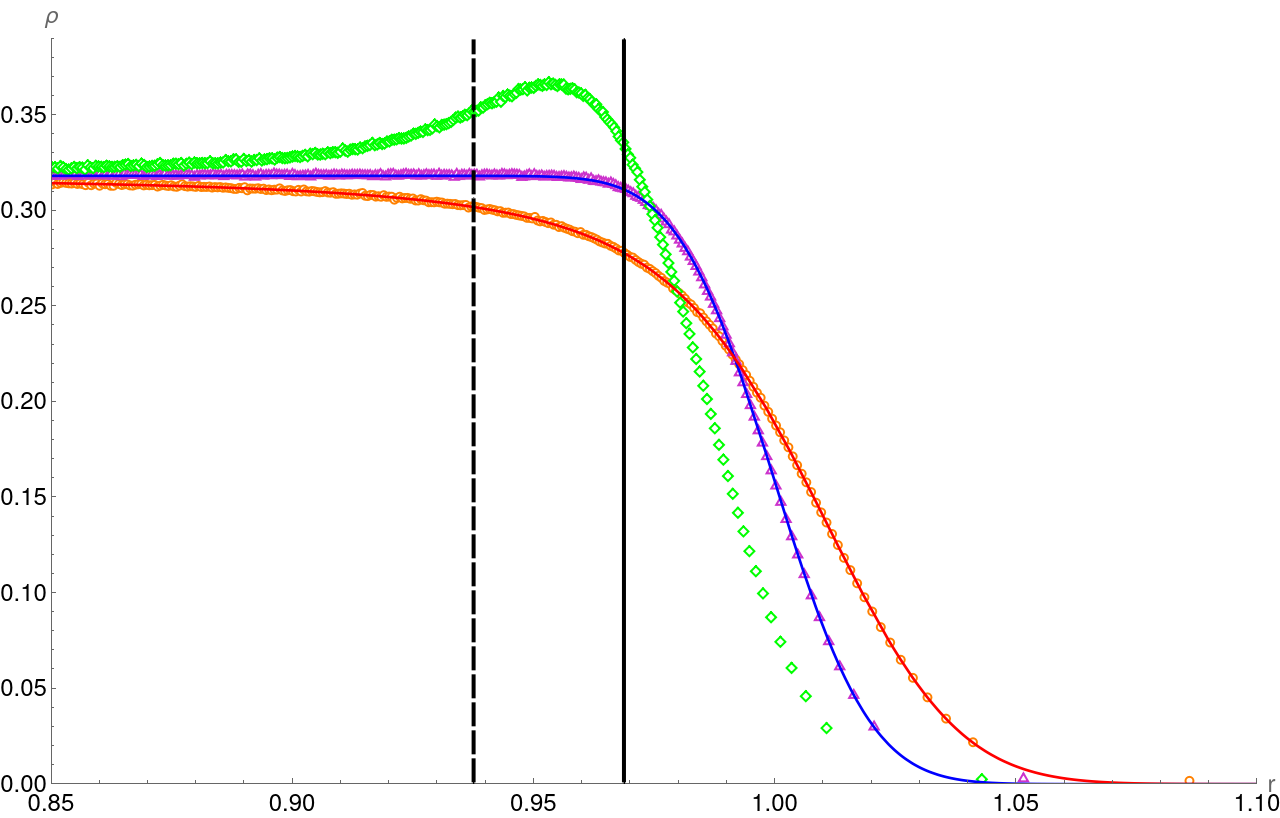}
	\includegraphics[width=0.49\linewidth]{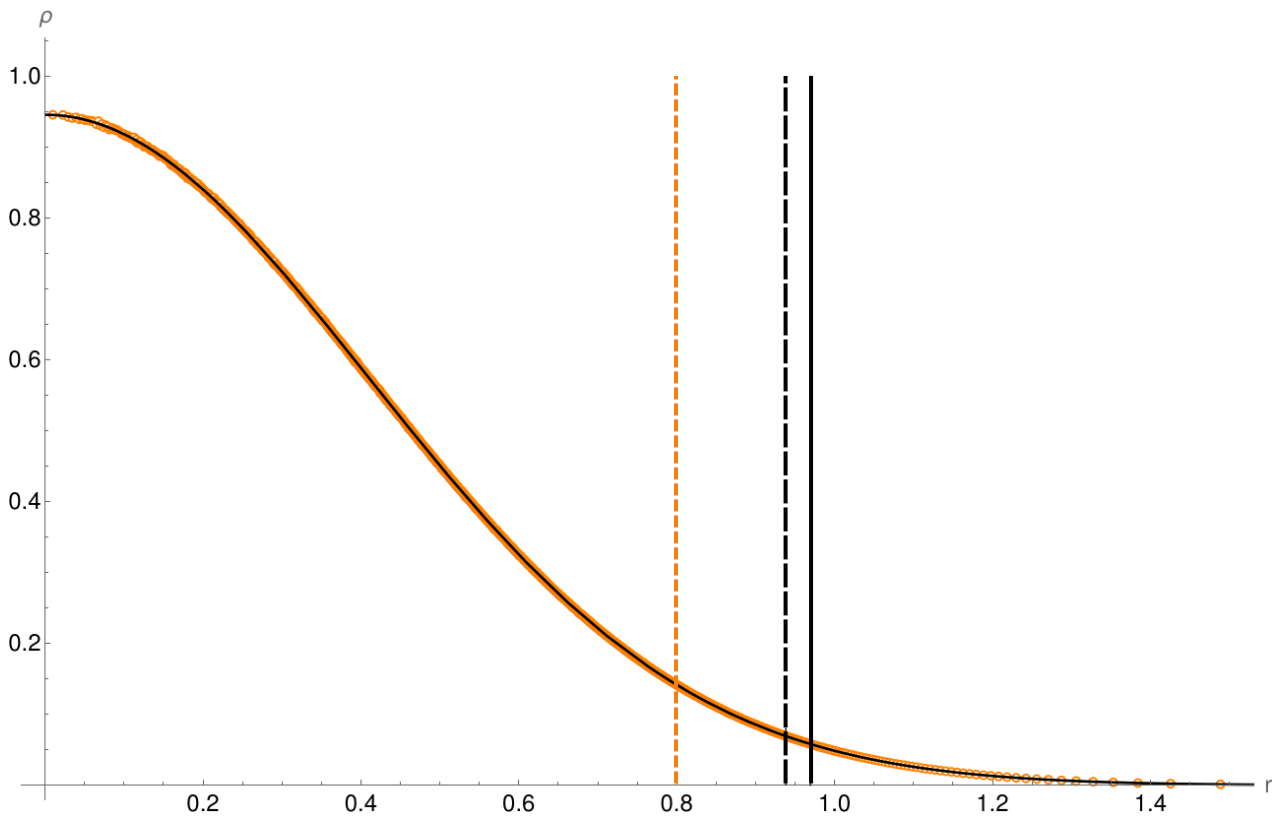}
	\caption[Part of Figure 4 and Figure 5 from \cite{AAAMPRW}, \textit{Radial spectral density of the symmetry classes A, AI$^\dagger$ and AII$^\dagger$ and the 2d Poisson process at the edge}]{\textbf{Left:} Zoom into the edge region for the radial spectral density of class A \eqref{eq-R1A} (blue full curve: analytical; purple triangle: numerical), class AI$^\dagger$ \eqref{eq-R1AI+} (red dashed curve: analytical; orange circle: numerical), both at $N=1024$, and of class AII$^\dagger$ (green diamond: numerical only) with $2N=1024$, with $720\,000$ samples for all ensembles. \textbf{Right:} The radial spectral density of the 2d Poisson process for $N=1024$, analytically (black full line) and from $720\,000$ ensembles (orange circles). \\
		The region to the right of the solid vertical line at radius $r_-=1-1/\sqrt{1024} = 0.96875$ defined as edge, and the dashed vertical line at $r_-^\prime=1-2/\sqrt{1024} = 0.9375$ as extended edge. \cite[Cf. Figures 4, 5]{AAAMPRW}
	}
	\label{fig-radial density plot symmetry classes edge and Poi}
\end{figure}\\
\noindent
Next, we have to define how to distinguish bulk and edge regions. Therefore, we revisit the plot of the radial spectral density in Figure \ref{fig-radial density plot symmetry classes} of our numerical analysis of a large but finite matrix size $N$ ($2N$). All three classes have a rotationally symmetric spectral density and share the same constant density in the bulk, with sharp drop at the edge. Thus, we can fully characterise the bulk and edge region by the respective radius. A zoom into the edge regions is given in Figure \ref{fig-radial density plot symmetry classes edge and Poi} left. The observed behaviour of the three global densities at the edge is consistent with the numerically obtained $\beta$-dependence of the edge behaviour of the density of a true 2dCG \cite[Figure 1]{Can et al}. There, for $\beta\leq2$ the radial density decreases monotonously towards the edge, where increasing $\beta$ makes the transition sharper. For $\beta>2$, a local maximum develops at the edge, with its height increasing with $\beta$.
For this reason, in the following we will always plot and discuss the three classes in the order AI$^\dag$, A and AII$^\dag$ of increasing effective $\beta=1.4,\ 2$ and 2.6, respectively.

For classes A and AI$^\dag$ \cite{complex sym Gernot} it is known analytically that the width of the edge region scales as $O(1/\sqrt{N})$. The same scaling has been found in a symmetry class related to AII$^\dag$ \cite{OK}, too. Therefore, we will make a simple choice for the edge region as $r_-=1-1/\sqrt{N}$, with $r=1$ being the limiting radius of support (for AII$^\dag$ we choose $r_-=1-1/\sqrt{2N}$, to have the same numerical value). 
In order to test our choice, how far the edge extends for the given value of $N$ ($2N$), we will also compare to a larger choice called extended edge, with a twice as large deviation from unity, $r_-^\prime=1-2/\sqrt{N}$. We indicate these choices by horizontal lines, from which radius $r_-=1-1/\sqrt{1024} = 0.96875$ (full line) onwards we count the eigenvalues with $|z_k|>r_-$ as edge eigenvalues in Figure \ref{fig-radial density plot symmetry classes edge and Poi}. For simplicity we take the same edge value for all three symmetry classes, as well as for the 2d Poisson point process of uncorrelated, Gaussian random variables.
In line with the argument of interpreting different classes as 2dCG at different effective values of $\beta$, our choice for 2d Poisson aims \eqref{eq-jpdf 2dCG} at $\beta=0$. We choose a Gaussian weight $\omega(z)=\exp[-|z|^2/(2 \sigma^2)]$, such that the probability inside the
typical edge radius ($r_- < 1-1/\sqrt{N}$) is the same as the ratio of areas of edge disc and unit disc. This results in the variance \cite{BScCR}
\begin{align}
2 \sigma^2 = -\frac{(1 - \frac{1}{\sqrt{N}})^2}{\log(1-(1 - \frac{1}{\sqrt{N}})^2)}.
\end{align}
The corresponding density is shown in Figure \ref{fig-radial density plot symmetry classes edge and Poi} right. Being given by a radially symmetric Gaussian, it is less obvious how to choose an appropriate value of $r_b$ for the bulk and of $r_-$ for the edge. In order to avoid any bias, we will choose the same values for $r_b$ and $r_-(r_-^\prime$) for the bulk and (extended) edge as for the other three ensembles.

An alternative choice for Poisson random variables in 2d would be independent points with uniform probability on the unit disc. 
Although one might expect a strong effect through this hard cut-off at the edge, this realisation of 2d Poisson leads to rather similar results, cf. \cite{BScCR}.\\
\\
\textbf{Complex Spacing Ratios in 2d: Edge vs. Bulk}
\\
\\
\noindent
In the following we will call the complex spacing ratio $\lambda_k$ to belong to the bulk, if the corresponding point satisfies $|z_k|<r_b$, having a radius less than what we chose to belong to the bulk. For our value of $N$ we define the bulk at radius $r_b=0.8$ (orange dotted vertical line in Figure \ref{fig-radial density plot symmetry classes}). On the other hand, it belongs to the (extended) edge, if it holds $|z_k|>r_-(r_-^\prime)$. Obviously, this removes some eigenvalues with $r_b\leq |z_k|\leq r_-(r_-^\prime)$ from our statistics, where it is not clear a priori, if they belong to the bulk or to the edge. Choosing two different values $r_-^\prime <r_-$ for the edge tests how far the edge actually extends.\clearpage
\begin{figure}[t]
	
	\begin{subfigure}{\textwidth}
		\hspace{80pt}{AI$^\dagger$} \hspace{130pt}{A}\hspace{130pt}{AII$^\dagger$} \hfill\\
		\begin{turn}{90}\hspace{60pt}{bulk}\end{turn}
		\includegraphics[width=0.3\linewidth]{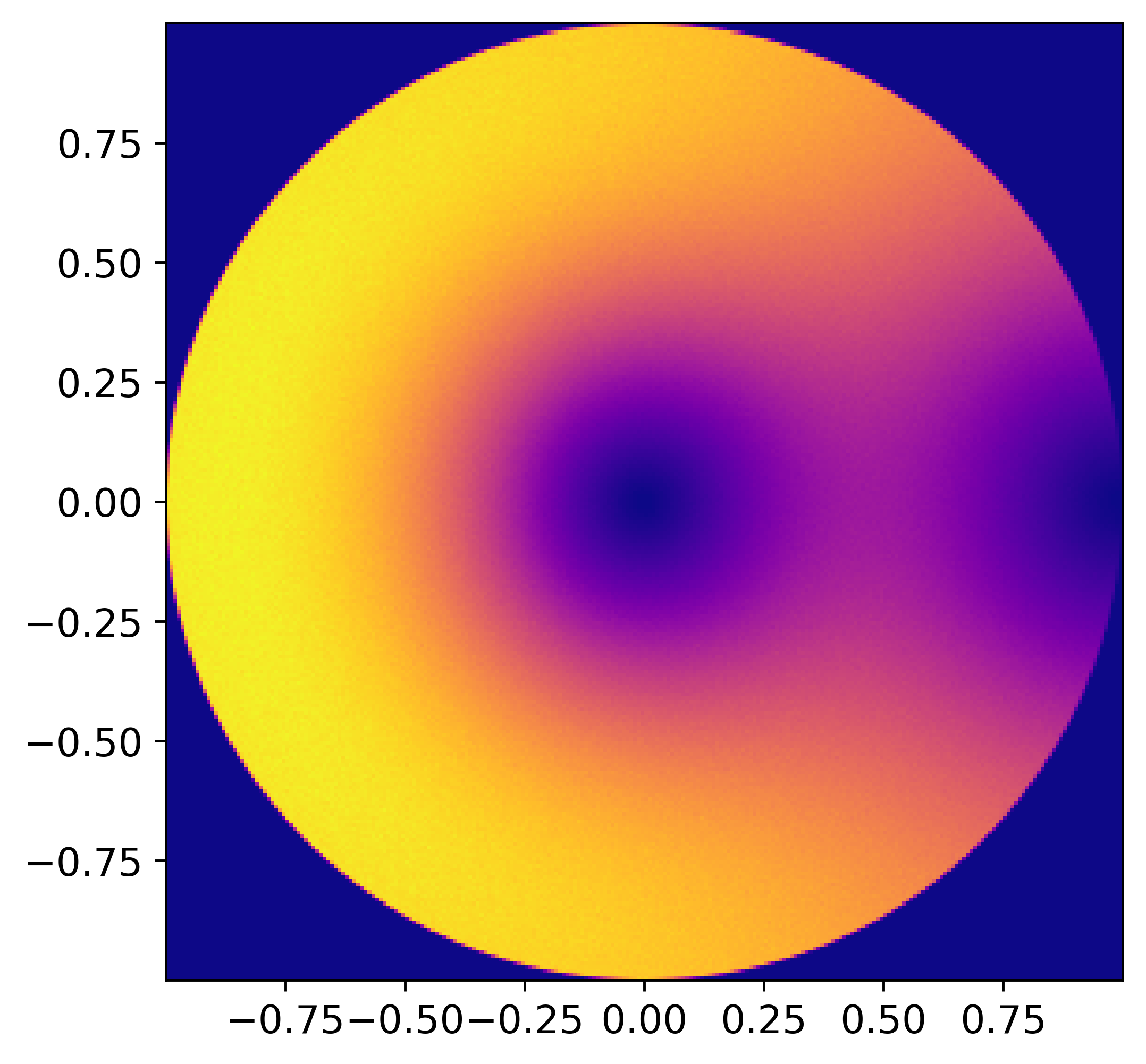} 
		\includegraphics[width=0.3\linewidth]{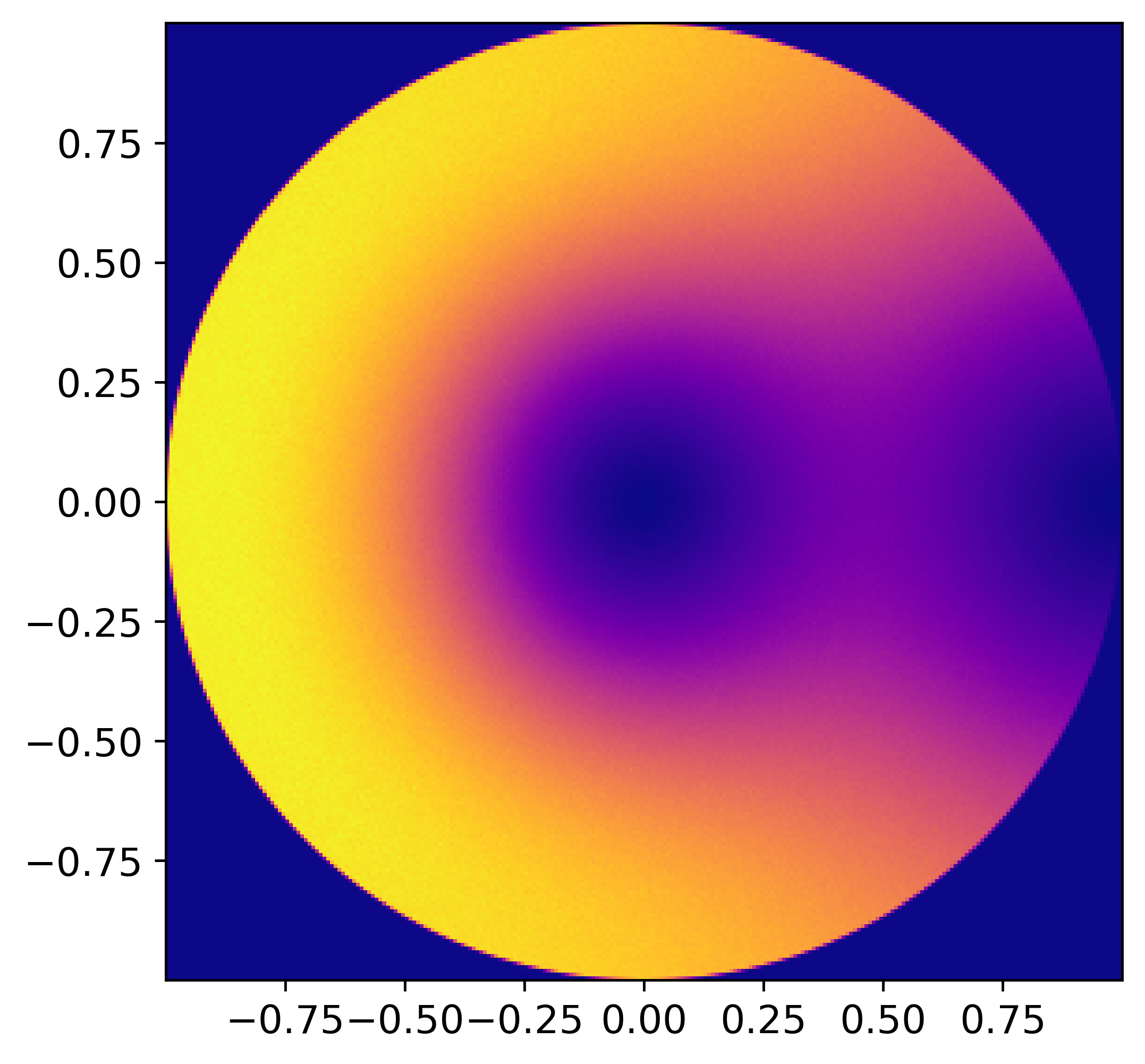}
		\includegraphics[width=0.3\linewidth]{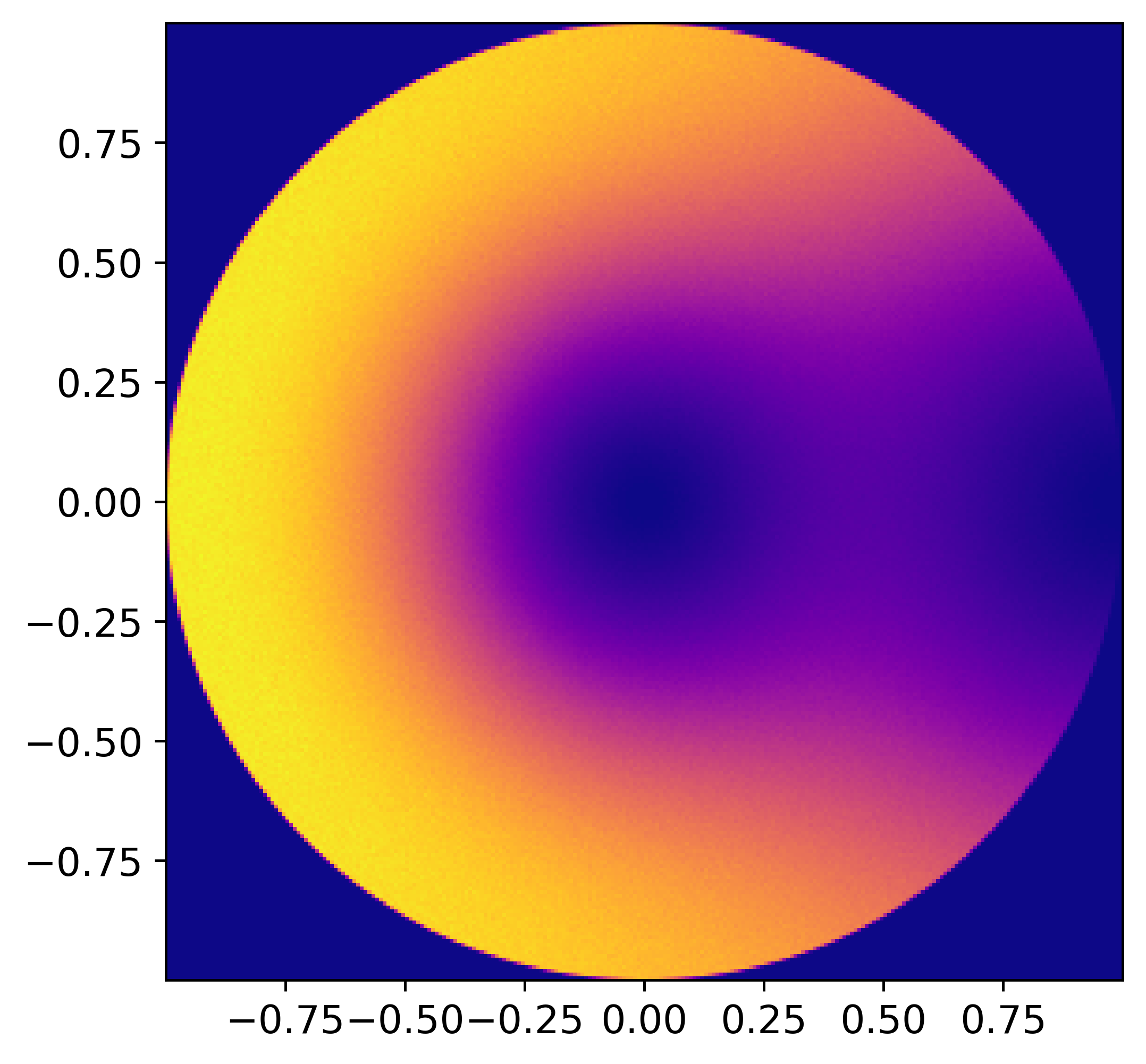}  
		\raisebox{5pt}{\includegraphics[scale=0.43]{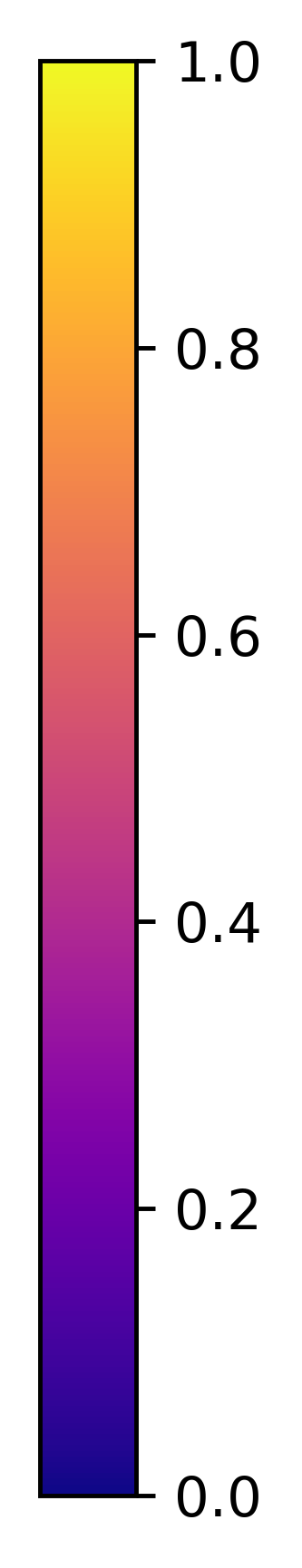}}
		\label{fig:spacing ratio bulk}
	\end{subfigure}
	\begin{subfigure}{\textwidth}
		\begin{turn}{90}\hspace{60pt}{edge}\end{turn}
		\includegraphics[width=0.3\linewidth]{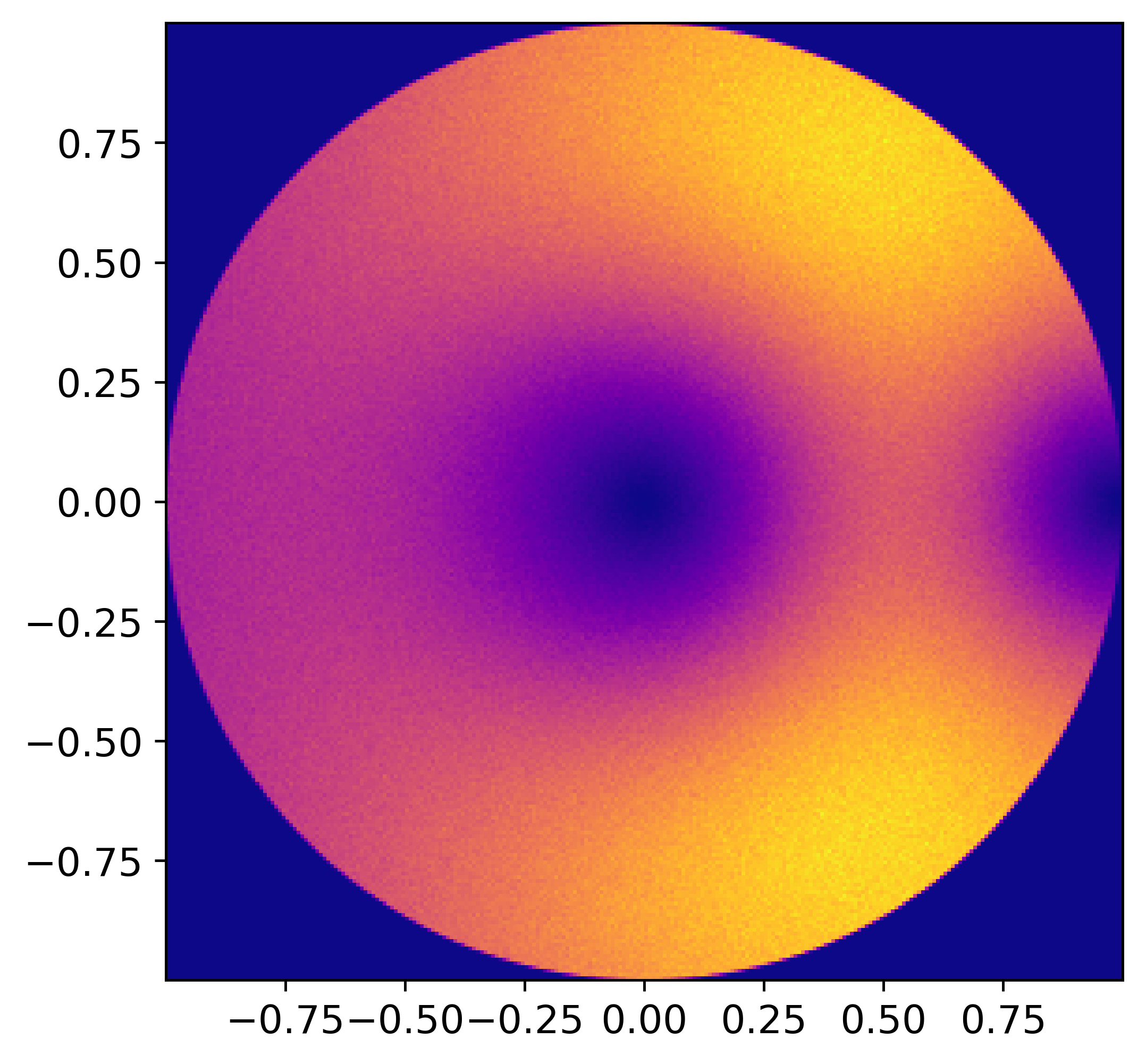}
		\includegraphics[width=0.3\linewidth]{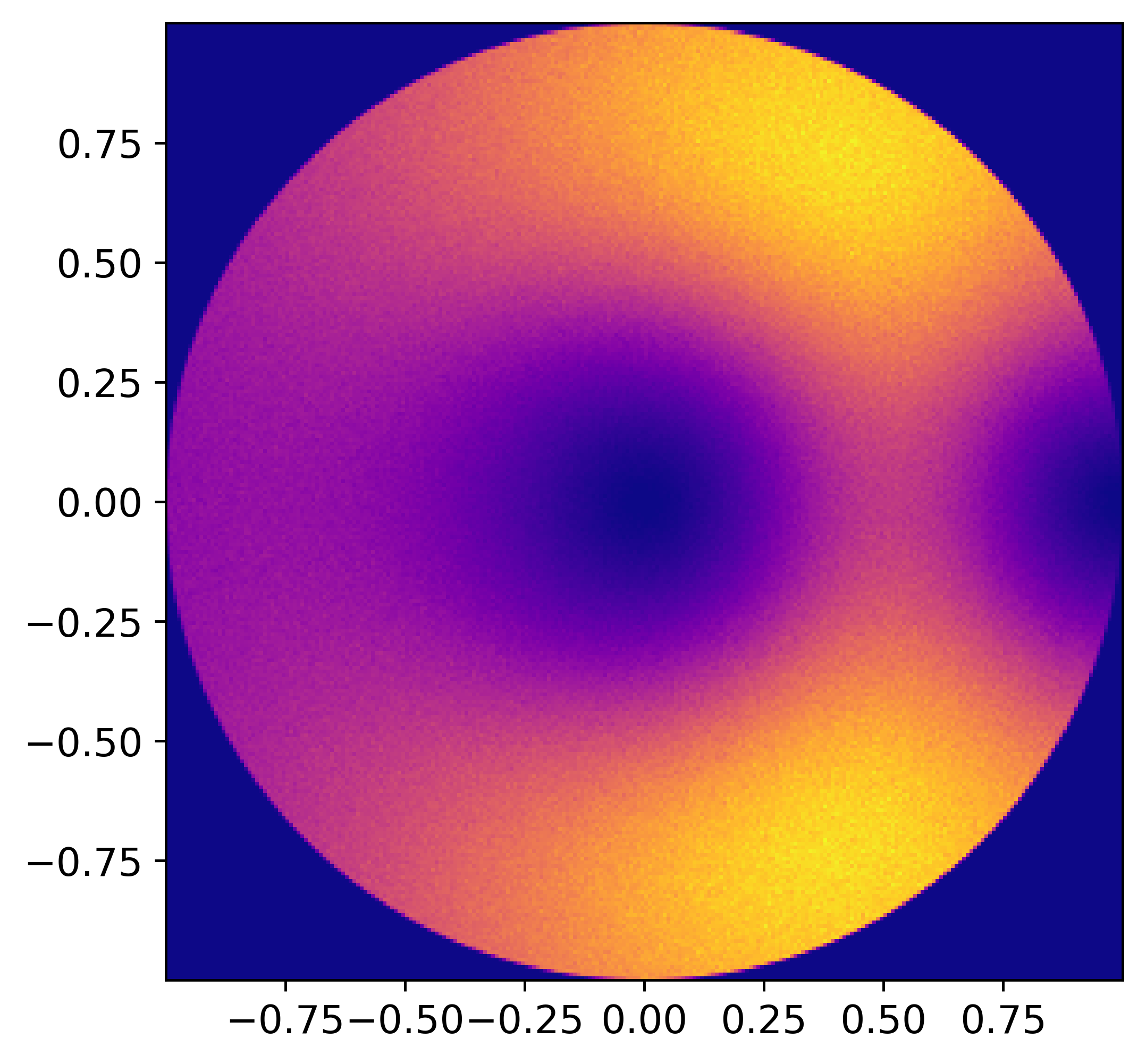}
		\includegraphics[width=0.3\linewidth]{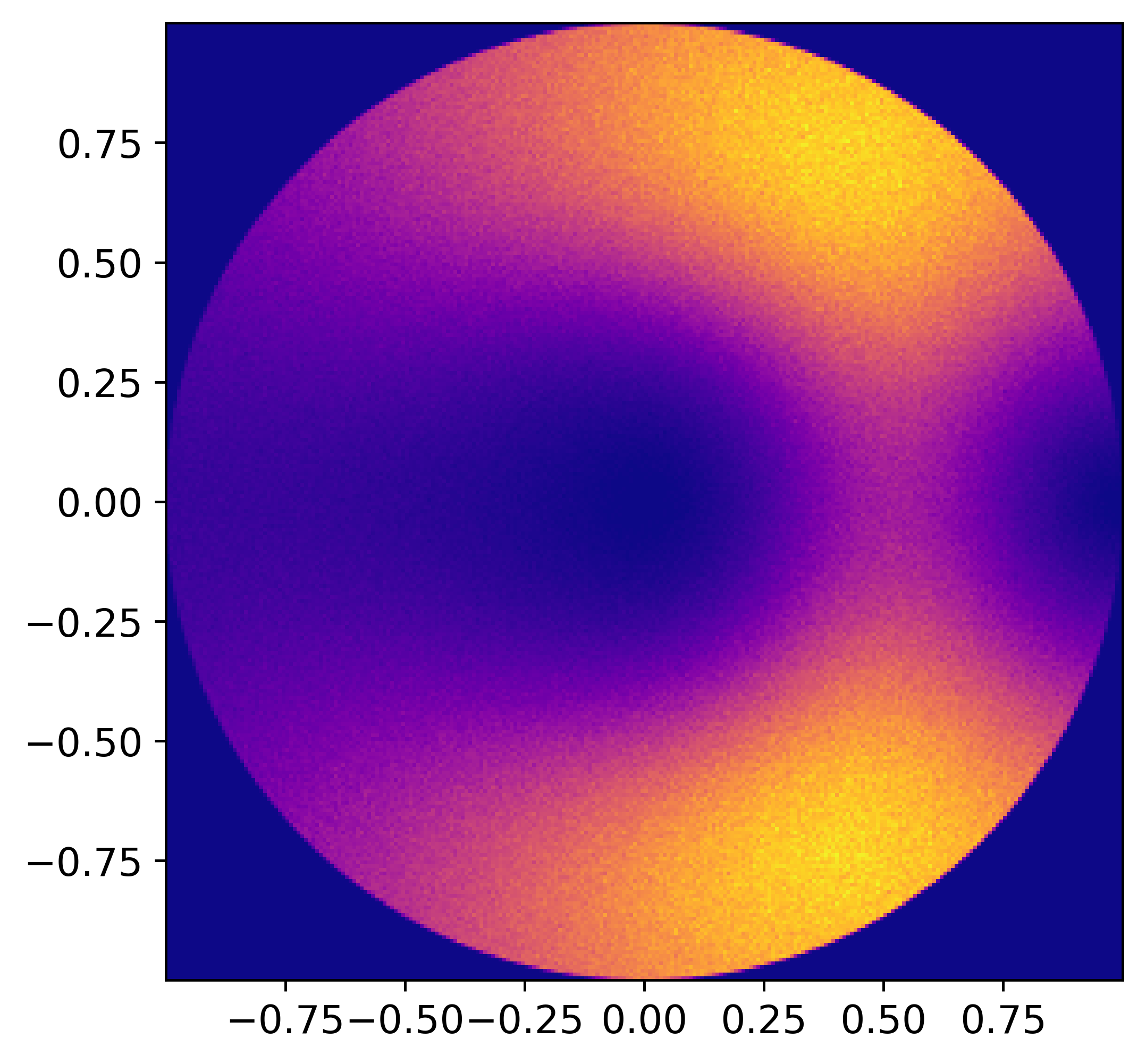}
		\raisebox{5pt}{\includegraphics[scale=0.43]{Figures_Sec4/colorbar.png}}
		\label{fig:spacing ratio edge}
	\end{subfigure}
	\caption[Figure 6 from \cite{AAAMPRW}, \textit{Density plots of the complex spacing ratio in the bulk and at the edge for classes AI$^\dagger$, A, AII$^\dagger$}]{Density plots of the complex spacing ratio in the bulk (top rows) and at the edge (bottom rows) for classes AI$^\dag$, A, AII$^\dag$, from left to right. 
		We show $720\,000$ ensemble realisations of size $N=1024$ (classes AI$^\dag$, A) and $2N=1024$ (class AII$^\dag$). The colour code gives the density from low (dark, blue) to high (bright, yellow).\\ \cite[Figure 6]{AAAMPRW}}
	\label{fig-spacing ratio symmetry classes}
\end{figure}\noindent
We now present the 2d density plots of the distribution of the complex spacing ratio \eqref{eq-def spacing ratio} for classes AI$^\dag$, A, AII$^\dag$, from left to right in Figure \ref{fig-spacing ratio symmetry classes}, both in the bulk (top) and the edge (bottom). Here, a brighter yellow intensity indicates a higher density as a function of $z=r e^{i \theta}=x+i y$.
Our results in the bulk agree with earlier studies by \cite{SaRibeiroProsen,DusaWettig}, where we add the following interpretation. 
The low density in the centre of the plot (dark central hole) represents the repulsion of the eigenvalues, i.e. the probability that two eigenvalues are getting very close to each other is going to zero. We observe that this region increases from class AI$^\dag$ (top left) over class A (top middle) to class AII$^\dag$ (top right).
This agrees with an effective CG description at $\beta=1.4$, $\beta=2$ and $\beta=2.6$ from left to right, as found in the local bulk statistics \cite{ABCPRS,AMP}. 

As a second feature, notice that, on the horizontal $x$-axis with the origin in the middle, points on the positive $x$-axis correspond to an angle $\theta=0$, that is NN and NNN are aligned. In contrast, points on the negative $x$-axis correspond to $\theta=\pi$, that is, NN and NNN are anti-aligned, cf. Figure \ref{fig-alignedspacings} below.
We observe, that due to the repulsion between points, the density for aligned NN and NNN with $\theta\approx 0$ is also suppressed, in particular close to radius $r=1$ (dark right hole). Again, the effect increases with the repulsion in $\beta$.
At the edge, in the bottom row in Figure \ref{fig-spacing ratio symmetry classes}, we first of all observe the same repulsion pattern and ordering in $\beta$ as in the bulk at the origin and small angle, with a weakening repulsion at $\theta\approx0$. In addition, also the case of anti-aligned NN and NNN at $\theta\approx \pm\pi$ is suppressed, again increasing with the repulsion in $\beta$. How can we understand this additional feature at the edge?
\begin{figure}[t]
	\includegraphics[width=0.3\linewidth]{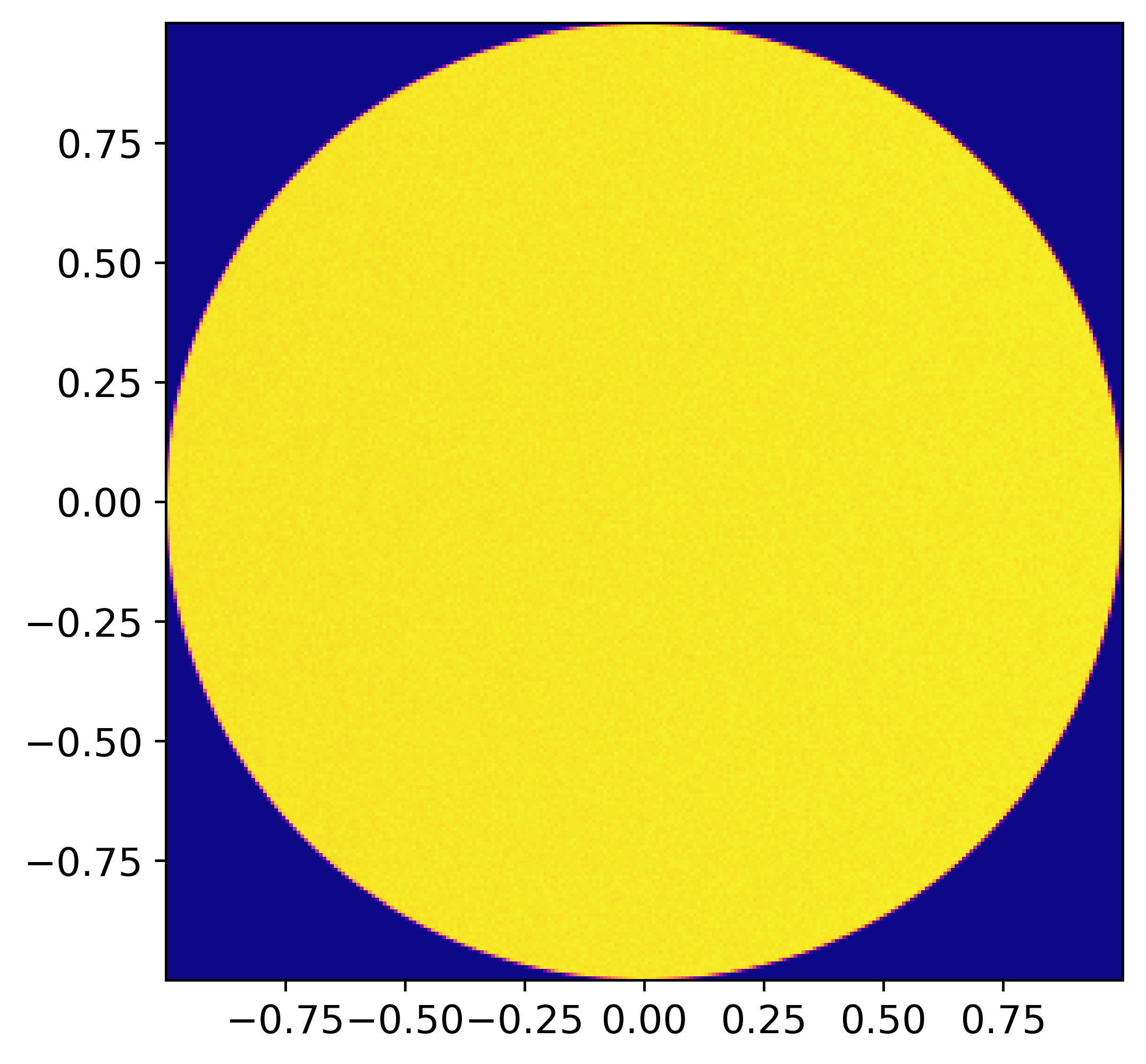} 
	\includegraphics[width=0.3\linewidth]{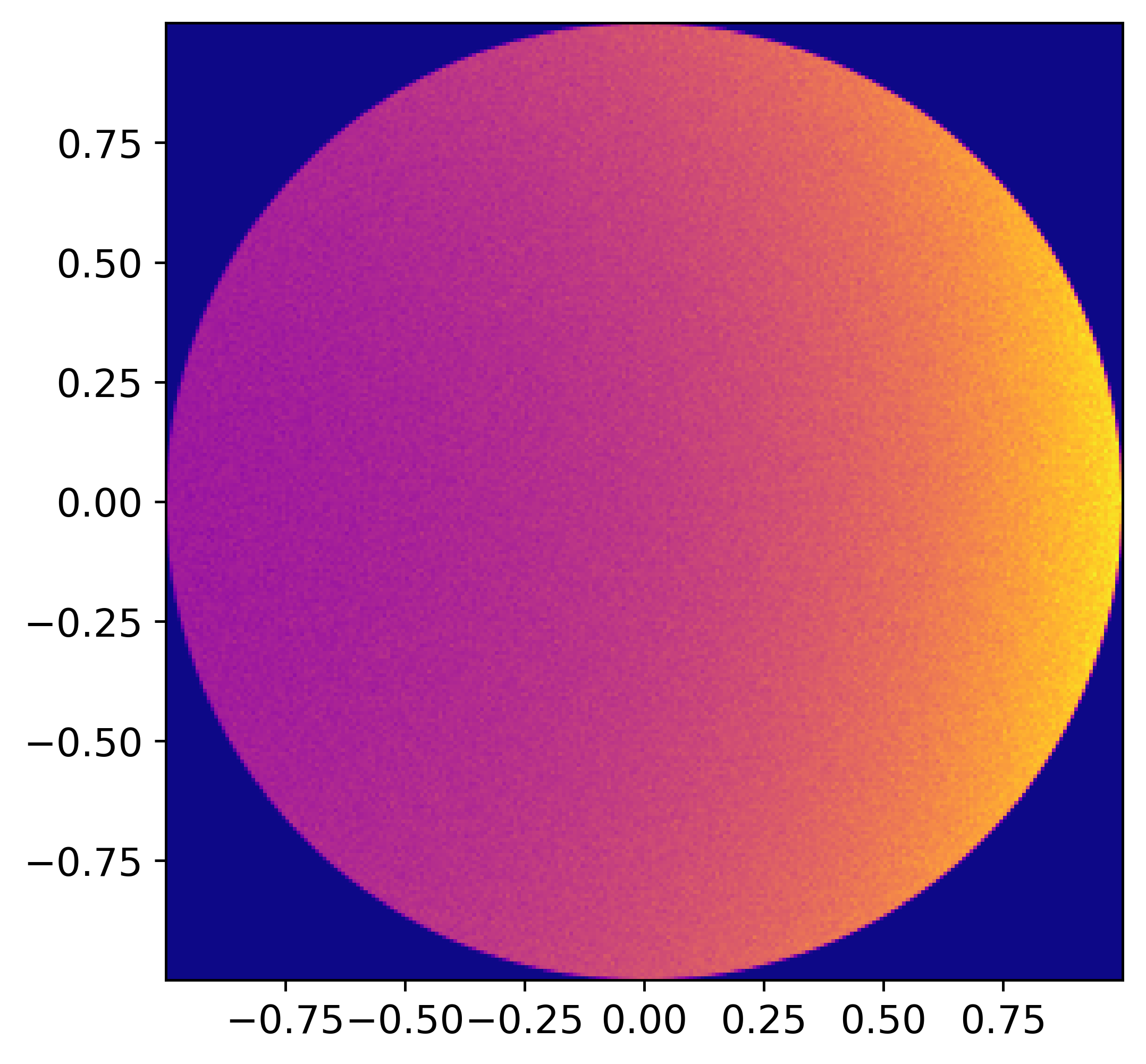}
	\raisebox{5pt}{\includegraphics[scale=0.43]{Figures_Sec4/colorbar.png}}
	\caption[Figure 7 from \cite{AAAMPRW}, \textit{Density plots of the complex spacing ratio for 2d Poisson}]{Density plots of the complex spacing ratio in the bulk (left) and at the edge (right) for the 2d Poisson ensemble. We consider $720\,000$ ensembles of size $N=1024$.\\ \cite[Figure 7]{AAAMPRW}
	}
	\label{fig-spacing ratio Gaussian Poisson}
\end{figure}

For comparison, let us first look at the edge effect in the complex spacing ratio of the 2d Poisson ensemble. In Figure \ref{fig-spacing ratio Gaussian Poisson} we present the bulk (left) and edge (right) spacing ratio distribution. Due to vanishing repulsion, the complex spacing ratio is flat in the bulk. 
At the edge (right), we observe a suppression except for small to intermediate angles close to radius $r\lessapprox 1$. 
As drawn schematically in Figure \ref{fig-alignedspacings},
the reason for that is when a point $z_1\ (\Delta)$ is in the edge region, it is favourable to have both NN $z_2\ (\bullet)$ and NNN $z_3\ (\circ)$ aligned in the direction of the increase of the density 
towards the bulk, as there is no repulsion between them. For anti-aligned points, most likely one of the points is at a larger radius, $r_k\gtrapprox1$, where the density is highly suppressed - unless they anti-align parallel to the edge. 

Coming back to the effect at the edge spacing ratio for classes AI$^\dag$, A, AII$^\dag$, in the bottom row in Figure \ref{fig-spacing ratio symmetry classes} there is a competition between repulsion (in favour of $\theta\approx \pm\pi$) and an almost vanishing density outside the edge (in favour of $\theta\approx0$). Consequently, an angle in around $\theta\lessapprox\pm\pi/2$ in the first and fourth quadrant is favoured in these plots, balancing repulsion and a vanishing spectral density outside the support (cf. sketch in Figure \ref{fig-alignedspacings} right).
This is most pronounced for class AII$^\dag$, with the largest effective value of $\beta$. 
The same behaviour can be seen much clearer when considering radial moments, see Figure \ref{fig-spacing ratio 1D dist} (bottom middle), discussed below.

The above comparison of the complex spacing ratio in the bulk and at the edge assumes, that the unfolding drops out also at the edge in the ratio of the difference with the NN and NNN eigenvalue \eqref{eq-def spacing ratio}, respectively. We notice that at the edge the density varies strongly on the scale of the mean level spacing $\sim O(1/\sqrt{N})$, see Figure \ref{fig-radial density plot symmetry classes edge and Poi} left.\clearpage

\begin{figure}
	\setlength{\unitlength}{1mm}
	\begin{picture}(150,15)
		\put(10,1){\circle{2}}
		\put(18,1){\circle*{2}}
		\put(31,0){$\Delta$}
		\put(8,6){$z_3$}
		\put(16,6){$z_2$}
		\put(31,6){$z_1$}
		\put(72,1){\circle*{2}}
		\put(78,0){$\Delta$}
		\put(94,1.5){\circle{2}}
		\put(70,6){$z_2$}
		\put(78,6){$z_1$}
		\put(92,6){$z_3$}
		\put(128,0){\circle*{2}}
		\put(132,-1){$\Delta$}
		\put(131.5,9){\circle{2}}
		\put(122,0){$z_2$}
		\put(137,0){$z_1$}
		\put(136,8){$z_3$}
	\end{picture}
	\caption[Figure 8 from \cite{AAAMPRW}, \textit{Schematic picture of eigenvalue positions}]{\textbf{Left:} \textit{aligned} NN $z_2\ (\bullet)$ and NNN $z_3\ (\circ)$ of point $z_1\ (\Delta)$ with angle $\theta\approx 0$ in the complex plane; \textbf{Middle:} \textit{anti-aligned} NN and NNN with angle $\theta\approx \pm\pi$; \textbf{Right:} NN and NNN with angle $\theta\approx \pi/2$. \cite[Figure 8]{AAAMPRW}}
	\label{fig-alignedspacings}
\end{figure}
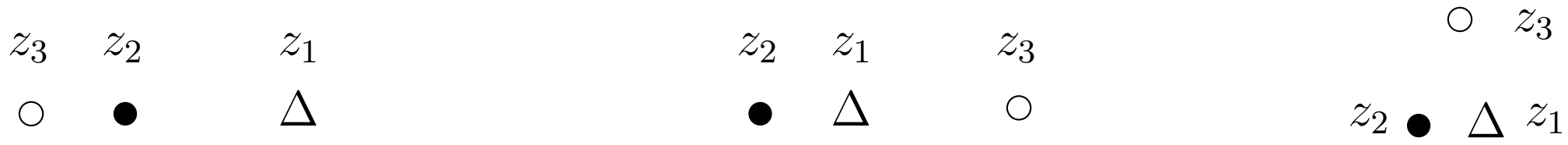
\noindent\textbf{Moments of Complex Spacing Ratios}\\
\\
We turn to the study of 1d and 0d moments of the complex spacing ratio, by integrating over the radial, respectively angular part of the distribution, or both. Such 1d curves or real values are much more favourable for a comparison to data. The radial respectively angular marginal for all symmetry classes is defined as follows:
\begin{align}
	\rho^{(N)}(r)&= \int_{-\pi}^{\pi}\! d \theta \ r \varrho^{(N)}(r\exp{[i \theta]}), 
	\label{eq-rad-marg n}\\
	\rho^{(N)}(\theta)  &= \int_{0}^{1}\! d r \ r \varrho^{(N)}(r\exp{[i \theta]}) .
	\label{eq-ang-marg n}
\end{align}
We integrate with respect to polar coordinates, $d ^2z=d r\ r d\theta$, \textit{including} the radial part $r$ in the angular integral. For a constant density $\varrho^{(N)}(r\exp{[i \theta]})=const.$, as in 2d Poisson, see Figure \ref{fig-spacing ratio Gaussian Poisson} (left), we thus obtain a linear radial density from \eqref{eq-rad-marg n}.

For the scalar, 0d moments to be discussed later, see Tables \ref{tab-scalar moments bulk} and \ref{tab-scalar moments edge} and Figure \ref{fig-scalar moments spacing ratio} below, we define
\begin{align}
	\langle r^{k} \cos(\phi)^l\rangle &:=  \int_{-\pi}^{\pi}\! d \theta\int_{0}^{1}\! d r \  r^{k+1}
	\cos(\theta)^l
	\varrho^{(N)}(r\exp{[i\theta]}) ,\quad k,l=0,1,2.
	\label{eq-r2mom}
\end{align}
\begin{figure}[h]
	\hspace{70pt}{bulk} \hspace{110pt}{edge}\hspace{110pt}{(bulk-edge)} \hfill\\
	\begin{turn}{90}\hspace{30pt}{radial}\end{turn}
	\includegraphics[width=0.31\linewidth]{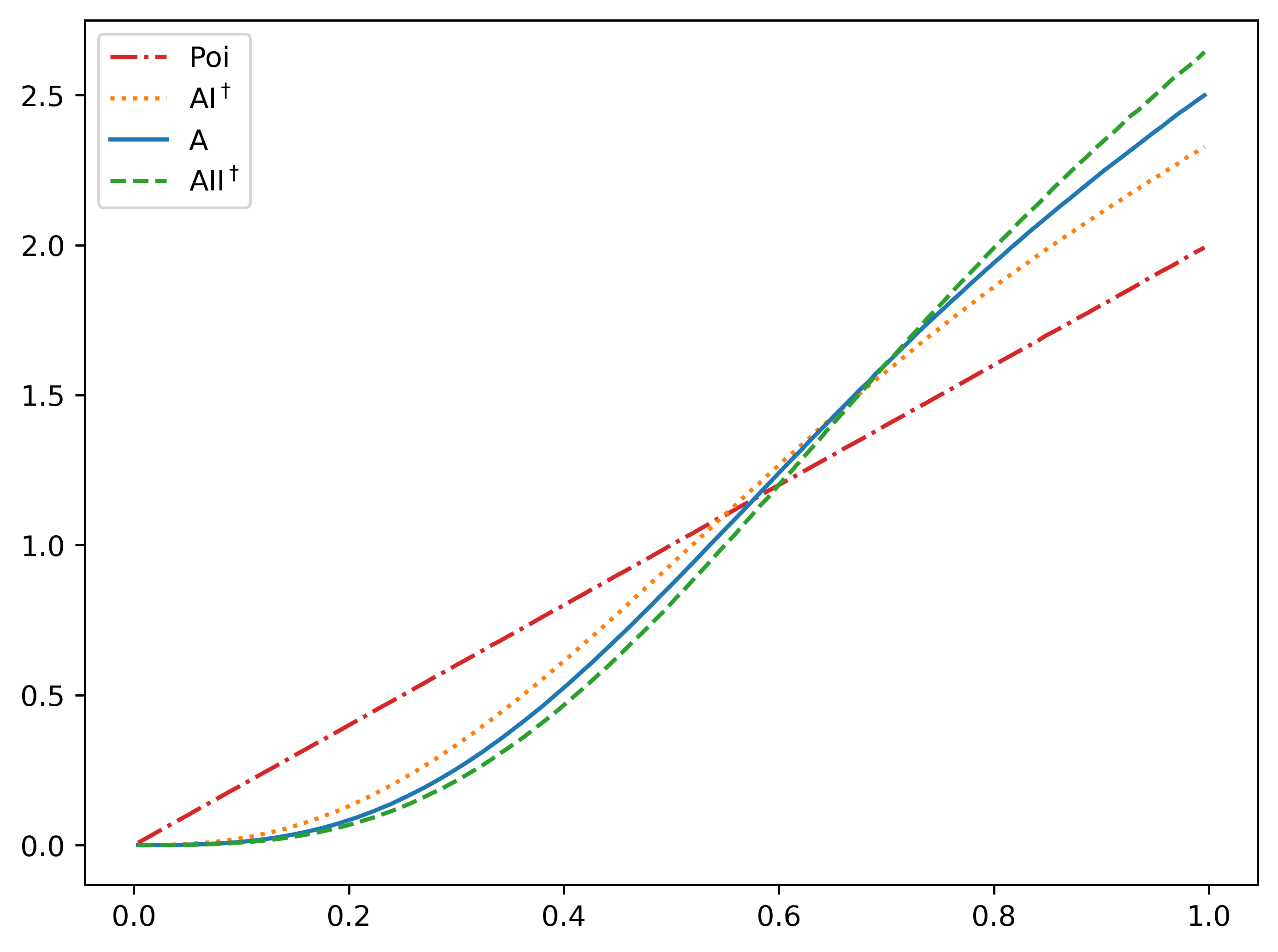} 
	\includegraphics[width=0.31\linewidth]{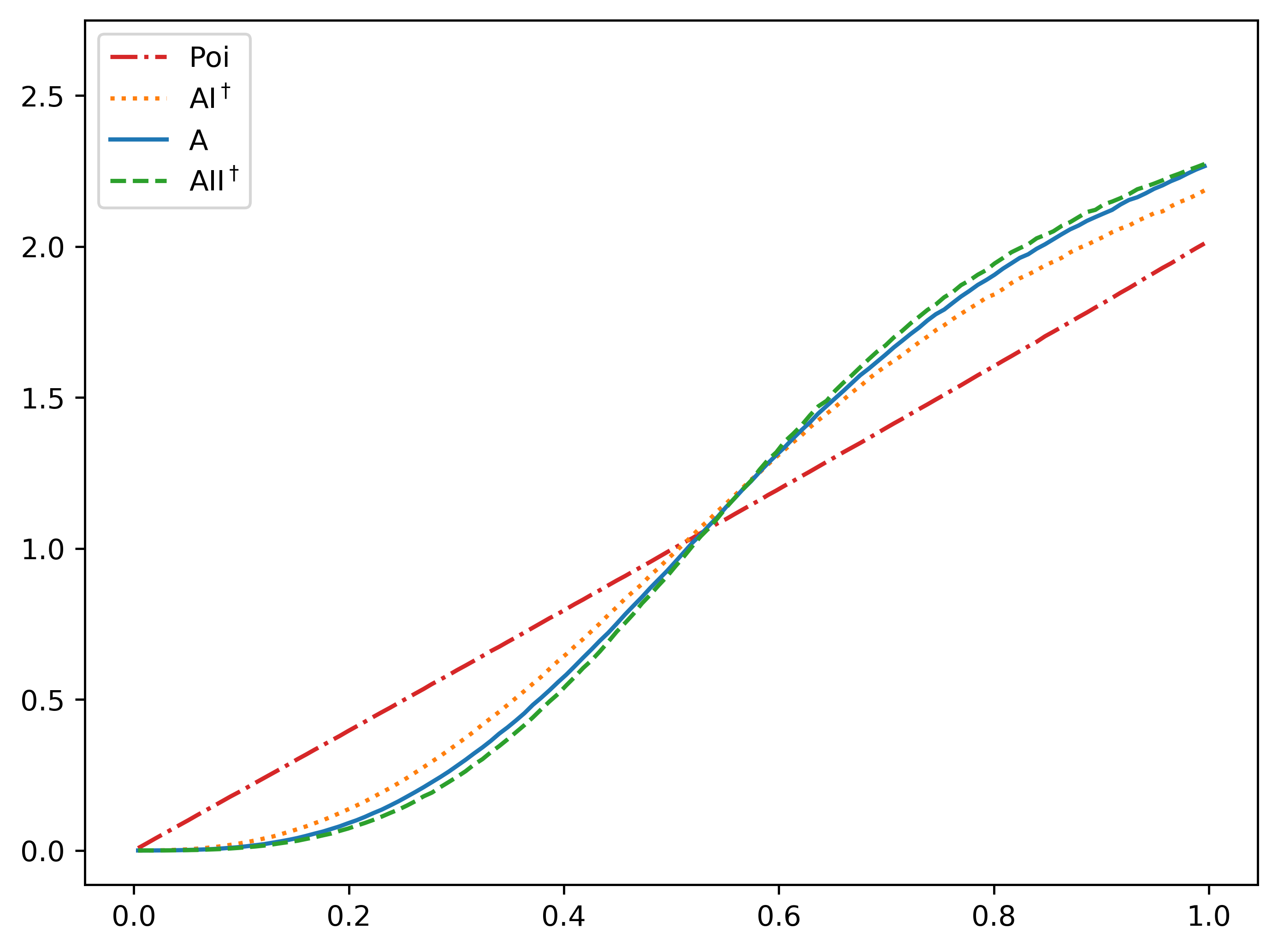}
	\includegraphics[width=0.31\linewidth]{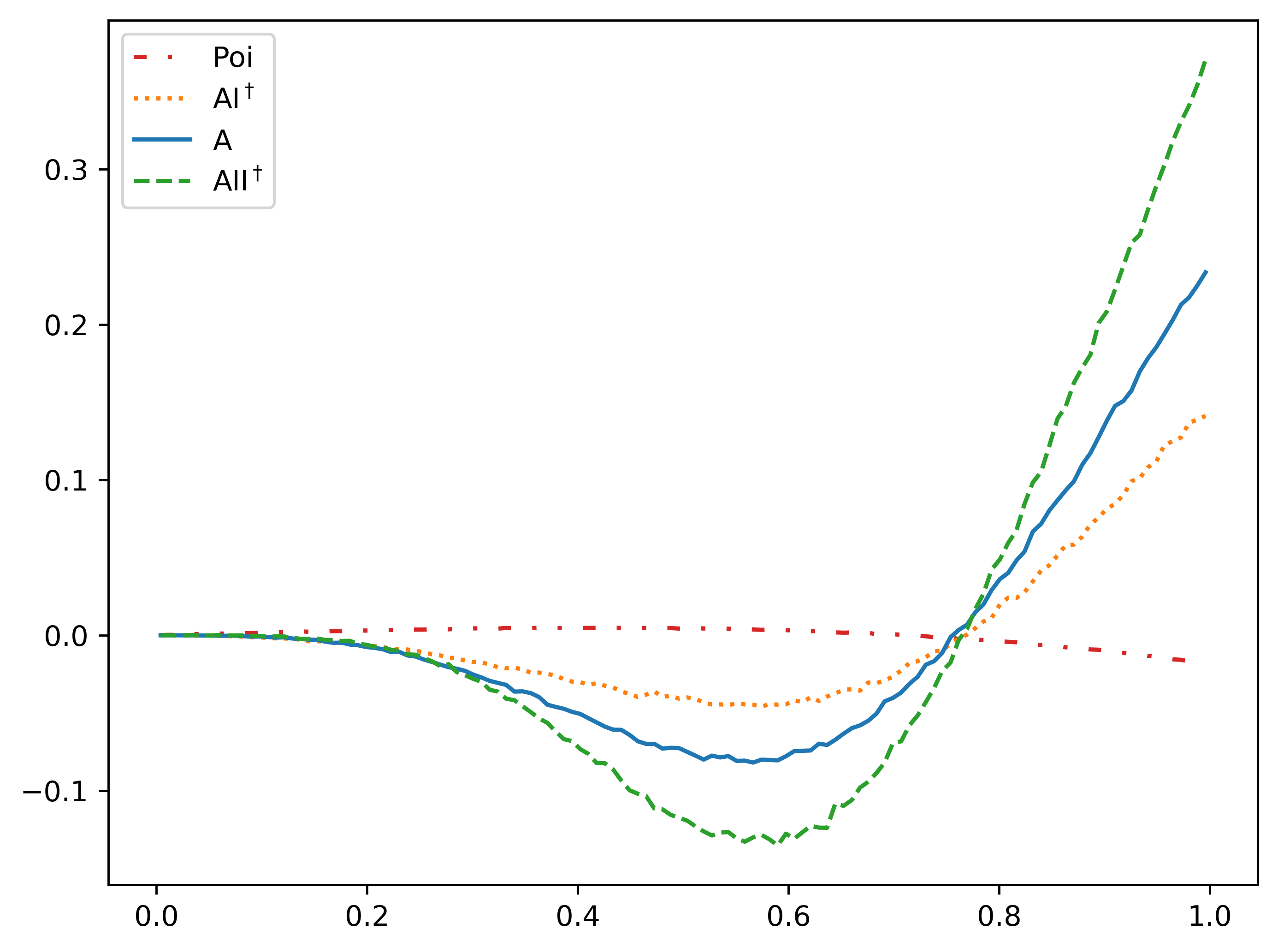}\\
	\begin{turn}{90}\hspace{30pt}{angular}\end{turn}
	\includegraphics[width=0.31\linewidth]{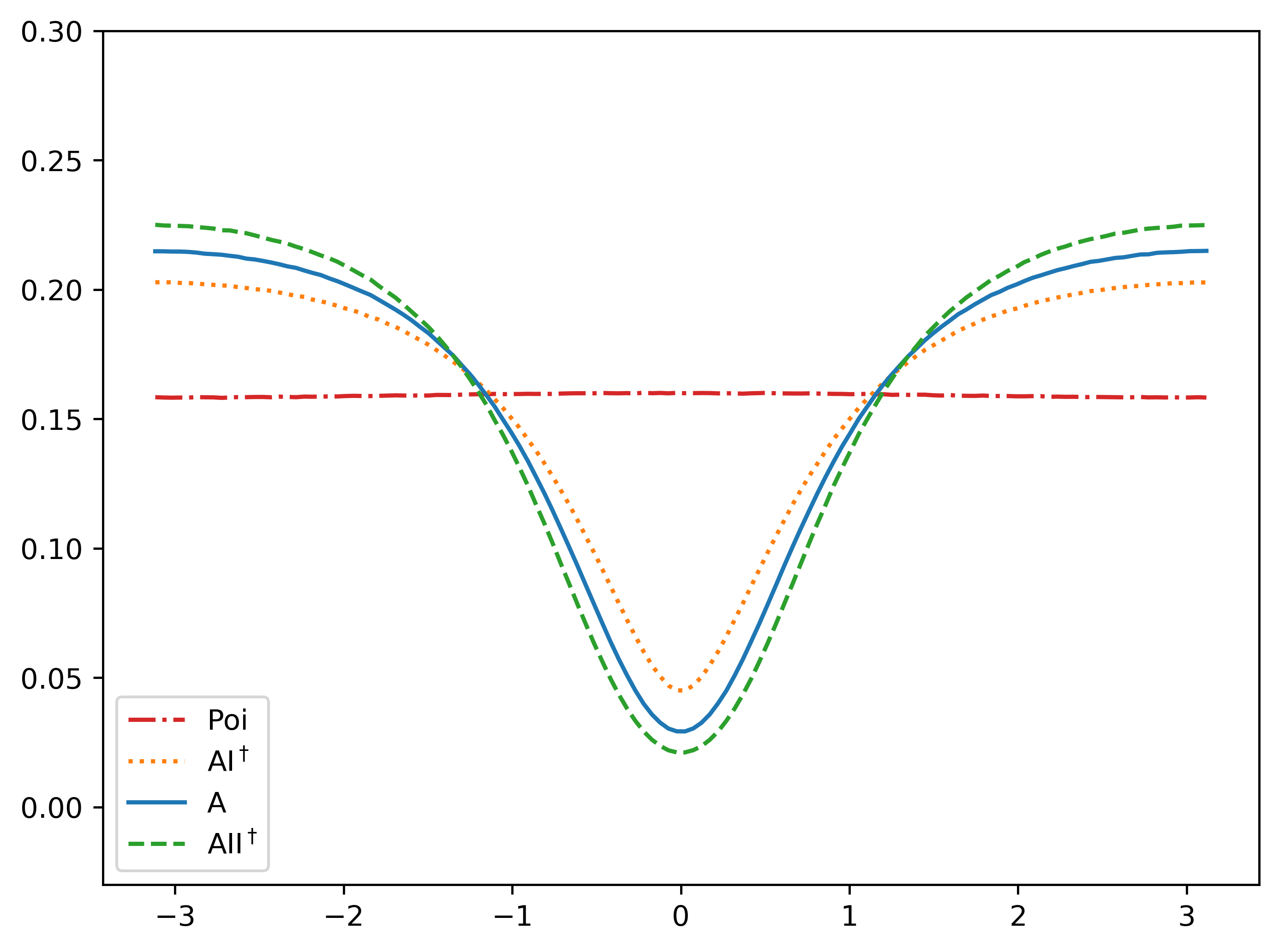}
	\includegraphics[width=0.31\linewidth]{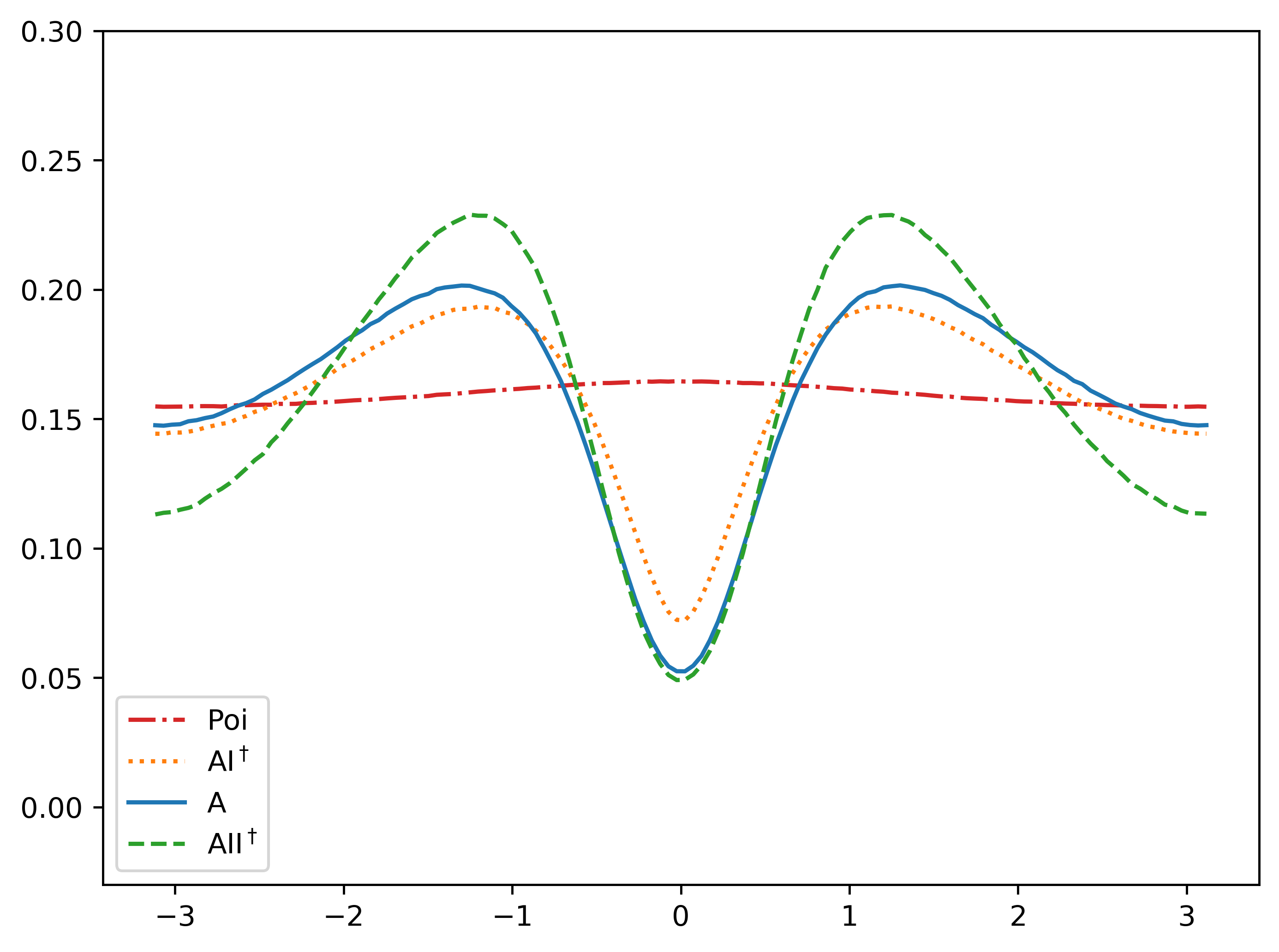}
	\includegraphics[width=0.31\linewidth]{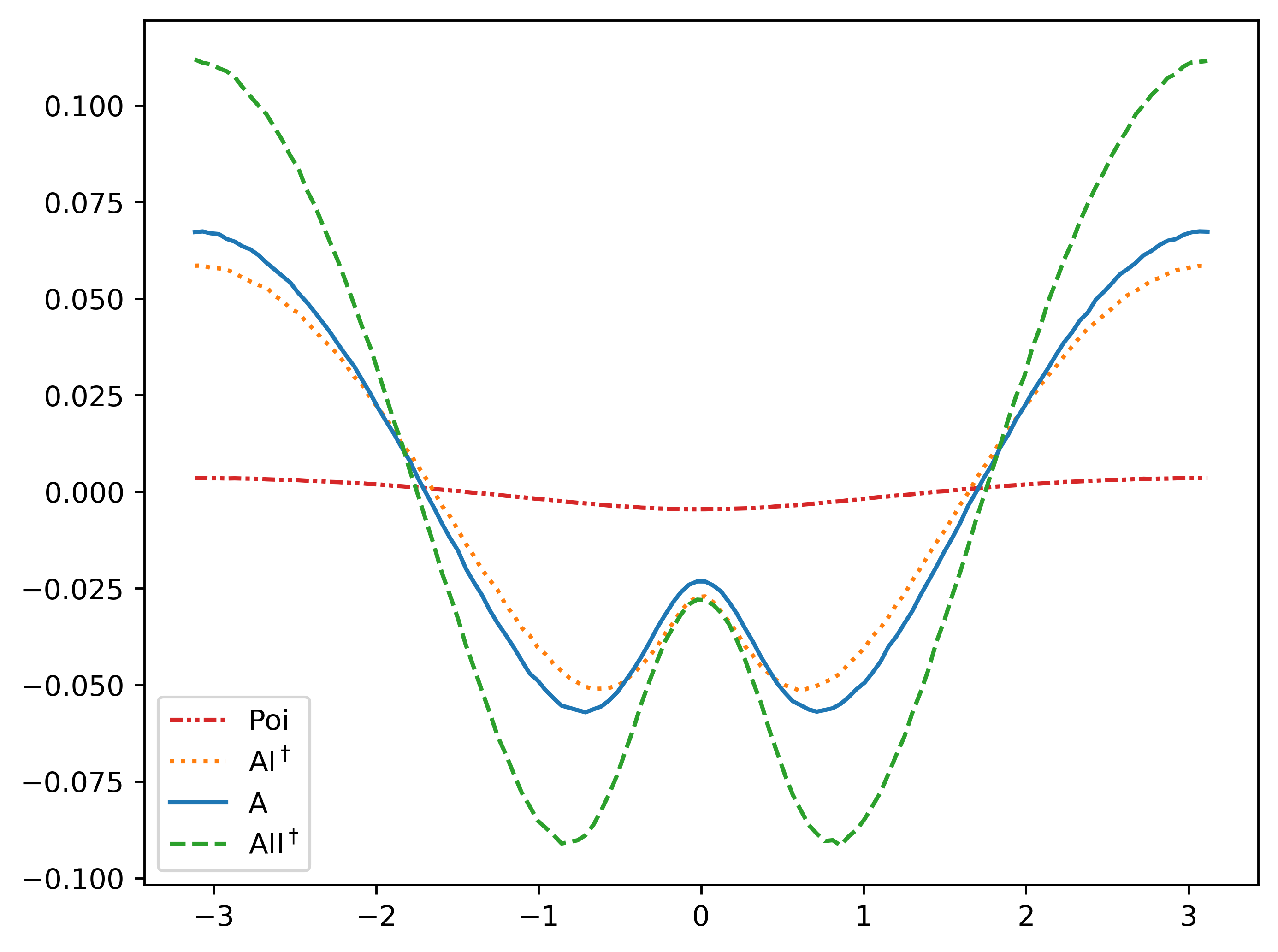}
	\caption[Figure 9 from \cite{AAAMPRW}, \textit{Angular and radial distributions of the complex spacing ratio in the bulk and at the edge for classes AI$^\dagger$, A, AII$^\dagger$}]{Radial (top row) and angular (bottom row) marginal distributions \eqref{eq-rad-marg n} and \eqref{eq-ang-marg n} in the bulk (left) for $r<r_b$, at the edge (middle) for $r>r_-$, and their difference bulk minus edge (right). The curves denote 2d Poisson (red dot-dashed), class AI$^\dag$ (orange dotted), A (blue full) and AII$^\dag$(green dashed), each with $720\,000$ copies of $N=1024$ matrices, respectively $2N=1024$ for class AII$^\dag$. \cite[Figure 9]{AAAMPRW}}
	\label{fig-spacing ratio 1D dist}
\end{figure}
The results for the marginal densities defined above are shown in Figure \ref{fig-spacing ratio 1D dist} below: 
The left plots give the radial (top) and angular (bottom) bulk distribution, agreeing with previous studies (cf. \cite{SaRibeiroProsen, KanazawaWettig}).
The same marginals at the edge are given in the middle column, and their difference to the bulk in the right column. In all plots the same coloured line code is used:
2d Poisson (red dot-dashed), complex symmetric AI$^\dag$ (orange dotted), complex Ginibre A (blue full) and complex self-dual AII$^\dag$ (green dashed). 
We briefly recall the findings in the bulk.
For the 2d Poisson process there is no correlation, with a linear rise in the radial (top) and flat curve in angle (bottom) - in agreement with the flat density in Figure \ref{fig-spacing ratio Gaussian Poisson} (left). 
The three symmetry classes AI$^\dag$, A, AII$^\dag$ show a repulsion increasing with the effective $\beta$ from the origin in radius (top) and from angle $\theta=0$ in the bulk plots.

Let us turn to the new results for the radial and angular distribution at the edge, 
in the second column of Figure \ref{fig-spacing ratio 1D dist}.  
While the radial distribution (top middle) is qualitatively very similar to the bulk results, see top right for the quantitative differences (in different units), the situation for the angular distribution differs significantly at the edge, see the bottom middle plot of Figure \ref{fig-spacing ratio 1D dist}. First, in 2d Poisson a small angle is now slightly favoured, as we already saw in Figure \ref{fig-spacing ratio Gaussian Poisson} right. 
For the other three symmetry classes, in addition to the minimum at $\theta=0$ (i.e. suppressed aligned NN and NNN),
there is a second minimum suppressing anti-aligned NN and NNN at $\theta=\pm\pi$. In between, the distributions develop two maxima at approximately $\theta\approx \pm\pi/2$, symmetrical with respect to $\theta \rightarrow -\theta$. The height/depth of the maxima/ minima increases with the repulsion in the effective $\beta$ values. 
This makes the picture we obtained from the previous paragraph from the 2d plots more quantitative.

The next step is to study the 0d scalar moments of the complex spacing ratio, see \eqref{eq-r2mom}.
These quantities certainly depend on the dimension and ensemble size of the studied class, as discussed in \cite[Figure 1]{DusaWettig}. This has to be kept in mind when comparing to the values in the literature. We choose a sufficiently large enough $N=1024$ (respectively $2N=1024$) 
to allow for a large number of ensembles within each class. This serves as a good compromise between large $N$ (which is computationally intensive) and the number of ensembles to minimise the statistical error.

\begin{table}[h]\centering
	\caption[Table 1 from \cite{AAAMPRW}, \textit{Scalar moments of the complex spacing ratio in the bulk for classes AI$^\dagger$, A, AII$^\dagger$ and the 2d Poisson process}]{Various scalar moments in the bulk (up to $r<r_b=0.8$), for $720\,000$ matrices of size $N=1024$ (respectively $2N=1024$). 
		We reproduce the known results \cite{SaRibeiroProsen, DusaWettig, KanazawaWettig} for 2d Poisson (analytical and numerical) and symmetry classes AI$^\dag$, A, AII$^\dag$. \\ \cite[Table 1]{AAAMPRW}
	}
	\begin{tabular}{|c|c || c | c | c | c | c|}
		\hline
		&\textbf{bulk moments}	 & $ \langle r \rangle $ & $ \langle r^2 \rangle $ & $ \langle \cos(\phi) \rangle $ & $ \langle \cos(\phi)^2 \rangle $ & $ \langle r\cos(\phi) \rangle $ \\ \hline
		& 2d Poisson exact & $2/3$ & $1/2$ & $0$ & $1/2$ & $0$\\ \hline
		\multirow{4}{*}{$r<0.8$}&2d Poisson & $0.66666 $ & $ 0.50000 $ & $\ 0.00273 $ & $ 0.50000 $ & $\ 0.00204 $  \\ \cline{2-7}
		& class AI$^\dag$ & $ 0.72231 $ & $ 0.56065 $ & $ -0.19532 $ & $ 0.45780 $ & $ -0.14868 $ \\ \cline{2-7}
		& class A & $ 0.73867 $ & $ 0.58086 $ & $ -0.24685 $ & $ 0.44991 $ & $ -0.18884 $\\ \cline{2-7}
		& class AII$^\dag$ & $ 0.74893 $ & $ 0.59431 $ & $ -0.28580 $ & $ 0.44684 $ & $ -0.21928 $\\ \hline
		
	\end{tabular}
	\label{tab-scalar moments bulk}
\end{table}
\begin{table}[h]\centering
	\caption[Table 2 from \cite{AAAMPRW}, \textit{Scalar moments of the complex spacing ratio at the (extended) edge for classes AI$^\dagger$, A, AII$^\dagger$ and the 2d Poisson process}]{The same scalar moments as in Table \ref{tab-scalar moments bulk} at the edge ($r>r_-\approx 0.96875$) (top) and the enlarged edge ($r>r_-'\approx0.9375$) (bottom) of the spectrum. \cite[Table 2]{AAAMPRW}
	}
	\begin{tabular}{|c|c || c | c | c | c | c|}
		\hline
		&\textbf{edge moments}	 & $ \langle r \rangle $ & $ \langle r^2 \rangle $ & $ \langle \cos(\phi) \rangle $ & $ \langle \cos(\phi)^2 \rangle $ & $ \langle r\cos(\phi) \rangle $ \\ \hline
		\multirow{4}{*}{$r>r_-$ }&
		2d Poisson & $ 0.68393 $ & $ 0.52171 $ & $ \ 0.17029 $ & $ 0.51219 $ & $\  0.13014 $\\ \cline{2-7}
		& class AI$^\dag$ & $ 0.71576 $ & $  0.55120 $ & $\  0.09769 $ & $ 0.43501 $ & $ \ 0.06779 $\\ \cline{2-7}
		& class A & $ 0.72735 $ & $ 0.56449 $ & $ \ 0.10872 $ & $ 0.40441 $ & $ \ 0.07473 $\\ \cline{2-7}
		& class AII$^\dag$ & $ 0.74438 $ & $ \ 0.58542 $ & $ 0.22132 $ & $ 0.35678 $ & $\  0.15712 $\\ \hline \hline
		\multirow{4}{*}{$r>r_-'$ }&
		2d Poisson & $ 0.68154 $ & $ 0.51870 $ & $ \ 0.14995 $ & $ 0.51025 $ & $\  0.11447 $\\ \cline{2-7}
		& class AI$^\dag$ & $ 0.71543 $ & $  0.55110 $ & $ -0.02321 $ & $ 0.44471 $ & $ -0.02115 $\\ \cline{2-7}
		& class A & $ 0.72735 $ & $ 0.56502 $ & $ -0.05943 $ & $ 0.42718 $ & $ -0.04899 $\\ \cline{2-7}
		& class AII$^\dag$ & $ 0.73233 $ & $ \ 0.57065 $ & $ 0.00686 $ & $ 0.39679 $ & $ -0.00022 $\\ \hline
	\end{tabular}\label{tab-scalar moments edge}
\end{table}

\begin{figure}[h!]
	\hspace{100pt}{$\langle r\rangle$} \hspace{210pt}{$\langle r^2 \rangle$} \hfill\\
	\includegraphics[width=0.49\linewidth]{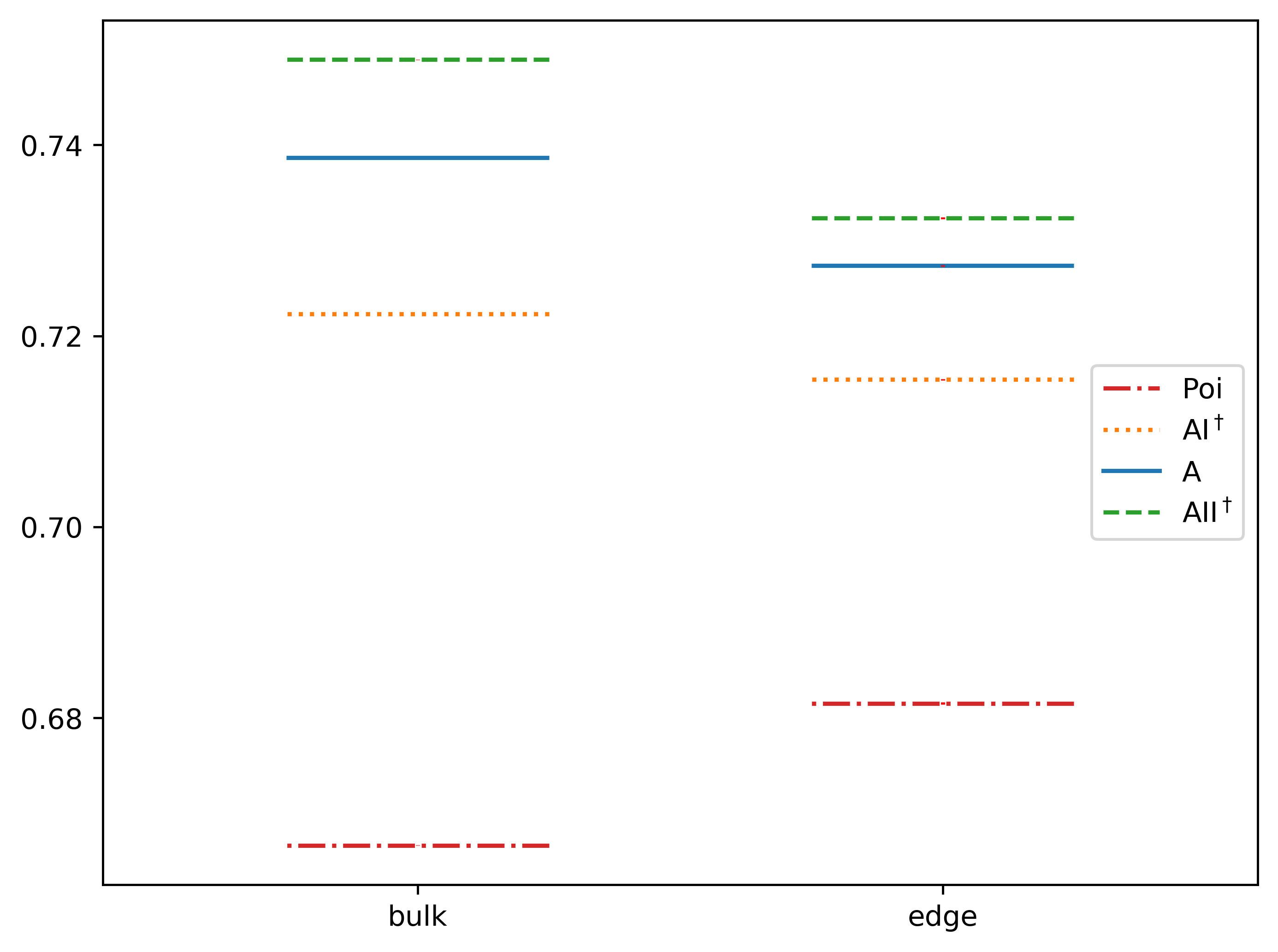} 
	\includegraphics[width=0.49\linewidth]{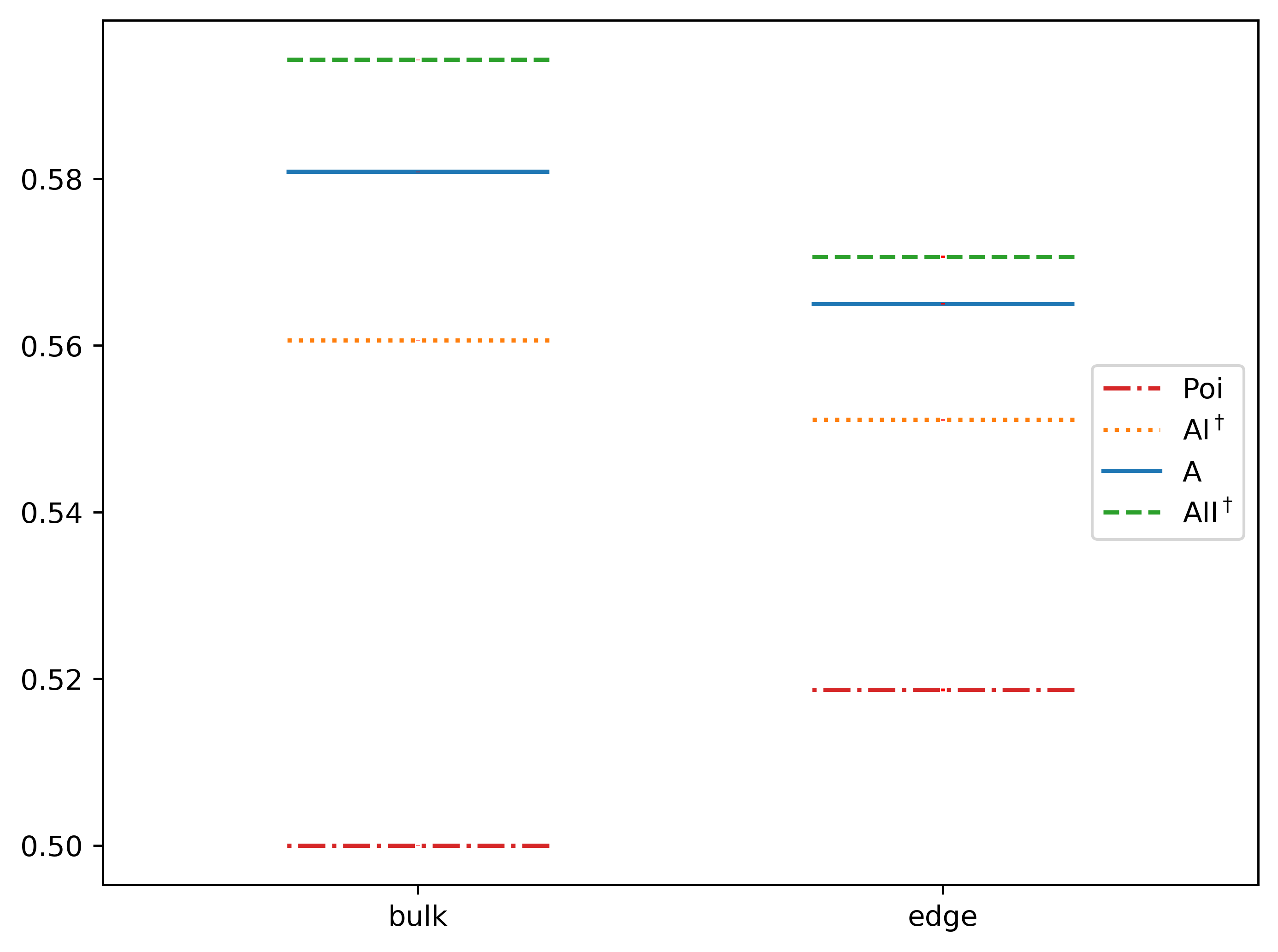}
	\includegraphics[width=0.49\linewidth]{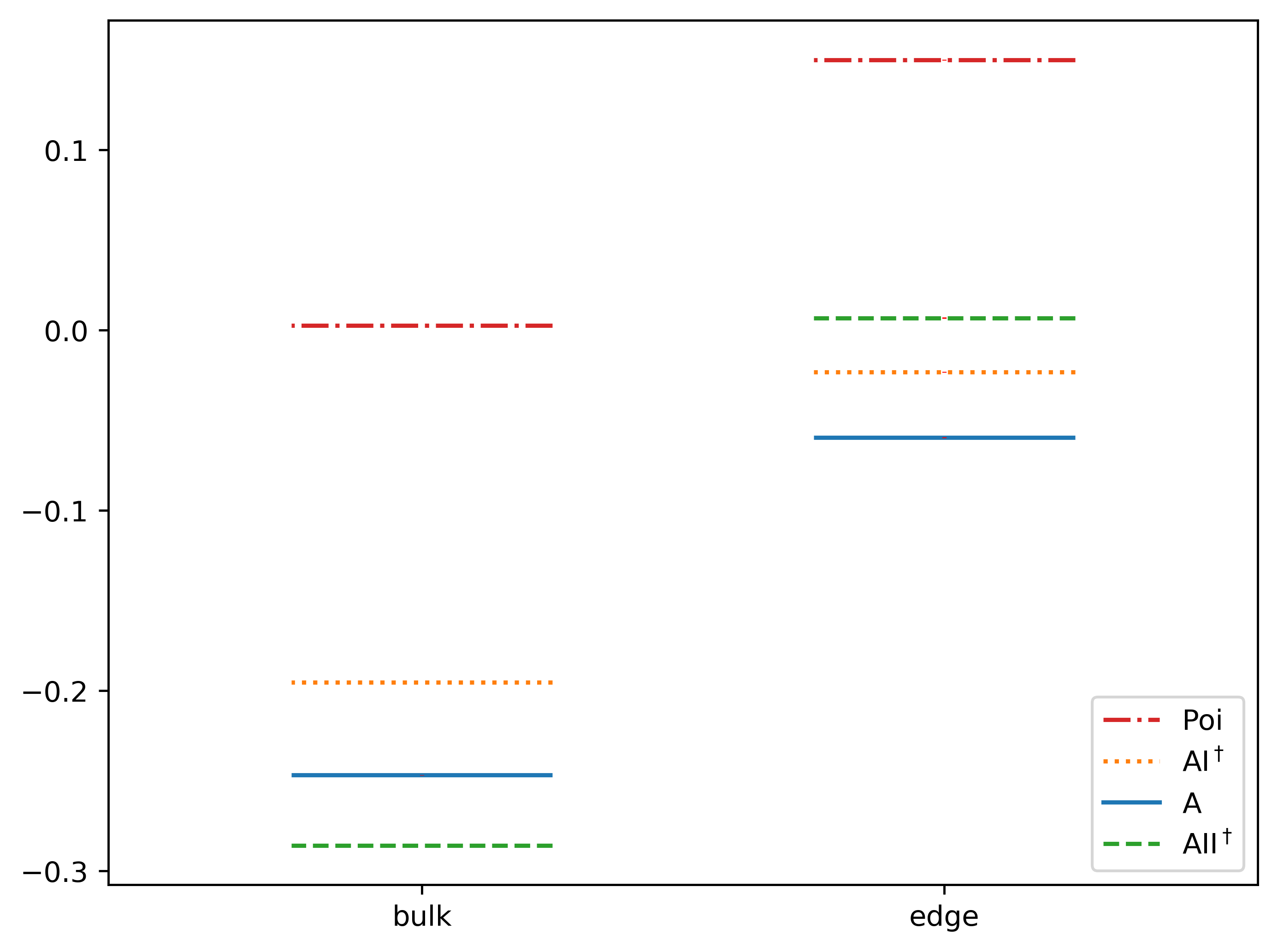}\hspace{8pt}
	\includegraphics[width=0.49\linewidth]{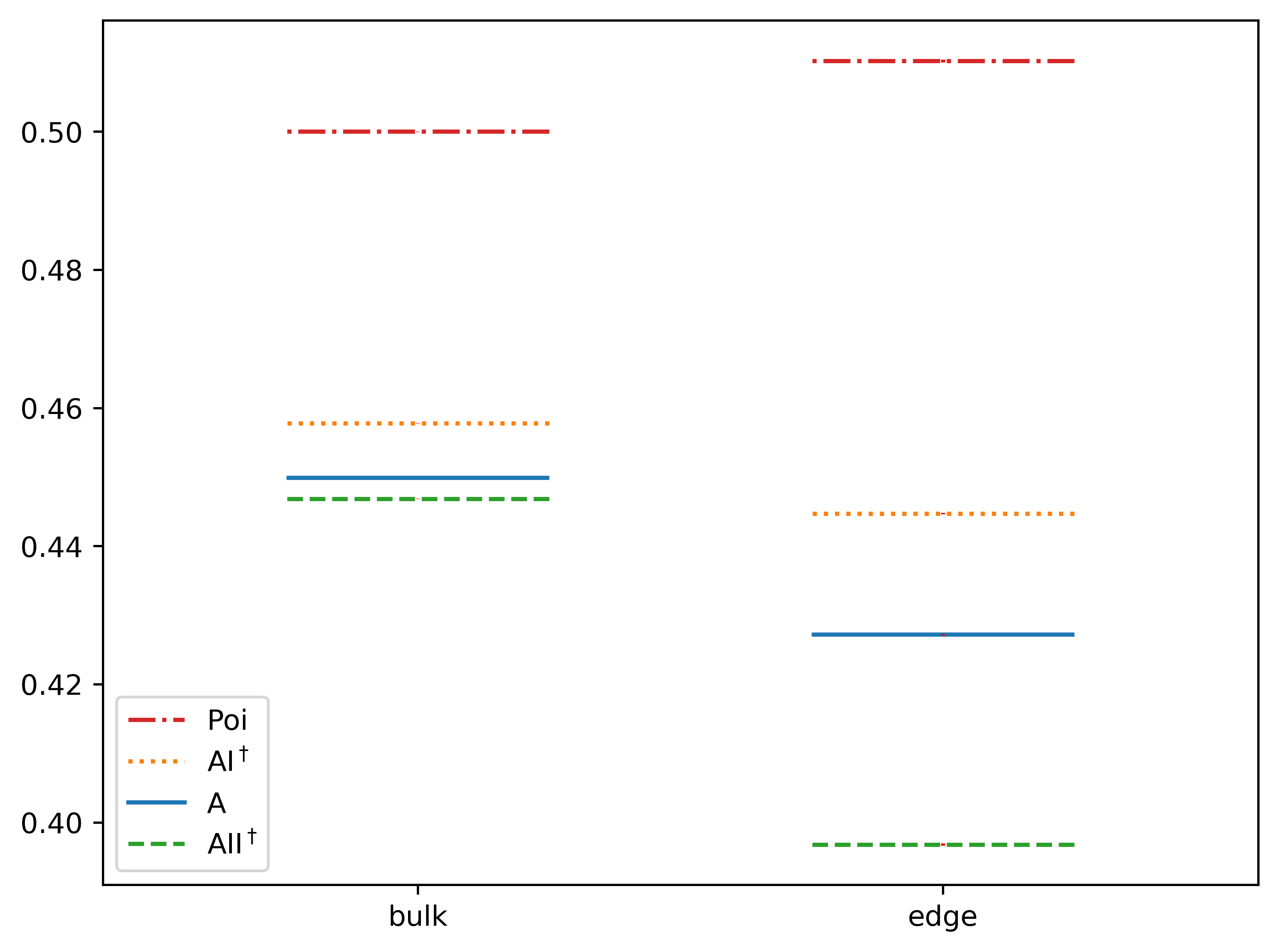}\\
	\hspace*{90pt}{$\langle \cos(\phi)\rangle$} \hspace*{190pt}{$\langle \cos(\phi)^2 \rangle$} \hfill
	\caption[Figure 10 from \cite{AAAMPRW}, \textit{Plot of the scalar moments of the complex spacing ratio in the bulk and at the edge for classes AI$^\dagger$, A, AII$^\dagger$ and the 2d Poisson process}]{\textbf{Top row:} Radial scalar moments $\langle r\rangle$ (left) and $\langle r^2 \rangle$ (right); \textbf{Bottom row:} Angular scalar moments $\langle \cos(\phi)\rangle$ (left) and $\langle \cos(\phi)^2 \rangle$ (right). We present the bulk for $r<r_b$ on the left sides, and the edge for $r>r_-$ on the right sides of each plot. We use $720\,000$ ensembles of size $N=1024$ (respectively $2N=1024$), denoting the 2d Poisson process (red dot-dashed), class AI$^\dag$ (orange dotted), A (blue full) and AII$^\dag$ (green dashed). While for radial moments $\langle r\rangle$, $\langle r^2\rangle$ and $\langle \cos(\phi)^2 \rangle$ the ordering with respect to $\beta$ is the same in the bulk (left sides) and the edge region (right sides), it is inverted at the edge for $\langle \cos(\phi) \rangle$. Also, relative gap sizes are more stable for radial moments. \cite[Figure 10]{AAAMPRW}
	}
	\label{fig-scalar moments spacing ratio}
\end{figure}

The results on the scalar moments $ \langle r \rangle,\  \langle r^2 \rangle,\  \langle \cos(\phi) \rangle, \  \langle \cos(\phi)^2 \rangle,$ and $ \langle r\cos(\phi) \rangle $ for $720\,000$ ensembles of size $N=1024$ (respectively $2N=1024$ for class AII$^\dag$) are given in Table \ref{tab-scalar moments bulk} for the bulk and in Table \ref{tab-scalar moments edge} in the edge region. Let us first discuss the bulk results, which were previously known (cf. \cite{SaRibeiroProsen, DusaWettig, KanazawaWettig}). For the 2d Poisson process, the moments of the complex spacing ratio can be calculated analytically in the bulk of the spectrum, cf. \cite{SaRibeiroProsen}. The results are given in the second row in Table \ref{tab-scalar moments bulk}. The numerically calculated results in the row below are in good agreement with the theory, up to the second or third digit. The vanishing moments $\langle \cos(\phi)\rangle$ and $\langle r\cos(\phi)\rangle$ reflect the rotational symmetry of the distribution, observed in Figures \ref{fig-spacing ratio Gaussian Poisson} left, and Figure \ref{fig-spacing ratio 1D dist} left. The radial moment $\langle r \rangle$ is a measure of distance between NN and NNN eigenvalue. Its value increases from top to bottom in the second column of Table \ref{tab-scalar moments bulk}, i.e. with the strength of the local repulsion in terms of $\beta$.
A similar increase (in absolute values) is observed for $\langle r^2\rangle,\ \langle \cos(\phi) \rangle$ and $\langle r\cos(\phi)\rangle$. Notice that the results for $\langle \cos(\phi) \rangle$ and $\langle r\cos(\phi)\rangle$ are negative in the bulk for all three symmetry classes AI$^\dag$, A and AII$^\dag$.

Next, we discuss the edge results, presented in Table \ref{tab-scalar moments edge} for the same ensemble sizes as before. 
We begin with the results for a smaller choice for the edge, with $r>r_-=1-1/\sqrt{N}$ in Table \ref{tab-scalar moments edge} top, where all spectral densities are in monotonous decay, see Figure \ref{fig-radial density plot symmetry classes edge and Poi} vertical black lines.
First, we note that the relative ordering of the radial moments $\langle r \rangle$ and $\langle r^2\rangle$ is consistent with the one in the bulk, i.e. we find an increase from top to bottom, with changes only in the second digit. Comparison with Table \ref{tab-scalar moments bulk} gives us that the mean radius of the complex spacing ratio is strictly smaller in the edge region than in the bulk, for all three ensembles under consideration. The confining potential in each matrix ensemble results into a finite support. When considering eigenvalues at the edge, these are more sensitive to boundary effects than in the bulk. This is reflected by the probability to find NN and NNN of an edge eigenvalue outside of the support to be significantly suppressed. Hence, the mean radius of the complex spacing ratio is smaller in the edge region than in the bulk. For the moments $\langle \cos(\phi) \rangle$ and $\langle r \cos(\phi)\rangle$, we find an inversion of the ordering of the symmetry classes compared to the bulk. Moreover, the values change sign for classes AI$^\dag$, A and AII$^\dag$.
For a better visual comparison, we present the moments given in Table \ref{tab-scalar moments bulk} and Table \ref{tab-scalar moments edge} top in Figure \ref{fig-scalar moments spacing ratio}. It summarises the effects we just described.

Let us turn to discuss the effect of a larger region for the edge, choosing $r>r_-^\prime=1-2/\sqrt{N}$, see Figure \ref{fig-radial density plot symmetry classes edge and Poi} vertical black dashed lines.
The corresponding values for the 0d moments are given in Table \ref{tab-scalar moments edge} bottom.
For 2d Poisson we see the smallest effect, where the angular moments $\langle \cos(\phi)\rangle$ and $\langle r\cos(\phi) \rangle$ which are odd powers of cosine are most sensitive (cf. Figure \ref{fig-radial density plot symmetry classes edge and Poi} (right)). While also for classes AI$^\dag$, A, AII$^\dag$ the radial moments are not very sensitive, the odd powers of cosine are very sensitive to the choice of $r_-^\prime$ and almost all change sign, moving towards their bulk value. In conclusion our more conservative choice for the edge region $r_-=1-1/\sqrt{N}$ (motivated from classes A and AI$^\dagger$, cf. Figure \ref{fig-radial density plot symmetry classes}), gives a clearer characterisation of the edge universality classes as visible in all complex spacing ratios and their moments.\\
We conclude that we verify the differences in the complex spacing ratio of the symmetry classes AI$^\dagger$, A and AII$^\dagger$ previously found in the literature, adding interpretation based on the effective description via 2dCG. Moreover, study the same quantities at the edge of the spectrum, where we assume that no unfolding is needed here. We observed that the results depend on the choice of the edge radius and moreover, the non-flat density of the complex spacing ratio for the 2d Poisson process in Figure \ref{fig-spacing ratio Gaussian Poisson} right, let us doubt that this assumption holds true at the edge. Again, we refer to \cite{AAAMPRW} to find a detailed study of the NN and NNN distributions for the three symmetry classes both in the bulk and at the edge, based on the same numerics \cite{data repository}, where an unfolding procedure is imposed.
\begin{Remark}
	We note that recently also the Krylov complexity of some of the 38 non-Hermitian symmetry classes was investigated \cite{krylov sv non-h}. While there, singular values are considered for the calculations, yielding a mapping onto the 10 Altland-Zirnbauer symmetry classes, a distinct behaviour of the three classes A, AI$^\dagger$ and AII$^\dagger$ was found. We will not discuss these results further, however, it is important to be aware that there exist other measures, which can be studied and it depends on the goal of each study, which one is the most convenient.
\end{Remark}
\subsection{Study of the Expectation Value of Characteristic Polynomials}\label{ssec-expectation of characteristic polynomials}
We set the stage for this section, by reminding the reader about the chronology of research on the universality of statistics in non-Hermitian RMT: A first conjecture based on numerics and heuristics arguments was published in \cite{Hamzaki et al}, where to that date only the $N=2$ spacing was known analytically for the classes AI$^\dagger$ and AII$^\dagger$.\footnote{Except for the spectral density at weak non-Hermiticity for class AI$^\dag$ in \cite{SFT}, which is given in terms of a triple integral.} Especially, only for class A the JPDF in terms of the complex eigenvalues was known at this time. Therefore, another technique came into flavour: The study of characteristic polynomials. See also \cite{Liu Zhang,Kulkarni Kawabata Ryu,Forrester24}, which appeared close in time. We will follow \cite{AAKP} closely, presenting only results for one pair of characteristic polynomials here, which were mainly contributed by the author. Most recently, the spectral density for class AI$^\dagger$ was derived in \cite{complex sym Gernot}.\\
The goal is to analytically confirm the conjecture that the two symmetry classes AI$^\dagger$ and AII$^\dagger$, do \textit{not} follow the local statistics of the complex Ginibre ensemble (class A). These ensembles are all Gaussian to keep the analysis simple. A direct computation based on their matrix representation for each symmetry class is feasible with the help of supersymmetric techniques, see \cite{Wegner} for an introduction of this method. \\
An alternative approach uses diffusion equations \cite{Liu Zhang}, which has been applied to the 9 classes considered in \cite{non-h classes}, including classes AI$^\dag$ and AII$^\dag$. In \cite{Liu Zhang}, a duality was derived expressing the average over $k$ pairs of characteristic polynomials with matrix size $N$ by matrix averages of dimension $k$, without taking the large $N$-limit though. We will recover these expressions using supersymmetry.

\subsubsection{The Setup}\label{sssec-setup expectation values}

We study the expectation value of a product of two characteristic polynomials, where one depends on the non-Hermitian random matrix $J$ and the other one on its complex conjugate $J^*$, i.e., the main object will be
\begin{align}
	D_N(z,w^*):=\langle \det[z\,\mathds{1}_{d_N}-J]\det[w^*\,\mathds{1}_{d_N}-J^*]\rangle , 
	\label{eq-def-char-pol}
\end{align}
with $z,w\in\mathbb{C}$ fixed, $\mathds{1}_{d_N}$ is the identity matrix of matrix dimension $d_N$, and $J$ is drawn from a Gaussian probability density
\begin{align}
	P(J)=\frac{1}{Z_N}e^{-\Tr (JJ^\dag)},
\end{align}
where the partition function $Z_N$ (normalisation constant) is specified below. The latter certainly strongly depends on from which non-Hermitian matrix space $J$ is taken. The average over an observable $\mathcal{O}(J)$ is denoted for all ensembles by 
\begin{align}
	\langle \mathcal{O}(J)\rangle
	:= \frac{1}{Z_N} 
	\int\![dJ]\,\mathcal{O}(J)
	e^{-\Tr (JJ^\dag)} ,
	\label{eq-vevdef}
\end{align}
where $[dJ]$ denotes the flat Lebesgue measure 
over all independent complex matrix elements. We underline that the measure $[dJ]$ looks different for the three ensembles when writing it out explicitly, as the possible symmetries relate the matrix entries. This leads to $N^2$ complex independent elements for class $A$, $N(N+1)/2$ for class AI$^\dag$ and $2N^2-N$ for class AII$^\dag$. The measure of a complex number $z = x + iy$ is chosen as $d^2z = dxdy$. We briefly revisit the three symmetry classes A, AI$^\dagger$, AII$^\dagger$ below.\footnote{Here, we choose again the ordering in such a way that the computations are of increasing complexity and not in increasing (effective) $\beta$ of the (assigned) 2dCG.}

\begin{itemize}
	\item For class A, which consist of independent Gaussian distributed entries, the normalization (partition function) of the JPDF in terms of the matrix elements $(J_{i,j})_{i,j=1}^N$, reads 
\begin{align}
	Z_N^\textrm{A} :=\int\![dJ]\ e^{-\Tr (JJ^\dag)} =\prod_{i,j=1}^N\int\! d^2J_{i,j}\ e^{-|J_{i,j}|^2} = \pi^{N^2}.
	\label{eq-ZA}
\end{align} 
\item For the class AI$^\dagger$ of complex symmetric matrices, i.e. $J=J^t$, the required invariance $J\to UJU^\dagger$ for any unitary matrix $U\in\textrm{U}(N)$ fixes the weight up to a scaling, when assuming that all independent matrix elements are Gaussian. This however also implies that the diagonal and off-diagonal matrix elements have a different variance. Its partition function is  
\begin{align}
	Z_N^{\textrm{AI}^\dag} &:=\!\int\![dJ]\ e^{-\Tr (JJ^\dag)} =
	\prod_{\ell=1}^N\int\! d^2J_{\ell ,\ell}\ e^{-|J_{\ell ,\ell}|^2} 
	\prod_{1\leq i<j\leq N}\int\! d^2J_{i,j}\ e^{-2|J_{i,j}|^2} \nonumber\\&= \pi^{N}\!\left( \frac{\pi}{2}\right)^{\frac{N(N-1)}{2}}\!\!\!\!. 
	\label{eq-ZCS}
\end{align} 
\item For the complex self-dual $2N\times 2N$ matrix $J$, we consider its representation via three complex non-Hermitian $N\times N$ matrices $A, B,C$, cf. Eq. \eqref{eq-para class AII dag}. The corresponding partition function, thus, reads 
\begin{align}
	Z_N^{\textrm{AII}^\dag} &:=\int\![dJ]\ e^{-\Tr (JJ^\dag)} 
	\label{eq-ZQS}\\
	&=\prod_{i,j=1}^N\int\! d^2A_{i,j}\ e^{-2|A_{i,j}|^2} 
	\prod_{1\leq n<\ell\leq N}\int\! d^2B_{n,\ell}d^2C_{n,\ell}\ e^{-2|B_{n,\ell}|^2-2|C_{n,\ell}|^2} \nonumber\\
	&= \left( \frac{\pi}{2}\right)^{2N^2-N}\nonumber
\end{align}
\end{itemize}

It is not difficult to see from the definition of the expectation value \eqref{eq-vevdef}, that for $zw^*\to\infty$ the RHS of 
\eqref{eq-def-char-pol} asymptotically becomes $\sim(zw^*)^N$ for the Ginibre and AI$^\dag$ ensembles (matrix dimension is $d_N=N$) while it will be $\sim(zw^*)^{2N}$ for the AII$^\dag$ ensemble because the matrices have dimension $d_N=2N$. 

In order to take the large $N$-limit later, we will adjust the normalisation, where we divide \eqref{eq-def-char-pol} by its value at vanishing arguments,
\begin{align}
	D_N(0,0)=\langle \det[J]\det[J^*] \rangle .
	\label{eq-norm-char-pol}
\end{align}
It is some natural choice of a reference point, despite that it will not cancel all $N$-dependent terms.

Below we will employ anticommuting complex Grassmann variables $\xi,\psi$, i.e., $\{\xi,\psi\}=\xi\psi+\psi\xi=0$, in order to write determinants as integrals.
We follow the conventions of Haake \cite{haake}, and we refer to \cite{Wegner} for an introduction into superalgebra and superanalysis.
Complex conjugation is chosen to satisfy the following relations
$(\xi^*)^*=-\xi$ for single and $(\xi\psi)^*=\xi^*\psi^*$ for a product of Grassmann variables. Complex conjugated variables anticommute equally, $\{\xi,\xi^*\}= \{\xi,^*\psi\}=0$. The integration convention of each Grassmann variable is chosen to be
\begin{align}
	\int\! d\xi =0\ , \quad \int \! d\xi\ \xi=1.
	\label{eq-Grassdef}
\end{align}
For a complex Grassmann vector of size $N$, $\xi^t=(\xi_1,\ldots,\xi_N)$ we, then, obtain for the determinant of an arbitrary $N\times N$ matrix $C$
\begin{align}
	\det[C ]=
	\int\! [d\xi]\ \exp[\xi^\dag C\xi] .
	\label{eq-Grassdet}
\end{align}
The integration measure is defined as $[d\xi]=\prod_{i=1}^N d\xi_id\xi_i^*$. 
A second identity is needed for the Pfaffian determinant of an arbitrary even dimensional antisymmetric matrix $D$ of size $2N\times 2N$, 
\begin{align}
	\Pf[D]=(-1)^{N(N+1)/2}
	\int\![d\xi]\ \exp\left[\frac12\begin{pmatrix}
	\xi^t,&\xi^{*\, t}\end{pmatrix} D
	\begin{pmatrix}
		\xi \\
		\xi^* \\
	\end{pmatrix}
	\right].
	\label{eq-GrassPf}
\end{align}

Finally, we would like to introduce some special functions that we will encounter repeatedly.
It is well-known \cite{ginibre} that the truncated exponential
\begin{align}
	E_N(x):= \sum_{j=0}^N\frac{x^j}{j!}=e^x \frac{\Gamma(N+1,x)}{\Gamma(N+1)}= e^x Q(N+1,x)\ ,
	\label{eq-def-eN}
\end{align}
plays an important role in the complex Ginibre ensemble (class A), cf. \eqref{eq-R1A}. It actually has its appearance for the other two ensembles studied by us.
The truncated exponential can be rewritten in terms of the upper incomplete Gamma function \cite[Eq. 8.4.10]{NIST}
\begin{align}
	\Gamma(a,x):= \int_x^\infty\! dt\  t^{a-1}e^{-t}\ ,
	\label{eq-def-Gax}
\end{align}
as well as its normalised version
\begin{align}
	Q(a,x):= \frac{\Gamma(a,x)}{\Gamma(a)}\ .
	\label{eq-def-Q}
\end{align}

At the edge of the complex Ginibre ensemble, 
one encounters the complementary error function
\begin{align}
	\erfc(x)= \frac{2}{\sqrt{\pi}}\int_x^\infty\! dt\  e^{-t^2} ,
	\label{eq-erfc-def}
\end{align}
whose relation to the error function $\erf$ is 
\begin{align}
	\erfc(y)=1-\erf(y)=1+\erf(-y)=2-\erfc(-y) ,
	\label{eq-erfc-rel}
\end{align}
and to the imaginary error function
\begin{align}
	\label{eq-erfi}
	\erfi(z)=-i\erf(iz)=\frac{2}{\sqrt{\pi}}\int_0^z \! ds\ e^{s^2}.
\end{align}
Those functions can be also found in the case of complex symmetric matrices (AI$^\dag$) while for the case of complex self-dual matrices (AII$^\dag$) we 
encounter Owen's $T$-function \cite{OwenT} defined as follows,
\begin{align}
	T(x,a):=\frac{1}{2\pi}\int_0^a \! dt\  \frac{e^{-\frac12 x^2(1+t^2)}}{1+t^2}, \quad x,a\in\mathbb{R}.
	\label{eq-OwenT-def}
\end{align}
We actually need the function analytically continued to the imaginary axis in its second argument,
\begin{align}
	\label{eq-OTi}
	T\left(x,ib\right)=\frac{i}{2\pi}\int_0^b\! du\ \frac{e^{-\frac12 x^2(1-u^2)}}{1-u^2} , \quad x\in\mathbb{R}\quad \textrm{and}\quad b\in(-1,1).
\end{align}
When going beyond the specified integral one needs to be aware of the emerging branch cuts on $b\in\mathbb{R}\setminus(-1,1)$. We will not run into this problem, though, as we avoid those cuts.

We will mention some properties of all these special functions where necessary in the computations.

\subsubsection{Results for One Pair of Characteristic Polynomials}\label{sssec-results 1 char poly}

We are now ready to state our first results for the expectation value of one pair of characteristic polynomials at finite matrix dimension $N$ for all three ensembles.

\begin{prop}
	[One pair of expected characteristic polynomials for finite $N$]
	\label{prop-char_exp}\

	Employing definitions \eqref{eq-def-char-pol} and \eqref{eq-norm-char-pol} as well as \eqref{eq-def-eN} and \eqref{eq-def-Q}, it holds for any finite $N\geq 1$
	\begin{align}
		\frac{D_N^{\textrm{A}}(z,w^*)}{D_N^{\textrm{A}}(0,0)}
		&=  E_N(zw^*)=e^{zw^*} Q(N+1,zw^*),
		\label{eq-DN-A}\\
		\frac{D_N^{\textrm{AI}^\dag}(z,w^*)}{D_N^{\textrm{AI}^\dag}(0,0)}&= E_N(2zw^*)
		-\frac{2zw^*}{N+1}E_{N-1}(2zw^*)\nonumber\\
		&= e^{2zw^*}\left(Q(N+1,2zw^*)-\frac{2zw^*}{N+1}Q(N,2zw^*)\right),
		\label{eq-DN-CS}\\
		\frac{D_N^{\textrm{AII}^\dag}(z,w^*)}{D_N^{\textrm{AII}^\dag}(0,0)}&= 
		\frac{N!}{(2N)!}\sum_{j=0}^N\frac{(2j)!}{j!}(4zw^*)^{N-j}E_{2j}(2zw^*)
		\nonumber\\
		&=e^{2zw^*} \frac{N!}{(2N)!}\sum_{j=0}^N\frac{(2j)!}{j!}(4zw^*)^{N-j}Q(2j+1,2zw^*).
		\label{eq-DN-QS}
	\end{align}
\end{prop}

The combination of truncated exponentials in \eqref{eq-DN-CS} for AI$^\dagger$ is precisely the one appearing in the computation of the eigenvector statistics for class A, see \cite[Eq. (2.16)]{ATTZ}. This may be a total coincidence, but it is worth mentioning.

The expectation \eqref{eq-DN-A} is well-known and only a special case of a more general theorem in \cite{AV03}. 
It was derived using the fact that the complex Ginibre ensemble (class A) is a determinantal point process, and planar orthogonal polynomials can be utilised, cf., Eq. \eqref{eq-K-DN-A} on the relation to the kernel.

We also mention in Eqs. \eqref{eq-DN-A} and \eqref{eq-DN-CS} the expression in terms of truncated exponentials \eqref{eq-def-eN} which become useful when taking the large $N$-limit in the bulk of the spectrum below. The second form of these expectation values in terms of the normalised incomplete Gamma function \eqref{eq-def-Q} will prove useful in the edge scaling limit. Last but not least there exists a double integral representation \eqref{eq-fNx-int2} together with \eqref{eq-QS-int} for the result \eqref{eq-DN-QS} of class AII$^\dag$, which is needed to derive the corresponding large $N$-limits at the edge and in the bulk.

\begin{proof} We omit the proof for the well-known result for class A, where we refer the reader to \cite{AAKP} and consider the classes AI$^\dagger$ and AII$^\dagger$ below.\\

\noindent
\textbf{Class AI$\boldsymbol{^\dag}$:} We show \eqref{eq-DN-CS} for the ensemble of complex symmetric matrices. Using \eqref{eq-Grassdet}, the expectation value \eqref{eq-def-char-pol} can be written as 
 \begin{align}
	D_N^{\textrm{AI}^\dag}(z,w^*)&=\int\!\frac{[dJ]}{Z_N^{\textrm{AI}^\dag}}\int\![d\xi][d\psi]\,
	\exp\left[ \sum_{j=1}^N(-|J_{j,j}|^2+\xi_j^*(z-J_{j,j})\xi_j+\psi_j^*(w^*-J_{j,j}^*)\psi_j)\right]
	\nonumber\\
	&\quad\times\exp\left[ -\sum_{1\leq j<\ell\leq N}2(|J_{j,\ell}|^2+J_{j,\ell}(\xi_j^*\xi_\ell+\xi_\ell^*\xi_j) +J^*_{j,\ell}(\psi_j^*\psi_\ell+\psi_\ell^*\psi_j))\right],
	\label{eq-CS-G1}
\end{align}
where we have taken the symmetry $J_{j,\ell}=J_{\ell ,j}$ into account. 
Splitting the matrix entries into real- and imaginary-part, $J_{j,\ell}=J^1_{j,\ell}+iJ^2_{j,\ell}$, the diagonal matrix elements in the exponent become,
\begin{align}
	\begin{split}
	\sum_{j=1}^N\Biggl(&-\left(J^1_{j,j}+\frac12(\xi_j^*\xi_j+\psi^*_j\psi_j)\right)^2+\frac14 (\xi_j^*\xi_j+\psi^*_j\psi_j)^2\\
	&-\left(J^2_{j,j}+\frac{i}{2}(\xi_j^*\xi_j-\psi^*_j\psi_j)\right)^2-\frac14 (\xi_j^*\xi_j-\psi^*_j\psi_j)^2\Biggr),
	\end{split}
	\label{eq-CS-G2}
\end{align}
while the upper triangular matrix elements take the form
\begin{align}
	\sum_{1\leq n<\ell\leq N}\left(-2\left(J^1_{n,\ell}+\frac14(\xi_n^*\xi_\ell+\xi_\ell^*\xi_n+\psi_n^*\psi_\ell+\psi_\ell^*\psi_n)\right)^2+\frac18 (\xi_n^*\xi_\ell+\xi_\ell^*\xi_n+\psi_n^*\psi_\ell+\psi_\ell^*\psi_n)^2\right.
	\nonumber\\
	\quad\quad\   \left.-2\left(J^2_{n,\ell}+\frac{i}{4}(\xi_n^*\xi_\ell+\xi_\ell^*\xi_n-\psi_n^*\psi_\ell-\psi_\ell^*\psi_n)\right)^2-\frac18 (\xi_n^*\xi_\ell+\xi_\ell^*\xi_n-\psi_n^*\psi_\ell-\psi_\ell^*\psi_n)^2\right).
	\label{eq-CS-G3}
\end{align}
Integrating out the shifted Gaussian matrix elements and multiplying out the quartic Grassmann terms, we obtain for \eqref{eq-CS-G1} after some simplifications
\begin{align}
	D_N^{\textrm{AI}^\dag}(z,w^*)&=\int\![d\xi][d\psi]\ 
	e^{\sum_{j=1}^N( \xi_j^*z\xi_j+\psi_j^*w^*\psi_j+\xi^*_j\xi_j\psi^*_j\psi_j)
		+\sum_{1\leq n<\ell\leq N}\frac12 (\xi_n^*\xi_\ell+\xi_\ell^*\xi_n)(\psi_n^*\psi_\ell+\psi_\ell^*\psi_n)
	}
	\nonumber\\
	&=\int\![d\xi][d\psi]
	\exp\!\left[ \sum_{j=1}^N (\xi_j^*z\xi_j+\psi_j^*w^*\psi_j) 
	+\frac12 \sum_{n,\ell=1}^N(\xi^*_n\psi_n \xi_\ell\psi^*_\ell-\xi^*_n\psi^*_n\xi_\ell\psi_\ell)
	\right].
	\label{eq-CS-G4}
\end{align}
In the second step, we have combined diagonal and upper triangular sums to obtain two fully factorised sums over all Grassmann variables, using their antisymmetry. In order to do the integrals, we linearise the quartic term in the Grassmann variables with an auxiliary variable $a\in\mathbb{C}$ integrated over the complex plane, i.e.,
\begin{align}
	1=\frac{1}{\pi}\int\! d^2a \,\exp\left[-\left(a-i\sum_{\ell=1}^N\xi_\ell\psi_\ell\right)\left(a^*-i\sum_{j=1}^N\xi^*_j\psi^*_j\right)\right].
	\label{eq-aux-int}
\end{align}

Applying \eqref{eq-aux-int} twice, though with different choices of factors, yields
\begin{align}
	D_N^{\textrm{AI}^\dag}(z,w^*)&=\int\![d\xi][d\psi]
	\int\! \frac{d^2a\, d^2b}{\pi^2}\exp\left[-|a|^2-|b|^2+\sum_{j=1}^N
	(\xi_j^*z\xi_j+\psi_j^*w^*\psi_j)\right]
	\nonumber\\
	&\quad\quad\quad\quad\times 
	\exp\left[\frac{1}{\sqrt{2}}\sum_{j=1}^N\left(i(a\xi_j\psi_j+a^*\xi_j^*\psi^*_j)+b\xi_j\psi^*_j+b^*\xi_j^*\psi_j\right)
	\right].
	\label{eq-CD-G5}
\end{align}
In this form we can identify the scalar product
\begin{align}
	\label{eq-CS-matrix}
	\begin{split}
		&\sum_{j=1}^N
		(\xi_j^*z\xi_j+\psi_j^*w^*\psi_j)+\frac{1}{\sqrt{2}}\sum_{j=1}^N\left(i(a\xi_j\psi_j+a^*\xi_j^*\psi^*_j)+b\xi_j\psi^*_j+b^*\xi_j^*\psi_j\right)\\
		=&
		\begin{pmatrix}
		\xi^\dag,&\xi^t,&\psi^\dag,&\psi^t\end{pmatrix}
		\begin{pmatrix}
			0&\frac{1}{2}z\mathds{1}_N&  \frac{i}{2\sqrt{2}}a^* \mathds{1}_N&  \frac{1}{2\sqrt{2}}b^* \mathds{1}_N\\
			-\frac{1}{2}z\mathds{1}_N& 0& \frac{1}{2\sqrt{2}}b \mathds{1}_N&  \frac{i}{2\sqrt{2}}a \mathds{1}_N\\
			\frac{-i}{2\sqrt{2}}a^* \mathds{1}_N&  \frac{-1}{2\sqrt{2}}b \mathds{1}_N&0&\frac{1}{2}w^*\mathds{1}_N  \\
			\frac{-1}{2\sqrt{2}}b^* \mathds{1}_N&  \frac{-i}{2\sqrt{2}}a \mathds{1}_N&-\frac{1}{2}w^*\mathds{1}_N& 0 \\
			\end{pmatrix}
		\begin{pmatrix}
			\xi^* \\
			\xi\\
			\psi^* \\
			\psi\\
		\end{pmatrix}
	\end{split}
\end{align}
This allows us to recognise and exploit the identity \eqref{eq-GrassPf} once we have changed $\xi\leftrightarrow
\xi^*$ and $\psi\leftrightarrow\psi^*$ such that
we arrive at
\begin{align}
	D_N^{\textrm{AI}^\dag}(z,w^*)=
	\int\! \frac{d^2a\, d^2b}{\pi^2}e^{-|a|^2-|b|^2}\Pf\begin{bmatrix}
		0&{z}\mathds{1}_N&  \frac{i}{\sqrt{2}}a^* \mathds{1}_N&  \frac{1}{\sqrt{2}}b^* \mathds{1}_N\\
		-{z}\mathds{1}_N& 0& \frac{1}{\sqrt{2}}b \mathds{1}_N&  \frac{i}{\sqrt{2}}a \mathds{1}_N\\
		\frac{-i}{\sqrt{2}}a^* \mathds{1}_N&  \frac{-1}{\sqrt{2}}b \mathds{1}_N&0&{w^*}\mathds{1}_N  \\
		\frac{-1}{\sqrt{2}}b^* \mathds{1}_N&  \frac{-i}{\sqrt{2}} a\mathds{1}_N&-{w^*}\mathds{1}_N& 0 \\
		\end{bmatrix}.
	\label{eq-CS-G6}
\end{align}
This duality formula with $k=1$ pair of integration variables over a Pfaffian determinant of size $4N$ was independently derived in \cite{Liu Zhang} using a diffusion equation rather than supersymmetry. Here, we further simplify the duality formula \eqref{eq-CS-G6}. 
Using the fact that the Pfaffian determinant of an antisymmetric matrix $G$ relates to the determinant as $\Pf[G]=\sqrt{\det[G]}$, as well as the evaluation of determinants with block matrices $A,B,C,D$ ($A$ must be invertible) via Schur complement,
\begin{align}
	\det\begin{bmatrix}
		A& B\\
		C& D\\
	\end{bmatrix}=\det\left[AD-ACA^{-1}B\right], 
	\label{eq-det-Id}
\end{align}
we can compute the Pfaffian in \eqref{eq-CS-G6} which is
\begin{align}
	\det
	\begin{bmatrix}
		-\frac12(2zw^*+|a|^2+|b|^2)\mathds{1}_N& 0\\
		0& -\frac12(2zw^*+|a|^2+|b|^2)\mathds{1}_N\\
\end{bmatrix}^{\frac12}=\frac{1}{2^N}(2zw^*+|a|^2+|b|^2)^N.
\end{align}
This leads to the following double integral expression for the expected characteristic polynomials
\begin{align}
	D_N^{\textrm{AI}^\dag}(z,w^*)=
	\int\! \frac{d^2a\, d^2b}{\pi^2}e^{-|a|^2-|b|^2}\frac{1}{2^N}(2zw^*+|a|^2+|b|^2)^N
	=\frac{1}{2^N}\int_0^\infty\! dt\ e^{-t}t(2zw^*+t)^N.
\end{align}
In the second equality we have employed spherical coordinates with $t=|a|^2+|b|^2$ the squared radius. Once we shift $t\to t-2zw^*$ we can anew identify the truncated exponential \eqref{eq-def-eN} and \eqref{eq-def-Gax} so that
\begin{align}
	D_N^{\textrm{AI}^\dag}(z,w^*)=\frac{N!}{2^N}\left( 
	(N+1)E_N(2zw^*)-2zw^*E_{N-1}(2zw^*)
	\right).
\end{align}
This result also implies the normalisation
\begin{align}
	\label{eq-DNAI00}
	D_N^{\textrm{AI}^\dag}(0,0)= \frac{N!}{2^N}(N+1)\ ,
\end{align}
finishing the derivation of \eqref{eq-DN-CS} for complex symmetric matrices class AI$^\dag$, 
after inserting \eqref{eq-def-eN}. 


\textbf{Class AII$\boldsymbol{^\dag}$:} Finally, we show \eqref{eq-DN-QS} for complex self-dual matrices. Because of the doubling of the dimension, we introduce two sets of $N$-dimensional Grassmann vectors $\xi_1^t=(\xi_{11},\ldots,\xi_{1N})$ and $\xi_2^t=(\xi_{21},\ldots,\xi_{2N})$ and likewise the vectors $\psi_1^t$ and $\psi_2^t$. Applying the relation \eqref{eq-Grassdet} and observing the block-structure of the complex self-dual matrix $J$ in \eqref{eq-para class AII dag} we thus obtain
\begin{align}
	D_N^{\textrm{AII}^\dag}(z,w^*)&=\int\!\frac{[dA][dB][dC]}{Z_N^{\textrm{AII}^\dag}}\!\int\![d\xi_1][d\xi_2][d\psi_1][d\psi_2]
	\nonumber\\
	&\times e^{\sum_{i,j=1}^N[-2|A_{i,j}|^2+\xi_{1i}^*(z\delta_{i,j}-A_{i,j})\xi_{1j}+\xi_{2i}^*(z\delta_{i,j}-A_{j,i})\xi_{2j}+\psi_{1i}^*(w^*\delta_{i,j}-A_{i,j}^*)\psi_{1j}+\psi_{2i}^*(w^*\delta_{i,j}-A_{j,i}^*)\psi_{2j}]}\nonumber\\
	&\times e^{-\sum_{1\leq \ell<n\leq N}[2|B_{n,\ell}|^2+(\xi_{1n}^*\xi_{2\ell}-\xi_{1\ell}^*\xi_{2n})B_{n,\ell}+(\psi_{1n}^*\psi_{2\ell}-\psi_{1\ell}^*\psi_{2n})B_{n,\ell}^*]}
	\nonumber\\
	&\times e^{-\sum_{1\leq \ell<n\leq N}[2|C_{n,\ell}|^2+(\xi_{2n}^*\xi_{1\ell}-\xi_{2\ell}^*\xi_{1n})C_{n,\ell}+(\psi_{2n}^*\psi_{1\ell}-\psi_{2\ell}^*\psi_{1n})C_{n,\ell}^*]}.
	\label{eq-QS-G1}
\end{align}
Splitting again the matrix elements $A_{n,\ell}=A^1_{n,\ell}+iA^2_{n,\ell}$ into real- and imaginary-part, we can write the exponent of all $A$-dependent parts as follows,
\begin{align}
	&-2\sum_{n,\ell=1}^N\left((A^1_{n,\ell})^2+(A^2_{n,\ell})^2 +\frac12 \Xi_{n,\ell}^+A^1_{n,\ell}+i\frac12 \Xi_{n,\ell}^-A^2_{n,\ell}\right)
	\nonumber\\
	&= -2\sum_{n,l=1}^N\left(\left( A^1_{n,\ell}+\frac14 \Xi_{n,\ell}^+\right)^2- \frac{1}{16}(\Xi_{n,\ell}^+)^2
	+\left( A^2_{n,\ell}+i\frac14 \Xi_{n,\ell}^-\right)^2+ \frac{1}{16}(\Xi_{n,\ell}^-)^2
	\right),
	\label{eq-QS-A}
\end{align}
where we have defined two bilinears of Grassmann variables
\begin{align}
	\Xi_{n,\ell}^+:= \xi_{1n}^*\xi_{1\ell}+\xi_{2\ell}^*\xi_{2n}+\psi_{1n}^*\psi_{1\ell}+\psi_{2\ell}^*\psi_{2n}\ ,
	\quad\Xi_{n,\ell}^-:=\xi_{1n}^*\xi_{1\ell}+\xi_{2\ell}^*\xi_{2n}-(\psi_{1n}^*\psi_{1\ell}+\psi_{2\ell}^*\psi_{2n}).\quad
	\label{eq-Xi-def}
\end{align} 
After integrating out all matrix elements of $A$, we find anew quartic terms in the Grassmann variables which are
\begin{align}
	&\frac18\sum_{n,\ell=1}^N\left( (\Xi_{n,\ell}^+)^2-(\Xi_{n,\ell}^-)^2\right) 
	=\frac12\sum_{n,\ell=1}^N(\xi_{1n}^*\xi_{1\ell}+\xi_{2\ell}^*\xi_{2n})(\psi_{1n}^*\psi_{1\ell}+\psi_{2\ell}^*\psi_{2n})\nonumber\\
	&=\frac12\sum_{n,\ell=1}^N(-\xi_{1n}^*\psi_{1n}^*\xi_{1\ell}\psi_{1\ell}+\xi_{1n}^*\psi_{2n}\xi_{1\ell}\psi_{2\ell}^*+\xi_{2n}\psi_{1n}^*\xi_{2\ell}^*\psi_{1\ell}-\xi_{2n}\psi_{2n}\xi_{2\ell}^*\psi_{2\ell}^*).
	\label{eq-QS-A4}
\end{align}
Likewise, all terms depending on the matrix elements of $B_{n,\ell}=B^1_{n,\ell}+iB^2_{n,\ell}$ in \eqref{eq-QS-G1} read
\begin{align}
	-2\sum_{1\leq \ell<n\leq N}\left(\left( B^1_{n,\ell}+\frac14 \Psi_{n,\ell}^+\right)^2- \frac{1}{16}(\Psi_{n,\ell}^+)^2
	+\left( B^2_{n,\ell}+i\frac14 \Psi_{n,\ell}^-\right)^2+ \frac{1}{16}(\Psi_{n,\ell}^-)^2
	\right),
	\label{eq-QS-B}
\end{align}
with 
\begin{align}
	\Psi_{n,\ell}^+:=\xi_{1n}^*\xi_{2\ell}-\xi_{1\ell}^*\xi_{2n}+\psi_{1n}^*\psi_{2\ell}-\psi_{1\ell}^*\psi_{2n},\quad
	\Psi_{n,\ell}^-:=\xi_{1n}^*\xi_{2\ell}-\xi_{1\ell}^*\xi_{2n}-(\psi_{1n}^*\psi_{2\ell}-\psi_{1\ell}^*\psi_{2n}) .\quad
	\label{eq-Psi-def}
\end{align} 
The integrals over all matrix elements of $B$, thus, yield 
\begin{align}
	\frac18\sum_{1\leq \ell<n\leq N}\left( (\Psi_{n,\ell}^+)^2-(\Psi_{n,\ell}^-)^2\right) =
	\frac12\sum_{1\leq \ell<n\leq N}(\xi_{1n}^*\xi_{2\ell}-\xi_{1\ell}^*\xi_{2n})(\psi_{1n}^*\psi_{2\ell}-\psi_{1\ell}^*\psi_{2n}).
	\label{eq-QS-B4}
\end{align}
Similarly, the terms depending on the matrix elements of 
$C_{n,\ell}=C^1_{n,\ell}+iC^2_{n,\ell}$ in \eqref{eq-QS-G1} give
\begin{align}
	-2\sum_{1\leq \ell<n\leq N}\left(\left( C^1_{n,\ell}+\frac14 \Phi_{n,\ell}^+\right)^2- \frac{1}{16}(\Phi_{n,\ell}^+)^2
	+\left( C^2_{n,\ell}+i\frac14 \Phi_{n,\ell}^-\right)^2+ \frac{1}{16}(\Phi_{n,\ell}^-)^2
	\right),
	\label{eq-QS-C}
\end{align}
with 
\begin{align}
	\Phi_{n,\ell}^+:= \xi_{2n}^*\xi_{1\ell}-\xi_{2\ell}^*\xi_{1n}+\psi_{2n}^*\psi_{1\ell}-\psi_{2\ell}^*\psi_{1n} ,
	\quad\Phi_{n,\ell}^-:=\xi_{2n}^*\xi_{1\ell}-\xi_{2\ell}^*\xi_{1n}-(\psi_{2n}^*\psi_{1\ell}-\psi_{2\ell}^*\psi_{1n}) ,\quad
	\label{eq-Phi-def}
\end{align} 
which leads to
\begin{align}
	\frac18\sum_{1\leq \ell<n\leq N}\left( (\Phi_{n,\ell}^+)^2-(\Phi_{n,\ell}^-)^2\right) =
	\frac12\sum_{1\leq \ell<n\leq N}(\xi_{2n}^*\xi_{1\ell}-\xi_{2\ell}^*\xi_{1n})(\psi_{2n}^*\psi_{1\ell}-\psi_{2\ell}^*\psi_{1n}) ,
	\label{eq-QS-C4}
\end{align}
after the integration over all matrix elements of $C$.

The sum of Eqs. \eqref{eq-QS-B4} and \eqref{eq-QS-C4} can be combined into
\begin{align}
	\frac12\sum_{\substack{1\leq n, \ell\leq N\\ n\neq \ell}}(-\xi_{1n}^*\psi_{1n}^*\xi_{2\ell}\psi_{2\ell}-\xi_{1n}^*\psi_{2n}\xi_{2\ell}\psi_{1\ell}^*- \xi_{2n}^*\psi_{2n}^*\xi_{1\ell}\psi_{1\ell}
	-\xi_{2n}^*\psi_{1n}\xi_{1\ell}\psi_{2\ell}^*
	) ,
	\label{eq-QS-BC4}
\end{align}
once we relabel indices.
Because at $n=\ell$ all terms cancel as the Grassmann variables anti-commute, we can remove the condition $n\neq \ell$ in the sum. 

Adding all quartic terms of Grassmann variables together, \eqref{eq-QS-A4} and \eqref{eq-QS-BC4}, we can write the exponential of the double sum in the following factorised form
\begin{align}
	&\!\!\!\exp\left[-\frac12\sum_{n,\ell=1}^N\left((\xi_{1n}^*\psi_{1n}^*+\xi_{2n}^*\psi_{2n}^*)(\xi_{1\ell}\psi_{1\ell}+\xi_{2\ell}\psi_{2\ell})
	+(\xi_{1n}^*\psi_{2n}-\xi_{2n}^*\psi_{1n})(\xi_{2\ell}\psi_{1\ell}^*-\xi_{1\ell}\psi_{2\ell}^*)\right)\right]
	\nonumber\\
	&=\int\!\frac{d^2p}{\pi}\,\exp\left[-|p|^2+\frac{i}{\sqrt{2}}p^*\sum_{n=1}^N(\xi_{1n}^*\psi_{1n}^*+\xi_{2n}^*\psi_{2n}^*)
	+\frac{i}{\sqrt{2}}p\sum_{\ell=1}^N(\xi_{1\ell}\psi_{1\ell}+\xi_{2\ell}\psi_{2\ell})
	\right]
	\nonumber\\
	&\times \int\!\frac{d^2q}{\pi}\,\exp\left[-|q|^2+\frac{i}{\sqrt{2}}q^*\sum_{n=1}^N
	(\xi_{1n}^*\psi_{2n}-\xi_{2n}^*\psi_{1n})
	+\frac{i}{\sqrt{2}}q\sum_{\ell=1}^N(\xi_{2\ell}\psi_{1\ell}^*-\xi_{1\ell}\psi_{2\ell}^*)
	\right].
	\label{eq-pq-int}
\end{align}
In the second step, we have again exploited the integral \eqref{eq-aux-int} to obtain bilinear Grassmann terms. These bilinear terms can be combined with the $z$- and $w^*$-dependent terms in \eqref{eq-QS-G1} to identify the scalar product
\begin{eqnarray}\small
	\label{eq-QS-matrix}
	\begin{pmatrix}
		\xi_1^* \\
		\xi_1\\
		\xi_2^* \\
		\xi_2\\
		\psi_1^* \\
		\psi_1\\
		\psi_2^* \\
		\psi_2\\
		\end{pmatrix}^t
	\begin{pmatrix}
		0&\frac{z}{2}\mathds{1}_N&0&0&  \frac{ip^*}{2\sqrt{2}} \mathds{1}_N& 0&0& \frac{iq^*}{2\sqrt{2}} \mathds{1}_N\\
		-\frac{z}{2}\mathds{1}_N&0&0&0& 0& \frac{ip}{2\sqrt{2}}p \mathds{1}_N&\frac{-iq}{2\sqrt{2}} \mathds{1}_N&0\\
		0&0&0&\frac{z}{2}\mathds{1}_N&0&\frac{-iq^*}{2\sqrt{2}}\mathds{1}_N&\frac{ip^*}{2\sqrt{2}} \mathds{1}_N&0\\
		0&0&-\frac{z}{2}\mathds{1}_N&0&  \frac{iq}{2\sqrt{2}} \mathds{1}_N& 0&0& \frac{ip}{2\sqrt{2}} \mathds{1}_N\\
		\frac{-ip^*}{2\sqrt{2}} \mathds{1}_N&0&0&\frac{-iq}{2\sqrt{2}} \mathds{1}_N&0&\frac{w^*}{2}\mathds{1}_N&0&0\\
		0&\frac{-ip}{2\sqrt{2}} \mathds{1}_N&\frac{iq^*}{2\sqrt{2}} \mathds{1}_N&0&-\frac{w^*}{2}\mathds{1}_N&0&0&0\\
		0&\frac{iq}{2\sqrt{2}} \mathds{1}_N&\frac{-ip^*}{2\sqrt{2}} \mathds{1}_N&0&0&0&0&\frac{w^*}{2}\mathds{1}_N\\
		\frac{-iq^*}{2\sqrt{2}}\mathds{1}_N&0&0&\frac{-ip}{2\sqrt{2}} \mathds{1}_N&0&0&-\frac{w^*}{2}\mathds{1}_N&0\\
		\end{pmatrix}
	\begin{pmatrix}
		\xi_1^* \\
		\xi_1\\
		\xi_2^* \\
		\xi_2\\
		\psi_1^* \\
		\psi_1\\
		\psi_2^* \\
		\psi_2\\
		\end{pmatrix}.\nonumber\\
\end{eqnarray}
The Grassmann integration can thus be performed using \eqref{eq-GrassPf} to arrive at
\begin{align}
	&D_N^{\textrm{AII}^\dag}(z,w^*)=
	\int\! \frac{d^2p\, d^2q}{\pi^2}\,\exp\left[-|p|^2-|q|^2\right]
	\nonumber\\
	&\times \Pf
	\begin{bmatrix}
		0&z\mathds{1}_N&0&0&  \frac{i}{\sqrt{2}}p^* \mathds{1}_N& 0&0& \frac{i}{\sqrt{2}}q^* \mathds{1}_N\\
		-z\mathds{1}_N&0&0&0& 0& \frac{i}{\sqrt{2}}p \mathds{1}_N&\frac{-i}{\sqrt{2}}q \mathds{1}_N&0\\
		0&0&0&z\mathds{1}_N&0&\frac{-i}{\sqrt{2}}q^* \mathds{1}_N&\frac{i}{\sqrt{2}}p^* \mathds{1}_N&0\\
		0&0&-z\mathds{1}_N&0&  \frac{i}{\sqrt{2}}q \mathds{1}_N& 0&0& \frac{i}{\sqrt{2}}p \mathds{1}_N\\
		\frac{-i}{\sqrt{2}}p^* \mathds{1}_N&0&0&\frac{-i}{\sqrt{2}}q \mathds{1}_N&0&w^*\mathds{1}_N&0&0\\
		0&\frac{-i}{\sqrt{2}}p \mathds{1}_N&\frac{i}{\sqrt{2}}q^* \mathds{1}_N&0&-w^*\mathds{1}_N&0&0&0\\
		0&\frac{i}{\sqrt{2}}q \mathds{1}_N&\frac{-i}{\sqrt{2}}p^* \mathds{1}_N&0&0&0&0&w^*\mathds{1}_N\\
		\frac{-i}{\sqrt{2}}q^* \mathds{1}_N&0&0&\frac{-i}{\sqrt{2}}p \mathds{1}_N&0&0&-w^*\mathds{1}_N&0\\
		\end{bmatrix}.\qquad
\end{align}
This duality formula for class AII$^\dag$ was also derived in \cite{Liu Zhang}.
Once again, we can exploit the block structure of the matrix and apply \eqref{eq-det-Id}, to obtain for the Pfaffian determinant
{\small
\begin{align}
		&\det\!\begin{bmatrix}
			-\!\left[zw^*\!+\!\frac{|p|^2+|q|^2}{2}\right]\!\mathds{1}_N &0&0& -pq^*\mathds{1}_N\\
			0& -\!\left[zw^*\!+\!\frac{|p|^2+|q|^2}{2}\right]\!\mathds{1}_N&p^*q\mathds{1}_N&0\\
			0&pq^*\mathds{1}_N&-\!\left[zw^*\!+\!\frac{|p|^2+|q|^2}{2}\right]\!\mathds{1}_N&0\\
			-p^*q\mathds{1}_N&0&0&-\!\left[zw^*\!+\!\frac{|p|^2+|q|^2}{2}\right]\!\mathds{1}_N\\
			\end{bmatrix}^{\frac12}\nonumber\\
		&=\left(\left(zw^*+\frac{|p|^2+|q|^2}{2}\right)^2-|p|^2|q|^2
		\right)^N\nonumber \\
		&=\frac{1}{4^N}\left((2zw^*)^2+4zw^*(|p|^2+|q|^2)+(|p|^2-|q|^2)^2
		\right)^N.
	\label{eq-QS-Pfresult}
\end{align}}
Thence, we are left with computing the following integral
\begin{align}
	D_N^{\textrm{AII}^\dag}(z,w^*)&=
	\int\! \frac{d^2p\, d^2q}{\pi^2}\ e^{-|p|^2-|q|^2}\frac{1}{4^N}\left((2zw^*)^2+4zw^*(|p|^2+|q|^2)+(|p|^2-|q|^2)^2\right)^N
	\nonumber\\
	&=\frac{1}{4^N}f_N(x)\ , \quad \mbox{with} \quad x=zw^*\quad \mbox{and}
	\label{eq-QS-int}\\
	f_N(x)&:=
	\int_0^\infty\! dr\int_0^\infty\! ds\ e^{-r-s}\left(4x^2+4x(r+s)+(r-s)^2\right)^N,
	\label{eq-fNx-def}
\end{align}
after changing to polar coordinates for $q$ and $p$.

It is useful to introduce so-called light cone variables at this stage, the sum and difference of the integration variables,
\begin{align}
	v:=r+s,\quad u:=r-s,\quad \mbox{with}\ v\geq0,\ u\in[-v,v].
	\label{eq-lightcone}
\end{align}
The Jacobian is $dr\,ds=\frac12 dv\,du$ so that we get, due to Fubini's theorem,
\begin{align}
	f_N(x)
	&= \frac12 \int_0^\infty\! dv\ e^{-v}\int_{-v}^v\!du\,
	\left(4x^2+4xv+u^2\right)^N.
	\label{eq-fNx-int1}
\end{align}
We note in passing that this integral satisfies 
an inhomogeneous first order differential equation as is explained in Appendix \ref{sec-app expec poly}.

The integral can be further simplified as follows. Using the evenness of the integrand with respect to variable $u\to-u$,
\begin{align}
	f_N(x)
	&= \int_0^\infty\! dv\int_0^v\!du\ e^{-v}
	\left(4x^2+4xv+u^2\right)^N\nonumber\\
	&= \int_0^\infty\! du\int_u^\infty\! dv \ e^{-v}
	\left(4x^2+4xv+u^2\right)^N
	\nonumber\\
	&= \int_0^\infty\! du\int_0^\infty\! dv \ e^{-v-u}
	\left(4xv+(2x+u)^2\right)^N.
	\label{eq-fNx-int2}
\end{align}
In the second equality we reordered the integrals and in the third one we shifted $v\to v'=v-u$.

This expression \eqref{eq-fNx-int2} leads to a simple summation formula at finite $N$, as well as to the large $N$ asymptotic limits performed in the next section. From a simple expansion of the binomial, we obtain 
\begin{align}
	f_N(x)
	&= \int_0^\infty\! du\int_0^\infty\! dv \ e^{-v-u}\sum_{j=0}^N\frac{N!}{j!(N-j)!}(4xv)^{N-j}(2x+u)^{2j}
	\nonumber\\
	&= \int_{2x}^\infty\! du' \ e^{-u'+2x} \sum_{j=0}^N\frac{N!}{j!}(4x)^{N-j}(u')^{2j}
	\nonumber\\
	&=e^{2x} \sum_{j=0}^N\frac{N!}{j!}(4x)^{N-j}\,\Gamma(2j+1,2x),
	\label{eq-fNx-sum}
\end{align}
after shifting again $u'=2x+u$ and using the definition of the incomplete Gamma function \eqref{eq-def-Gax}. 

At $x=zw^*=0$ we only have a contribution from $j=N$, and hence with $\Gamma(2j+1,0)=(2j)!$ we obtain
\begin{align}
	f_N(0)=(2N)!
\end{align}
and, thus, together with \eqref{eq-QS-int}
\begin{align}
	\label{eq-DNAII00}
	D_N^{\textrm{AII}^\dag}(0,0)=\frac{(2N)!}{4^N}.
\end{align}
Inserting this into the definition on the LHS of \eqref{eq-DN-QS} together with \eqref{eq-def-Q} and \eqref{eq-def-eN}, this proves both RHS of \eqref{eq-DN-QS}.
\end{proof}

\subsubsection{Asymptotic Global and Local Large $\boldsymbol{N}$ Limits}\label{sssec-asymptot 1 char poly}
Recall that we are looking for an analytical distinction of the three classes A, AI$^\dagger$ and AII$^\dagger$ to prove universality. Therefore, we study the results from Proposition \ref{prop-char_exp} in the large $N$-limit and discuss the correct scaling in this regime. We start with the latter and discuss the global limit for the derived results for a single pair of characteristic polynomials comparing with the (analytically) known density shape from Figure \ref{fig-radial density plot symmetry classes}. For the complex Ginibre ensemble (class A) we recall that there exists a direct relation of $D_N^{\textrm{A}}(z,w^*)/D_N^{\textrm{A}}(0,0)$ to the spectral density due to its determinantal structure. Next, we study local edge and bulk scaling limits of the results from Proposition \ref{prop-char_exp} confirming distinct behaviours within the three symmetry classes.\\
\\
\textbf{Scaling with $\boldsymbol{N}$ and Global Limit for One Pair}\\
\\
In this paragraph, we start exploring the possible large $N$-limits.
First, we present plots for the rescaled characteristic polynomials at different values of $N$ over the entire range of the spectrum, helping us in identifying different scaling regimes. In particular, we present the global limit for all three classes, before zooming into the local edge and bulk below. To deal with those in a unifying way, it is useful to introduce the dimension
\begin{align}\label{eq-def.dim}
	d_N=N\quad\textrm{for}\quad \textrm{A}\ \textrm{and}\ \textrm{AI}^\dag\qquad\textrm{and}\qquad d_N=2N\quad\textrm{for}\quad \textrm{AII}^\dag\ ,
\end{align}
and the scaling
\begin{align}\label{eq-def.scale}
	s=1\quad\textrm{for}\quad \textrm{A} \qquad\textrm{and}\qquad s=\frac{1}{\sqrt{2}}\quad\textrm{for}\quad \textrm{AI}^\dag\ \textrm{and}\  \textrm{AII}^\dag.
\end{align}
The scaling $s$ properly normalises the limiting support of the expectation value \eqref{eq-def-char-pol} as well as the eigenvalues to the unit disc for all three ensembles. The latter we have checked numerically which also implies that the local mean level spacing of the eigenvalues is normalised to the same distance so that the local statistics become comparable.

Our results for the global limit can be summarised as follows.
\begin{figure}[t!]
	\centering
	\includegraphics[width=\textwidth]{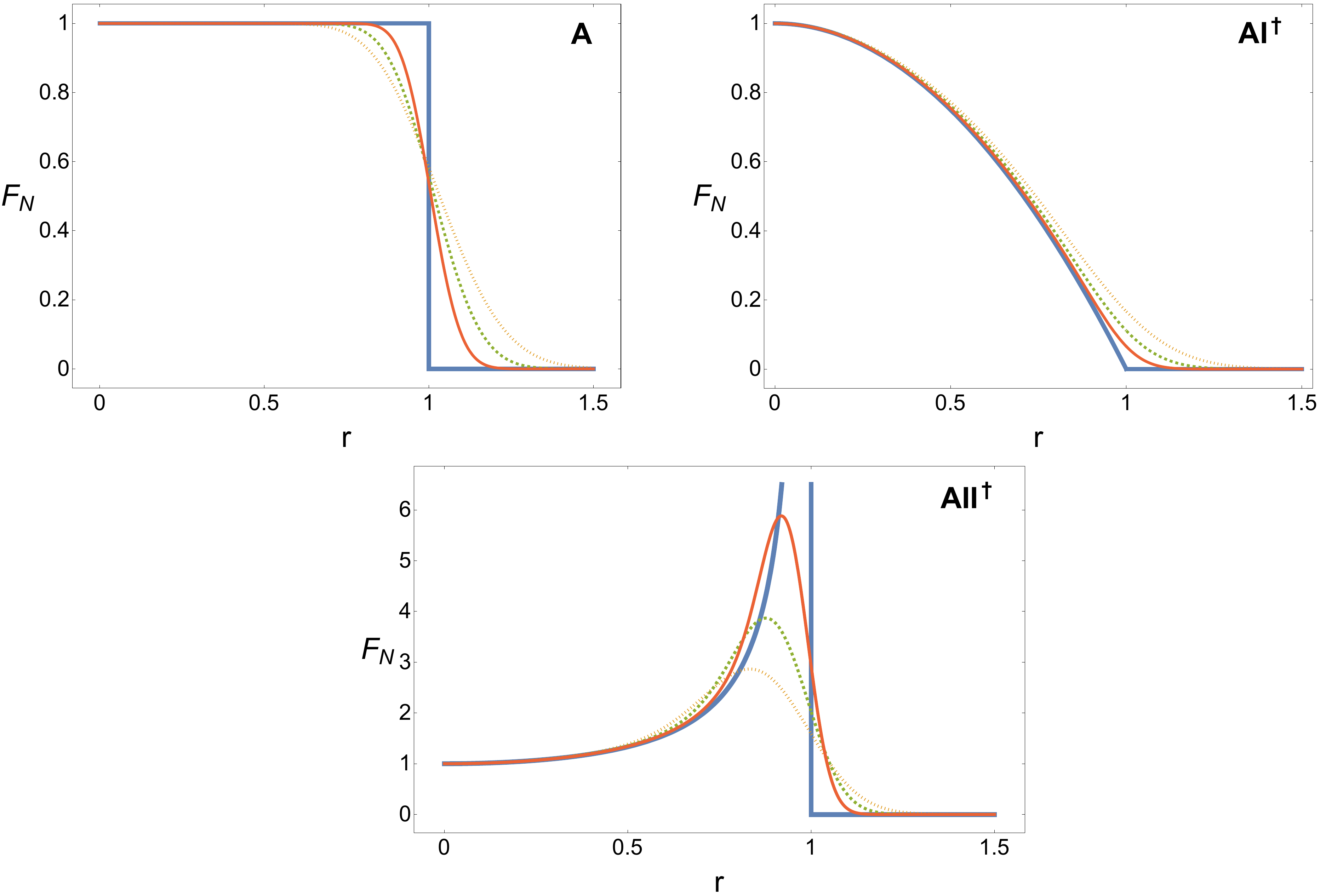}
	\caption[Figure 1 from \cite{AAKP}, \textit{Plots of rescaled characteristic polynomials of the symmetry classes A, AI$^\dagger$ and AII$^\dagger$}]{Plots of the rescaled characteristic polynomials for finite $N$ as a function of $r=zw^*\geq0$ for $N=10,20,50$ (dotted orange, dashed green, solid red, respectively) and the global limit $N\to\infty$ (thick blue). \textbf{Top left:} Rescaled densities \eqref{eq-Gin-NF} of class A; \textbf{Top right:} Rescaled characteristic polynomials \eqref{eq-CS-NF} of class AI$^\dagger$; \textbf{Bottom:} Rescaled characteristic polynomials \eqref{eq-QS-NF} of class AII$^\dagger$. The position of the edge of the global limit is rescaled to unity. \cite[Figure 1]{AAKP}}
	\label{fig-rescaled poly}
\end{figure}

\begin{prop}
	[Global limit for one pair of characteristic polynomials]\label{prop-global}\
	
	Let $r\geq0$. For the rescaled characteristic polynomials 
	\begin{align}
		\label{eq-F-def}
		F_N(r):=\left.\exp[-d_N zw^*] \frac{D_N\left(\sqrt{d_N}sz,\sqrt{d_N}s w^*\right)}{D_N(0,0)}\right|_{zw^*=r^2},
	\end{align}
	cf., Proposition \ref{prop-char_exp}, we obtain the following global limits 
	\begin{align}
		D_\textrm{global}^\textrm{A}(r)&:=\lim_{N\to\infty}F_N^\textrm{A}(r)\ \ \ =\Theta(1-r)\ ,\label{eq-Dgobal-A}\\
		D_\textrm{global}^{\textrm{AI}^\dag}(r)&:=\lim_{N\to\infty}F_N^{\textrm{AI}^\dag}(r)\ =(1-r^2)\Theta(1-r)\ ,
		\label{eq-Dgobal-AI dag}\\
		D_{\textrm{global}}^{\textrm{AII}^\dag}(r)&:=\lim_{N\to\infty}F_N^{\textrm{AII}^\dag}(r)
		=\frac{1}{1-r^2}\Theta(1-r)\ ,
		\label{eq-Dgobal-AII dag}
	\end{align}
	where $\Theta(1-r)$ is the Heaviside step-function restricting the support to the interval $[0,1]$.
\end{prop}
\begin{proof}
\textbf{Class A:} We begin with the derivation for the complex Ginibre ensemble, where it is well-known that it converges to the circular law, since $F_N^\textrm{A}(r)$ is up to a normalisation factor the global spectral density of the eigenvalues in the complex plane when $z=w$, see \cite{ginibre}. The rescaled quantity \eqref{eq-F-def} can be rewritten in terms of the ratio \eqref{eq-def-Q} due to \eqref{eq-DN-A}, i.e.,
\begin{align}
	F_N^\textrm{A}(r)=Q(N+1,Nr^2).
	\label{eq-Gin-NF}
\end{align}
The exponential factor in \eqref{eq-DN-A} is removed, as it is well-known that the remaining normalised incomplete Gamma function has a finite limit. 
The global limit of this function
is obtained from the convergence of the normalised lower incomplete Gamma function \cite[Eqs. 8.11.6, 8.11.7]{NIST}
\begin{align}
	D_\textrm{global}^\textrm{A}(r)=\Theta(1-r).
	\label{eq-Dglobal-Gin}
\end{align}
The corrections are exponentially suppressed except when $r\approx 1$ where it becomes algebraic due to \cite[Eq. 8.11.12]{NIST}.
Later, in the local edge scaling limit the step of the function $F_N^\textrm{A}(r)$ at $r=1$ will be studied at a finer resolution, cf., Eqs. \eqref{eq-EN-P-relation} and \eqref{eq-P-lim}. 
The results \eqref{eq-Gin-NF} and \eqref{eq-Dglobal-Gin} are plotted in Figure \ref{fig-rescaled poly} (top left plot) for $N=10,20,50$ (dotted orange, dashed green, solid red, respectively) and in the limit $N\to\infty$ (thick blue).

\textbf{Class AI$\boldsymbol{^\dag}$:} We turn to the derivation for complex symmetric matrices. Due to the very similar form, cf. Eqs. \eqref{eq-DN-A} and \eqref{eq-DN-CS}, the computation goes along the same lines. We would like to emphasise that the JPDF of the eigenvalues for this ensemble is unknown, and no relation between the expected characteristic polynomials and the eigenvalue density can be expected a priori. We discuss a comparison with the recently derived spectral density \cite{complex sym Gernot} below.

To have a similar situation as in the complex Ginibre ensemble, we also remove the exponential prefactor in \eqref{eq-DN-CS}. Once we use the result \eqref{eq-DN-CS} in \eqref{eq-F-def}, we consider
\begin{align}
	F_N^{\textrm{AI}^\dag}(r)= Q(N+1,Nr^2)-\frac{Nr^2}{N+1}Q(N,Nr^2),
	\label{eq-CS-NF}
\end{align}
As for class $A$, it is clear from the asymptotic of
the incomplete Gamma function \cite[Eqs. 8.11.6, 8.11.7]{NIST} that we obtain for the global limit 
\begin{align}
	D_\textrm{global}^{\textrm{AI}^\dag}(r)
	=(1-r^2)\Theta(1-r) .
	\label{eq-Dglobal-CS}
\end{align}
As in class A the correction terms are exponentially suppressed when staying away from the vicinity about $r=1$. 
Plots of \eqref{eq-CS-NF} for $N=10,20,50$ and \eqref{eq-Dglobal-CS} for $N\to\infty$ can be found in Figure \ref{fig-rescaled poly} (top right plot).

\textbf{Class AII$\boldsymbol{^\dag}$:} The case of the complex self-dual ensemble is more involved. 
We start like before with properly scaling the spectral variables $z$ and $w^*$ and then removing the exponential prefactor in \eqref{eq-DN-QS},  
\begin{align}
	F_N^{\textrm{AII}^\dag}(r)=\sum_{j=0}^N\frac{N!(2N-2j)!}{(2N)!(N-j)!}(4Nr^2)^{j}Q(2N-2j+1,2Nr^2).
	\label{eq-QS-NF}
\end{align}
We have renamed the summation index $j\to N-j$ as the main contribution for $N\to\infty$ comes from a neighbourhood where $j$ is of order one. The summands have indeed a point wise limit
\begin{align}
	\lim_{N\to\infty}\frac{N!(2N-2j)!}{(2N)!(N-j)!}(4Nr^2)^{j}Q(2N-2j+1,2Nr^2)=r^{2j}\Theta(1-r)
\end{align}
for any fixed $j$, due Stirling's approximation for the factorials and the asymptotic of the incomplete Gamma function \cite[Eqs. 8.11.6, 8.11.7]{NIST}. One can even relax the fixed condition of $j$ and allow any $j=o(\sqrt{N})$ when considering the Taylor expansion of the summands. We omit here to show that there is actually a sequence of $N$-independent summands whose series converges. This would be needed to be allowed to apply Lebesgue's dominated convergence theorem.

Summarising, we obtain a geometric series
\begin{align}
	D_\textrm{global}^{\textrm{AII}^\dag}(r)
	&\approx \sum^{\infty}_{j=0}r^{2j} \Theta(1-r)=\frac{1}{1-r^2}\Theta(1-r),
	\label{eq-Dglobal-QS}
\end{align}
which is drawn in Figure \ref{fig-rescaled poly} (bottom plot) as well as for the values $N=10,20,50$ of the finite $N$ expression \eqref{eq-QS-NF}.
\end{proof}
Let us remark again that in both classes AI$^\dag$ and AII$^\dag$ the rescaled characteristic polynomials do not directly relate to the eigenvalue density as it is the case in class A, though some kind of relation will be still present. The quantity $D_\textrm{global}(r)$ is not normalised nor it is guaranteed to be normalisable as we can see in the class AII$^\dag$ for which \eqref{eq-Dglobal-QS} is not integrable over its support. Indeed, the integrals over their finite $N$ expressions depend on the matrix dimension.

The question remains how we can interpret the different shapes of the curves in Figure \ref{fig-rescaled poly} above. Therefore, we compare with the plot of the spectral densities in Figure \ref{fig-radial density plot symmetry classes} and find similar behaviours at the edge. Class AI$^\dagger$ shows a damping near the edge while class AII$^\dagger$ evolves a peak, which is again in agreement with the effective description of the NN and NNN bulk distributions by 2dCG at $\beta= 1.4$, $\beta= 2.6$ respectively \cite{AMP, ABCPRS}.

For this reason, we expect that the expectation value of two characteristic polynomials contains important information about the spectrum, too, at least at the edge of the spectrum. 
\\
\\
\textbf{Complex Ginibre Ensemble: Edge and Bulk Scaling of the Density Revisited}
\\
\\
We briefly recall the edge and bulk scaling limit of the complex Ginibre ensemble, which is a determinantal point process with kernel $K_N(z,w^*)$. We choose the convention that $K_N(z,w^*)$ denotes the polynomial part of the kernel, without the corresponding weight functions, which is also known as pre-kernel. In this form it equals the normalised expectation value of two conjugate characteristic polynomials.

The following relation is known between this kernel at finite $N$ and the characteristic polynomials, e.g., see \cite{ginibre,AV03},
\begin{align}
	K_N(z,w)=
	\frac{D_N^\textrm{A}(z,w^*)}{D_N^\textrm{A}(0,0)}
	=
	E_N(zw^*)= e^{
		zw^*}Q(N+1,zw^*).
	\label{eq-K-DN-A}
\end{align}
The eigenvalue density \eqref{eq-R1A} is given by the kernel at equal arguments, times the weight function of the Ginibre ensemble, $w(z)=\exp[-|z|^2]$. We would like to point out that this density is normalised to $N$, and not to 1 yet. 

In the global large $N$-limit it converges to the circular law on a disc of radius $\sqrt{N}$ in this normalisation, 
\begin{align}
	R_{N}^{\textrm{A}}(z)\sim\Theta(\sqrt{N}-|z|).
	\label{eq-Gin-density}
\end{align}
A further rescaling $z\to\sqrt{N} z$ maps the support to the unit disc.

The local bulk scaling limit is defined as follows. We pick a point $\sqrt{N}z_0$ in the bulk of the spectrum, that is with $|z_0|<1$. The vicinity of this bulk point is then parametrised as follows
\begin{align}
	z&=\sqrt{N}z_0+\chi, \quad w^*=\sqrt{N}z_0^*+\eta^*,
	\nonumber\\
	x&:=zw^*=N|z_0|^2+\sqrt{2N}\,y+\chi\eta^*,\quad y:=(\chi z_0^*+\eta^*z_0)/\sqrt{2}\ .
	\label{eq-zoom}
\end{align}
The limiting kernel in the bulk is, then, obtained by replacing the truncated sum $E_N(zw^*)$ in \eqref{eq-K-DN-A} by an exponential, leading to
\begin{align}
	K_\textrm{bulk}^\textrm{A}(\chi,\eta^*)&=\lim_{N\to\infty}K_N(\sqrt{N}z_0+\chi,\sqrt{N}z_0^*+\eta^*)\Big|_{|z_0|<1}\sim \exp\left[ N|z_0|^2+\sqrt{2N}\,y+\chi\eta^*\right].\qquad
	\label{eq-Gin-bulk-kernel}
\end{align}
Strictly speaking, the pre-kernel does not have a limit. 
Only after multiplying with the weight function, $e^{-\frac12(|z|^2+|w|^2)}K_N(z,w^*)$ and after cancelling cocycles that drop out in the determinant of the $k$-point correlation function, it does lead to the limiting Ginibre kernel. 
This is why below we will remove the exponential factor in \eqref{eq-K-DN-A} when taking limits. In Appendix B in \cite{AAKP}, the bulk \eqref{eq-Gin-bulk-kernel} and edge \eqref{eq-Gin-edge-kernel} asymptotic are used for a meaningful comparison of the results for averages over more than one pair of characteristic polynomials.

Turning to the local edge scaling limit, we will choose the same rescaling as in \eqref{eq-zoom} but require now that the point we zoom in is located on the unit circle, that is $|z_0|=1$.
For the edge scaling we rewrite the truncated exponential function in the following way,
\begin{align}
	E_N(x)=e^xQ(N+1,x)
	=e^x(1-P(N+1,x)),
	\label{eq-EN-P-relation}
\end{align}
where 
\begin{align}
	P(N+1,x)=\frac{\gamma(N+1,x)}{\Gamma(N+1)}
	\label{eq-Pdef}
\end{align}
is the regularised lower incomplete Gamma function
$\gamma(a,x)=\Gamma(a)-\Gamma(a,x)$. It enjoys the following asymptotic \cite[Eq. 8.11.10]{NIST} for large $N$, with the edge rescaling \eqref{eq-zoom} at $|z_0|=1$,
\begin{align}
	P(N+1,x)=\frac12\erfc(-y)+\mathcal{O}(N^{-\frac12}), 
	\label{eq-P-lim}
\end{align}
with the complementary error function \eqref{eq-erfc-def}. Once, we employ the identity \eqref{eq-erfc-rel}, we find
\begin{align}
	Q(N+1,x)=\frac12\erfc(y)+\mathcal{O}(N^{-\frac12}), 
	\label{eq-Q-lim}
\end{align}
and hence the edge scaling asymptotic becomes
\begin{align}
	K_\textrm{edge}^\textrm{A}(\chi,\eta^*)&=\left.\lim_{N\to\infty}K_N(\sqrt{N}z_0+\chi,\sqrt{N}z_0^*+\eta^*)\right|_{|z_0|=1}\nonumber\\
	&\sim\frac12\exp\left[ N+\sqrt{2N}\,y+\chi\eta^*\right]
	\erfc(y).\quad
	\label{eq-Gin-edge-kernel}
\end{align}
After multiplying it with the weight function and removing cocycles, it leads to the universal complementary error function kernel (sometimes called Faddeyeva plasma-kernel \cite{FT61}).
\begin{Remark}
For complex symmetric matrices from class AI$^\dagger$ there is no direct relation of the results from Proposition \ref{prop-char_exp} with the recently derived spectral density \eqref{eq-R1AI+} \cite{complex sym Gernot}. While the expression \eqref{eq-DN-CS} for $D_N^{\textrm{AI}^\dagger}(z,z^*)/D_N^{\textrm{AI}^\dagger}(0,0)$ at $w=z$ consists of two functions $Q(a,2|z|^2)$, the functional form for spectral density \eqref{eq-R1AI+} is much more evolved. Therefore, it is not surprising that also the results in the global limit of the two quantities does not agree. For the radial density $\rho_N(r)$ it was found in \cite{complex sym Gernot} for $N\gg 1$:
\begin{align}
	\rho_{N\gg1}(r)&\propto \Theta\left(\sqrt{N}(r^2-1)\right)\nonumber\\
	&\propto \Theta(r-1),
\end{align}
where we choose the support up to $r=1$, compared to $r=\sqrt{2}$ in \cite{complex sym Gernot}, and used the fact that $r>0$. Moreover, we do not include the radial factor from the polar coordinates in the density. In contrast we have a prefactor of $(1-r^2)$ in front of the Heaviside-function in \eqref{eq-Dgobal-AI dag}. Similarly, the results in Theorem \ref{thm-lim_char_exp} below at the spectral edge do not represent the edge behaviour of the density derived in \cite{complex sym Gernot}. 
\end{Remark}
\noindent \textbf{Local Edge Scaling Limit for One Pair of Characteristic Polynomials}\\
\\
\noindent
Because only one of the three ensembles we consider is a determinantal point process, the complex Ginibre ensemble, we do not have the freedom to remove $N$-dependent cocycles from the limiting characteristic polynomials which are proportional to the kernel, nor can we argue why to multiply with a weight function as in the Ginibre class A, which removes the leading contribution $\exp[N|z_0|^2]$ from the expected characteristic polynomials in that case. 

Therefore, we choose a different route in defining the limiting characteristic polynomials, by removing the exponential prefactor $\exp[zw^*]$ that appears in all three ensembles, see Eqs. \eqref{eq-DN-A}, \eqref{eq-DN-CS} and \eqref{eq-DN-QS}, respectively. Furthermore, as in the expressions for finite matrix dimension $N$ given in Proposition \ref{prop-char_exp}, we divide by the value of the respective characteristic polynomials at the origin, $D_N(0,0)$. It turns out that for classes AI$^\dag$ and AII$^\dag$ an additional rescaling in powers of $N^{\,\pm1/2}$ is necessary. 

\begin{theorem}[Local limits of one pair of characteristic polynomials at the spectral edge]\label{thm-lim_char_exp}\
	
	We recall the definitions \eqref{eq-def.dim}, \eqref{eq-def.scale} as well as Owen's $T$-function \eqref{eq-OTi} and the results of Proposition \ref{prop-char_exp}. The vicinity of edge points is defined as 
	\begin{align}
		z=\sqrt{d_N}sz_0+s\chi\ , \quad
		w^*=\sqrt{d_N}sz_0^*+s\eta^*, \quad zw^*=s^2(d_N+\sqrt{2d_N}\,y
		+\chi\eta^*),
		\label{eq-zoom1}
	\end{align}
	with $|z_0|=1$ and $y=(z_0^*\chi+z_0\eta^*)/\sqrt{2}$. With this rescaling we have 
	\begin{align}
		D_\textrm{edge}^\textrm{A}(\chi,\eta^*)&:=\lim_{N\to\infty}
		e^{-zw^*}\frac{D_N^\textrm{A}(z,w^*)}{D_N^\textrm{A}(0,0)}=\frac12\erfc(y),
		\label{eq-Dedge-A}\\
		D_\textrm{edge}^{\textrm{AI}^\dag}(\chi,\eta^*)&:=\lim_{N\to\infty}
		\sqrt{N}\,e^{-2zw^*}
		\frac{D_N^{\textrm{AI}^\dag}\left(z,w^*\right)}{D_N^{\textrm{AI}^\dag}(0,0)}
		=\frac{1}{\sqrt{2\pi}}e^{-y^2}-\frac{y}{\sqrt{2}}
		\erfc(y),
		\label{eq-Dedge-CS}
		\\
		D_\textrm{edge}^{\textrm{AII}^\dag}(\chi,\eta^*)&:=\lim_{N\to\infty}
		\frac{1}{\sqrt{2N}}\,e^{-2zw^*}\frac{D_N^{\textrm{AII}^\dag}(z,w^*)}{D_N^{\textrm{AII}^\dag}(0,0)}\nonumber\\
		&=
		\left(-\frac{\sqrt{\pi}}{4}\erfc\left(y\right)\erfi(y/\sqrt{2})-i\sqrt{\pi}\,T\left(\sqrt{2}y,i/\sqrt{2}\right)\right) e^{-y^2/2}.
		\label{eq-Dedge-QS}
	\end{align}
	The corresponding left and right tail asymptotic read (for real $y$)
	\begin{align}
		\label{eq-Dtails-A}
		D_\textrm{edge}^\textrm{A}(\chi,\eta^*)&\sim
		\left\{
		\begin{array}{cl}
			\displaystyle\frac{1}{2\sqrt{\pi}y}\, e^{-y^2}, & \textrm{for}\ y\to+\infty,\\
			\displaystyle1+\frac{1}{2\sqrt{\pi}y}\,e^{-y^2}, & \textrm{for}\ y\to-\infty,
			\\
		\end{array}
		\right.
		\\
		\label{eq-Dtails-CS}
		D_\textrm{edge}^{\textrm{AI}^\dag}(\chi,\eta^*)&\sim
		\left\{
		\begin{array}{cl}
			\displaystyle\frac{1}{\sqrt{8\pi}\,y^2}\, e^{-y^2}, & \textrm{for}\ y\to+\infty,\\
			\displaystyle-\sqrt{2}\,y+\frac{1}{\sqrt{8\pi}\,y^2}\, e^{-y^2}, & \textrm{for}\ y\to-\infty,
			\\
		\end{array}
		\right.
		\\
		\label{eq-Dtails-QS}
		D_\textrm{edge}^{\textrm{AII}^\dag}(\chi,\eta^*)&\sim
		\left\{
		\begin{array}{cl}
			\displaystyle\frac{1}{\sqrt{8\pi}\, y^2}e^{-y^2}, & \textrm{for}\ y\to+\infty,\\
			\displaystyle-\frac{1}{\sqrt{2}\,y}-\frac{1}{\sqrt{2}\,y^3}, & \textrm{for}\ y\to-\infty.\\
		\end{array}
		\right.
	\end{align}
	
\end{theorem}

The asymptotic \eqref{eq-Dedge-CS} also follows from \cite[Corollary 2]{ATTZ} for the eigenvector statistics as the initial expression \eqref{eq-DN-CS} has been the same.

We start with a brief discussion of the results from Theorem \ref{thm-lim_char_exp} and present the proof afterwards. From Eqs. \eqref{eq-Dedge-A}-\eqref{eq-Dedge-QS} we can readily read off the value of these normalised expectation values at $\chi=\eta^*=0$ which are
\begin{align}\label{eq-edge.origin}
	D_\textrm{edge}^\textrm{A}(0,0)=\frac{1}{2},\qquad D_\textrm{edge}^{\textrm{AI}^\dag}(0,0)=\frac{1}{\sqrt{2\pi}}\qquad\textrm{and}\qquad D_\textrm{edge}^{\textrm{AII}^\dagger}(0,0)=\frac{\textrm{arctanh}(\frac{1}{\sqrt{2}})}{\sqrt{4\pi}}.
\end{align}
For AII$^\dag$, we need Owen's $T$-function at $x=0$ in its first argument, see \eqref{eq-OTi}, which is an elementary integral.

We would like to highlight that the edge limits of all three ensembles are translation invariant along the boundary since they only depend on $y$ and not on $\chi$ and $\eta^*$ independently. Indeed, when shifting $(\chi,\eta^*)\to(\chi,\eta^*)+t(z_0,-z_0^*)$ the variable $y$ remains unchanged, see \eqref{eq-zoom}. This could be expected for a smooth boundary of the global support because in the local limit the boundary will be well approximated by its tangent at this point so that the edge will follow this straight line.

\begin{figure}[t!]
	\centering
	\includegraphics[width=\textwidth]{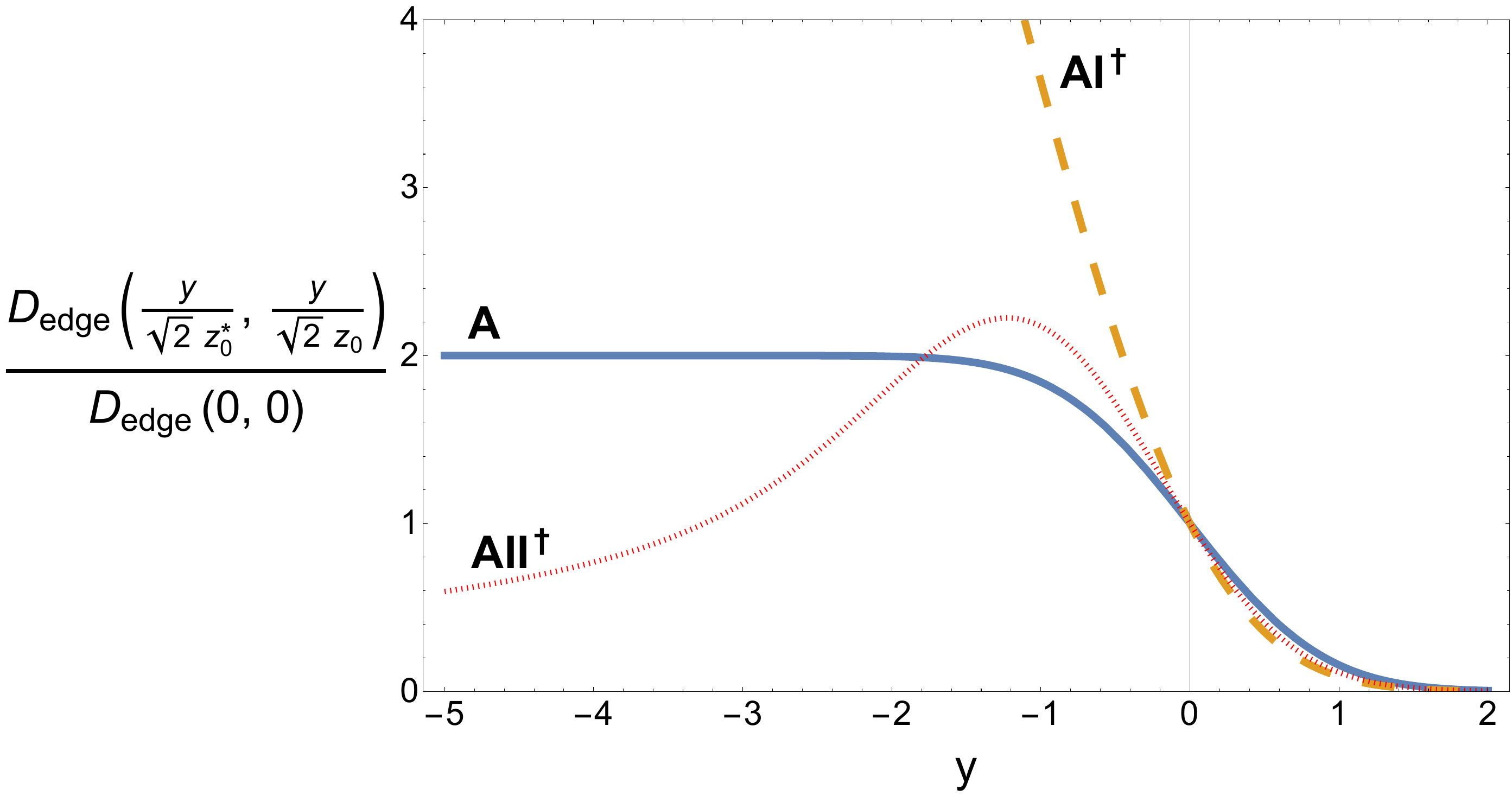}
	\caption[Figure 2 from \cite{AAKP}, \textit{Plot of the expectation value of one pair of characteristic polynomials in the edge scaling limit for the symmetry classes A, AI$^\dagger$, AII$^\dagger$}]{Plots of the expectation value of one pair of characteristic polynomials in the edge scaling limit as a function of a real $y=(z_0^*\chi+z_0\eta^*)/\sqrt{2}$, which can be achieved by setting $\chi=y/\sqrt{2}z_0^*$ and $\eta^*=y/\sqrt{2}z_0$. We normalised the expectation value by its value at $\chi=\eta^*=0$ for comparability which is given by \eqref{eq-edge.origin}. The curve for class A (solid blue) with $D_\textrm{edge}^\textrm{A}(\chi,\eta^*)$ from \eqref{eq-Dedge-A} normalised by $D_\textrm{edge}^\textrm{A}(0,0)$ saturates at $2$. Class AI$^\dag$ (dashed orange) given by \eqref{eq-Dedge-CS} has a clearly visible linear rise, and class AII$^\dag$ (dotted red) following \eqref{eq-Dedge-QS} shows the decay $\sim-1/\sqrt{2}y$ which starts to be visible for large negative values of $y$. \cite[Figure 2]{AAKP}}
	\label{fig-edge char poly}
\end{figure}

Theorem \ref{thm-lim_char_exp} implies that at the edge of the spectrum we indeed found three different classes of limiting characteristic polynomials. This is in agreement with the different behaviour of the three classes A, AI$^\dagger$ and AII$^\dagger$ we observed already in the numerics of the NN and NNN distributions in the bulk (cf. Section \ref{sssec-NN NNN classes}) and in the complex spacing ratio in the bulk and the edge (cf. Section \ref{sssec-complex spacing ratio}). We recall that they are conjectured to be universal based on numerics and symmetry arguments \cite{Hamzaki et al}. We plot the results for the characteristic polynomial at the edge from Theorem \ref{thm-lim_char_exp} in Figure \ref{fig-edge char poly} for illustration and comparison.

The difference between the three ensembles is obvious. It is perhaps simplest to see in the left tail asymptotic, $y\to-\infty$ for which we have given not only the leading but also the next to leading order. For the complex Ginibre ensemble the limiting characteristic polynomial saturates to unity, the mean value of the spectral density, whereas for the classes AI$^\dag$ and AII$^\dag$ a linear rise and an algebraic decay dominates, respectively, cf. Figure \ref{fig-rescaled poly}. 
The right tail in the complex Ginibre ensemble was noted first in \cite{Kanzieper replica} for the decay of the density at the edge. 

We can only speculate why a different scaling is needed in all three ensembles in the edge scaling limit. For comparison, in the real and quaternion Ginibre ensembles the corresponding skew-symmetric kernel, that relates to the spectral density, is obtained from the expectation value of two characteristic polynomials after multiplying, respectively dividing by the difference of the arguments. When rescaling them with $\sqrt{N}$ as it is done here, this will also give rise to extra powers in $\sqrt{N}$. Our choice of the numerical part of the scaling is natural as we multiply with $\sqrt{d_N}=\sqrt{N}$ for AI$^\dagger$ and divide by $1/\sqrt{d_N}=1/\sqrt{2N}$ for AII$^\dagger$.

\begin{proof}
\textbf{Class A:} The edge limit of the complex Ginibre ensemble is well known and has been summarised in the previous paragraph, especially one needs to insert the expansion \eqref{eq-P-lim} together with \eqref{eq-EN-P-relation} into the definition \eqref{eq-Dedge-A} and \eqref{eq-DN-A} to arrive at the claim. The tail asymptotic \eqref{eq-Dtails-A} follows from the expansion of the complementary error function \cite[Eq. 7.12.1]{NIST}
\begin{align}
	\label{eq-erfc-asympt}
	\erfc(z)&\sim 
	\left\{
	\begin{array}{r}
		\displaystyle \frac{1}{\sqrt{\pi}z}e^{-z^2}\left( 1-\frac{1}{2z^2}+\mathcal{O}(z^{-4})\right), \quad \mbox{for}\ z\to+\infty,\\
		\displaystyle 2+\frac{1}{\sqrt{\pi}z}e^{-z^2}\left( 1-\frac{1}{2z^2}+\mathcal{O}(z^{-4})\right), \quad \mbox{for}\ z\to-\infty,\\
	\end{array}
	\right.
\end{align}
where we give the next-to-leading order, as well.

\textbf{Class AI$\boldsymbol{^\dag}$:} For the complex symmetric ensemble, it is preferable to write $E_{N-1}(x)=\partial_xE_N(x)= 
\partial_x(e^xQ(N+1,x))$ in \eqref{eq-DN-CS}. Let us mention again that the scaling with $s=1/\sqrt{2}$ guarantees the unit disc as the limiting support. Proceeding as for the Ginibre ensemble, we multiply \eqref{eq-DN-CS} with $e^{-2zw^*}$ to cancel the exponential and set $x=2zw^*=N+\sqrt{2N}y+\chi\eta^*$,
\begin{align}
	e^{-2zw^*}
	\frac{D_N^{\textrm{AI}^\dag}\left(z,w^*\right)}{D_N^{\textrm{AI}^\dag}(0,0)}=Q(N+1,x)-\frac{x e^{-x}}{N+1}\partial_x\left(e^xQ(N+1,x) \right).
\end{align}
With help of the edge rescaling \eqref{eq-zoom1} at $|z_0|=1$ and the asymptotic \eqref{eq-Q-lim} with $y=(z_0^*\chi+z_0\eta^*)/\sqrt{2}$, we arrive at
\begin{align}
	e^{-2zw^*}
	\frac{D_N^{\textrm{AI}^\dag}\left(z,w^*\right)}{D_N^{\textrm{AI}^\dag}(0,0)}=& \frac12\erfc(y)+\mathcal{O}(N^{-\frac12})\nonumber\\
	&-\frac{N+\sqrt{2N}y+\chi\eta^* }{N+1}\left(1+\frac{1}{\sqrt{2N}}\partial_y\right)\left(\frac12\erfc(y)+\mathcal{O}(N^{-\frac12}) \right)
	\nonumber\\
	=&-\frac{y}{\sqrt{2N}}\erfc(y)+\frac{1}{\sqrt{2\pi N}}e^{-y^2}+\mathcal{O}\left(\frac{1}{N}\right).
\end{align}
The two order one terms cancel after expanding the ratio in $1/N$. Eventually, we multiply with $\sqrt{N}$ to find the result \eqref{eq-Dedge-CS}. 

For the tail asymptotic we insert \eqref{eq-erfc-asympt} into \eqref{eq-Dedge-CS} and see that the leading order Gaussian decay cancels for the right tail, $y\to+\infty$ 
\begin{align}
	D_\textrm{edge}^{\textrm{AI}^\dag}(\chi,\eta^*)&\sim -y\frac{1}{\sqrt{2\pi}y}e^{-y^2}\left(1-\frac{1}{2y^2}+\mathcal{O}(y^{-4})\right)+\frac{1}{\sqrt{2\pi}}e^{-y^2}\nonumber\\ & = \frac{1}{\sqrt{8\pi}\,y^2}e^{-y^2}\left(1+\mathcal{O}(y^{-2})\right).
\end{align}
The errors are uniform for any compact set where $y$ can be drawn from so that the derivative can indeed be interchanged with the limit $N\to\infty$.

For the left tail with $y\to-\infty$, we obtain from the corresponding asymptotic \eqref{eq-erfc-asympt} of the complementary error function, 
\begin{align}
	D_\textrm{edge}^{\textrm{AI}^\dag}(\chi,\eta^*)&\sim 
	\frac{1}{\sqrt{2\pi}}e^{-y^2}-\frac{y}{\sqrt{2}}\left(2+\frac{1}{\sqrt{\pi}y}e^{-y^2}
	\left(1-\frac{1}{2y^2}+\mathcal{O}(y^{-4})\right)
	\right)
	\nonumber\\
	&=-\sqrt{2}y+\frac{1}{\sqrt{8\pi}\,y^2}e^{-y^2}\left(1+\mathcal{O}(y^{-2})\right),
\end{align}
a linear rise in $y$ (as $y<0$) as the leading term.

\textbf{Class AII$\boldsymbol{^\dag}$:} For the remaining ensemble of complex self-dual matrices, we cannot use standard asymptotic results. Our starting point is the integral representation \eqref{eq-fNx-int2} of \eqref{eq-DN-QS} in the derivation for finite $N$, that we repeat for the readers' convenience:
\begin{align}\label{eq-DNAII+fNrel}
\begin{split}
	&\frac{D_N^{\textrm{AII}^\dag}(z,w^*)}{D_N^{\textrm{AII}^\dag}(0,0)}\biggl |_{x=zw^*}=\frac{1}{(2N)!}f_N(x),\quad 
	f_N(x)= \int_0^\infty\! dv\int_0^\infty\! du \ e^{S(u,v)},\quad\\
	&S(u,v):=-u-v+N\log\left(4xv+(2x+u)^2\right).
	\end{split}
\end{align}
This calls for a Laplace approximation at large $N$. Any critical point $(u_*,v_*)$ satisfies
\begin{align}
	\begin{split}
	0&=\partial_v S(u_*,v_*)= -1+\frac{4Nx}{4xv_*+(2x+u_*)^2}\quad\textrm{and}\quad\\
	 0&=\partial_u S(u_*,v_*)= -1+\frac{2N(2x+u_*)}{4xv_*+(2x+u_*)^2}.\quad
	\label{eq-SPeqs}
	\end{split}
\end{align}
Inserting the first into the second equation yields $u_*=0$ and reinserted into the first equation gives $v_*=N-x$. 

To figure out whether the point $(u_*,v_*)$ is the global maximum of $S(u,v)$ we need to compare it with the extrema along the two boundaries. For the boundary $u=0$ and $v\geq0$, 
\begin{align}
	0=\partial_v S(0,v)= -1+\frac{N}{v+x},
\end{align}
we find the same point $(u_*,v_*)$, whereas along the boundary $u\geq0$ and $v=0$ 
\begin{align}
	0=\partial_uS(u_0,0)= -1+\frac{2N}{2x+u_0},
\end{align}
we find $(u_0,v_0)=(2(N-x),0)$. Taking the second derivative confirms that both extrema are indeed local maxima. 
However, when inserting the scaling limit \eqref{eq-zoom1} with $x=N+\sqrt{N}y+\chi\eta^*/2$, since $d_N=2N$ and $s=1/\sqrt{2}$, we see that both maxima $(v_*,u_*)$ and $(u_0,v_0)$ are only within the integration domain $v,u\geq0$ for negative values $y<0$. This can be expected as this is the direction into the bulk of the spectrum which becomes immediate when setting $z=w$.

What is more important is that the leading order of both extrema is $\sqrt{N}$ which is the same scale as the standard deviations of the Gaussians once we would expand about these points. This means that even an expansion about the origin would cover all relevant contributions from this integral. To pursue this direction, we rescale $v=\sqrt{N}v'$ and $u=2\sqrt{N}u'$,
\begin{align}
	f_N(x)=(2N)^{2N+1}\int_0^\infty\! dv'\int_0^\infty\! du'\ 
	e^{-\sqrt{N}v'-2\sqrt{N}u'+N\log\left(\frac{xv'}{N^{3/2}}+ \left(\frac{x}{N}+\frac{u'}{\sqrt{N}}\right)^2\right)
}.
	\label{eq-fNexp}
\end{align}
The logarithm can be expanded as follows,
\begin{align}
	\begin{split}
		&\log\left[\frac{xv'}{N^{3/2}}+ \left(\frac{x}{N}+\frac{u'}{\sqrt{N}}\right)^2\right]\\
		=&\log\left[1+\frac{v'+2u'+2y}{\sqrt{N}}+\frac{1}{N}\left(yv'+(y+u')^2+\chi\eta^*\right)+\mathcal{O}\left(\frac{v'+u'}{N^{3/2}}\right)\right]\\
		=&1+\frac{1}{\sqrt{N}}(v'+2u'+2y)-\frac{1}{N}\left(\frac{(v')^2}{2}+(u')^2+y^2+2v'u'+v'y+2u'y-\chi\eta^*\right)\\
		&+\mathcal{O}\left(\frac{(v')^3+(u')^3}{N^{3/2}}\right).
	\end{split}
\end{align}
This expansion can be stopped at the second order as long as $(v')^3+(u')^3\ll \sqrt{N}$ because the logarithm is multiplied with $N$. This restriction is, nonetheless, not a problem since already for $v'+u'\geq N^{\epsilon}$ for any fixed $\epsilon>0$ the integrand in \eqref{eq-fNexp} times $e^{-2\sqrt{N}y}$ is exponentially small which we omit to show here. It is essentially due to the fact that the maximum of the integrand in the region $v'+u'\geq N^{\epsilon}$ is attained on the boundary $v'+u'= N^{\epsilon}$, because the directional derivative of $S(u,v)$ is always negative when the direction vector is non-zero and has non-negative entries, cf. the partial derivatives \eqref{eq-SPeqs} which are negative in this region.

Summarising, we find asymptotically
\begin{align}
	f_N(x)\sim(2N)^{2N+1}e^{2\sqrt{N}y-y^2+\chi\eta^*}\int_0^\infty\! dv\int_0^\infty\! du\,
	e^{-u^2-\frac12v^2-y(v+2u)-2vu},
\end{align}
when $N\to\infty$.
Consequently, the full expression together with the scaling \eqref{eq-zoom1} reads to leading order
\begin{align}
	\frac{e^{-2zw^*}}{\sqrt{2N}}\frac{D_N^{\textrm{AII}^\dag}(z,w^*)}{D_N^{\textrm{AII}^\dag}(0,0)}
	\sim\frac{(2N)^{2N+1}}{\sqrt{2N}(2N)!}
	e^{-2N-y^2}\int_0^\infty\! dv\int_0^\infty\! du\,
	e^{-u^2-\frac12v^2-y(v+2u)-2vu}.
\end{align}
After we exploit the Stirling formula $(2N)!\approx \sqrt{2\pi}(2N)^{2N+1/2}e^{-2N}$, we are allowed to take the large $N$-limit leading to
\begin{align}
	D_\textrm{edge}^{\textrm{AII}^\dag}(\chi,\eta^*)= \frac{1}{\sqrt{2\pi}}e^{-y^2}\int_0^\infty\! dv\int_0^\infty\! du\,
	e^{-u^2-\frac12v^2-y(v+2u)-2vu}.
\end{align}

We can simplify this expression further when substituting $v'=(v+y)/\sqrt{2}$ and $u'=u+v+y$
\begin{align}
	D_\textrm{edge}^{\textrm{AII}^\dag}(\chi,\eta^*)&=\frac{1}{\sqrt{\pi}}e^{-y^2/2}\int_{y/\sqrt{2}}^\infty\! dv'\,e^{+\frac12v'^2}\int_{\sqrt{2}v'}^\infty\! du'\,e^{-u'^2}\nonumber \\
	&=\frac{1}{2}e^{-y^2/2}\int_{y/\sqrt{2}}^\infty\! dv\,e^{v^2}\erfc(\sqrt{2}v),
	\label{eq-DN-AII-int1}
\end{align}
where we employed the definition \eqref{eq-erfc-def} of the complementary error function. To get a closed expression as stated in \eqref{eq-Dedge-QS}, we make the ansatz  
\begin{align}
	\label{eq-gdiff}
	g'(v)=-\frac12\erfc(\sqrt{2}v)\, e^{v^2},
\end{align}
cf. Appendix \ref{sec-app expec poly}, which would simplify the integral \eqref{eq-DN-AII-int1} as follows
\begin{align}
	D_\textrm{edge}^{\textrm{AII}^\dag}(\chi,\eta^*)=e^{-y^2/2}(g(y/\sqrt{2})-g(+\infty)).
	\label{eq-DAII-g}
\end{align}
This approach is motivated by the fact that the original integral \eqref{eq-fNx-int1} satisfies an inhomogeneous first order differential equation, as derived in the Appendix \ref{sec-app expec poly}. Here, we directly solve the differential equation \eqref{eq-gdiff} for $g(v)$. 

For this purpose, we make use of Owen's $T$-function with an imaginary number in its second entry \eqref{eq-OTi}. Its derivative in its first entry is explicitly given by
\begin{align}
	\partial_x T\left(x,ib\right)=-\frac{ix}{2\pi}\int_0^b\! du\ e^{-\frac12 x^2(1-u^2)}=-\frac{i}{\sqrt{8\pi}}e^{-x^2/2}{\erfi}\left[\frac{bx}{\sqrt{2}}\right].
\end{align}
Thence, when choosing $x=2v$ and $b=1/\sqrt{2}$, the derivative of
\begin{align}
	g(v)=-\frac{\sqrt{\pi}}{4}\erfc\left(\sqrt{2}\, v\right)\erfi(v)-i\sqrt{\pi}T\left(2v,\frac{i}{\sqrt{2}}\right)
	\label{eq-g-sol}
\end{align}
solves the differential equation \eqref{eq-gdiff}.

We still have to determine $g(+\infty)$.
While the asymptotic of the complementary error function is given in \eqref{eq-erfc-asympt}, for the error function at imaginary argument we have \cite{Dingle}
\begin{align}
	\erfi(v)=\frac{1}{\sqrt{\pi}\,v}e^{+v^2}\left(1+\frac{1}{2v^2}+\mathcal{O}(1/v^4)\right),
\end{align}
which is even valid for both $y\to\pm\infty$.
Therefore, for the first term in \eqref{eq-g-sol} we obtain
\begin{align}
	-\frac{\sqrt{\pi}}{4}\erfc\left(\sqrt{2}\, v\right)\erfi(v)
	\sim
	\left\{
	\begin{array}{r}
		\displaystyle-\frac{1}{4\sqrt{2\pi}v^2}e^{-v^2}\left( 1+\frac{1}{4v^2}+\mathcal{O}(v^{-4})\right), \quad \mbox{for}\ v\to+\infty,\\
		\displaystyle-\frac{1}{2v}e^{+v^2}\left( 1+\frac{1}{2v^2}+\mathcal{O}(v^{-4})\right), \quad \mbox{for}\ v\to-\infty,\\
	\end{array}
	\right.
\end{align}
implying that it vanishes for $v\to+\infty$.
For the tail asymptotic, we also give the expansion for $v\to-\infty$.

We did not find the asymptotic of Owen's $T$-function in the literature. Hence, we present a short derivation. Starting from \eqref{eq-OTi}, the substitution $z=(1-u^2)v^2$ leads to 
\begin{align}
	-i\sqrt{\pi}T\left(2v,i/\sqrt{2}\right)&=\frac{1}{\sqrt{8\pi}\,v^2}e^{-v^2}\int_0^{v^2}\! dz\,{e^{-z}}\left(1+\frac{z}{v^2}\right)^{-1}\left(1-\frac{z}{v^2}\right)^{-1/2}
	\nonumber\\
	&=\frac{1}{\sqrt{8\pi}\,v^2}e^{-v^2}\int_0^{v^2}\! dz\,{e^{-z}}\left[1 -\frac12 \frac{z}{v^2}+\mathcal{O}\left(\frac{z^2}{v^4}\right)\right]
	\nonumber\\
	&=\frac{1}{\sqrt{8\pi}\,v^2}e^{-v^2}\left[1 -\frac{1}{2v^2}+\mathcal{O}\left(\frac{1}{v^4}\right)\right],
\end{align}
where in the last step we used elementary integrals. The expansion of the integrand is only uniform when cutting off at $z\leq v^{\epsilon}\ll v^2$ for any constant $\epsilon\in(0,2)$. Like before this is not a problem as the integrand is uniformly bounded from above by $e^{-v^\epsilon}/\sqrt{1-z/v^2}$ when $z\in( v^\epsilon,v^2)$, whose integral thus vanishes exponentially for $v\to\infty$.

Let us remark that the expansion in $1/v$ holds for both $y\to\pm\infty$; in particular, it vanishes in both limits.
As a result, the asymptotic yields $g(+\infty)=0$ so that the full solution for the edge scaling limit from \eqref{eq-DAII-g} is given by \eqref{eq-Dedge-QS}.

After having evaluated the asymptotic of $g(v)$ already in great detail, we can immediately give the tail estimates. For $y\to+\infty$, it is 
\begin{align}
	D_\textrm{edge}^{\textrm{AII}^\dag}(\chi,\eta^*)\sim e^{-y^2}\frac{1}{\sqrt{8\pi}\,y^2}\left( 1-\frac{5}{2y^2}+\mathcal{O}(y^{-4})\right),
\end{align}
where both terms in \eqref{eq-Dedge-QS} contribute to the same order. For $y\to-\infty$, the asymptotic of the complementary error function \eqref{eq-erfc-asympt} gives the leading contribution while Owen's $T$-function can be neglected. Thence, it is
\begin{align}
	D_\textrm{edge}^{\textrm{AII}^\dag}(\chi,\eta^*)\sim -\frac{1}{\sqrt{2}y}\left( 1+\frac{1}{y^2}+\mathcal{O}(y^{-4})\right)
	+ e^{-y^2}\frac{1}{\sqrt{2\pi}\,y^2}\left( 1+\mathcal{O}(y^{-2})\right),
\end{align}
which completes the derivation of\eqref{eq-Dtails-QS}. We note that for negative $y$ this is a positive function. 
\end{proof}

\noindent \textbf{Bulk Scaling Limit for One Pair of Characteristic Polynomials}\\
\\
\noindent
The limiting expressions for the rescaled characteristic polynomials in the bulk can be defined in a similar way as at the edge.

\begin{prop}
	[Local limits of one pair of characteristic polynomials in the spectral bulk]
	\label{prop-lim_char_exp}
	
	Recalling Proposition \ref{prop-char_exp}, the local scaling in the bulk, defined as 
	\begin{align}
		z=\sqrt{d_N}sz_0+s\chi\ , \quad
		w^*=\sqrt{d_N}sz_0^*+s\eta^*, \quad zw^*=s^2(d_N|z_0|^2+\sqrt{2d_N}\,y
		+\chi\eta^*)
		\label{eq-zoom2}
	\end{align}
	with $|z_0|<1$ and $y=({z_0^*\chi+z_0\eta^*})/{\sqrt{2}}$ as well as the definitions \eqref{eq-def.dim} and \eqref{eq-def.scale}, leads to
	\begin{align}
		D_\textrm{bulk}^\textrm{A}(\chi,\eta^*)&:=\lim_{N\to\infty}
		e^{-zw^*}\frac{D_N^\textrm{A}(z,w^*)}{D_N^\textrm{A}(0,0)}=1 ,
		\label{eq-Dbulk-A}\\
		D_\textrm{bulk}^{\textrm{AI}^\dag}(\chi,\eta^*)&:=\lim_{N\to\infty}
		e^{-2zw^*}\frac{D_N^{\textrm{AI}^\dag}\left(z,w^*\right)}{D_N^{\textrm{AI}^\dag}(0,0)} =1-|z_0|^2  ,
		\label{eq-Dbulk-CS}
		\\
		D_\textrm{bulk}^{\textrm{AII}^\dag}(\chi,\eta^*)&:= 
		\lim_{N\to\infty}e^{-2zw^*}\frac{D_N^{\textrm{AII}^\dag}(z,w^*)}{D_N^{\textrm{AII}^\dag}(0,0)}=\frac{1}{1-|z_0|^2} .
		\label{eq-Dbulk-QS}
	\end{align}
\end{prop}

All three results are consistent with the respective global limit in Eqs. \eqref{eq-Dglobal-Gin}, \eqref{eq-Dglobal-CS} and \eqref{eq-Dglobal-QS} respectively, with $r=|z_0|$. In particular, they give unity at the origin, by setting $|z_0|=0$ on the RHS, which is consistent with their definition. This agreement and its independence of $\chi$ and $\eta^*$ reflect that all three ensembles satisfy locally translation invariance and isotropy as it is well known for the three Ginibre ensembles.

\begin{proof}
\textbf{Class A:} Looking at \eqref{eq-DN-A}, the result for the Ginibre ensemble is trivial, as the corrections to $e^{-x}E_N(x)\to 1$ are exponentially suppressed in this kind of large $N$-limit.

\textbf{Class AI$\boldsymbol{^\dag}$:} Likewise, the result \eqref{eq-DN-CS} for complex symmetric matrices yields for the limit \eqref{eq-Dbulk-CS}
\begin{align}
	\lim_{N\to\infty} e^{-2zw^*}\left(E_N(2zw^*)-\frac{2zw^*}{N+1}E_{N-1}(2zw^*)\right)
	&=-\frac{N|z_0|^2+\sqrt{2N}\,y+\chi\eta^*}{N+1}
	\nonumber\\
	&= 1-|z_0|^2-\sqrt{\frac{2}{N}}\,y + \mathcal{O}(1/N),
\end{align}
as both truncated exponentials give unity.
Evidently, the sub-leading term contains non-trivial information about the local spectral fluctuation $y$, that deviates from a constant behaviour.

\textbf{Class AII$\boldsymbol{^\dag}$:} Let us turn to the ensemble of complex self-dual matrices, where we return to \eqref{eq-SPeqs} determining the critical points. We note that this time the bulk scaling \eqref{eq-zoom2} is equal to $x=zw^*=N|z_0|^2+\sqrt{N}y+\chi\eta^*/2$, with $0<|z_0|<1$. Because of that, the maximum at $(u_*,v_*)=(0,N-x)$ (and maximum at the boundary $(0,v>0)$) as well the maximum $(u_0,v_0)=(2(N-x),0)$ at the boundary $(u>0,0)$ are both inside the integration domain $u,v\geq0$.

We need to compare the integrand in $f_N(x)$, see \eqref{eq-fNx-int2}, at these two points to decide which of those is the global maximum. For this purpose, we ask for the leading asymptotic of the integrand $\exp[S(u,v)]$, with
\begin{align}
	S(u,v)=-u-v+N\log[4xv+(2x+u)^2],
\end{align}
at these two points which is 
\begin{align}\begin{split}
	e^{S(u_*,v_*)}&\sim (2N|z_0|)^{2N}\exp[-N(1-|z_0|^2)]
	\quad \textrm{and}\quad \\ e^{S(u_0,v_0)}&\sim(2N)^{2N}\exp[-2N(1-|z_0|^2)].
	\end{split}
\end{align}
Thus, we obtain for the difference
\begin{align}
	S(u_*,v_*)-S(u_0,v_0)\sim N(1-|z_0|^2+2\log|z_0|)
\end{align}
The function $h(a)=1-a+\log a$ is strictly increasing for $a=|z_0|^2\in(0,1)$ and takes its maximum at $a=1$ which is $h(1)=0$. Therefore, it is $S(u_*,v_*)<S(u_0,v_0)$ for suitable large $N$ and $(u_0,v_0)=(2(N-x),0)$ is the position of the global maximum.

Consequently, we substitute $u=2N(1-|z_0|^2)+2\sqrt{N}u'$ while the variable $v$ remains unchanged. Plugging this into the integral \eqref{eq-DNAII+fNrel} we arrive at
{\small
\begin{align}
	f_N(x)\!=& 2\sqrt{N}\!\int_0^\infty\!\! dv \!\int_{-\sqrt{N}(1-|z_0|^2)}^\infty\!\!\! du'
	e^{-2N(1-|z_0|^2)-2\sqrt{N}u'-v} 
	\!\left[ 4xv \!+\! (2x \!+\! 2N(1-|z_0|^2)\!+\!2\sqrt{N}u')^2\right]^N 
	\nonumber\\
	=& \frac{(2N)^{2N+1}}{\sqrt{N}}e^{-2N(1-|z_0|^2)}\int_0^\infty\! dv\int_{-\sqrt{N}(1-|z_0|^2)}^\infty\! du'\, e^{-2\sqrt{N}u'-v} 
	\left[1+\frac{2}{\sqrt{N}}(u'+y)
	\right.
	\nonumber\\
	&\left.
	+\frac{1}{N}\left(v|z_0|^2+\chi\eta^*+(u'+y)^2\right)
	+\frac{1}{N^{\frac32}}\left(vy+\chi\eta^*(u'+y)\right)
	+\mathcal{O}\left(\frac{v+(u')^2}{N^{2}}\right)\right]^N.
\end{align}}
Here, we have inserted the bulk scaling \eqref{eq-zoom2} and expanded far enough to keep the first sub-leading order in the final result. 
In the next step, we have to exponentiate the bracket to the $N$th power, and then expand the logarithm up to third order, $\log[1+z]=z-\frac{z^2}{2}+\frac{z^3}{3}+\mathcal{O}(z^4)$. In addition, from the definition \eqref{eq-Dbulk-QS} we have to appropriately normalise according to \eqref{eq-DNAII+fNrel}. Inserting all together a lot of terms cancel, and we arrive at 
{\small
\begin{align}
	e^{-2x}\frac{f_N(x)}{(2N)!}=&
	\frac{(2N)^{2N+1}e^{-2N}}{\sqrt{N}(2N)!}\int_0^\infty \! dv\int_{-\sqrt{N}(1-|z_0|^2)}^\infty\! du'\,\exp\left[-v(1-|z_0|^2)-(u'+y)^2\right]
	\nonumber\\
	&\times\left[1+\frac{1}{\sqrt{N}}\left( vy-2(u'+y)\left(vy+\frac12\chi\eta^*-\frac13(u'+y)^2\right)\right)+\mathcal{O}\left(\frac{v^2+(u')^4}{N}\right)\right]. \qquad
\end{align}}
Applying the Stirling formula simplifies the prefactor to $1/\sqrt{\pi}$ when $N\to\infty$. The error in the integral is uniform when we restrict the integration to $v^2+(u')^4\ll N$. As before, one can show that this does not cause any problems since the integrand is exponentially small compared to the main contribution at the origin for any choice $\epsilon>0$ of $v+|u'|\geq N^\epsilon$, since the maximum of $S(u,v)$ in this region lies on the boundary $v+|u'|= N^\epsilon$. Therefore, Lebesgue's dominated convergence theorem is applicable.

A similar statement would also allow us to extend the integral to $-\infty$ for which we skip the rigorous mathematical details. Once done, we can exploit that all the odd moments in the Gaussian variable $u'+y$ vanish, i.e., 
\begin{align}
	e^{-2x}\frac{f_N(x)}{(2N)!}=
	\int_0^\infty\! dv \,\exp\left[-v(1-|z_0|^2)\right]\left[1+\frac{vy}{\sqrt{N}}+\mathcal{O}\left(\frac{v^2}{N}\right)\right].
\end{align}
Here, we carried out the integral over $u'$ yielding $\sqrt{\pi}$ up to order $1/N$. We are left with the elementary integral over the variable $v$ which is
\begin{align}
	e^{-2x}\frac{f_N(x)}{(2N)!}= \frac{1}{1-|z_0|^2} +\frac{1}{\sqrt{N}}\frac{y}{(1-|z_0|^2)^2}+\mathcal{O}(N^{-1}). 
\end{align}
This ends the derivation of the bulk limit for class AII$^\dag$ in \eqref{eq-Dbulk-QS}, including the first $y$-dependent correction term.
\end{proof} 
Let us briefly summarise this Section \ref{ssec-expectation of characteristic polynomials}. We provided analytical results for three ensembles of non-Hermitian random matrices classes A, AI$^\dagger$ and AII$^\dagger$ via a method which is applicable to each of them. Choosing
expectation values of one pair of conjugate characteristic polynomials as correlations functions, we proved that indeed all three ensembles lead to different local bulk and edge statistics for the complex eigenvalues. The explicit expressions for one pair in classes AI$^\dag$ and AII$^\dag$, see Proposition \ref{prop-char_exp}, show a remarkably similar structure to the well-known respective answer for the complex Ginibre ensemble for finite $N$, being given in terms of sums of truncated exponentials. This allowed us to derive asymptotic results in the local bulk and edge scaling limit, as well as for the global large $N$-limit, with different results in all three limits for each ensemble, cf. Propositions \ref{prop-global} and \ref{prop-lim_char_exp} and Theorem \ref{thm-lim_char_exp}. 
In particular, the resulting properties at the edge are compatible with the expected behaviour of the spectral density based on an approximate CG description with three different values of inverse temperature $\beta$, discussed in the previous Section \ref{ssec-numerics symmetry classes}.

\subsection{Application: Spin Chain with Dissipative Disorder}\label{ssec-application spin chain}
As remarked throughout this thesis, applications of random matrices play always an important role in the motivation of new studies as well as probing results. Therefore, we come now to an example, where we probe the statistics of a quantum spin chain against the statistics of various ensembles, we encountered already. We follow \cite{ABCPRS} closely, where the author mainly contributed to the numerical analysis of the local statistical behaviour in the bulk of the spectrum. This section illustrates the applicability of non-Hermitian random matrix statistics to quantum systems and thus strengthens the importance of the presented numerical and analytical effort in finding new results and matrix ensembles in this research field. We briefly revisit the motivation of such models: Quantum systems are in general not perfectly isolated but rather interact with their surrounding environment. When the timescales of the environment are much shorter than the system's ones and under the Born approximation, the reduced system's dynamics can be effectively described by a Schrödinger equation with a non-Hermitian Hamiltonian \cite{MMCN 19, NAJM 19,RPCB 22} and thus generically has a complex spectrum. 
We recall that the GHS conjecture (cf. the beginning of Section \ref{sec-non Herm beta}) indicates the NN distribution as a good measure to distinguish chaotic from integrable behaviour \cite{GHS}. Here comes the universal statistical behaviour of the non-Hermitian symmetry classes into play. It depends on the symmetry of the Hamiltonian which random matrix statistics are reached. We will encounter class AI$^\dagger$ in the model discussed below.\\
The crossover between integrable and chaotic systems has currently been studied in systems that are either Hermitian---it then corresponds to going from a 1d Poisson process to a Hermitian random matrix ensemble---or non-Hermitian---the transition is then from a 2d Poisson to one of the non-Hermitian symmetry classes. However, the interplay between Hermiticity breaking and integrability breaking has never been investigated. We do it here, by studying a strongly-interacting many-body system with dissipative local disorder. In absence of dissipation, disordered interacting quantum models have been the focus of much attention recently, due to the presence of many-body localized phases \cite{Basko2006Metal,OH,Znidaric2008Many,Pal2010Many,Bardarson2012Unbounded,DeLuca2013Ergodicity,Abanin2019Colloquium,Sierant2024Many}. These are robust, non-ergodic phases of quantum matter, that violate the eigenstate thermalization hypothesis \cite{Deutsch1991Quantum,Srednicki1994Chaos} and display an emergent integrability via local integrals of motion \cite{Serbyn2013Local,Ros2015Integrals,Imbrie2016Diagonalization}. Similar non-chaotic regimes have been conjectured to be present also in an open setting, especially in the context of non-Hermitian Hamiltonians \cite{hamazaki2019non,Zhai2020Many,unfolding Suthar et al,Ghosh2022Spectral,detomasi2023stable}. 
\\We consider a model, which was introduced in reference \cite{roccati2024diagnosing}. It has the peculiarity to have a purely imaginary disorder, breaking at the same time both integrability and Hermiticity. This model is motivated by a recent experimental work \cite{lapp2019engineering}, showing that local tunable loss terms can be engineered in synthetic lattices. Moreover, the model allows us to investigate whether the breaking of Hermiticity happens on a different scale than the transition between chaotic and integrable behaviour. We quantitatively characterize the crossovers by comparing the NN (and NNN) level statistics of complex eigenvalues to that of a static 2dCG. The latter allows us to reproduce the 2d Poisson and approximate the random matrix class AI$^\dag$, thus enabling to capture a transition from integrable ($\beta=0)$ to chaotic ($\beta=1.4$) behaviour, respectively. The breaking of Hermiticity is captured through a $D$-dimensional Poisson point process (cf. Section \ref{ssec-point processes}), with $D$ transitioning from a value of 1 for a purely Hermitian spectrum to 2 for strongly non-Hermitian systems. 
\\In the next Section \ref{sssec-model spin} we describe our model, the XXZ spin chain with purely imaginary disorder, which is Hermitian and integrable at vanishing disorder, and its symmetries. 
Section \ref{sssec-results spin} contains our results, namely the transition between random matrix and 2d Poisson statistics in $\beta$ of the 2dCG, and, for very small disorder, the transition from 1d to 2d Poisson statistics, indicating a Hermiticity breaking crossover. The transition in terms of the complex spacing ratio is also investigated in \cite{ABCPRS}. The observed distributions are similar to the ones discussed in Section \ref{sssec-complex spacing ratio} for the 2d Poisson process and class AI$^\dagger$, such that we omit the discussion here and refer the reader to \cite[Section III.D]{ABCPRS} for details.
\subsubsection{The Setup}\label{sssec-model spin}
In this section, we describe the dissipative model under consideration and discuss some aspects of its symmetries and spectrum. We consider a chain of $L$ interacting spins with local dissipative disorder, described by the non-Hermitian Hamiltonian
\begin{align}
	\label{eq-XXZ_iW_Hamiltonian}
	H = H_{\textrm{XXZ}} - i \, \Gamma/2,
\end{align}
with the XXZ spin chain Hamiltonian 
\begin{align}\label{eq-H_XXZ}
	H_{\textrm{XXZ}} = \sum_{a=1}^L \left( S_a^x S_{a+1}^x + S_a^y S_{a+1}^y + \mathcal{D} S_a^z S_{a+1}^z \right),
\end{align}
and 
\begin{align}
	\label{eq-Gamma}
	\Gamma = \sum_{a=1}^L \gamma_a \left(S^z_a + \frac{1}{2} \mathds{1}\right) 
\end{align}
representing site-dependent, local, \textit{dissipative} disorder, because of the imaginary $i$ in \eqref{eq-XXZ_iW_Hamiltonian}. We choose the anisotropy factor $\mathcal{D}=1$ in our simulations.
The site-dependent losses $\gamma_a$ can be experimentally implemented, as demonstrated in a synthetic lattice \cite{lapp2019engineering}. Here, they are sampled from a uniform distribution $\mathcal{U}(0,\gamma)$, where $\gamma$ represents the `strength' of the disorder. For each value of $\gamma$, we generate spectral data from a number of realizations of the model, each realization having a different randomly chosen set of local disorder terms, $\{\gamma_a \}_{a=1}^L$. We consider the case of periodic boundary conditions.

Note that in this model, all the disorder and its implications to the spectral statistics come from the openness of the system---the spin chain Hamiltonian \eqref{eq-H_XXZ} is always fixed and is integrable according to the Bethe Ansatz \cite{XXZ_bethe_ansatz}. For the study of spectral statistics of non-Hermitian models with real disorder see e.g. \cite{hamazaki2019non, huang2020anderson, roccati2024diagnosing}.
In \cite{roccati2024diagnosing}, it was shown that increasing the range $\gamma$, the statistics of the \textit{singular values} exhibit a crossover from chaotic to integrable. A similar observation is made here from the eigenvalue spacing statistics. We complete the story by extending the analysis to very small disorder, which reveals a richer behaviour coming from Hermiticity breaking. 
For some aspects of thermalization and transport in this model, see \cite{Mahoney:2024sgu}. 
This model preserves the total spin in the $z$ direction, $\mathcal{S}_z = \sum_{a=1}^L S_a^z$. We thus work in a specific total $\mathcal{S}_z$ magnetization sector, usually the largest one with $\mathcal{S}_z=0$ which occurs for even $L$ and is of dimension $\binom{L}{L/2}$. Without disorder ($\gamma=0$), additional symmetries need to be considered to avoid exact degeneracies in the spectrum (see e.g. \cite{joel2013introduction}). Specifically, each $\mathcal{S}_z$ sector then 
splits into two: one sector with states of even parity and another with states of odd parity.

The Hamiltonian \eqref{eq-XXZ_iW_Hamiltonian} falls into the universality class AI$^\dagger$ of non-Hermitian random matrices \cite{Hamzaki et al}. 
In addition, the complex spectrum of the Hamiltonian \eqref{eq-XXZ_iW_Hamiltonian} has reflection symmetry along a line parallel to the real axis of the complex plane at $(-i/4)\sum_{a=1}^L\gamma_a$, i.e. the eigenvalues come in complex conjugate pairs above and below this line. This reflection symmetry in the complex spectrum is a result of the matrix being \textit{pseudo-Hermitian} \cite{pauli1943dirac, feinberg2022metrics, melkani2023degeneracies}. We recall from \eqref{eq-pseudo}, that a non-Hermitian matrix is pseudo-Hermitian if there exists a Hermitian matrix $S$ such that\footnote{Note that we add the condition that $S$ is Hermitian here compared to \cite{ABCPRS}. It can be seen in various references, that this is a necessary condition for a proper definition of pseudo-Hermicitiy for non-Hermitian Hamiltonians, cf. \cite{feinberg2022metrics,melkani2023degeneracies,Mostafazadeh, Bender Boettcher}. }
\begin{align}
	\label{eq-Pseudo-Hermitian}
	H = SH^\dag S^{-1} ~.
\end{align}
Comparing to the discussion in Section \ref{ssec-intro symmetry classes}, here $S$ is not assumed to be unitary as in Eq. \eqref{eq-pseudo}. However, $S$ can be chosen to be additionally unitary as each Hermitian matrix $S$ generates a unitary matrix as $e^{iS}$, which leaves \eqref{eq-Pseudo-Hermitian} invariant, cf. \cite{Mostafazadeh}. Indeed, if such a similarity transformation exists, 
and $z$ is an eigenvalue of $H$ with right-eigenvector $\mathbf{R}$, $H\mathbf{R} = z \mathbf{R}$, then also $z^*$ is an eigenvalue of $H$. This can be shown by taking the Hermitian conjugate $H^\dag\mathbf{R}^* = z^* \mathbf{R}^*$ and using \eqref{eq-Pseudo-Hermitian}, giving $HS\mathbf{R}^* = z^* S\mathbf{R}^*$, which means that $z^*$ is an eigenvalue of $H$ with eigenvector $S\mathbf{R}^*$.

In our case, such a $S$ would need to flip the sign of $\Gamma$ in Eq. \eqref{eq-Gamma},\footnote{After removing the term proportional to the identity, which is responsible for the constant imaginary shift in the complex plane.} while keeping $H_{\textrm{XXZ}}$ unchanged. It turns out that this can be achieved by the operator
\begin{align}
	S = \prod_{a=1}^L S_a^x ~,
\end{align}
which acts as $S S^x_a S^{-1}=+S^x_a$, $S S^y_a S^{-1}=-S^y_a$ and $S S^z_a S^{-1}=-S^z_a$ for any site $a$. In $H_{\textrm{XXZ}}$, the spin operators appear in pairs and so the sign change is cancelled. In turn, the $S^z_a$ operators appear once in each term of $\Gamma$, leading to $S \Gamma S^{-1} = -\Gamma$, as needed to fulfil the symmetry \eqref{eq-Pseudo-Hermitian}. 
\subsubsection{Hermiticity and Integrability Breaking}\label{sssec-results spin}
We now characterize the spectral statistics of the spin chain with dissipative disorder described by the non-Hermitian Hamiltonian \eqref{eq-XXZ_iW_Hamiltonian}. We do so by comparing the distributions of its eigenvalue spacings with that of the 2d spectral statistics of the reference models. Here, we use again the methods described in Section \ref{ssec-numerics methods}.
We focus on the NN and NNN spacing distributions for values of the disorder strength from $\gamma=0-20$, fitting the numerical distribution to either the 2dCG distributions and finding the corresponding $\beta$, or to a $D$-dimensional Poisson point process with $1\leq D \leq 2$, cf. Section \ref{ssec-point processes}. We focus here on the study for $L=16$ corresponding to an eigenvalue space of dimension 12870 for over 450 disorder realizations. We recall that for the purpose of universality, we need a large system size and we checked in \cite[Section III.B]{ABCPRS} that we find the best agreement for $L=16$, comparing also with $L=12,14$.

We expect the spectral statistics of our model to behave similarly to those of a XXZ spin-chain Hamiltonian with real disorder, which exhibits a crossover of the spectral statistics from chaos to integrability (for non-zero disorder strength) as the disorder strength is increased. In such a case, the NN spectral statistics changes from random matrix statistics approximated by the Wigner surmise $\propto s e^{-a s^2}$ for the GOE (cf. Eq. \eqref{eq-pWigner} at $\beta=1$) in this model to 1d Poisson, $\propto e^{-s}$ (cf. Eq. \eqref{eq-Poid-NN} for $D=1$), $s$ being the spacing between NN eigenvalues. 
In 1d, there is universal level repulsion between the eigenvalues only in the first, chaotic regime, while in the second, 1d Poisson displays uncorrelated eigenvalues---a universal signature of generic integrable (deterministic) systems.

In the non-Hermitian (2d) case with imaginary disorder, where the eigenvalues lie in the complex plane, we expect the spectral statistics to exhibit a crossover between random matrix statistics of class AI$^\dag$ in our case to that of the 2d Poisson point process---which has a linear repulsion from the area measure, cf. Eq. \eqref{eq-Poid-NN} at $D=2$. 
We follow the proposal \cite{AKMP} and use the spectral statistics of a 2dCG with some value for $\beta\geq0$, which includes 2d Poisson at $\beta=0$ as a special case and approximates the AI$^\dag$ class for $\beta = 1.4$ \cite{AMP}. Fitting $\beta$ thus allows us to capture this crossover. 
However, the transition has more structure than initially expected, showing two different crossovers.
\begin{figure}[t]
	\centering
	\includegraphics[width=0.8\linewidth]{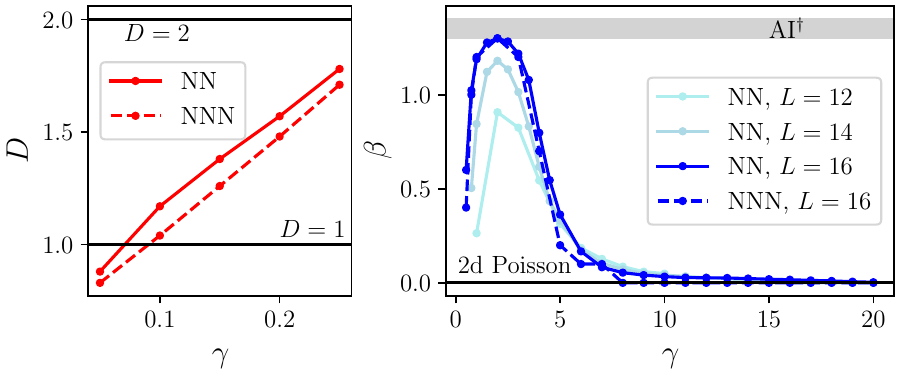}
	\caption[Figure 3 from \cite{ABCPRS}, \textit{Hermiticity and integrability breaking in the XXZ spin chain with dissipative disorder}]{\textbf{Left:} For $0.05\leq\gamma<0.25$ we find that the NN and NNN spectral statistics changes from those of 1d Poisson to 2d Poisson, thus \textit{Hermiticity is broken first}. Shown is the fitted effective dimension $D$ for NN (full curve) and NNN (dashed curve). Data is shown for $L=16$ spins in the $\mathcal{S}_z = 0$ magnetization sector. \textbf{Right:} For larger $\gamma$, i.e. $0.5<\gamma<20$ we find a very good fit to $\beta$ of the 2dCG. At $\gamma=2$ the NN (full curve) and NNN statistics (dashed curve) are very close to those of AI$^\dagger$ corresponding to $\beta=1.4$ (horizontal band), representing a \textit{complete breaking of integrability}. As $\gamma$ is increased further, there is a crossover to the statistics of 2d Poisson (horizontal line at $\beta=0$). \cite[Figure 3]{ABCPRS}}
	\label{fig-summary spin chain}
\end{figure}

On a technical level, the spin chain data need to be drawn away from the edges of the eigenvalue distributions, where the statistics may be different or not even universal, as discussed in Section \ref{sssec-unfolding}. To do so, we focus on the bulk given by the denser inner part of the spectrum. We systematically select only that part of the spectrum for which the eigenvalue density is not less than some fraction of the maximal density. For $L=16$ spins with $\gamma=0.75 - 20$, we use $1/2$ of the maximal density. For spins with $\gamma < 0.75$ we choose a cut-off equal or smaller to $1/4$ the maximal density because, as we discuss below, the spectrum is very peaked around a line parallel to the real line.\\

\begin{figure}[t]
	\centering
	\includegraphics[width=\linewidth]{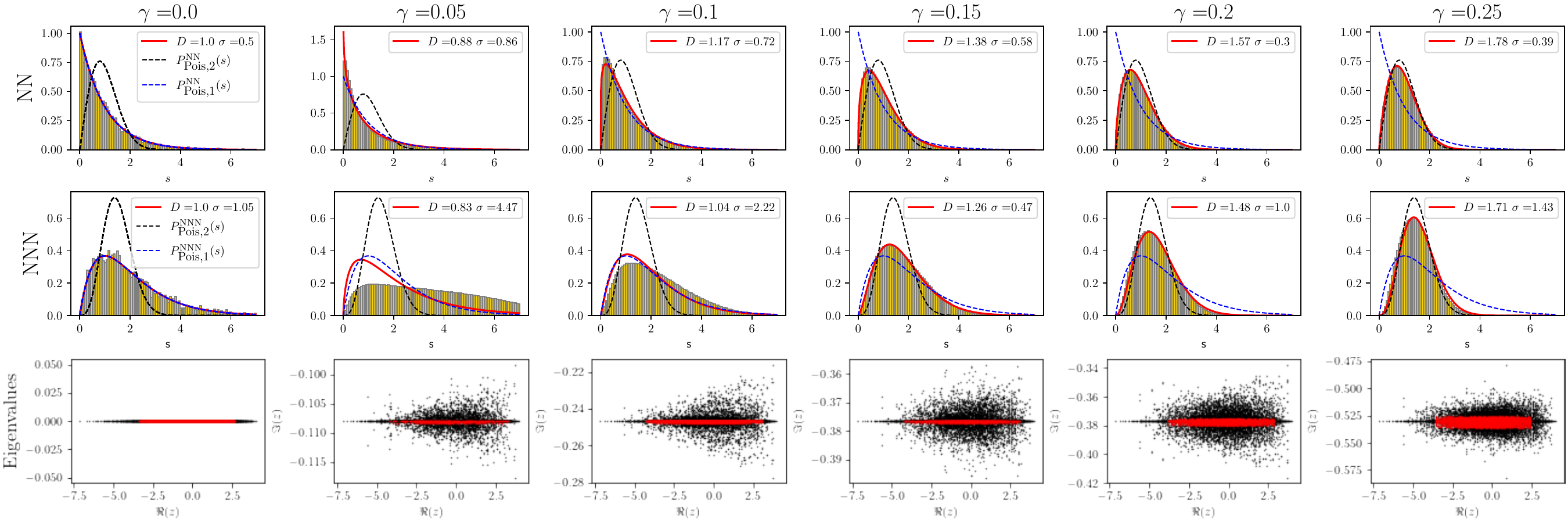}
	\caption[Figure 4 from \cite{ABCPRS}, \textit{NN and NNN distributions and eigenvalue scatter plots for the XXZ spin chain with small dissipative disorder}]{The NN (top), NNN spacing distributions (middle) of at least 490 disorder realizations, and scatter plots with selected eigenvalues highlighted in red for a single realization of the spectrum (bottom) for $\gamma=0,0.05,0.1,0.15,0.2,0.25$. In red (full line) we show the analytics of the Poisson process, i.e. Eqs. \eqref{eq-Poid-NN} and \eqref{eq-Poid-NNN}, for the NN and NNN distribution, respectively, for the best fitted value for the effective dimension $D$, with corresponding standard deviation $\sigma$ (in units of $10^{-2})$. For comparison we also show the results for $D=1$ (blue dashed) and for $D=2$ (black dashed line) as an orientation. \cite[Figure 4]{ABCPRS}
	}
	\label{fig-smallga}
\end{figure}

We recall from Section \ref{sssec-comparison numerical data}, Figure \ref{fig-2d-statistics} the following observations: The NN distribution of the 2dCG has its leftmost and lowest maximum at $\beta=0$, see Figure \ref{fig-2d-statistics}(b) left (grey curve). Increasing $\beta$ leads to a monotonous increase of the maximum in height and in position to the right (grey to blue). The random matrix class AI$^\dag$ expected at maximal chaos is the one with the lowest maximum of the three distinct random matrix classes (see Figure \ref{fig-NN NNN symmetry classes}), reached at approximately $\beta= 1.4$. As such, it is approached by the 2dCG monotonously from below, starting at $\beta=0$. 
What happens if the maximum of the NN spacing distribution of our spin chain data---which is normalised and has first moment of unity---is to the \textit{left} of the 2d Poisson distribution (at $\beta=0$)? Obviously the 2dCG will no longer provide a good fit then. Only the $D$-dimensional Poisson distribution might provide a good fit then for $D<2$, see Figure \ref{fig-2d-statistics}(a) left. There, it is shown that the maximum of the distribution goes from left to right as the effective dimension is increased.

Our findings are summarised by Figure \ref{fig-summary spin chain} above and described as follows. In the small disorder regime, there is a crossover between 1d Poisson at $\gamma=0$---where our XXZ spin chain is integrable--- towards 2d Poisson, that is well described by an increase in effective dimension $1\leq D<2$. We reach up to $D\approx 1.7$ at $\gamma=0.25$ for both NN and NNN spacing distributions. 
Therefore, Hermiticity is broken first on a smaller scale. This is summarised in Figure \ref{fig-summary spin chain} left.

For intermediate to large values of $\gamma=0.5 - 20$, we see a rise from $\beta\approx0.6$ at $\gamma=0.5$ to $\beta\approx 1.4$ at $\gamma=2$ where class AI$^\dag$ is located, to which we also compare. We see that integrability is broken as we approach the fully chaotic behaviour characterised by random matrix class AI$^\dag$. For larger $\gamma$, the disorder term \eqref{eq-Gamma} leads to an alignment of spins and thus again to a deterministic behaviour, approaching 2d Poisson with $\beta=0$ around $\gamma\approx 8$, see Figure \ref{fig-summary spin chain} right. We also use the surmise from Section \ref{ssec-surmise}, cf. Eq. \eqref{eq-pbetasurmise}, for a finer scaling in the NN distribution in an asymptotic approach to $\beta=0$ between $\gamma=8-20$. \\
\\
\noindent
\textbf{Hermiticity Breaking at Small Disorder $\boldsymbol{\gamma}$: From 1d to 2d Poisson}\\
\\
\noindent
We study the regime of small $\gamma$, with range $\gamma=0-0.25$ and compare our model to Poisson statistics with varying dimension, Eqs. \eqref{eq-Poid-NN} and \eqref{eq-Poid-NNN} for the NN and NNN spacing distribution, respectively, see Figure \ref{fig-smallga}. The corresponding spectrum is also shown---notice the different scales on the $y$-axis---with the part of the spectrum that is used, representing the bulk, highlighted in red. As explained in Section \ref{sssec-model spin}, the spectrum is symmetric with respect to a line parallel to the real axis. This symmetry line changes with $\gamma$. 
\\Starting with $\gamma=0$ (left plots in Figure \ref{fig-smallga}), the spectrum is real, the XXZ model integrable, and we obtain a very good agreement with 1d Poisson statistics for both NN and NNN as expected. Notice that at $\gamma=0$ the energies are doubly degenerate, which is why we only show spacings with $\mathcal{S}_z = 0$ and parity $P=+1$ (for $L=16$, this sector is of dimension 6470). The plots for $\mathcal{S}_z = 0$ with $P=-1$ (for $L=16$, this sector is of dimension 6400) look similar. For the smallest non-zero value of $\gamma=0.05$, we obtain a somewhat reasonable fit for the NN, with $D\approx 0.88$ consistent with $D=1$. In contrast, the distribution for the NNN does not yield a good fit at all, a feature that persists for $\gamma=0.1$. A possible explanation is that for these small values of $\gamma$ the spectrum is almost singular, being concentrated on a line parallel to the real line, with very few eigenvalues in the complex plane (see Figure 5 in \cite{ABCPRS}). Thus, unfolding procedure described in Section \ref{sssec-unfolding} for a 2d complex spectrum may then not be completely suitable. We recall that following \cite{AKMP}, the average spectral density is approximated by the Gaussian curve.
For $\gamma=0$, we used a 1d version of the unfolding method described in Section \ref{sssec-unfolding}.\footnote{We found that for 1d, $\Sigma = 1.8 \bar{s}$ gives good results.}

From $\gamma=0.1$ both NN and NNN start to deviate from $D=1$, and the best fits in $D$ reach values up to $D\approx 1.7$ at $\gamma=0.25$. Comparing NN and NNN leads to consistent values within a small margin. 
We thus believe that the fits to a one-parameter family of distributions are not a coincidence but rather capture properties of an underlying effective Poisson interaction. 
\\This leads to the interpretation that Hermiticity is broken first, while maintaining an (almost) integrable system, the characteristic of which is given by a Poisson distribution at intermediate dimension $1<D<2$, depicted in the summary plot in Figure \ref{fig-summary spin chain} above. Increasing to the next higher value of $\gamma=0.5$, detailed below, the maximum of the NN crosses to the right of the 2d Poisson distribution (at $\beta=0$), which then fits the 2dCG. 
The fact that $D=2$ is not fully reached by increasing $\gamma$ here can be interpreted as a competition between Hermiticity and integrability breakings around the scale $\gamma=0.25-0.5$.
\clearpage

\begin{figure}[t]
	\includegraphics[width=\linewidth]{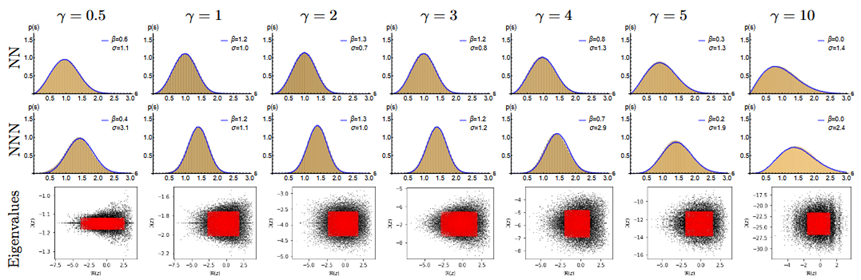}	
	\caption[Figure 6 from \cite{ABCPRS}, \textit{NN and NNN distributions and eigenvalue scatter plots for the XXZ spin with intermediate dissipative disorder}]{Comparison of the model \eqref{eq-XXZ_iW_Hamiltonian} with increasing dissipative order $\gamma$, with size $L=16$ and spins in the $\mathcal{S}_z = 0$ magnetization sector (histograms) for at least 450 realizations, and the best fit to the 2dCG (blue curves) of size $N=5000$. The NN (NNN) distribution is given in the first (second) row. Additionally, we give the standard deviation $\sigma$ in units $10^{-2}$ for each comparison of the spin chain data with the 2dCG. We show an example of the complex eigenvalue data for a single realization in the bottom row, with the points used in the statistics highlighted in red. \cite[Figure 6]{ABCPRS}}
	\label{fig-intermedg}
\end{figure}


\begin{table}[t]
	\centering 
	\caption[Table I from \cite{ABCPRS}, \textit{List of best fit $\beta$ for the comparison of the NN and NNN spacings of the XXV spin chain with dissipative disorder to 2dCG statistics}]{List of best fit $\beta$ for the comparison of the NN and NNN spacing of the spin chain with $L=16$ and $\gamma=0.5,0.75,1,2,\ldots 20$ to the 2dCG and the corresponding standard deviation $\sigma$ in units of $10^{-2}$. Notice that apart from $\gamma=0.5$ the values obtained for $\beta$ from NN and NNN agree or differ by $0.1$ only. \cite[Table I]{ABCPRS}}\vspace{3mm}
	\begin{tabular}[width=\linewidth]{|c|c|c|c|c|c|c|c|c|c|c|c|c|}
		\hline
		&$\boldsymbol{\gamma}$& \textbf{0.5} & \textbf{0.75} &\textbf{1.0}  & \textbf{2.0} &\textbf{3.0} & \textbf{4.0} & \textbf{5.0} &\textbf{6.0}&\textbf{7.0}  &\textbf{8.0}&\textbf{9.0}\\
		\hline\hline
		\multirow{2}{*}{NN}&	$\beta$ & 0.6& 1.0& 1.2 & 1.3 & 1.2& 0.8 & 0.3 & 0.2 &0.1 &0.1 &0.0\\
		\cline{2-13}
		&$\sigma$ & 1.1& 1.3& 1.0 &0.7 & 0.8& 1.3 & 1.3 & 1.3 & 1.3 & 1.7 &1.6       \\
		\hline
		\hline
		\multirow{2}{*}{NNN}&	$\beta$ & 0.4& 1.0& 1.1,$\ $1.2 & 1.3 & 1.2& 0.7 & 0.2 & 0.1 &0.1 &0.0 &0.0 \\
		\cline{2-13}
		&$\sigma$ &3.1& 1.5& 1.1 & 1.0 & 1.2& 2.9 & 1.9 & 1.6 & 2.6 & 2.8 &2.6    \\
		\hline
		\multicolumn{13}{c}{}\\[-0.5em]
		\hline
		&$\boldsymbol{\gamma}$&\textbf{10.0} &\textbf{11.0}  & \textbf{12.0} &\textbf{13.0} & \textbf{14.0} & \textbf{15.0} &\textbf{16.0}& \textbf{17.0}  &\textbf{18.0} &\textbf{19.0}&\textbf{20.0}\\ \hline\hline
		\multirow{2}{*}{NN}&	$\beta$ & 0.0& 0.0 & 0.0 & 0.0 & 0.0  & 0.0 & 0.0 &0.0 &0.0 &0.0 & 0.0 \\
		\cline{2-13}
		&$\sigma$&1.4 & 1.4 & 1.3 & 1.3& 1.2 & 1.2 & 1.0 & 0.9 & 0.9 &0.8 &0.8 \\
		\hline
		\hline
		\multirow{2}{*}{NNN}&	$\beta$ &  0.0& 0.0 & 0.0 & 0.0 & 0.0 & 0.0 & 0.0 &0.0 &0.0 &0.0 & 0.0\\
		\cline{2-13}
		&$\sigma$ &2.4& 2.4 & 2.3 & 2.2& 2.1 & 2.0 & 1.9 & 1.7 & 1.6 &1.4 &1.3\\
		\hline
	\end{tabular}
	\label{tab-spin chain gamma}
\end{table}
\begin{table}[t]
	\centering 
	\caption[Table III from \cite{ABCPRS}, \textit{List of best fit $\beta$ for the comparison of the NN spacing of the XXV spin chain with dissipative disorder to the NN 2dCG surmise}]{List of best fit $\beta$ for the comparison of the NN spacing of the spin chain with $L=16$ and $\gamma=8,\ldots 20$ to the NN 2dCG surmise \eqref{eq-pbetasurmise} and the corresponding standard deviation $\sigma$ in units of $10^{-2}$. \cite[Table III]{ABCPRS} }\vspace{3mm}
	\begin{tabular}{|c|c|c|c|c|c|c|c|c|}
		\hline
		&$\boldsymbol{\gamma}$ & \textbf{8} & \textbf{9} &\textbf{10}  & \textbf{11} &\textbf{12} & \textbf{13} & \textbf{14} \\
		\hline\hline
		\multirow{2}{*}{NN} &	$\beta$ & 0.0547 & 0.0413 & 0.0337 & 0.0283 & 0.0273 & 0.026 & 0.023  \\
		\cline{2-9}
		&$\sigma$ & 0.36 & 0.35 & 0.33 & 0.35 & 0.33 & 0.33 & 0.31     \\
		\hline
		\multicolumn{9}{c}{}\\[-0.5em]
		\hline
		&$\boldsymbol{\gamma}$ &\textbf{15}& \textbf{16}  &\textbf{17}&\textbf{18}&\textbf{19} &\textbf{20} &\\
		\hline\hline
		\multirow{2}{*}{NN} &	$\beta$ & 0.0196 &
		0.0169 & 0.0135 & 0.0108 & 0.0066 & 0.0035 &\\
		\cline{2-9}
		&$\sigma$& 0.3 & 0.27 & 0.24 & 0.24 &
		0.21 & 0.2 &\\
		\hline
	\end{tabular}
	\label{tab-g8-20}
\end{table}

\noindent
\textbf{Integrability Breaking at Intermediate to Large Disorder $\boldsymbol{\gamma}$: From 2d Poisson to Class AI$\boldsymbol{^\dag}$ and Back} \\
\\
\noindent
Now, we discuss the transition from 2d Poisson to random matrix statistics class AI$^\dag$ and back, witnessed for intermediate ($\gamma=0.5$) to large ($\gamma=20$) disorder dissipation strength. 

The smallest value of disorder for which a meaningful fit with a 2dCG 
at $\beta\geq0$ can be made is $\gamma=0.5$, when, $\beta=0.6\, (0.4)$ for NN (NNN), see Figure \ref{fig-intermedg} left. At this disorder strength, both Hermiticity and integrability appear to be broken. The system is in between the integrable ($\beta=0$) and the fully chaotic case ($\beta= 1.4$, equivalent to random matrix symmetry class AI$^\dag$). The latter is then attained at $\gamma=2$, where the maximal value for $\beta$ (and height of the maximum) is reached, see Figure \ref{fig-summary spin chain} right. We compare here the NN and NNN spacings with the 2dCG rather than AI$^\dagger$. Furthermore, random matrix statistics is approached in $\beta$ from below and the $\sigma$-value gives a measure of how close it is.

\begin{figure}[t]
	\centering
	\includegraphics[width=\linewidth]{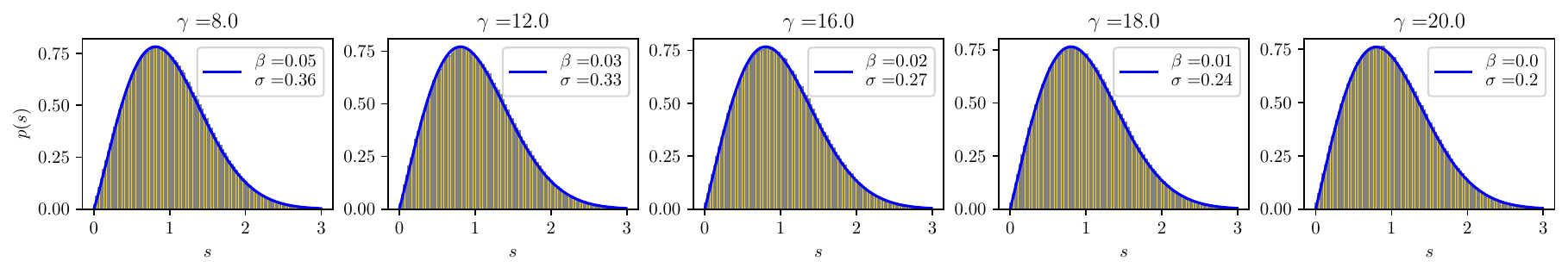}
	\caption[Figure 9 from \cite{ABCPRS}, \textit{Comparison of the NN distribution of the XXZ spin chain with dissipative disorder to the 2dCG surmise}]{Asymptotic approach to 2d Poisson for large disorder strength $\gamma$. The histograms for the NN spacings of at least 450 realizations of the spin chain with $L=16$ in the $\mathcal{S}_z = 0$ magnetization sector are compared with the best $\beta$ fit of the surmise \eqref{eq-pbetasurmise} shown as a smooth blue line. The legend shows the value of $\beta$ and the standard deviation (in units of $10^{-2}$) of the fit. \cite[Figure 9]{ABCPRS}}
	\label{fig-beta0_approach}
\end{figure}

Increasing $\gamma$ further leads to a rapid decrease in the best fitted value for $\beta$, see also Table \ref{tab-spin chain gamma}. From $\gamma \simeq 8-10$ onwards we consistently obtain $\beta=0$. This is because the disorder is so strong that the spins align with the disorder term and the statistics becomes again deterministic. The best fits for $\beta$ for NN and NNN and corresponding values, including standard deviations, are collected in Table \ref{tab-spin chain gamma}. Here, we only fit to $\beta$ in the 2dCG 
where the step width is 0.1, but both NN and NNN distribution are available. Next, we present a much finer fit in $\beta$ for $\gamma\geq 8$, possible for the NN only thanks to the surmise \eqref{eq-pbetasurmise}, which is a very good approximation so close to $\beta=0$, cf. Section \ref{ssec-surmise}. The approach to zero can also be seen well in the summary Figure \ref{fig-summary spin chain} right. 
Similar findings for Hermitian Hamiltonians comparing to 1d Poisson and corresponding classical random matrices, using both NN and NNN spacings, were made very recently \cite{KKSG}.\\
\\
\noindent
\textbf{Asymptotic Approach to 2d Poisson for Large Disorder $\boldsymbol{\gamma}$}
\\
\\
We detail the asymptotic approach to 2d Poisson statistics $\beta\to0$ for large values of $\gamma=8-20$. Because for the 2dCG we have only discrete steps in $\beta$ of size 0.1 available, we will use the surmise \eqref{eq-pbetasurmise} instead here. It is an excellent approximation for the local statistics close to $\beta=0$, but it only allows fitting the NN spacing distributions. Figure \ref{fig-beta0_approach} shows fits in $\beta$ of the surmise to the NN spacing distribution for various values of $\gamma=8-20$ in steps of size 2 (or more). 
The fitted $\beta$-values for all values of $\gamma\geq8$, together with the corresponding standard deviations, are provided in Table \ref{tab-g8-20}. \clearpage

\noindent
Let us briefly summarise our findings. Using the NN and NNN distributions, we show that the XXZ model with dissipative disorder exhibits two spectral crossovers as a function of disorder strength $\gamma$ summarised in Figure \ref{fig-summary spin chain}. At $\gamma=0$, the model is a pure integrable XXZ spin chain and its spectral statistics (once restricted to an appropriate symmetry sector to avoid exact degeneracies) corresponds to the 1d Poissonian statistics. As the dissipative disorder strength is increased, but still small, the spectrum begins spreading into the complex plane and the statistics, approximately still integrable, are close to Poisson with an effective dimension $D>1$ (Figure \ref{fig-summary spin chain} left). As the dissipative disorder is increased further, integrability begins to break and both the NN and NNN distributions are well fitted by the 2dCG with $\beta>0$ (Figure \ref{fig-summary spin chain} right). At $\gamma=2$, the model becomes fully chaotic with NN and NNN spectral statistics close to those of random matrices of class AI$^\dagger$, which is in the same non-Hermitian symmetry class as our model. Beyond $\gamma>2$,
we observe a crossover from chaos to integrability. The same crossover from chaos to integrability was detected in \cite{roccati2024diagnosing} through real singular value statistics and is now validated also from complex spectral analysis. In a similar manner to the Hermitian case, increasing the dissipative disorder induces a crossover of the spectral statistics from chaotic---described by random matrix symmetry class AI$^\dag$ or the 2dCG statistics with $\beta= 1.4$, to integrable---described by the statistics of the 2d Poisson point process (or equivalently by the 2dCG 
with $\beta=0$). We tracked the crossover by fitting the NN and NNN spectral distributions to the corresponding value of $\beta$ in the 2dCG, obtaining a consistent value for $\beta$ from both fits. \cleardoublepage
\section{Summary and Open Problems}\label{sec-summary}
We recall the definition at the very beginning of this thesis in Section \ref{sec-intro}: "A random matrix contains elements which follow a certain probability distribution." Starting from this simple statement, we revisited and developed diverse matrix ensembles as well as applications of them. We investigated the subject of (non-)Hermitian matrices from two perspectives: First, by application to various examples based on universality of local statistics, following the spirit of Wigner \cite{Wigner57} strengthening the importance of the field. Second, we turned our view back to the probability distributions of matrix entries, modifying them to step into the direction of a random matrix ensemble for the two-dimensional Coulomb gas (2dCG). We view the subject of non-Hermitian matrices in a broader context, by distinguishing symmetry classes, discussing especially universality of local statistics bringing full circle to the starting point.\\
After a brief introduction into the field in Section \ref{sec-intro}, we discuss Hermitian $\beta$-ensembles in Section \ref{sec-Hermitian beta ensembles}. We learned already about their importance and wide applicability before, especially for the three classical Gaussian ensembles GO/U/SE which share the statistics of a Dyson gas in 1d at $\beta=1,2,4$ respectively. This set the course for the first two parts in Sections \ref{ssec-Householder}, \ref{ssec-Anderson Dixon}: We revisit the proof techniques, used in \cite{Dumitriu Edelman} and \cite{Forrester Rains 05} to establish a tridiagonal matrix ensemble, corresponding to a Dyson gas in 1d at $\beta>0$. While the first method of Householder reflections relies heavily of inner (Hermitian) symmetry of the matrix, the second one is intertwined to the understanding and knowledge of the Selberg-integral. While we notice the problems of an application to non-Hermitian matrices through the lines, we use the insight to introduce and discuss a Hermitian tridiagonal matrix-valued process with resetting at $\beta>0$ in Section \ref{ssec-tridiag with resetting}. The key ingredient is here the replacement of the unit standard deviation in the proof of Dumitriu and Edelman \cite{Dumitriu Edelman} by a time-dependent quantity $\sigma(t)$. We study the ensemble under simultaneous and independent resetting of the matrix elements finding distinct limiting spectral densities. Although the latter can be studied analytically for simultaneous resetting (cf. \cite{BMS2025}), the analytical computation for independent resetting remains a challenging open problem. Finding the Cox-Ingersoll-Ross process, which has applications for example in finance \cite{CIR85}, in the off-diagonal elements of the established matrix ensemble, we are looking forward to see further relations to processes in diverse research areas in the future.\\
We turn to the subject of non-Hermitian $\beta$-ensembles in Section \ref{sec-non Herm beta}. In order to set the stage, we revisit the $D$-dimensional Poisson point process and the 2dCG in Section \ref{ssec-point processes}. They serve as reference processes to distinguish integrable and chaotic behaviour. In order to quantify the comparisons, we follow the Grobe-Haake-Sommers conjecture \cite{GHS} and study the nearest-neighbour (NN) as well as the next-to-nearest-neighbour (NNN) distributions. The standard deviation and the Kolmogorov-Smirnov distance are introduced in Section \ref{ssec-numerics methods} and used repeatedly as measures of (dis-)agreement of two distributions. We discuss that the unfolding is an important step in the analysis of numerics in Section \ref{sssec-unfolding}. Another measure, which is conjectured to be independent of unfolding in the bulk of the spectrum, is the complex spacing ratio, which we discuss in detail in the later Section \ref{sssec-complex spacing ratio}. In Section \ref{ssec-tridiag beta} we introduce and study three tridiagonal non-Hermitian $\beta$-ensembles. Their joint probability density function (JPDF) in terms of the complex eigenvalues includes a pairwise logarithmic repulsion at inverse temperature $\beta>0$ and a non-trivial coupling to eigenvector components of varying complexity for the three ensembles. The key idea is here, to start from a tridiagonal ensemble, rather than reducing a full matrix to it as in the Hermitian case and compute the Jacobian etc. explicitly. The functional form of the JPDF shows partly the Coulomb repulsion, we are looking for. However, the additional coupling part, makes the analysis of the limiting density via standard techniques challenging. Nevertheless, this pushed us to the introduction of an alternative proof technique, via first the explicit study of the characteristic polynomial of centred Jacobi matrices in Section \ref{sssec-characteristic polynomial}, which we find in the low temperature limit $\beta \gg 1$ in Section \ref{sssec-low temperature}, and second by the application of a result from free probability in Section \ref{sssec-equilibrium density}. This yields to a novel logarithmic density for two of the three ensembles and to a Dirac delta in the third. We discuss the stability of the tridiagonal ensembles and study their NN distributions numerically in Section \ref{sssec-numerics}. We find that two of the three ensembles show Poisson like statistics of integrable systems in 2d independent of the choice of $\beta$, while the third ensemble deviates both from Poisson as well as from Ginibre statistics and shows a $\beta$-dependence. Currently, we have no understanding of this behaviour, but it seems that the coupling to the eigenvector components have a significant impact to the eigenvalue statistics. It would be interesting whether this could be quantified from the recursive structures given in the proof of Theorem \ref{thm-generalised Jacobian}. Alternatively, one can look for a matrix ensemble, with less free parameters, e.g. a bidiagonal matrix, which results in less free spectral parameters that are carried by eigenvector components. The distribution of the matrix elements will probably look more "complicated" (than the normal and the Chi-distribution), in order to produce a Coulomb-type interaction but the evolved coupling terms which arise from the triangular part in the Schur decomposition could become tractable.\\
In the next Section \ref{sec-universality symmetry classes}, we take a step back from the explicit study of matrix ensembles via their JPDF in terms of the eigenvalues. First, we revisit the classification of non-Hermitian matrices into 38 classes by \cite{non-h classes,Bernard LeClair}. We catch up the key word universality in the following sections. Initiated by Hamazaki et al \cite{Hamzaki et al}, who presented via numerics and heuristic arguments, that the classes A of complex Ginibre, AI$^\dagger$ of complex symmetric and AII$^\dagger$ of complex self-dual matrices are the simplest representatives of local statistics in the bulk, we investigated an effective description of the latter two. We established a correspondence with the NN and NNN statistics of a 2dCG at $\beta=1.4$ for class AI$^\dagger$, respectively at $\beta=2.6$ for class AII$^\dagger$. Although the explicit JPDF in terms of the complex eigenvalues is not known for each of these two classes until today, this effective description enabled us to interpret and detail the complex spacing ratio statistics in Section \ref{sssec-complex spacing ratio}. We revisit the conjecture that no unfolding is needed for this spectral measure in the bulk, at the edge of the spectrum, finding significant doubts. It remains as an open problem to quantify the applicability of the study of complex spacing ratios without unfolding depending on the local and global scales of the underlying eigenvalue density. While we only briefly mentioned analytical results in this section, it is certainly of interest to investigate this in future research. Ensembles, which interpolate between real symmetric (GOE) and complex symmetric (AI$^\dagger$) or Hermitian self-dual (GSE) and complex self-dual (AII$^\dagger$) may serve as good tool to reach, first in weak non-Hermiticity regimes, a better understanding of the two classes AI$^\dagger$ and AII$^\dagger$. We recall that the elliptic Ginibre ensemble, which interpolates between Hermitian (GUE) and Ginibre (GinUE, class A), is well-studied. However, it should be noted that in contrast to class A, where a determinantal point process is well-established, it is until today not clear whether there are underlying point processes for classes AI$^\dagger$ and AII$^\dagger$. Circumventing this lack of knowledge, we investigated the expectation value of one pair of characteristic polynomials in Section \ref{ssec-expectation of characteristic polynomials}. Here, we make use of the Gaussian structure of the matrix ensembles in each of the classes A, AI$^\dagger$ and AII$^\dagger$, presenting results in global, local bulk and local edge limits. A rigorous study of all 38 symmetry classes, resulting in comparability in terms of the expectation value of characteristic polynomial would be a first step into the analytical classification of local statistics. We end in Section \ref{ssec-application spin chain} with the study of a quantum spin chain with dissipative disorder, presenting applications for the surmise from Section \ref{ssec-surmise}, the 2dCG introduced in Section \ref{ssec-point processes} and the symmetry class AI$^\dagger$ of the Hamiltonian. This model enables us to investigate Hermiticity and integrability breaking via the analysis of the NN and NNN local bulk statistics. It would be interesting whether the eigenvector statistics become tractable in the future.\\
We end with the observation that there is still a lot work to do and we hope that the present thesis may serve as a building block for future work, by revisiting the various possibilities as well as (and maybe most important) the elimination of methods from Hermitian RMT which are (so far) not applicable to non-Hermitian RMT.\cleardoublepage
\begin{appendix}
	\section{Overview on Hermitian $\boldsymbol{\beta}$-Ensembles}\label{app-overview Hermitian beta ensembles}
	We give a brief overview on the Hermitian-$\beta$-ensembles known in the literature and comment on the used proof-strategy.
	\begin{itemize}
		\item \textbf{Gaussian-}$\boldsymbol{\beta}$\textbf{-Ensemble:}
		\begin{itemize}
			\item Also known as the Hermite-$\beta$-ensemble
			\item Consists of Hermitian $N\times N$ matrices, with eigenvalue distribution match the one of particles in a 1d Dyson gas at inverse temperature $\beta>0$. We repeat the JPDF, given in Eq. \eqref{eq-jpdf 1d Dyson}
			\begin{align*}
				P(\lambda_1,\ldots,\lambda_N)&=\frac{1}{C_\beta}e^{-\frac{\beta}{2}\sum_{j=1}^N\lambda_j^2}\prod_{1\leq i < j\leq N}|\lambda_i-\lambda_j|^\beta, \text{ with } \\
				C_\beta&=(2\pi)^{N/2}\beta^{-N/2-\beta N (N-1)/4}\frac{\prod_{j=1}^N\Gamma(1+\beta j/2)}{\left(\Gamma(1+N\beta/2)\right)^N}.
			\end{align*}
			\item Tridiagonal matrix representation due to \cite{Dumitriu Edelman}
			\begin{align*}
				H_\beta \sim \frac{1}{\sqrt{2}}\begin{pmatrix}
					\mathcal{N}(0,2)& \chi^{}_{(N-1)\beta}& & &\\
					\chi^{}_{(N-1)\beta}&\mathcal{N}(0,2)& \chi^{}_{(N-2)\beta}& &\\
					&\ddots&\ddots&\ddots&\\
					& &\chi^{}_{2\beta}&  \mathcal{N}(0,2) & \chi^{}_\beta\\
					& & &\chi^{}_\beta&\mathcal{N}(0,2)
				\end{pmatrix}.
			\end{align*}
			\item Proof strategies: via direct computation of the Jacobian in \cite{Dumitriu Edelman}, or via the Anderson-Dixon integral in \cite{Forrester Rains 05}.
		\end{itemize}
		\item \textbf{Wishart-}$\boldsymbol{\beta}$\textbf{-Ensemble:}
		\begin{itemize}
			\item Also known as the Laguerre-$\beta$-ensemble
			\item Consists of $m\times m$ matrices $L=BB^t$, where $B$ a rectangular $m\times N$ matrix with $N>m-1$. The JPDF in terms of the $m$ eigenvalues $\{\lambda_i\}_{i=1}^m$ of the Wishart-$\beta$-matrix $L_\beta$ reads for $\beta>0$
			\begin{align*}
				P(\lambda_1,\ldots,\lambda_m)=\frac{1}{c_{\beta,a}}\prod_{1\leq i < j\leq m}|\lambda_i-\lambda_j|^\beta\prod_{i=1}^m\lambda_i^{a-p} e^{-\frac{1}{2}\sum_{j=1}^m\lambda_j}\\
				c_{\beta,a}=2^{ma}\prod_{j=1}^m \frac{\Gamma\left(1+\frac{\beta}{2}j\right)\Gamma\left(a-\frac{\beta}{2}(m-j)\right)}{\Gamma\left(1+\frac{\beta}{2}\right)},
			\end{align*}
			with $a=\frac{\beta}{2}N $, $p=1+\frac{\beta}{2}(m-1)$.
			\item Bidiagonal parametrization of $L_\beta$ due to \cite{Dumitriu Edelman}
			\begin{align*}
				L_\beta=B_\beta B_\beta ^t,\text{ where }
				B_\beta \sim \begin{pmatrix}
					\chi_{2a}& &\\
					\chi_{\beta(m-1)}& \chi_{2a-\beta}&\\
					& \ddots & \ddots &\\
					& & \chi_\beta & \chi_{2a-\beta(m-1)}
				\end{pmatrix}
			\end{align*}
			\item Proof strategy: via direct computation of the Jacobian in \cite{Dumitriu Edelman}
		\end{itemize} 
		\item \textbf{Jacobi-}$\boldsymbol{\beta}$\textbf{-Ensemble:}
		\begin{itemize}
			\item Also known as MANOVA-ensemble
			\item Listed in \cite{Dumitriu Edelman} as an open problem
			\item Consists of Hermitian $N\times N$ matrices, with eigenvalues distributed in the interval $\left[-2,2\right]$. Let $a,b>-1$ be real parameters, the JPDF of the eigenvalues $\{\lambda_i\}_{i=1}^N$ of the Jacobi-$\beta$-matrix $J_\beta$ reads for $\beta>0$
			{\small
			\begin{align*}
				P(\lambda_1,\ldots,\lambda_N)&=\frac{1}{c_{\beta,a,b}}\prod_{1\leq i < j\leq N}|\lambda_i-\lambda_j|^\beta\prod_{k=1}^N(2-\lambda_k)^a(2+\lambda_k)^b\\
				c_{\beta,a,b}&=2^{2N\left[(N-1)\frac{\beta}{2}+a+b+1\right]}\prod_{j=0}^{N-1} \frac{\Gamma\left(j\frac{\beta}{2}+a+1\right)\Gamma\left(j\frac{\beta}{2}+b+1\right)\Gamma\left((j+1)\frac{\beta}{2}\right)}{\Gamma\left(\frac{\beta}{2}+1\right)\Gamma\left((N-j+1)\frac{\beta}{2}+a+b+2\right)}.
			\end{align*}}
			\item Tridiagonal parametrization of $J_\beta$ due to \cite{Killip Nenciu}
			\begin{align*}
				J_\beta=\begin{pmatrix}
					b_1& a_1& & &\\
					a_1&b_2& a_2& &\\
					&\ddots&\ddots&\ddots&\\
					& &a_{N-2}&  b_{N-1} & a_{N-1}\\
					& & &a_{N-1}&b_N
				\end{pmatrix},
			\end{align*}
			with parameters
			\begin{align*}
				b_{k+1}&=(1-\alpha_{2k-1})\alpha_{2k}-(1+\alpha_{2k-1})\alpha_{2k-2}\\
				a_{k+1}&=\left[(1-\alpha_{2k-1})(1-\alpha_{2k}^2)(1+\alpha_{2k+1})\right]^{1/2}
			\end{align*}
			for $k=0,\ldots,N-1$. The random variables $\alpha_k$ are distributed as
			\begin{align*}
				\alpha_k\sim \begin{cases}
					B\left(\frac{2N-k-2}{4}\beta +a+1,\frac{2N-k-2}{4}\beta +b+1\right)& \text{ for }k \text{ even,}\\
					B\left(\frac{2N-k-3}{4}\beta +a+b+2,\frac{2N-k-1}{4}\beta\right)& \text{ for }k \text{ odd,}
				\end{cases}
			\end{align*}
			for $k=0,\ldots, 2N-2$ and $\alpha_{2N-1}=\alpha_{-1}=-1$. $B(s,t)$ denotes the beta-distribution, with probability density for $x\sim B(s,t)$ 
			\begin{align*}
				f_{s,t}(x)=\frac{2^{1-s-t}\Gamma(s+t)}{\Gamma(s)\Gamma(t)}(1-x)^{s-1}(1+x)^{t-1},\text{ for } x\in \left[-1,1\right].
			\end{align*}
			\item Proof via direct computation of the Jacobian in \cite{Killip Nenciu,Forrester Rains unitary} or via CS-decomposition in \cite{Edelman Sutton}
			\item There exists a circular Jacobi-$\beta$-ensemble via a circular analogue of the Anderson-Dixon density \cite{book forrester}
		\end{itemize}
		
		\item \textbf{Anti-Symmetric-Gaussian-}$\boldsymbol{\beta}$\textbf{-Ensemble:}
		\begin{itemize}
			\item Consists of Hermitian anti-symmetric $N\times N$ matrices, i.e. $H=\frac{i}{2}(X-X^t)$ with the real matrix $X$. It is convenient to study the non-Hermitian analogue $A=\frac{1}{i}H$. The eigenvalues of $A$ come in pairs $\{\pm i\lambda_j\}_{j=1}^{\lfloor \frac{N}{2}\rfloor}$ with $\lambda_j>0$ for $N$ even and an additional zero eigenvalue for $N$ odd.\footnote{The imaginary eigenvalues of $A$ are parametrized fully by the real quantities $\lambda_j$.} The JPDF reads for $\beta>0$
			{\small
			\begin{align*}
				&P(\lambda_1,\ldots,\lambda_N)\\
				&=\frac{1}{C_{N,\beta}^{\mathrm{antisym}}}\begin{cases}
					\prod_{1\leq j<k\leq N/2}(\lambda_j^2-\lambda_k^2)^{\beta}\prod_{i=1}^{N/2}\lambda_i^{\beta/2-1}e^{-\lambda_i^2}& \text{, for } N \text{ even,}\\
					\prod_{1\leq j<k\leq (N-1)/2}(\lambda_j^2-\lambda_k^2)^{\beta}\prod_{i=1}^{(N-1)/2}\lambda_i^{3\beta/2-1}e^{-\lambda_i^2}& \text{, for }N \text{ odd,}
				\end{cases}
			\end{align*}}
			with \begin{align*}
				C_{N,\beta}^{\mathrm{antisym}}=\begin{cases}
					\prod_{j=1}^{N/2}\frac{\Gamma(2j\beta /4)\Gamma((2j-1)\beta/4)}{2\Gamma(\beta/2)}& \text{, for }N \text{ even,}\\
					\frac{\Gamma(N\beta/4)}{\Gamma(\beta/4)}C_{(N-1),\beta}& \text{, for }N \text{ odd}.
				\end{cases}
			\end{align*}
			\item Bidiagonal parametrization of $A_\beta$ due to Dumitriu and Forrester \cite{Dumitriu Forrester}\footnote{The factor $1/\sqrt{2}$ in front of the bidiagonal matrix is in agreement with the result in \cite{Dumitriu Forrester} taking the definition of distribution $\widetilde{\chi}_k$ used there into account.}
			\begin{align*}
				A_\beta \sim \frac{1}{\sqrt{2}}\begin{pmatrix}
					0& \chi^{}_{(N-1)\beta/2}& & &\\
					-\chi^{}_{(N-1)\beta/2}&0& \chi^{}_{(N-2)\beta/2}& &\\
					&\ddots&\ddots&\ddots&\\
					& &-\chi^{}_{\beta}&  0 & \chi^{}_{\beta/2}\\
					& & &-\chi^{}_{\beta/2}&0
				\end{pmatrix}.
			\end{align*}
			\item Three proofs are presented in \cite{Dumitriu Forrester}: using Anderson-Dixon-integral, i.e. study of conditional probability of minor processes, via direct computation of the Jacobian and by the mapping to the Laguerre-$\beta$-ensemble
		\end{itemize}
	\end{itemize}\cleardoublepage
\section{Additional Information to the Study of Birds of Prey}\label{app-birds of prey}
We provide supplementary material for the study of interaction within and between birds of prey, discussed in Section \ref{ssec-birds of prey}. In Section \ref{apps-random displacement}, we probe whether a random displacement of the positions of nests can explain the repulsion, which we described in Section \ref{sssec-interact}. For completeness, we give the data of nests per year of each species between 2000 and 2020 in Section \ref{apps-number of nests}. We follow \cite{ACKMOP} closely.
\subsection{Effect of Random Displacement}\label{apps-random displacement}
\begin{center}
	\begin{figure}[h]
		\centering
		\hspace{0.04\linewidth} Displaced Buzzard \hspace{0.08\linewidth}
		Displaced Goshawks\\
		\hspace{-4pt}
		\includegraphics[width=0.295\linewidth,angle=0]{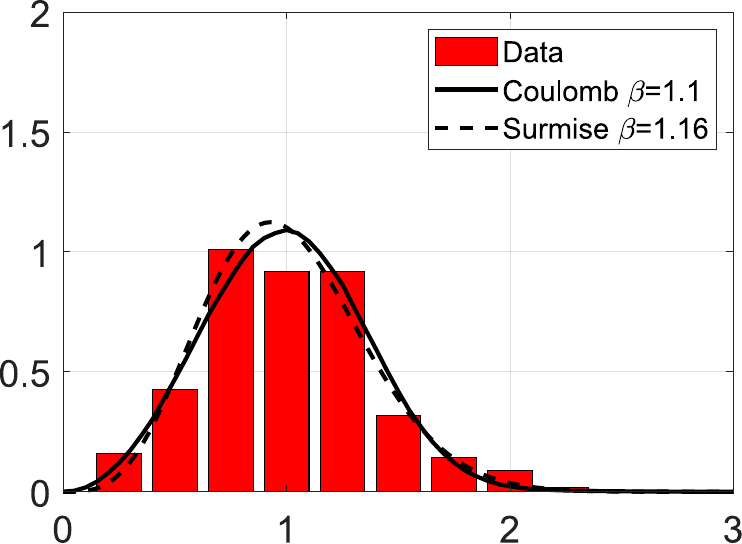}
		\hspace{12pt}
		\includegraphics[width=0.295\linewidth,angle=0]{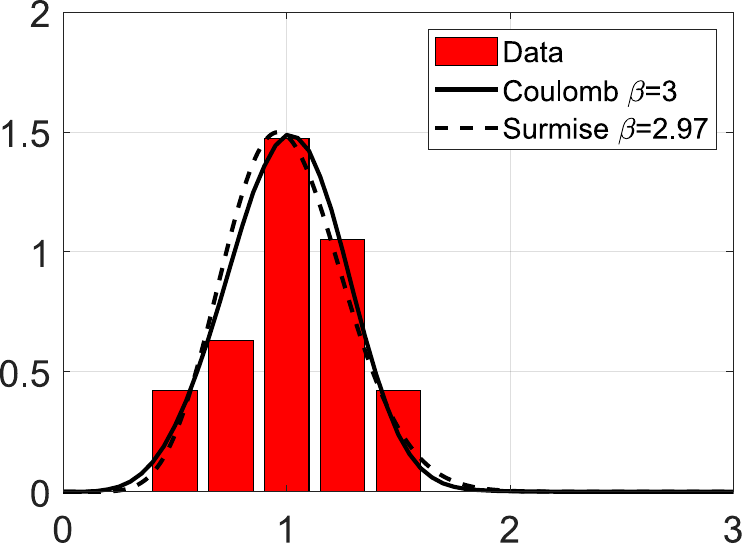}
		\\		
		\includegraphics[width=0.3\linewidth,angle=0]{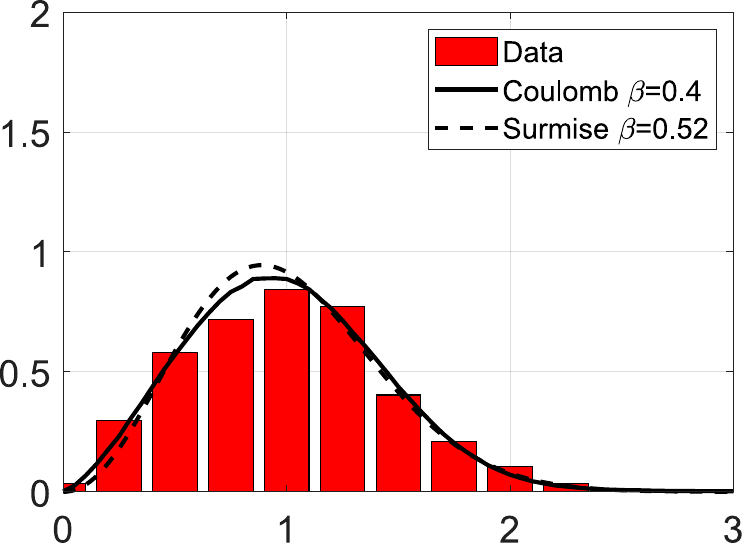}
		\hspace{10pt}
		\includegraphics[width=0.3\linewidth,angle=0]{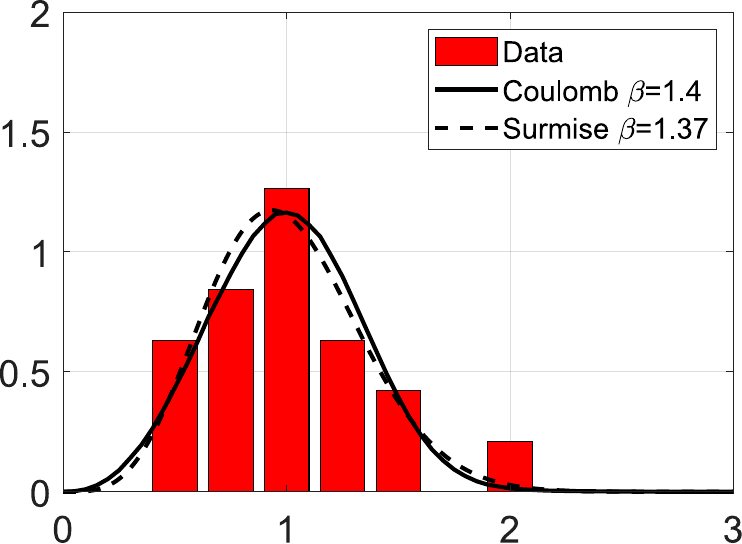}
		\\
		\includegraphics[width=0.3\linewidth,angle=0]{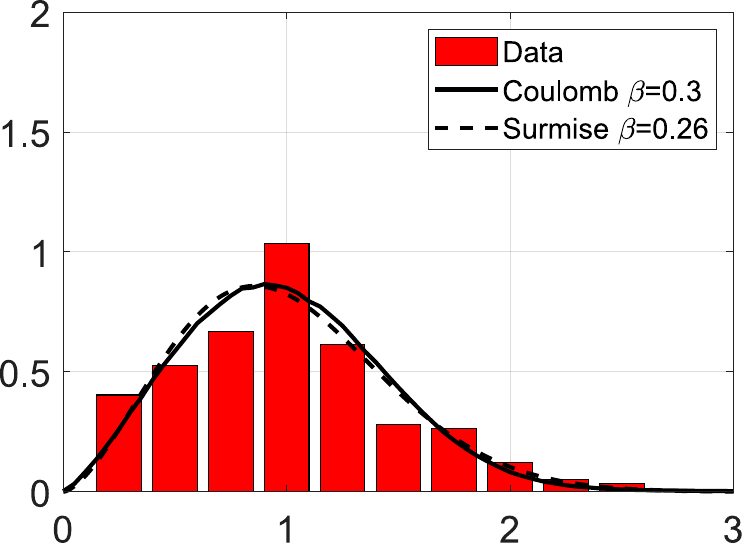}
		\hspace{10pt}
		\includegraphics[width=0.3\linewidth,angle=0]{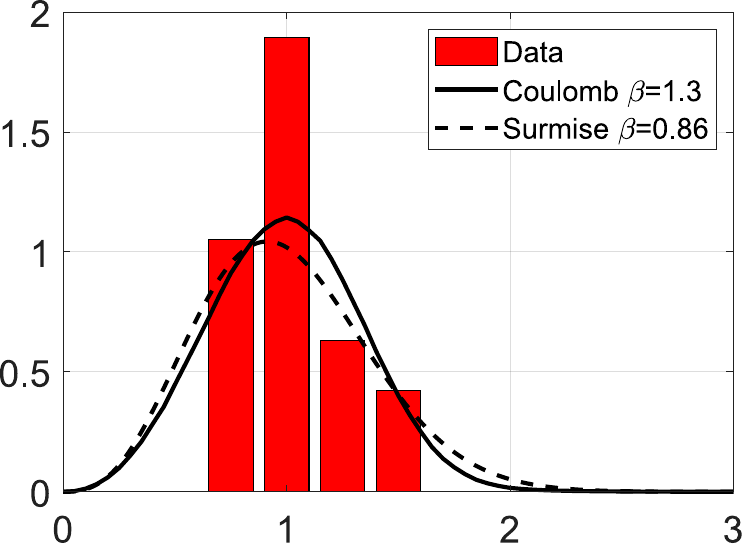}
		\\		\includegraphics[width=0.3\linewidth,angle=0]{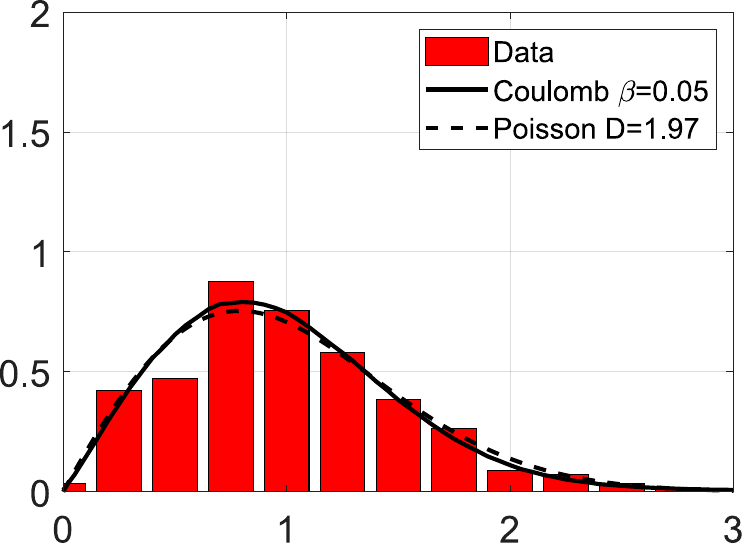}
		\hspace{10pt}
		\includegraphics[width=0.3\linewidth,angle=0]{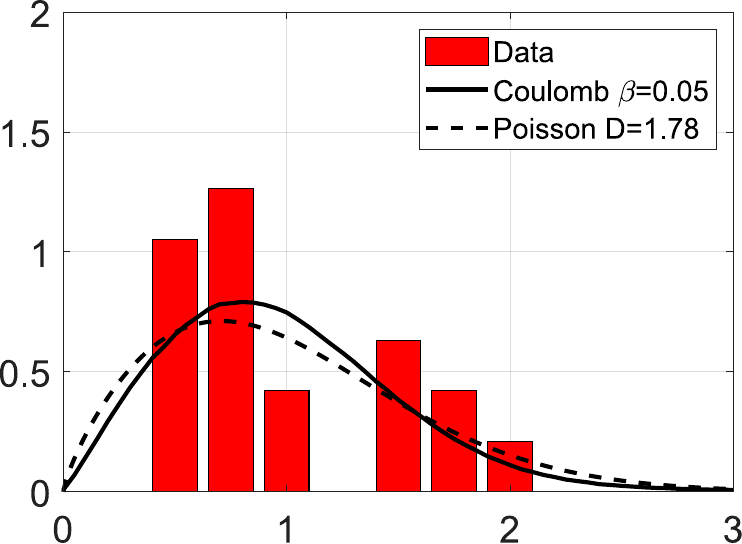}
		\caption[Figure 12 from \cite{ACKMOP}, \textit{Probing effects of displacement on fitted $\beta$ for Common Buzzards and Goshawks}]{
			\textbf{Left row:} Fitted $\beta$ values for NN for randomly displaced Buzzard nests by $Xm_B$: $\beta=1.1\ (X=0)$, $\beta=0.4\ (X=0.5)$, $\beta=0.3\ (X=1.0)$, $\beta=0.05\ (X=1.5)$ from top to bottom. \textbf{Right row:} Fitted $\beta$ values for NN for randomly displaced Goshawk nests by $Xm_G$: 
			$\beta=3\ (X=0)$, $\beta=1.4\ (X=0.5)$, $\beta=1.3\ (X=1.0)$, $\beta=0.05\ (X=1.5)$, from top to bottom. \cite[Figure 12]{ACKMOP}.
		}
		\label{fig-Displace}
	\end{figure}
\end{center}

In this appendix we test the sensitivity of our assumption to fit the NN spacings of nests with $\beta$ from a 2d Coulomb distribution to a random displacement of the positions of the nests. 
Because this is an issue in particular for a small number of spacings available, we perform this test for two examples of spacings of a single year 2020 in Figure \ref{fig-Displace} for Buzzards (left) and for Goshawks (right). Each year represents an ensemble for each species. We have picked these examples as first, they contain a moderate (228) respectively small number (19) of spacings, see Figure \ref{fig-GroupingYears} top and middle plot of the left column for the undisplaced data. These are reproduced in the top row in Figure \ref{fig-Displace}. Second, the two chosen examples give a fitted value of $\beta=1.1$ for the Buzzards and $\beta=3$ for the Goshawks, that is significantly away from the Poisson value of $\beta=0$.
If we were to perform this test on data that are already close to that it would be difficult to see an effect.

The procedure we apply is as follows. First, we determine the mean spacing between the nests of both species separately in 2020, cf. Figure \ref{fig-PlotBirds}. We obtain $m_B=0.63$ km for Buzzards (dark blue points) and $m_G=2.62$ km for Goshawks (yellow points), before unfolding. 
Next, we pick a random direction for each nest of the ensemble of a given species, say Buzzard. The nest is then displaced by $Xm_B$ in that direction, where we choose $X=0.5,1.0,1.5$ for that ensemble in three steps. For each chosen value of $X$ we do a fit of $\beta$ to the displaced nests.

The result for the Buzzards in shown Figure \ref{fig-Displace} left column with $X$ increasing from top to bottom: the top row corresponds to $X=0$, second row to $X=0.5$, third row to $X=1.0$ and bottom row to $X=1.5$. 
For comparison we give the resulting $\beta$ values (insets) and densities for the Coulomb gas (full line) and for the surmise (dashed line).
The same random displacements and fits for $\beta$ are then made for the ensemble of Goshawks nests, shifted by $Xm_G$ in the random direction and choosing the same values of $X=0.5,1.0,1.5$, see Figure \ref{fig-Displace} right column. As in all previous plots we fit to the cumulative distribution rather than to the density, which would depend on the choice of histogram width.

For the Buzzards, clearly the fitted $\beta$ value rapidly decreases towards Poisson at $\beta=0$. It makes sense that Poisson is only reached once the average mean spacing is exceeded by $X>1$, as only then on average points can become very close.
For the Goshawks, the statistics of 19 spacings in 2020 is (deliberately) poor. Nevertheless, also here the fitted value for $\beta$ consistently moves down towards Poisson, even if the quality of each fit is very low. \\
\noindent
We conclude from this numerical experiment of an increasing random displacement, that our initial data do contain valid information about the repulsion by fitting $\beta$, even down to low statistics with the order of 20 spacings as observed here. For an order of magnitude higher, with about 200 spacings, also the quality of the fits from $\beta$ of order unity down to Poisson becomes acceptable. Poisson is reached when a random displacement larger than the mean spacing is allowed.
\clearpage
\subsection{Number of Nests per Species per Year}\label{apps-number of nests}
\begin{table}[t]
	{ 
		\begin{tabular}{c||c|c|c|c|c}
			Year & Buzzards & Goshawks & Eagle Owls & Eagle Owls (F) & Eagle Owls (P)\\
			\hline
			\hline
			2000 & 63 & 12 & 1 & 1 & 0\\
			\hline
			2001 & 85 & 19 & 1 & 1 & 0\\
			\hline
			2002 & 78 & 15 & 2 & 2 & 0\\
			\hline
			2003 & 101 & 17 & 3 & 3 & 0\\
			\hline
			2004 & 99 & 17 & 6 & 6 & 0\\
			\hline
			2005 & 121 & 17 & 7 & 7 & 0\\
			\hline
			2006 & 106 & 18 & 9 & 9 & 0\\
			\hline
			2007 & 107 & 19 & 9 & 9 & 0\\
			\hline
			2008 & 130 & 21 & 7 & 7 & 0\\
			\hline
			2009 & 105 & 20 & 5 & 5& 0\\
			\hline
			2010 & 195 & 22 & 8 & 8 & 0\\
			\hline
			2011 & 161 & 27 & 10 & 8 & 2\\
			\hline
			2012 & 242 & 29 & 12 & 11 & 1\\
			\hline
			2013 & 147 & 22 & 16 & 10 & 6\\
			\hline
			2014 & 220 & 23 & 10 & 10 & 0\\
			\hline
			2015 & 240 & 22 & 15 & 12 & 3\\
			\hline
			2016 & 217 & 22 & 15 & 12 & 3\\
			\hline
			2017 & 267 & 21 & 19 & 13 & 6\\
			\hline
			2018 & 214 & 22 & 16 & 10 & 6\\
			\hline
			2019 & 251 & 19 & 20 & 14 & 6\\
			\hline
			2020 & 228 & 19 & 23 & 16 & 7\\
			\hline
			\hline
			Sum & 3377 & 423 & 214 & 174 & 40\\
		\end{tabular}
		\vspace{10pt}
		\caption[Table 2 from \cite{ACKMOP}, \textit{Collection of numbers of birds of prey from 2000 to 2020}]{The number of nests per each species per year, including the split of the Eagle owls into (F) and (P) in the last two columns. The sum of all nests per species over all years is given in the last row. \cite[Table 2]{ACKMOP}}
		\label{tab-numbersbirds}}
\end{table}
In Table \ref{tab-numbersbirds} we collect the number of observed nests per species per year that are depicted in Figure \ref{fig-AllBirds} bottom left for the Buzzards, in Figure \ref{fig-AllBirds} bottom right for the Goshawks, and in Figure \ref{fig-EagleOwlGroup} bottom left respectively right for the Eagle Owls in the forest (F) respectively plain (P). We also give the total number of Eagle Owls per year in the middle column.\cleardoublepage
\section{Results for the Tridiagonal $\boldsymbol{\beta}$-Ensembles for $\mathbf{N=2}$} \label{app- tridiag N=2}
\begin{figure}[h]
	\begin{subfigure}{0.49\textwidth}
		\includegraphics[width=\textwidth]{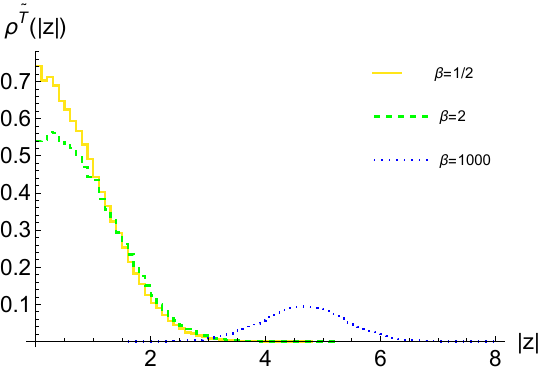}
		\subcaption{$\widetilde{T}_\beta$ \label{fig-radial dist T tilde n=2}}	
	\end{subfigure}
	\begin{subfigure}{0.49\textwidth}
	\includegraphics[width=\textwidth]{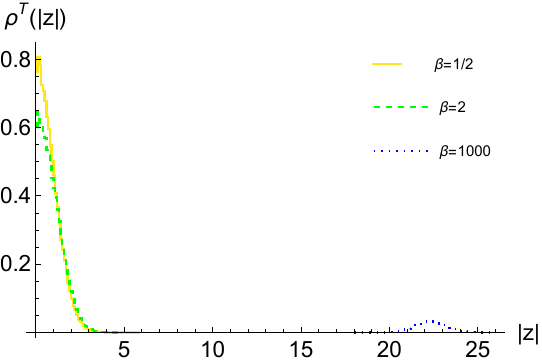}
	\subcaption{$T_\beta$\label{fig-radial dist T n=2}}	
	\end{subfigure}
	\caption[Cf. Figure 6 from \cite{AMPT}, \textit{Radial density of complex eigenvalues for the tridiagonal ensembles $\widetilde{T}_\beta$ and $T_\beta$ with $N=2$}]{Histograms of the radial density of complex eigenvalues for $50000$ matrices of size $N=2$ of the ensemble $\widetilde{T}_\beta$ (left) and $T_\beta$ (right) at $\beta=1/2$ (yellow full), $\beta=2$ (green dashed) and $\beta=1000$ (blue dotted). \cite[Cf. Figure 6]{AMPT} \label{fig-radial dist T and T tilde beta 1/2,2,1000 n=2}}
\end{figure}\noindent
As discussed in Section \ref{ssec-surmise}, it is well known that the $N=2$ statistics are for non-Hermitian matrices not a good approximation of the large $N$-behaviour. This is also what we observe for the three tridiagonal non-Hermitian matrix ensembles, introduced in Section \ref{ssec-tridiag beta}. We present the $N=2$ results here. For the general ensemble $T_\beta$, it is clear from Figure \ref{fig-radial dist T n=2} that we do not have $\beta$-universality here, i.e. we see a distinct behaviour of the radial distribution for various $\beta$ at $N=2$. For the non-symmetric tridiagonal ensemble $\widetilde{T}_\beta$ we see in Figure \ref{fig-radial dist T tilde n=2} that for small $\beta=1/2,2$ we have a relatively large support of the eigenvalues where a large amount concentrates near the origin. For $\beta=1000$ we have one large peak which is not placed at the origin and the eigenvalues are supported in a small neighbourhood of this radius. As $N=2$ is the simplest case of these non-Hermitian ensembles where we can study the JPDF and average spectral density, we will show the results and comment on the differences compared to the previously derived large $N$-results. 
\begin{lemma}\label{lemma-n=2 jpdf}
	In the low temperature limit $\beta \gg 1$, the JPDF of the eigenvalues of the $2\times 2$ matrix $J=\widetilde{T},S,T$, is given by
	\begin{align}
		P_\beta^{\widetilde{T}}(\mathbf{z})&\approx C_\beta^{\widetilde{T}}|z_1-z_2|^\beta e^{\frac{|z_1-z_2|^2}{4}-\frac{1}{32}|z_1-z_2|^4-\frac{|z_1|^2}{2}-\frac{|z_2|^2}{2}}\text{ , with }C_\beta^{\widetilde{T}}=\frac{2^{1-\beta}\pi}{\beta Z_\beta^{\widetilde{T}}}\label{eq-n=2 jpdf tilde T},\\
		P_\beta^S(\mathbf{z})&\approx C_\beta^S |z_1-z_2|^\beta e^{-\frac{|z_1|^2}{2}-\frac{|z_2|^2}{2}} \text{ , with }C_\beta^S=\frac{2^{1-\beta}\pi}{\beta Z_\beta^S},\label{eq-n=2 jpdf S}\\
			P_\beta^T(\mathbf{z})&\approx C_\beta ^T |z_1-z_2|^\beta e^{\frac{|z_1-z_2|^2}{4}-\frac{|z_1|^2}{2}-\frac{|z_2|^2}{2}}K_0\left(\frac{|z_1-z_2|^2}{4}\right)\text{ , with }
		C_\beta^T=\frac{2^{4-\beta}\pi^2}{\beta Z_\beta^T},\label{eq-n=2 jpdf T}
	\end{align}
	where $K_0(x)$ denotes the modified Bessel function.
\end{lemma}
\begin{Remark}
	We observe in the symmetric ensemble with $N=2$ that the function $g_S(\mathbf{z},\mathbf{r})$ contributes only as a normalization factor. This can be understood because a $2\times2$ matrix which is complex symmetric can be made normal by fixing only one phase. As the Frobenius norm is invariant under this change one can assume that a $2\times 2$ complex symmetric matrix is also normal, i.e. it can be diagonalized via a unitary transformation.\\
	Let 
	\begin{align}
		S=\begin{pmatrix}
			a_1 & c_1\\
			c_1 & a_2
		\end{pmatrix}
	\end{align}
	be a complex symmetric matrix with $a_1, a_2, c_1\in \mathbb{C}$. The normality condition $SS^\dagger =S^\dagger S$ yields one equation:
	\begin{align}
		a_1c_1^*+c_1a_2^*=a_1^*c_1+c_1^*a_2.
	\end{align}
	We choose polar coordinates $a_1=r_1e^{i\Theta_1}$, $a_2=r_2e^{i\Theta_2}$ and $c_1=se^{i\varphi}$ with $r_1,r_2,s\in [0,\infty)$, $\Theta_1,\Theta_2, \varphi\in [0,2\pi)$. After some basic algebraic computations, one finds the condition
	\begin{align}
		e^{2i\varphi}=\frac{r_2e^{i\Theta_2}-r_1e^{i\Theta_2}}{r_2e^{-i\Theta_2}-r_1e^{-i\Theta_1}},
	\end{align}
	i.e. the normality condition for a $2\times 2$ complex symmetric matrix can be expressed by fixing one phase.
\end{Remark}
\begin{proof}
	We apply the method of steepest descent for large $\beta \gg 1$ to calculate the function $f_J(Z)$ for $J=\widetilde{T},S,T$. As the $\beta$-dependent part in the expression of $f_J(Z)$ is the same for all three ensembles under consideration, the saddle point analysis can be performed before specifying the ensemble. We denote the $\beta$-dependent part of $f_J(Z)$ as
	\begin{align}
		f_J^\beta(Z,\mathbf{R}_1)=\int_{\mathbb{C}}\! d^2r_1\, \exp(-g_J(Z,\mathbf{r},\mathbf{R}_1))|r_1|^{\frac{\beta}{2}-2}|1-r_1|^{\frac{\beta}{2}-2},
	\end{align}
	where the $\mathbf{R}_1$-dependence drops out for the ensembles $S$ and $\widetilde{T}$. Note that the function $g_J(Z,\mathbf{r},\mathbf{R}_1)$ explicitly depends on the choice of the ensemble via its definition in Eq. \eqref{eq-def g J} and the expressions of the matrix elements in terms of the spectral parameters. We use the following formula for a two-dimensional integration via steepest descent
	\begin{align} \label{eq-method of steepest descent}
		\int\!d\mathbf{x}\  h(\mathbf{x})e^{Mp(\mathbf{x})}\approx \frac{2\pi}{M}\frac{h(\mathbf{x_0})e^{Mp(\mathbf{x_0})}}{\det(-H_p(\mathbf{x_0}))^{1/2}} \text{ as }M\rightarrow \infty,
	\end{align}
	where $H_p(\mathbf{x_0})$ is the Hessian matrix of $p$ evaluated at $\mathbf{x_0}$, which is a global maximum/ minimum of $p(\mathbf{x})$. We denote the real- and imaginary-parts of $r_1$ as $(x,y)$ and rewrite $f_J^\beta(Z,\mathbf{R}_1)$ as follows
	\begin{align}
		f_J^\beta(Z,\mathbf{R}_1)=\int_{-\infty}^\infty \! dx\int_{-\infty}^\infty\! dy \frac{\exp(-g_J(Z,\mathbf{r},\mathbf{R}_1))}{(x^2+y^2)((1-x)^2+y^2)}e^{\frac{\beta}{4}\log\left[(x^2+y^2)((1-x)^2+y^2)\right]}.
	\end{align}
	Hence, we are looking for the maximum/ minimum of the function 
	\begin{align*}
		p(x,y)=\frac{1}{4}\log\left[(x^2+y^2)((1-x)^2+y^2)\right].
	\end{align*}
	We find that only $(x,y)=(1/2,0)$ yields $\frac{\partial p}{\partial x}=\frac{\partial p}{\partial y}=0$. The determinant of the Hessian at $\mathbf{x_0}$ is negative. Hence, we have a saddle point. Moreover, we observe that $p(x,y)\rightarrow -\infty$ for $(x,y)=(0,0)$ and $(x,y)=(1,0)$. Hence, the contribution of $p(x,y)$ vanishes at these points. For $x,y\rightarrow \infty$ we have a logarithmic divergence of $p(x,y)$. It follows that
	\begin{align}
		f_J^\beta(Z,\mathbf{R}_1)\approx \frac{2^{4-\beta}\pi}{\beta}\exp(-g_J(Z,1/2,\mathbf{R}_1)),
	\end{align}
	for $\beta \gg1$. Now, we have to specify the function $g_J(Z,\mathbf{R}_1)$ for each matrix ensemble $J=\widetilde{T},S,T$, where the $\mathbf{R}_1$ dependence only appears for the general ensemble $T$, which makes another integration over the complex variable $R_2$ necessary.
	\begin{enumerate}
		\item[I.] For $\mathbf{J=\widetilde{T}}$, we use the recurrence relations detailed in the proof of Theorem \ref{thm-roughbijection}, Eqs. \eqref{eq-first recurrence step T tilde} and find at $r_1=1/2$
		\begin{align}
			g_{\widetilde{T}}(Z,1/2)=-\frac{|z_1-z_2|^2}{4}+\frac{|z_1-z_2|^4}{32},
		\end{align}
		which yields the joint probability density in Eq. \eqref{eq-n=2 jpdf tilde T}.
		\item[II.] Similarly, we find for the symmetric matrix $\mathbf{J=S}$, via the Eqs. \eqref{eq-first recurrence step S} at $r_1=1/2$ that $g_S(Z,\mathbf{r})=1$. Hence, the factor $f_S(Z,\mathbf{r})$ gives only a constant contribution yielding the result in Eq. \eqref{eq-n=2 jpdf S}.
		\item[III.] For $\mathbf{J=T}$, we conclude via the recurrence relations \eqref{eq-first recurrence step T} at $r_1=1/2$
		\begin{align}
			g_T(Z,1/2,R_2)=-\frac{|z_1-z_2|^2}{4}+\frac{|z_1-z_2|^2}{2}\left(|R_2|^2+\frac{1}{16|R_2|^2}\right),
		\end{align}
	with $\mathbf{R}_1^t=(1/2,R_2)^t$.
		We obtain the function $f_T(Z)$ (cf. Eq. \eqref{eq-factor f T}) as follows
		\begin{align}
			f_T(Z)&=\int_{\mathbb{C}\backslash \{0\}} \! d^2R_2 \frac{f_T^\beta(Z,\mathbf{R}_1)}{|R_2|^2}\nonumber\\
			&\approx \frac{2^{5-\beta}\pi^2}{\beta}e^{\frac{|z_1-z_2|^2}{4}}\int_{0}^\infty \! d |R_2|\  |R_2|^{-1}e^{-\frac{|z_1-z_2|^2}{2}(|R_2|^2+\frac{1}{16}|R_2|^{-2})}\nonumber \\
			&= \frac{2^{5-\beta}\pi^2}{\beta}e^{\frac{|z_1-z_2|^2}{4}}K_0\left(\frac{|z_1-z_2|^2}{4}\right),
		\end{align}
		where we used the integral representation \cite[Eq. 8.432(7)]{gradshteyn} of the modified Bessel function $K_0(x)$. The JPDF in Eq. \eqref{eq-n=2 jpdf T} follows now from Eq. \eqref{eq-jpdf eig tridiag}.
	\end{enumerate}	
\end{proof}
Using the results of Lemma \ref{lemma-n=2 jpdf}, we compute the average spectral density, which is defined as
\begin{align}
	\rho^J(Z)=\int_\mathbb{C} \! dz_2\ P_J(z,z_2).
\end{align}
\begin{prop}
	In the low temperature limit $\beta \gg 1$, the spectral density of the $2\times 2$ matrix $J=\widetilde{T},S,T$, is given by
	\begin{align}
		\rho^{\widetilde{T}}(z)&=\frac{e^{-|z|^2+\frac{1}{4}}\sum_{k=0}^\infty\frac{\Gamma\left(\frac{\beta}{2}+k+1\right)}{\Gamma(k+1)^2}|z|^{2k}D_{-\frac{\beta}{2}-k-1}(1)}{\pi \Gamma\left(\frac{\beta}{2}+1\right)D_{-\frac{\beta}{2}-1}(0)}\label{eq-n=2 density tilde T}\\
		\rho^S(z)&=\frac{1}{\pi 2^{\frac{\beta}{2}+1}}{}_1F_1\left(\frac{\beta}{2}+1,1,\frac{|z|^2}{2}\right)e^{-|z|^2}\label{eq-n=2 density S}\\
		\rho^T(z)&=\frac{1}{\pi{}_2F_1\left(1+\frac{\beta}{2},1+\frac{\beta}{2}; \frac{3}{2}+\frac{\beta}{2};\frac{1}{2}\right)}e^{-|z|^2}{}_2F_2\left(1+\frac{\beta}{2},1+\frac{\beta}{2}; 1,\frac{3}{2}+\frac{\beta}{2};\frac{|z|^2}{2}\right)\label{eq-n=2 density T}
	\end{align}
	for $z\in \mathbb{C}$. $D_\nu(x)$ denotes the parabolic cylinder function.
\end{prop}
\begin{Remark}
	We observe that spectral densities of the general ensemble $T$ and the symmetric one $S$ agree as $\beta/2+1\approx \beta/2+3/2$ in the low temperature limit $\beta \gg 1$, which reduces the hypergeometric function ${}_2F_2$ in Eq. \eqref{eq-n=2 density T} to the same ${}_1F_1$ as in Eq. \eqref{eq-n=2 density S}.
\end{Remark}
\begin{proof}
	To perform the complex integral, we choose the new coordinate $w=z_2-z$ and take polar coordinates for $z$ and $w$, i.e. $z=r_1e^{i\theta_1}$ and $w=r_2e^{i\theta_2}$ with $r_1,r_2\in [0,\infty)$ and $\theta_1,\theta_2 \in [0,2\pi)$. It follows that the angular-depending part of the integral is of the same form for all three ensembles $J=\widetilde{T},S,T$, i.e. we have to perform the following integral
	\begin{align}
		\int_0^{2\pi}\! d\Theta_2 \, e^{-r_1r_2\cos(\Theta_1-\Theta_2)}.
	\end{align}
	Therefore, we apply the addition theorem for the cosine and use \cite[Eq. 3.338(4)]{gradshteyn}. The result is given in terms of the modified Bessel function of the first kind $I_0(x)$.
	\begin{enumerate}
		\item[I.] For $\mathbf{J=\widetilde{T}}$, we start from the radial integration of the form 
		\begin{align}\label{eq-int form n=2 non sym}
			\rho_\beta^{\widetilde{T}}(z)&=2\pi K_\beta^{\widetilde{T}} \int_0^\infty\! dr_2\  r_2^{\beta+1}e^{-r_1^2-\frac{r_2^2}{4}-\frac{r_2^4}{32}}I_0(r_1r_2).
		\end{align}
		We again apply the summation formula Eq. \eqref{eq-sum I_0}, change to the variable $w=r_2^2$ and apply \cite[Eq. 3.462(1)]{gradshteyn}. We find
		\begin{align}
			\rho_\beta^{\widetilde{T}}(z)	=\pi K_\beta^{\widetilde{T}} e^{-r_1^2}\sum_{k=0}^\infty \frac{r_1^{2k}}{4^k\Gamma(k+1)^2}2^{\beta +2k+2}\Gamma\left(\frac{\beta}{2}+k+1\right)e^{\frac{1}{4}}D_{-\frac{\beta}{2}-k-1}(1).\label{eq-unnormalised spectral density T tilde model n=2}
		\end{align}
		In order to calculate the normalization, we start from the integral representation in \eqref{eq-int form n=2 non sym} and apply integration over $z\in \mathbb{C}$. Using \cite[Eq. 6.614(3)]{gradshteyn} and \cite[Eqs. 9.220(2), 9.215]{gradshteyn} to evaluate the resulting Whittaker function, we find
		\begin{align}
			\int_\mathbb{C}\! d^2 z\  \rho^{\widetilde{T}}(z)&=\pi^2 K_\beta^{\widetilde{T}}\int_0^\infty\! dr_2\  r_2^{\beta+1}e^{-\frac{r_2^4}{32}}\\
			&=2^{2+2\beta}\pi^2K_\beta^{\widetilde{T}}\Gamma(1+\frac{\beta}{2})D_{-(1+\frac{\beta}{2})}(0),
		\end{align}
		where we applied \cite[Eq. 3.462(1)]{gradshteyn} and $D_\mu(x)$ denotes the parabolic cylinder function.
		\item[II.] As the function $g_S(Z)$ is constant for $\mathbf{J=S}$, the calculation agrees with the one of the $2\times 2$ matrix model describes in Section \ref{ssec-surmise}, cf. \cite{AMP}.
		\item[III.] For $\mathbf{J=T}$, we are left with the following radial integral
		\begin{align}
			\rho^T(z)&=2\pi C_\beta^T \int_0^\infty\! dr_2\  r_2^{\beta+1}e^{-r_1^2-\frac{r_2^2}{4}}I_0(r_1r_2)K_0\left(\frac{r_2^2}{4}\right).
		\end{align}
		We use the summation formula \cite[Eq. 8.445]{gradshteyn}
		\begin{align}\label{eq-sum I_0}
			I_0(x)=\sum_{k=0}^\infty \frac{x^{2k}}{4^k\Gamma(k+1)^2}.
		\end{align}
		Next, we swap the summation and the integration. Therefore, we apply Tonelli's theorem (we use the counting measure over $\mathbb{N}$ to identify the summation as an integration), as the function is positive for all $r_2\in [0,\infty)$ and $k\in \mathbb{N}$.
		\begin{align}
			\Rightarrow \rho^T(z)&=2\pi C_\beta^T e^{-r_1^2}\sum_{k=0}^\infty \frac{r_1^{2k}}{4^k\Gamma(k+1)^2}\int_0^\infty\! dr_2\  r_2^{\beta+2k+1}e^{-\frac{r_2^2}{4}}K_0\left(\frac{r_2^2}{4}\right)
		\end{align}
		We change variables to $t=r_2^2/4$ and find
		\begin{align}
			\Rightarrow \rho_\beta^T(z)&=2\pi C_\beta^T e^{-r_1^2}\sum_{k=0}^\infty \frac{r_1^{2k}2^{\beta+2k+1}}{4^k\Gamma(k+1)^2}\int_0^\infty\! dt\  t^{\frac{\beta}{2}+k}e^{-t}K_0(t)\\
			&=2\pi C_\beta^T e^{-r_1^2}\sum_{k=0}^\infty \frac{r_1^{2k}2^{\beta+2k-2k+1}}{\Gamma(k+1)^2}\sqrt{\pi}2^{-\frac{\beta}{2}-k-1}\frac{\Gamma(\frac{\beta}{2}+k+1)^2}{\Gamma(\frac{\beta}{2}+k+\frac{3}{2})}\\
			&=2^{\frac{\beta}{2}+1}\pi^{\frac{3}{2}}\frac{\Gamma(\frac{\beta}{2}+1)^2}{\Gamma(\frac{\beta}{2}+\frac{3}{2})} C_\beta^T e^{-r_1^2}{}_2F_2\left(1+\frac{\beta}{2},1+\frac{\beta}{2}; 1,\frac{3}{2}+\frac{\beta}{2};\frac{r_1^2}{2}\right),
		\end{align}
		where we used \cite[Eq. 6.621(3)]{gradshteyn} and ${}_2F_1(a,b;c;0)=1$ in the first step.	We are now left with the calculation of the normalization such that
		\begin{align}
			1=\int_\mathbb{C}\! dz \ \rho^T (z).
		\end{align}
		Using a change of variables $t=r_1^2$ and \cite[Eq. 7.522(5)]{gradshteyn}, we find
		\begin{align*}
			\int_0^\infty\! dr_1\  r_1 e^{-r_1^2}{}_2F_2\left(1+\frac{\beta}{2},1+\frac{\beta}{2}; 1,\frac{3}{2}+\frac{\beta}{2};\frac{r_1^2}{2}\right)
			=\frac{1}{2}{}_2F_1\left(1+\frac{\beta}{2},1+\frac{\beta}{2}; \frac{3}{2}+\frac{\beta}{2};\frac{1}{2}\right).
		\end{align*}
		This proves Eq. \eqref{eq-n=2 density T}.
	\end{enumerate}
\end{proof}\clearpage
\section{Differential Equation for Class AII$\boldsymbol{^\dagger}$ and the Edge Scaling Limit}\label{sec-app expec poly}
In this appendix, we derive a differential equation for the integral representation for the expected characteristic polynomials in class AII$^\dag$ at finite $N$, see Section \ref{sssec-results 1 char poly},
\begin{align}
	D_N^{\textrm{AII}^\dag}(z,w^*)\Big |_{x=zw^*}=\frac{1}{4^N}f_N(x),\qquad 
	f_N(x)=\frac{1}{2}\int_0^\infty\! dv\ e^{-v}\int_{-v}^v \! du\,
	\left(4x^2+4xv+u^2\right)^{N},
\end{align}
in the form \eqref{eq-fNx-int1} which we need here.
We will also show how this could be employed to derive the edge scaling limit in this class.  

From the above integral we obtain for $x\neq0$ 
\begin{align}
	\frac{\partial}{\partial x}f_N(x)&= \frac{1}{2}\int_0^\infty\! dv\ e^{-v}\int_{-v}^v \! du
	N(8x +4v)\left(4x^2+4xv+u^2\right)^{N-1}
	\nonumber\\
	&= \frac{1}{2}\int_0^\infty \! dv\ e^{-v}\int_{-v}^v \! du
	N\left(4x +\frac{4x^2+4vx+u^2}{x} - \frac{u^2}{x}\right)\left(4x^2+4xv+u^2\right)^{N-1}
	\nonumber\\
	&= \frac{1}{2}\int_0^\infty \! dv\ e^{-v}\int_{-v}^v \! du\left( \frac{\partial}{\partial v}+\frac{N}{x}-\frac{u}{2x}\frac{\partial}{\partial u}\right) \left(4x^2+4xv+u^2\right)^{N}.
	\label{eq-partialfN}
\end{align}
While the middle term immediately gives back $\frac{N}{x}f_N(x)$, the two terms with derivatives with respect to $v$ and $u$ can be simplified using integration by parts. 

For the third term differentiated with respect to $u$ we obtain
\begin{align}
	&\frac{1}{2}\int_0^\infty \! dv\ e^{-v}\left\{\left[-\frac{u}{2x}\left(4x^2+4xv+u^2\right)^{N}\right]_{-v}^v +\int_{-v}^v \! du\frac{1}{2x}\left(4x^2+4xv+u^2\right)^{N}\right\}
	\nonumber\\
	&=-\frac{1}{2x}\int_0^\infty \! dv\ e^{-v}v(2x+v)^{2N}+ \frac{1}{2x}f_N(x)
	\nonumber\\
	&=-\frac{1}{2x}e^{2x}\int_0^\infty\! dv\ e^{-v-2x}\left((2x+v)^{2N+1}-2x(2x+v)^N\right)+ \frac{1}{2x}f_N(x)\nonumber\\
	&=-\frac{1}{2x}e^{2x}\left(\Gamma(2N+2,2x)-2x\Gamma(2N+1,2x)\right)+ \frac{1}{2x}f_N(x).
	\label{eq-partf1}
\end{align}
For the first term in \eqref{eq-partialfN} we consider the following total derivative:
{\small
\begin{align}
	&\frac{d}{dv}\int_{-v}^v \! du\ e^{-v}(4x^2+4xv+u^2)^N
	\\
	&= 2e^{-v}(4x^2+4xv+v^2)^N-\int_{-v}^v \! du\ e^{-v}(4x^2+4xv+u^2)^N+\int_{-v}^v \! du\ e^{-v}\frac{\partial}{\partial v}(4x^2+4xv+u^2)^N.\nonumber
\end{align}}
Integrating this expression over $v$ from zero to infinity, the boundary terms on the LHS obviously vanish, and we arrive at
{\small
\begin{align}
	0&=\int_0^\infty \! dv\ e^{-v}(2x+v)^{2N} - f_N(x)+ \frac{1}{2}\int_0^\infty\!  dv\int_{-v}^v \! du\ e^{-v}\frac{\partial}{\partial v}(4x^2+4xv+u^2)^N\nonumber\\
	&= e^{2x}\Gamma(2N+1,2x)- f_N(x)+ \frac{1}{2}\int_0^\infty \! dv\int_{-v}^v \! du\ e^{-v}\frac{\partial}{\partial v}(4x^2+4xv+u^2)^N.
	\label{eq-partf2}
\end{align}}
Inserting this expression together with \eqref{eq-partf1} into \eqref{eq-partialfN} we obtain as a final result
\begin{align}
	\frac{\partial}{\partial x}f_N(x)= \left(1+\frac{N}{x}+\frac{1}{2x}\right) f_N(x)-\frac{1}{2x}e^{2x}\Gamma(2N+2,2x) .
\end{align}
We note that two terms with the incomplete Gamma function $\Gamma(2N+1,2x)$ cancel.

This can be translated into an equation for the rescaled quantity
\begin{align}
	F_N(x)= \frac{1}{(2N)!}\ e^{-2x} f_N(x) = e^{-2zw^*}\frac{D_N^{\textrm{AII}^\dag}(z,w^*)}{D_N^{\textrm{AII}^\dag}(0,0)}\biggl|_{x=zw^*},
	\label{eq-FNdef}
\end{align}
see \eqref{eq-F-def}, in order to take the edge (and bulk) large $N$-limit. It 
gives the following differential equation 
\begin{align}
	\frac{\partial}{\partial x}F_N(x)= \left(\frac{2N+1}{2x}-1\right) F_N(x)-\frac{1}{2x}\frac{\Gamma(2N+2,2x)}{\Gamma(2N+1)}.
	\label{eq-FdiffN}
\end{align}

Let us briefly indicate how \eqref{eq-FdiffN} can be used to alternatively derive the edge scaling limit \eqref{eq-zoom1}. Employing the scaling
\begin{align}
	x=N+\sqrt{N}y+\chi\eta^*/2,
\end{align}
while keeping $y=(z_0^*\xi+z_0\eta^*)/\sqrt{2}$ fixed, the partial derivatives are related like
\begin{align}
	\frac{\partial}{\partial x}=\frac{1}{\sqrt{N}}\frac{\partial}{\partial y}.
\end{align}
This implies for \eqref{eq-FdiffN} that
\begin{align}
	\frac{1}{\sqrt{N}}\frac{\partial}{\partial y}F_N(x)= \left[-\frac{y}{\sqrt{N}}+O(N^{-1})\right] F_N(x)
	-\left[1+O(N^{-1/2})\right]\frac{\Gamma(2N+2,2x)}{\Gamma(2N+2)}.
	\label{eq-diffFN}
\end{align}
Due to \eqref{eq-def-Q} and its asymptotic \eqref{eq-Q-lim} we have 
\begin{align}
	\frac{\Gamma(2N+2,2x)}{\Gamma(2N+2)}=Q(2N+2,2x)=\frac12\erfc(y)+O(N^{-1/2}). 
\end{align}
From this we deduce that the limit $F(y)=\lim_{N\to\infty}F_N(x)/\sqrt{2N}$ satisfies the differential equation
\begin{align}
	\frac{\partial}{\partial y}F(y)= -yF(y)-\frac{1}{\sqrt{8}}\erfc(y).
	\label{eq-diffF}
\end{align}
The solution of the homogeneous solution $F_h(y)$ is given by 
\begin{align}
	F_h(y)=Ce^{-y^2/2}.
	\label{eq-Fhom}
\end{align}
In order to find a special solution $F_s(y)$ to the inhomogeneous equation \eqref{eq-diffF}, we vary the integration constant, setting $F_s(y)=g(y/\sqrt{2})e^{-y^2/2}$ which reduces \eqref{eq-diffF} to
\begin{align}
	g'(v)=-\frac12\erfc(\sqrt{2}\,v)\, e^{+v^2}\qquad\textrm{with}\quad v=y/\sqrt{2}.
\end{align}
It is identical to \eqref{eq-gdiff} and further motivates why we have introduced this function. From here onwards we can proceed as in Section \ref{sssec-asymptot 1 char poly} to arrive at \eqref{eq-Dedge-QS}. The bulk scaling limit can be derived from \eqref{eq-diffFN} in a similar fashion.
\clearpage

	\cleardoublepage
	\listoffigures\clearpage
	
	\listoftables \vfill \footnotesize{Gedruckt auf alterungsbeständigem Papier °°ISO 9706}\clearpage
	
\end{appendix}
\end{document}